 \documentclass[trackchanges,modern]{aastex63}
\usepackage[utf8]{inputenc}
\usepackage{threeparttable}
\usepackage{amsmath}
\usepackage{amssymb}
\usepackage{xspace}
\submitjournal{ApJ}
\usepackage{multirow}
\usepackage[all]{nowidow}

\usepackage{soul}
\usepackage{import}
\usepackage{hyperref}
\usepackage{calculator}
\usepackage{chngcntr}
\usepackage{multirow}
\usepackage{array}
\usepackage{tablefootnote}
\usepackage[nomessages]{fp}

\usepackage{printlen}
\usepackage{xcolor}
\usepackage{xfp}
\usepackage{pgfplots}
\usepackage{gensymb}

\makeatletter
\newcommand\footnoteref[1]{\protected@xdef\@thefnmark{\ref{#1}}\@footnotemark}
\newcommand{\as}{$^{\prime\prime}$}
\newcommand{\bapp}{$\beta_{app}$}
\newcommand{\siml}{$\sim$}

\newcommand{\newref}[2]{\hyperref[#1]{\ref*{#1}(#2)}}

\edef\writeimagedesc#1,#2,#3,#4.#5{The VLA #1 band #2 config contours are overlaid on the X-ray image binned to a size of #3 and smoothed with a Gaussian kernel of FWHM=#4.#5\as. }

\newcommand{\chdr}{\textit{Chandra}\xspace}
\usepackage[maxfloats=256]{morefloats}
\makeatletter
\def\RemoveSpaces#1{\zap@space#1 \@empty}

\newcommand{\ntitl}[1]{\textbf{#1}~(Figure \ref{fig:results-\RemoveSpaces{#1}}):~}
\newcommand{\z}[1]{($z$=#1)}
\newcommand*{\MyNum}[1]{\pgfmathprintnumber[precision=2]{#1}}
\newcommand{\ascale}[2]{#1$^{''}$~(\MyNum{\fpeval{round(#1*#2,2)}}~kpc) }
\newcommand{\bappb}[1]{$\beta_{app}$=#1}
\newcommand{\lsep}{($< 1^{''}$)~}
\defcitealias{2021ApJS..253...37R}{I}
\newcommand{\comment}[1]{}
\newcommand{\pval}{$p$-value}
\makeatother
\newcommand{\papri}{Paper \citetalias{2021ApJS..253...37R}\xspace}
\usepackage{nicefrac}

\received{...}
\revised{...}
\accepted{...}

\begin{document}
\title{Offsets between X-ray and Radio Components in X-ray Jets: The AtlasX}

\author[0000-0001-9018-9553]{Karthik Reddy}
\affiliation{Department of Physics, University of Maryland, Baltimore County, 1000 Hilltop Circle, Baltimore, MD, 21250, USA}
\author[0000-0002-2040-8666]{Markos Georganopoulos}
\affiliation{Department of Physics, University of Maryland, Baltimore County, 1000 Hilltop Circle, Baltimore, MD, 21250, USA}
\affil{NASA Goddard Space Flight Center, Code 663, Greenbelt, MD 20771, USA}
\author[0000-0002-7676-9962 ]{Eileen T. Meyer}
\affiliation{Department of Physics, University of Maryland, Baltimore County, 1000 Hilltop Circle, Baltimore, MD, 21250, USA}
\author{Mary Keenan}
\affil{ ADNET Systems, Inc., Bethesda, MD 20817 USA}
\author[0000-0003-4004-2451]{Kassidy E. Kollmann}
\affil{Department of Physics, Princeton University, Princeton, NJ 08544, USA}
\begin{abstract}
The X-ray emission mechanism of powerful extragalactic jets, which has important implications for their environmental impact, is poorly understood. The X-ray/radio positional offsets in individual features of jets provide important clues. Extending the previous work in \citealt{2021ApJS..253...37R}, we present a detailed comparison between X-ray maps deconvolved using the Low Count Image Reconstruction and Analysis (LIRA) tool and radio maps of 164 components from 77 Chandra-detected X-ray jets.
We detect 94 offsets (57\%), with 58 new detections. In FR II-type jet knots, the X-rays peak and decay before the radio in about half the cases, disagreeing with the predictions of one-zone models. While a similar number of knots lack statistically significant offsets, we argue that projection and distance effects result in offsets below the detection level. Similar de-projected offsets imply that X-rays could be more compact than radio for most knots, and we qualitatively reproduce this finding with a `moving-knot' model. The bulk Lorentz factor ($\Gamma$) derived for knots under this model is consistent with previous radio-based estimates, suggesting kpc-scale jets are only mildly relativistic. An analysis of X-ray/radio flux ratio distributions does not support the commonly invoked mechanism of X-ray production from inverse Compton scattering of the cosmic microwave background but does show a marginally significant trend of declining flux ratio as a function of distance from the core. 
Our results imply the need for multi-zone models to explain the X-ray emission from powerful jets. We provide an interactive list of our X-ray jet sample at \url{http://astro.umbc.edu/Atlas-X}. 
\end{abstract}
\section{Introduction}

It is now generally agreed that practically all galaxies host a supermassive black hole (SMBH, $M\sim 10^6-10^9\; M_\odot $) at their centers \citep[e.g.,][]{richstone1998supermassive}.
A small fraction of SMBHs accrete matter and produce intense radiation that covers up to 20 orders of magnitude in frequency. These Active Galactic Nuclei (AGN) sometimes produce collimated double-sided relativistic jets \citep[e.g.][]{Padovani_2017}, which are detected in large numbers in the radio. These jets transport energy from sub-parsec scales of the AGN out to kpc and frequently to  Mpc scales, well beyond the host galaxy, and play a crucial role in driving the AGN feedback on the host galaxy and cluster \citep[e.g.,][]{fabian2012}, which may affect how structure in the Universe evolves \citep[e.g.,][]{blandford2019relativistic}. However, the total power carried by jets is still not well known.

Despite hundreds of known resolved (arcsecond-scale/kpc-scale) radio jets, there were very few resolved jets detected in the optical and X-rays prior to the advent of the \emph{Hubble} and \emph{Chandra} observatories \citep[e.g.,][]{schreier1979einstein,butcher1980optical,biretta1991radio,harris1994x}.
Since the launch of \textit{Chandra} in 1999, over 150 jets have been detected in the X-rays \citep[e.g.,][]{sambruna2004survey,marshall2005chandra,massaro2015chandra} showing in many cases X-ray knot structure roughly coincident with the radio. These detections of jet knots and terminal `hot spots' stimulated optical observations with the Hubble Space Telescope (HST) resulting in over 3 dozen optical jet detections \citep[e.g.,][]{2004ApJ...608..698S,tavecchio2007chandra,2012ApJ...748...81K,}

For most low-power jets, with the typical `plumey' and edge-darkened \citet{Fanaroff1974TheLuminosity} class I (FR I) morphology, the level and spectrum of the knot X-ray emission seems to be a smooth continuation of the radio to optical synchrotron spectrum (e.g., \citealt{Marshall_2002,worrall2003x,Hardcastle-2003}, but see \cite{Meyer_2018} for a counter-example).  In contrast, the powerful, highly-collimated and edge-brightened FR II-type jets exhibit puzzling X-ray characteristics, seen first in the powerful source PKS 0637-752. The latter was the first source
\textit{Chandra} detected \citep{chartas2000chandra,schwartz2000chandra}. Its bright and spectrally hard knot X-ray emission is not the high energy tail of knot radio-to-optical synchrotron spectrum, but rather part of a second, high energy spectral component. Synchrotron self-Compton (SSC) and inverse Compton emission by up scattering cosmic microwave background (IC/CMB) photons in equipartition conditions and  assuming non-relativistic flow speeds were found to severely under-produce the observed X-rays \citep{chartas2000chandra,schwartz2000chandra}.


\citet{celotti2001large} and \citet{tavecchio2000x} independently noted that at pc scales (probed with very long baseline radio interferometry or VLBI) the jet of PKS 0637-752 exhibits superluminal motion with apparent speed (in units of $c$) of $\beta_{app}=13.8$ \citep{2000aprs.conf..215L}, requiring a minimum bulk Lorentz factor $\Gamma_{min}=\beta_{app}$. By assuming that the bulk Lorentz
factor $\Gamma$ is the same at pc and kpc scales and that the jet is oriented close to our line of sight ($\theta\sim 1/\Gamma $, where $\theta$ is the angle between the jet and the line of sight), they showed that
the X-rays could be explained as IC/CMB due to the much higher Doppler boosting\footnote{$\delta=\frac{1}{\Gamma(1-\beta \cos\theta)}$, where $\beta$ is the speed of the emitting plasma in units of the speed of light, and $\theta$ is the angle of the jet to the line of sight} ($\delta$) than previously adopted for this mechanism. 

The IC/CMB model was quickly adopted as the standard explanation for the high X-ray emission from many FR-II type jets  \citep[e.g.,][]{sambruna2004survey,marshall2005chandra,jorstad2006x,tavecchio2007chandra,miller2006x,2012ApJ...748...81K}. 
Despite the popularity of the IC/CMB model, it failed to explain some features of multi-wavelength observations: ({\sl i}) For a given X-ray spectrum this model predicts a unique gamma ray flux \citep{georganopoulos2006quasar} that in most cases is not detected  \citep[e.g.,][but see \citet{Meyer_2019} for
two notable exceptions]{meyer2013fermi,meyer2015ruling,breiding2017fermi}
({\sl ii}) High polarization (on the order of 30\%) is observed in the optical spectrum of the jet knots of PKS 1136-135 \citep{cara2013polarimetry}; in this source the optical samples the low-energy tail of the high-energy (X-ray) emission component. A high level of polarization is in disagreement with the expected low polarization of IC/CMB emission \citep{2007ApJ...661..719U}.
({\sl iii})
If we assume that the emission comes
from a moving feature, we expect to see proper motions on the order of $\beta_{app}\sim 10$. Contrary to this expectation, an HST proper motion study of the jet of 3C 273 found $\beta_{app}\lesssim 1$
\citep{meyer2016hst}.
({\sl iv}) 
If we assume that the emission comes
from a stationary feature, the cooling time of the X-ray emitting electrons is \siml$10^6$ years \citep{2007RMxAC..27..188H}, much larger than that of the radio-emitting electrons \citep{worrall2009x}. One would thus expect X-ray jets to extend past the radio, while the opposite is seen to happen in many jets \citep[e.g.,][]{sambruna2004survey,jester2006new}.
({\sl v}) X-ray variability has been detected in a knot in the jet of Pictor A on the order of a few years \citep{marshall2010flare}, and possibly even a few weeks \citep{2016MNRAS.455.3526H}. Such short time scales are incompatible with the enormously longer  cooling time of the X-ray emitting electrons under IC/CMB.

Some workers \citep[e.g.,][]{jester2006new,hardcastle2006testing,Clautice:2016zai}, motivated by the problems in the IC/CMB model, favored synchrotron emission from an ad-hoc additional population of energetic electrons ($\sim$ 30-100 TeV). 
Testing this interpretation requires high-resolution X-ray polarimetry, which is possible with the recently launched Imaging X-ray Polarimetry Explorer \citep[IXPE~][]{weisskopf2016imaging}. 
In general, both double-synchrotron and IC/CMB models can fit the observed radio to X-ray spectral energy distributions (SEDs), as \cite{cara2013polarimetry} demonstrated for PKS 1136-135.
It is essential to understand the nature of the knot X-ray emission, as this has a major consequence on the implied total power in the jet. While the IC/CMB model requires Eddington or super-Eddington jet powers, the synchrotron model is sub-Eddington by at least an order of magnitude \citep{dermer2004nonthermal,atoyan2004synchrotron}.

Almost all the discussion in the literature has been implicitly framed in the context of one-zone models, which requires the emission in different frequency bands to be co-spatial. This assumption has been seen not to hold in some cases,
where displacements and/or size differences between radio and X-ray knots have been noted
\citep[e.g.,][]{sambruna2004survey,jester2006new}. X-rays peaking and decaying before the radio were previously reported in some knots of FR-II sources  \citep[e.g.,][]{kataoka2008chandra,Clautice:2016zai,Harris-2017}. Similar offsets between the peaks of radio and X-rays are also seen in FR-I sources like M87 \citep{Marshall_2002}, though this is generally attributed to synchrotron cooling effects \citep[e.g.,][]{hardcastle2001chandra}. 

The majority of the known radio/X-ray offsets were only noted individually in their respective detection papers, and for the majority of X-ray jets the low number of X-ray counts attributed to the knots and the resulting statistics limited attempts to detect offsets. We addressed this problem using a novel approach based on a statistical tool called Low count Image Reconstruction and Analysis \citep[LIRA, ][]{esch2004image,stein2015detecting} to detect offsets from  low-count jets in \citet[][hereafter Paper \citetalias{2021ApJS..253...37R}]{2021ApJS..253...37R}. In that study we detected offsets in roughly half of the analyzed 22 jets; however the overall frequency of occurrence in the full X-ray jet sample is unknown, and it is not yet clear whether these offsets are exceptions or the norm. In this work, we extend our analysis to the data of all the remaining X-ray jets available in the \textit{Chandra} archive, and compare them to high-resolution radio images, to look for radio/X-ray offsets, where possible. We aim to find whether these offsets represent a norm, and examine any statistical correlations that could illuminate the physical nature of the knots and hotspots. 

The rest of this document is organized as follows. Section \ref{sec:methods} describes the data sample, reduction and analysis methods. Section~\ref{sec:results} describes the results we discuss them in section \ref{sec:discussion}. Appendix \ref{sec:src-notes} provides brief descriptions for offsets in each source. 

The  adopted cosmology follows the NASA Extragalactic Database (NED): $H_0=67.8$ km/s/Mpc, $\Omega_m=0.308$ and $\Omega_\Lambda=0.692$. Spectral index $\alpha$ is given by the flux density, $F_\nu=\nu^{-\alpha}$ and the photon index by $\Gamma=\alpha+1$.


\section{Methods\label{sec:methods}}
\subsection{X-ray Jet Sample\label{subsec:xjet-sample}}
The initial sample of X-ray jets was taken from a broad literature and archive search, as previously described in \citet{2021ApJS..253...37R}. These include the over 100 sources listed in the XJET\footnote{\url{https://hea-www.harvard.edu/XJET/}} database, the 3CR snapshot surveys \citep[e.g.,][]{massaro2010chandra,massaro2015chandra,Massaro_2018,Stuardi_2018} and other serendipitous discoveries, totaling 189 sources. In addition, for this paper, we added four detections from \citet{snios2021discovery}, three from \citet{2018ApJ...856...66M}, and one each from \citet{2020MNRAS.497..988W}, \citet{connor2021enhanced} and \citet{2022AnA...659A..93I}, bringing the total count to 199 sources. These are listed in Table \ref{table:list_of_sources} in the Appendix, and is to the best our of knowledge, the complete list of all published X-ray jets detected by \chdr at the time of writing this paper.  The table lists the common name of the source in column 1, IAU name in column 2, J2000 Right Ascension and Declination in column 3, redshift in column 4, angular scale (kpc/\arcsec) in column 5, class of the source (e.g., FR-I, FR-II, quasar) in column 6, and a reference to the previous X-ray observation publication in column 7.  In what follows, unless otherwise stated, we group sources classified as FR II or `quasars' (i.e., lobe-dominated quasars or LDQ and core-dominated quasars or CDQ) collectively as FR II-type and all FR I together with BL Lacertae (BL Lac) objects as FR I-type.

In \papri, we examined 69 of these jets identified as `low counts' jets, where at least one feature was detected with less than 20 counts, in order to apply a specialized low-counts localization analysis in these cases. In that work we were not able to measure offsets for 47/69 due to various shortcomings of the observations or source characteristics (such as lacking a radio core or having no compact X-ray emission). In this paper we re-consider one source from the previous paper with a high-count hotspot (3C~275.1).
Of the remaining 130 jets left for this paper, we exclude 52 jets from the analysis for a variety of reasons. (In the Appendix we further describe these sources and the impact of their removal on our conclusions.) In particular, in 17 of these there is no point-like or spatially correlated X-ray/radio feature in the jet, and in 2 sources the short exposure leads to an under-exposed `core' which prevents accurate X-ray/radio alignment. In two cases it is clear that the X-ray emission originates from a thermal medium surrounding the jet rather than from the jet itself -- these are 3C 171 \citep{10.1111/j.1365-2966.2009.15855.x} and 3C 305 \citep{2012MNRAS.424.1774H}. We also exclude 13 jets at high redshift ($z>2.2$) because in these cases it is difficult to ascertain whether the X-ray origin is from the jet, a hotspot, or an extended lobe due to the reduced scale and greater foreshortening of the jet. Following \papri, we only considered jet components without neighboring components in a radius of two native ACIS-S pixels (0.492\arcsec) to construct regions large enough to measure their centroids. We excluded ten sources with no component satisfying this criterion. Finally, we did not re-analyze ten sources with known (published) offsets similar to or larger than a single ACIS-S pixel, where instrumental effects are unlikely to produce an offset (see \ref{subsec:measuring-offsets} for details), or sources where they are already accounted for while measuring offsets. However, these sources are included in our general analysis. In total we newly analyzed 164 components from 77 jets in this paper, joining the 37 features in 22 jets from paper I. With the literature cases, this comes to 226 components from 107 X-ray jets.

\subsection{Radio Data\label{subsec:radio-data}}
We retrieved radio data mainly from archives of the Very Large Array (VLA) and, in a few cases, from the Australian Telescope Compact Array (ATCA). As the intent is to detect offsets using \chdr X-ray images at a resolution of $\approx$0.25\arcsec, we also selected the radio data to have a similar resolution ($\approx$0.2-0.4\arcsec). Table \ref{table:radio-observations-mrt} provides the details of these observations with the name of the source in column 1, project name in column 2, date of observation in column 3, frequency in column 4, beam size in column 5, and the RMS of the final image in column 6.  
\startlongtable
\begin{deluxetable*}{lllccl}
    \tablecaption{Details of Radio Observations  \label{table:radio-observations-mrt}}
    \tabletypesize{\footnotesize}
    \tablehead{
    \colhead{Name} &
    \colhead{Program}&
    \colhead{Date}&
    \multicolumn{1}{p{1.5cm}}{\centering Frequency\\GHz} &
    \multicolumn{1}{p{2cm}}{\centering Beam \\\as$\times$\as, PA\textdegree}&
    \multicolumn{1}{p{1cm}}{\centering RMS\\$10^{-5}$\nicefrac{Jy}{beam}}
    }
    \tablecolumns{6}
    \startdata
    3C 9 &     15A-357 &  2019 Jun 15 &           14.99 & 0.15$\times$0.12, -55.58 & \phantom{11}0.47 \\
        3C15 &      AB0534 &  1990 May 25 & \phantom{1}8.40 &  0.35$\times$0.26, 51.76 & \phantom{11}2.21 \\
        3C 31 &      AL0405 &  1996 Nov 13 & \phantom{1}8.46 & 0.24$\times$0.22, 167.00 & \phantom{11}1.04 \\
    4C +01.02 &      AR0197 &  1989 Jan 08 & \phantom{1}4.80 &  0.47$\times$0.39, 18.26 & \phantom{1}11.80 \\
        3C 33 &      AR0148 &  1986 Jul 13 & \phantom{1}4.86 & 1.54$\times$1.49, -32.53 &           124.00 \\
        3C 47 &      AB0796 &  1996 Nov 07 & \phantom{1}4.84 &  0.40$\times$0.38, 61.81 & \phantom{11}2.19 \\
PKS 0144-522 &   ATCA(C890) &  2004 May 08 &           17.73 &   0.40$\times$0.40, 0.00 & \phantom{1}12.10 \\
    4C +35.03 &      AM0221 &  1987 Aug 15 & \phantom{1}4.86 & 0.39$\times$0.36, -20.48 & \phantom{11}5.26 \\
PKS 0208-512 &  ATCA(C890) &  2002 Feb 01 & \phantom{1}8.64 & 1.33$\times$1.21, -26.06 & \phantom{1}27.40 \\
    3C 66B &      AA0128 &  1991 Aug 19 & \phantom{1}8.21 &   0.24$\times$0.20, 9.80 & \phantom{11}2.66 \\
\enddata
\tablecomments{Table 1 is published in its entirety in the machine-readable format.
      A portion is shown here for guidance regarding its form and content.}
\end{deluxetable*}

The VLA observations were calibrated and imaged using the Common Astronomy Software Applications \citep[CASA,][]{mcmullin2007astronomical} toolbox. For VLA data, 3C 286, 3C 48, or 3C 137 were used as flux calibrators,  while the sources were bright enough themselves in several cases to be used as phase calibrators. The initial bandpass, and amplitude and phase calibrations were performed using \texttt{bandpass} and \texttt{gaincal} tasks in CASA, respectively. The calibrated data are mostly imaged using the \texttt{clean} task in CASA with the ``Briggs'' weighting scheme and where we set \texttt{robust=0.5}. In a few cases the newer \texttt{tclean} task was used with equivalent parameters. Additionally, for JVLA data, we set \texttt{nterms=2} to capture the spectral curvature over the wideband. For ATCA data, the initial calibration was performed using the MIRIAD \citep{1995ASPC...77..433S} toolbox following the standard procedures described in the ATCA cookbook. 1934-638 served as the flux calibrator for all the observations. The RFI  was flagged using \texttt{pgflag} and \texttt{blflag} tasks, and amplitude and phase were initially calibrated using \texttt{mfcal} and \texttt{gpcal} tasks in MIRIAD. The calibrated data were imaged using the Difmap program \citep{1997ASPC..125...77S} following the same procedure used for the VLA data. Natural weighting was applied in nearly all cases to produce the images. Finally, we measured the radio flux from each component by fitting an elliptical Gaussian model using the 2D fitting tool available in CASA. If the fit did not converge, we estimated the flux by summing the pixel values within a chosen region of interest (ROI) around the component.
\subsection{Chandra X-ray Data\label{subsec:chandra-xray-data}}
We obtained X-ray data from the \chdr~archives and re-processed them using the standard methods described in \chdr~data analysis threads. We used calibration information from CALDB 4.9.4 and the Chandra Interactive Analysis of Observations \citep[CIAO, ][]{2006SPIE.6270E..1VF} v4.13 suite for reducing the data. Table \ref{table:xray-observations-mrt} summarizes details of the observations used in our analysis with the source name in column 1, \chdr~ObsID in column 2, the effective exposure time in column 3, observation mode in column 4, and the count rate in column 5.
    \begin{deluxetable*}{l>{\raggedright}p{3cm}clc}
      \tablecaption{Details of \textit{Chandra} X-ray Observations \label{table:xray-observations-mrt}}
      \tabletypesize{\footnotesize}
      \tablewidth{0pt}
      \tablehead{
        \colhead{Name} &
        \multicolumn{1}{p{3cm}}{\centering Chandra\\ObsIDs} &
        \multicolumn{1}{p{2cm}}{\centering Effective\\exposure\textsuperscript{a} (ks)} &
        \colhead{Modes\textsuperscript{b}}&
        \colhead{\nicefrac{counts}{s}\textsuperscript{c}}
      }
      \tablecolumns{5}
      \startdata
      3C 9 &                          1595, 17088, 18700-1 & \phantom{1}89.00 & F, V & 0.021 \\
      3C15 &                                              17128 &           116.34 &    V & 0.045 \\
     3C 31 &                                               2147 & \phantom{1}44.00 &    F & 0.027 \\
 4C +01.02 &                                 9281, 10380, 10799 & \phantom{1}70.30 &    V & 0.163 \\
     3C 33 &                                         6910, 7200 & \phantom{1}36.40 &    F & 0.068 \\
     3C 47 &                                               2129 & \phantom{1}34.70 &    F & 0.222 \\
PKS 0144-522 &                                              10366 & \phantom{11}5.41 &    V & 0.091 \\
 4C +35.03 &                                                856 & \phantom{11}8.20 &    V & 0.020\\
\enddata
             \tablenotetext{a}{After background flare removal, if any.}
             \tablenotetext{b}{The observation mode for each obsID where F=FAINT, V=VFAINT. A single value is provided if all the current and subsequent obsIDs have the same mode.  }
             \tablenotetext{c}{Average of the count rate from each epoch.}
             \tablecomments{Table 1 is published in its entirety in the machine-readable format.
      A portion is shown here for guidance regarding its form and content.}
            \end{deluxetable*}

We inspected the time-binned light curve for each observation and any periods with count rates above the 2$\sigma$-level were excluded. Only the events falling between 0.5-7.0 keV are considered, while those falling in regions on the detector with less than 2\% of the total exposure time are excluded. For sources with multiple observations, the centroid of the core in each observation, measured using \texttt{dmstat}, is aligned to the observation with the maximum of exposures. The aligned observations are merged using \texttt{merge\_obs} to produce the final merged observation. Following \papri, we bin all the events files on half-pixel boundaries (bin factor: 0.5) of ACIS-S pixels and set the image size to either 64x64 or 128x128 to comply with LIRA input size requirements, further explained below.

\subsection{Measuring X-ray/Radio Offsets with LIRA\label{subsec:measuring-offsets}}  
Sub-pixel level offsets ($\lesssim$0.3\arcsec) in high-count jets are generally detected by comparing longitudinal radio brightness profiles against their X-ray counterparts, constructed using sub-pixel images \citep[e.g.,][]{2007ApJ...670...74S,2007ApJ...657..145S,2011ApJ...739...65P,2012ApJ...748...81K,Harris-2017} or by measuring X-ray/radio centroid differences \citep[e.g.,][]{2004ApJ...608..698S,2005MNRAS.360..926W,2011ApJ...729...26M}, or visually in simple cases where offsets are similar to or larger than \textit{Chandra's} resolution \citep[e.g.,][]{kataoka2008chandra,2007ApJ...669..893H,erlund2007luminous,Perlman_2009}. The brightness profile-based method, which is usually applied to the entire jet at once, ignores the possibility that curvature in the jet that may reduce any observable offset below a detectable level at different points along the jet. Although centroid-difference can alleviate this problem by measuring the offset individually for each component independent of the jet morphology,  \textit{Chandra's} point spread function (PSF) and background fluctuations introduce additional uncertainties in offsets, which are unaccounted by this method. 
Moreover, emission from the PSF of the bright core can add excess emission to nearby components and move their centroid closer to the core, thereby creating false offsets.
 The same effects can additionally induce false detections of X-ray components themselves in the case of low-count jets, and LIRA was used to examine this possibility for the jets analyzed in \papri. We extend the approach used in \papri to high-count jets in this work to estimate the uncertainties in offsets and thereby the significance of their detection.
 
 We refer the reader to \papri for full details on the method and \citet{esch2004image} and \citet{stein2015detecting} for a detailed statistical treatment of LIRA. Briefly, LIRA models the observed image as a sum of two model images. The baseline model comprises all user-specified features assumed to be present in the observation. For our case, we use a point source (representing the bright core) overlaid with a flat background for the baseline model. In addition to this there is the ``added'' model, containing any emission required in excess of the baseline to explain the observation. LIRA applies a Bayesian methodology to infer the posterior distribution of the added model by fitting a superposition of the baseline and the added models to the observation with a user-supplied PSF. Put another way, LIRA reconstructs the observation by convolving the sum of the baseline and the added models with the PSF. It samples the output posterior distribution using Markov Chain Monte Carlo (MCMC) method to produce a series of images of the added model; they represent the brightness distribution of the observation after accounting for effects of the core, background, and the PSF. We compute an offset from each MCMC image using the X-ray/radio centroid difference method to derive its posterior distribution and thereby its associated uncertainty. In this work, we discard the first 1000 iterations as \textit{burn-in}, and generate 2000 MCMC images of the added model. These 2000 images are averaged to produce a core and background subtracted deconvolved image.

\subsection{Preparing PSFs and baseline images}
We follow a similar procedure used in \papri{}~to generate PSFs and baseline models for each source. We used the published values of spectral parameters for the core where available. For the rest, we extracted their spectra using \texttt{specextract} command in CIAO and initially fit them using a power-law model in SHERPA \citep{2001SPIE.4477...76F}, with two absorption components, one each for the host galaxy and the Milky Way. If the hydrogen absorption column density (nH) of the host galaxy was poorly constrained in the fit, we only used the Milky Way nH\footnote{Retrieved from WebPIMMS service (\url{https://heasarc.gsfc.nasa.gov/cgi-bin/Tools/w3nh/w3nh.pl})} for the absorption component and re-fit the spectra. We used the final spectral model with \texttt{trace-nest} command from SAOTrace\footnote{\url{https://cxc.cfa.harvard.edu/cal/Hrma/Raytrace/SAOTrace.html}} v2.0.5 to simulate 100 ray traces through \textit{Chandra}'s optics. We projected these ray traces onto the ACIS-S detector using \texttt{simulate\_psf} command from CIAO, which used MARX v5.4.0 \citep[][]{2012SPIE.8443E..1AD} as the backend, and generated a merged event file without applying any sub-pixel event repositioning.

Since the default value of 0.07\arcsec for the \texttt{blur} parameter in \texttt{simulate\_psf} is known to produce a PSF narrower\footnote{See \url{https://cxc.cfa.harvard.edu/ciao/why/aspectblur.html} for details} than the observation, we generated multiple  PSFs by varying \texttt{blur} between 0.07\arcsec and 0.35\arcsec in steps of 0.07\arcsec (see also discussion in \papri). We compared each simulated PSF's enclosed counts fraction (ECF) with that of the observation and selected the PSF with the least squared error. We binned the final PSF events file to match the pixel boundaries of the observed image. For sources with multiple observations, we summed the PSF of all individual observations to obtain the final PSF. We loaded this PSF in SHERPA and fit a 2D Gaussian model (\texttt{gauss2d}) with a flat background (\texttt{const2d}) to the observation to generate the baseline model.  In addition, as we discuss in the following section, we also generated a baseline model with only a flat background to estimate uncertainties in the location of the X-ray core. 

\subsection{X-ray/Radio Image Registration\label{subsec:xray-radio-alignment}}
The radio core locations were measured by fitting with a 2D elliptical-Gaussian model in CASA and were localized to within 0.1 mas for all sources. We use the \texttt{dmstat} command in CIAO to estimate the centroid of the X-ray core using a circular region. The centroid was computed multiple times by varying the radius and location of the region to minimize any bias from manual selection of regions. In addition, we measured the core's centroid on MCMC images generated using LIRA with a flat-background-only baseline model (i.e., without subtracting the core). The major axis of the standard deviational ellipse \citep[e.g.,][]{10.2307/490885}, which encloses 98\% of the LIRA derived centroids was as always less than 0.03\arcsec. Furthermore, for all the sources, the offset between \texttt{dmstat}-measured and LIRA-measured centroids was always less than 0.05\arcsec.  To estimate the uncertainty in the alignment, we conservatively take the sum of the maximum uncertainty in X-ray (\texttt{dmstat}+LIRA) and radio core locations in quadrature yielding a value of $\sim$0.06\arcsec. We note the intrinsic dispersion in the X-ray core's location was absent in the slightly smaller uncertainty estimate of 0.05\arcsec used in \papri. However, its conclusions remain unaltered even with the slightly larger estimate as all offsets claimed in \papri{}~are larger than 0.15\arcsec.
\subsection{Measuring X-ray Flux and Spectral Index\label{subsec:measuring-flux-spec-index}}
Similar to the case of cores, we use the published values of spectral indices and fluxes where available. These values are listed in table \ref{table:offset-results} along with their respective publications. For the remaining sources, 
we extracted the emission spectrum using the \texttt{specextract} command in CIAO. We fit each spectrum with an absorbed power-law model over the 0.5-7.0 keV energy band and fixed nH  to Milky Way value. The counts were grouped to a minimum of one count per bin and WStat statistic was used with Monte-Carlo (\texttt{moncar}) optimizer. Once the fit converged, we used the normalization value to estimate the spectral flux density at 1 keV. For sources with poor fits (with no spectral indices listed in table \ref{table:offset-results}), we measured the flux using exposure-corrected LIRA deconvolved images. The aspect histogram and instrument map required to create the exposure map were generated using \texttt{asphist} and \texttt{mkinstmap} commands in CIAO, respectively. The instrument map was weighted using an absorbed power-law model with $\Gamma=2$~and appropriate galactic nH value. We summed the individual exposure maps in the case of sources with multiple observations. We divided the averaged LIRA image by the final exposure map to create a flux image with the units of counts sec$^{-1}$ pixel$^{-1}$. The pixel values inside each chosen region were summed to estimate the count flux and were converted to energy flux density at 1 keV using the same spectral model adopted for the instrument map.
\section{Results\label{sec:results}}
We show in Figure \ref{fig:results-3C9} an example figure for 3C 9 showing the \chdr observations (left panel) and LIRA-deconvolved (average of 2000 MCMC images) image (right panel), with radio contours overlaid in green. The dashed-gray circles indicate the regions used to measure the centroids for each knot. In each LIRA deconvolved image, we normalize any residuals in the core region to improve the contrast among features in the jet. The mean centroid for each region is marked with a red \texttt{X}. Similar figures for the remaining 76 sources we have analyzed are given in the Appendix.

\begin{figure*}[ht]
    \gridline{
        \fig{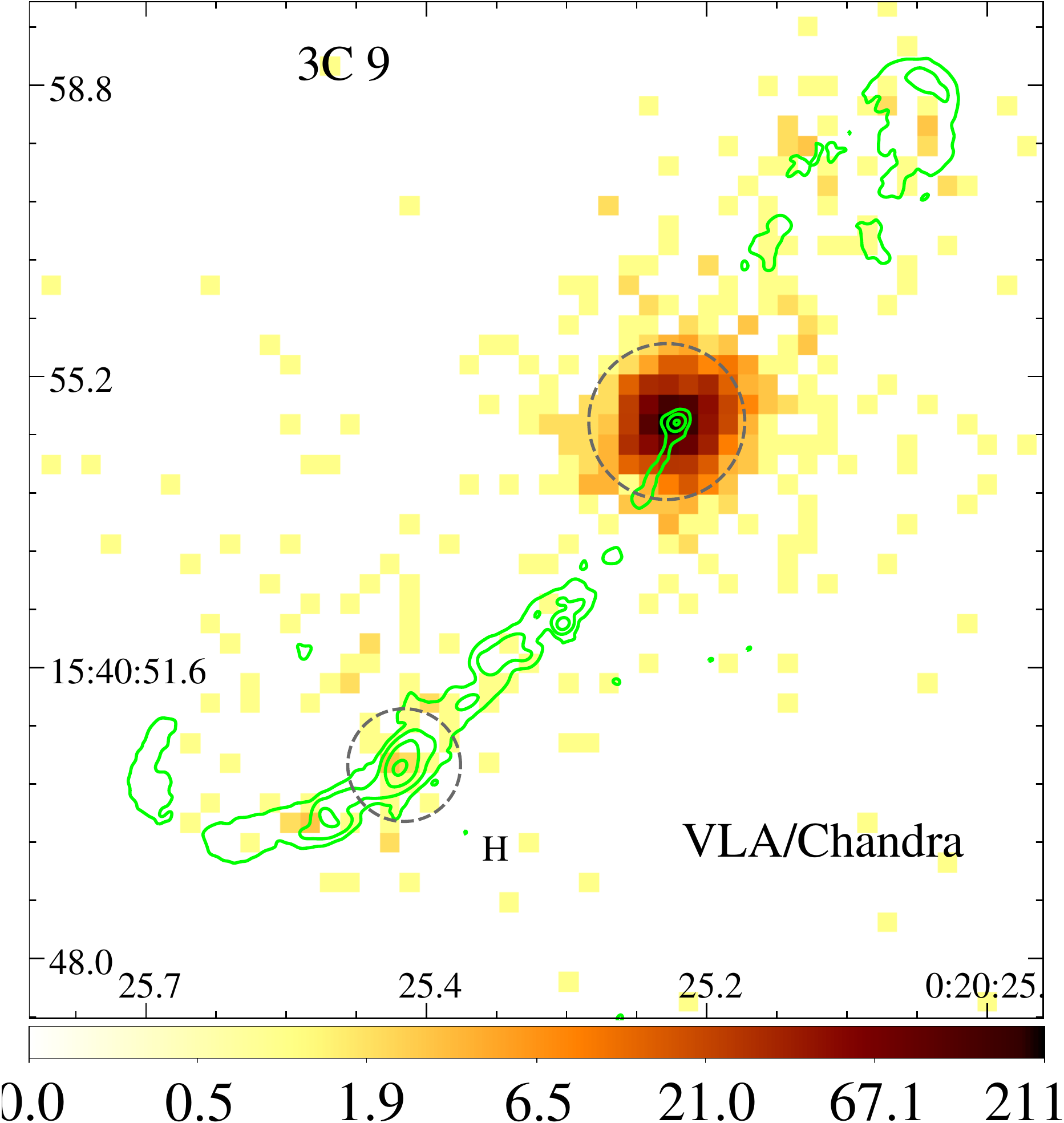}{0.5\textwidth}{(a)}
        \fig{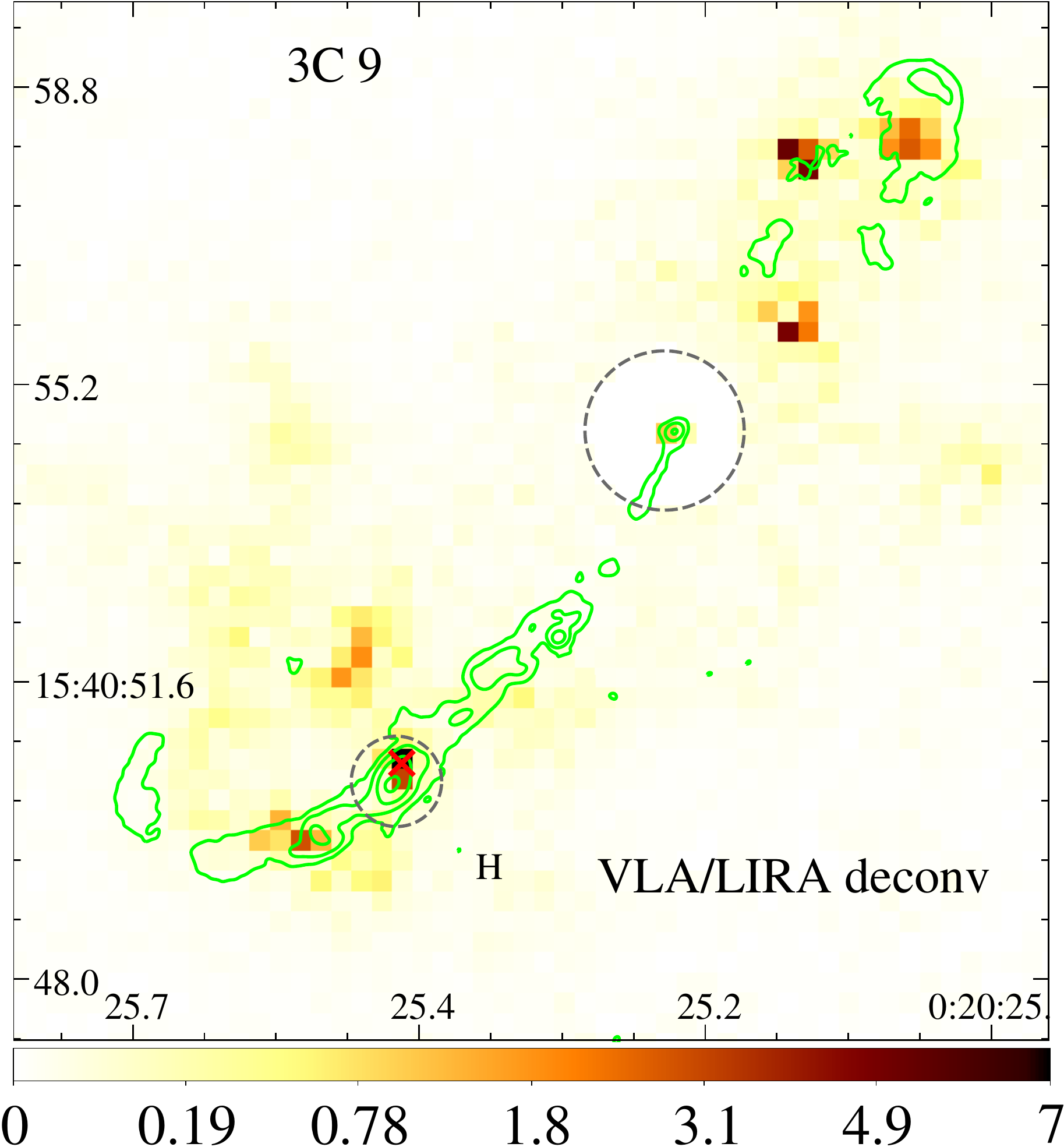}{0.5\textwidth}{(b)}
    }
    \caption{Results for the offset analysis of 3C 9. With the radio contours overlaid on both panels, (a) shows the \chdr X-ray image and (b) shows the LIRA deconvolved image. The dashed-gray circles indicate the regions used to measure the centroid for each feature and their respective mean centroids are marked with a red \texttt{X}. The radio contours are given by 0.03, 0.4, 2.0, 6.0, 8.0 mJy beam$^{-1}$. The complete figure set (75 images) is available in the online journal.\label{fig:results-3C9}}
\end{figure*}



 %

 In our analysis, an offset was considered to be significant only when the mean of the distribution of offsets measured using individual MCMC images from LIRA was at least 1$\sigma$~level above 0.15\as, which is a threshold  $\sim$2~times the estimated X-ray/radio astrometric registration error  (taken as $\approx$0.07\as). If multiple unresolved radio peaks existed within a hotspot region with only a single source of X-ray emission, the offset is measured as the distance between the X-ray centroid and the brightest radio peak. With this criterion, we find 94 (from 61 jets) offsets in our sample of 164 features in the 77 jets newly analyzed in this paper. Table \ref{table:source_stats} summarizes the source and offset counts from \papri and this work. In \papri we found offsets in 19/37 features of 15/22 jets analyzed there.
 
 Of the  94 offsets detected in this work, we newly detect 58 offsets from 43 jets, while 36 offsets from 21 jets confirm what was previously reported in the literature. The previously reported offsets were generally identified either visually or based on X-ray/radio centroid differences or using longitudinal brightness profiles although without any proper uncertainty estimates, which we now provide. In addition, we include 15 sources from 9 jets with offsets on the order of an ACIS-S pixel (0.492\as); we also include 4 offsets in 3C 273 from \citet{Marchenko_2017}, which were measured after accounting for the effects of PSF. In addition, we include 6 other features from 3 of these jets under the Co-s category, where within the published uncertainty, estimates, the offsets are consistent with $
 \leq$0.15\as.
 Furthermore, although our selection criteria excluded components surrounded by any other feature within 2 ACIS-S pixels, in 18 of such features, we find radio knots co-spatial with distinct X-ray peaks. We tentatively include them under the no-offset category, and are indicated with gray labels on their individual images. This comes to 132 offsets in a total of 226 components where a total of 201 components are analyzed in \papri and this work.
 
\begin{center}
    \begin{table}[h]
        \caption{Summary of the sources considered in \papri and this work.\label{table:source_stats}}
        \begin{threeparttable}[b]
        \begin{tabular}{|l|c|c|c|}
            \hline
            \multicolumn{2}{|c|}{}                                    & \papri   & This work       \\
            \hline
            \multirow{2}{4em}{Analyzed}                               & Jets     & 22        & 77\tnote{a}  \\

            \cline{2-4}
                                                                      & Features & 37        & 164 \\
            \hline
            Excluded                                                  & Jets     & 47        & 52  \\\hline
            \multirow{2}{4em}{Offsets found}                          & Jets     & 15 (68\%)        & 61 (79\%)  \\

            \cline{2-4}
                                                                      & Features & 19        & 94  \\
            \hline
            \multirow{2}{10em}{Previously identified\newline offsets} & Jets     & -         & 21  \\
            \cline{2-4}
                                                                      & Features & -         & 36  \\\hline
            \multirow{2}{10em}{Newly identified\newline offsets} & Jets     & 15         & 43  \\
            \cline{2-4}
                                                                      & Features & 19         & 58  \\\hline
            \multirow{2}{10em}{Offsets used\newline from reference}   & Jets     & -         & 10   \\
            \cline{2-4}
                                                                      & Features & -         & 19  \\\hline
        \end{tabular}
        \begin{tablenotes}
            \centering
            \item [a] Total count including the common source, 3C 275.1.
          \end{tablenotes}
        \end{threeparttable}
    \end{table}
\end{center}

 The default prior for LIRA used in this work is optimized to detect compact sources that causes smooth diffuse (e.g., M87) or resolved sources to appear with a speckled texture in their deconvolutions. That means diffuse emission around point or point-like sources, if present, may introduce errors in the source's mean location. However, the associated uncertainty in the localization also increases due to increased fluctuation in the measured offset between individual MCMC draws, thereby suppressing the detection of a false offset. For example, even with a mean offset of 0.69\as, knot K45 in 3C 111 is classified as Co-s because it is still within 1$\sigma$~of 0.15\as.
 
 For two reasons, diffuse emission cannot solely be responsible for the offsets observed in our study. Most of the analyzed components are isolated where we expect minimal diffuse (inter-knot) emission. Two, diffuse emission must induce offsets in random directions contrary to the observed predominance of a specific offset type (see section \ref{subsec:fr2-offsets}). That means a diffuse component must exist in jet features that always brightens in a preferred orientation to reproduce the observations. However, this is equivalent to a spatially separated emission zone configuration, whose detection is the main focus of this work. Hence we conclude that diffuse emission has little influence on the overall conclusions of this study.

Table \ref{table:offset-results} summarizes the results of our offset analysis on 77 sources with 164 components. Additionally, we provide the details of 25 sources we included from the literature. The name of the source is provided in column 1, the X-ray feature's name in column 2, the angular X-ray/radio offset in column 3, along with any reference that noted an offset previously, the sky-projected offset in column 4, and the type of the offset in column 5. We adopt the offset-type nomenclature from \papri~where ``Xf'' indicates X-ray-first or the X-rays peaking upstream of the radio, ``Rf'' indicates radio-first or the radio peaking upstream of the X-rays, and ``Co-s'' or co-spatial indicates the lack of any significant evidence for an offset. Tentative Co-s-types are indicated with a (T). ``Amb'' or ambiguous offset-type is used for components with an unclear direction of the jet (e.g., 3C 351, Figure \ref{fig:results-3C351}) or those that lack unambiguous X-ray/radio spatial correlation (e.g., knot or large-scale hot gas). Furthermore, we specify an additional attribute for the offset-type based on the structure of the component. ``Bnd'' indicates a knot lying at a bend in the jet, ``Flr'' indicates the so-called \textit{flaring point}, commonly found in FR I-type sources \citep[e.g.,][]{2001MNRAS.326.1499H}, and ``Vnsh'' if the emission from an FR II-type jet quickly diminishes at the specified location before re-emerging at a hotspot further downstream \citep[e.g., 4C+19.44, ][]{Harris-2017}. ``Jet-HS'' indicates the hotspot presumably produces X-rays when the jet enters the turbulent hotspot region, generally coinciding with a faint radio peak followed by a much brighter radio hotspot. In addition to the results on offsets, we also compile spectral data in Table \ref{table:offset-results} with radio spectral index ($\alpha_r$) in column 6, X-ray spectral index ($\alpha_X$) in column 7, radio frequency in column 8, the corresponding radio flux in column 9, X-ray energy flux density at 1 keV in column 10 and the X-ray/Radio flux ratio ($R=\log_{10}(\frac{\nu_XF_X}{\nu_rF_r})$, flux ratio, hereafter) in column 11.  
\startlongtable
\tablecolumns{11}
\centerwidetable 
\begin{deluxetable*}{ll|LlC|LLCCCC}
  \tabletypesize{\scriptsize}
  \tablecaption{Offset analysis results with spectral data\label{table:offset-results}}
  \tablehead{
    \colhead{Name} &
    \colhead{Component}&
    \multicolumn{3}{c}{Offset} &
    \colhead{$\alpha_r$}&
    \colhead{$\alpha_X$}&
    \colhead{$\nu_r$} &
    \colhead{$f_r$} &
    \colhead{$f_{1 keV}$}&
    \colhead{$R$}\\[-5pt]
    \colhead{}&
    \colhead{}&
    \colhead{$\arcsec$}&
    \colhead{Type\textsuperscript{b}}&
    \colhead{kpc\textsuperscript{c}}&
    \colhead{}&
    \colhead{}&
    \colhead{GHz}&
    \colhead{mJy}&
    \colhead{nJy}&
    \colhead{$\log_{10}(\frac{\nu_XF_X}{\nu_rF_r})$}\\
  }
    \startdata
    3C 9 & H & 0.29\pm0.09  & Xf+Bnd & 2.46\pm0.77 &  &  & 14.99 & 47.20 & 0.30  & 0.10 \\
    3C15 & B & 0.66\pm0.29  & Xf+Flr & 0.92\pm0.40 &  &  & 8.40 & 33.20 & 0.64  & 0.56 \\
     & C & 0.14\pm0.04  & Co-s & 0.20\pm0.06 & 0.90 & 0.70\pm0.36 ~(32) & 8.40 & 8.30 & 1.10  & 3.82 \\
    3C 31 & B & 0.77\pm0.05  & Xf+Flr & 0.26\pm0.02 & 0.55 & 1.30\pm0.20  & 8.46 & 6.83 & 2.45  & 10.25 \\
    4C +01.02 & B & 0.12\pm0.07  & Co-s+Bnd & 0.97\pm0.56 & 1.02 & 0.80\pm0.30 ~(1) & 4.80 & 10.04 & 4.55 ~(1) & 22.85 \\
     & SHS & 0.40\pm0.06 ~(1) & Xf & 3.41\pm0.50 & 1.02 & 0.90\pm0.60 ~(1) & 4.80 & 114.60 & 1.72 ~(1) & 0.76 \\
\enddata
\tablecomments{Table 1 is published in its entirety in the machine-readable format.
      A portion is shown here for guidance regarding its form and content.}
\tablerefs{(1) \citet{2012ApJ...748...81K}. (2) \citet{2007ApJ...659.1008K}. (3) \citet{2004ApJ...612..729H}. (4) \citet{2011ApJ...739...65P}. (5) \citet{2001MNRAS.326.1499H}. (6) \citet{Clautice:2016zai}. (7) \citet{harris2004x}. (8) \citet{hardcastle2001chandra}. (9) \citet{2016MNRAS.455.3526H}. (10) \citet{2007ApJ...669..893H}. (11) \citet{Stanley_2015}. (12) \citet{2006ApJ...641..717S}. (13) \citet{2002ApJ...581..948H}. (14) \citet{2012ApJ...745...84W}. (15) \citet{2018ApJ...860....9M}. (16) \citet{jester2006new}. (17) \citet{Marshall_2002}. (18) \citet{worrall2016x}. (19) \citet{Harris-2017}. (20) \citet{gelbord2005knotty}. (21) \citet{2005MNRAS.360..926W}. (22) \citet{2001ApJ...561L.157B}. (23) \citet{kataoka2008chandra}. (24) \citet{2005ApJ...622..149K}. (25) \citet{wilson2000chandra}. (26) \citet{2008ApJ...684..862S}. (27) \citet{2010MNRAS.404..629E}. (28) \citet{2012ApJ...755..174G}. (29) \citet{2012MNRAS.424.1346W}. (30) \citet{Perlman_2009}. (31) \citet{tavecchio2007chandra}. (32) \citet{2003AA...410..833K}. (33) \citet{siemiginowska2003chandra}. (34) \citet{2012ApJ...750..124S}. (35) \citet{2002ApJ...568..133W}. (36) \citet{jorstad2006x}. (37) \citet{kataoka2003chandra}. (38) \citet{2004ApJ...608..698S}. (39) \citet{2004ApJ...614..615J}. (40) \citet{2011ApJ...729...26M}. (41) \citet{2007ApJ...657..145S}. (42) \citet{Perlman-2011}. (43) \citet{Marchenko_2017}. (44) \citet{pyrzas2015multiwavelength}. (45) \citet{worrall2012jet}.}
\end{deluxetable*}

\begin{figure}
    \includegraphics[width=\textwidth]{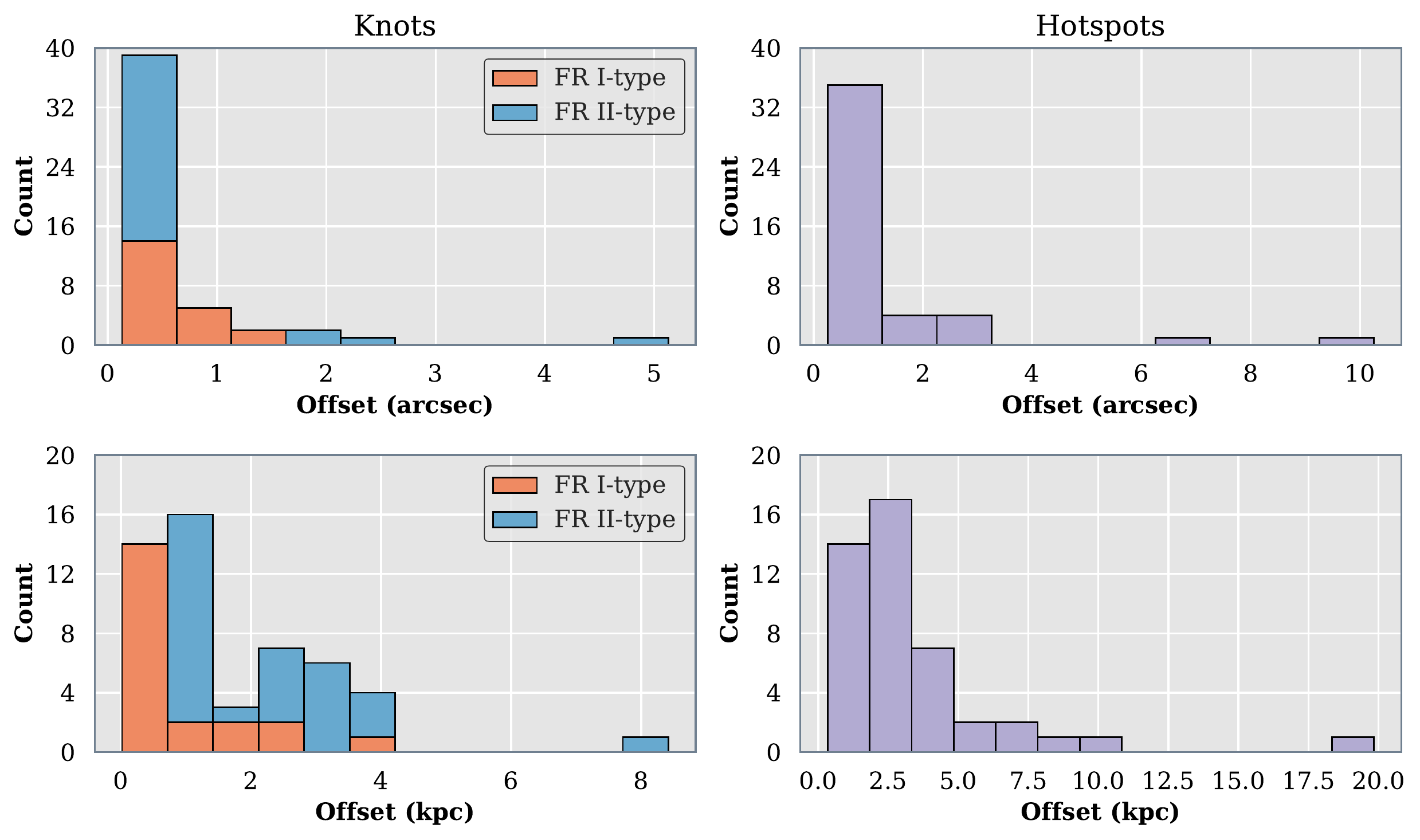}
    \caption{Histograms of the sky-projected offsets in knots and hotspots. Left panels show histograms of offsets in knots for the FR-I and FR-II types and the right panels show histograms for offsets in hotspots. The units of offsets are arc second and kpc in the top and bottom panels, respectively.    \label{fig:offset-histograms}}
\end{figure}

Of the 94 detected offsets, we find 75  Xf-type offsets, 13 with the Bnd-type, 4 with the Flr-type, 11 with Jet-HS type; 9 features show the Rf-type offset, 2 with the Bnd-type, 2 with the Vnsh-type. Figure \ref{fig:offset-histograms} shows the histogram of sky-projected offsets in our sample for knots (left panels) and hotspots (right panels), where the top and bottom panels show the offsets in arcseconds and kpc, respectively. The average magnitude of sky-projected knot-offsets is 0.59$\pm$0.79\as~(1.79$\pm$1.44 kpc) in FR II-type sources, 0.58$\pm$0.29\as~(0.81$\pm$1.04 kpc) in FR I-type sources, and 1.13$\pm$1.77\as~(3.17$\pm$3.21 kpc) in hotspots. The observed high dispersion in the offsets is expected as the jets in our sample span a wide range of orientations (see section \ref{subsec:xjet-sample}) and redshifts. Put another way, even an identical intrinsic offset can observationally produce a wide range of offsets due to projection and distance effects.

\subsection{New X-ray Feature Detections\label{subsec:new_detections}}
Combined with the high resolution of \chdr, our X-ray image deconvolution using LIRA has allowed us to quantify offsets while accounting for the effects of core and background, which is unavailable to traditionally used longitudinal profile or centroid-based methods. For example, we detected an offset in the hotspot of 3C~454.3, located at $\approx$5\as~from its extremely bright blazar core. Here, the wings of the core's PSF inevitably introduce an uncertainty in the centroid of the hotspot, but our method is able to overcome this interference. Besides offsets, our deconvolutions have also revealed new features in jets, undetected in their previous analyses. In a few cases, we also detect features buried within the PSF of the core.
 Below we briefly comment on the new detections in individual sources.
\\
\\
\ntitl{3C 66B} In addition to an X-ray peak associated with the radio knot A, we also detect two distinct X-ray peaks between knots x and B, which, however, lack clear radio counterparts. The one closer to the core lies between knot x and the upstream edge of knot A. The other peak, further downstream, lies on the eastern edge of knot B, where the radio contours appear gently deflected around it, suggesting it may be a stationary obstacle shocked by the jet \citep[e.g.,][]{kraft2000chandra, Hardcastle-2003}. This detection is consistent with the need for two or more spectral components to explain knot B's complex X-ray spectra \citep{2001MNRAS.326.1499H}.
\\
\ntitl{PKS 1229-02} We detect a new X-ray knot roughly upstream of the radio knot C, which lies at an apparent bend in the jet. Interestingly, the upstream region of radio C also shows a spike in the rotation measure \citep[see Fig. 3 in ][]{russell2012x}. This spike suggests a bend-induced shock presumably produces the X-rays in the jet \citep[e.g., ][]{2005MNRAS.360..926W}. A relatively fainter X-ray feature is also detected further down the jet, beyond which the jet appears disrupted. This morphology suggests a stationary obstacle possibly interacts with the jet and gently disrupts it \citep[for example, see ][]{2008ApJ...675.1057E}, while presumably producing the X-rays.
\\
\ntitl{PKS 1335-127} We detect the X-ray counterpart to the radio knot A, which lies $\sim$1\as~to the southeast of the core. 
\\
\ntitl{4C +19.44}Besides all the knots except S2.0, 
We also detect distinct X-ray features between knots S4.0-S5.3, S5.3-S8.3, and S10.0-S11.2, along the outer edges of the jet where the radio contours curve inside. This structure suggests these features are possibly stationary obstacles, for example, hot gas clouds, gently \textit{pinching} the jet. These features match the locations where the transverse X-ray brightness profile of the jet deviates from its mean position angle, presented in Figure 2 of \citet{Harris-2017}. We newly detect two X-ray features in the jet region (previously referred to as S25.7 in \citet{Harris-2017}) upstream of the southern hotspot. We label the northern feature as S25.7, and the southern feature as SHS-a where the jet presumably enters the turbulent hotspot region.
\\
\ntitl{3C 78}We detect the X-ray counterpart to the radio knot C.
\\
\ntitl{4C +21.35}We detect an X-ray knot $\sim$0.2\as~upstream of the radio knot C2, which lies at an apparent bend in the jet.
\\
\ntitl{4C +28.07}We detect two X-ray knots co-spatial with two radio knots, C and D, respectively. A new X-ray feature is also detected upstream of NHS-a, where the jet presumably enters the turbulent hotspot region.
\\
\ntitl{PKS 1202-262}We detect a clear knotty structure in the X-rays with three knots co-spatial with radio knots B, C, D, respectively. Although we detect two more X-ray features further downstream of knot D, they lie midway between the radio knots, making their radio association ambiguous. This knotty structure is contrary to the observed relatively flat longitudinal brightness profile, presented in Figure 9 of \citet{Perlman-2011}, which originally led the authors to believe IC/CMB primarily produces the X-rays.
\\
\ntitl{3C 270.1}Our deconvolved X-ray image shows a \siml1\as~(8.7 kpc)~bar-like structure upstream of the southern radio hotspot, previously seen in the western hotspot of Pictor A \citep{2016MNRAS.455.3526H}. This structure presumably indicates the location where the jet impacts the turbulent hotspot region and produces X-rays along its entire cross-section. 

%
\subsection{Are X-ray knots resolved?\label{subsec:knot_variance}}
Throughout the literature on X-ray jets, the sizes of knots required to calculate various jet parameters, for example, equipartition magnetic fields, have been inferred using radio data (in the image or UV planes) mainly due to their well-characterized PSFs \citep[e.g.,][]{kataoka2005x}. A few attempts have been made to infer the sizes of knots in the X-rays, for example, by comparing the radial profiles of observations and simulated PSFs \citep[e.g.,][]{kataoka2008chandra,Marchenko_2017}. However, firm conclusions could not be made due to the imperfect knowledge of \textit{Chandra's} PSF. 

\begin{figure}
    \fig{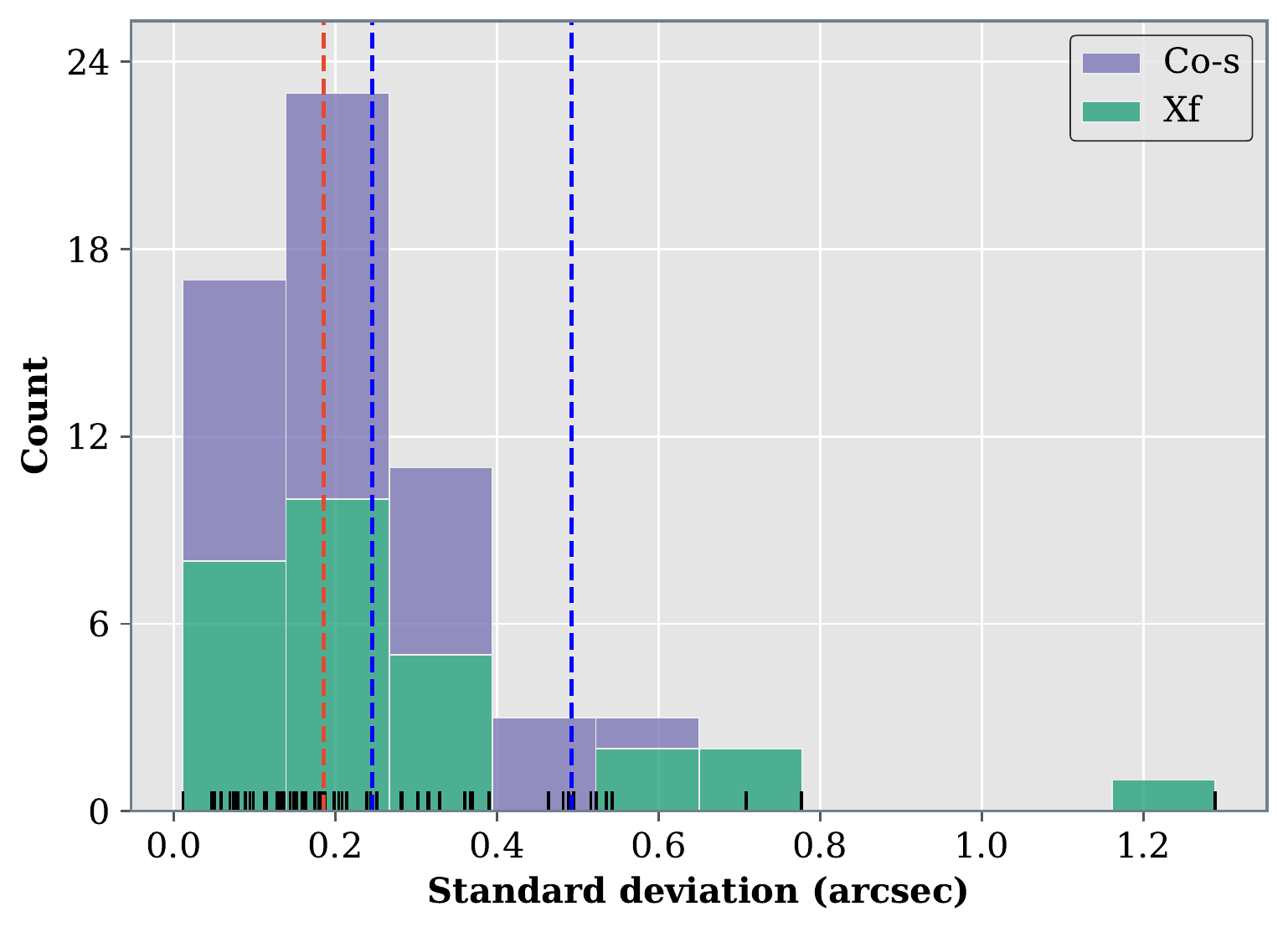}{0.5\textwidth}{}
    \caption{Histogram of the standard deviation of knot regions on deconvolved images for Xf (green) and Co-s-type (gray) knots. The two blue-dashed lines locate the half and single ACIS-S pixel sizes,  and the red-dashed line indicates the median of the distribution. \label{fig:std_dist_knot_size}}
\end{figure}

Despite these imperfections, we can examine whether the X-ray knots are resolved or not by studying the standard deviation of knot regions in LIRA-deconvolved images. We estimate the standard deviation as:
\begin{equation}
    \sigma = \sqrt{\frac{1}{N}\sum_{pixel~i}n_i(x_i-\bar{x})^2}
\end{equation}
where $x_i$~is the location of the $i^{th}$~pixel, $n_i$~is its pixel value, and $N$~is the sum of all pixel values. Here we only consider knots in FR-II-type jets as they are the main focus of our work. Figure \ref{fig:std_dist_knot_size} shows the histogram of standard deviation of all the knot regions.
The median of the distribution (red-dashed line) is less than the half ACIS-S pixel size (0.246\as), and about 90\% of the knots lie below single ACIS-S pixel size (0.492\as), both indicated with blue-dashed lines. That means emission from the majority of the knots is mostly confined to a single image pixel, which supports our assumption that we can generally treat knots as point (unresolved) or point-like sources. For knots with values greater than a single pixel, the source may be intrinsically extended. Larger standard deviations may also result in sources with low signal-to-noise ratios (SNR) or with mismatched PSFs. For comparison, the typical standard deviation for the cores with high SNRs is about 0.03\as, consistent with a point source. However, due to the spatio-spectral dependence of \textit{Chandra's} PSF, the PSF simulated for the core and input to LIRA may not always match the emission from knots. Hence, their emission may be smeared in the deconvolved images, resulting in a larger variance even for actual point sources.

\section{Discussion\label{sec:discussion}}
We analyzed a large sample of X-ray jets with the main aim of constraining the nature of X-ray emission from FR II-type jets on kpc scales, and obtained several lines of constraints in the form of offsets. While offsets argue for two or multi-zone models, the formation mechanism of knots and their internal structure are other important uncertainties in understanding how they emit X-rays. Although a few multi-zone models have been considered in the literature to reconcile the predictions of IC/CMB with the observations \citep[e.g.,][]{jester2006new,tavecchio2021constraining}, they are generally found to require unrealistic conditions \citep[e.g.,][]{jester2006new} in the jet, and lack a mechanism to explain offsets. Below we discuss the statistics of the observed offsets and explain them using a knot description based on previous works \citep[][and references therein]{stawarz2004multiwavelength,kataoka2008chandra} that suggest knots represent separate moving parts of the jet, and estimate their bulk Lorentz factor under this scenario. We then use X-ray/radio flux ratios as a means to test whether IC/CMB can be the dominant mechanism with the limits of the current sample.
\subsection{Knot-Offsets in FR II-type jets\label{subsec:fr2-offsets}}
Including offsets from \papri~and those directly taken from the literature, we find that the Xf-type offset is the majority type of offset observed in knots of FR II-type sources (48 out of 114). The Rf-type offset occurs only in a few cases (8 out of 114). These Xf-type offsets clearly contradict the X-ray/radio co-spatiality required by {\sl any one-zone} IC/CMB interpretation, including the IC/CMB one.

To first order, Rf-type offsets in knots of FR II jets are compatible with a multizone  IC/CMB scenario, where the X-rays (produced by electrons with $\gamma\sim10$) can extend past the radio ($\gamma\sim10^4$) due to their larger radiative lifetimes. However, their peculiar morphology combined with additional evidence suggests that in these cases the X-ray emission is not from the jet itself. For example, the Rf-type offset found in knot C of 4C+11.45 (Figure in 4.15 \papri) appears at an apparent bend in the jet and is surrounded by Ly$\alpha$~clouds. The X-rays may be produced by the surrounding clouds, shock-heated by the jet which is being deflected, leading to an Rf-type offset. Similarly, Knot C in PKS 1030-357 (Figure \ref{fig:results-PKS1030-357}) and knot S17.7 in 4C+19.44 (Figure \ref{fig:results-4C+19.44}) show an Rf-type offset with an unusual structure: the radio and X-ray emission from the jet rapidly declines with the X-ray emission persisting slightly further downstream than the radio (indicated with ``+Vnsh'' in Table \ref{table:offset-results}). A previously cold gas may be deflecting the jet away from our line of sight, thereby diminishing its observed emission \citep[e.g.,][]{mendoza2001deflection}, while the jet shock-heats the cloud, which produces the X-rays \citep[e.g., ][]{worrall2016x}. Such shock-heated clouds presumably exhibit a thermal X-ray spectrum which can be tested with deeper observations.

\begin{figure*}
    \includegraphics[width=\textwidth]{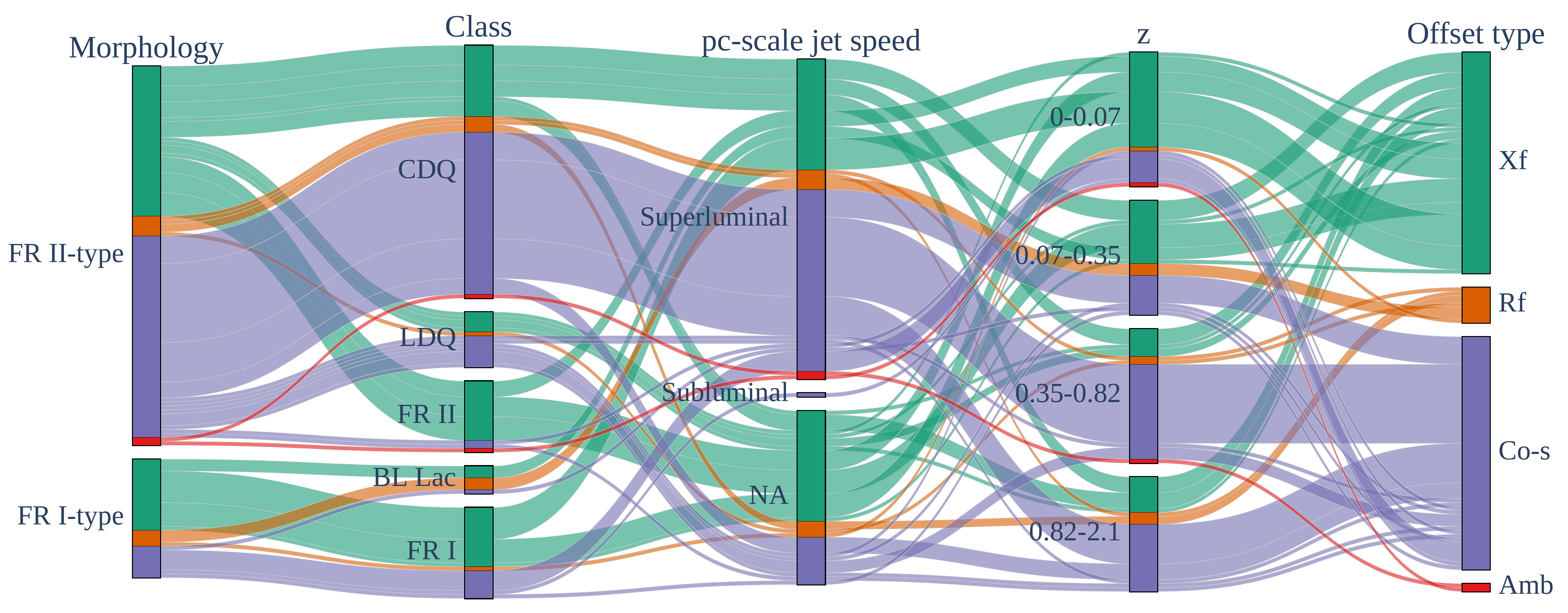}
    \caption{Categorical parallel-coordinates plot for the offsets in knots of FR II-type and FR I-type sources. Each stripe constitutes a group of knots with identical value in each coordinate and are colored according to the offset-type. Starting from the left, the coordinates are source type, sub-classification, pc-scale jet speed, redshift range, type of the offset.\label{fig:parallel-coord-knots}}
\end{figure*}

Although many knots show significant offsets, a large fraction  (56 out of 114) have radio \& X-ray positions consistent to within errors (within 0.15$''$), which may appear to satisfy the co-spatiality condition that one-zone models require. However, using these knots as evidence for one-zone emission is limited by several factors, including projection and distance effects, and limited instrumental resolution. To examine the importance of these effects, we construct a \textit{categorical-parallel-coordinates} plot, shown in Figure \ref{fig:parallel-coord-knots} for all the knot-offsets in FR II-type sources (as well as FR I knots for completeness). Each ``stripe'' represents a group of one or more knots with the same attributes across all the coordinates and are color-coded based on the offset type. Starting from the left, the coordinates are namely, ({\sl i}) the source type (FR I or FR II), ({\sl ii}) the sub-class (e.g., CDQ or LDQ), ({\sl iii}) the pc-scale jet speed (superluminal for \bapp$>1$ and subluminal for \bapp$\leq1$), ({\sl iv}) the redshift, divided into four quartiles, and ({\sl v}) the offset-type.

The class-coordinate shows that the CDQ-knots, which come from closely aligned jets, comprise the majority of the Co-s-type knots. The LDQ class, although less populous than the CDQ class, has a similar fraction of Xf/Co-s-types. Conversely, the knots in FR II-jets, which come from misaligned jets, predominantly possess the Xf-type offset. Furthermore, the z-coordinate indicates that the fraction of Co-s-type knots increases with redshift, which is expected as the CDQs mostly populate the two higher-redshift quartiles. That means Co-s-type knots possibly belong to the Xf-type class, and a combination of projection and distance effects may be foreshortening their observable offsets to below \textit{Chandra}'s resolution.

On the other hand, the Co-s-type knots may also be intrinsically different from their Xf-type counterparts and may result in different spectral properties. To examine this scenario, we adopted the 2-sample Anderson-Darling (AD)  test \citep[]{scholz1987k}, to test for any differences in the distributions of spectral indices and flux ratios between the Xf-type and Co-s-type knots. For both the parameters, we are unable to detect any statistically significant differences between the two offset classes. Specifically, the \pval s are 0.16 and 0.25 for spectral index and flux ratio, respectively, with the null hypothesis being that the two distributions belong to the same parent population.


\subsection{Jet orientation scheme\label{subsec:angle_scheme}}
To further interpret the measured offsets (expressed in a sky-projected form in Table \ref{table:offset-results}) we require an estimate of orientation angle in order to compute de-projected offsets. A few methods exist to measure the orientations for kpc-scale jets  or pc-scale jets. For example, \cite{marin2016robust} use the ratio of radio core to lobe luminosity to determine the orientation of kpc-scale jets while \cite{drouart2012jet} use it for pc-scale jets. A few authors have proposed using optical emission lines from the central AGN to determine the pc-scale jet orientations \citep[e.g.,][]{risaliti2011iii,matthews2017quasar,yong2020determining}. However, most of these methods only apply to restricted redshift ranges. Moreover, the optical emission line data is absent for several sources in our sample and the proposed methods measure much larger angles for pc-scale jets than what their superluminal motions imply.

Hence, for 29 sources with a measured value of superluminal speed, we first estimate the upper limit on its pc-scale jet angle. We then convert it to an upper limit on the kpc-scale jet by numerically solving the conversion equations given in \citet{conway1993helical,osti_22679542}. The required inputs are {\sl (i)} the position angle difference between kpc jets, and pc-scale jets, which we measure using VLBI images obtained from NED\footnote{NASA Extragalactic Database, \url{https://ned.ipac.caltech.edu/}}, and {\sl (ii)} the intrinsic pc-kpc jet bend, which we take as 5\textdegree. We omitted three superluminal sources, 3C 66B, B2 0737+313 and 4C -03.79, from this analysis due to lack of numerical convergence. For these three and the remaining sources, following \papri we adopt a spectral class-based orientation scheme to assign representative upper limits on the jet's angle to the line of sight. In this scheme, CDQs are assigned 15\textdegree, LDQs 30\textdegree, and FR-II (BLRG) 60\textdegree and FR II (NLRG) 90\textdegree.  Furthermore, for all features where the jet displays a projected bend larger than 20\textdegree, we increment the derived/assumed angle by 5\textdegree~for CDQs, 10\textdegree~for LDQs and 20\textdegree~for FR-IIs.

\subsection{De-projected offsets\label{subsec:deproj-offsets}}
\begin{figure*}[!ht]
    \includegraphics[width=\textwidth]{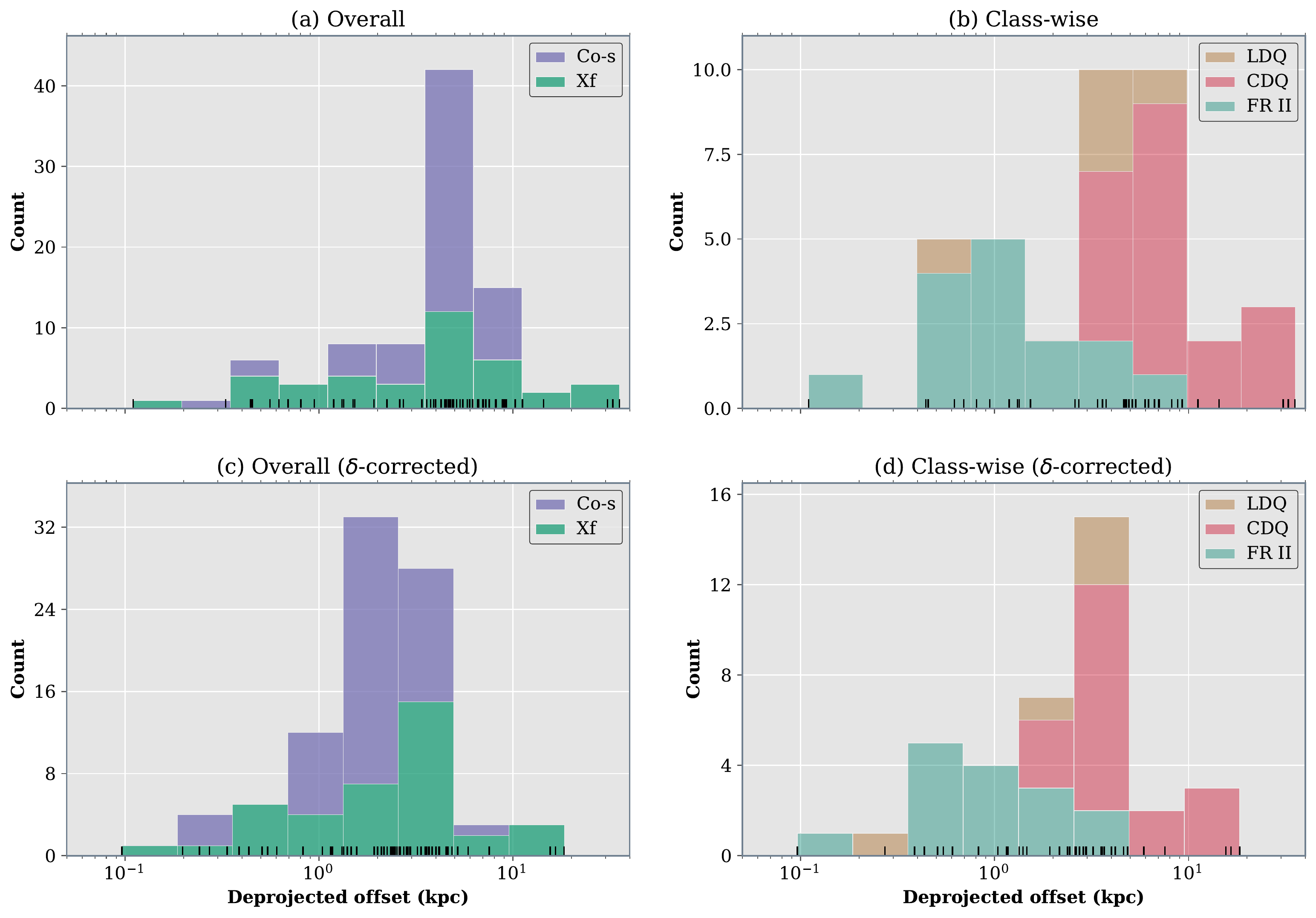}
    \caption{Stacked-histograms for de-projected offsets (kpc), plotted on a log scale, for knots in FR II-type sources. (a)  shows the histograms for Xf-type and Co-s-type offsets for all the FR II-type sources. Co-s type knots are plotted by assuming a 0.15\arcsec~offset, also distributed like the Xf-types. The histograms roughly describe a unimodal profile peaking at $\approx$4 kpc, suggesting similar knot sizes and formation mechanisms. (b)  shows only the Xf-type offsets, grouped into their respective subclasses; CDQs and LDQs peak at higher values than FR-IIs, which may suggest an intrinsic difference between knots in quasars and FR-IIs. Alternatively, we can interpret this difference as a result of moving knots where their relativistic motion increases the observed offset by a factor of $\delta$. (c) and (d) show the $\delta$-corrected counterparts of the top panels where we assume $\Gamma=1.25$~to estimate $\delta$. The corrected-offsets are distributed more uniformly  across classes and peak at $\approx$2 kpc, which is a better lower-limit on the sizes of knots (see section \ref{subsec:deproj-offsets} for discussion).\label{fig:deproj_offsets}}
\end{figure*}
Figure \ref{fig:deproj_offsets}a shows the histogram for de-projected Xf-type offsets,de-projected using the upper limits from section \ref{subsec:angle_scheme}. We also stack the histograms for Co-s-type knots, estimated from a constant angular offset of 0.15\as, which is an upper limit to any offset present in them. Both histograms peak at $\approx$4 kpc. Interestingly, this value is similar to the typical sizes of radio knots found in nearby FR II jets \citep[e.g.,][]{kataoka2008chandra,Clautice:2016zai}. If not a coincidence, it suggests that most of the observed knots have similar structures, and presumably, a single process produces them. Put another way, the radio knots at lower-redshifts, if resolved along the jet, generally appear extended \citep[e.g.,][]{kataoka2008chandra,Clautice:2016zai}, while their structure at higher-redshifts remains indeterminate due to limited instrumental resolution.
Furthermore, as discussed in \ref{subsec:knot_variance}, nearly all the deconvolved X-ray knots show no evidence for extended structures at any redshift. Therefore, the similarity of offsets in resolved and unresolved radio knots, and with magnitudes comparable to the resolved low-redshift knot sizes, indicates a similarity in their morphology: extended structure in the radio with a relatively more compact structure in the X-rays.

\begin{figure*}
    \gridline{
        \fig{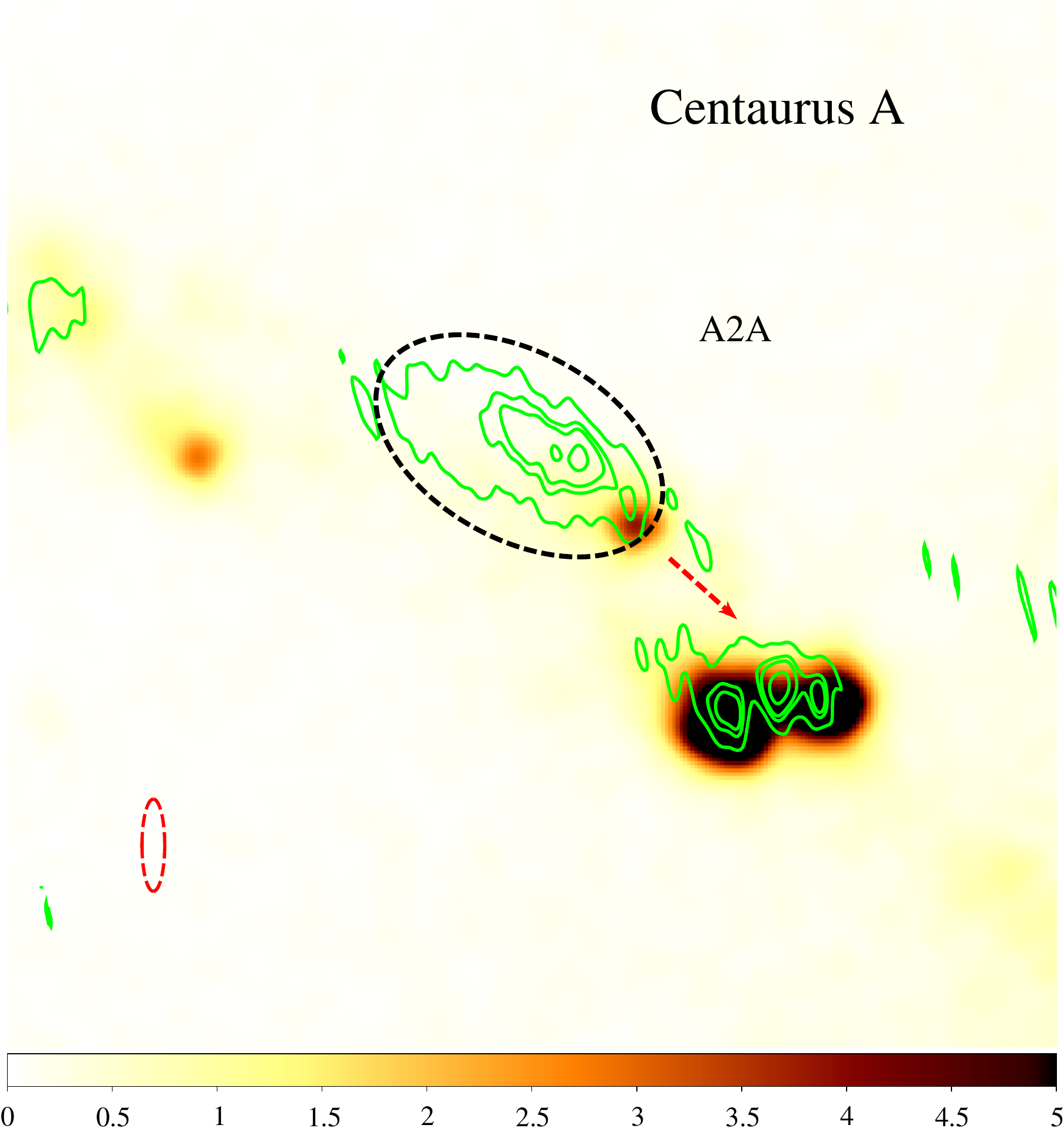}{0.5\textwidth}{(a)}
        \fig{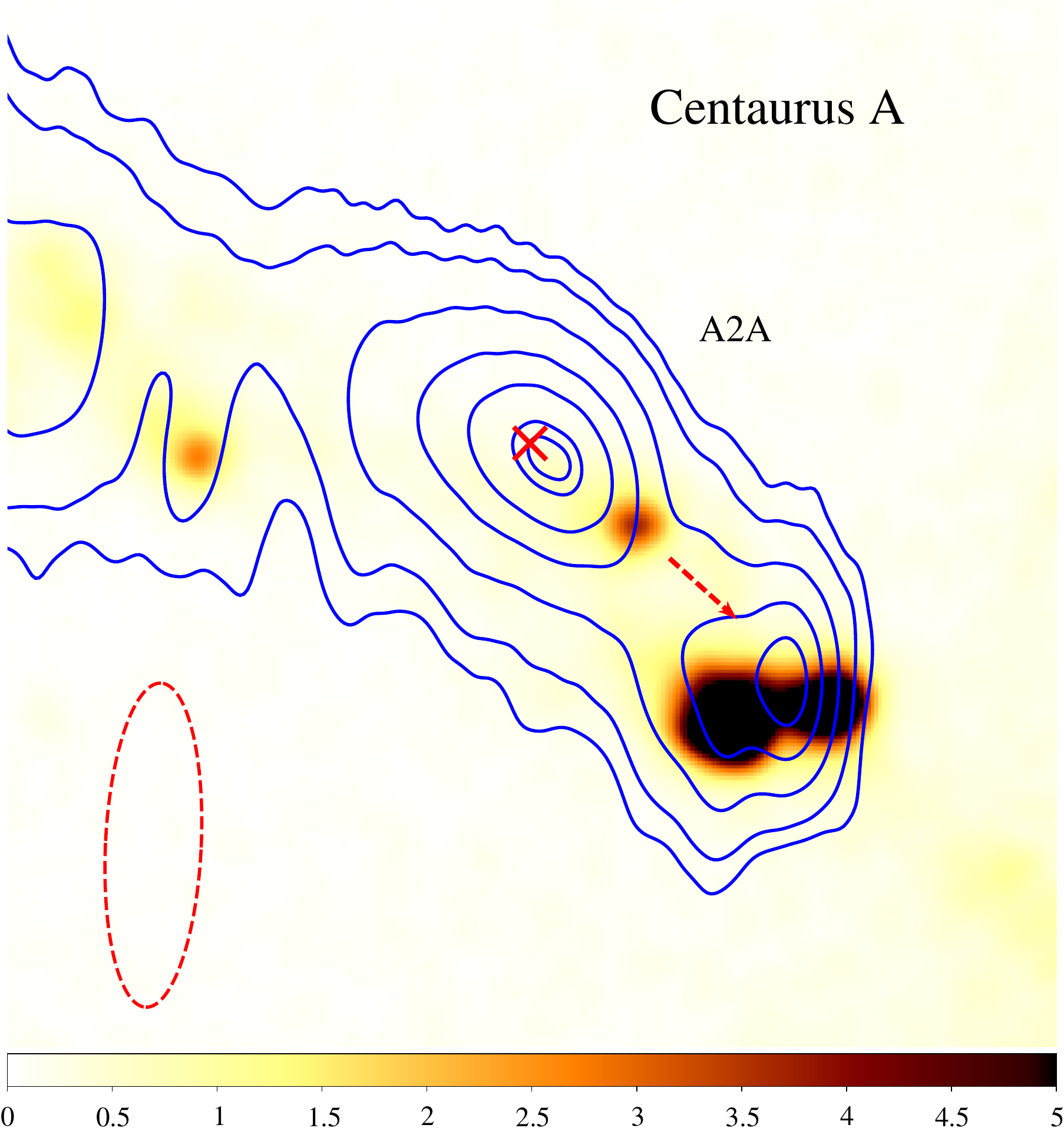}{0.5\textwidth}{(b)}
    }
    \caption{Demonstrating how similar offsets imply similar linear sizes using Centaurus A. With the smoothed \chdr X-ray image in the background, the VLA 8.4 GHz A and B-config contours are overlaid on the left and right panels, respectively. The beam size is shown at the bottom left on each image and the red-dashed arrow leads to the core. A-config resolves the radio knot A2A while the B-config produces a point-like knot with its peak located near the weighted centroid of the resolved knot (indicated with a red X), measured using the black-dashed ellipse, producing an X-ray/radio peak-to-peak offset. That means when a knot like A2A--with extended radio and compact X-ray emitting regions--becomes unresolved, it produces an offset, which is a lower limit on the knot's size. Hence, similar offsets imply similar structures.  \label{fig:cena_offset_demo}}
\end{figure*}

We demonstrate this inference in Figure \ref{fig:cena_offset_demo} with the \chdr~X-ray image and the VLA 8.4 GHz radio images of a nearby FR-I type jet, Centaurus A.
The left panel shows the smoothed X-ray image of knot A2A \citep{Hardcastle-2003} overlaid with the A-config radio contours (beam size: 0.77\as$\times$0.19\as, -0.3\textdegree~PA) while the right panel shows the B-config radio contours (beam size: 2.76\as$\times$0.80\as, -3.3\textdegree~PA). A red-dashed arrow leads to the core in both the panels. The A-config radio observation resolves the knot across and along the jet, while the B-config observation, with a lower resolution than its A-config counterpart, shows a point-like structure, thereby producing a typical peak-to-peak X-ray/radio offset. Furthermore, the radio peak in the B-config image lies near the centroid of the knot's brightness distribution (indicated with a red ``X'' in the right panel), measured using the dashed-ellipse in the A-config image. That means, when an extended radio knot becomes unresolved (while the X-ray knot remains compact), and unless the radio knot's brightness varies significantly along its length, its centroid would lie roughly at its center, creating an offset roughly half its length. Hence, similar offsets would imply similar spatial extents.

Despite a roughly unimodal profile of the overall de-projected offset distribution, the offsets in quasars (CDQ and LDQ) appear to be slightly larger than FR-II type sources (see Fig.~\ref{fig:deproj_offsets}b). The assumed angles for quasars in our orientation scheme can be smaller than their actual angles, which would result in larger offsets for quasars. Alternatively, the quasar knots in our sample, which lie at higher redshifts, could be intrinsically larger than their lower-redshift, misaligned counterparts, FR-II knots. However, it is unclear with the current understanding how the sizes of knots could increase with redshift. One physically motivated explanation could be that these knots are separate moving portions of plasma moving with at least mildly-relativistic speeds, embedded in an outer flow \citep[e.g.,][]{stawarz2004multiwavelength,kataoka2008chandra}, which we detail in the next section. In this case, relativistic effects would magnify the observed offsets by a factor of $\delta$ \citep[see e.g.,][]{jester2008retardation}.
To check the applicability of the moving-knots scenario to the observed offsets,  we correct the de-projected offsets by dividing by an estimated value of $\delta$ for each source. Here we set $\Gamma$=1.25 ($\beta=$0.6) based on the limits on kpc-scale bulk Lorentz factor ($\Gamma\approx 1.18-1.49$) derived from radio data \citep{wardle1997fast,mullin2009bayesian}. We show the resulting histogram of corrected offsets in the bottom panels of Fig.~\ref{fig:deproj_offsets}. The overall distribution becomes less dispersed and peaks at $\approx$2 kpc, which is roughly half of the typical knot sizes of $\approx$4 kpc \citep[e.g.,][]{kataoka2008chandra}. The $\delta$-corrected class-wise histograms also indicate similar offsets across all the spectral classes, which, within the errors of the assumed angles and the bulk Lorentz factor,  can be read as evidence for the moving-knots scenario. 

\subsection{Knot formation mechanism\label{subsec:knot_formation_mechanism}}
Initially, the knots in FR II-type jets were thought to be produced by stationary re-confinement shocks \citep[e.g.,][]{komissarov1998large}, which, however, lack a natural way to explain offsets. \citet{stawarz2004multiwavelength} suggested the knots are separate faster-moving parts of the jet, presumably produced by a possible intermittent activity in the central engine \citep[e.g.,][]{bridle1986collimation,bridle1989unusual,clarke1992origin}; see \citet{godfrey2012periodic} in this context. A forward-reverse shock between faster and slower moving parts of the jet would produce the observed multiwavelength emission. However, \citet{kataoka2008chandra} note in a detailed study of 3C 353, that it is unclear why the two shocks produce different electron energy populations. They suggest the knots are slower and heavier moving blobs of plasma (produced by intermittent activity) embedded in a faster and lighter outer flow. A forward-reverse shock develops at the upstream end of the blob, where the reverse shock travels in the faster outer flow producing X-rays, while the radio emission coincides with the blob. 
\comment{
\begin{figure*}[!ht]
    \gridline{
        \fig{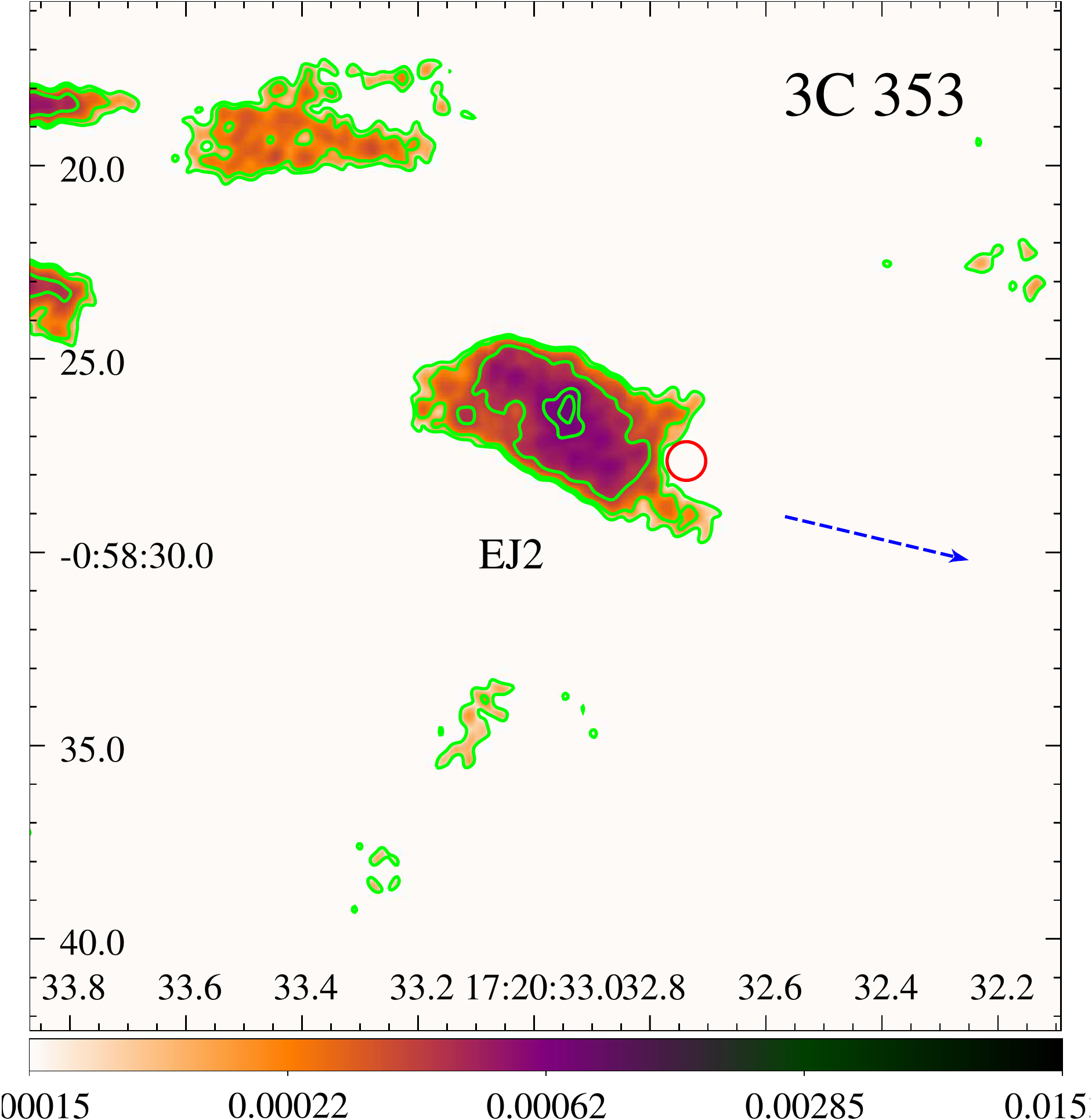}{0.5\textwidth}{(a)}
        \fig{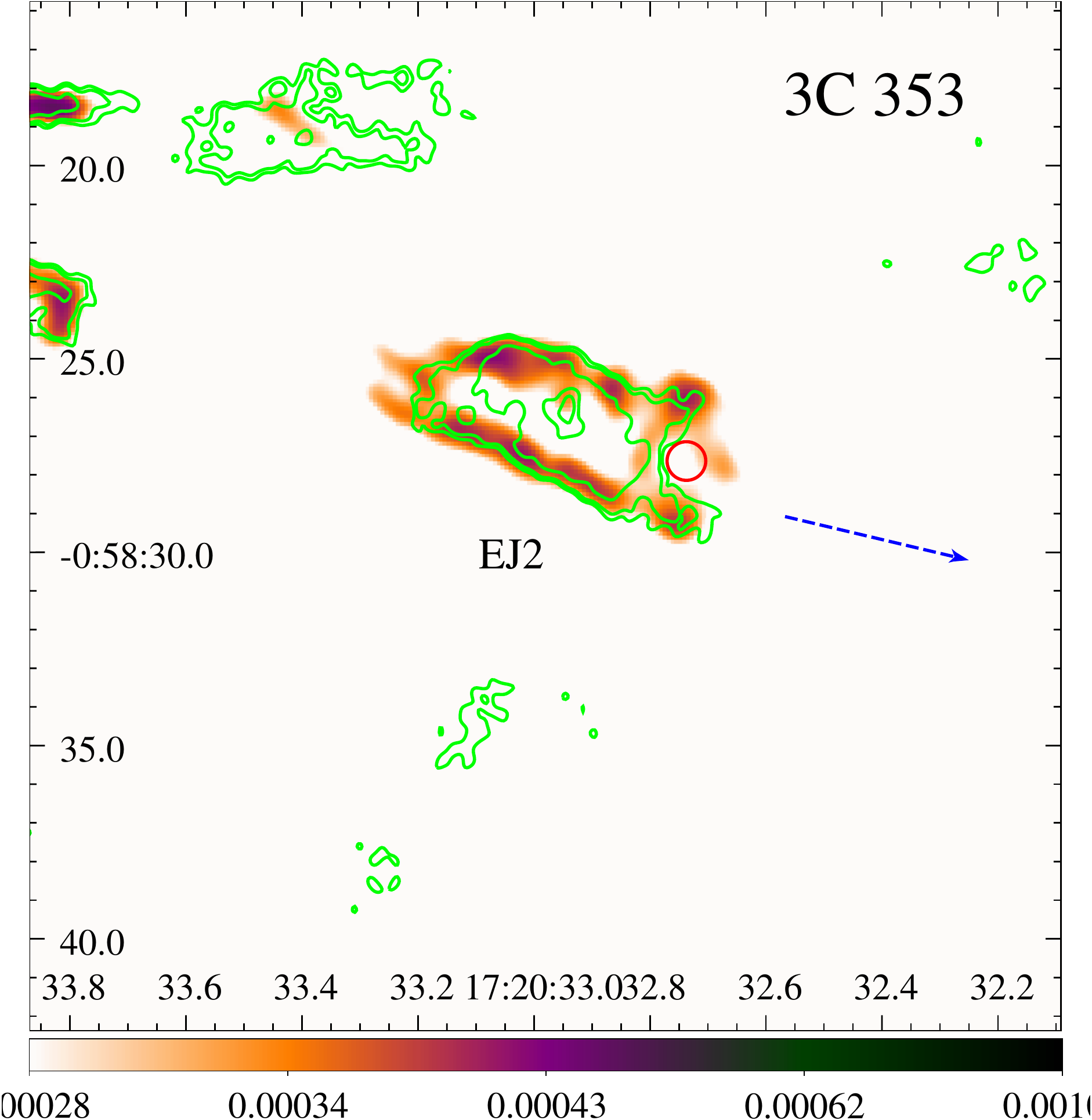}{0.5\textwidth}{(b)}
    }
    \caption{Indication of a shear layer in 3C 353. The left panel shows the VLA 4.8 GHz radio image of knot EJ2; the contours, shown in green, start from 0.1 mJy and increase by a factor of 2. The red circle (0.5\as~radius) marks the centroid of X-ray emission, and the blue-dashed line leads to the nucleus. The right panel shows the residual image of EJ2, obtained after several steps of sequentially removing a portion of the peak flux at each step (i.e., \textit{clean} iterations), and shows an edge-brightened structure, suggesting the radio emission originates from a surface layer.
     \label{fig:3c353_radio_rail}}
\end{figure*}
}

Although this model explains a few features of knots in FR II-type jets, including Xf-type offsets, it 
fails to explain why the transverse brightness profiles of the radio knots in 3C 353 imply emission from a surface instead of a blob \citep{swain1997multifrequency}. \added{This can be clearly seen in Figure 5.12 in \citet{1996PhDT.........6S}, which shows an edge-filtered radio image of 3C 353, where emission from all knots peaks on jet boundaries. Furthermore, nearly all of the knots show a bridge-like structure on their upstream end that connects the edges of the jet. Their X-ray emission is observed upstream of the bridges \citep{kataoka2008chandra}, where the jet's outer flow presumably shocks the blob's upstream end. This may suggest that we need a different picture.}


One possibility is that the X-ray-emitting electrons accelerated at the upstream edge of the plasma blob flow along its surface blob, presumably forming a shear layer, while emitting synchrotron radiation progressively at longer wavelengths. This simple advection, however, leads to offsets of $\sim$100 kpc \citep{bai2003radio} contrary to the observed offsets of a few kpc. Instead, accelerated plasma undergoes multiple interactions (compression or turbulence) while re-joining the outer flow, which would alter the brightness distribution and result in the observed morphology. Moreover, if synchrotron electrons produce the observed X-rays, their shorter lifetimes than their radio counterparts can also explain the relatively smaller emitting regions in X-rays than in the radio. Alternatively, the radio plasma can also diffuse into an already existing shear layer.

This advection through a shear layer scenario also fits two more observed properties of jets:
\begin{enumerate}
    \item Magnetic fields are parallel (i.e., negligible radial component) to the jet in their outer layers while perpendicular in the inner regions of several jets \citep[e.g.,][]{bridle1984extragalactic,bridle1986collimation,swain1998internal}. It is possible for a shock at the upstream end of the blob to compress the plasma and produce a magnetic field parallel to the shock front \citep[for a recent review, see][]{tavecchio2021probing} in the inner parts, while it can be parallel to the jet in the outer parts due to the presumed shear layer \citep[e.g.,][]{urpin2006generation}.
    \item Optical polarization studies on knots of nearby jets like 3C 273 reveal high polarization either on jet-edges or near their maxima \citep{perlman2020unraveling} in the central parts of the jet. Depending on the mass density of the blob, most of the observed optical emission could be produced in a few different ways. For example, by a forward shock in the blob \citep[e.g.,][]{stawarz2004multiwavelength}, or by a shear layer between the outer flow and the blob \citep[e.g.,][]{ostrowski2002radiation}, or by the synchrotron cooled X-ray electrons. Interestingly, similar features are also observed in FR I jets \citep[e.g., ][and the references therein]{perlman2020unraveling} where the knots could be produced by stationary obstacles obstructing the jet flow \citep[e.g.,][]{Hardcastle-2003,perlman2006optical,avachat2016multi,2019ApJ...871..248S}.
\end{enumerate}

\subsubsection{Constraints on the pattern speed of knots}
\citet{Reynolds-1997} suggested that an intermittent activity on timescales of $\geq$10$\textsuperscript{5}$ yrs in the central engine with active jet production on periods of $\sim$10$\textsuperscript{4}$ yrs can explain the observed distribution of radio-jet sizes. Accretion disk models with thermal-viscous instability \citep[e.g,. ][]{Siemiginowska-1997} also indicate similar timescales for jet activity. Based on these works, \citet{stawarz2004multiwavelength} suggested modulated jet activity produces the observed knotty morphology in radio jets. If the moving-knot model is viable at all, we can use the inter-knot distances to constrain their bulk Lorentz factor.
\begin{figure*}
    \gridline{
        \fig{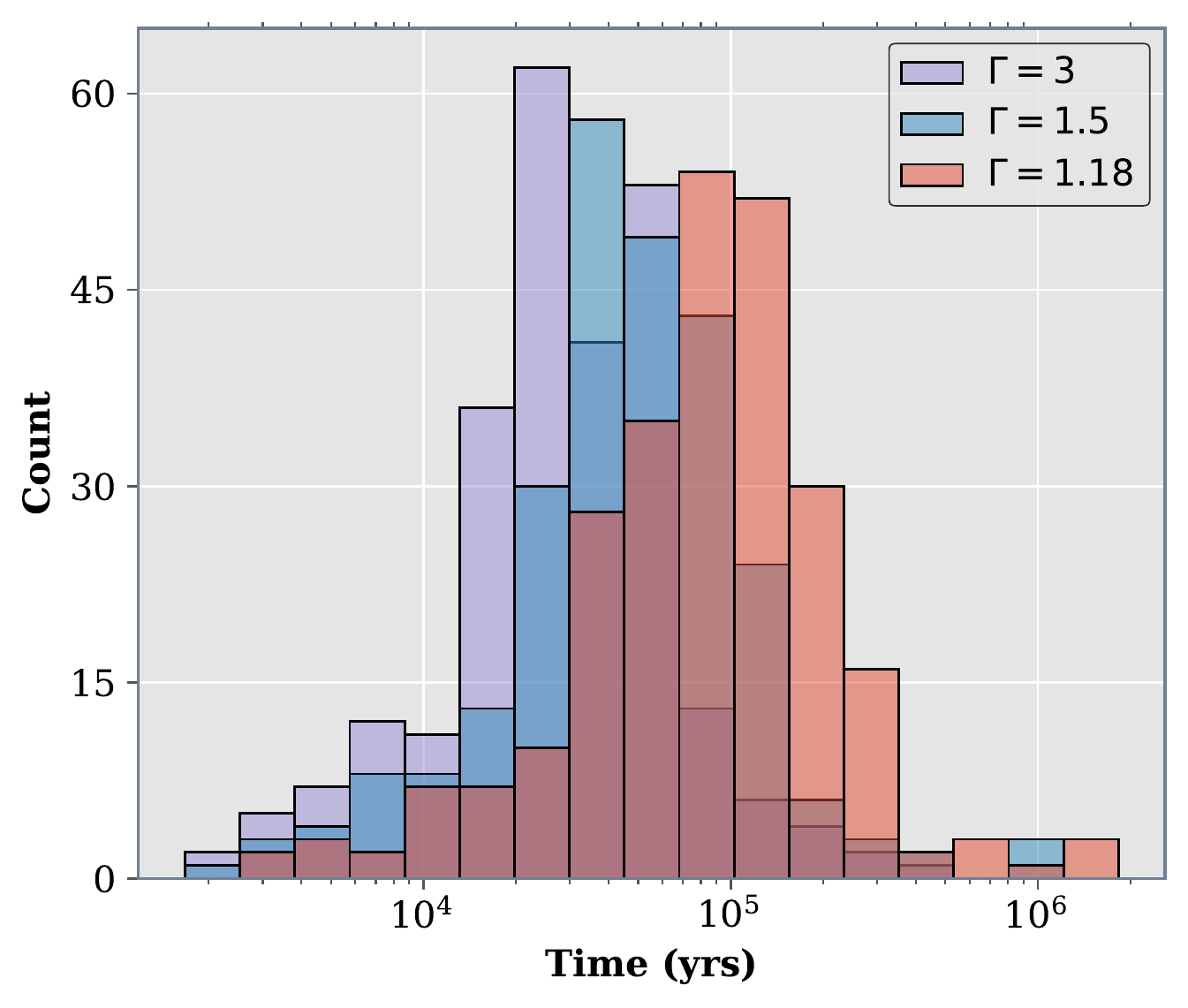}{0.5\textwidth}{(a)}
        \fig{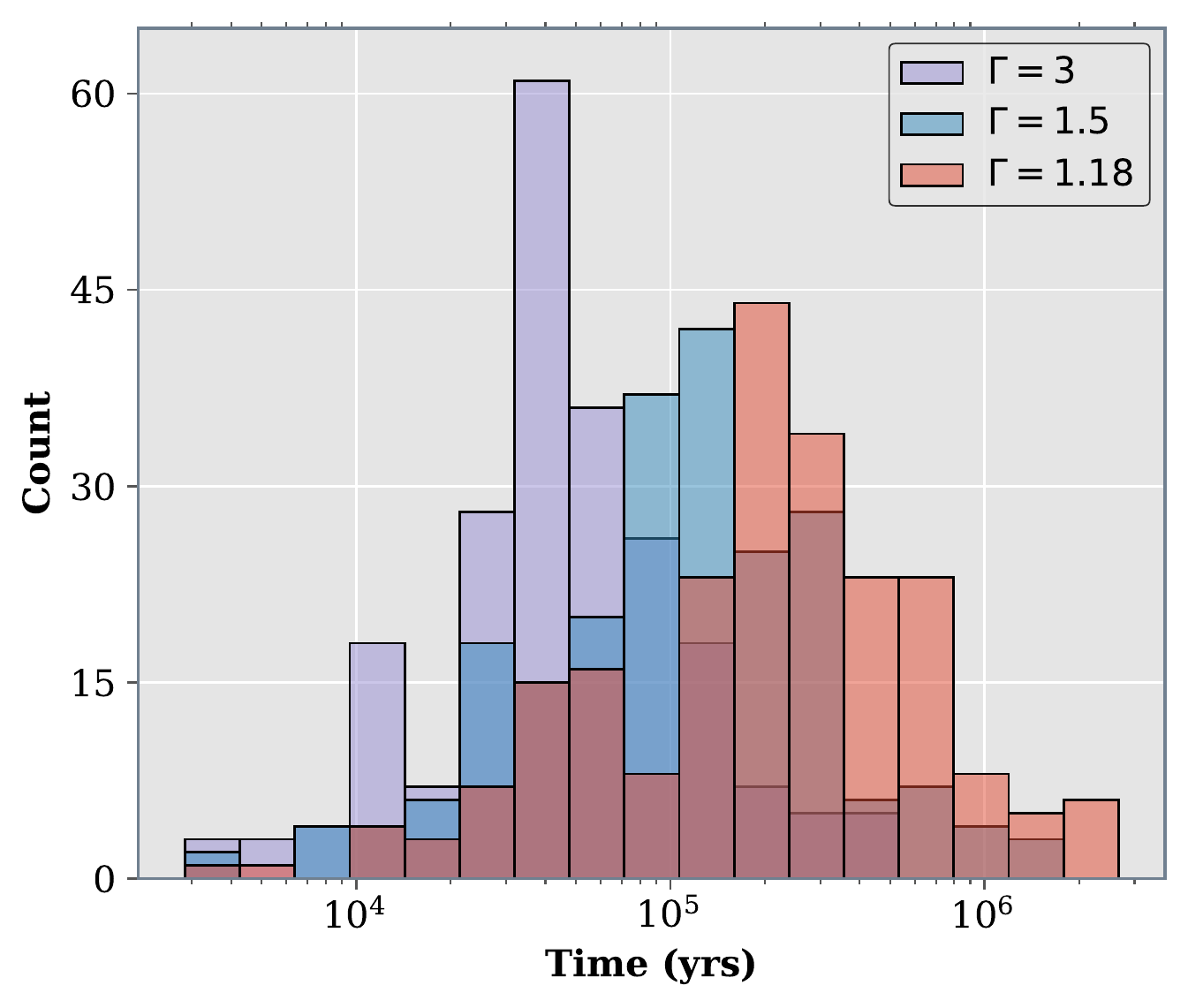}{0.5\textwidth}{(b)}
    }
    \caption{Histograms of knot-ejection timescales derived using inter-knot distances for different bulk Lorentz factors. (a) shows the histogram for timescales measured using the orientation scheme adopted in section \ref{subsec:angle_scheme} where the mean is $\approx10^5$~yrs for $\Gamma=1.18$, while smaller by a factor of 3 for $\Gamma=1.5,3$. (b) shows the same but with all the angles of quasars set to an unrealistic value of 5\textdegree. The distribution peaks above $10^5$~yrs for $\Gamma=1.18, 1.5$, while it remains roughly similar for $\Gamma=3$, suggesting the knots can at most be mildly relativistic if an intermittent activity in the central engine is producing them. \label{fig:knot_timescales}}
\end{figure*}

To derive this constraint, we measure the length of the jet segment between any two adjacent radio-knots for each source in our sample and de-project them using the orientation scheme discussed in section \ref{subsec:angle_scheme}. Each segment is $\delta$-corrected and converted into a timescale using a set of three bulk Lorentz factors. Figure \ref{fig:knot_timescales}a plots the resulting distribution for each of $\Gamma=$1.18, 1.5 and 3 where it peaks about $\approx$10$\textsuperscript{5}$~yrs for $\Gamma=1.18$, while it peaks at about 3$\times$10$\textsuperscript{4}$ yrs for $\Gamma=$ 1.5 and 3, which is about three times smaller than the expected value. Because knots from CDQs and LDQs form the majority in our sample, our orientation scheme may be underestimating the de-projected lengths leading to smaller periods. To test if that is the case, we fix the inclinations of all the quasar knots to 5\textdegree, which is highly unlikely, and re-plot them in Figure \ref{fig:knot_timescales}b. The distributions now become wider with peaks above $\gtrsim$10$\textsuperscript{5}$~yrs for $\Gamma=$1.18, 1.5. However, it still peaks at about 3$\times$10$\textsuperscript{4}$~yrs for $\Gamma=3$. That means if the radio knots represent slow-moving blobs in the jet produced by modulated jet-activity, they can at most be mildly-relativistic ($\Gamma\lesssim1.5$). This limit is consistent with the bulk Lorentz factors derived from radio data \citep{wardle1997fast,mullin2009bayesian}, and also with the upper limits on proper motions ($\Gamma<2.9$) of the kpc-scale optical knots in 3C 273 \citep{meyer2017proper}. Moreover, if the X-ray emitting plasma is roughly at rest in the blob (knot) frame, it questions the large bulk Lorentz factors invoked to explain the observed X-ray emission in one-zone models.

\subsubsection{Offsets vs. distance from the core}
\begin{figure}
    \gridline{
        \fig{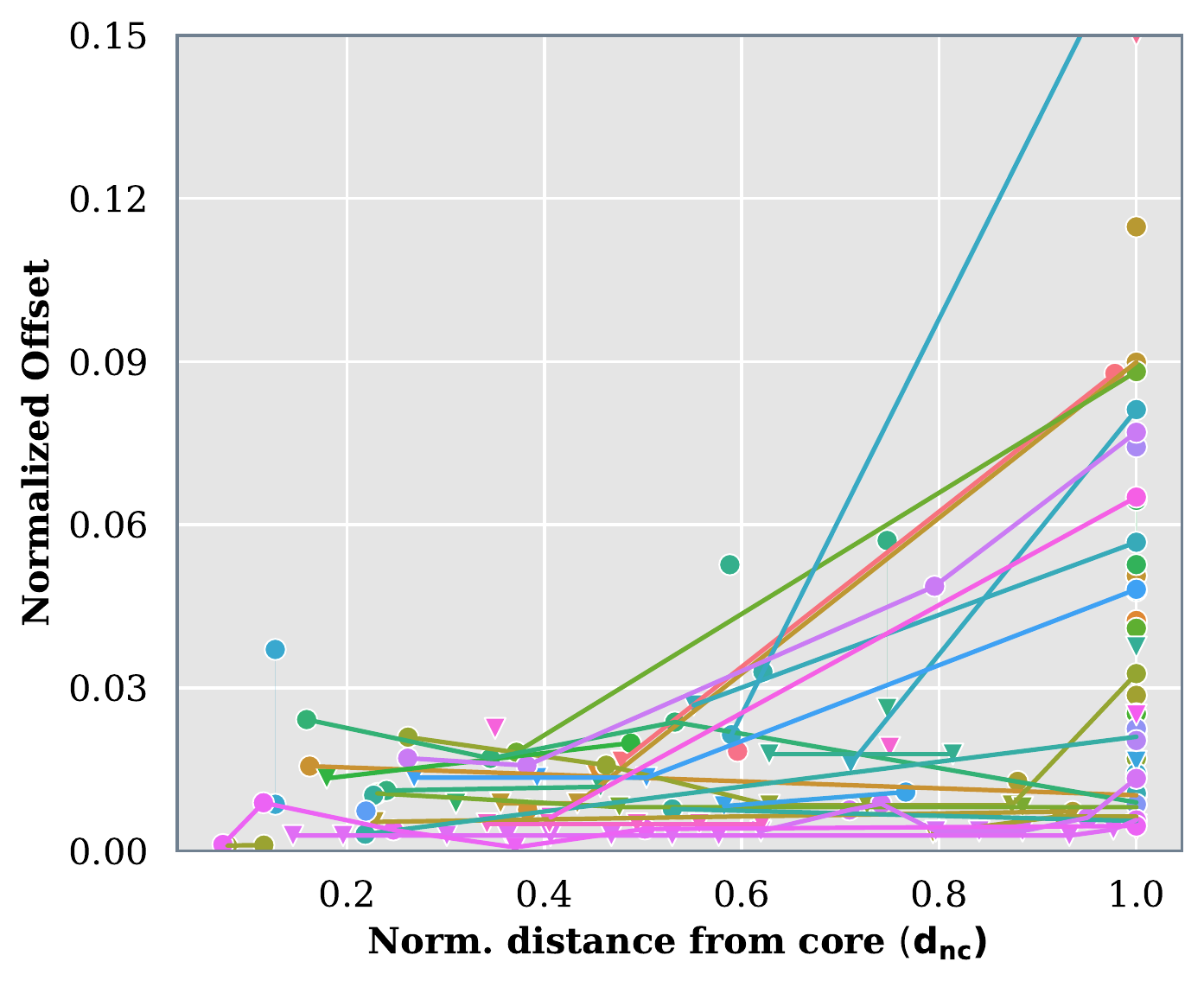}{0.5\textwidth}{(b)}
    }
    \caption{Normalized offsets in knots and hotspots of FR II-type jets plotted against the normalized distance from the core, showing a loose trend of offsets increasing with distance. The moving-knot model can produce this increase if a forward-moving shock determines the radio peak in the blob.  
     \label{fig:offsets_v_d_from_core}}
\end{figure}
Figure \ref{fig:offsets_v_d_from_core} shows offsets in knots and hotspots (discussed in section\ref{subsec:hotspots_offsets}) plotted against distance from the core, both normalized with the total length of the jet. Each jet is assigned a different color, and Xf-type and Co-s-type components are indicated with circles and triangles, respectively. Similar to \papri~we only observe a loose trend of offsets increasing with distance from the core, even with a larger sample. In 16 of 24 jets with at least one knot and a hotspot, the offset in the hotspot exceeds the maximum offset in knots; the count increases to 20 if we consider $\delta$-corrected knot-offsets (see section \ref{subsec:hotspots_offsets}). Suppose the offsets increase with distance from the core, and environmental effects presumably (e.g., jet-deflection) modify this trend. In that case, the radio peak must be moving further away from the X-ray peak as observed in the resolved knots of 3C 353 \citep{kataoka2008chandra}, with intrinsically larger offsets in hotspots. This trend can be reproduced in the moving-knot model if a forward shock propagating within the blob establishes the radio peak. As a result, the radio peak would move further away from the blob's upstream edge with time, increasing the offset. Despite the qualitative match of the moving-blob model with several observed properties of jets, further theoretical investigations and more advanced numerical simulations are clearly necessary to confirm this and other speculations.
\subsection{Flux Ratio\label{subsec:flux-ratio}}
The observed offsets challenge one-zone models and clearly indicate the need for multi-zone models. However, offsets alone are insufficient to establish the exact emission mechanism in multi-zone models. One way to differentiate between synchrotron and IC/CMB modes is by studying X-ray/radio flux ratios. Because of the differences in beaming patterns, IC/CMB mode experiences enhanced beaming at smaller viewing angles \citep[e.g.,][]{dermer1995beaming,harris2002c} than the synchrotron radiation. Hence, closely aligned knots may show larger flux ratios than their misaligned counterparts. Moreover, due to the stronger dependence of IC/CMB on the Doppler factor compared to synchrotron, in a simplified scenario, the flux ratio may also smoothly decline along a decelerating jet. Below we test for any such systematic trends in the flux ratios of FR II-type knots in our sample.



\begin{figure*}[h!]
    \epsscale{0.6}
    \plotone{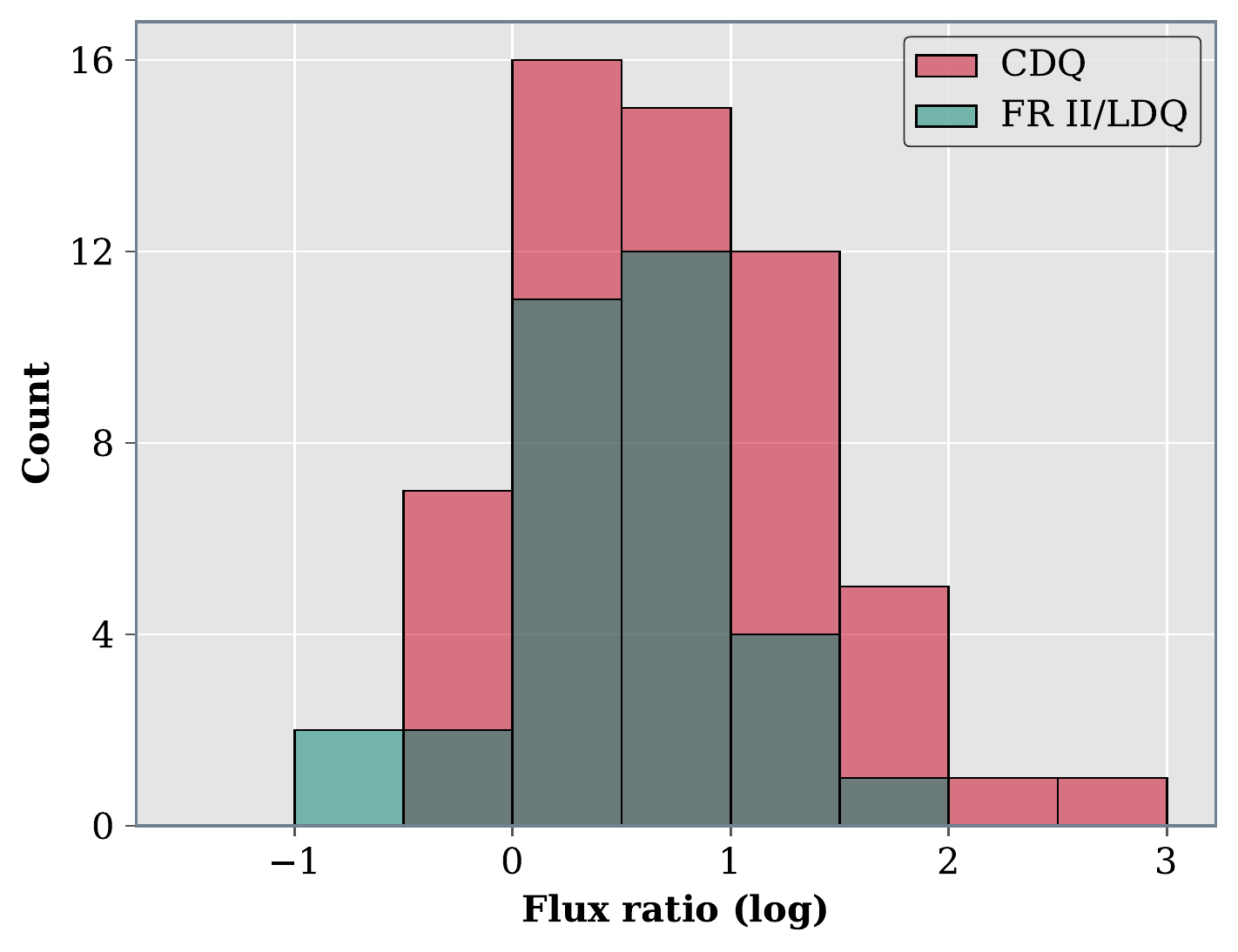}
    \caption{Histograms for log-scaled flux ratios of knots in CDQs and LDQ/FR II sources. A 2-sample AD test between the two groups reveals no statistically significant difference, suggesting that IC/CMB's contribution to the X-ray emission may be minimal.}
    \label{fig:cdq_vs_ldq_fr2_flux_ratio}
\end{figure*}
\subsubsection{Test for differences in flux ratio distributions of aligned and misaligned sources}
We first test for any differences in flux ratios of knots by dividing them into two groups. One with knots from CDQs, which are commonly believed to be viewed at small angles to our line of sight, and the other from their misaligned counterparts, the FR IIs and LDQs. Under this division, we might expect a significant difference in flux ratios between these two groups for simple one-zone IC/CMB mode of X-ray emission. Table \ref{table:offset-results} lists the X-ray flux density, evaluated at 1 keV, and the radio flux density along with its respective frequency for individual sources.

Figure \ref{fig:cdq_vs_ldq_fr2_flux_ratio} shows the flux ratio distributions for the two groups on a log-scale. A {\sl 2}-sample AD test is unable to detect a difference between the two distributions ($p$-value$>$0.25) at a high significance. Here the null hypothesis states that the two distributions are drawn from a single population. However, a large imbalance between the number of samples in each group (CDQ: 64, LDQ/FR II: 32) results in low statistical power [$<$0.3 \citep[see Table 1 in][]{razali2011power}]. Put another way, this test has a $>$70\% chance to incorrectly accept the null hypothesis even when there is an intrinsic difference. At least twice the number of present observations of knots in both the groups are required to confirm the presence (or absence) of any flux ratio differences with high statistical power.

\subsubsection{Evolution of flux ratio along the jet}

\begin{figure*}
    \gridline{
        \fig{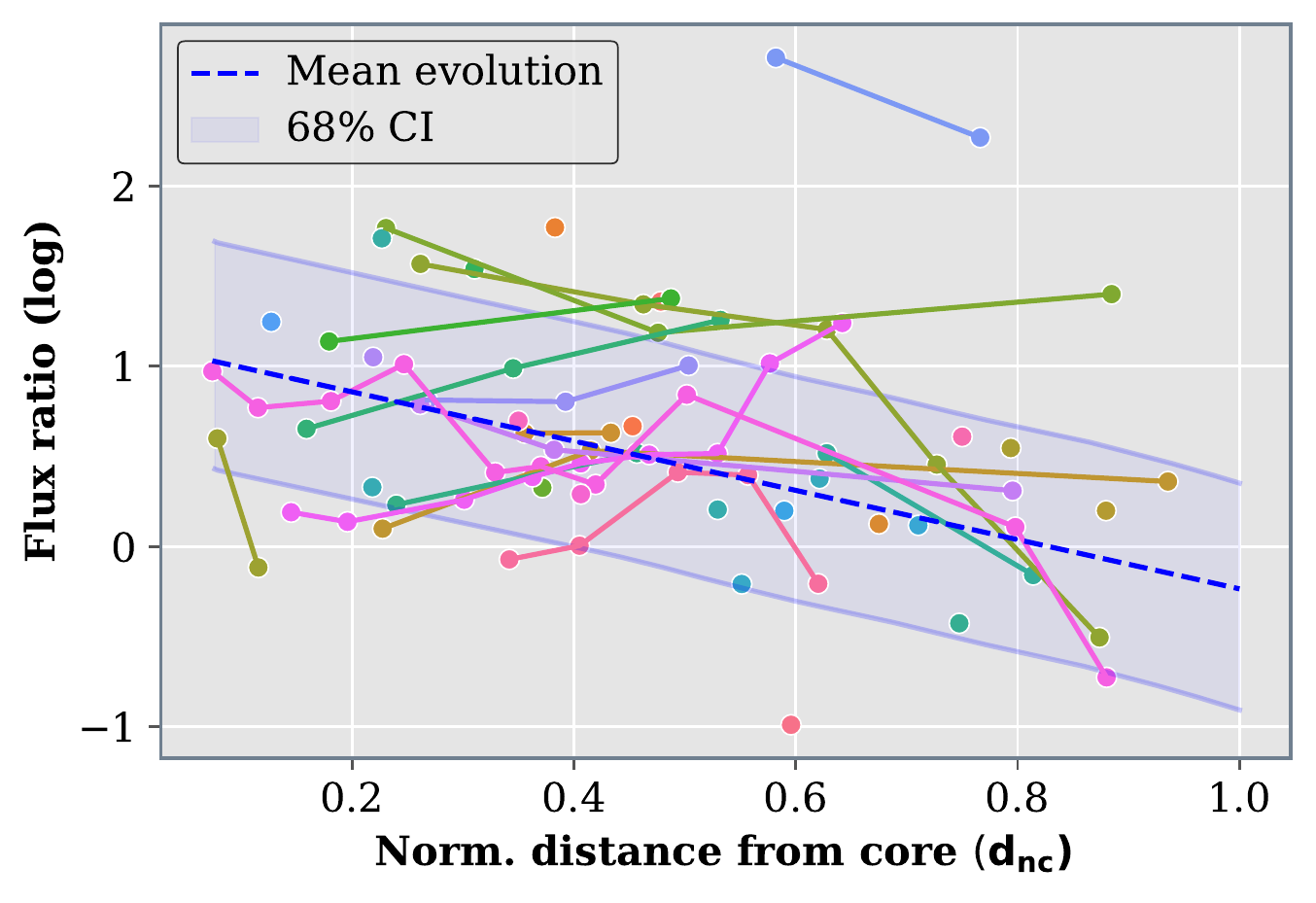}{0.5\textwidth}{(a)}
        \fig{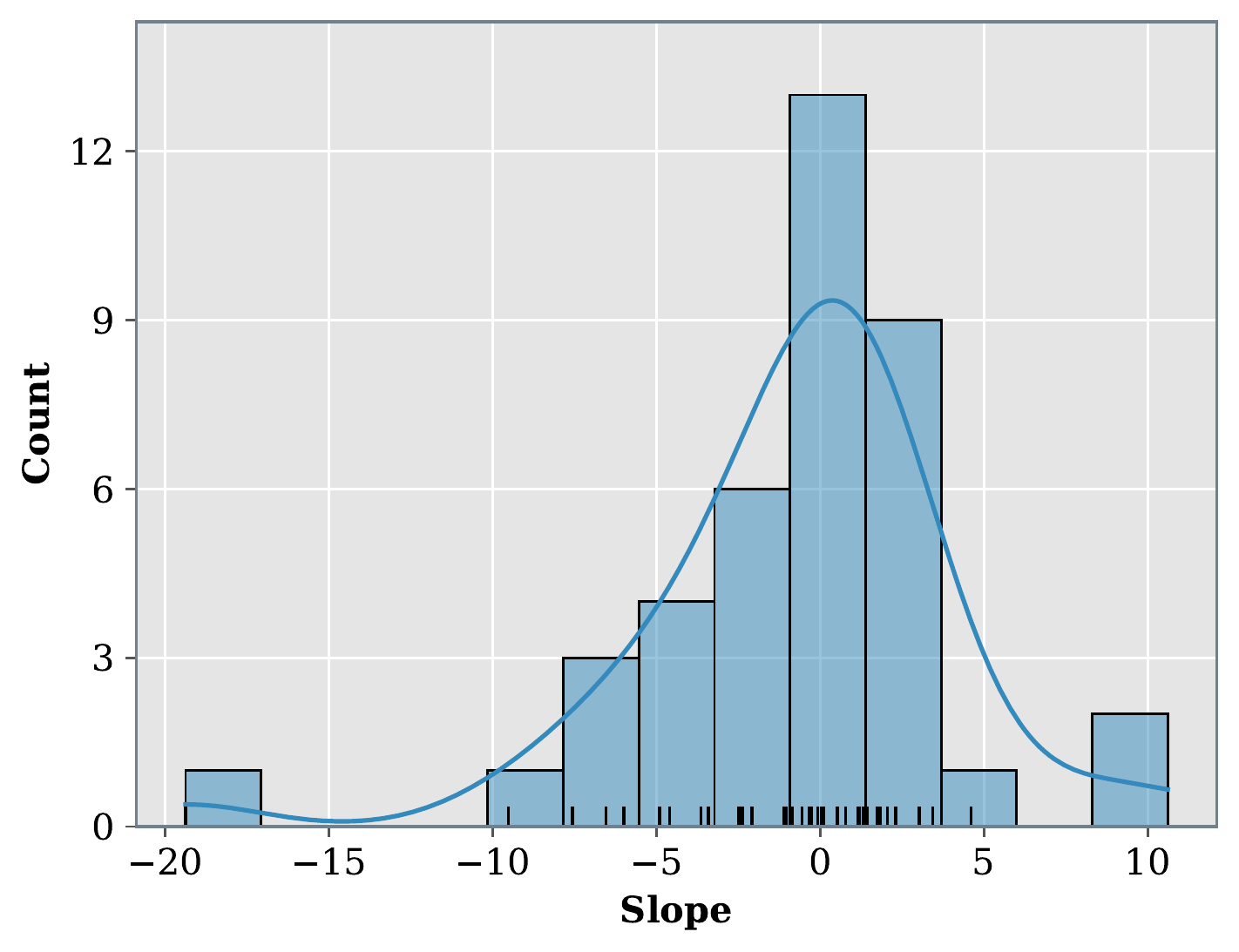}{0.5\textwidth}{(b)}
    }
    \caption{Trends in flux ratios. (a) shows flux ratio plotted against normalized distance from the core. Components from each jet are assigned the same color and joined with a solid line. The blue-dashed line shows the linear fit, which describes a general trend where the flux ratio decreases along the jet. The blue-shaded region indicates its 68\% confidence interval. (b) shows the histogram of slopes for each line segment plotted in (a), with a peak near zero. Only the slopes for 16 out of the 34 segments lie below zero, indicating the declining trend is only marginally significant.}
    \label{fig:flux_ratio_vs_dist}
\end{figure*}

We next search for any trends associated with the flux ratio. To understand the general evolution of flux ratio along X-ray jets while accounting for the wide range of orientations and jet-lengths in our sample, we consider the relation between flux ratio and distance from the core, normalized by the jet's total length. The left panel in Figure \ref{fig:flux_ratio_vs_dist} shows the flux ratio plotted against the normalized distance, and the data points for individual jets are joined with solid lines. We perform a linear regression on the flux ratios of knots and obtain the following equation:
\begin{equation}
    \log_{10} R=1.14^{+0.13}_{-0.13}-1.37_{-0.17}^{+0.17}\times d_{nc}
    \label{eq:flux_ratio_fit}
\end{equation}
where $R$~is the flux ratio and $d_{nc}$~is the normalized distance from the core. The uncertainties correspond to the 1$\sigma$~values. The blue dashed line indicates the fitted model in Figure \ref{fig:flux_ratio_vs_dist} and the blue-shaded region its 68\% confidence interval (CI). The slope of Eq.~\ref{eq:flux_ratio_fit} indicates a mean flux ratio decline by a factor of $\sim23_{-8}^{+11}$ between the inner and outer regions of the jet, which is slightly higher than what is observed in representative FR II sources like 3C 353 \citep[][]{kataoka2008chandra}. To estimate the significance of this trend, we measured the slope of each line segment (59 total segments) in Figure \ref{fig:flux_ratio_vs_dist}a. Their histogram, shown in Figure \ref{fig:flux_ratio_vs_dist}b, describes a unimodal distribution peaking at $\approx$0 with only 21 out of 40 (52\%) segments lying below zero, indicating the measured declining trend is only marginally significant. It is possible, however, that the sample is contaminated by sources with non-jet sources of X-ray emission (for example those with shock-heated gas), which will modify the flux ratios. Deeper observations of jets would allow us to confirm via spectral fitting whether the X-rays are likely from the jet or another source.

\subsubsection{Does a decelerating jet explain a decreasing flux ratio?}
\begin{figure}
    \epsscale{0.6}
    \plotone{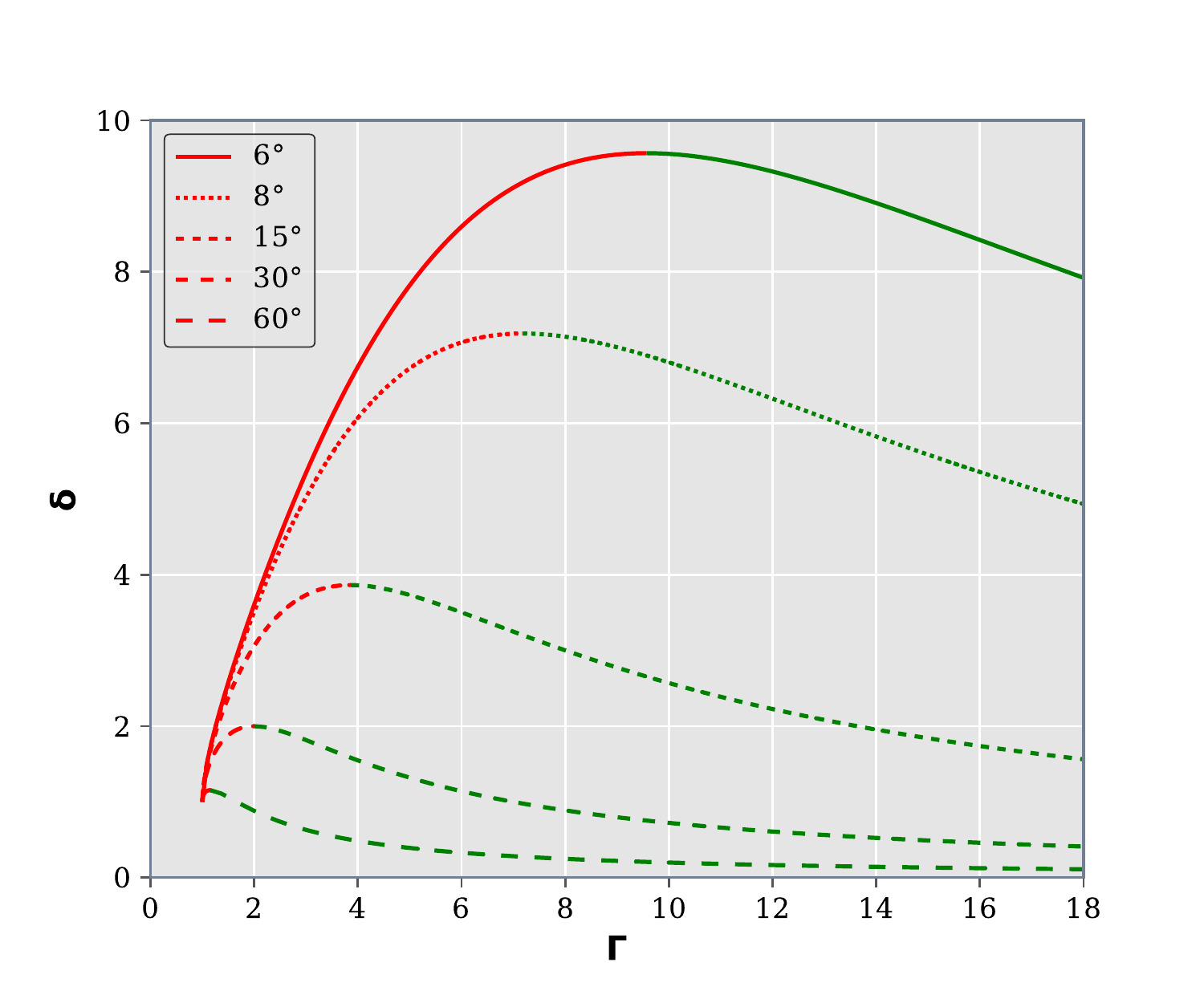}
    \caption{$\delta$ vs. $\Gamma$ for different inclination angles. A jet decelerating in the red zone of each curve experiences reduced $\delta$ while it rises in the green zone.\label{fig:delta_vs_gamma}}
\end{figure}

Under the IC/CMB scenario for X-ray production, a decreasing $\delta$ along the jet could result in a declining X-ray/Radio flux ratio since IC/CMB has a stronger dependence on $\delta$ than synchrotron due to differences in their beaming patterns. However, this is only true for  jets with small inclination angles and fails to hold for misaligned jets. Specifically, for a jet aligned at an angle $\theta$~to our line of sight,  deceleration can lower $\delta$ only when $\Gamma$~falls below $\Gamma<1/\sin\theta$ while the opposite happens when it remains above $\Gamma>1/\sin\theta$. To see this visually, in Figure \ref{fig:delta_vs_gamma} we have plotted the theoretical curves of $\delta$ vs. $\Gamma$~for different alignment angles ($\theta$). Any deceleration decreases $\delta$ in the red zone while the converse happens in the green zone.

For a jet with $\theta$=6\textdegree, if a knot requires, for example,  $\Gamma=$15 to explain the observed X-ray flux, the jet must decelerate to below $\Gamma\lesssim 5$~at the next knot before any deceleration can reduce the flux ratio. However, the cold matter required to decelerate the jets by such large values can sometimes be implausibly high \citep{hardcastle2006testing}. Although a smaller deceleration in the red zone can reduce $\delta$ by an equal amount, the range of $\Gamma$~within these zones is generally insufficient to reproduce the observed X-ray flux under IC/CMB. Furthermore, traditional one-zone IC/CMB interpretations of X-rays assume $\theta$ remains constant throughout the jet's extent. At $\Gamma=10$, $\delta$ decreases from about 10 to 7 even with a small change in $\theta$, for instance, from $\theta$=4\textdegree~to~$\theta$=6\textdegree, which we may falsely perceive as deceleration. Hence, fine-tuned angles are unreasonable given the absence of any \textit{straight} jets in our entire sample. Finally, in the case of highly-misaligned jets (e.g., $\theta$=60\textdegree~curve in Fig.~\ref{fig:delta_vs_gamma}) where any deceleration practically always increases $\delta$, no evidence for smoothly rising X-ray and radio fluxes along such jets has been reported so far. The absence of such observations independently rules out simple one-zone IC/CMB models and warrants the need for a multi-zone emission model, as previously suggested in different contexts \citep[e.g.,][]{stawarz2004multiwavelength,jester2006new,2007ApJ...657..145S,kataoka2008chandra}.


\subsection{Is there a wavelength-based beaming?}
The previous two sections have evaluated flux ratios in the context of a one-zone IC/CMB model. Alternatively, to reconcile the small bulk Lorentz factors inferred from radio observations with large values required in IC/CMB models, a ``spine-sheath'' structure in the jet, with an X-ray-emitting fast spine surrounded by a radio-emitting slow sheath, has been suggested \citep[e.,g][]{hardcastle2006testing,jester2006new}. Although this model explained the observed X-ray emission in a few jets, it sometimes required unrealistically large bulk Lorentz factors \citep[e.g.,][]{jester2006new}. In a different context, \citep{10.1093/mnras/stt1676} find a fast spine ($\Gamma\gtrsim10$) is necessary to explain similar injection spectral indices during different jet-production  episodes of nearby double-double radio galaxies (DDRG).

These spine-sheath models were only evaluated in the case of a few jets, and it is unclear in general whether X-rays and radio can come from a fast spine and a slow sheath, respectively.
Below we test whether the bulk Lorentz factors of X-ray and radio-emitting plasmas differ using the luminosity distributions for resolved knots.
\begin{figure*}
    \gridline{
        \fig{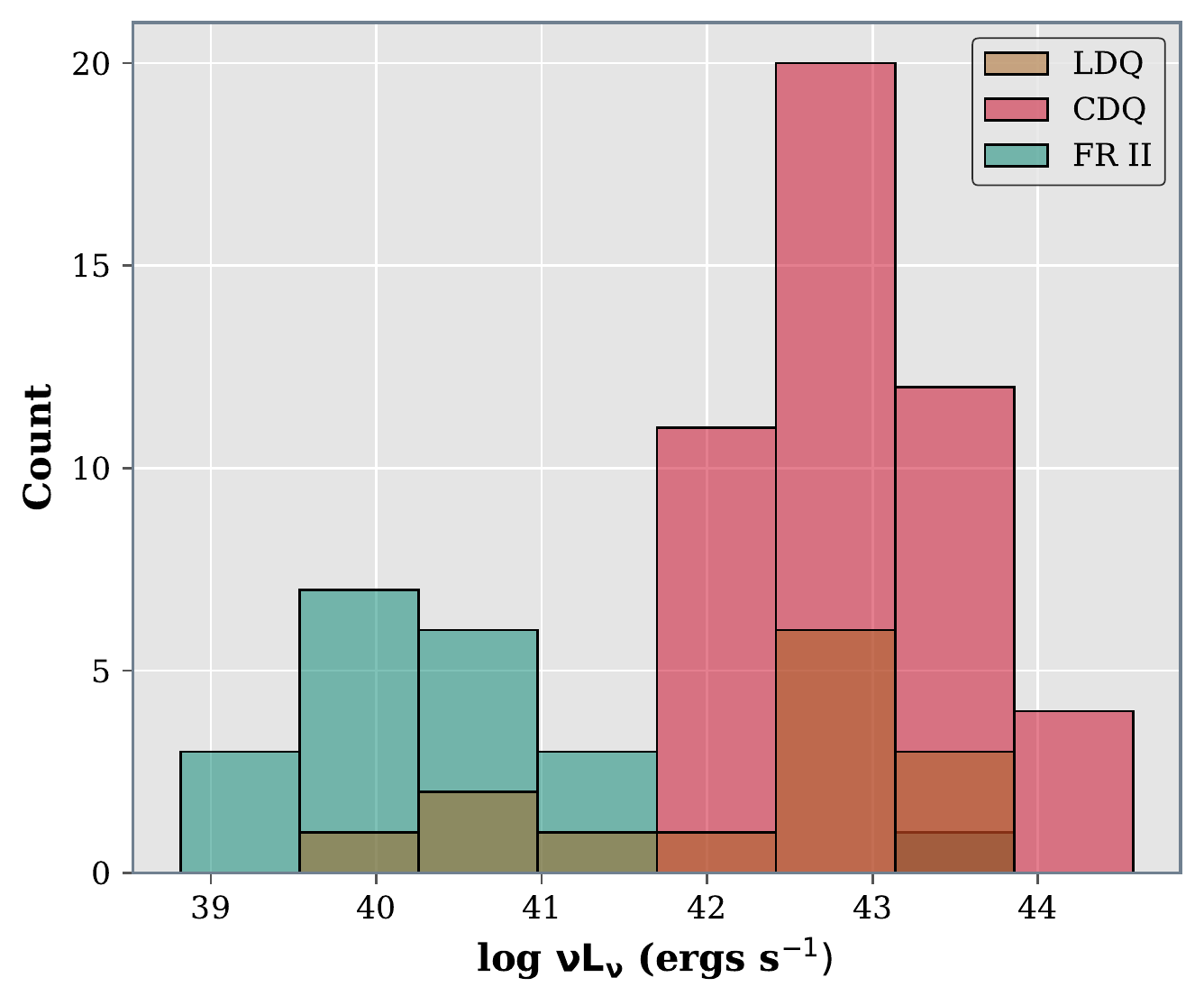}{0.5\textwidth}{(a) X-rays}
        \fig{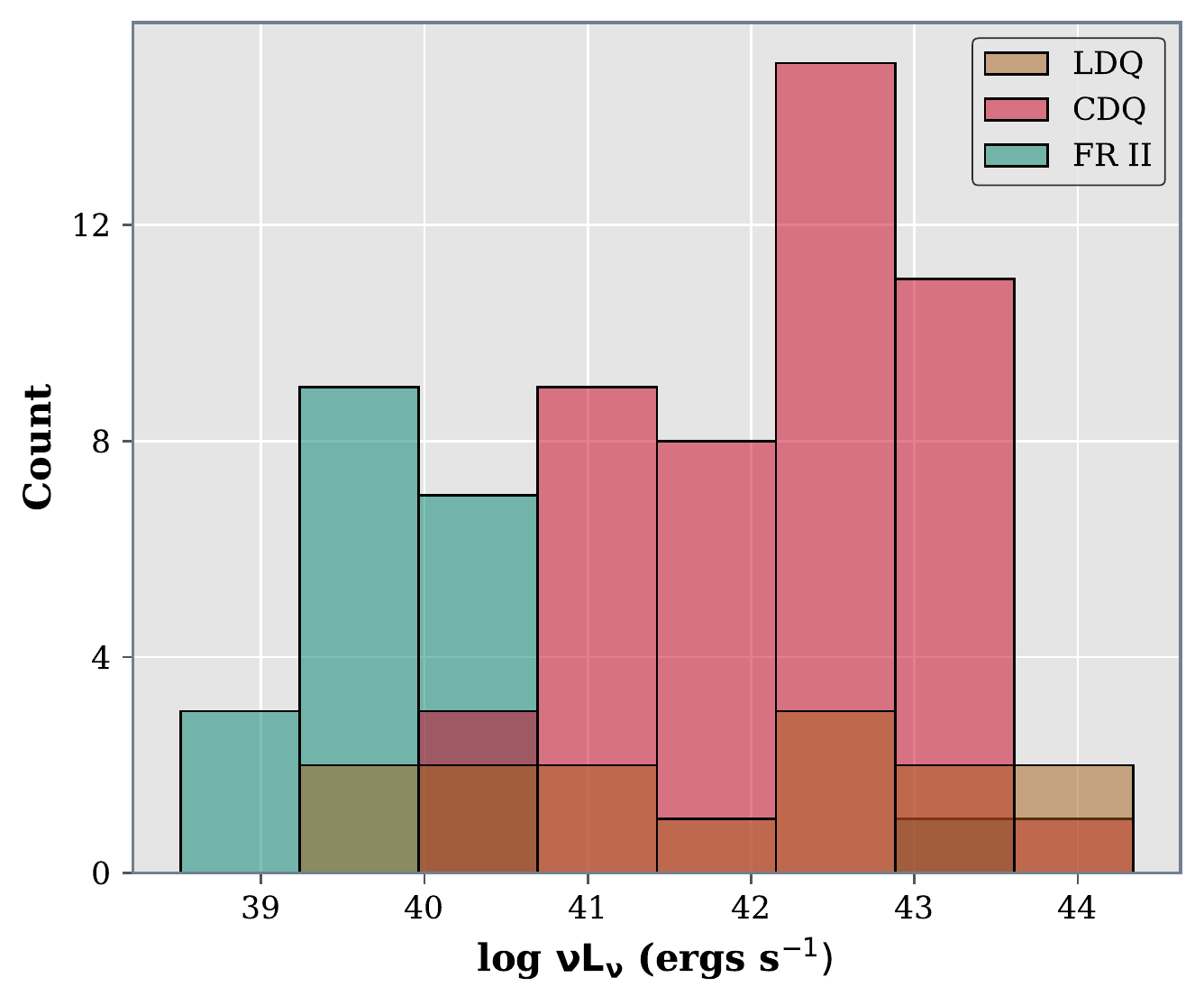}{0.5\textwidth}{(b) Radio}
    }
    \gridline{\fig{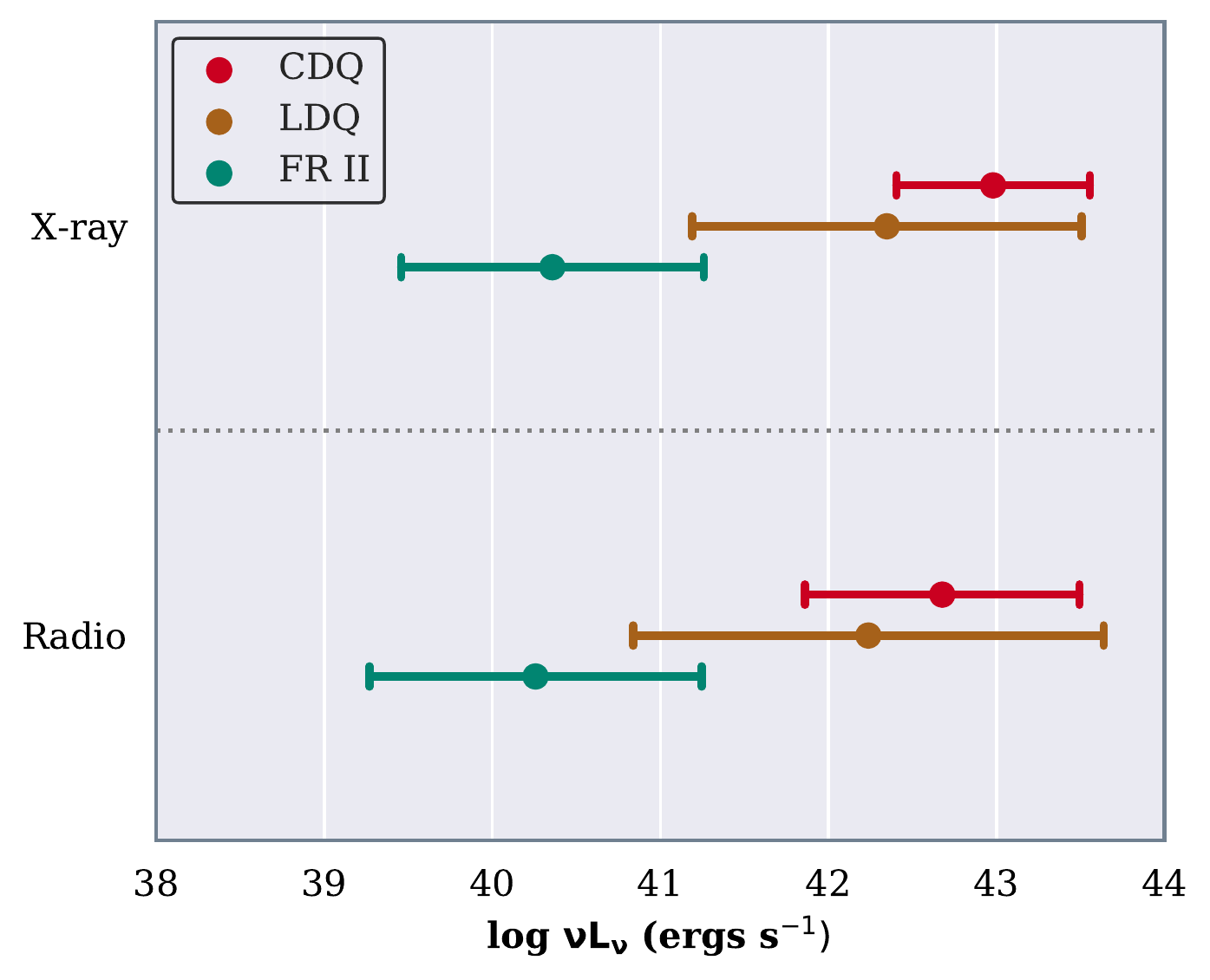}{0.5\textwidth}{(c) }}
    \caption{Class-wise histograms for (a) X-ray and (b) radio (right panel) luminosity of knots in FR II-type sources. (c) shows the mean values of each class at each wavelength, where the radio luminosity values are nudged to the right by 0.4 units (multiplied by 1.26) for easier comparison. The similarity in inter-class distance between the distributions (measured using the AD test statistic) at X-ray and radio wavelengths suggests that bulk Lorentz factors for both the emitting plasmas are similar. \label{fig:xray_radio_lum_dist}}
\end{figure*}

The observed luminosity ($L_\nu^{obs}$) from an isotropically emitting optically thin source with rest-frame luminosity $L_\nu^{in}\propto\nu^{-\alpha}$, moving with a bulk Lorentz factor $\Gamma$~along an angle $\theta$~to our line of sight is given by \citep[e.g.,][]{blandford1979relativistic,jester2008retardation}
\begin{equation}
    L_\nu^{obs}=
    \begin{cases}
        L_\nu^{in}\delta^{3+\alpha}(1+z)^{(1-\alpha)}           & \text{(Discrete blob)}   \\
        L_\nu^{in}\delta^{2+\alpha}(1+z)^{(1-\alpha)} & \text{(Continuous flow)}
    \end{cases}
    \label{eq:lum_cases}
\end{equation}
Because radio knots show extended structures in a wide range of redshifts \citep[e.g.,][]{kataoka2008chandra,2007ApJ...657..145S}, we approximate them as cylindrical blobs and use the first expression in Eq. \ref{eq:lum_cases} for radio in the following discussion. However, we consider both expressions for X-rays due to uncertainty in their emission geometry.

Figure \ref{fig:xray_radio_lum_dist} shows the X-ray and radio luminosity distributions in the left and right panels, respectively, and are color-coded based on their spectral class. The bottom panel shows the mean and 1$\sigma$~limits for each class at each wavelength. The radio luminosity values are multiplied by a factor of 1.26 to enable easier comparison. Let us first consider the distributions in radio. The mean luminosity for FR II knots is $\sim$10$^{40}$ ergs~s$^{-1}$ and $\sim$10$^{42}$ ergs~s$^{-1}$ for CDQ and LDQ knots, both indicating a difference of about 2 decades between FR II and quasar knot populations. If we assume all the FR-II knots align at 60\textdegree~to our line of sight~and all the quasar knots at 5\textdegree, the luminosity of an FR II knot with $\alpha=0.8$~and $\Gamma=1.8$ when it aligns at 5\textdegree, i.e., when it is \textit{observed} as a quasar knot,  increases by about 2 orders of magnitude. That means if the rest-frame luminosity of all the radio knots belong to a single parent population, beaming effects alone are sufficient to explain most of the observed dispersion in luminosity distributions. We quantify the magnitude of its effect as the sum of AD test-statistic measured between all the two-class combinations from the three spectral classes--the test statistic increases with a rise in the population's average bulk Lorentz factor. We measure a value of 34.96$\pm$0.35 for the current sample, estimating errors via bootstrapping and assuming 30\% error on the flux ratios.

On the other hand, the mean X-ray luminosity for FR II knots is about $\sim$10$^{40.5}$ ergs~s$^{-1}$, while it is about $\sim$10$^{42.5}$ ergs~s$^{-1}$ and $\sim$10$^{43}$ ergs~s$^{-1}$ for CDQ and LDQ knots, respectively. These values place the mean luminosity difference between FR II and quasar knots in the range of 2-2.5 decades. Similar to radio, the X-ray luminosity of a knot with  $\alpha=0.8$, $\Gamma=2.1$, when approximated as a discrete blob, increases by $\sim$2.5 orders of magnitude when its alignment changes from 60\textdegree~to 5\textdegree; continuous flow-like geometry requires $\Gamma\sim 1.8$. The magnitude of the beaming effect as quantified by the AD test-statistic for the X-ray knot population is 36.35$\pm$0.65 and differs from the radio population's statistic 34.96$\pm$0.35 only at the 1.1$\sigma$-level. That means the bulk Lorentz factors of X-ray and radio-emitting plasmas must be similar contrary to the large differences required in spine-sheath IC/CMB models.

The conclusions drawn here are independent of the assumed orientations. For example, if we assume quasar knots align at 15\textdegree~and FR II-type knots at 90\textdegree, $\Gamma\sim 1.5$~is required in the radio while it becomes $\Gamma\sim 1.5$~(discrete blob) and $\Gamma\sim 1.75$~(continuous flow) in the X-rays.

We emphasize that the bulk Lorentz factors are similar only for the emitting plasmas and not the jet material that carries it. For example, the shocked plasma in the moving-knot model can be at rest in the slow-moving blob's frame, while the outer flow producing the shock can be much faster. Hence, this result is still consistent with the requirement for a faster spine in DDRGs.

The mean redshift for FR II knots is 0.13$\pm$0.31 while it is 0.81$\pm$0.41 and 0.75$\pm$0.67 for CDQ and LDQ knots, respectively. Our analysis ignores any redshift-related enhancements in the intrinsic X-ray luminosity of knots. For instance, an IC/CMB component could become comparable to an already existing synchrotron component at higher redshift \citep[e.g.,][]{2019MNRAS.482.2016Z}, further reducing the beaming requirement and the bulk Lorentz factor. Finally, if the X-ray/radio bulk Lorentz factors and spectral indices are similar, and their emitting their emission geometries are similar, the flux ratio only weakly depends on beaming ($R\propto \delta^{\alpha_x-\alpha_r}$). That means differences in the X-ray/radio intrinsic luminosity ratio must provide the major contribution to the observed 2-3 orders of magnitude spread in the flux ratio's distribution (see Fig. \ref{fig:cdq_vs_ldq_fr2_flux_ratio}). This spread is consistent with the observed offsets that indicate spatially separate X-ray/radio-emitting regions and, therefore, independent plasma conditions, producing a large scatter in intrinsic luminosity ratios.
\subsection{Offsets in hotspots\label{subsec:hotspots_offsets}}
\begin{figure}
    \gridline{
        \fig{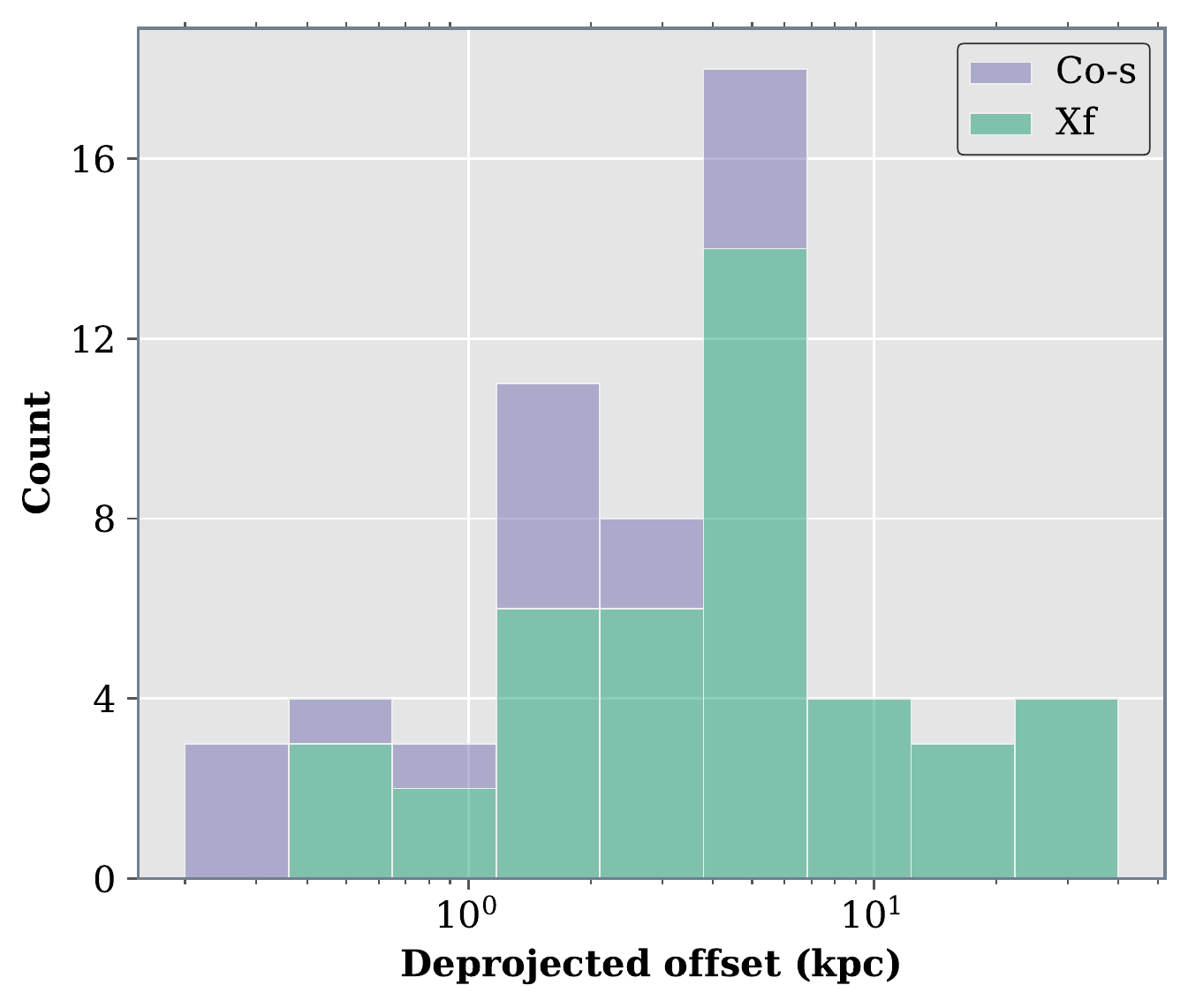}{0.5\textwidth}{(a)}
    }
    \caption{Histograms of de-projected Xf-type offsets in hotspots, stacked with Co-s-type hotspots assuming a 0.15\as~angular offset. Unlike offsets in knots, the Xf-type and Co-s-type hotspots may indicate different structures (see section \ref{subsec:hotspots_offsets}).
     \label{fig:frI_hotspot_offsets}}
\end{figure}

Although we mainly focus on studying offsets in knots of FR II-type sources, we briefly discuss the offsets in hotspots for completeness. Figure \ref{fig:frI_hotspot_offsets} shows the histograms of de-projected Xf-type offsets is hotspots, stacked with Co-s-type offsets by assuming a 0.15\as~angular offset. The Xf-type offsets are also the predominant type of offsets in hotspots, similar to the knots in FR II-type jets. 43 out of the 63 hotspots in our sample show Xf-type offsets, ruling out simple one-zone IC (CMB or SSC) or synchrotron models, while 17 show Co-s-type offsets. The remaining three hotspots also show offsets, however, we cannot determine the type of offset, i.e., Xf or Rf-type, due to the uncertainty in the direction of their jets at this point. Hence, we indicate them as ``Amb'' offset-type in table \ref{table:offset-results}.


Upper limits on offsets in Co-s-type hotspots, unlike in the case of knots, are distributed on the left side of the Xf-type's roughly unimodal distribution. Projection and distance-related effects may be reducing any offsets in Co-s-type hotspots to below the detectable level. A 2-sample AD test also reveals no difference ($p$-value=0.25) between the flux ratios of the two types of hotspots, suggesting a possible intrinsic structural similarity between them. However, the redshift distributions for both kinds of hotspots are similar with means of 0.71$\pm$0.55 and 0.65$\pm$0.37, respectively; each offset-type group also contains roughly equal proportions of FR II and quasar jets. These common traits suggest projection and distance-related effects are probably minimal, and possibly different mechanisms produce Xf and Co-s-type hotspots.

Interestingly, the mean de-projected offset in hotspots and knots are similar ($\approx$4 kpc, see sec. \ref{subsec:deproj-offsets}). Moreover, a 2-sample AD test reveals no difference between their distributions ($p$-value $> 0.25$), suggesting a possible similarity in their multi-wavelength morphology. However, with the current physical understanding, it is unclear why the two offset distributions are similar when we expect entirely different processes to produce them. One possibility is that the orientations of knots and hotspots differ in most if not all the jets due to the larger likelihood of jets getting deflected in the turbulent hotspot region. To test this possibility, we modified the alignment angle for each hotspot by adding a random angle between -15\textdegree~and+15\textdegree~while keeping the knot-angles unchanged and re-performed the AD test 10,000~times. None of the cases reported a $p$-value less than 0.14, ruling out the jet-deflection scenario. Alternatively, if we use the $\delta$-corrected knot-offsets (with $\Gamma=1.25$), we find the expected difference at the 0.2\% significance level (although with low statistical power), further supporting the moving-knot model.

\section{Summary and Conclusions}
We used a statistical tool called LIRA to analyze a sample of 77 \chdr-detected X-ray jets to detect X-ray/radio offsets in individual features. 
We required an offset to be at least 1$\sigma$-level above 0.15\as~to be statistically significant. Our analysis of 164 components revealed 94 offsets with this criterion, where 58 of these are newly reported. 
Of the 114 FR II-type knots considered in this study as well as \papri, we find 48 X-ray-first (Xf-type) offsets, 8 Radio-first (Rf-type) offsets, and 56 knots, which were co-spatial to within 0.15$''$ (Co-s-type). Considering hotspots (bright terminal knots) separately, out of 69 considered in this paper and \papri together, 49 are Xf-type offsets, 20 are Co-s-type offsets, and none are Rf-type. In the case of FR I-type sources, in total, we analyzed 30 knots and detected 20 Xf-type, 5 Rf-type, and 8 Co-s-type knots.

The predominance of Xf-type offsets in FR II-type knots questions the applicability of one-zone models. Although Co-s-type knots occur in similar numbers, they are mostly found in closely-aligned and higher redshift jets. A 2-sample AD test reveals no difference between the distributions of flux ratio and X-ray spectral index for Xf-type and Co-s-type knots. These considerations suggest both kinds of knots are intrinsically similar, and projection and distance effects may be reducing any observable offsets below the detectable level. 

The distribution of offsets in FR II-type knots, de-projected using a simple orientation scheme based on the spectral classification of sources, roughly peak at $\sim$4 kpc; the Co-s-type knots with assumed angular offset of 0.15\as~also shows a similar distribution. These offsets are also similar to the sizes of low-redshift radio knots, suggesting most if not all knots possess a similar morphology with an extended radio and compact X-ray structures. Adopting a simple `moving-knot' model further reduces the effective mean offset to $\sim$2 kpc and inter-class dispersion in the detected offsets. 

We speculate that a modified version of the moving-knot model, previously proposed by \citet{kataoka2008chandra}, can explain the observed offsets and multiple other observed properties of jets. The bulk Lorentz factor of radio knots, as implied by inter-knot distances, can at most be mildly relativistic ($\lesssim 1.5$) even with extreme jet-angle constraints. This speed is consistent with the limits derived using radio observations \citep[e.g.,][]{mullin2009bayesian}. Finally, we find a loose trend of offsets increasing with distance from the core, which can be explained with the moving-knot model by invoking a forward shock in the blob.

We find no difference between the flux ratio distributions of closely aligned (CDQs) and misaligned (LDQs and FR IIs) sources, indicating IC/CMB may be less likely to be the dominant X-ray emission mechanism. The flux ratio appears to decline with distance from the core in general (with marginal significance). Although a decelerating jet with IC/CMB can explain this trend, it generally requires unrealistic conditions. Furthermore, we measure the beaming effect on the dispersion of knot luminosity distributions as the sum of AD test statistic computed between the distributions of CDQ, LDQ, and FR II classes. Here we assume the intrinsic luminosity of all the knots in X-rays and radio belong to the same individual parent populations. The test statistic differs between X-rays and radio only at the 1.1$\sigma$-level, suggesting similar bulk Lorentz factors for the emitting plasmas.  

These results indicate the need for two or multi-zone emission models that, unfortunately, increase the number of model parameters but provide more flexibility in explaining the observations. We summarize below the general properties of knots that future models must account for:
\begin{enumerate}
    \item The X-rays peak and decay before the radio in most knots, requiring spatially separated emitting regions.
    \item The radio appears to be more extended than the X-ray emission, which is generally unresolved in \emph{Chandra} observations, and both the emitting plasmas have similar bulk Lorentz factors.
    \item There is evidence that the radio-emitting plasma occupies a surface layer surrounding the jet instead of the entire cross-section \citep{swain1998internal}. On the other hand, the X-rays appear in the central regions of transversely-resolved jets  \citep[e.g.,][]{kataoka2008chandra}.
    \item Radio polarization measurements indicate magnetic fields parallel to the jet in the outer regions while perpendicular to the jet in the inner regions \citep[e.g.,][]{swain1998internal,bridle1994deep}.
\end{enumerate}

Hotspots predominantly show Xf-type offsets (43 out of 69). Although we find no difference between the flux ratio distributions of Xf-type and Co-s-type hotspots, hotspots in both groups are located at similar redshifts and comprise similar mixtures of aligned and misaligned sources. These similarities suggest that projection or distance-related effects are unlikely to reduce any offsets to produce Co-s-type hotspots, and the two types of hotspots possibly have different structures. The distributions of Xf-type offsets in knots and hotspots of FR II-type sources are similar; however, they differ for the moving-knot scenario.






\pagebreak
\appendix
\begin{longrotatetable}
\begin{deluxetable*}{lcccclccl}
    \tablecaption{List of all \textit{Chandra} detected X-ray jets\label{table:list_of_sources}}
    \tabletypesize{\scriptsize}
    \tablehead{
        \colhead{Name} &
        \colhead{IAU Name} &
        \multicolumn{1}{p{1.5cm}}{\centering R.A. (J2000)\\hh:mm:ss} &
        \multicolumn{1}{p{1.5cm}}{\centering decl. (J2000)\\dd:mm:ss} &
        \colhead{z} &
        \colhead{$\beta_{app}$\textsuperscript{a}}&
        \multicolumn{1}{p{0.75cm}}{\centering Scale\\(kpc/\arcsec) } &
        \colhead{Class\textsuperscript{b}}&
        \multicolumn{1}{p{3cm}}{\centering Reference }
    }
    \tablecolumns{9}
    \startdata
    \cutinhead{Sources analyzed in this work}
    \object{3C9} & 0017+154 & +00:20:25.0 & +15:40:54.7 & 2.012 &       & 8.57 &          LDQ &      \citet{2003MNRAS.338L...7F} \\
           \object{3C15} & 0034-014 & +00:37:04.0 & -01:09:08.5 & 0.073 &       & 1.40 &   FRI (NLRG) &      \citet{2003AA...410..833K}  \\
           \object{3C31} & 0104+321 & +01:07:25.0 & +32:25:45.0 & 0.017 &       & 0.34 &   FRI (LERG) &   \citet{hardcastle2002chandra}  \\
       \object{4C+01.02} & 0106+013 & +01:08:39.0 & +01:35:00.3 & 2.099 & 26.83 & 8.43 &          CDQ &      \citet{2011ApJ...730...92H} \\
           \object{3C47} & 0133+207 & +01:36:24.0 & +20:57:27.4 & 0.425 &  6.41 & 5.50 &          LDQ &      \citet{2004ApJ...612..729H} \\
   \object{PKS 0144-522} & 0144-522 & +01:46:48.6 & -52:02:33.5 & 0.098 &       & 1.86 &            Q &      \citet{2018ApJ...856...66M} \\
       \object{4C+35.03} & 0206+355 & +02:09:39.0 & +35:47:50.9 & 0.037 &       & 0.72 &    FR I (RG) &     \citet{2001MNRAS.326L...7W}  \\
    \object{PKS0208-512} & 0208-512 & +02:10:46.0 & -51:01:02.9 & 0.999 &       & 8.04 &          CDQ &      \citet{2005ApJS..156...13M} \\
          \object{3C66B} & 0220+427 & +02:23:11.0 & +43:00:31.2 & 0.022 &  1.03 & 0.43 &   FRI (LERG) &      \citet{2001MNRAS.326.1499H} \\
       \object{4C+28.07} & 0234+285 & +02:37:52.0 & +28:48:08.9 & 1.213 & 27.33 & 8.36 &          CDQ &            \citet{marshall2011x} \\
           \object{3C78} & 0305+039 & +03:08:26.2 & +04:06:39.3 & 0.029 &  0.13 & 0.58 &  FR I (LERG) &    \citet{harwood2012determines} \\
         \object{3C83.1} & 0315+416 & +03:18:16.0 & +41:51:27.9 & 0.025 &       & 0.50 &   FRI (LERG) &     \citet{2005ApJ...633..165S}  \\
           \object{3C88} & 0325+023 & +03:27:54.2 & +02:33:42.0 & 0.030 &       & 0.61 &  FR I (NLRG) &      \citet{2009ApJ...704.1586S} \\
     \object{PKS0405-12} & 0405-113 & +04:07:48.0 & -12:11:36.6 & 0.574 &       & 6.50 &          CDQ &       \citet{sambruna2004survey} \\
     \object{PKS0413-21} & 0413-210 & +04:16:04.0 & -20:56:27.5 & 0.808 &       & 7.54 &          CDQ &      \citet{2005ApJS..156...13M} \\
          \object{3C111} & 0415+379 & +04:18:21.0 & +38:01:35.8 & 0.049 &  8.42 & 0.95 & FR II (BLRG) &      \citet{2011ApJ...730...92H} \\
          \object{3C120} & 0430+052 & +04:33:11.0 & +05:21:15.6 & 0.033 &  9.08 & 0.65 &  FR I (BLRG) &              \citet{harris2004x} \\
          \object{3C123} & 0433+295 & +04:37:04.0 & +29:40:13.7 & 0.218 &       & 3.49 & FR II (LERG) &    \citet{hardcastle2001chandra} \\
    \object{PKS 0454-46} & 0454-463 & +04:55:51.0 & -46:15:58.7 & 0.858 &       & 7.70 &          CDQ &            \citet{marshall2011x} \\
     \object{PKS0605-08} & 0605-085 & +06:08:00.0 & -08:34:49.9 & 0.870 & 32.81 & 7.70 &          CDQ &       \citet{sambruna2004survey} \\
    \object{PKS0637-752} & 0637-752 & +06:35:46.0 & -75:16:16.8 & 0.651 & 11.40 & 6.90 &          CDQ &       \citet{chartas2000chandra} \\
     \object{B20738+313} & 0738+313 & +07:41:11.0 & +31:12:00.2 & 0.631 & 11.09 & 6.90 &          CDQ & \citet{siemiginowska2003chandra} \\
        \object{OJ +248} & 0827+243 & +08:30:52.0 & +24:10:59.8 & 0.939 & 24.67 & 7.90 &          CDQ &      \citet{2004ApJ...614..615J} \\
       \object{4C+29.30} & 0836+299 & +08:40:02.0 & +29:49:00.5 & 0.064 &       & 1.20 &       FRI RG &       \citet{sambruna2004survey} \\
          \object{3C207} & 0837+133 & +08:40:47.0 & +13:12:23.0 & 0.680 & 13.32 & 7.10 &          LDQ &       \citet{2002AA...381..795B} \\
          \object{OJ287} & 0851+202 & +08:54:49.0 & +20:06:30.6 & 0.306 & 20.06 & 4.48 &       BL Lac &      \citet{2011ApJ...729...26M} \\
    \object{PKS0920-397} & 0920-397 & +09:22:46.4 & -39:59:35.0 & 0.591 & 30.79 & 8.50 &          CDQ &     \citet{2005ApJS..156...13M}  \\
          \object{3C228} & 0947+145 & +09:50:10.8 & +14:20:00.6 & 0.552 &       & 6.62 & FR II (NLRG) &      \citet{2004ApJ...612..729H} \\
      \object{Q0957+561} & 0957+561 & +10:01:21.0 & +55:53:56.5 & 1.410 &       & 8.50 &          LDQ &  \citet{chartas2002constraining} \\
    \object{PKS1030-357} & 1030-357 & +10:33:08.0 & -36:01:56.8 & 1.455 &       & 8.53 &          CDQ &      \citet{2005ApJS..156...13M} \\
    \object{PKS1045-188} & 1045-188 & +10:48:07.0 & -19:09:35.7 & 0.595 & 10.86 & 6.64 &          CDQ &      \citet{2011ApJ...730...92H} \\
    \object{PKS1055+201} & 1055+201 & +10:58:18.0 & +19:51:50.9 & 1.110 & 10.46 & 8.20 &          LDQ &    \citet{schwartz2006discovery} \\
          \object{3C254} & 1111+408 & +11:14:38.0 & +40:37:20.3 & 0.734 &       & 7.30 &          LDQ &   \citet{donahue2003constraints} \\
    \object{PKS1127-145} & 1127-145 & +11:30:07.0 & -14:49:27.4 & 1.180 & 31.62 & 8.30 &          CDQ & \citet{siemiginowska2002chandra} \\
    \object{PKS1136-135} & 1136-135 & +11:39:11.0 & -13:50:43.6 & 0.554 &       & 6.40 &          LDQ &       \citet{sambruna2002survey} \\
          \object{3C263} & 1137+660 & +11:39:57.0 & +65:47:49.4 & 0.656 &  2.31 & 7.00 &          LDQ &      \citet{2002ApJ...581..948H} \\
          \object{3C265} & 1142+318 & +11:45:29.0 & +31:33:49.4 & 0.811 &       & 7.55 & FR II (NLRG) &      \citet{2004MNRAS.354L..43B} \\
       \object{4C+49.22} & 1150+497 & +11:53:24.0 & +49:31:06.9 & 0.334 & 18.17 & 4.80 &          CDQ &       \citet{sambruna2002survey} \\
    \object{PKS1202-262} & 1202-262 & +12:05:33.0 & -26:34:04.5 & 0.789 & 10.97 & 7.48 &          CDQ &      \citet{2005ApJS..156...13M} \\
        \object{3C270.1} & 1218+339 & +12:20:33.9 & +33:43:12.1 & 1.528 &       & 8.69 &            Q &     \citet{2012ApJ...745...84W}  \\
            \object{M84} & 1222+131 & +12:25:04.0 & +12:53:13.1 & 0.003 &       & 0.08 &   FRI (LERG) &             \citet{harris2002b}  \\
      \object{4C +21.35} & 1222+216 & +12:24:55.0 & +21:22:46.4 & 0.432 & 27.95 & 5.60 &          CDQ &             \citet{jorstad2006x} \\
            \object{M87} & 1228+216 & +12:30:49.0 & +12:23:28.0 & 0.004 &  6.30 & 0.08 &   FRI (NLRG) &            \citet{Marshall_2002} \\
     \object{PKS1229-02} & 1229-021 & +12:32:00.0 & -02:24:05.3 & 1.045 &       & 8.10 &          CDQ &     \citet{tavecchio2007chandra} \\
        \object{3C277.3} & 1251+278 & +12:54:12.0 & +27:37:34.0 & 0.085 &       & 1.67 & FR II (HERG) &             \citet{worrall2016x} \\
       \object{4C+52.27} & 1317+520 & +13:19:46.0 & +51:48:05.8 & 1.060 &       & 8.15 &          CDQ &            \citet{jorstad2006x}  \\
        \object{3C287.1} & 1330+022 & +13:32:53.3 & +02:00:45.7 & 0.216 &       & 3.62 & FR II (BLRG) &   \citet{balmaverde2012extended} \\
    \object{PKS1335-127} & 1334-127 & +13:37:40.0 & -12:57:24.7 & 0.539 & 22.67 & 6.33 &          CDQ &      \citet{2011ApJ...730...92H} \\
       \object{4C+19.44} & 1354+195 & +13:57:04.0 & +19:19:07.4 & 0.720 &  9.84 & 7.20 &          CDQ &       \citet{sambruna2004survey} \\
          \object{3C294} & 1404+344 & +14:06:44.0 & +34:11:25.1 & 1.786 &       & 8.50 & FR II (NLRG) &      \citet{2003MNRAS.341..729F} \\
          \object{3C295} & 1409+524 & +14:11:21.0 & +52:12:09.0 & 0.450 &       & 5.70 & FR II (NLRG) &       \citet{massaro2013chandra} \\
          \object{3C299} & 1419+419 & +14:21:05.6 & +41:44:48.5 & 0.367 &       & 5.25 & FR II (NLRG) &       \citet{massaro2010chandra} \\
    \object{PKS1421-490} & 1421-490 & +14:24:32.0 & -49:13:54.0 & 0.662 &       & 6.98 &            Q &        \citet{gelbord2005knotty} \\
          \object{3C303} & 1441+522 & +14:43:03.0 & +52:01:37.3 & 0.141 &       & 2.50 & FR II (BLRG) &       \citet{kataoka2003chandra} \\
     \object{PKS1510-08} & 1510-089 & +15:12:50.0 & -09:05:59.7 & 0.361 & 47.00 & 5.00 &          CDQ &       \citet{sambruna2004survey} \\
          \object{3C325} & 1549+628 & +15:49:58.4 & +62:41:21.7 & 1.135 &       & 8.44 &        FR II &             \citet{salvati2008x} \\
       \object{4C+00.58} & 1603+001 & +16:06:13.0 & +00:00:27.0 & 0.059 &       & 1.10 &          LDQ &      \citet{2010ApJ...717L..37H} \\
          \object{3C330} & 1609+660 & +16:09:37.0 & +65:56:43.6 & 0.550 &       & 6.40 & FR II (NLRG) &      \citet{2002ApJ...581..948H} \\
       \object{4C+15.55} & 1622+158 & +16:25:14.4 & +15:45:21.9 & 1.406 &       & 8.66 &        FR II &    \citet{stockton2006extremely} \\
          \object{3C346} & 1641+173 & +16:43:49.0 & +17:15:49.0 & 0.161 &       & 2.74 &   FRI (NLRG) &      \citet{2005MNRAS.360..926W} \\
          \object{3C345} & 1641+399 & +16:42:59.0 & +39:48:37.0 & 0.594 & 24.64 & 6.60 &          CDQ &       \citet{sambruna2004survey} \\
       \object{4C+69.21} & 1642+690 & +16:42:08.0 & +68:56:39.7 & 0.751 & 24.90 & 7.30 &          CDQ &       \citet{sambruna2004survey} \\
          \object{3C351} & 1704+605 & +17:04:41.0 & +60:44:30.7 & 0.372 &       & 5.10 &          LDQ &      \citet{2001ApJ...561L.157B} \\
\object{[HB89] 1800+440} & 1800+440 & +18:01:32.0 & +44:04:21.9 & 0.663 & 21.85 & 6.98 &          CDQ &      \citet{2011ApJ...730...92H} \\
       \object{4C+56.27} & 1823+568 & +18:24:07.1 & +56:51:01.5 & 0.664 & 26.98 & 7.19 &            Q &      \citet{2018ApJ...856...66M} \\
          \object{3C380} & 1828+487 & +18:29:32.0 & +48:44:46.6 & 0.692 & 15.36 & 7.11 &        FR II &      \citet{2005ApJS..156...13M} \\
          1849+670        & 1849+670 & +18:49:16.0 & +67:05:41.7 & 0.657 & 31.60 & 6.95 &          CDQ &      \citet{2011ApJ...730...92H} \\
       \object{4C+73.18} & 1928+738 & +19:27:48.0 & +73:58:01.6 & 0.302 & 22.37 & 4.40 &          CDQ &       \citet{sambruna2004survey} \\
          \object{3C403} & 1949+023 & +19:52:16.0 & +02:30:24.4 & 0.059 &       & 1.13 & FR II (NLRG) &      \citet{2005ApJ...622..149K} \\
     \object{S52007+777} & 2007+777 & +20:05:31.0 & +77:52:43.0 & 0.342 &  4.40 & 4.80 &       BL Lac &      \citet{2008ApJ...684..862S} \\
     \object{PKS2101-49} & 2101-490 & +21:05:01.0 & -48:48:46.5 & 1.040 &       & 8.12 &          CDQ &      \citet{2005ApJS..156...13M} \\
     \object{PKS2152-69} & 2152-699 & +21:57:06.0 & -69:41:23.7 & 0.028 &       & 0.56 &     FR II RG &          \citet{ly2005discovery} \\
       \object{4C+08.64} & 2209+080 & +22:12:02.0 & +08:19:16.5 & 0.485 &       & 5.98 &          CDQ &             \citet{jorstad2006x} \\
       \object{4C-03.79} & 2216-038 & +22:18:52.0 & -03:35:36.9 & 0.901 &  6.91 & 7.81 & (FRI/II) CDQ &      \citet{2011ApJ...730...92H} \\
          \object{3C452} & 2243+394 & +22:45:49.0 & +39:41:15.7 & 0.081 &       & 1.50 & FR II (NLRG) &      \citet{2004ApJ...612..729H} \\
        \object{3C454.3} & 2251+158 & +22:53:58.0 & +16:08:53.6 & 0.859 & 25.40 & 7.70 &          CDQ &      \citet{2005ApJS..156...13M} \\
    \cutinhead{Sources from reference}
    3C33 & 0106+130 & +01:08:53.0 & +13:20:13.8 & 0.060 &       & 1.14 & FR II (NLRG) &   \citet{2007ApJ...659.1008K} \\
 PictorA & 0518-458 & +05:19:50.0 & -45:46:44.5 & 0.035 &       & 0.69 & FR II (BLRG) &   \citet{2001ApJ...547..740W} \\
   3C227 & 0945+076 & +09:47:45.0 & +07:25:20.6 & 0.086 &       & 1.59 & FR II (BLRG) &   \citet{2007ApJ...669..893H} \\
   3C273 & 1226+023 & +12:29:07.0 & +02:03:08.6 & 0.158 & 19.19 & 2.70 &          CDQ & \citet{marshall2001structure} \\
   3C327 & 1600+021 & +16:02:27.0 & +01:57:56.2 & 0.104 &       & 1.89 &        FR II &   \citet{2007ApJ...669..893H} \\
   3C353 & 1717+009 & +17:20:28.0 & -00:58:47.0 & 0.030 &       & 0.60 &        FR II &    \citet{kataoka2008chandra} \\
 3C390.3 & 1845+797 & +18:42:09.0 & +79:46:17.2 & 0.056 &  2.57 & 1.08 & FR II (BLRG) &   \citet{2007ApJ...669..893H} \\
    CygA & 1957+405 & +19:59:28.0 & +40:44:01.9 & 0.056 &  1.20 & 1.07 &     FR II RG &    \citet{wilson2000chandra}  \\
4C+74.26 & 2043+749 & +20:42:37.0 & +75:08:02.5 & 0.104 &       & 1.90 &          LDQ &    \citet{erlund2007luminous} \\
   3C445 & 2221-020 & +22:23:50.0 & -02:06:12.4 & 0.056 &       & 1.10 &     FR II RG &          \citet{Perlman_2009} \\
    \cutinhead{Remaining sources}
    3C6.1 &  0013+790 & +00:16:31.0 & +79:16:49.9 & 0.840 &       & 7.84 &    FR II (NLRG) &        \citet{2004ApJ...612..729H}  \\
    3C13 &  0031+391 & +00:34:14.6 & +39:24:16.6 & 1.351 &       & 8.63 &    FR II (NLRG) &        \citet{wilkes2013revealing}  \\
    3C16 &  0035+130 & +00:37:44.6 & +13:19:55.0 & 0.405 &       & 5.57 &           FR II &          \citet{massaro2013chandra} \\
    3C17 &  0035-024 & +00:38:21.0 & -02:07:40.7 & 0.220 &       & 3.64 &    FR II (BLRG) &               \citet{Massaro_2009}  \\
    3C19 &  0038+328 & +00:40:55.0 & +33:10:08.0 & 0.482 &       & 6.16 &    FR II (LERG) &          \citet{massaro2015chandra} \\
    3C29 &  0055+016 & +00:57:34.9 & -01:23:27.9 & 0.045 &       & 0.89 &     FR I (LERG) &          \citet{massaro2012chandra} \\
  NGC315 &  0055+300 & +00:57:49.0 & +30:21:08.8 & 0.016 &  0.81 & 0.32 &         FR I RG &                \citet{worrall2003x} \\
    3C41 &  0123+239 & +01:26:44.4 & +33:13:11.2 & 0.795 &       & 7.70 &    FR II (HERG) &                \citet{Massaro_2018} \\
    3C52 &  0145+532 & +01:48:29.0 & +53:32:35.4 & 0.290 &       & 4.47 &    FR II (HERG) &          \citet{massaro2010chandra} \\
    3C54 &  0152+435 & +01:55:30.2 & +43:45:55.4 & 0.827 &       & 7.80 &    FR II (HERG) &                \citet{Massaro_2018} \\
  3C61.1 &  0210+860 & +02:22:35.0 & +86:19:06.2 & 0.188 &       & 3.32 &    FR II (HERG) &          \citet{massaro2010chandra} \\
    3C65 &  0220+397 & +02:23:43.2 & +40:00:52.5 & 1.176 &       & 8.49 &    FR II (HERG) &         \citet{wilkes2013revealing} \\
  3C68.2 &  0231+313 & +02:34:23.9 & +31:34:17.5 & 1.575 &       & 8.69 &    FR II (HERG) &         \citet{wilkes2013revealing} \\
WISE J031552.09-190644.2 &  0313-192 & +03:15:52.0 & -19:06:44.3 & 0.067 &       & 1.27 &         FR I RG &         \citet{2006AJ....132.2233K} \\
   3C105 &  0404+033 & +04:07:16.5 & +03:42:25.8 & 0.089 &       & 1.71 &            FR II &        \citet{massaro2010chandra}\\
   3C109 &  0410+110 & +04:13:40.0 & +11:12:13.8 & 0.306 &       & 4.50 &    FR II (BLRG) &         \citet{2004ApJ...612..729H} \\
   3C129 &  0445+449 & +04:49:09.0 & +45:00:39.3 & 0.021 &       & 0.42 &         FR I RG &                 \citet{harris2002c} \\
   3C133 &  0459+252 & +05:02:58.5 & +25:16:25.1 & 0.278 &       & 4.35 &    FR II (HERG) &          \citet{massaro2010chandra} \\
PKS0521-36 &  0521-365 & +05:22:58.0 & -36:27:30.8 & 0.055 &       & 1.06 &          BL Lac &             \citet{birkinshaw2002x} \\
WISE J053239.00+073243.2 &  0529+075 & +05:32:39.0 & +07:32:43.3 & 1.254 & 23.26 & 8.40 &             CDQ &         \citet{2011ApJ...730...92H} \\
   3C154 &  0610+260 & +06:13:50.1 & +26:04:36.7 & 0.580 &       & 6.77 &               Q &                \citet{Massaro_2018} \\
   3C171 &  0651+542 & +06:55:14.8 & +54:09:00.1 & 0.238 &       & 3.89 &    FR II (HERG) &          \citet{massaro2010chandra} \\
 3C173.1 &  0702+749 & +07:09:18.0 & +74:49:31.8 & 0.292 &       & 4.30 &    FR II (LERG) &         \citet{2004ApJ...612..729H} \\
   3C179 &  0723+679 & +07:28:12.0 & +67:48:47.5 & 0.846 &  8.99 & 7.70 &             LDQ &          \citet{sambruna2004survey} \\
   3C181 &  0725+147 & +07:28:10.3 & +14:37:36.2 & 1.382 &       & 8.65 &               Q &         \citet{wilkes2013revealing} \\
B3 0727+409 &  0727+409 & +07:30:51.3 & +40:49:50.8 & 2.500 &  6.61 & 8.27 &               Q & \citet{simionescu2016serendipitous} \\
4C+25.21 &  0730+257 & +07:33:08.8 & +25:36:25.1 & 2.686 &       & 8.14 &               Q &               \citet{McKeough_2016} \\
   DA240 &   0745+56 & +07:48:37.0 & +55:48:58.0 & 0.036 &       & 0.70 &            - &                                - \\
QSO B0748+126 &  0748+126 & +07:50:52.0 & +12:31:04.8 & 0.889 & 18.84 & 7.98 &               Q &         \citet{2018ApJ...856...66M} \\
   3C189 &  0755+379 & +07:58:28.0 & +37:47:11.8 & 0.043 &       & 0.83 &         FR I RG &          \citet{worrall2001chandra} \\
   3C191 &  0802+013 & +08:04:48.0 & +10:15:23.7 & 1.956 &       & 8.60 &               Q &          \citet{erlund2006extended} \\
4C +05.34 &  0805+046 & +08:07:57.5 & +04:32:34.5 & 2.877 &       & 7.99 &               Q &               \citet{McKeough_2016} \\
   3C200 &  0824+294 & +08:27:25.4 & +29:18:44.9 & 0.458 &       & 6.00 &    FR II (LERG) &         \citet{2004ApJ...612..729H} \\
QSO J0833+0959  &  0830+101 & +08:33:22.5 & +09:59:41.1 & 3.713 &       & 7.33 &               Q &         \citet{2018ApJ...856...66M} \\
TXS0833+585 &  0833+585 & +08:37:22.4 & +58:25:01.8 & 2.101 & 14.17 & 8.53 &               Q &               \citet{McKeough_2016} \\
   3C212 &  0855+143 & +08:58:41.0 & +14:09:43.9 & 1.049 &       & 8.10 &             LDQ &         \citet{2003ApJ...597..751A} \\
   3C210 &  0855+280 & +08:58:10.0 & +27:50:51.6 & 1.169 &       & 8.48 &    FR II (HERG) &     \citet{gilmour2009distribution} \\
 3C213.1 &  0858+292 & +09:01:05.3 & +29:01:46.9 & 0.194 &       & 3.33 &    FR II (LERG) &          \citet{massaro2010chandra} \\
4C+47.29 &  0859+470 & +09:03:04.0 & +46:51:04.1 & 1.465 & 16.41 & 8.68 &               Q &         \citet{2018ApJ...856...66M} \\
  3c 215 &  0903+169 & +09:06:31.9 & +16:46:11.9 & 0.412 &       & 5.64 &               Q &         \citet{2004ApJ...612..729H} \\
PKS0903-573 &  0903-573 & +09:04:53.0 & -57:35:05.8 & 0.695 &       & 7.12 &             CDQ &         \citet{2005ApJS..156...13M} \\
6C0905+39 &  0905+399 & +09:08:16.9 & +39:43:26.0 & 1.883 &       & 8.63 &    FR II (HERG) &         \citet{2008MNRAS.386.1774E} \\
PMN J0909+0354 &  0906+041 & +09:09:15.9 & +03:54:43.0 & 3.431 &       & 7.66 &               Q &         \citet{2022AnA...659A..93I} \\
   3C219 &  0917+458 & +09:21:09.0 & +45:38:57.4 & 0.174 &       & 2.90 & FR II RG (HERG) &         \citet{2003MNRAS.340L..52C} \\
 3C220.2 &  0927+362 & +09:30:33.5 & +36:01:24.2 & 1.157 &       & 8.47 &               Q &                \citet{Stuardi_2018} \\
  3C225B &  0939+139 & +09:42:15.4 & +13:45:50.5 & 0.580 &       & 6.77 &    FR II (HERG) &                \citet{Massaro_2018} \\
4C +55.17 &  0954+556 & +09:57:38.2 & +55:22:57.8 & 0.900 &       & 8.01 &               Q &        \citet{tavecchio2007chandra} \\
   3C234 &  0958+290 & +10:01:49.5 & +28:47:09.0 & 0.185 &       & 3.21 &           FR II &         \citet{2012ApJS..203...31M} \\
4C+13.41 &  1004+130 & +10:07:26.0 & +12:48:56.2 & 0.240 &       & 3.76 &             LDQ &         \citet{2006ApJ...652..163M} \\
   3C238 &  1008+066 & +10:11:00.4 & +06:24:39.7 & 1.405 &       & 8.66 &    FR II (HERG) &                \citet{Stuardi_2018} \\
  3C 239 &  1008+467 & +10:11:45.3 & +46:28:18.8 & 1.781 &       & 8.60 &    FR II (HERG) &         \citet{2020ApJS..250....7J} \\
7C 1013+2053  &  1013+208 & +10:16:44.3 & +20:37:47.3 & 3.114 &       & 7.80 &               Q &          \citet{snios2021discovery} \\
   3C245 &  1040+123 & +10:42:44.6 & +12:03:31.3 & 1.028 &  0.60 & 8.28 &               Q &          \citet{sambruna2004survey} \\
PKS1046-409 &  1046-409 & +10:48:38.0 & -41:13:59.6 & 0.620 &       & 6.78 &             CDQ &         \citet{2005ApJS..156...13M} \\
  3C 249 &  1059-009 & +11:02:03.8 & -01:16:16.7 & 1.554 &       & 8.61 &             LDQ &         \citet{2020ApJS..250....7J} \\
  3C 257 &  1120+057 & +11:23:09.4 & +05:30:18.5 & 2.474 &       & 8.24 &           FR II &         \citet{2020ApJS..250....7J} \\
   3C264 &  1142+195 & +11:45:05.0 & +19:36:22.7 & 0.022 &       & 0.43 &     FR I (LERG) &         \citet{2010ApJ...708..171P} \\
 3C268.1 &  1157+732 & +12:00:19.2 & +73:00:45.7 & 0.970 &       & 8.17 &    FR II (NLRG) &         \citet{massaro2015chandra}  \\
 3C268.2 &  1158+318 & +12:00:59.1 & +31:33:27.9 & 0.362 &       & 5.21 &    FR II (HERG) &          \citet{massaro2012chandra} \\
 NGC4261 &  1216+061 & +12:19:23.0 & +05:49:30.8 & 0.007 &       & 0.15 &         FR I RG &         \citet{chiaberge2003hubble} \\
B3 1239+376 &  1239+376 & +12:42:09.8 & +37:20:05.7 & 3.819 &       & 7.25 &               Q &               \citet{McKeough_2016} \\
 3C275.1 &  1241+266 & +12:43:58.0 & +16:22:53.2 & 0.555 &       & 6.40 &             LDQ &        \citet{crawford2003extended} \\
   3C280 &  1254+476 & +12:56:57.0 & +47:20:19.9 & 0.996 &       & 8.00 &    FR II (HERG) &      \citet{donahue2003constraints} \\
3C 280.1 &  1258+404 & +13:00:33.4 & +40:09:07.3 & 1.667 &       & 8.62 &             LDQ &         \citet{2020ApJS..250....7J} \\
   3C281 &  1305+069 & +13:07:54.0 & +06:42:14.3 & 0.602 &       & 6.70 &             LDQ &        \citet{crawford2003extended} \\
PKS1311-270 &  1311-270 & +13:13:47.4 & -27:16:49.3 & 2.260 &       & 8.44 &               Q &               \citet{McKeough_2016} \\
4C+11.45 &  1318+113 & +13:21:18.8 & +11:06:50.0 & 2.179 &       & 8.49 &               Q &               \citet{McKeough_2016} \\
    CenA &  1322-428 & +13:25:28.0 & -43:01:08.8 & 0.002 &       & 0.02 &          FRI RG &            \citet{kraft2000chandra} \\
4C+65.15 &  1323+655 & +13:25:30.0 & +65:15:13.5 & 1.625 &       & 8.60 &             LDQ &                 \citet{Miller_2009} \\
    CenB &  1343-601 & +13:46:49.0 & -60:24:29.9 & 0.013 &       & 0.26 &          FRI RG &         \citet{2005ApJS..156...13M} \\
   3C293 &  1350+316 & +13:52:17.8 & +31:26:46.5 & 0.045 &       & 0.93 &     FR I (LERG) &          \citet{massaro2010chandra} \\
QSO B1402+044  &  1402+044 & +14:05:01.1 & +04:15:35.8 & 3.215 &       & 7.73 &               Q &          \citet{snios2021discovery} \\
   3C296 &  1414+110 & +14:16:53.0 & +10:48:27.7 & 0.024 &       & 0.47 &     FR I (LERG) &      \citet{hardcastle2005chandrab} \\
   3C297 &  1414-037 & +14:17:24.0 & -04:00:47.5 & 1.406 &       & 8.66 &               Q &                \citet{Stuardi_2018} \\
PKS 1418-064 &  1418-064 & +14:21:07.8 & -06:43:56.4 & 3.689 &       & 7.35 &               Q &               \citet{McKeough_2016} \\
B3 1428+422 &  1428+422 & +14:30:23.7 & +42:04:36.5 & 4.715 &       & 6.62 &               Q &                 \citet{Cheung_2012} \\
   3c305 &  1448+634 & +14:49:21.6 & +63:16:14.0 & 0.042 &       & 0.85 &            FRII &         \citet{2012MNRAS.424.1774H} \\
   3C313 &  1508+080 & +15:11:00.0 & +07:51:50.0 & 0.461 &       & 6.02 &    FR II (HERG) &          \citet{massaro2013chandra} \\
TXS1508+572 &  1508+572 & +15:10:03.0 & +57:02:44.0 & 4.300 &       & 6.90 &               Q &         \citet{siemiginowska2003x}  \\
  AP Lib &  1514-241 & +15:17:42.0 & -24:22:19.5 & 0.049 &       & 0.94 &          BL Lac &       \citet{kaufmann2013discovery} \\
   3C321 &  1529+242 & +15:31:44.0 & +24:04:19.0 & 0.096 &       & 1.80 &    FR II (NLRG) &         \citet{2004ApJ...612..729H} \\
  3C 322 &  1533+557 & +15:35:01.3 & +55:36:52.3 & 1.681 &       & 8.61 &           FR II &         \citet{2020ApJS..250....7J} \\
   3C324 &  1547+215 & +15:49:49.0 & +21:25:38.1 & 1.206 &       & 8.40 &    FR II (HERG) &         \citet{2004ApJ...612..729H} \\
3C 326.1 &  1553+202 & +15:56:10.2 & +20:04:20.7 & 1.825 &       & 8.59 &              RG &         \citet{2020ApJS..250....7J} \\
 3C327.1 &  1602+014 & +16:04:45.3 & +01:17:51.0 & 0.462 &       & 6.02 &    FR II (HERG) &          \citet{massaro2013chandra} \\
QSO B1607+1819  &  1607+183 & +16:10:05.3 & +18:11:43.5 & 3.122 &       & 7.80 &               Q &          \citet{snios2021discovery} \\
   3C334 &  1618+177 & +16:20:21.8 & +17:36:24.0 & 0.555 &  3.36 & 6.63 &               Q &         \citet{2004ApJ...612..729H} \\
   3C341 &  1626+738 & +16:28:04.0 & +27:41:43.0 & 0.448 &       & 5.92 &    FR II (HERG) &          \citet{massaro2013chandra} \\
 NGC6251 &  1637+826 & +16:32:32.0 & +82:32:16.5 & 0.025 &       & 0.49 &   FRI/II (LERG) &           \citet{evans2005chandra}  \\
   3C349 &  1658+371 & +16:59:29.5 & +47:02:44.1 & 0.205 &       & 3.47 &    FR II (HERG) &          \citet{massaro2010chandra} \\
4C+62.29 &  1745+624 & +17:46:14.0 & +62:26:54.7 & 3.889 &       & 7.20 &             LDQ &                 \citet{Cheung_2006} \\
7C 1754+6737 &  1754+678 & +17:54:22.3 & +67:37:34.6 & 3.600 &       & 7.42 &               Q &               \citet{McKeough_2016} \\
   3C368 &  1802+110 & +18:05:06.5 & +11:01:35.1 & 1.131 &       & 8.44 &    FR II (HERG) &          \citet{massaro2015chandra} \\
   3C371 &  1807+698 & +18:06:51.0 & +69:49:28.1 & 0.051 & 18.17 & 0.98 &          BL Lac &          \citet{pesce2001detection} \\
TXS1834+612 &  1834+612 & +18:35:19.7 & +61:19:40.0 & 2.274 &       & 8.43 &               Q &               \citet{McKeough_2016} \\
   3C402 &  1940+504 & +19:41:46.0 & +50:35:48.4 & 0.026 &       & 0.53 &            FR I &          \citet{massaro2012chandra} \\
  3C 418 &  2037+511 & +20:38:37.0 & +51:19:12.4 & 1.686 &  6.86 & 8.61 &               Q &         \citet{2021ApJS..253...37R} \\
4C+23.56 &  2104+233 & +21:07:14.8 & +23:31:45.0 & 2.483 &       & 8.29 &               Q &               \citet{blundell2011x} \\
PKs2123-463 &  2123-463 & +21:26:31.0 & -46:05:47.9 & 1.670 &       & 8.56 &             CDQ &               \citet{marshall2011x} \\
   3C436 &  2141+279 & +21:44:11.7 & +28:10:19.0 & 0.214 &       & 3.58 &    FR II (HERG) &          \citet{massaro2010chandra} \\
   3C437 & 2145+151  & +21:47:25.1 & +15:20:37.5 & 1.480 &       & 8.68 &           FR II &          \citet{massaro2015chandra} \\
PKS2155-152 &  2155-152 & +21:58:06.0 & -15:01:09.3 & 0.672 & 21.46 & 7.02 &             CDQ &         \citet{2011ApJ...730...92H} \\
PKS2201+044 &  2201+044 & +22:04:18.0 & +04:40:02.0 & 0.027 &       & 0.54 &          BL Lac &         \citet{2007ApJ...670...74S} \\
4C+31.63 &  2201+315 & +22:03:15.0 & +31:45:38.3 & 0.295 &  8.81 & 4.37 &             CDQ &         \citet{2011ApJ...730...92H} \\
 NGC7385 &  2247+313 & +22:49:54.6 & +11:36:30.8 & 0.028 &       & 0.52 &            FR I &       \citet{10.1093/mnras/stv1501} \\
PKS 2255-282 &  2255-282 & +22:58:06.0 & -27:58:21.3 & 0.926 &  6.15 & 7.88 &             CDQ &               \citet{marshall2011x} \\
   3C458 &  2310+050 & +23:12:54.4 & +05:16:46.0 & 0.289 &       & 4.46 &    FR II (HERG) &          \citet{massaro2012chandra} \\
   3C465 &   2335+46 & +23:38:29.0 & +27:01:55.9 & 0.029 &       & 0.58 &            FR I &       \citet{hardcastle2005chandra} \\
   2345-167     &  2345-167 & +23:48:03.0 & -16:31:12.0 & 0.576 & 13.50 & 6.54 &             CDQ &         \citet{2011ApJ...730...92H} \\
   3C470 &  2356+437 & +23:58:35.9 & +44:04:45.6 & 1.653 &       & 8.69 &           FR II &         \citet{2011ApJ...730...92H} \\
PKSJ2310-437 &           & +23:10:42.0 & -43:47:34.3 & 0.089 &       & 1.60 &            FR I &         \citet{2009ApJ...698.1061B} \\
PSO J352.4034-15.3373  &           & +23:29:36.8 & -15:20:14.5 & 5.840 &       &      &               Q &         \citet{2021ApJ...911..120C} \\
PSO J047.4478+27.2992 &           & +03:09:47.4 & +27:17:57.6 & 6.100 &       & 5.79 &               Q &          \citet{snios2021discovery} \\
    \enddata
    \tablenotetext{a}{The maximum apparent superluminal motions are compiled from \citet{2021MNRAS.505.4726K}}
    \tablenotetext{b}{The class column gives a radio jet morphology descriptor: Fanaroff-Riley I or II, CDQ (core dominated quasar), LDQ (Lobe dominated quasar), BL Lac, and an optical spectroscopic designation: LERG (low-excitation radio galaxy), HERG (high-excitation radio galaxy), NLRG (Narrow-line radio galaxy),BLRG (broad-line radio galaxy)  }
\end{deluxetable*}
\end{longrotatetable}

\section{Excluded sources\label{sec:excluded_sources}}
\begin{figure*}
    \gridline{
        \fig{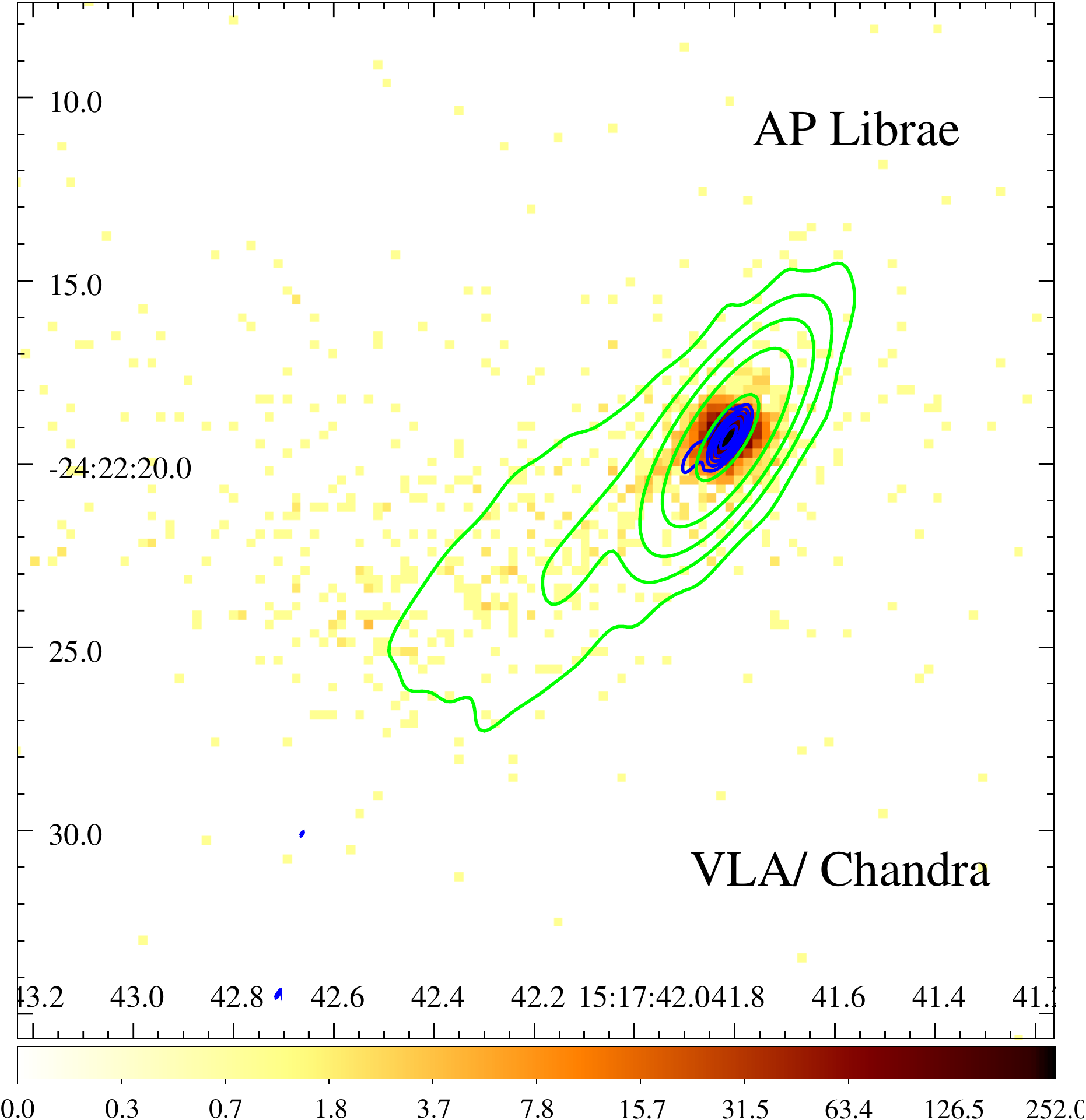}{0.5\textwidth}{(a)}
        \fig{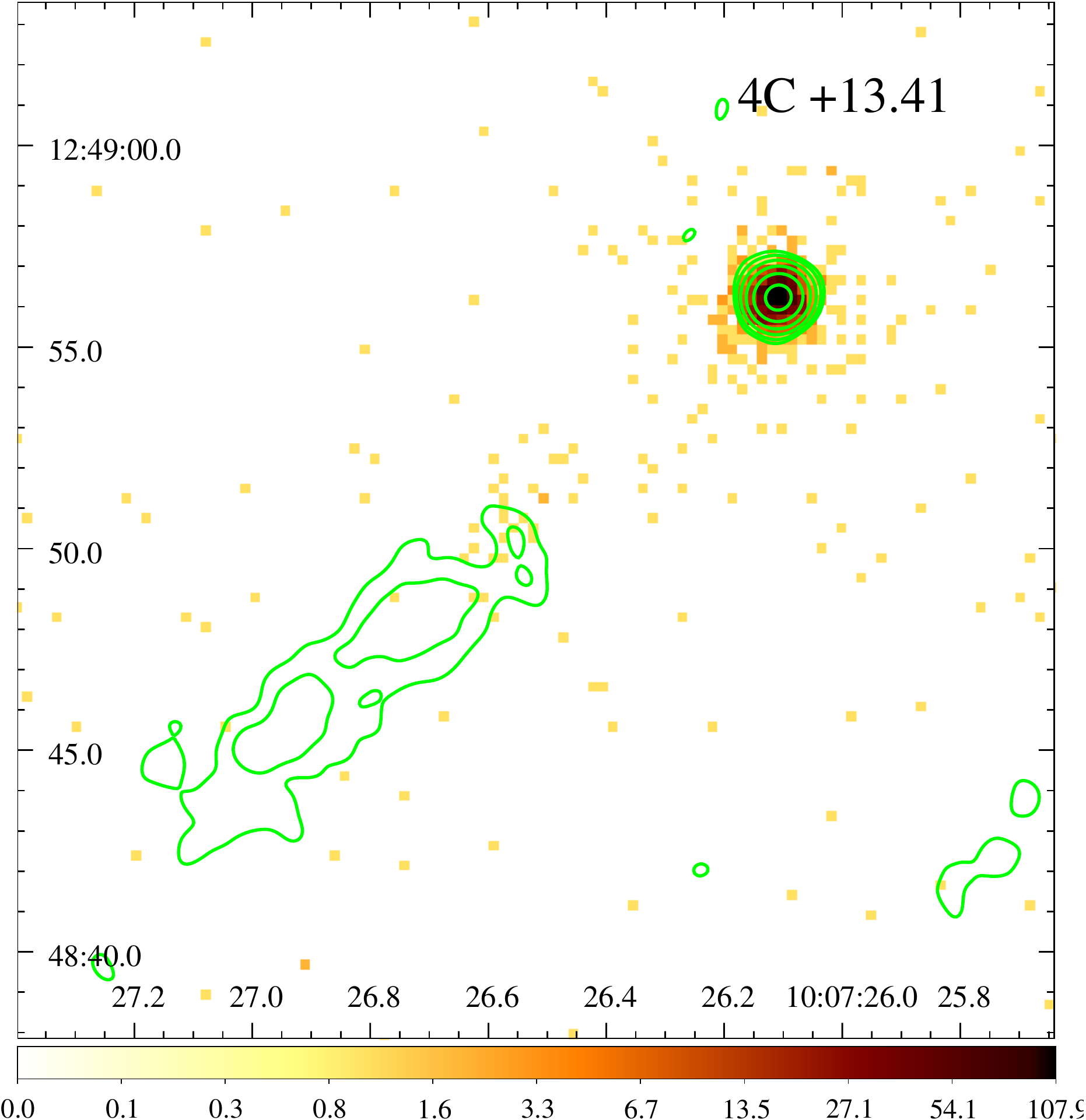}{0.5\textwidth}{(b)}
    }
    \gridline{
        \fig{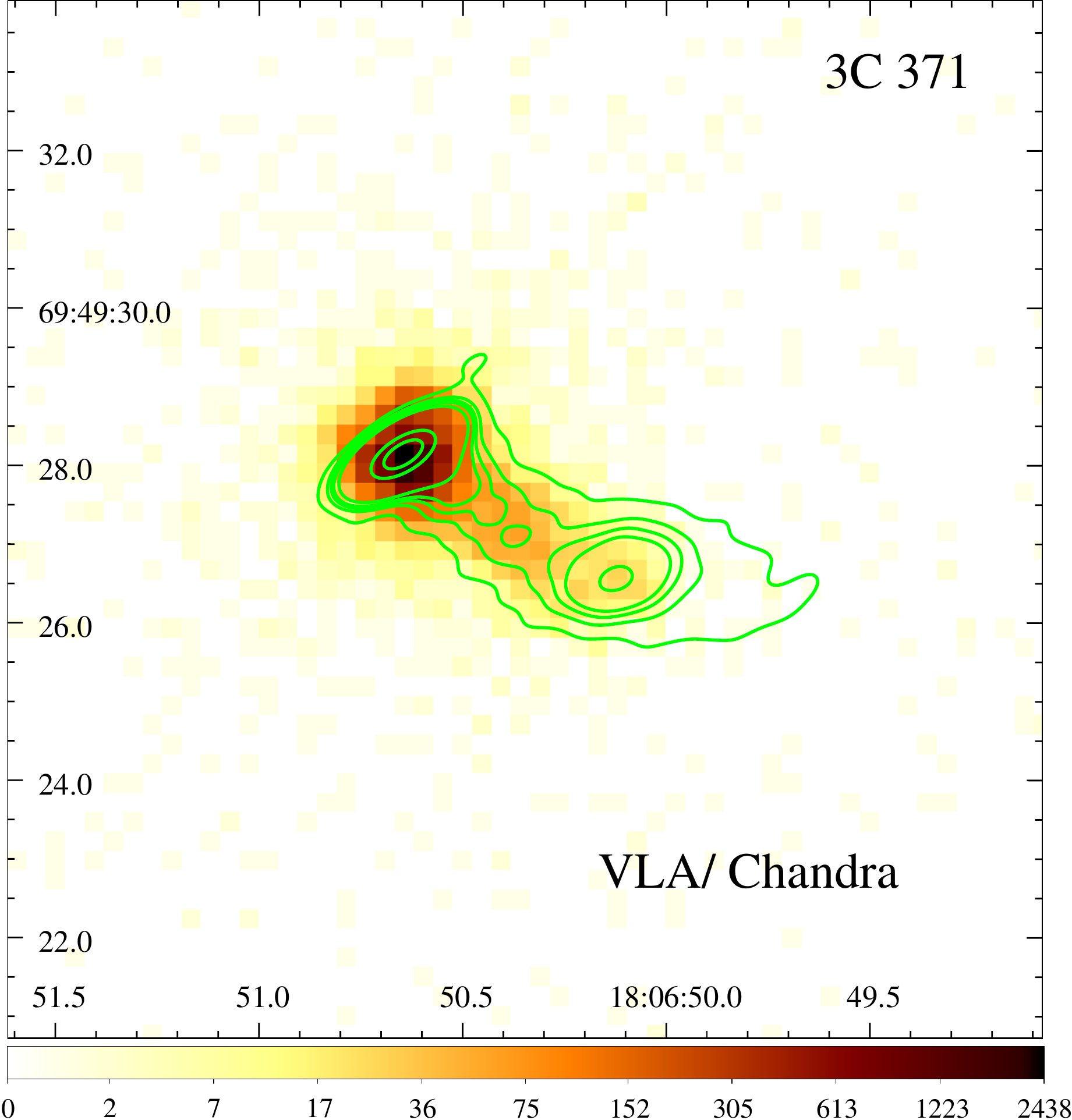}{0.5\textwidth}{(c)}
    }
    \caption{Examples of sources excluded from analysis. The top panels show examples for jets with no spatially-correlated X-ray radio features. (a) X-ray image (bin: 0.5) of AP Librae overlaid with 1.4 GHz VLA radio contours, where both the jets lie along the same position angle but show not correlated features (b) X-ray image of 4C +13.41 with the VLA 8.4 GHz radio contours overlaid. The X-ray and radio jets are mostly spatially disjoint. (c) X-ray image of 3C 371 with the VLA 4.8 GHz radio contours overlaid. The small inter-knot spacing (separation$<1$\as) prevents accurately measuring the centroids on LIRA images.\label{fig:excluded_examples}}
\end{figure*}

Besides high-z jets, lack of spatially-correlated features and jets with closely-spaced knots are the two main reasons for exclusion. Here we present examples of sources from the two categories in Figure \ref{fig:excluded_examples} where each panel shows an X-ray image with the corresponding VLA radio contours overlaid on top. The top panels show examples for the first type where (a) shows AP Librae with spatially-overlapping radio and X-ray jets but with no correlation in their brightnesses, and (b) shows 4C+13.41 with X-ray and radio jets but mostly non-overlapping. Figure \ref{fig:excluded_examples}(c) shows 3C 371 as an example for the second type where all the features are closely-spaced (separation$<1$\as) that preclude constructing large enough ROIs around each component to accurately measure offsets.

\begin{figure}
    \gridline{
        \fig{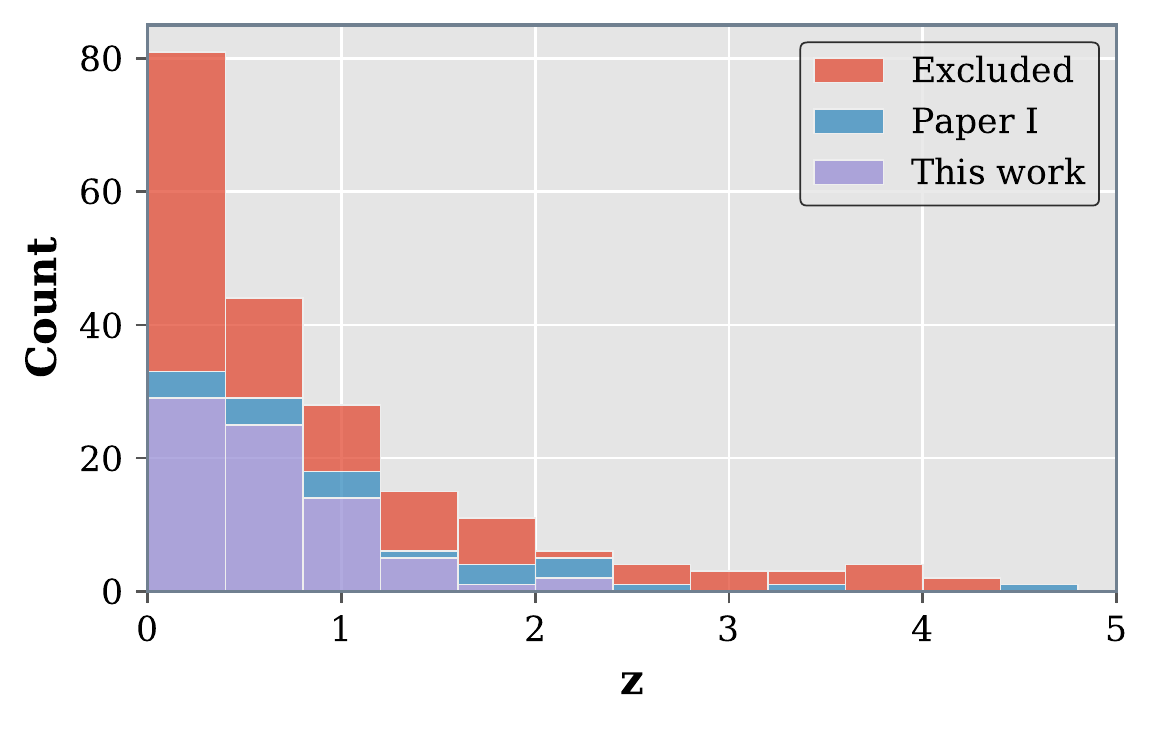}{0.5\textwidth}{(a) All X-ray jets}
        \fig{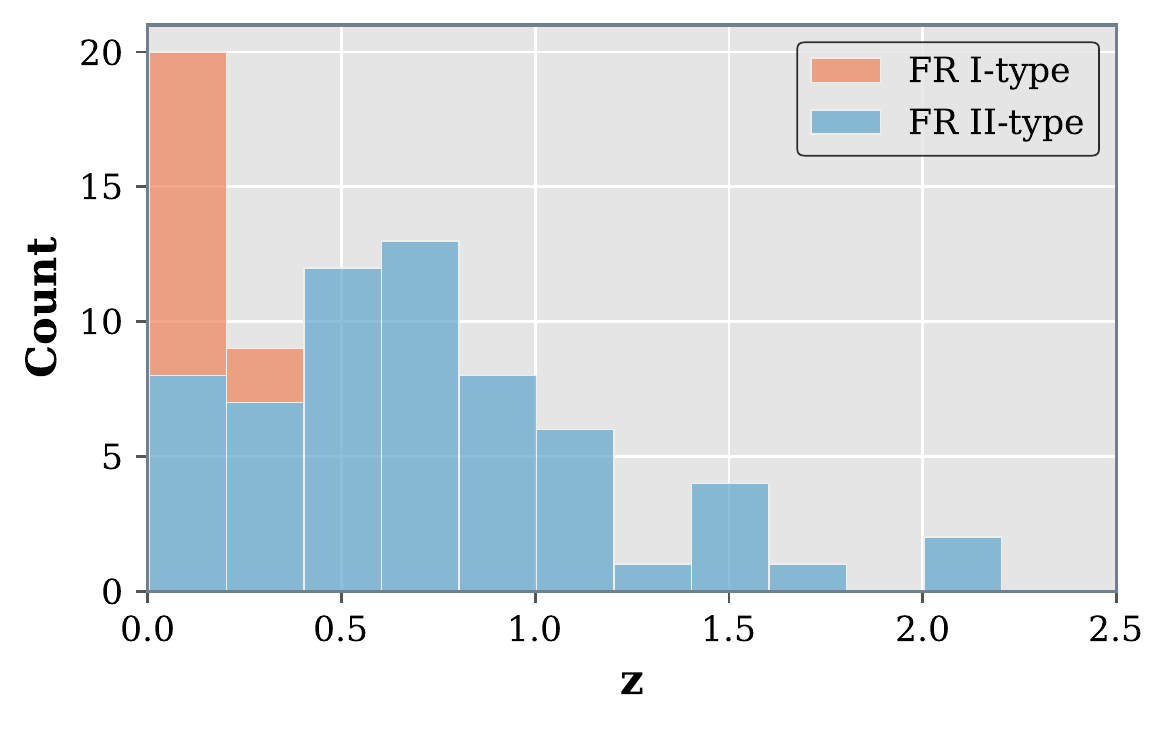}{0.5\textwidth}{(b) Analyzed in this work}
    }
    \gridline{
        \fig{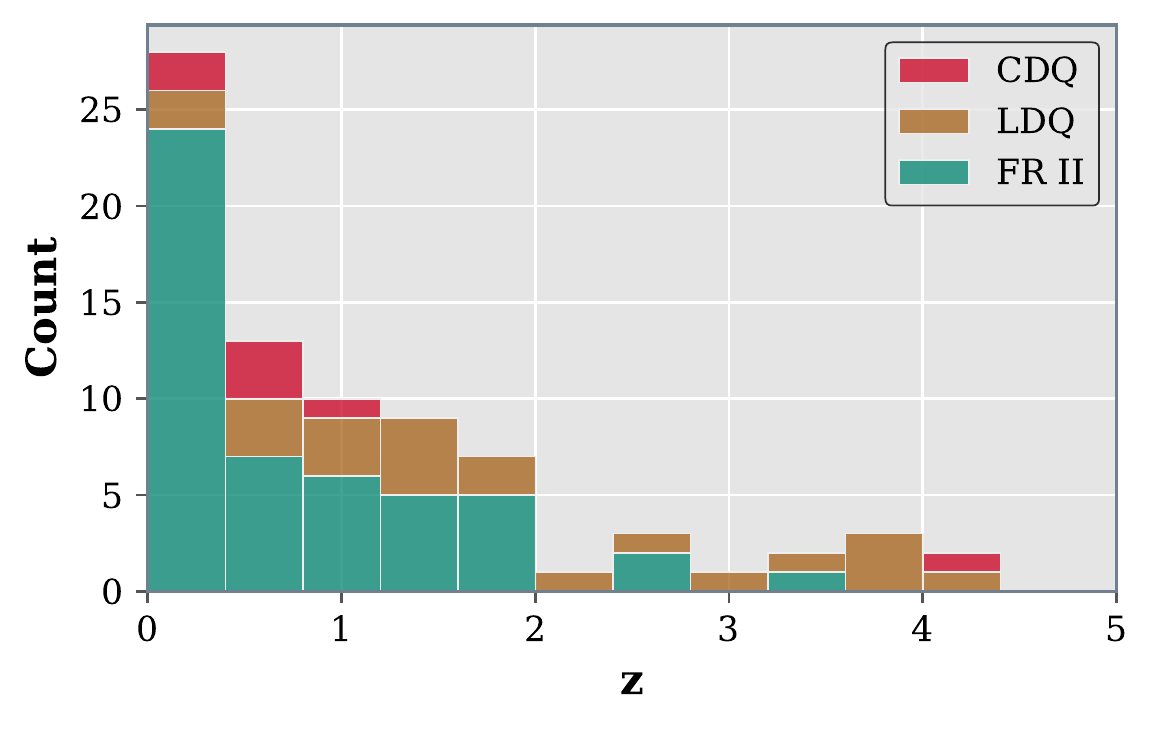}{0.5\textwidth}{(c) All excluded FR II-types}
    }
    \caption{Redshift distribution for X-ray jets. (a) shows the stacked histogram of redshift for all the 199  X-ray jets compiled in this work; sources included and excluded from the offset analysis are colored blue and gray, respectively. (b) shows the class-wise stacked histogram of redshift for the sources considered for offset analysis; FR I-type and FR II-type are colored orange and blue, respectively. (c) shows the stacked redshift distributions of excluded FR II-type sources, divided into individual spectral classes. \label{fig:redshift-histograms}}
\end{figure}

Figure \ref{fig:redshift-histograms}(a) shows the redshift distribution for all the 199 X-ray jets with the sources considered for offset analysis in \papri and this paper. (b) shows the stacked histogram of redshift only for the sources considered for offset analysis with the FR I-type colored orange and the FR-II type colored blue.

Because we mainly focus on understanding the nature of X-ray emission from FR II-type sources, to understand the impact of several excluding sources, we plotted their redshift distributions in \ref{fig:redshift-histograms}(c). As can be seen, the FR II radio galaxies form the major portion of the excluded sources. In section  \ref{subsec:flux-ratio}, we find no significant evidence for difference between the flux ratios of FR II/LDQ and CDQ knots, although with a low statistical power due to the small sample size. In section \ref{subsec:deproj-offsets}, we find similar offsets in FR II and quasar knots that become more uniform if the knots are moving. Hence, unless the offsets and flux ratios of any knots in the excluded FR II are significantly lower than their included counterparts, the trends found in Section \ref{sec:discussion} will remain unaltered.



\comment{\begin{figure*}
    \gridline{
        \fig{3C9_counts_ax-crop.pdf}{0.5\textwidth}{(a)}
        \fig{3C9_lira_ax-crop.pdf}{0.5\textwidth}{(b)}
    }
    \caption{Same as in Fig. \ref{fig:results-3C9} but for 3C 9. The radio contours are given by 0.03, 0.4, 2.0, 6.0, 8.0 mJy beam$^{-1}$.\label{fig:results-3C9}}
\end{figure*}
}

\begin{figure*}[ht]
    \gridline{
        \fig{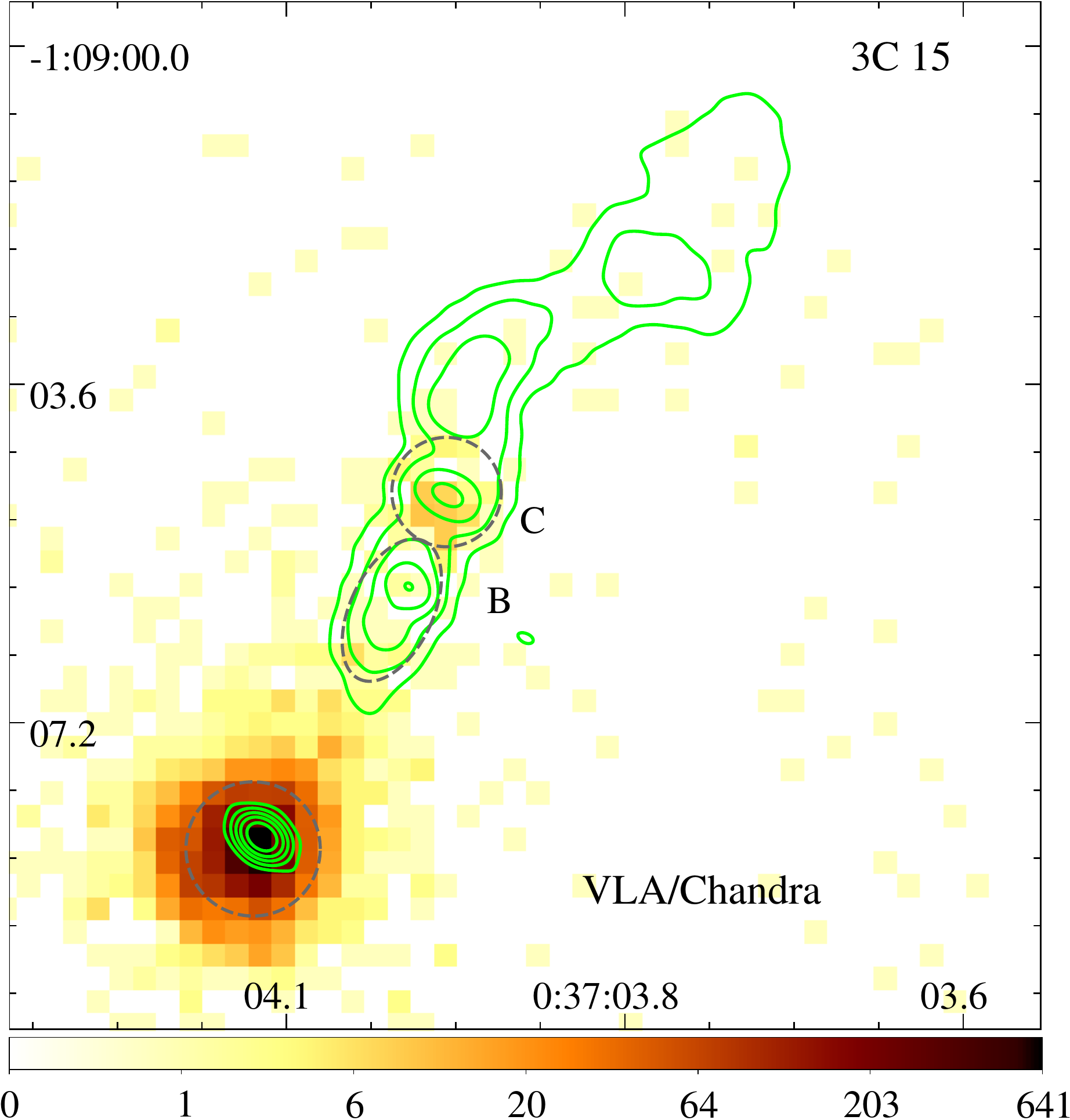}{0.5\textwidth}{(a)}
        \fig{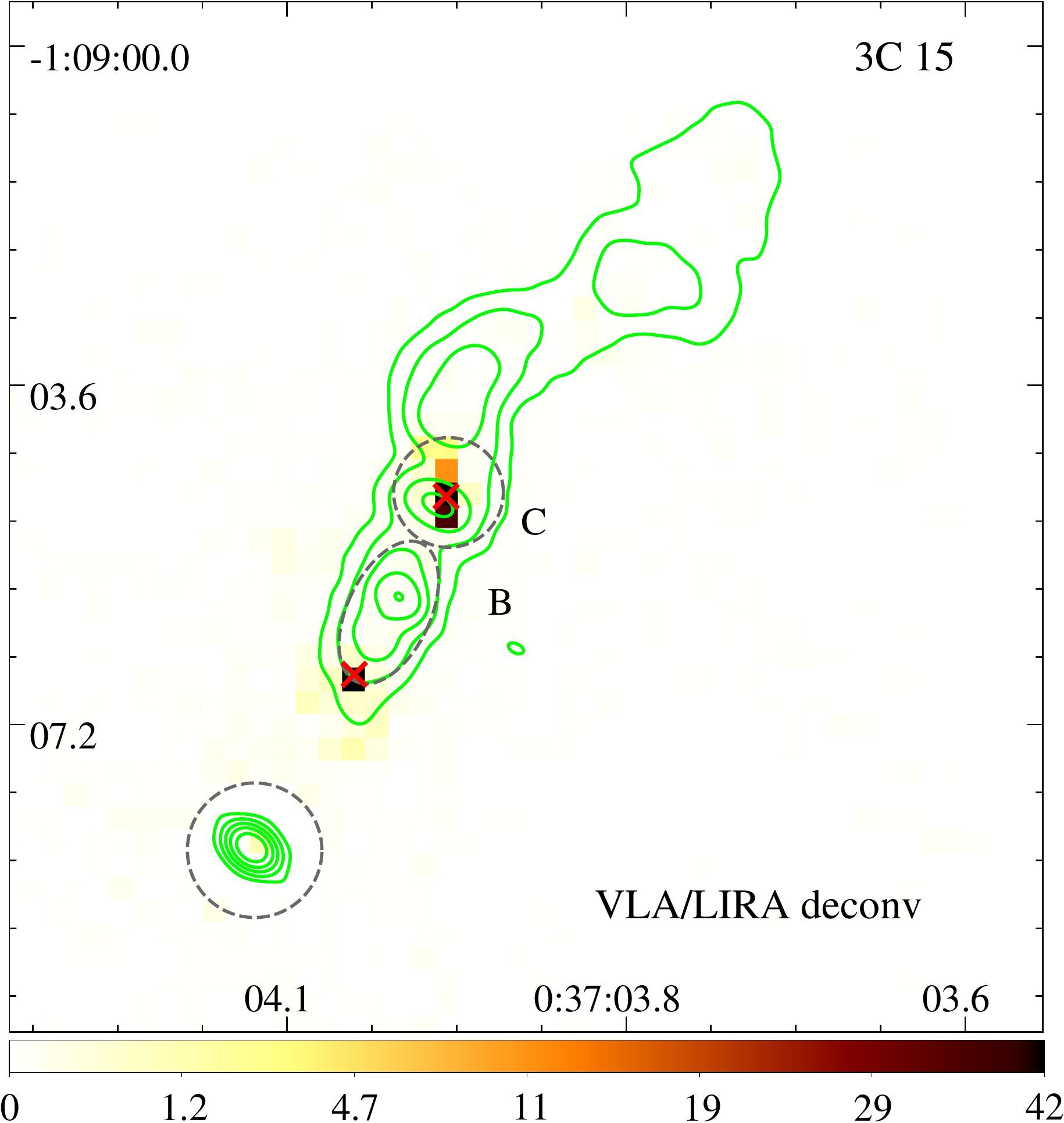}{0.5\textwidth}{(b)}
    }
    \caption{Same as in Fig. \ref{fig:results-3C9} but for 3C15. The radio contours are given by 0.2, 0.8, 2.0, 4.0, 8.0 mJy beam$^{-1}$.\label{fig:results-3C15}}
\end{figure*}

\begin{figure*}[ht]
    \gridline{
        \fig{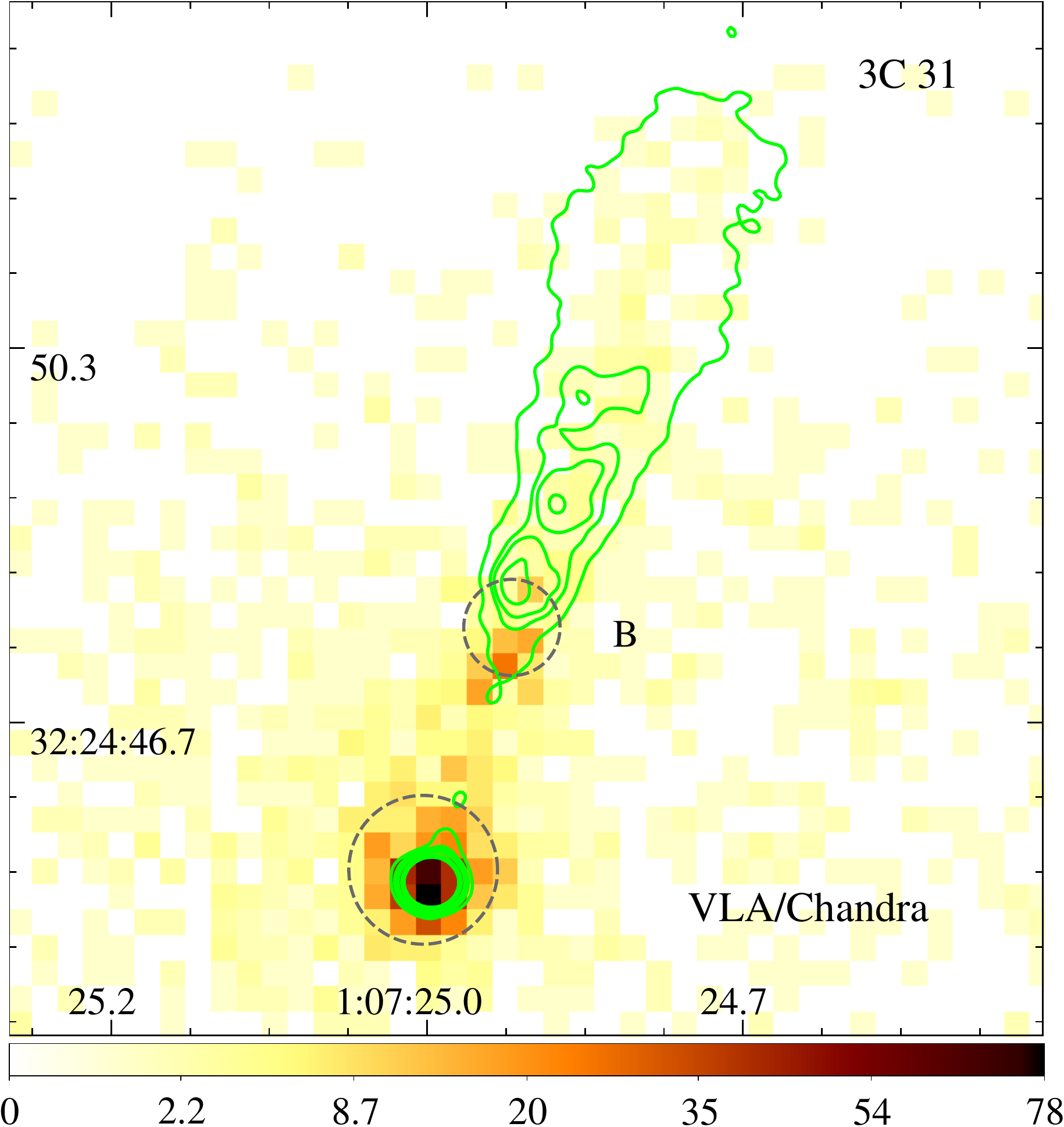}{0.5\textwidth}{(a)}
        \fig{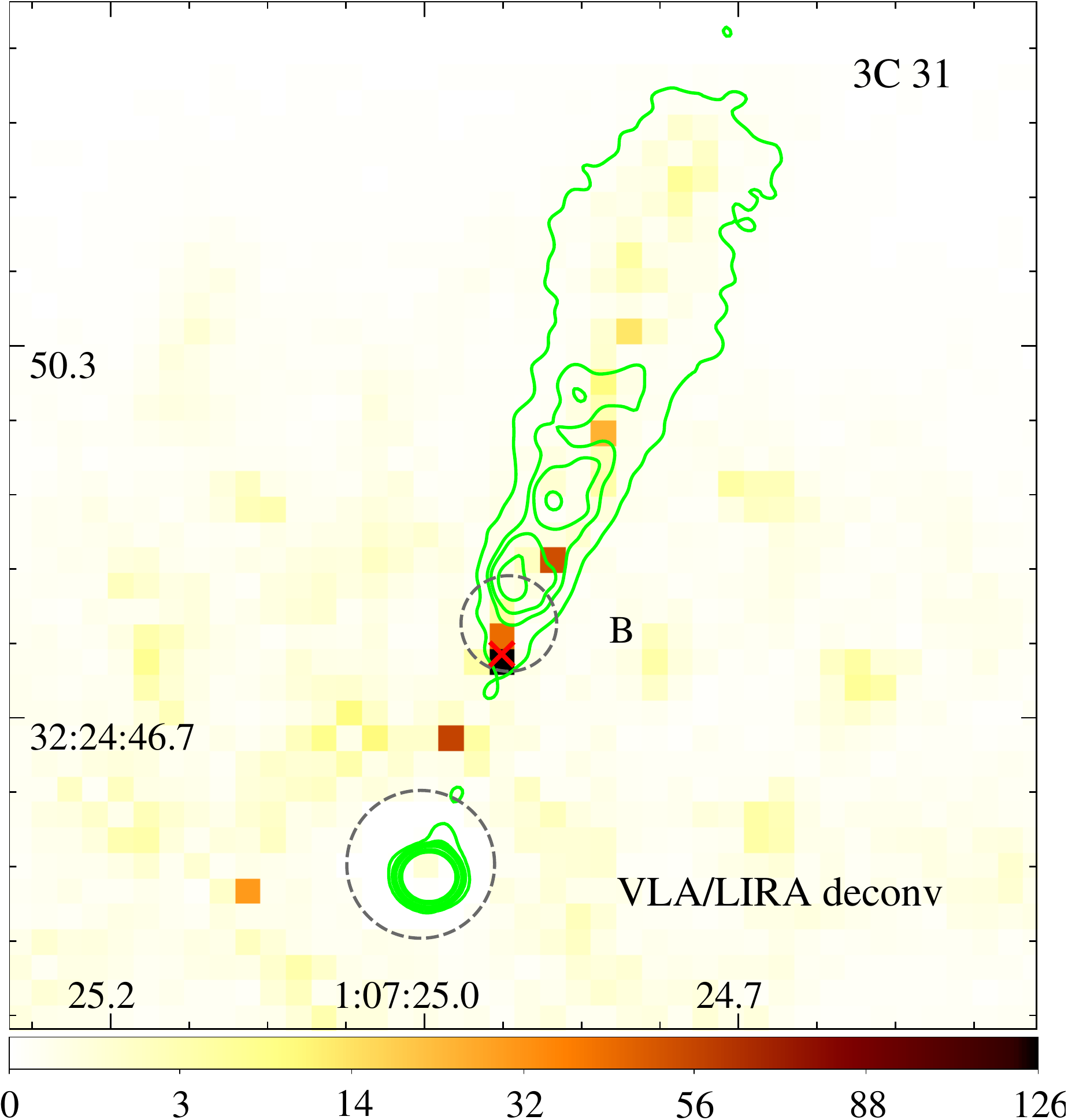}{0.5\textwidth}{(b)}
    }
    \caption{Same as in Fig. \ref{fig:results-3C9} but for 3C 31. The radio contours are given by 0.1, 0.3, 0.4, 0.6, 0.8, 2.0, 3.5 mJy beam$^{-1}$.\label{fig:results-3C31}}
\end{figure*}

\begin{figure*}[ht]
    \gridline{
        \fig{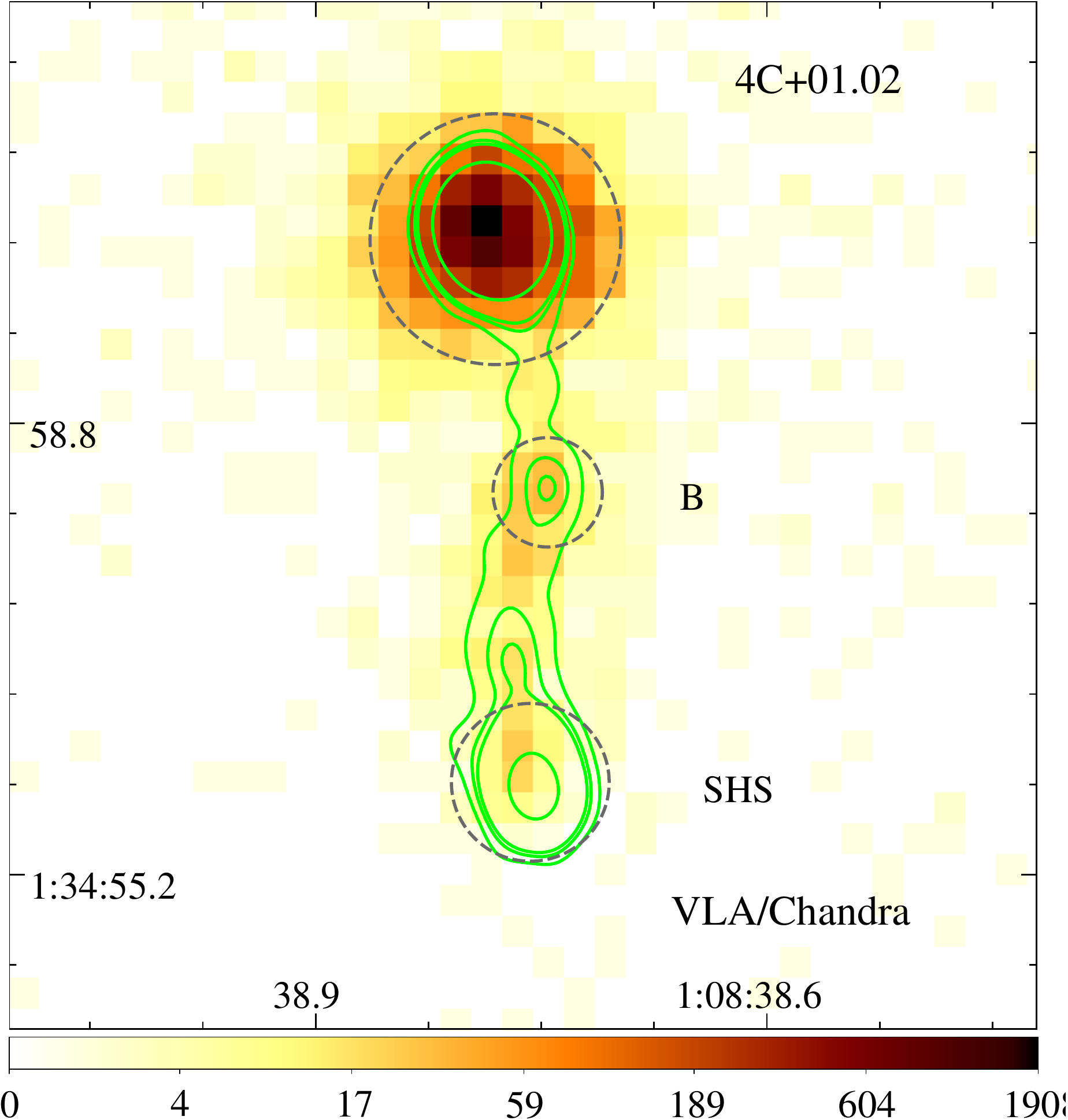}{0.5\textwidth}{(a)}
        \fig{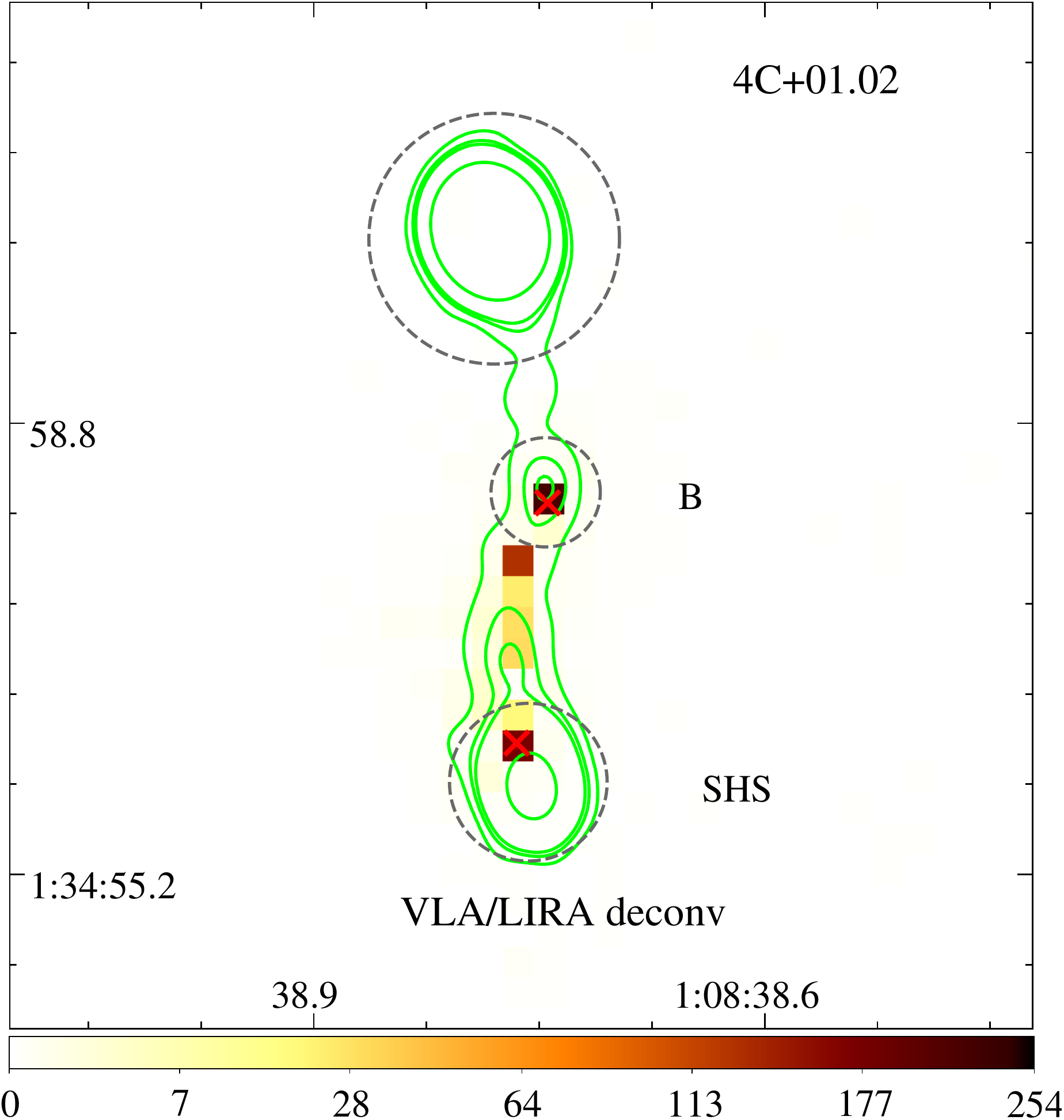}{0.5\textwidth}{(b)}
    }
    \caption{Same as in Fig. \ref{fig:results-3C9} but for 4C +01.02. The radio contours are given by 1.5, 4.0, 6.0, 50.0 mJy beam$^{-1}$.\label{fig:results-4C+01.02}}
\end{figure*}

\begin{figure*}[ht]
    \gridline{
        \fig{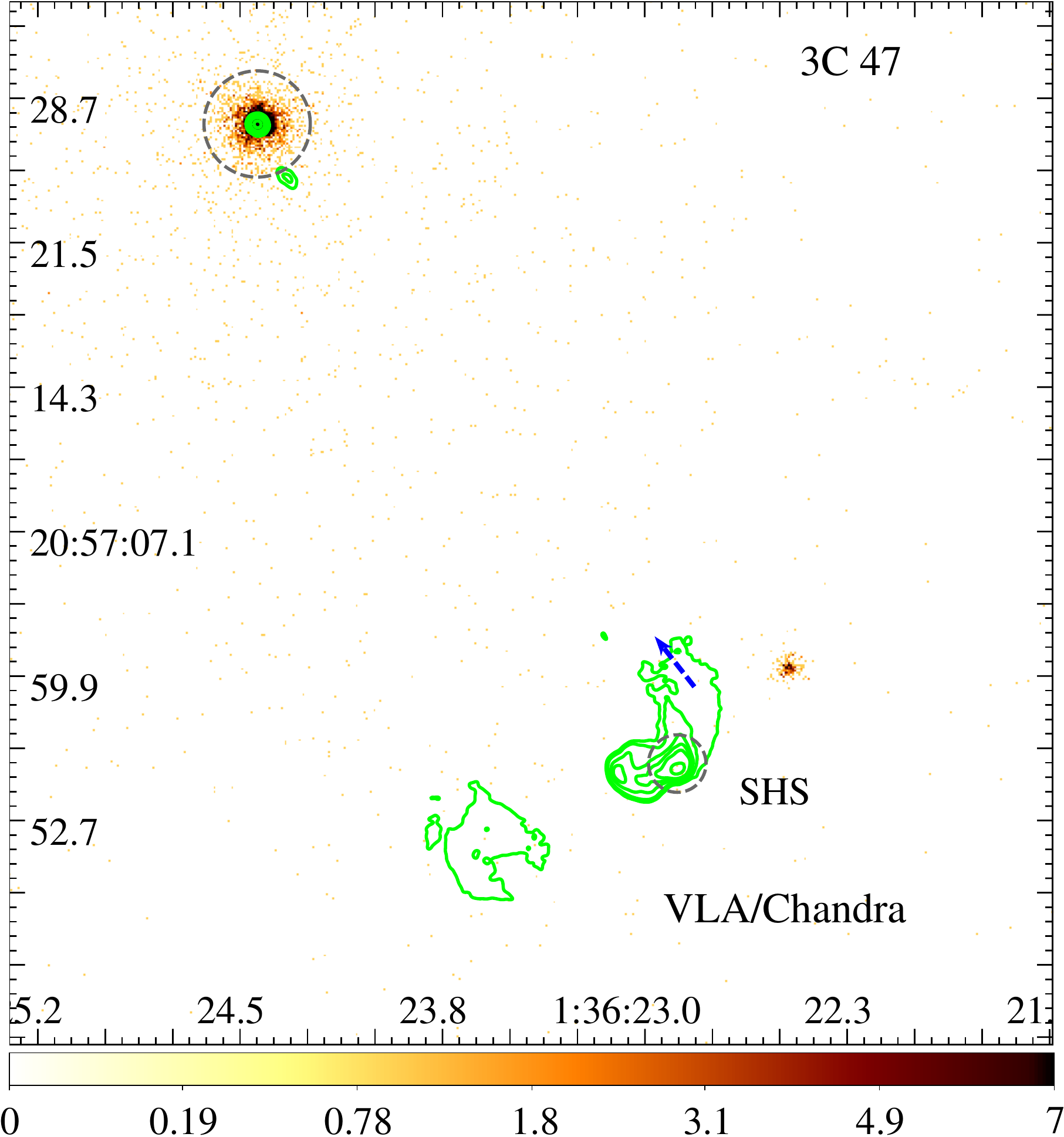}{0.5\textwidth}{(a)}
    }
    \gridline{
        \fig{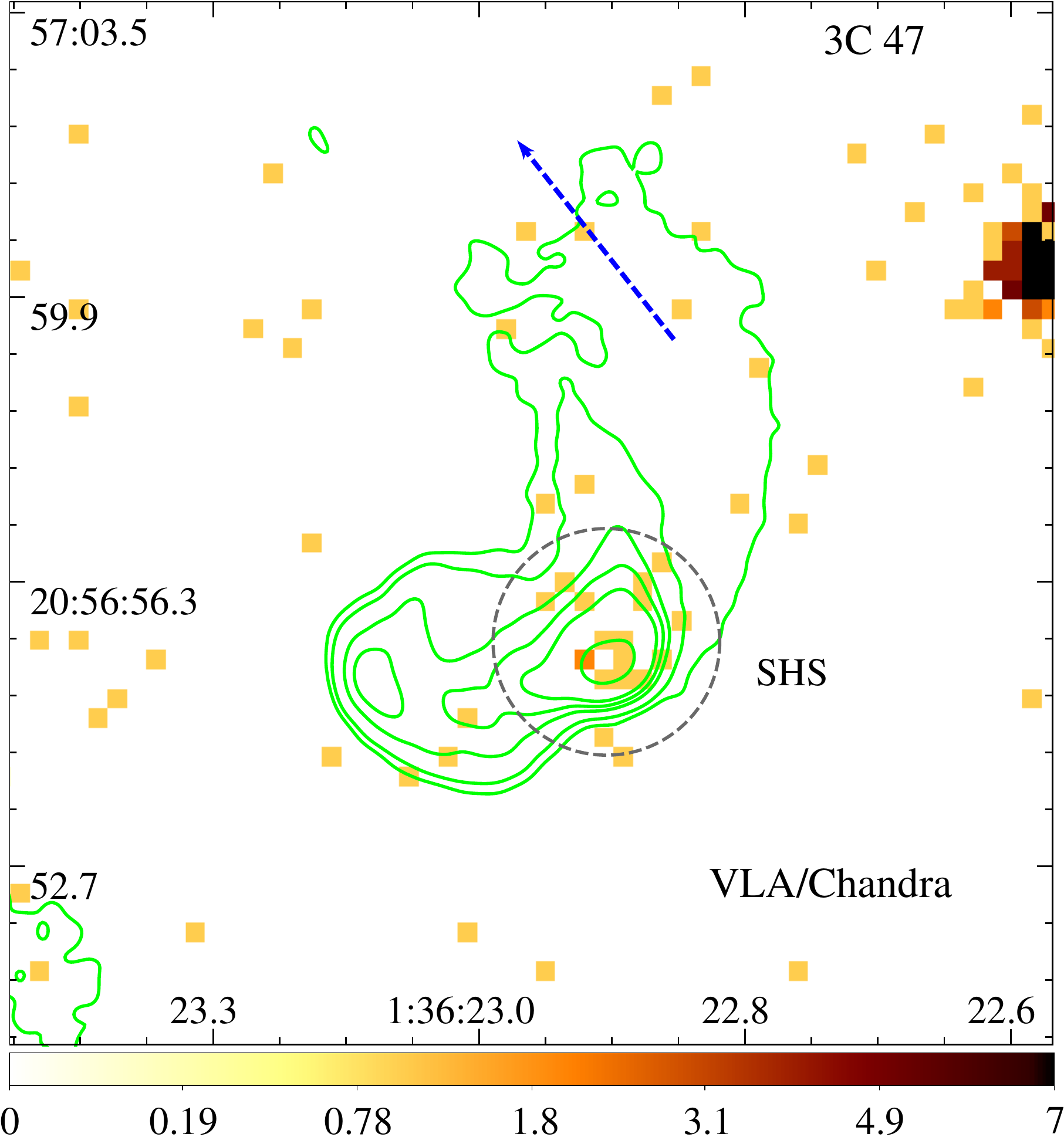}{0.5\textwidth}{(b)}
        \fig{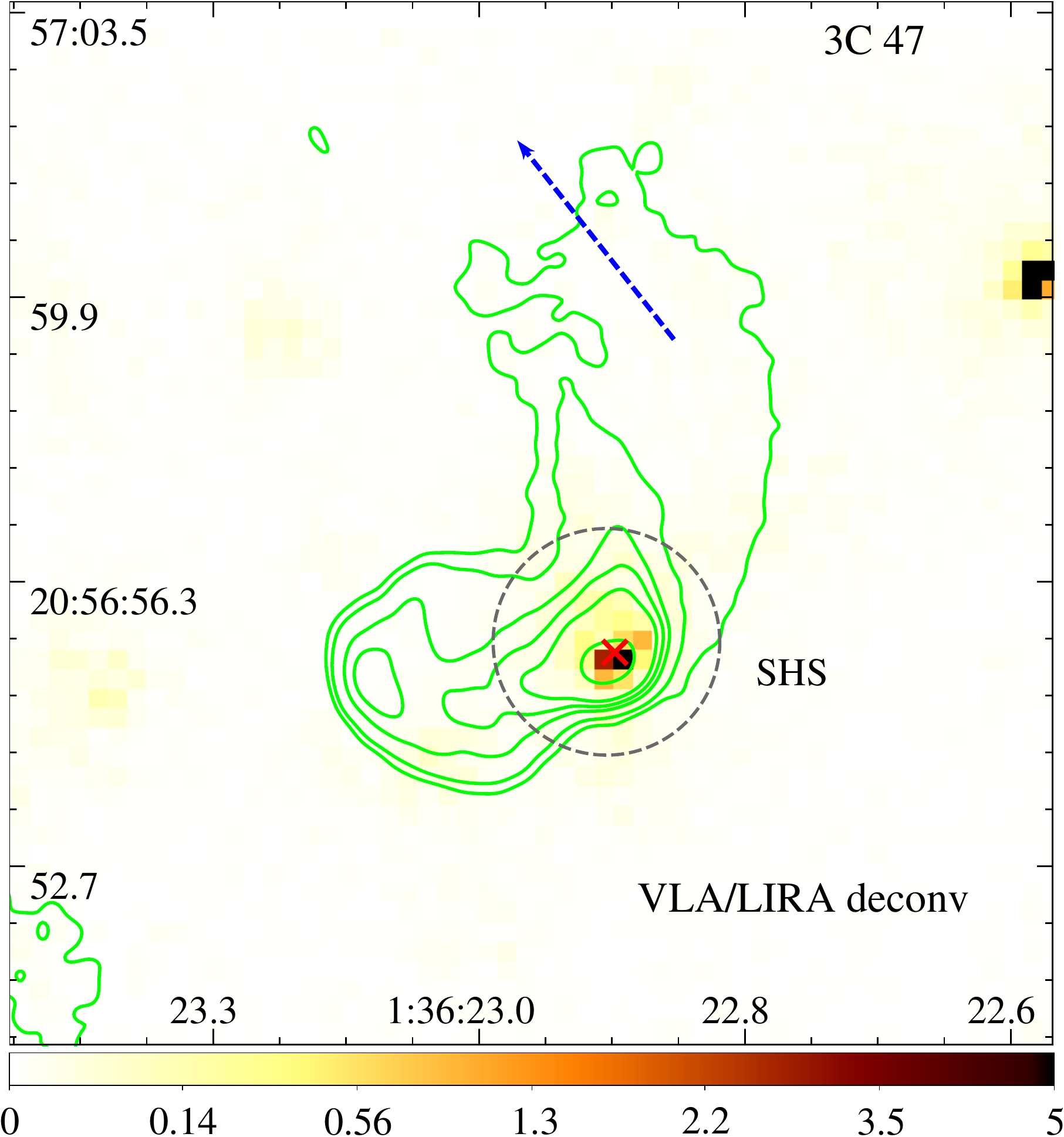}{0.5\textwidth}{(c)}
    }
    \caption{Same as in Fig. \ref{fig:results-3C9} but for 3C 47. The radio contours are given by 0.2, 0.5, 2.0, 4.0, 8.0, 40.0 mJy beam$^{-1}$.\label{fig:results-3C47}}
\end{figure*}

\begin{figure*}[ht]
    \gridline{
        \fig{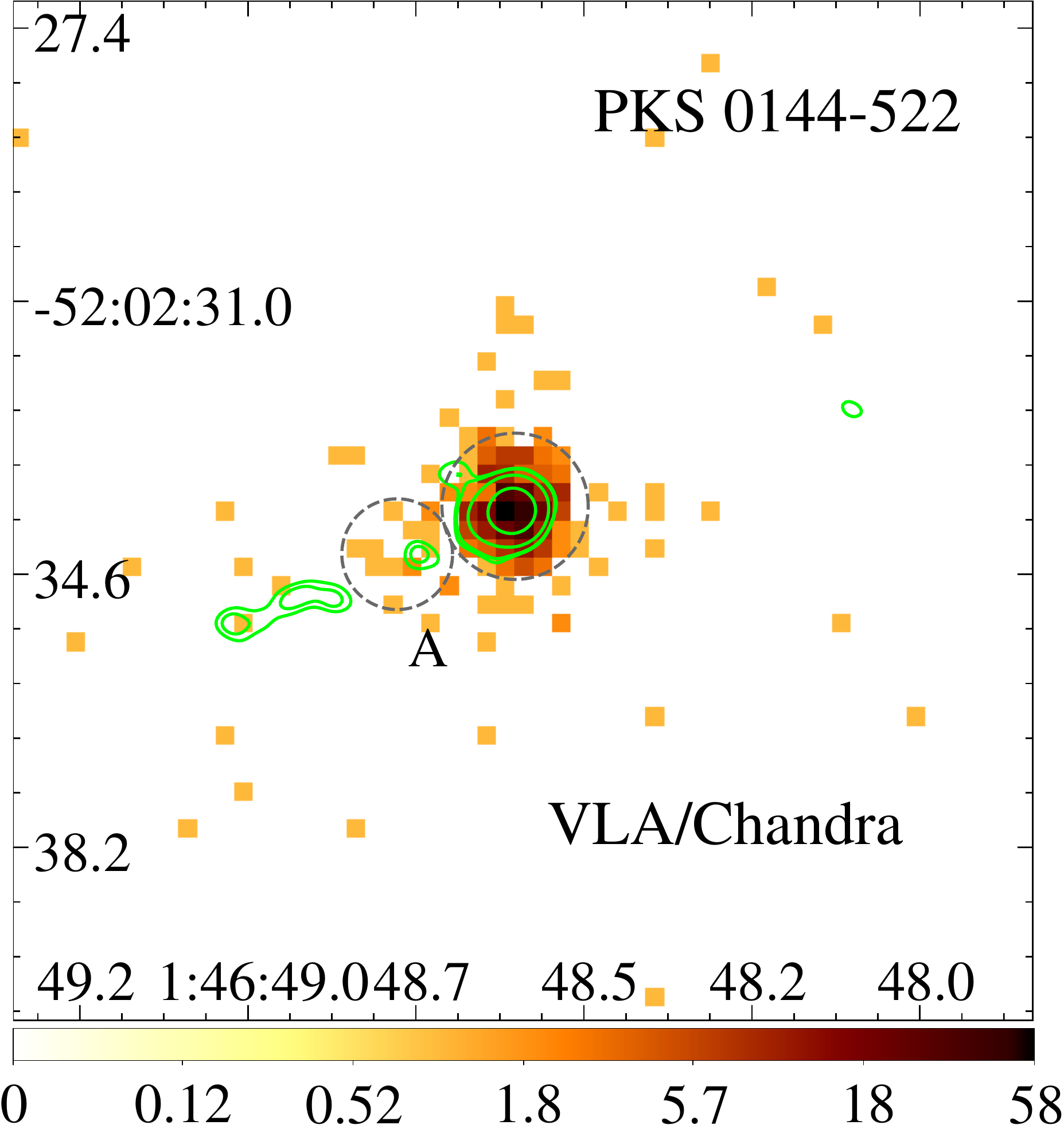}{0.5\textwidth}{(a)}
        \fig{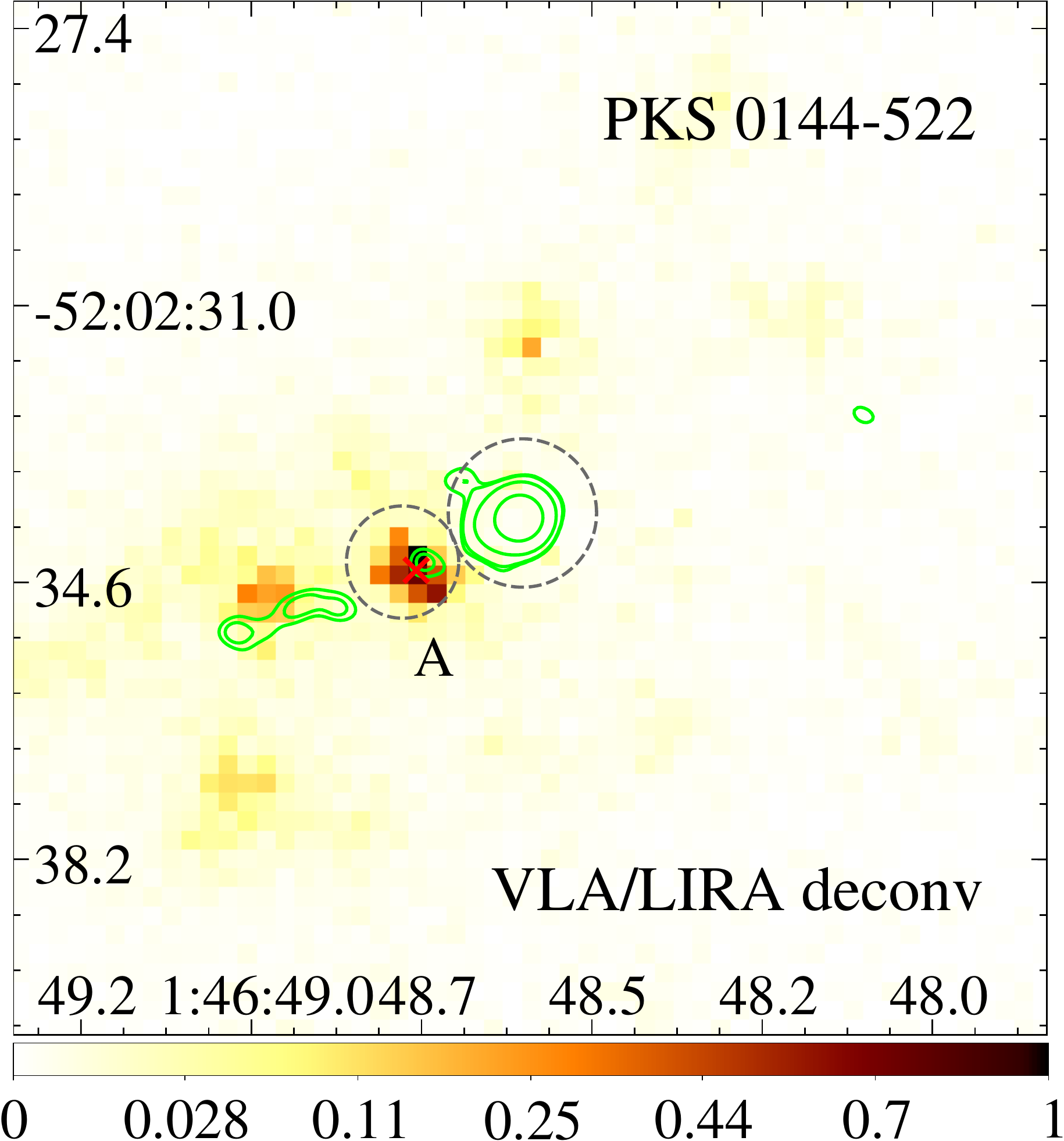}{0.5\textwidth}{(b)}
    }
    \caption{Same as in Fig. \ref{fig:results-3C9} but for PKS 0144-522. The radio contours are given by 0.5, 0.7, 4.0, 40.0 mJy beam$^{-1}$.\label{fig:results-PKS0144-522}}
\end{figure*}

\begin{figure*}[ht]
    \gridline{
        \fig{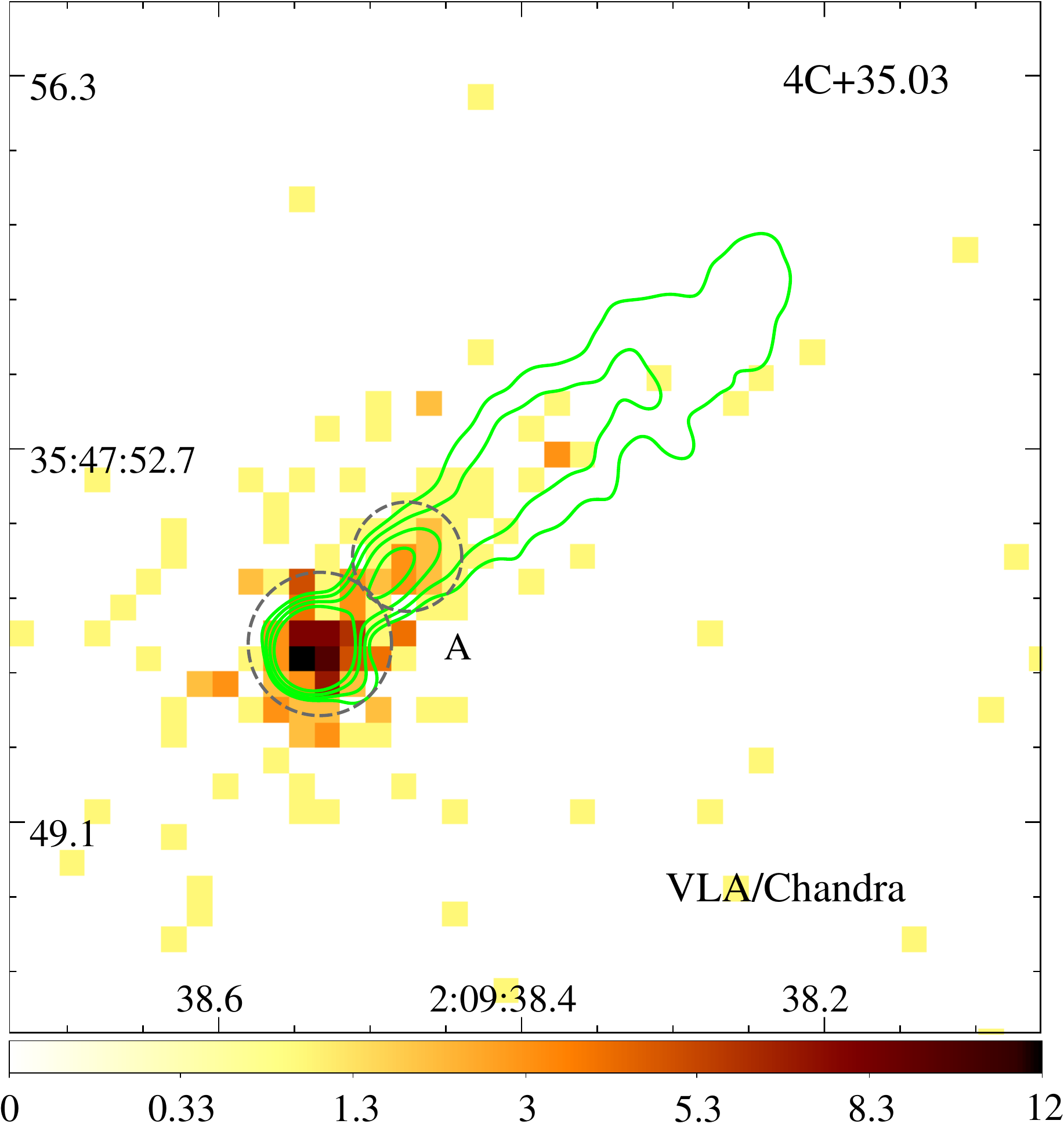}{0.5\textwidth}{(a)}
        \fig{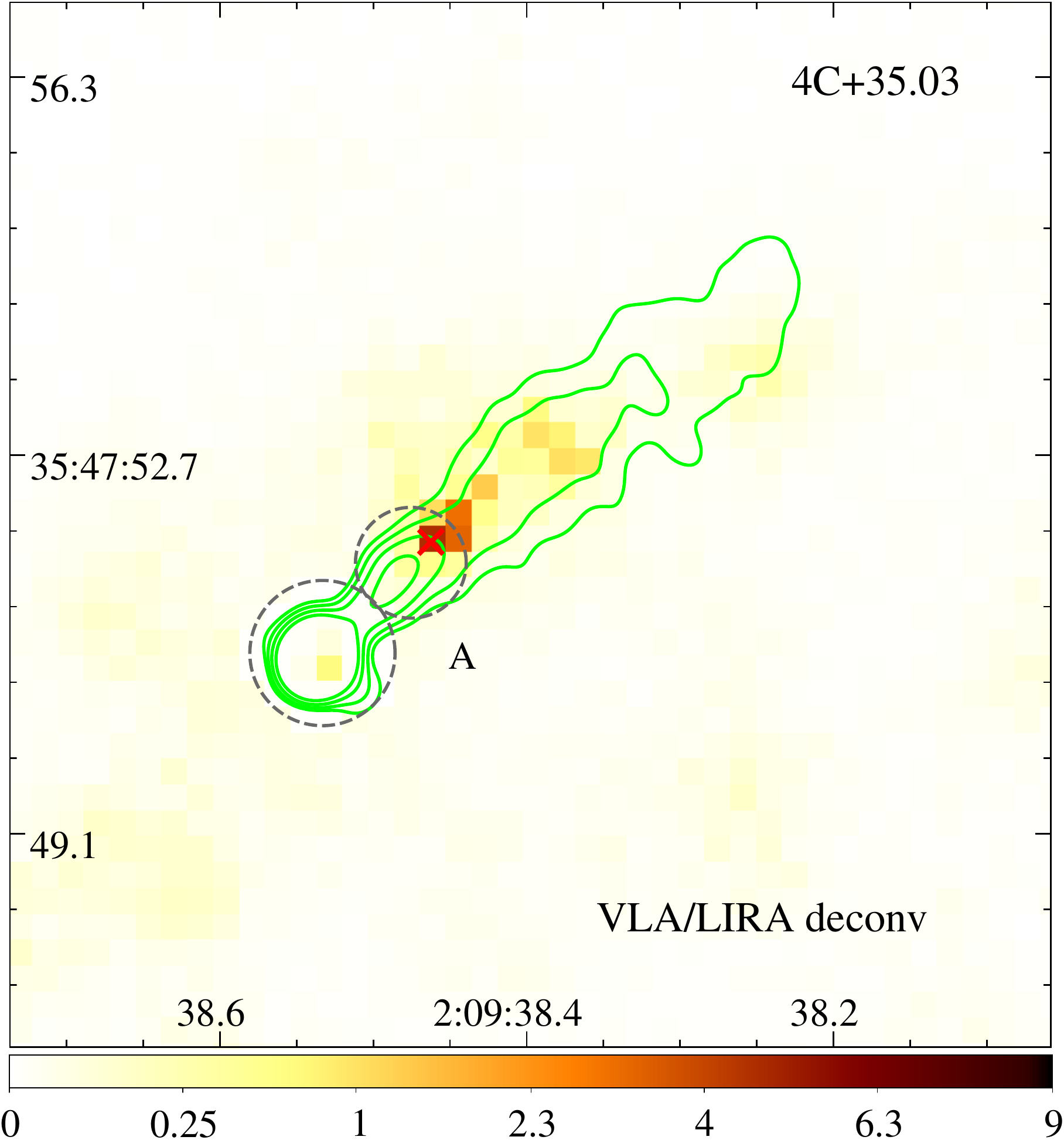}{0.5\textwidth}{(b)}
    }
    \caption{Same as in Fig. \ref{fig:results-3C9} but for 4C +35.03. The radio contours are given by 0.5, 1.0, 2.0, 4.0 mJy beam$^{-1}$.\label{fig:results-4C+35.03}}
\end{figure*}

\begin{figure*}[ht]
    \gridline{
        \fig{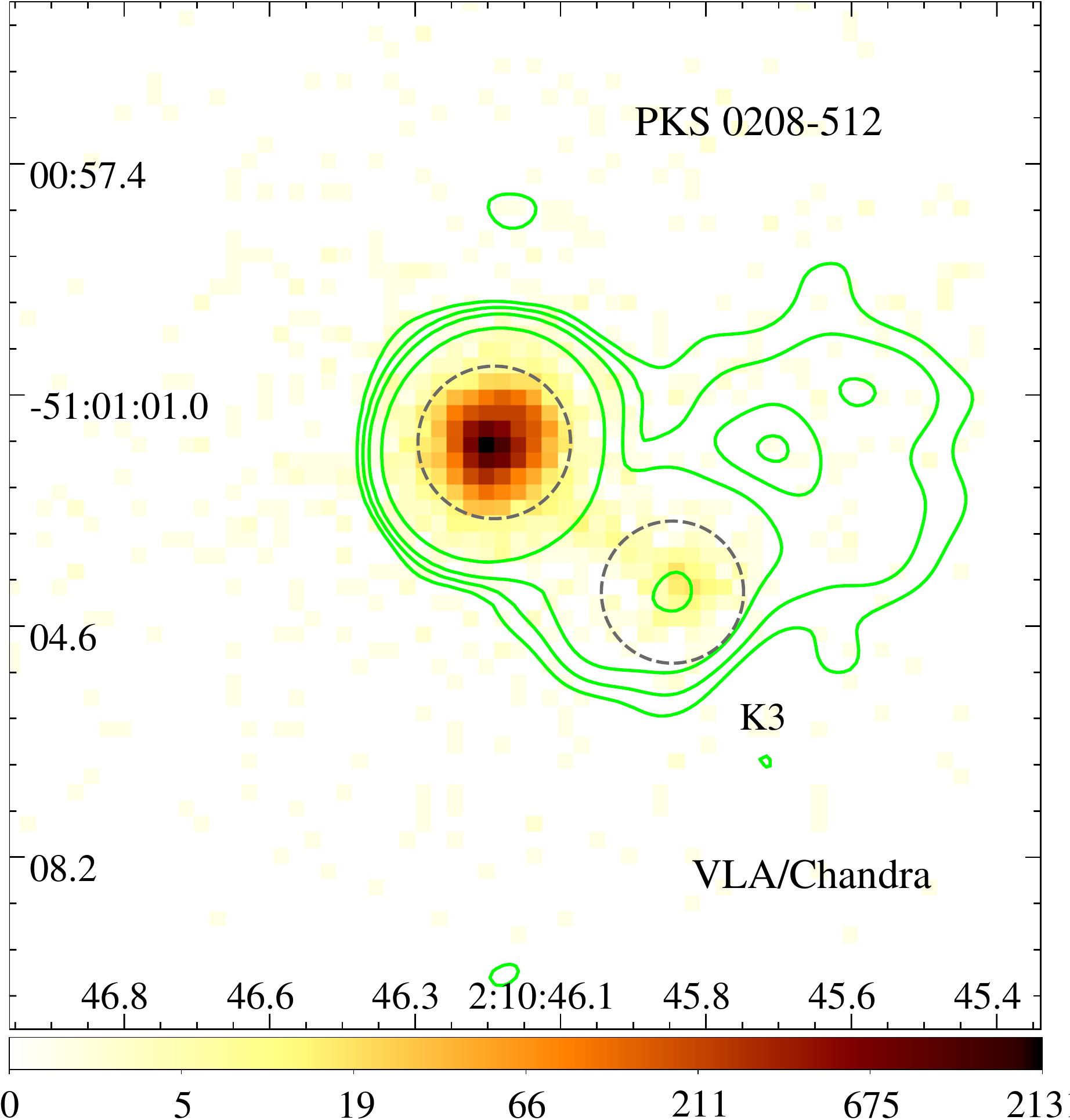}{0.5\textwidth}{(a)}
        \fig{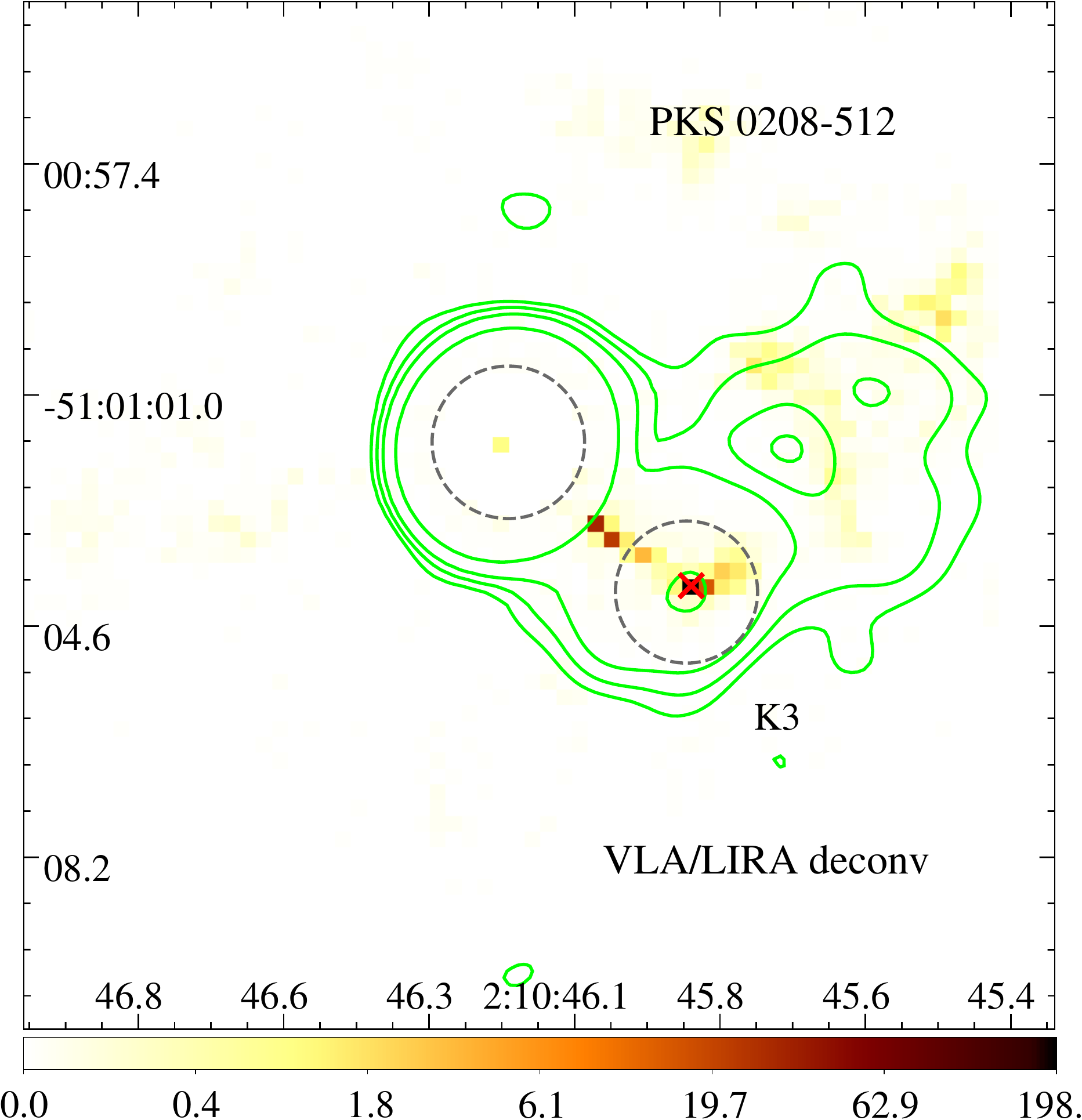}{0.5\textwidth}{(b)}
    }
    \caption{Same as in Fig. \ref{fig:results-3C9} but for PKS 0208-512. The radio contours are given by 1.0, 2.0, 4.0, 15.0 mJy beam$^{-1}$.\label{fig:results-PKS0208-512}}
\end{figure*}

\begin{figure*}[ht]
    \gridline{
        \fig{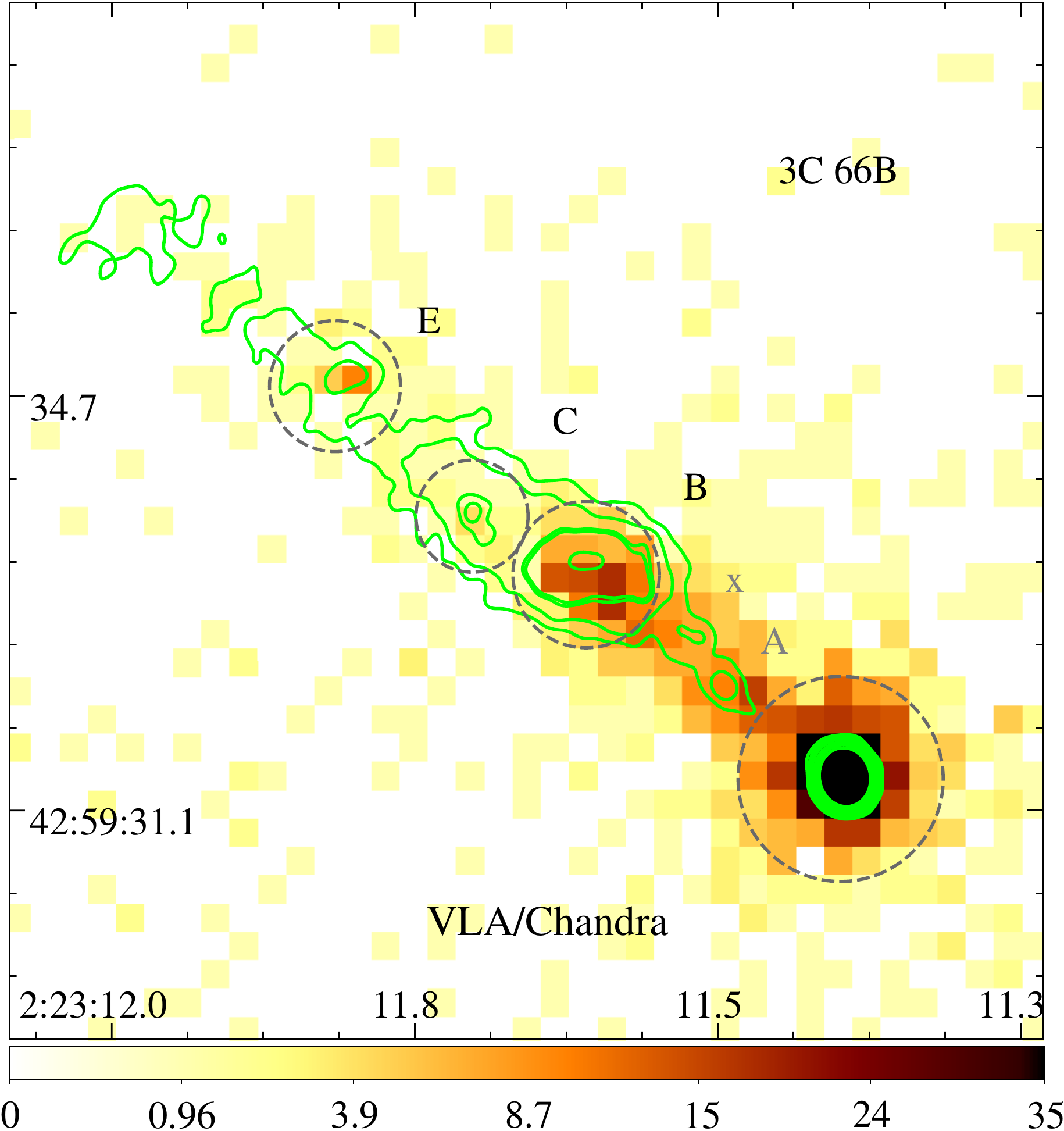}{0.5\textwidth}{(a)}
        \fig{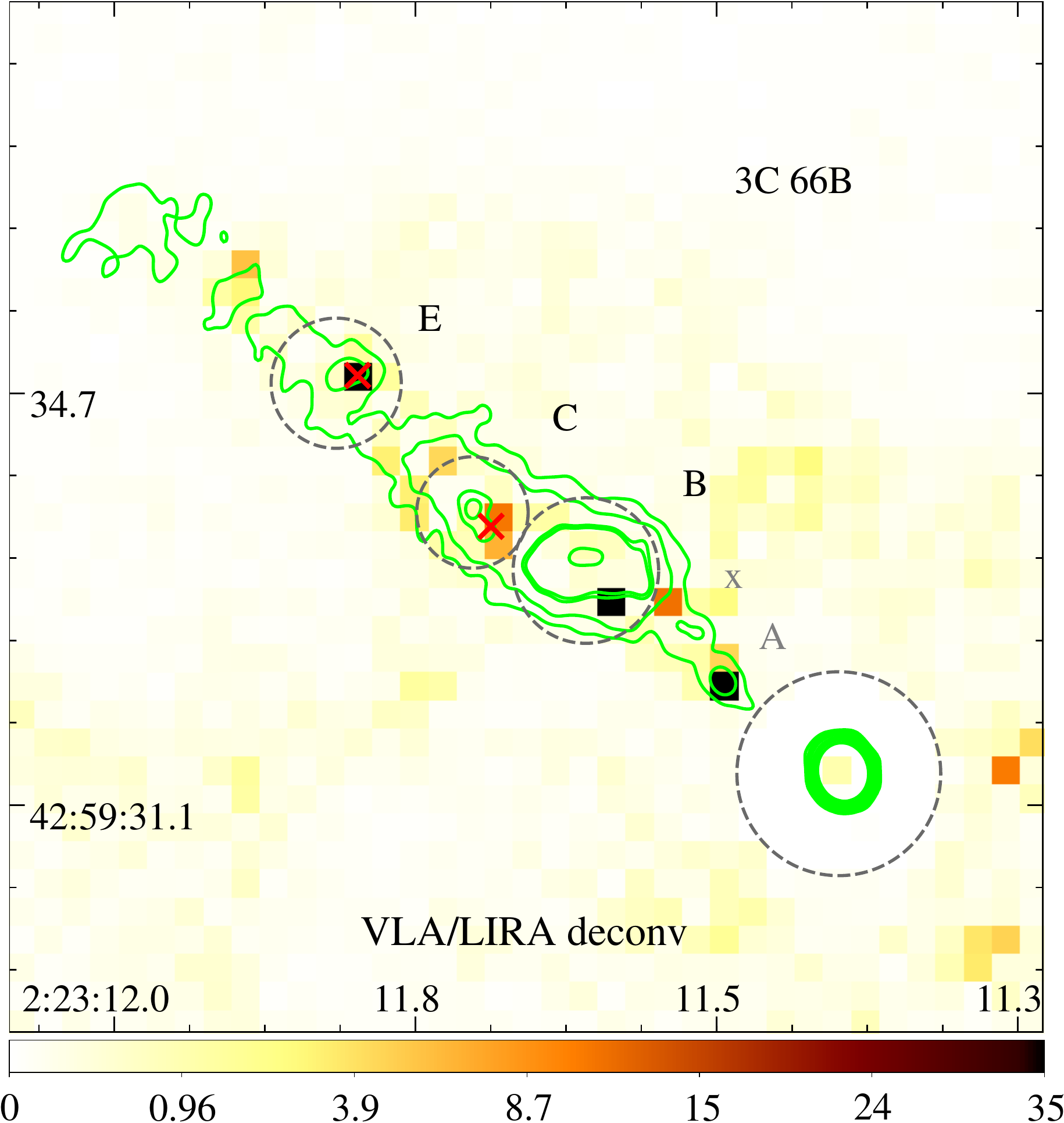}{0.5\textwidth}{(b)}
    }
    \caption{Same as in Fig. \ref{fig:results-3C9} but for 3C 66B. The radio contours are given by 0.2, 0.4, 0.8, 0.9, 2.0, 4.0, 8.0 mJy beam$^{-1}$.\label{fig:results-3C66B}}
\end{figure*}

\begin{figure*}[ht]
    \gridline{
        \fig{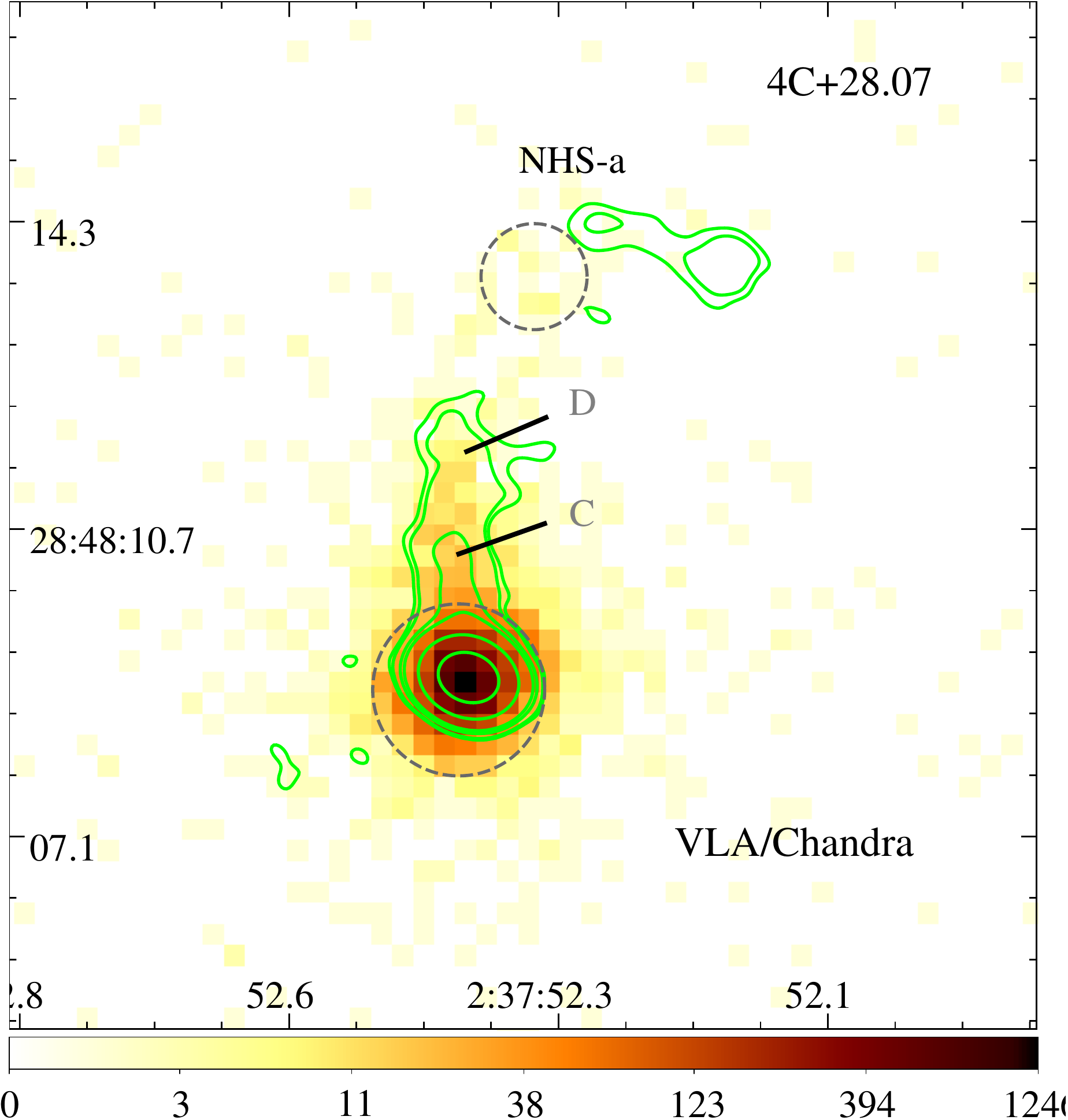}{0.5\textwidth}{(a)}
        \fig{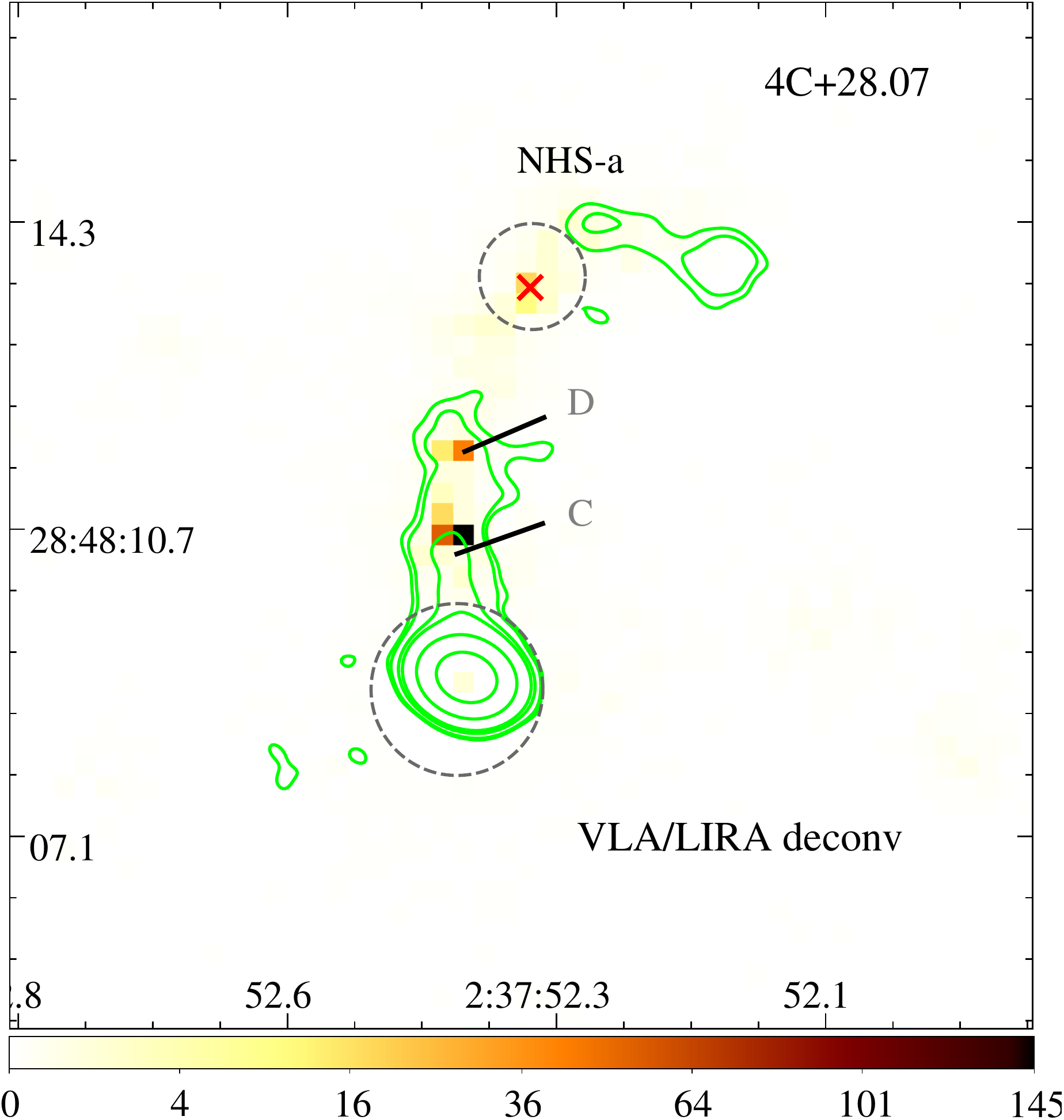}{0.5\textwidth}{(b)}
    }
    \caption{Same as in Fig. \ref{fig:results-3C9} but for 4C +28.07. The radio contours are given by 1.0, 1.5, 5.0, 10.0, 100.0, 1000.0 mJy beam$^{-1}$.\label{fig:results-4C+28.07}}
\end{figure*}

\begin{figure*}[ht]
    \gridline{
        \fig{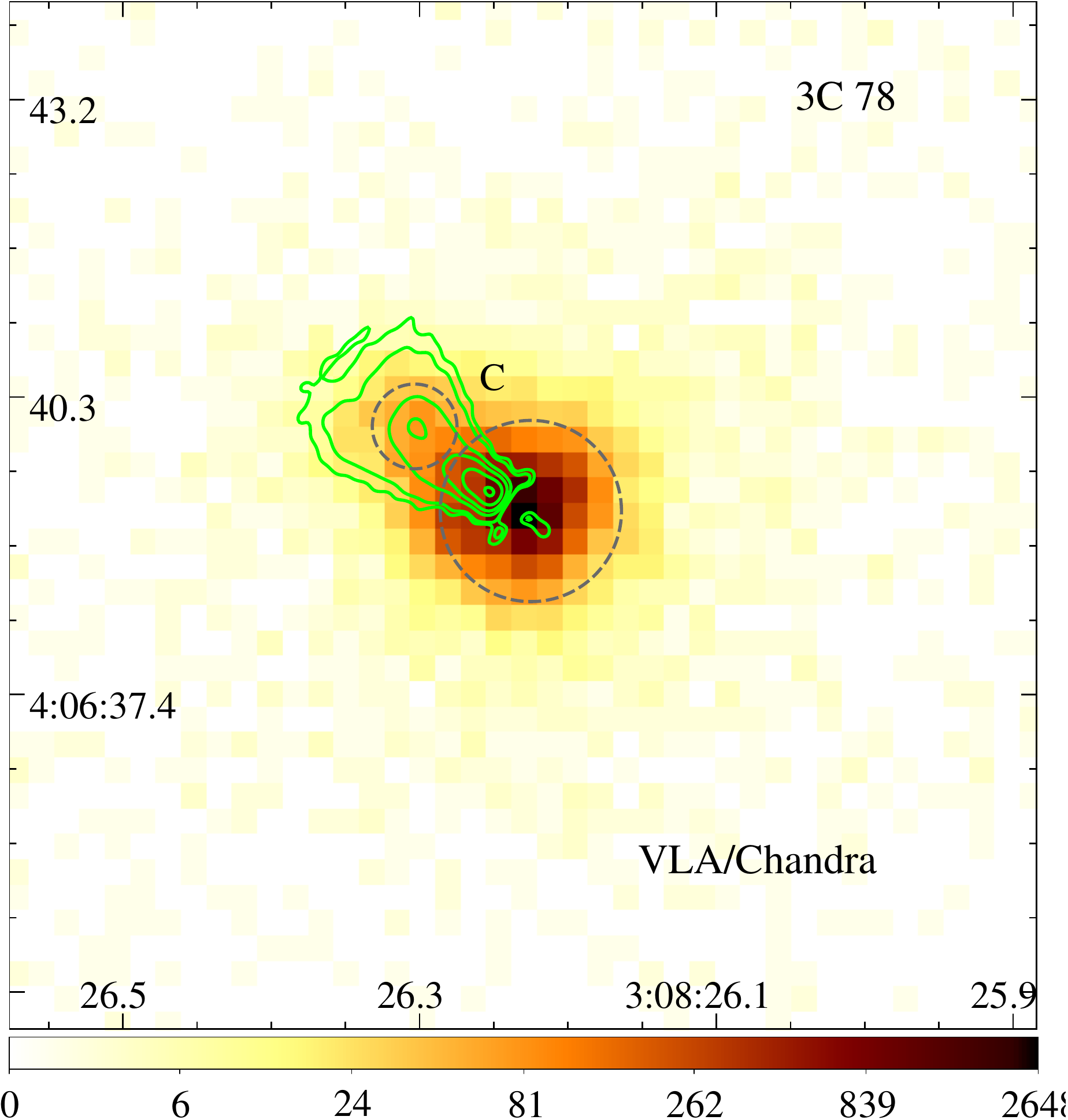}{0.5\textwidth}{(a)}
        \fig{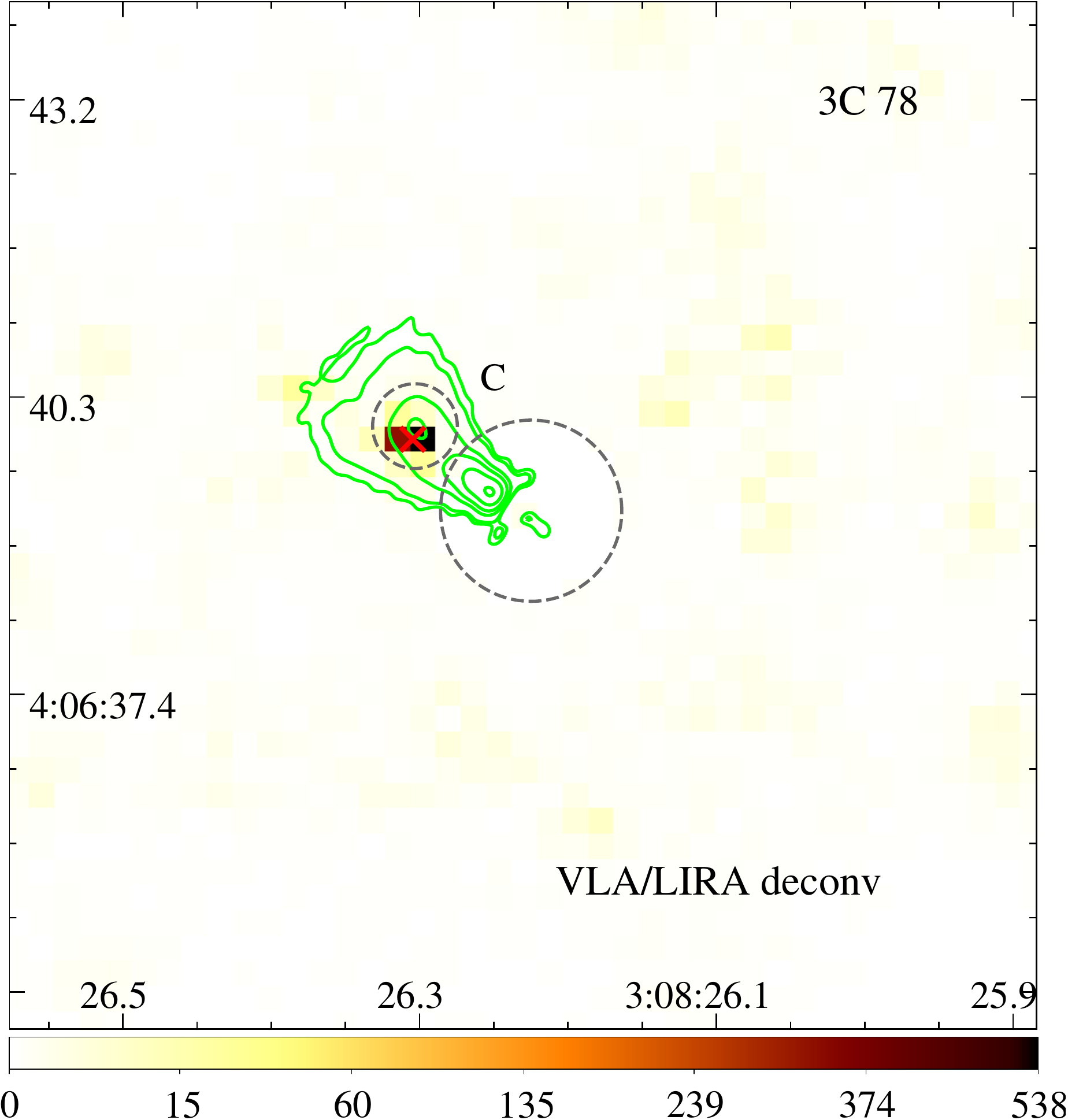}{0.5\textwidth}{(b)}
    }
    \caption{Same as in Fig. \ref{fig:results-3C9} but for 3C 78. The radio contours are given by 0.5, 1.0, 5.0, 10.0, 20.0, 35.0 mJy beam$^{-1}$.\label{fig:results-3C78}}
\end{figure*}

\begin{figure*}[ht]
    \gridline{
        \fig{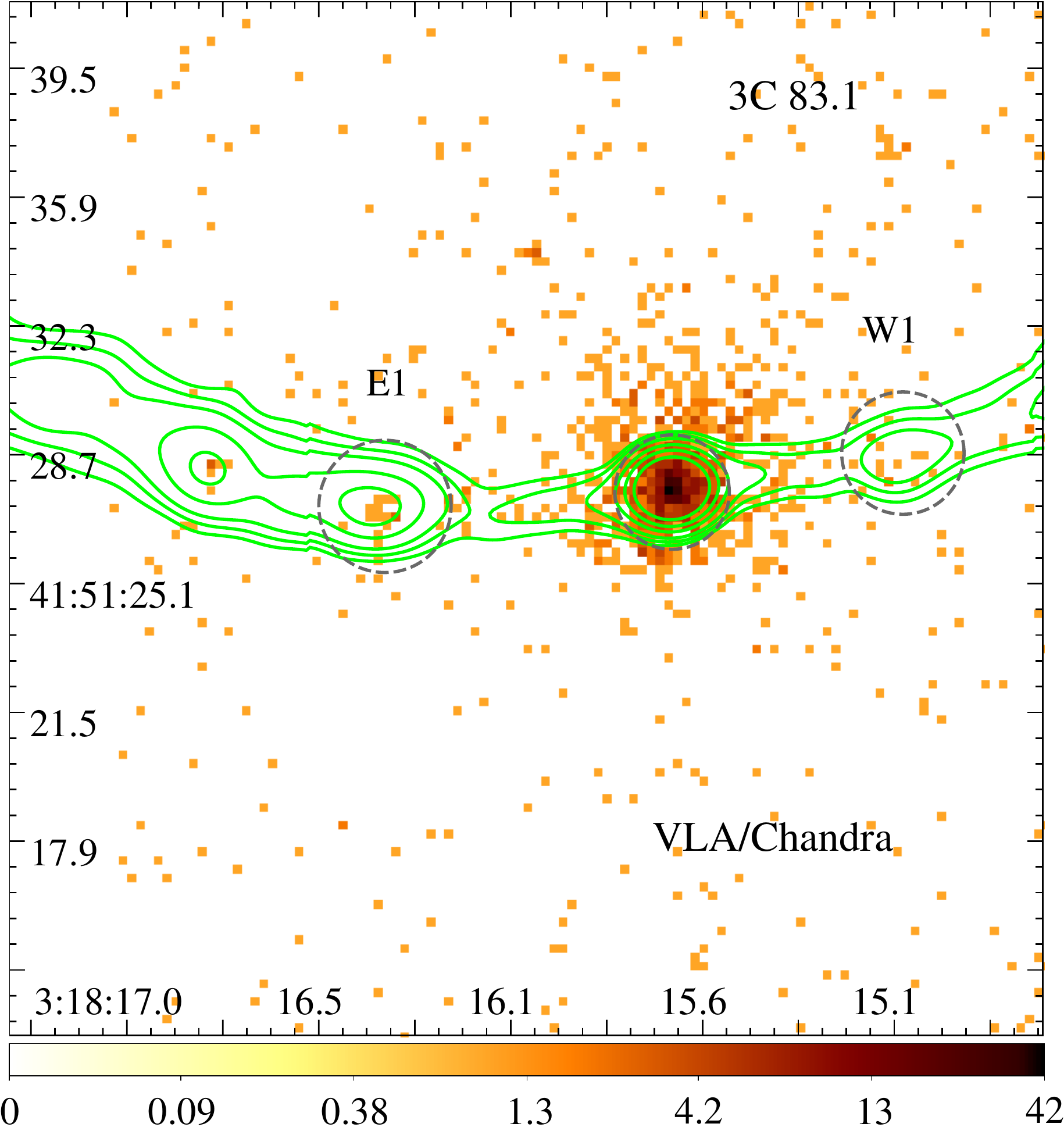}{0.5\textwidth}{(a)}
        \fig{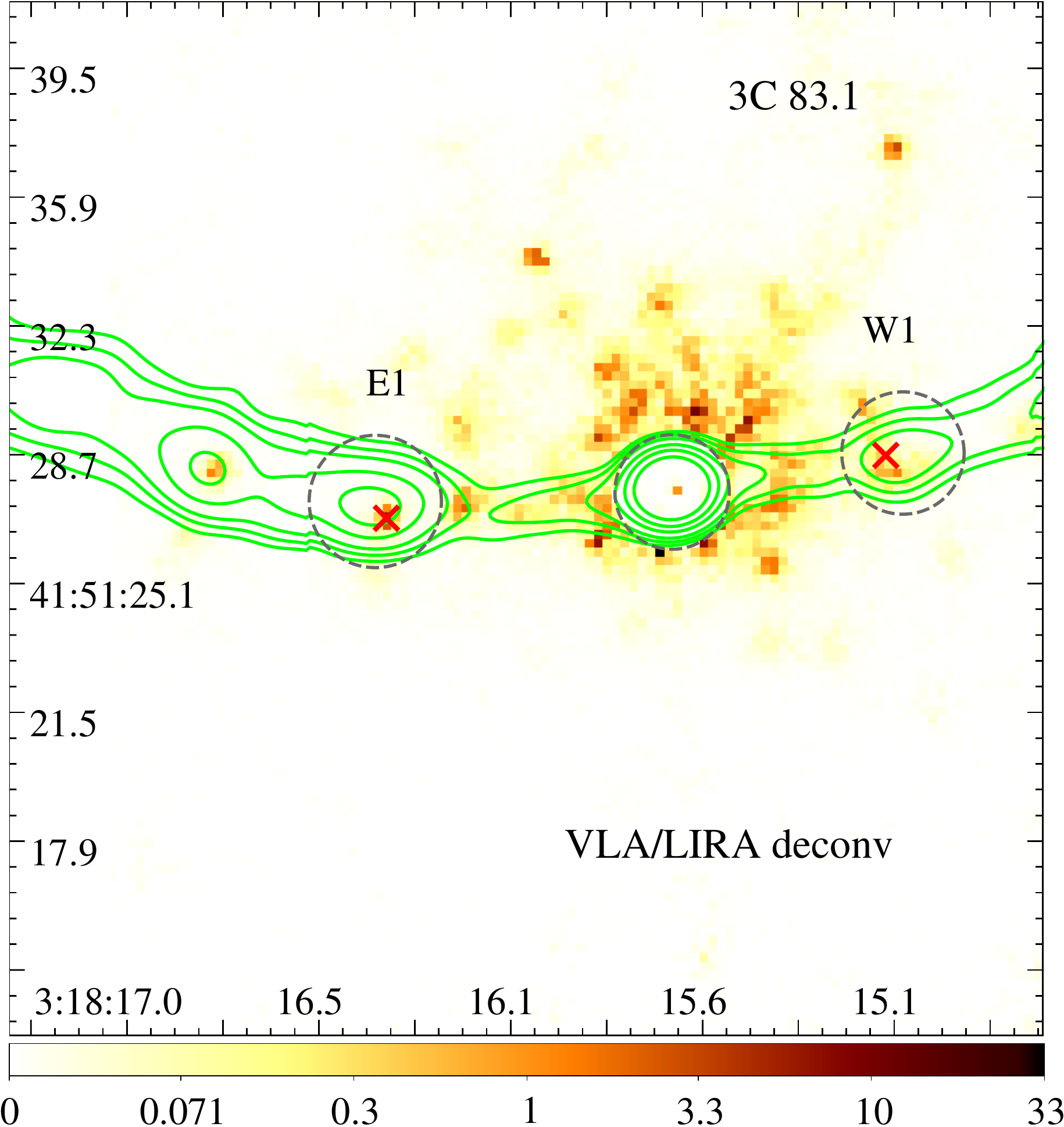}{0.5\textwidth}{(b)}
    }
    \caption{Same as in Fig. \ref{fig:results-3C9} but for 3C 83.1. The radio contours are given by 0.17, 0.4, 0.8, 2.0, 4.0, 8.0 mJy beam$^{-1}$.\label{fig:results-3C83.1}}
\end{figure*}

\begin{figure*}[ht]
    \gridline{
        \fig{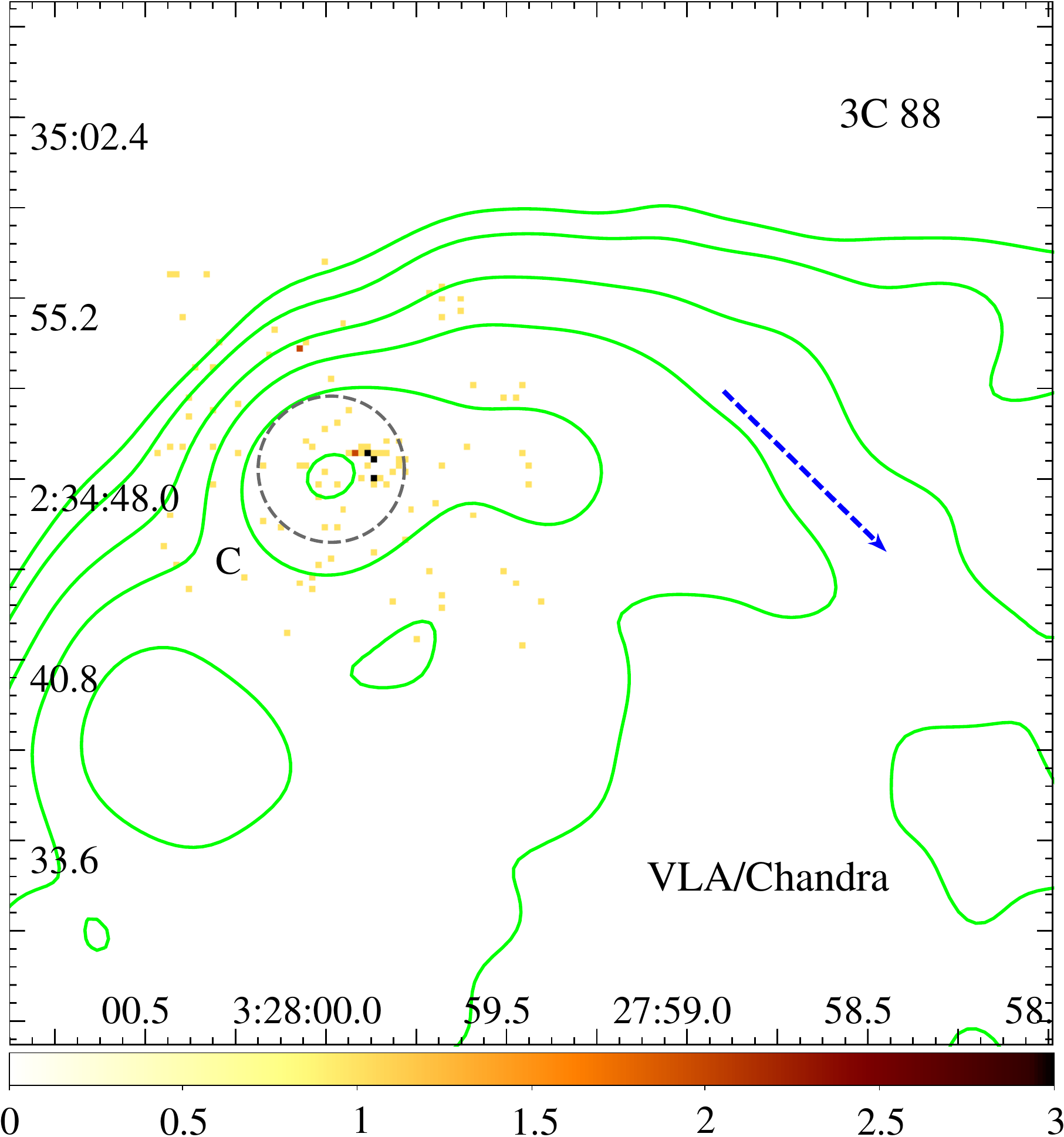}{0.5\textwidth}{(a)}
        \fig{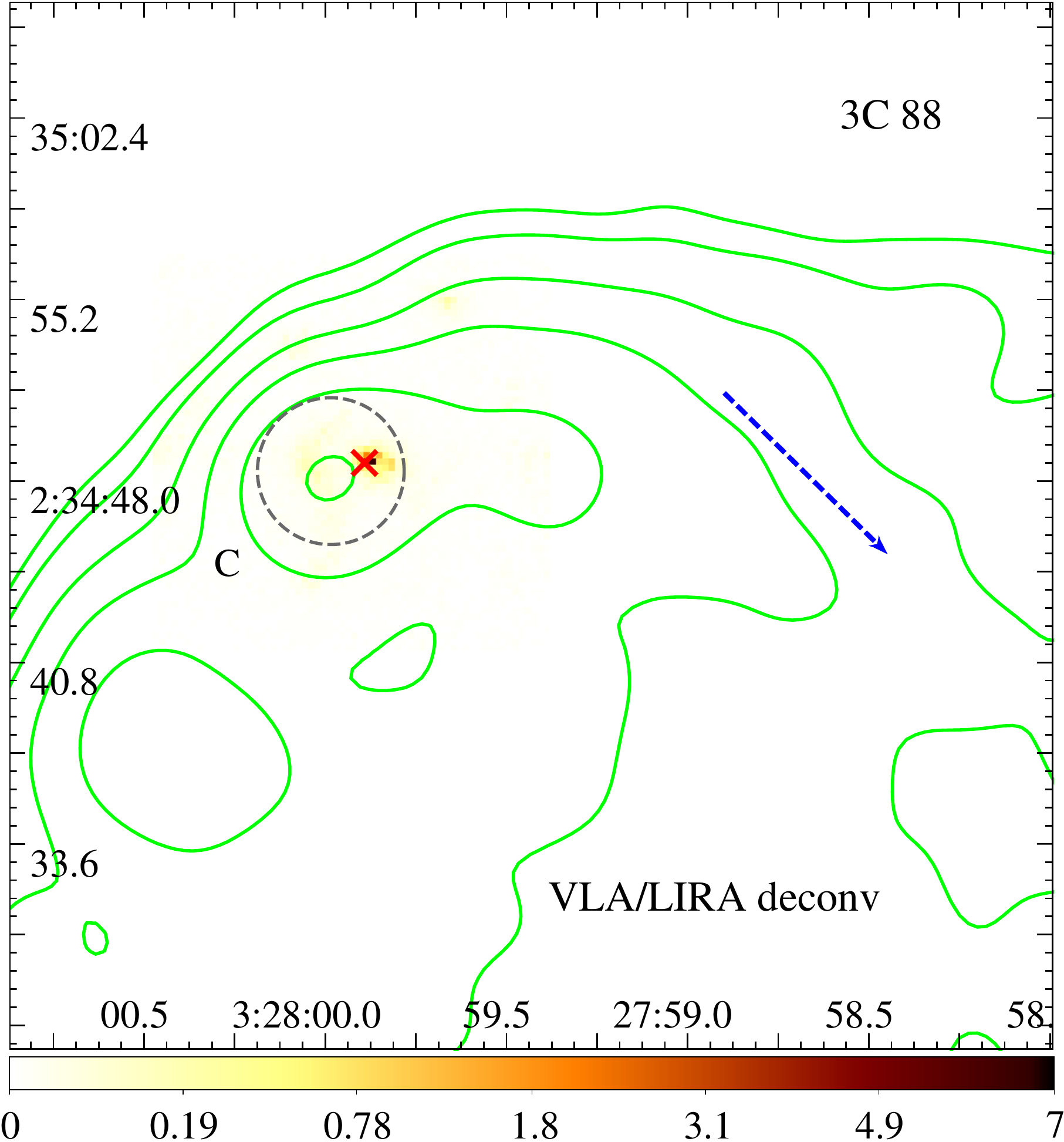}{0.5\textwidth}{(b)}
    }
    \caption{Same as in Fig. \ref{fig:results-3C9} but for 3C 88. The radio contours are given by 0.4, 0.8, 2.0, 4.0, 8.0, 18.0 mJy beam$^{-1}$.\label{fig:results-3C88}}
\end{figure*}

\begin{figure*}[ht]
    \gridline{
        \fig{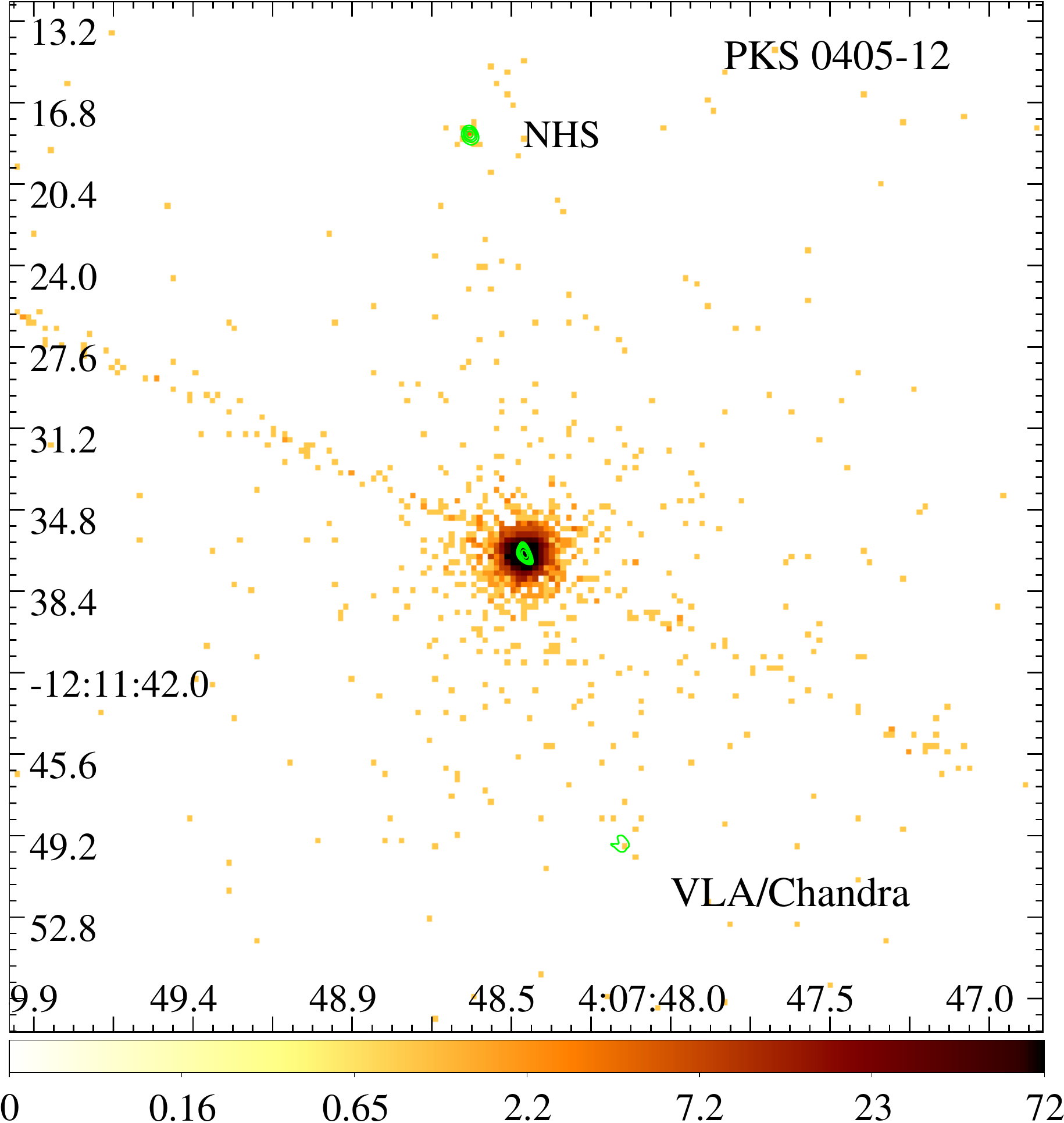}{0.5\textwidth}{(b)}
    }
    \gridline{
        \fig{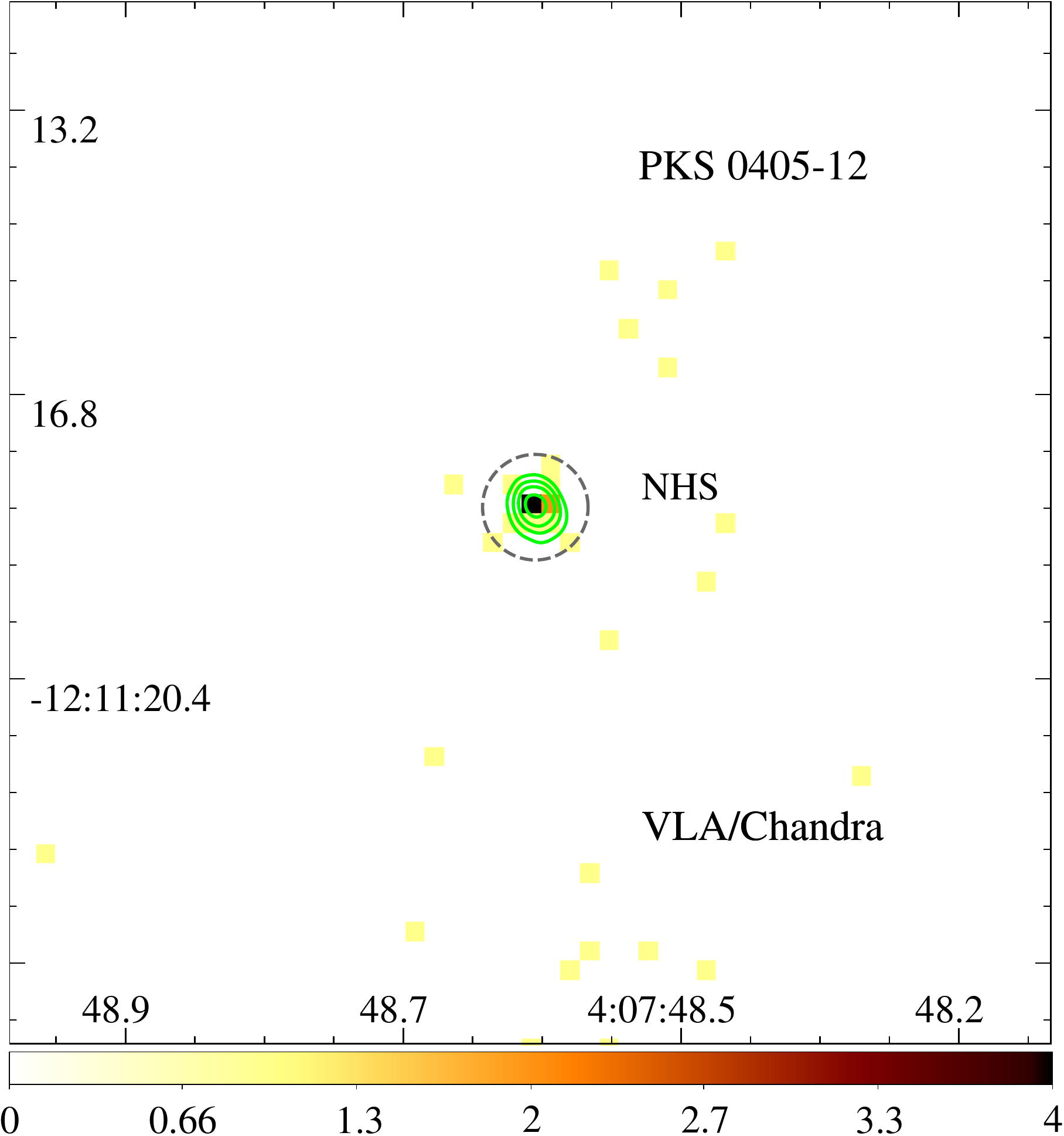}{0.5\textwidth}{(b)}
        \fig{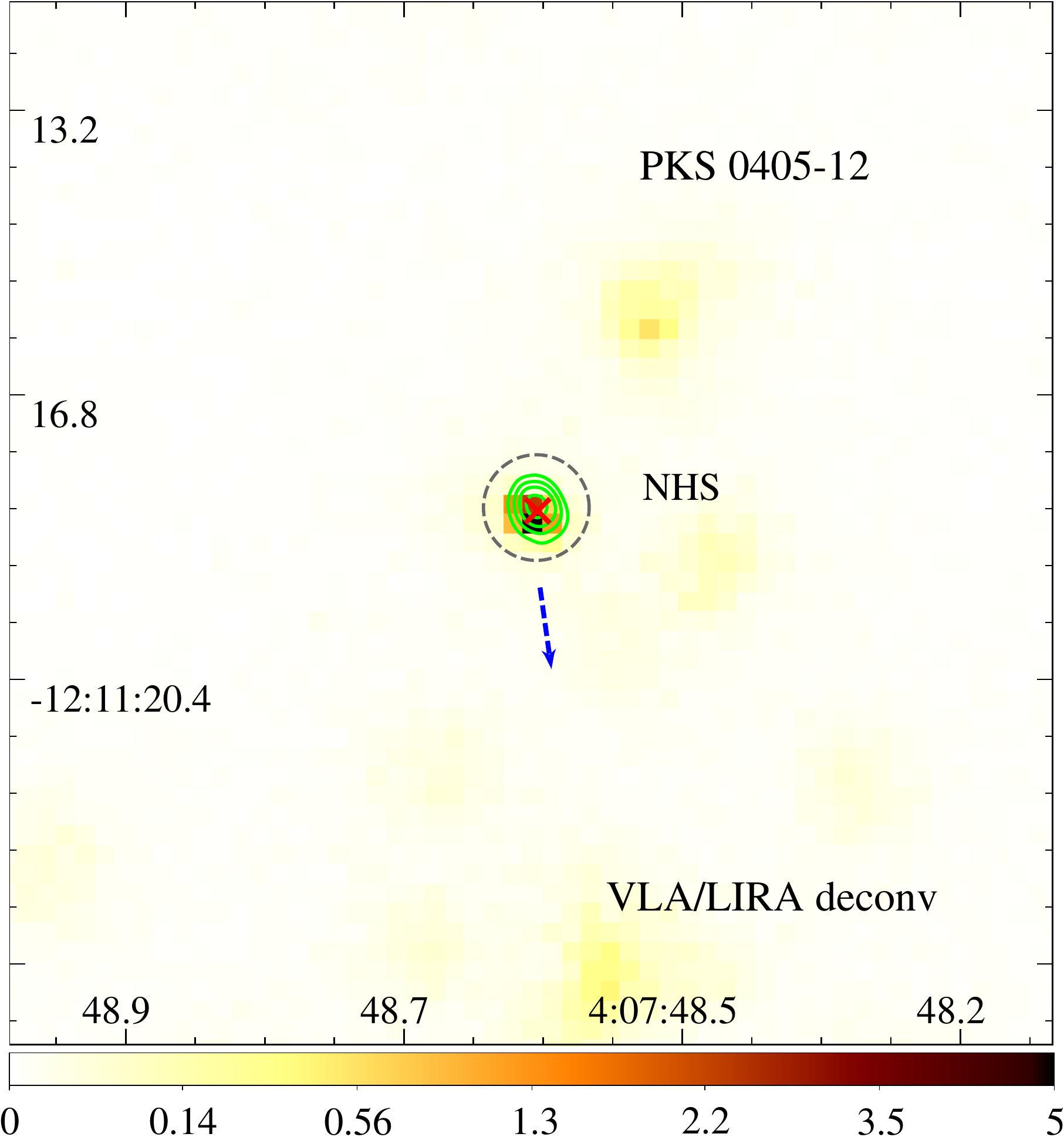}{0.5\textwidth}{(c)}
    }
    \caption{Same as in Fig. \ref{fig:results-3C9} but for PKS 0405-12. (a) shows the full image while (b) and (c) show zoomed-in region around the northern hotspot (NHS). The radio contours are given by 4.0, 10.0, 20.0, 40.0, 100.0, 400.0 mJy beam$^{-1}$.\label{fig:results-PKS0405-12}}
\end{figure*}
\comment{
\begin{figure*}[ht]
    \gridline{
        \fig{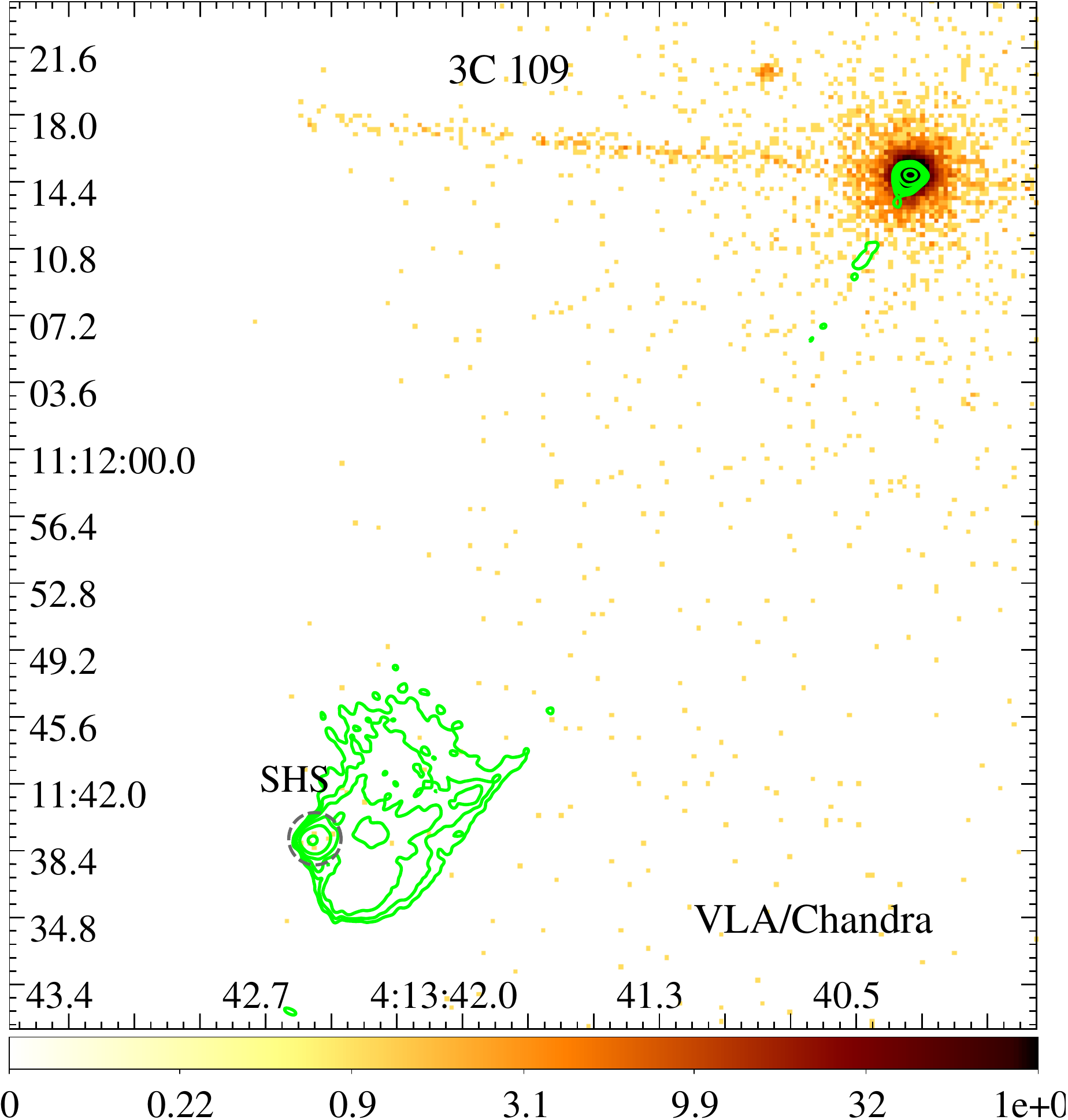}{0.5\textwidth}{(b)}
    }
    \gridline{
        \fig{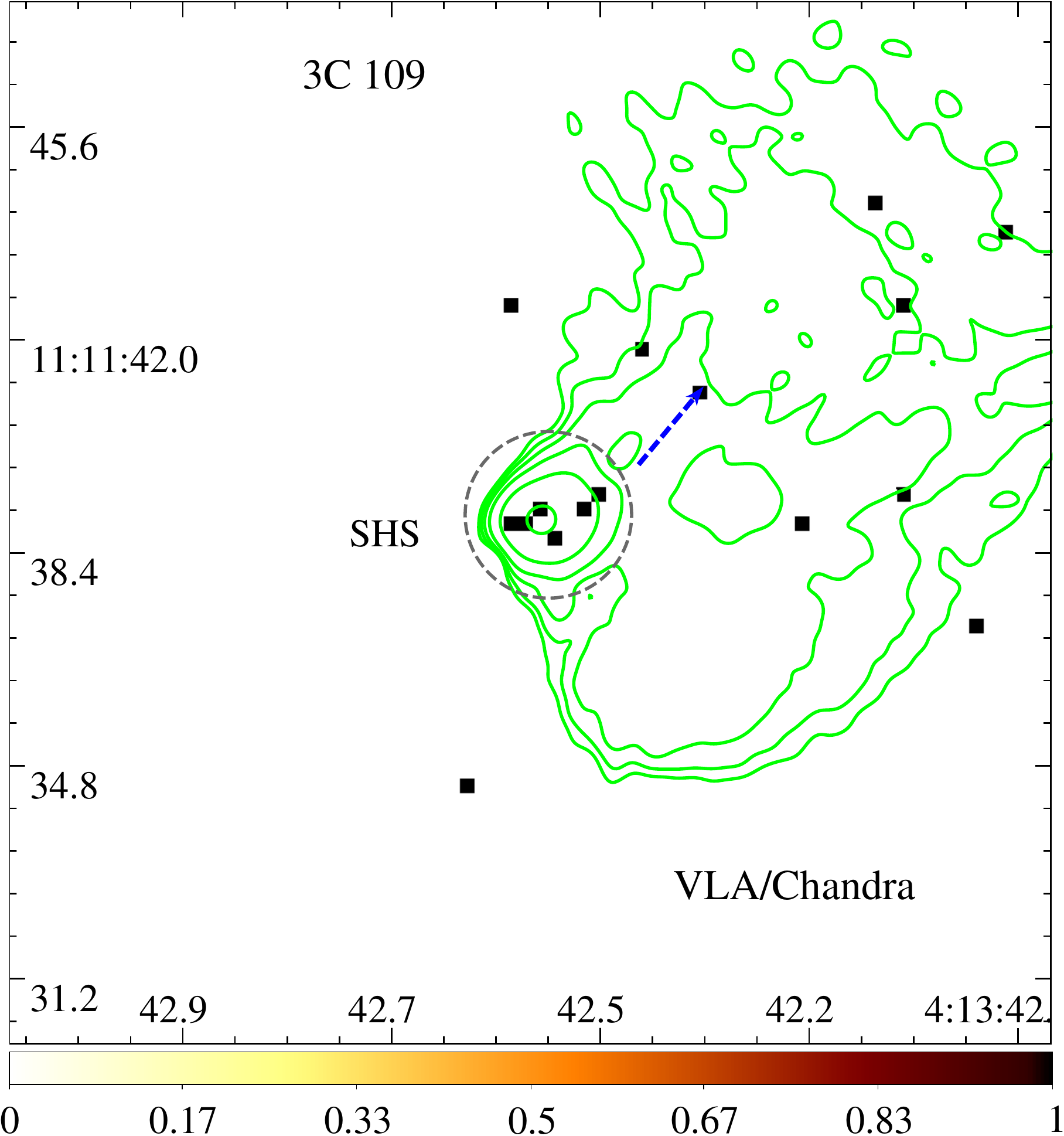}{0.5\textwidth}{(b)}
        \fig{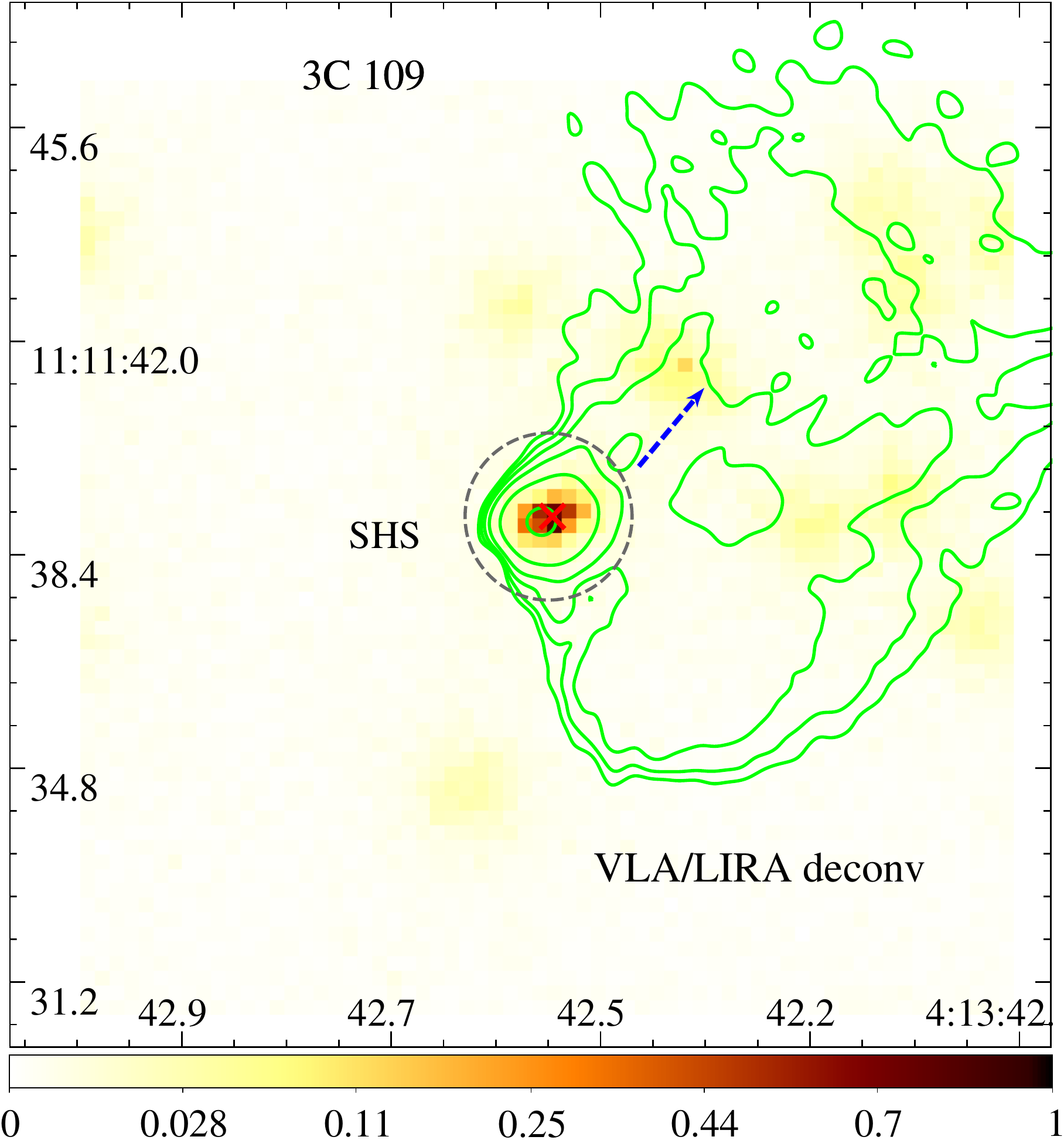}{0.5\textwidth}{(c)}
    }
    \caption{Same as in Fig. \ref{fig:results-3C9} but for 3C 109. (a) shows the full image while (b) and (c) show zoomed-in region around the southern hotspot (SHS). The radio contours are given by 0.2, 0.4, 0.8, 2.0, 10.0, 100.0 mJy beam$^{-1}$.\label{fig:results-3C109}}
\end{figure*}
}

\begin{figure*}[ht]
    \gridline{
        \fig{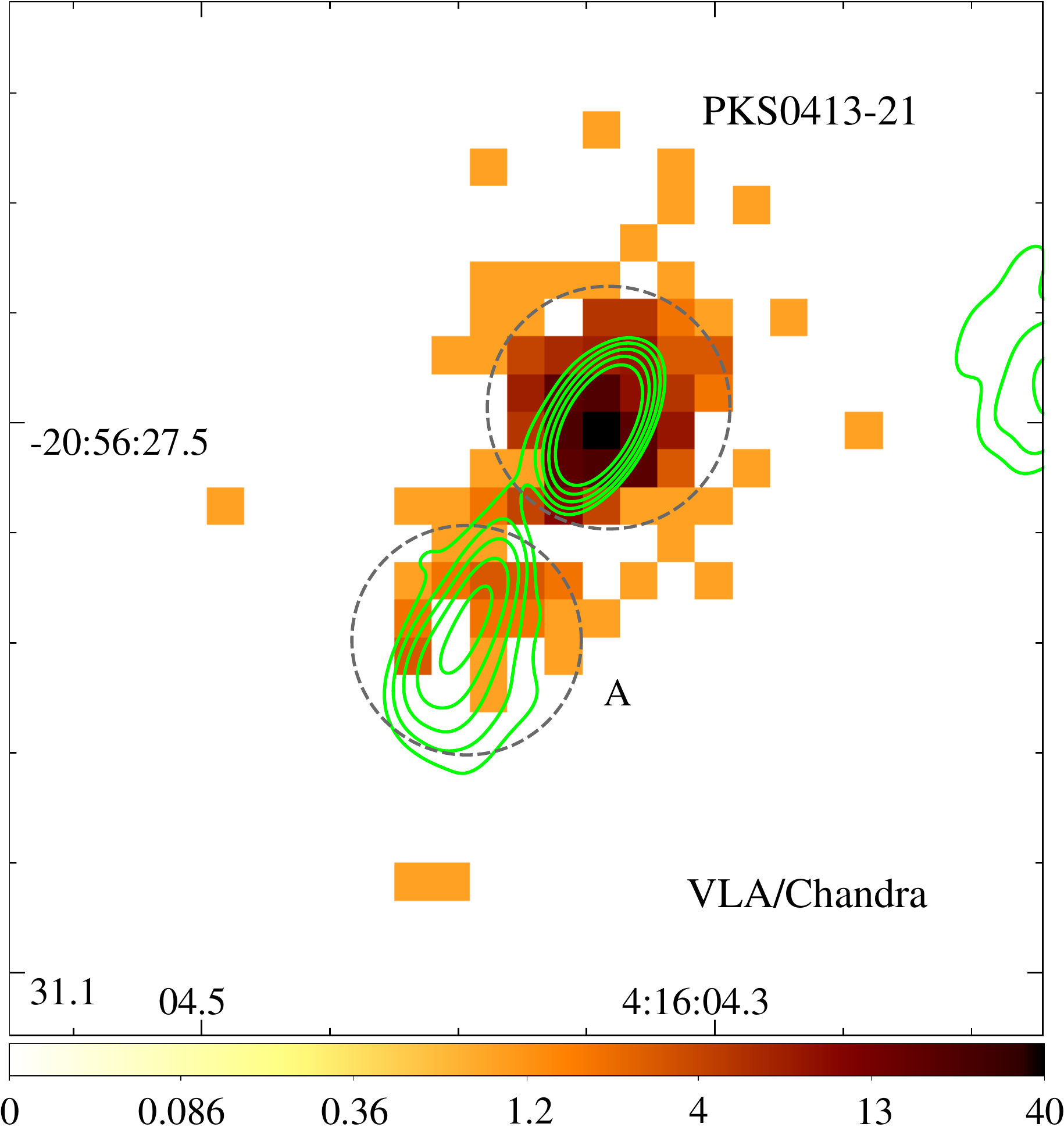}{0.5\textwidth}{(a)}
        \fig{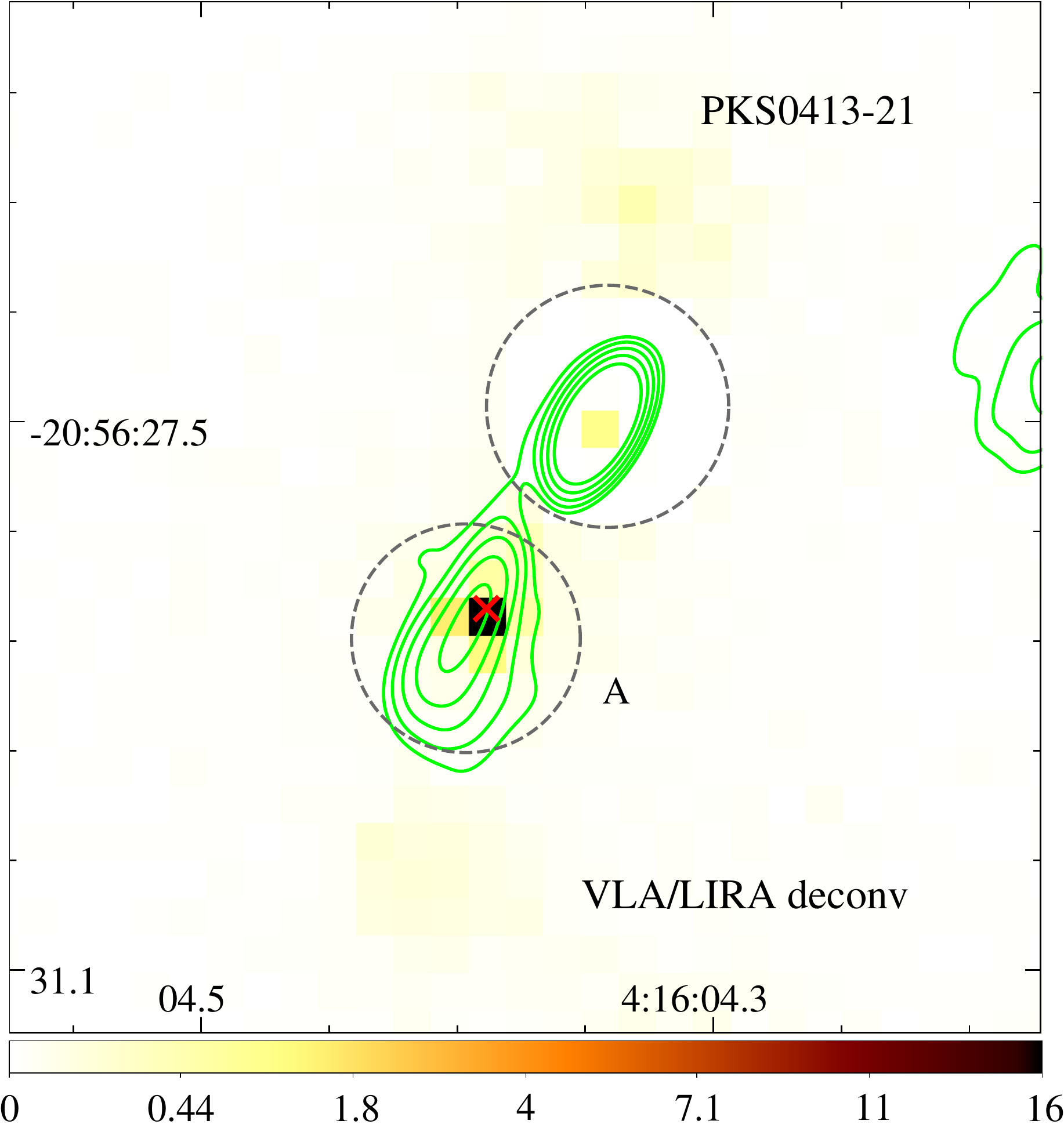}{0.5\textwidth}{(b)}
    }
    \caption{Same as in Fig. \ref{fig:results-3C9} but for PKS 0413-21. The radio contours are given by 5.0, 10.0, 20.0, 40.0, 80.0 mJy beam$^{-1}$.\label{fig:results-PKS0413-21}}
\end{figure*}

\begin{figure*}[ht]
    \gridline{
        \fig{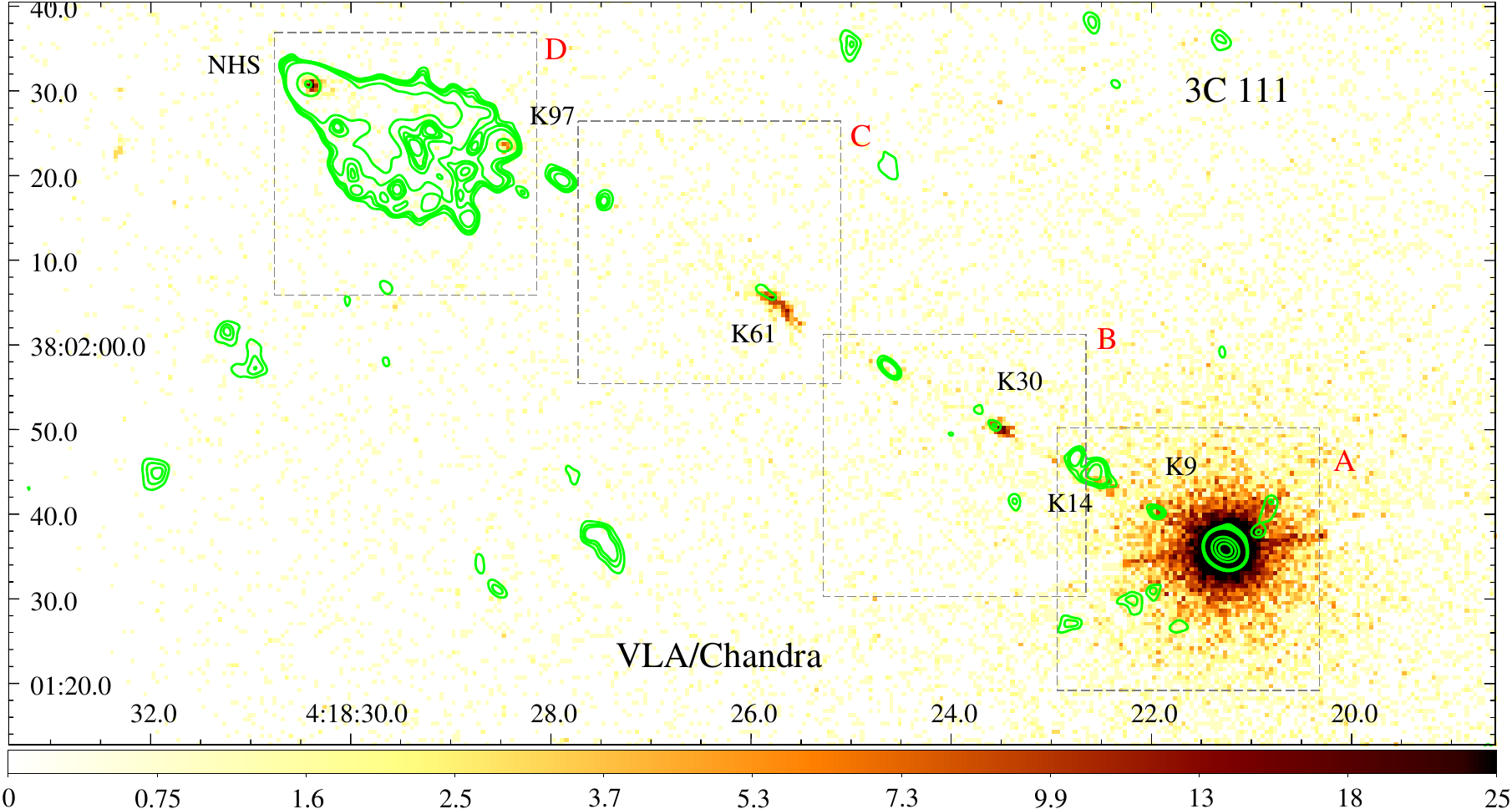}{\textwidth}{(a) Full image}
    }
    \gridline{
        \fig{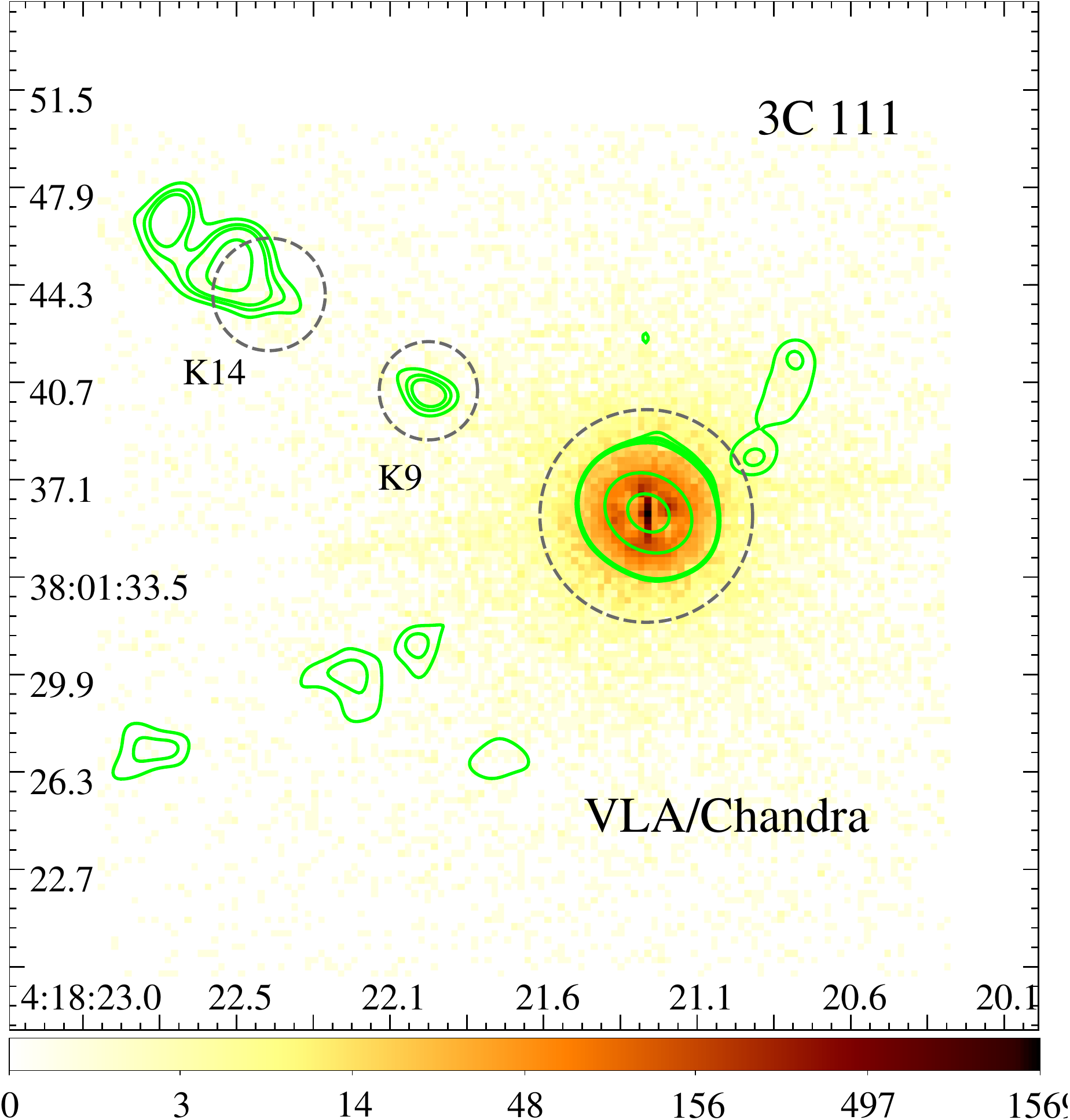}{0.5\textwidth}{(a)}
        \fig{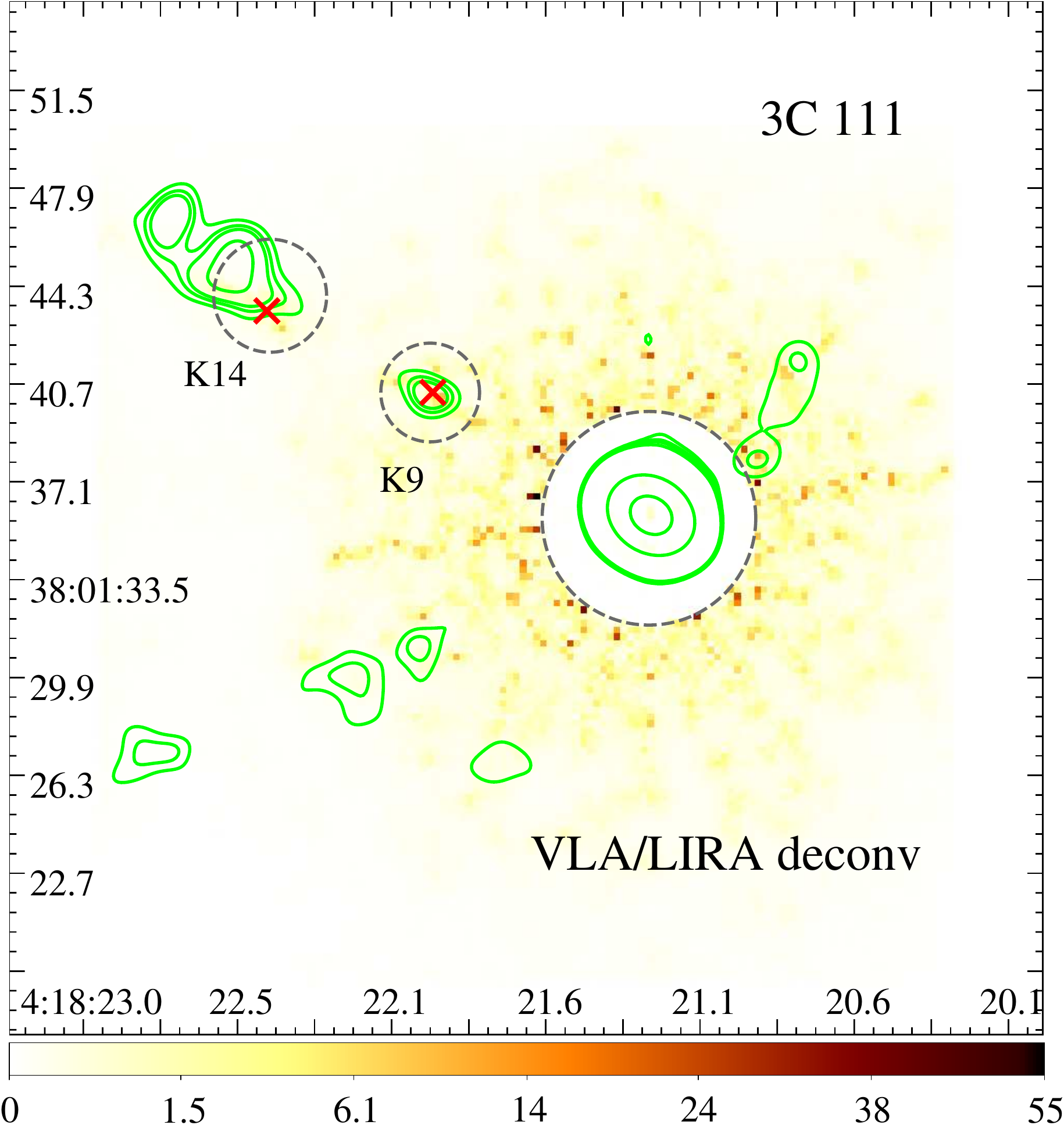}{0.5\textwidth}{(b)}
    }
    \caption{Same as in Fig. \ref{fig:results-3C9} but for 3C 111 (part A). The radio contours are given by 1.5, 1.8, 2, 2.5, 100, 500 mJy beam$^{-1}$.\label{fig:results-3C111-A}}
\end{figure*}

\begin{figure*}[ht]
    \gridline{
        \fig{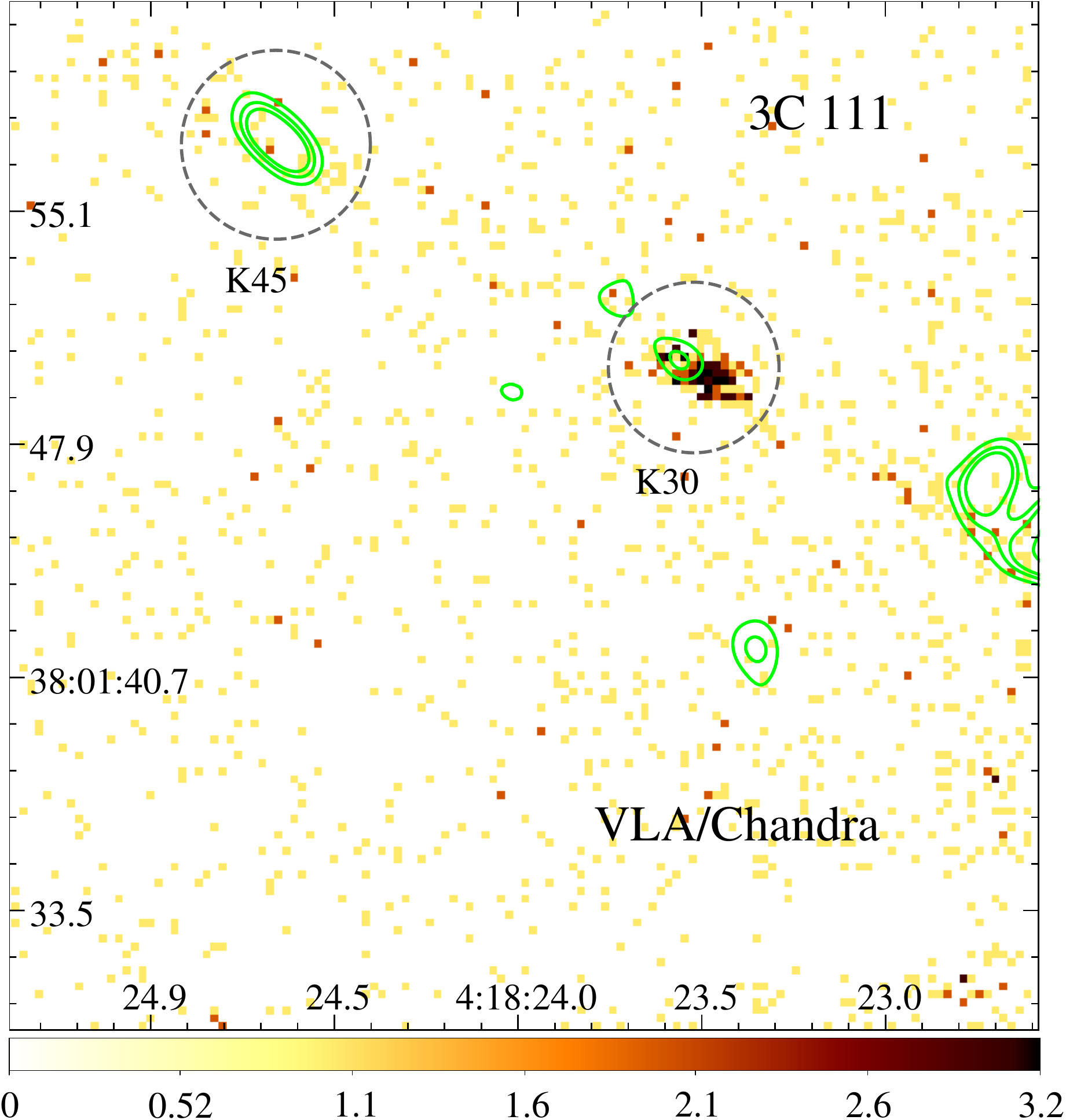}{0.5\textwidth}{(a)}
        \fig{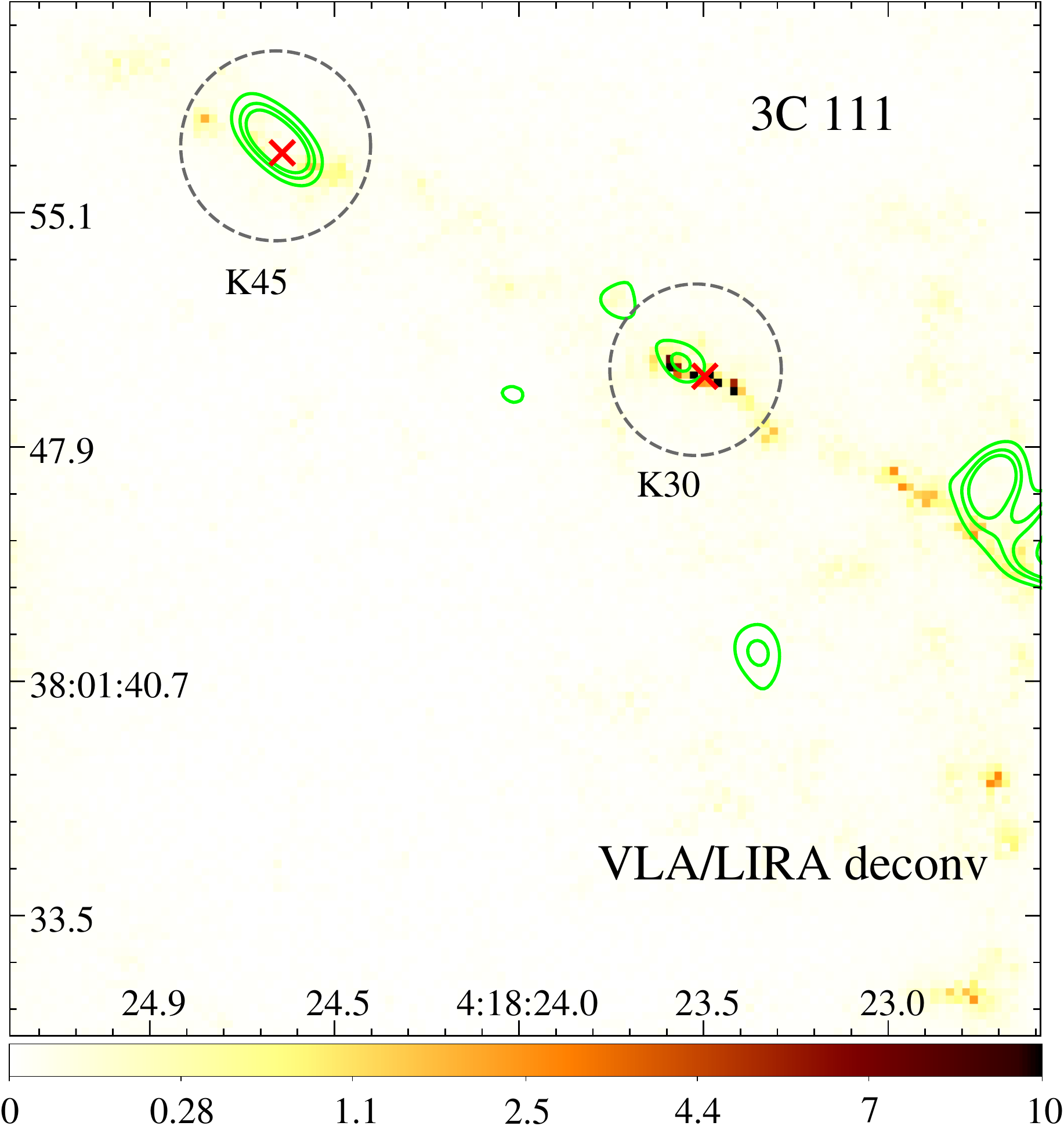}{0.5\textwidth}{(b)}
    }
    \caption{Same as in Fig. \ref{fig:results-3C9} but for 3C 111 (part B). The radio contours are given by 1.5, 1.8, 2, 2.5, 100, 500 mJy beam$^{-1}$.\label{fig:results-3C111-B}}
\end{figure*}

\begin{figure*}[ht]
    \gridline{
        \fig{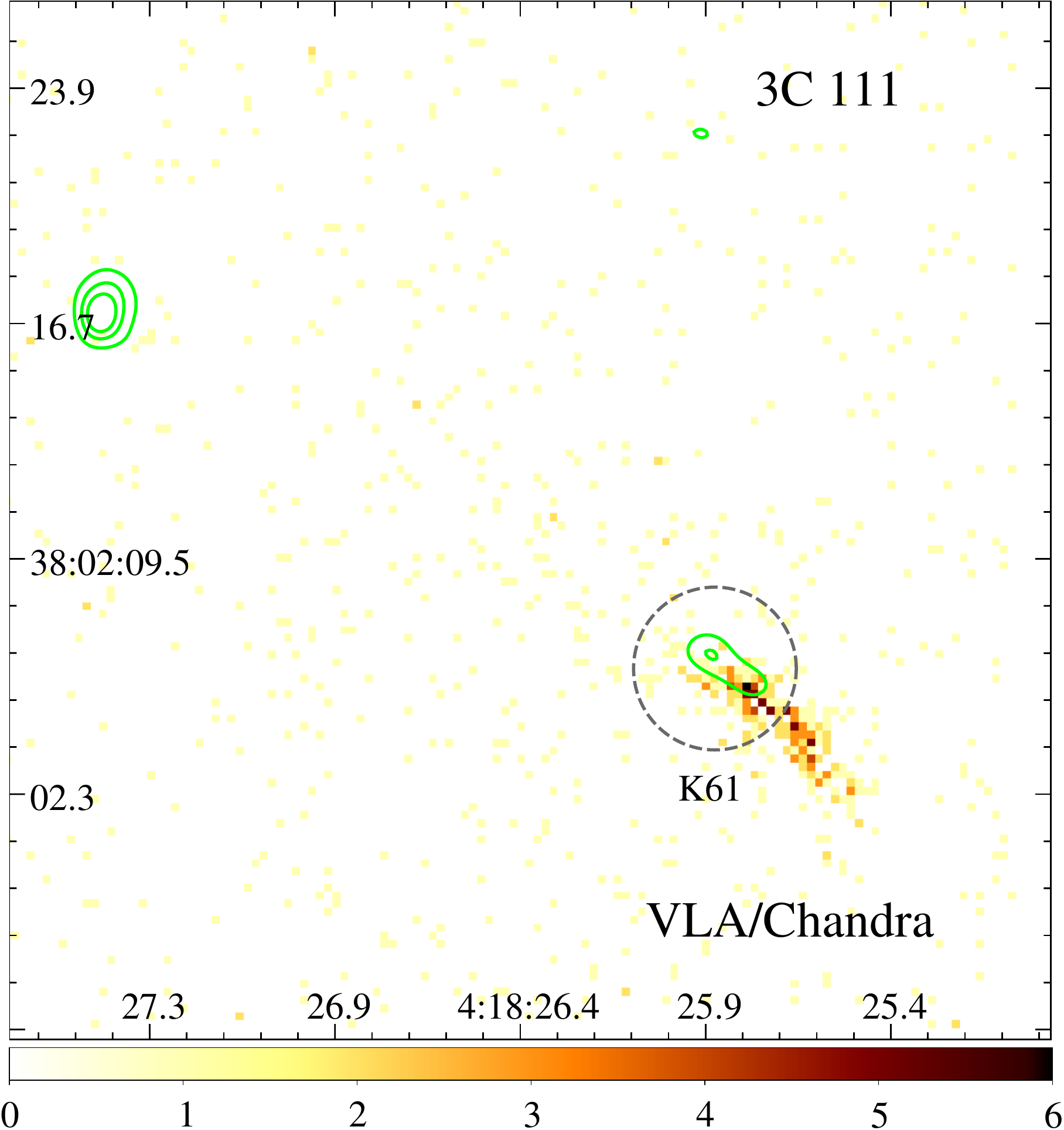}{0.5\textwidth}{(a)}
        \fig{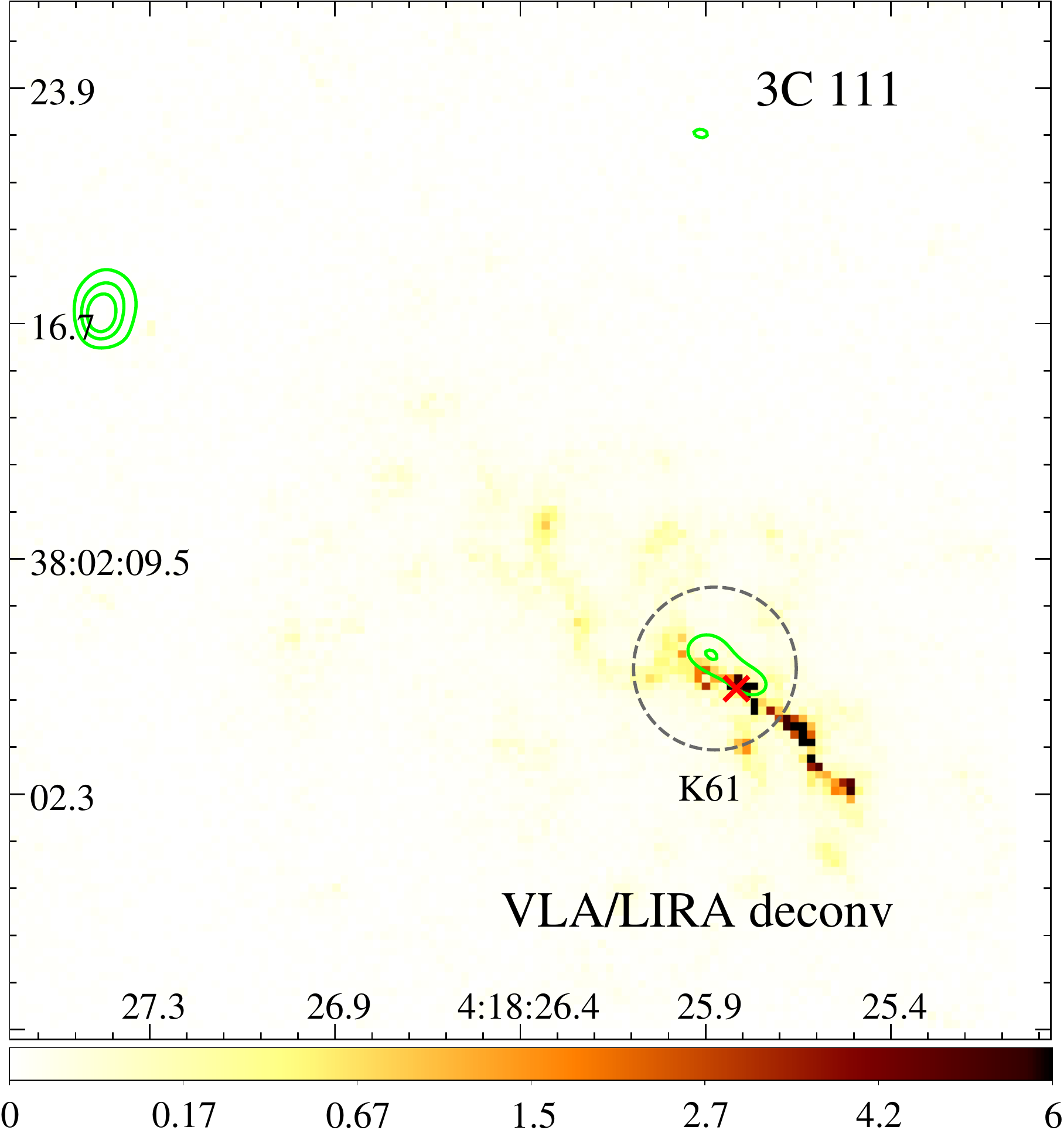}{0.5\textwidth}{(b)}
    }
    \caption{Same as in Fig. \ref{fig:results-3C9} but for 3C 111 (part C). The radio contours are given by 1.5, 1.8, 2, 2.5, 100, 500 mJy beam$^{-1}$.\label{fig:results-3C111-C}}
\end{figure*}

\begin{figure*}[ht]
    \gridline{
        \fig{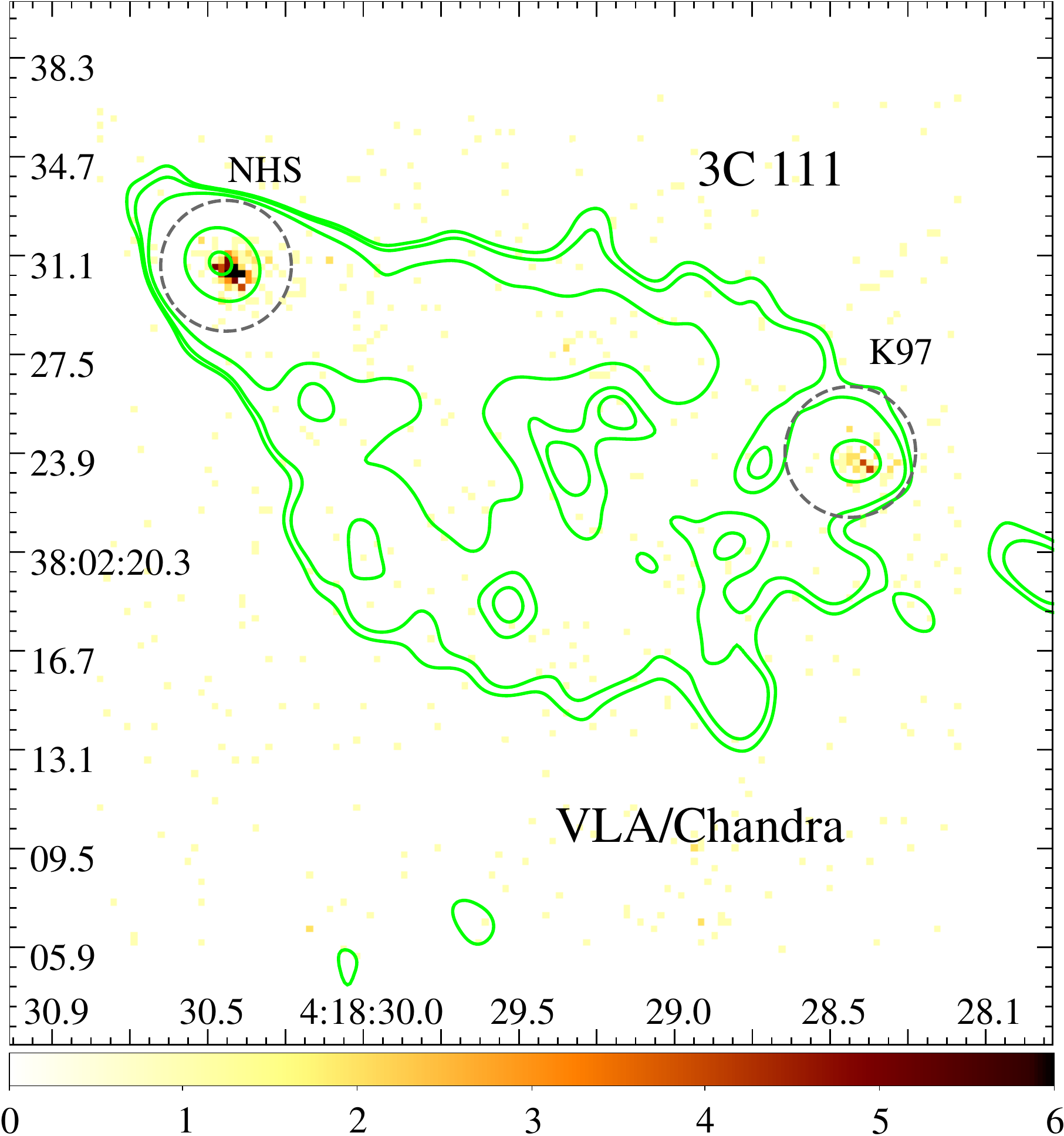}{0.5\textwidth}{(a)}
        \fig{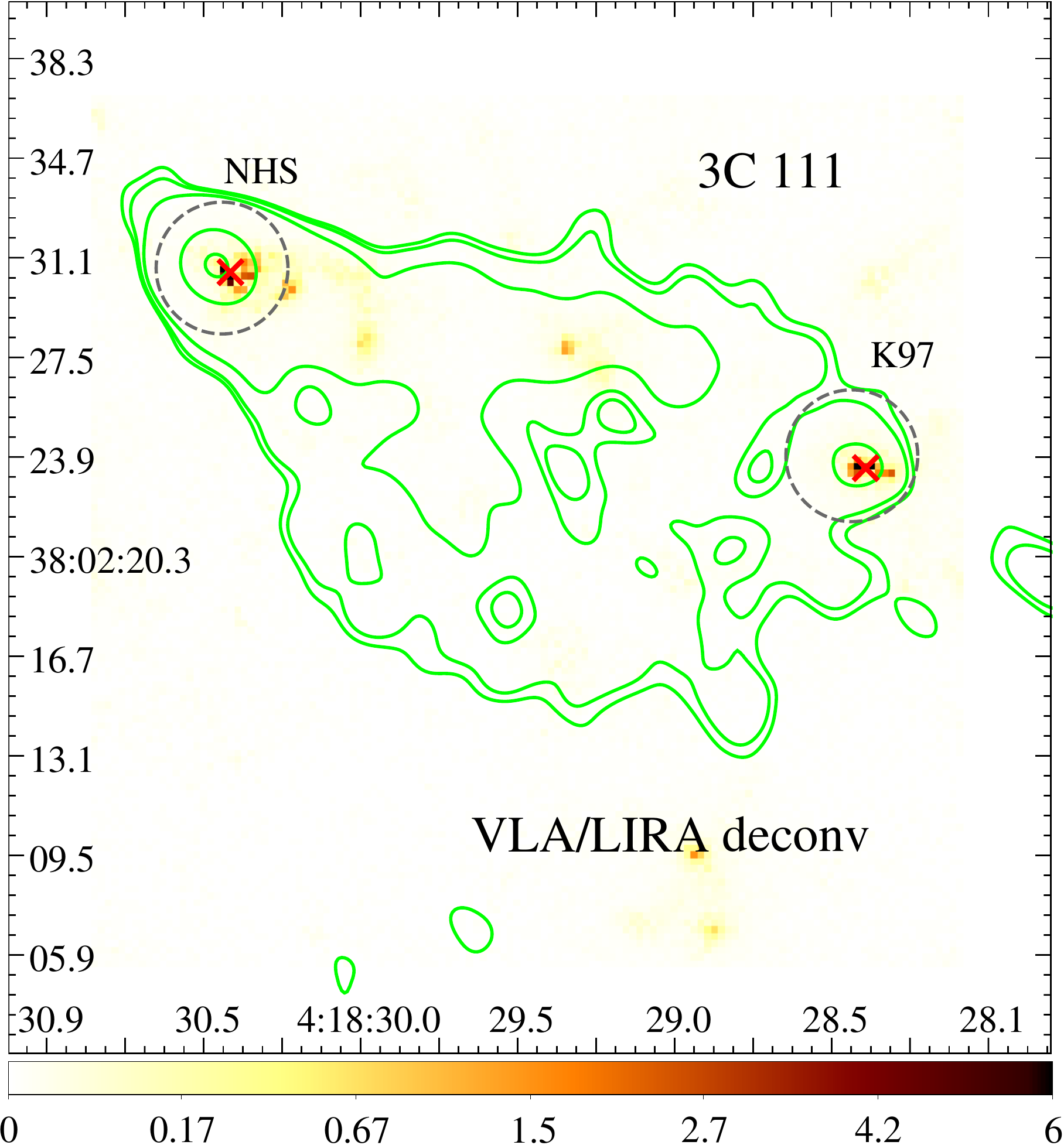}{0.5\textwidth}{(b)}
    }
    \caption{Same as in Fig. \ref{fig:results-3C9} but for 3C 111 (part D). The radio contours are given by 1.5, 1.8, 2, 2.5, 100, 500 mJy beam$^{-1}$.\label{fig:results-3C111-D}}
\end{figure*}

\begin{figure*}[ht]
    \gridline{
        \fig{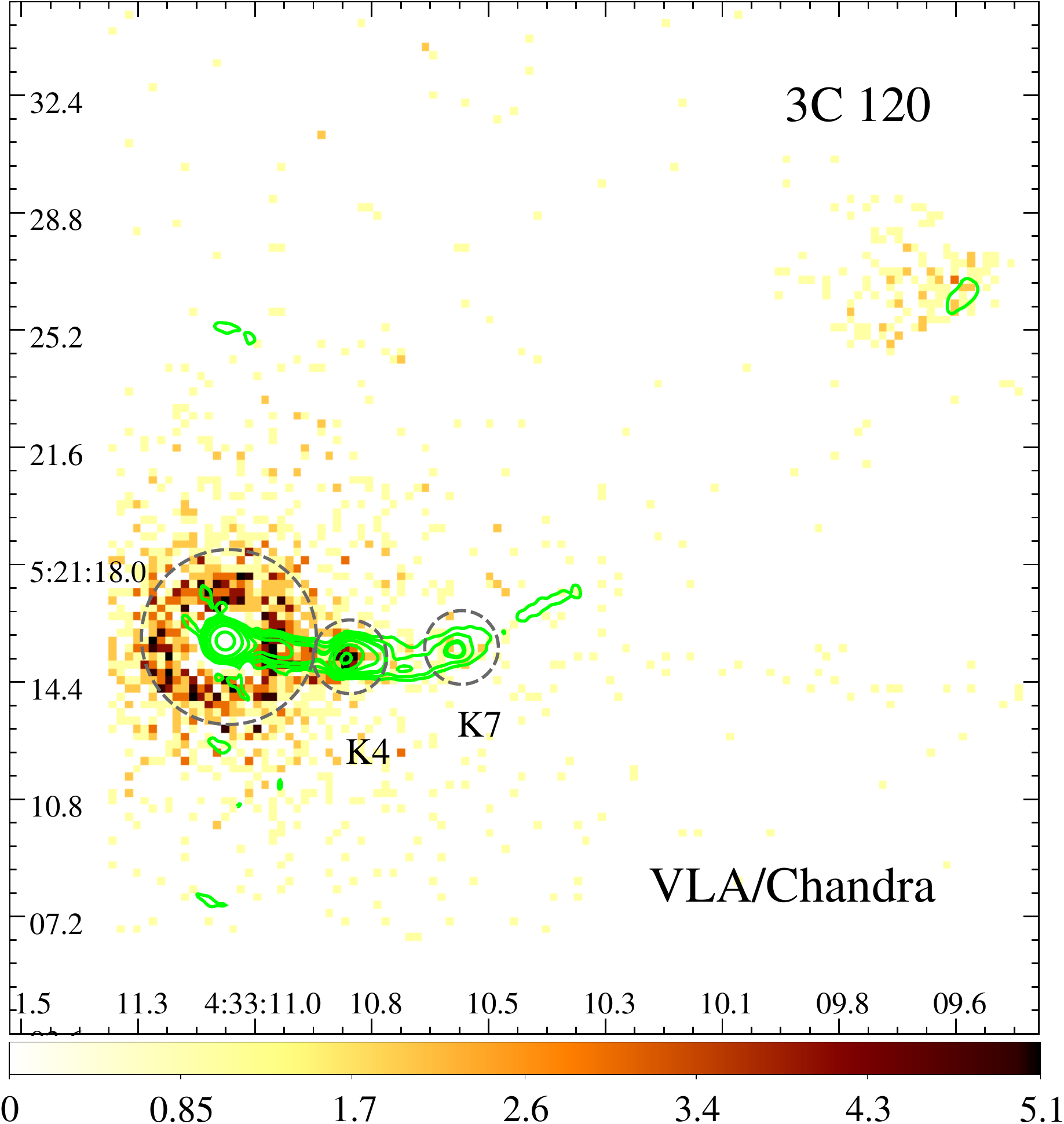}{0.5\textwidth}{(a)}
        \fig{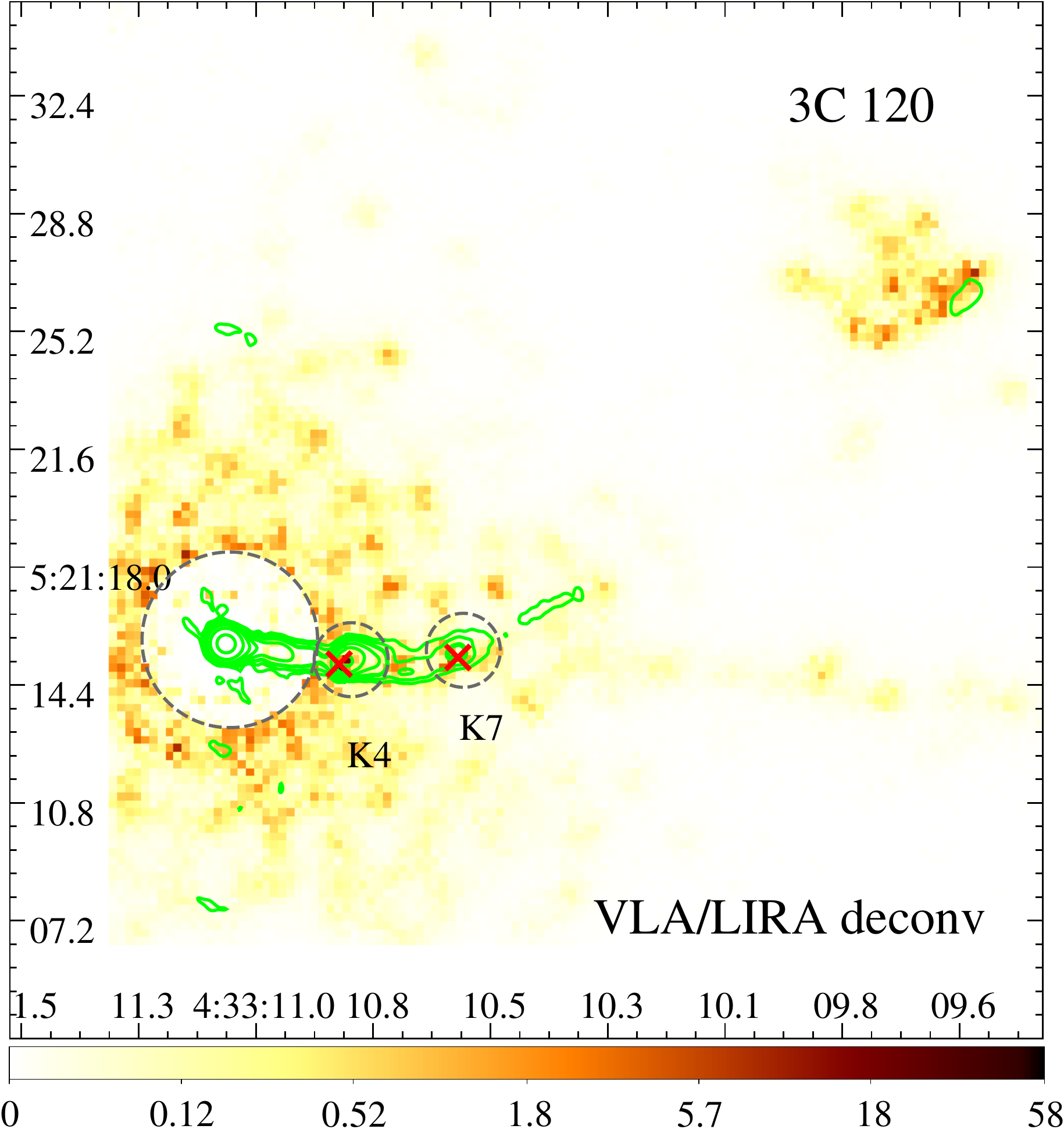}{0.5\textwidth}{(b)}
    }
    \caption{Same as in Fig. \ref{fig:results-3C9} but for 3C 120. The radio contours are given by 0.5, 1.0, 1.8, 2.0, 4.0, 8.0, 20.0, 100.0, 1000.0 mJy beam$^{-1}$.\label{fig:results-3C120}}
\end{figure*}

\begin{figure*}[ht]
    \gridline{
        \fig{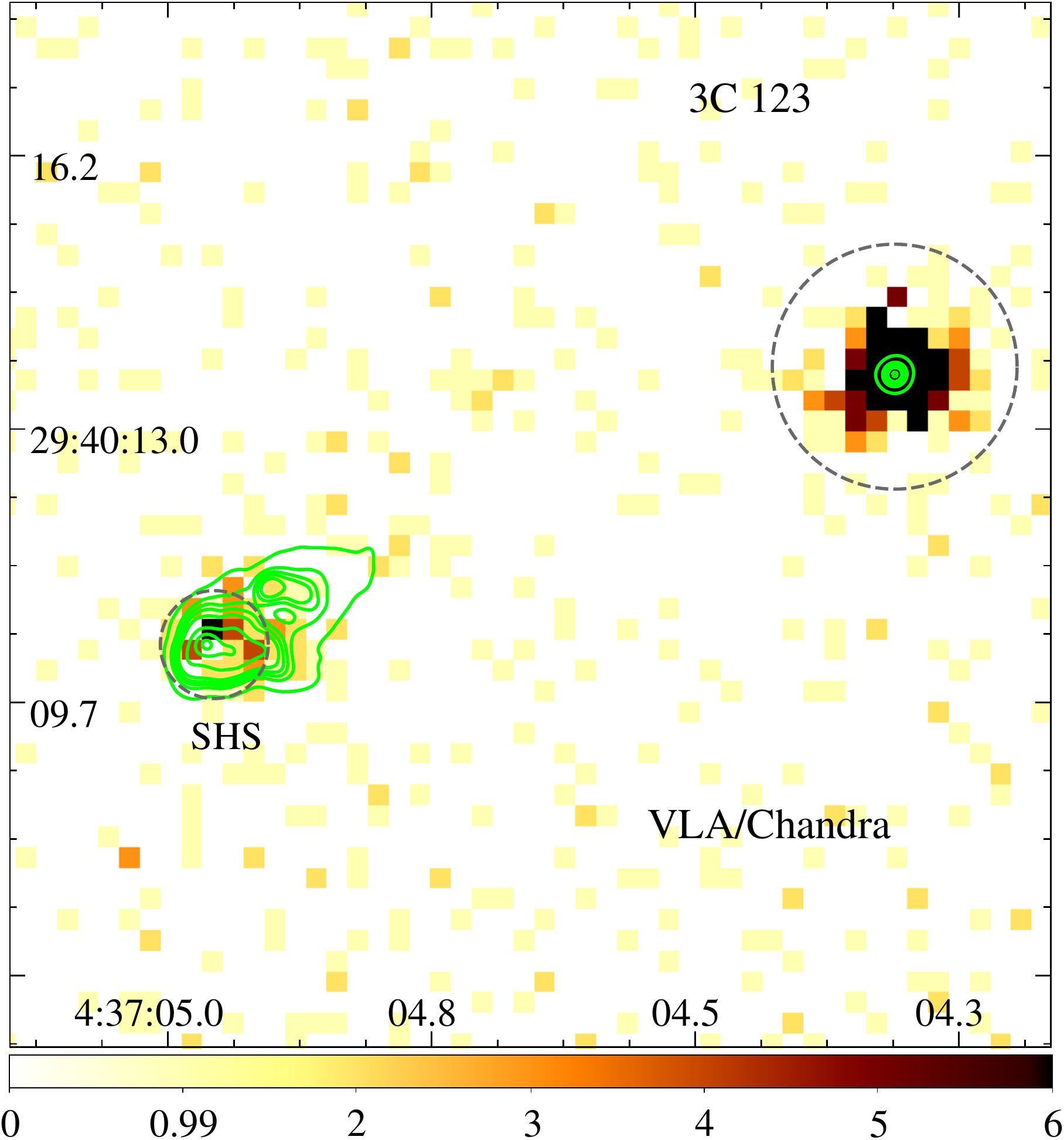}{0.5\textwidth}{(a)}
        \fig{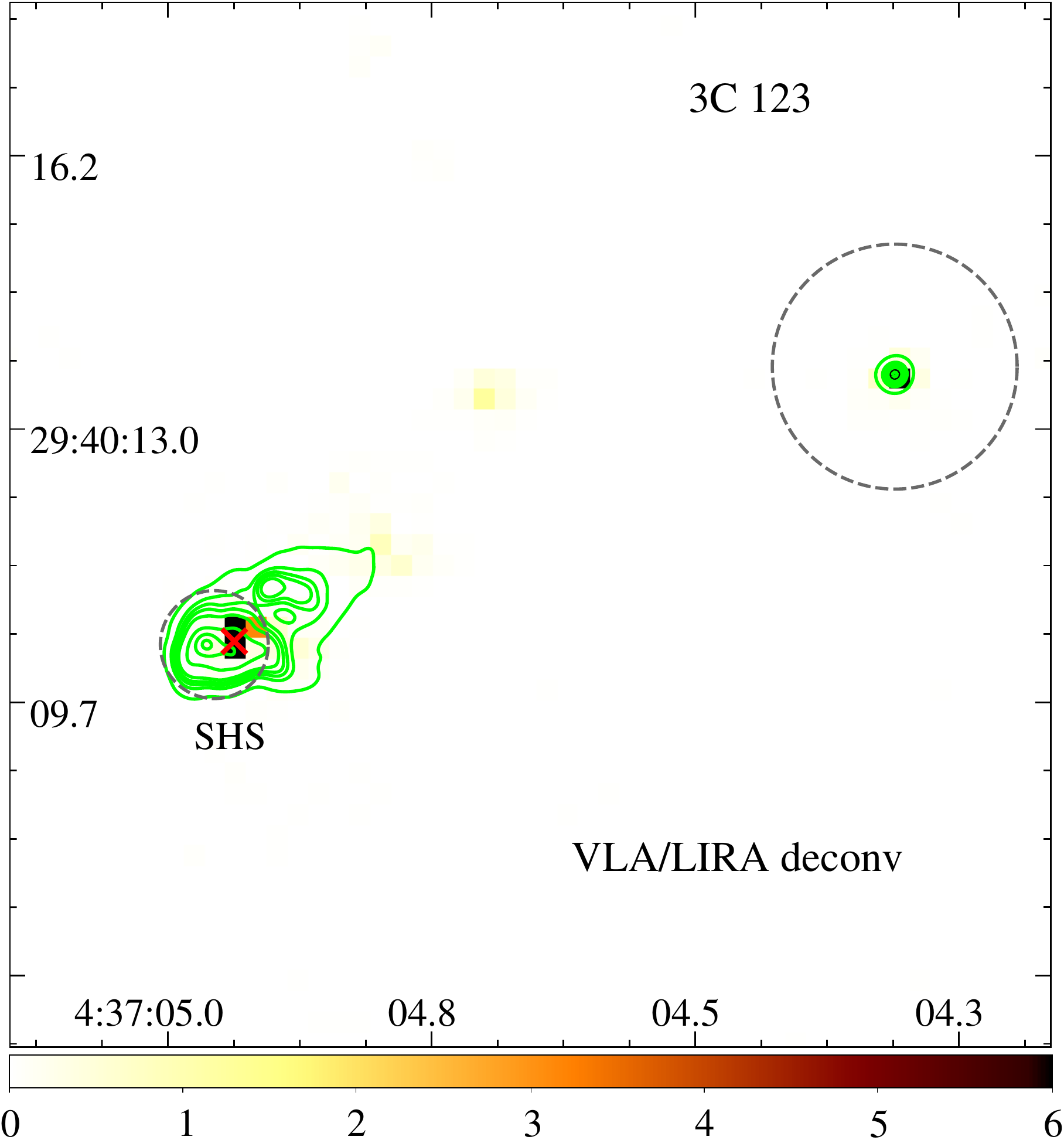}{0.5\textwidth}{(b)}
    }
    \caption{Same as in Fig. \ref{fig:results-3C9} but for 3C 123. The radio contours are given by 5.0, 30.0, 50.0, 70.0, 100.0, 200.0, 300.0, 380.0 mJy beam$^{-1}$.\label{fig:results-3C123}}
\end{figure*}

\begin{figure*}[ht]
    \gridline{
        \fig{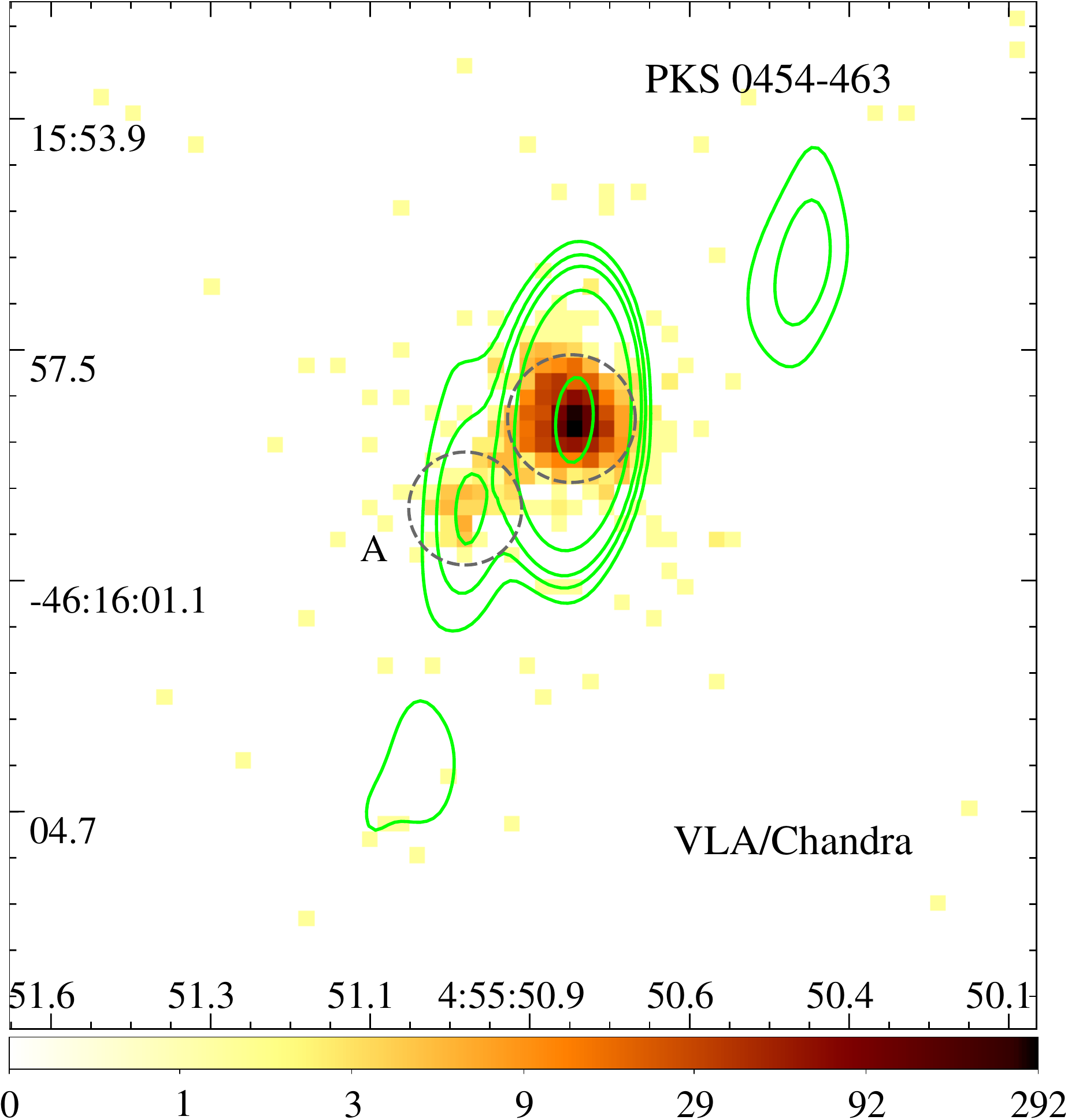}{0.5\textwidth}{(a)}
        \fig{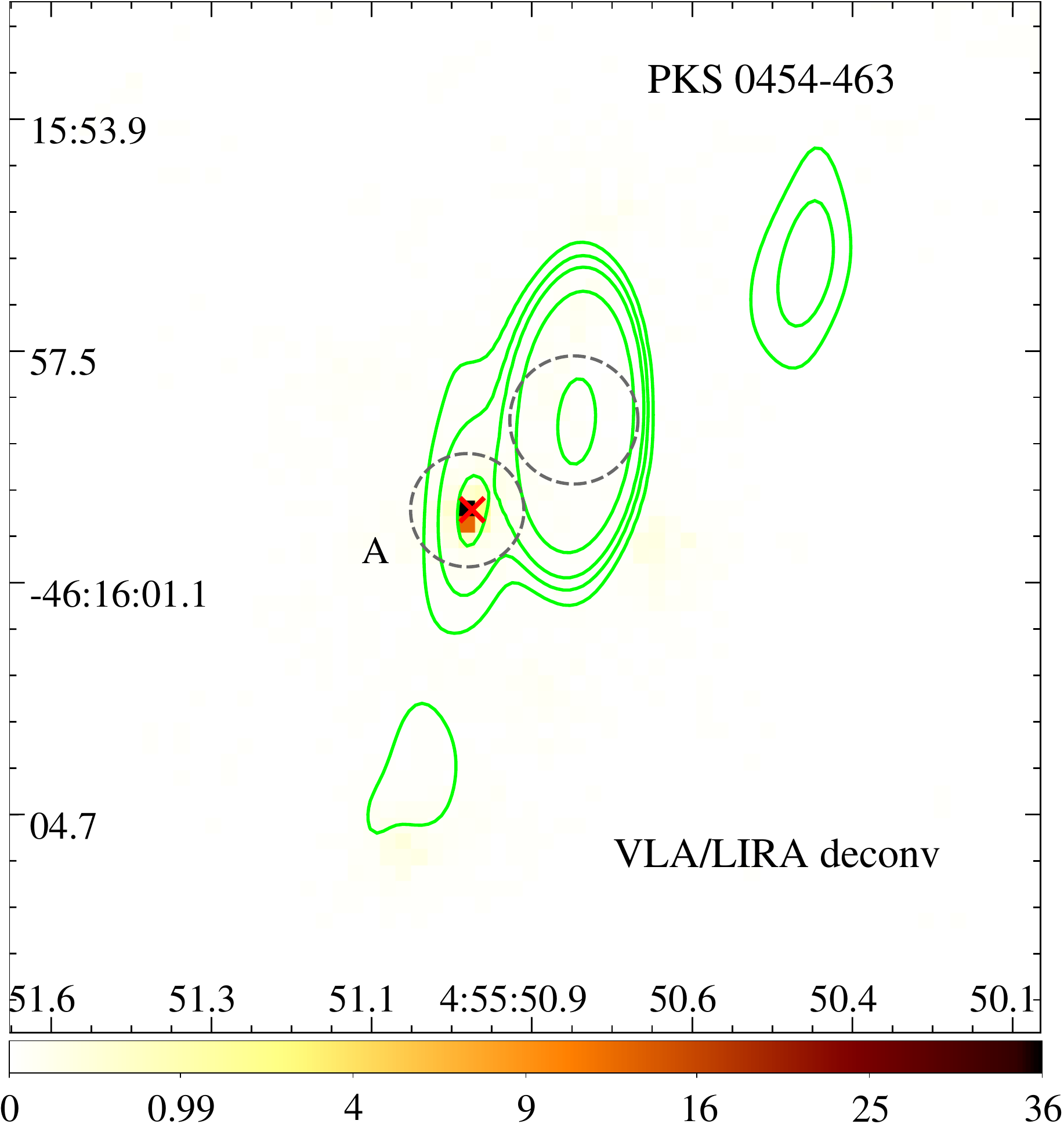}{0.5\textwidth}{(b)}
    }
    \caption{Same as in Fig. \ref{fig:results-3C9} but for PKS 0454-46. The radio contours are given by 10.0, 20.0, 35.0, 100.0, 1000.0 mJy beam$^{-1}$.\label{fig:results-PKS0454-463}}
\end{figure*}

\begin{figure*}[ht]
    \gridline{
        \fig{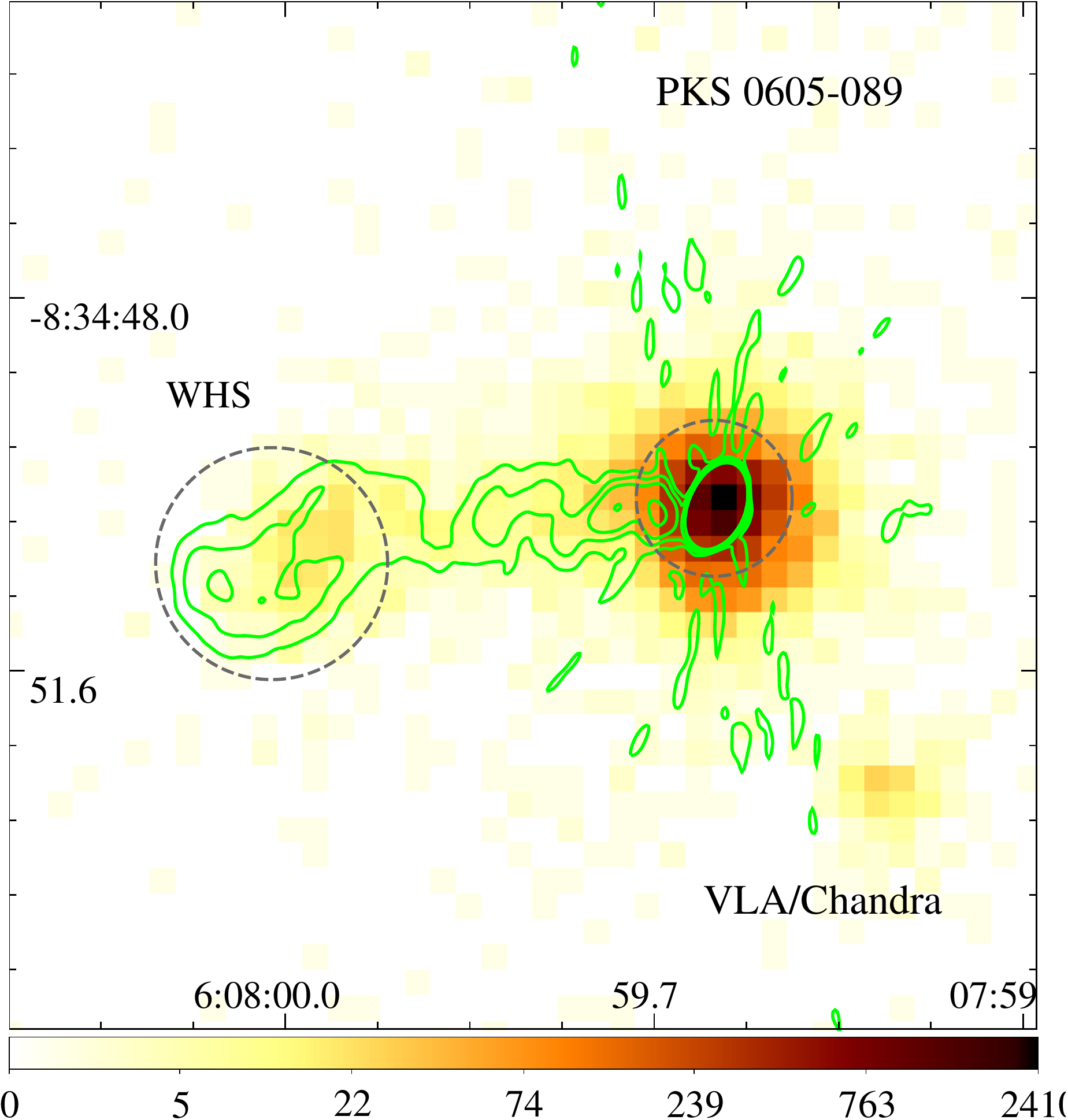}{0.5\textwidth}{(a)}
        \fig{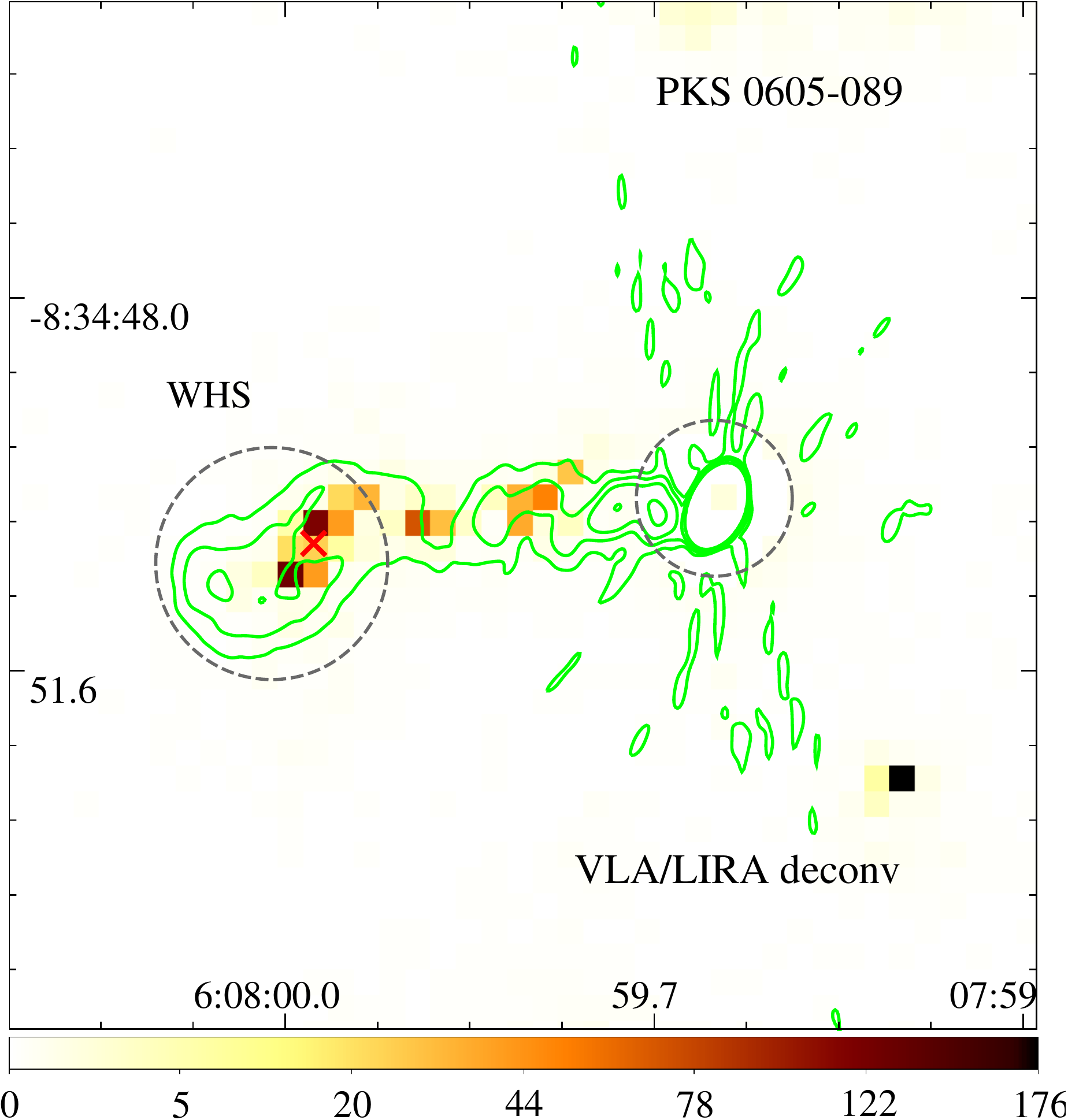}{0.5\textwidth}{(b)}
    }
    \caption{Same as in Fig. \ref{fig:results-3C9} but for PKS 0605-089. The radio contours are given by 0.1, 0.3, 0.6, 1.5 mJy beam$^{-1}$.\label{fig:results-PKS0605-089}}
\end{figure*}

\begin{figure*}[ht]
    \gridline{
        \fig{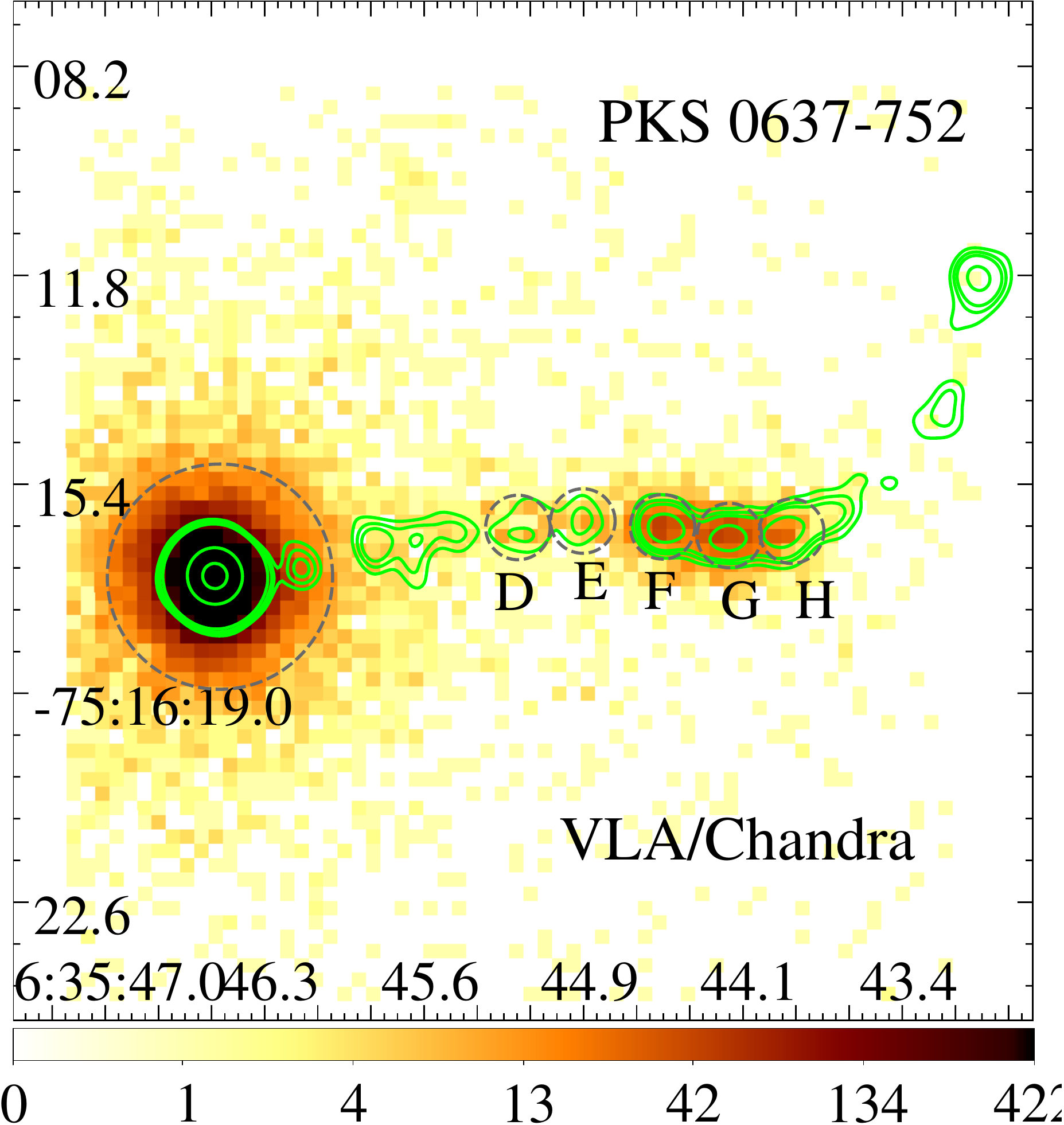}{0.5\textwidth}{(a)}
        \fig{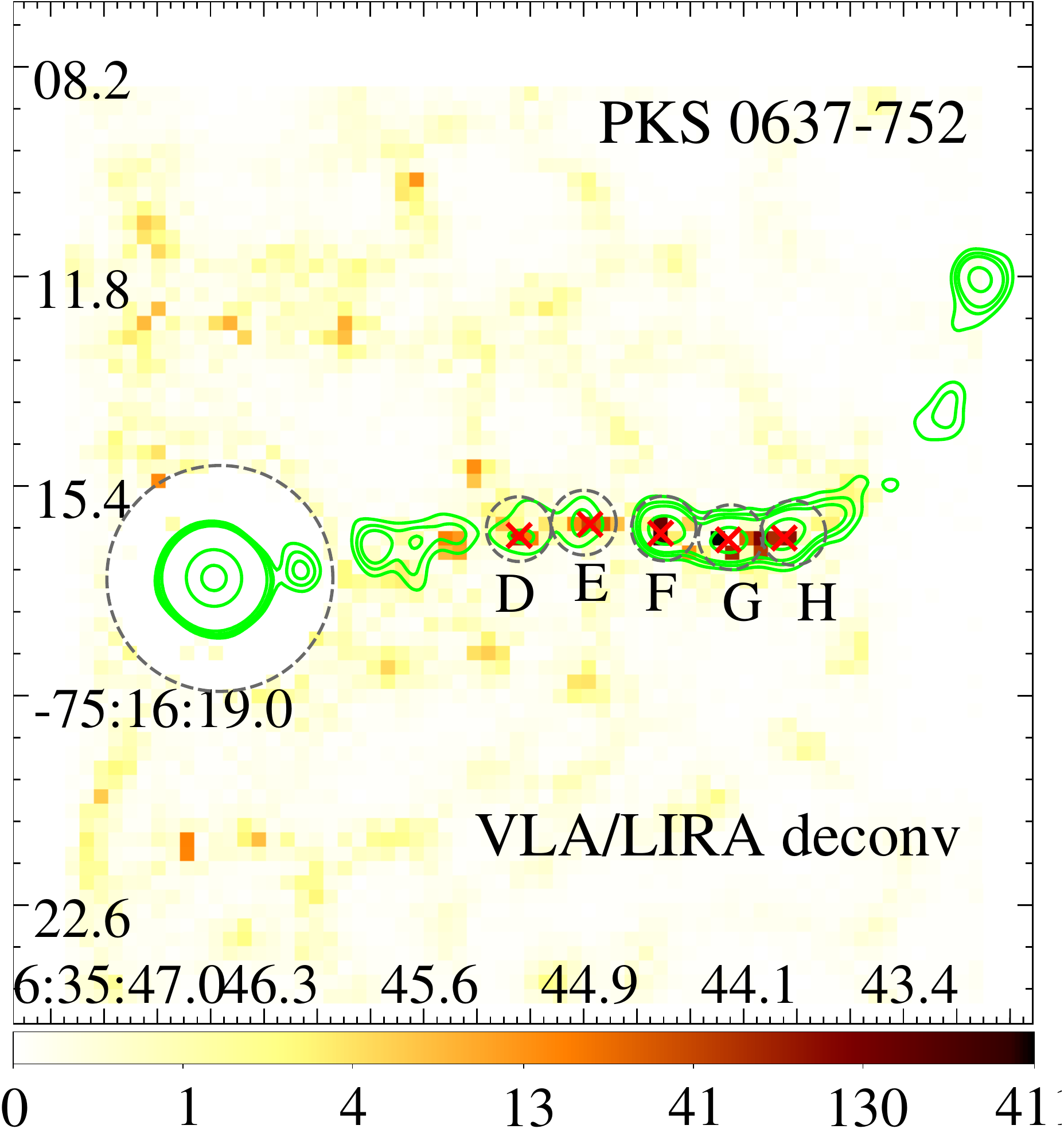}{0.5\textwidth}{(b)}
    }
    \caption{Same as in Fig. \ref{fig:results-3C9} but for PKS 0637-752. The radio contours are given by 4.0, 6.0, 8.0, 1.5, 1000.0, 4000.0 mJy beam$^{-1}$.\label{fig:results-PKS0637-752}}
\end{figure*}

\begin{figure*}[ht]
    \comment{\gridline{
        \fig{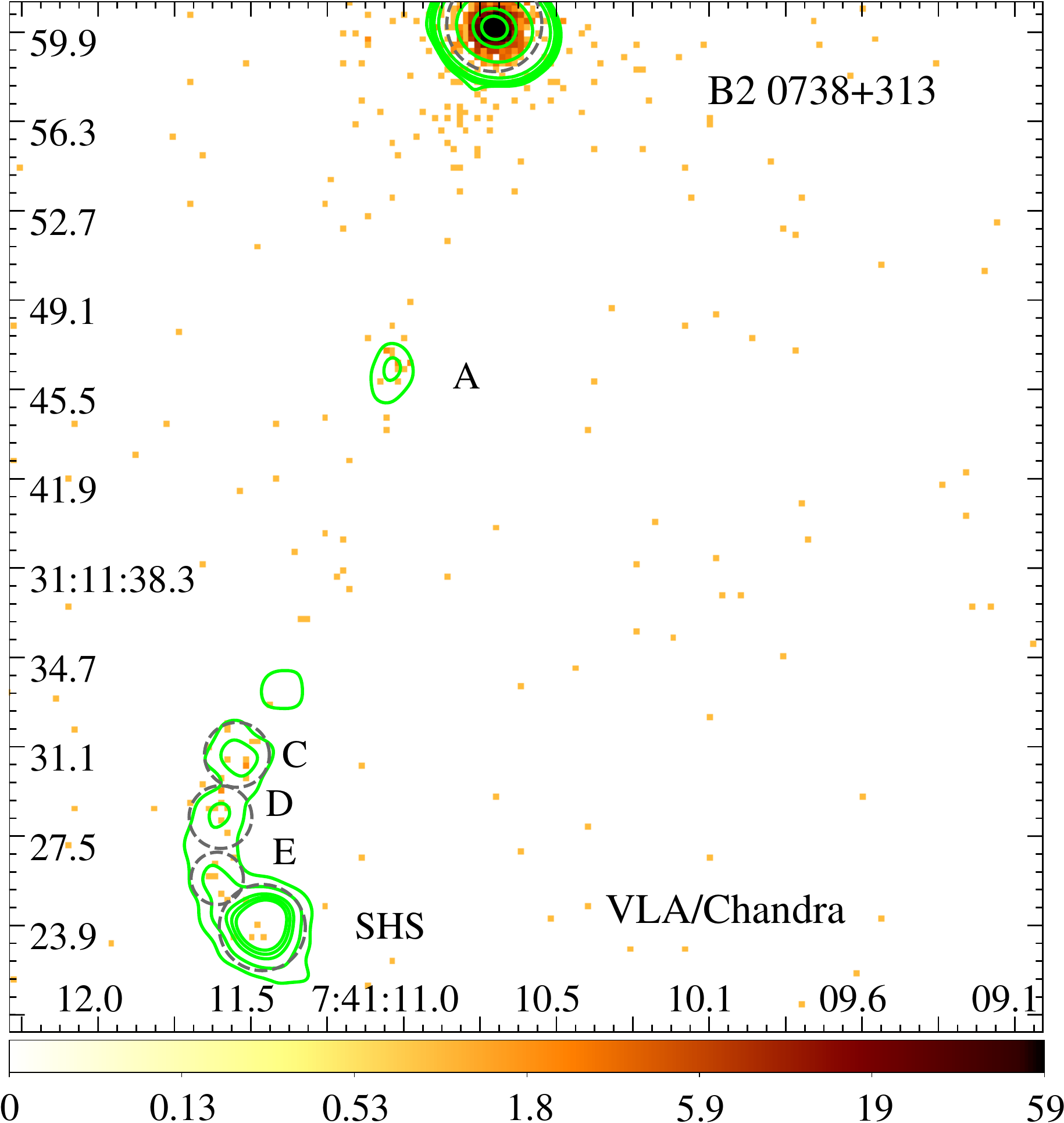}{0.5\textwidth}{(a)}
    }}
    \gridline{
        \fig{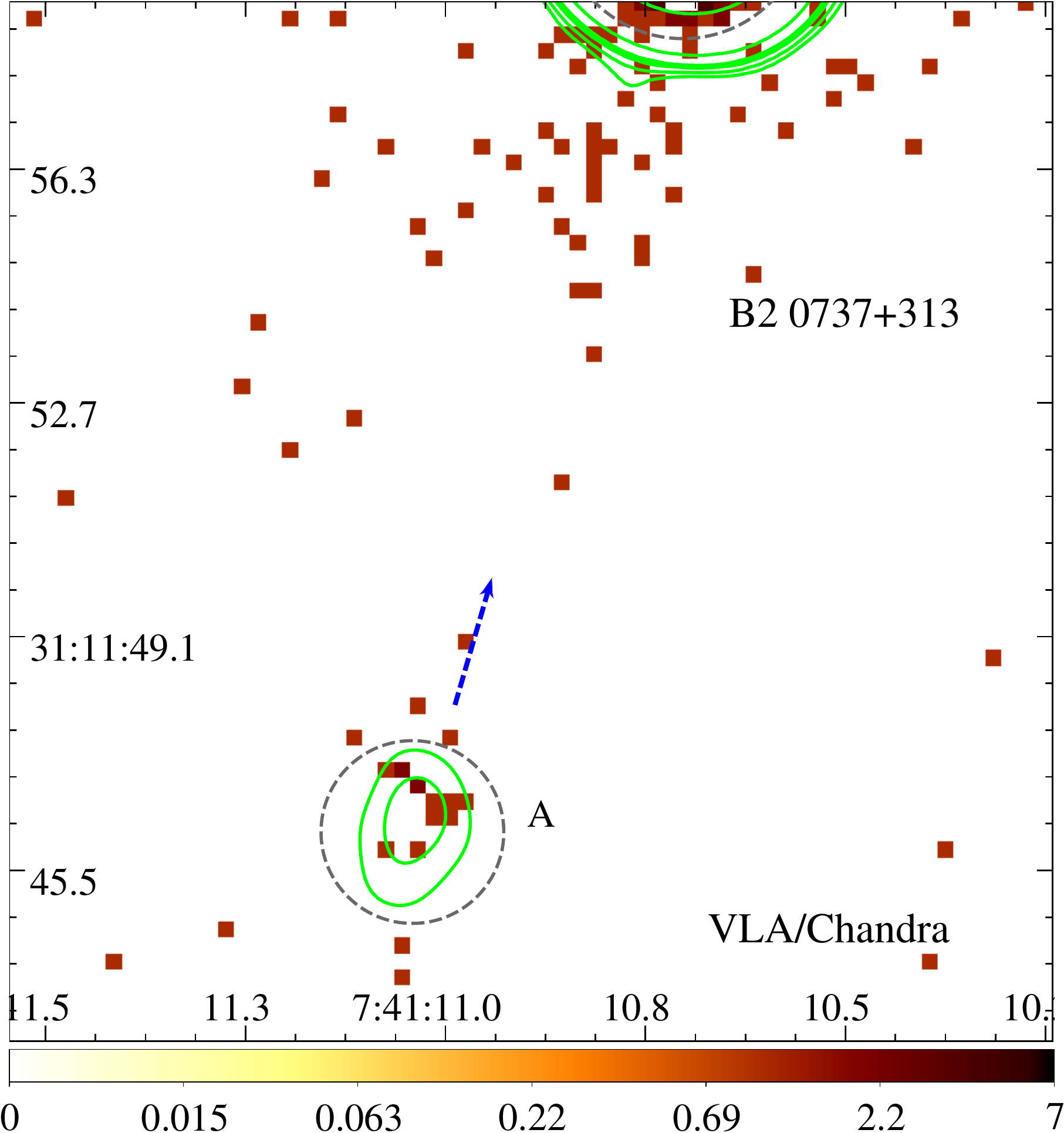}{0.5\textwidth}{(b)}
        \fig{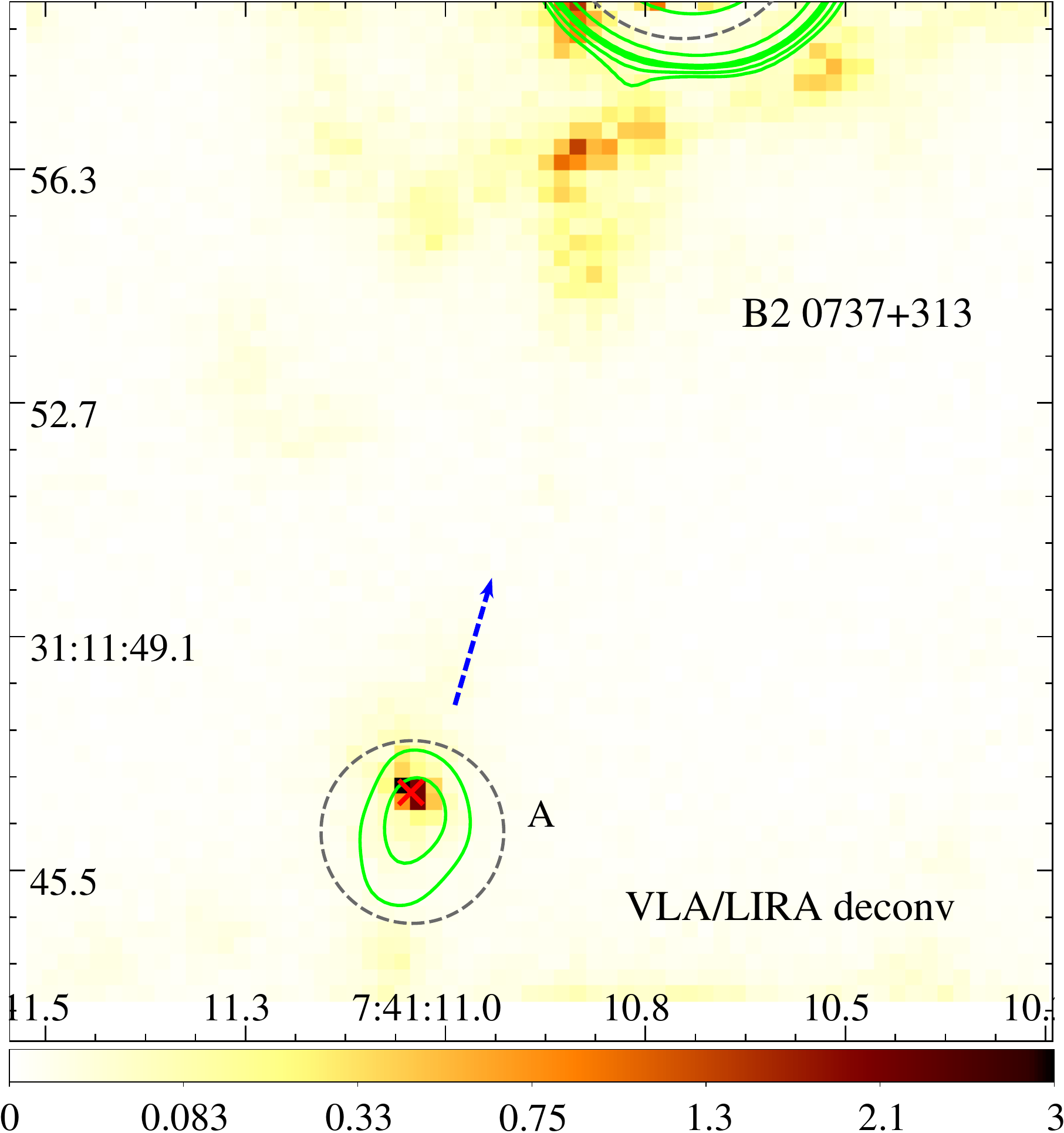}{0.5\textwidth}{(c)}
    }
    \comment{
    \gridline{
        \fig{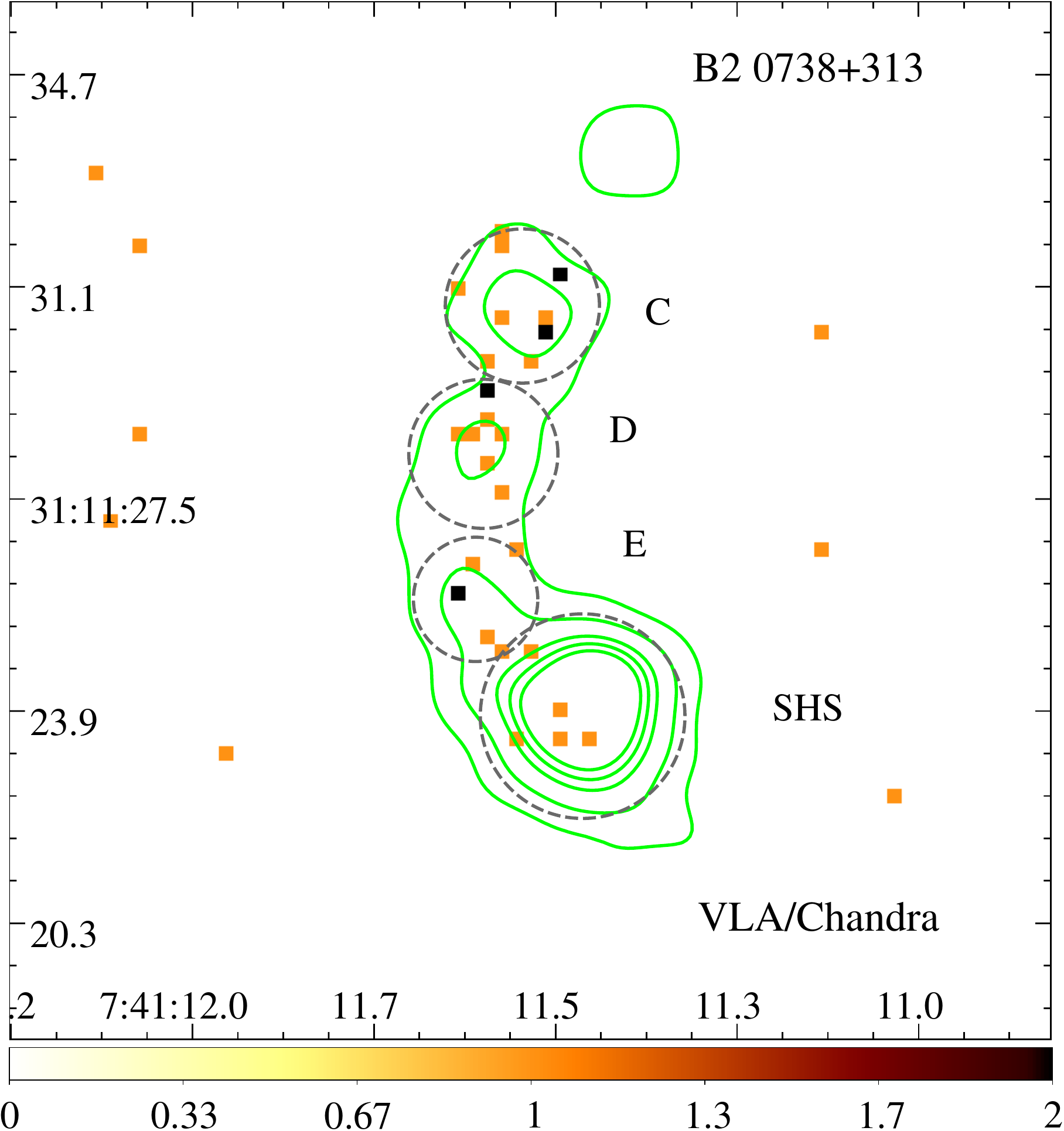}{0.5\textwidth}{(d)}
        \fig{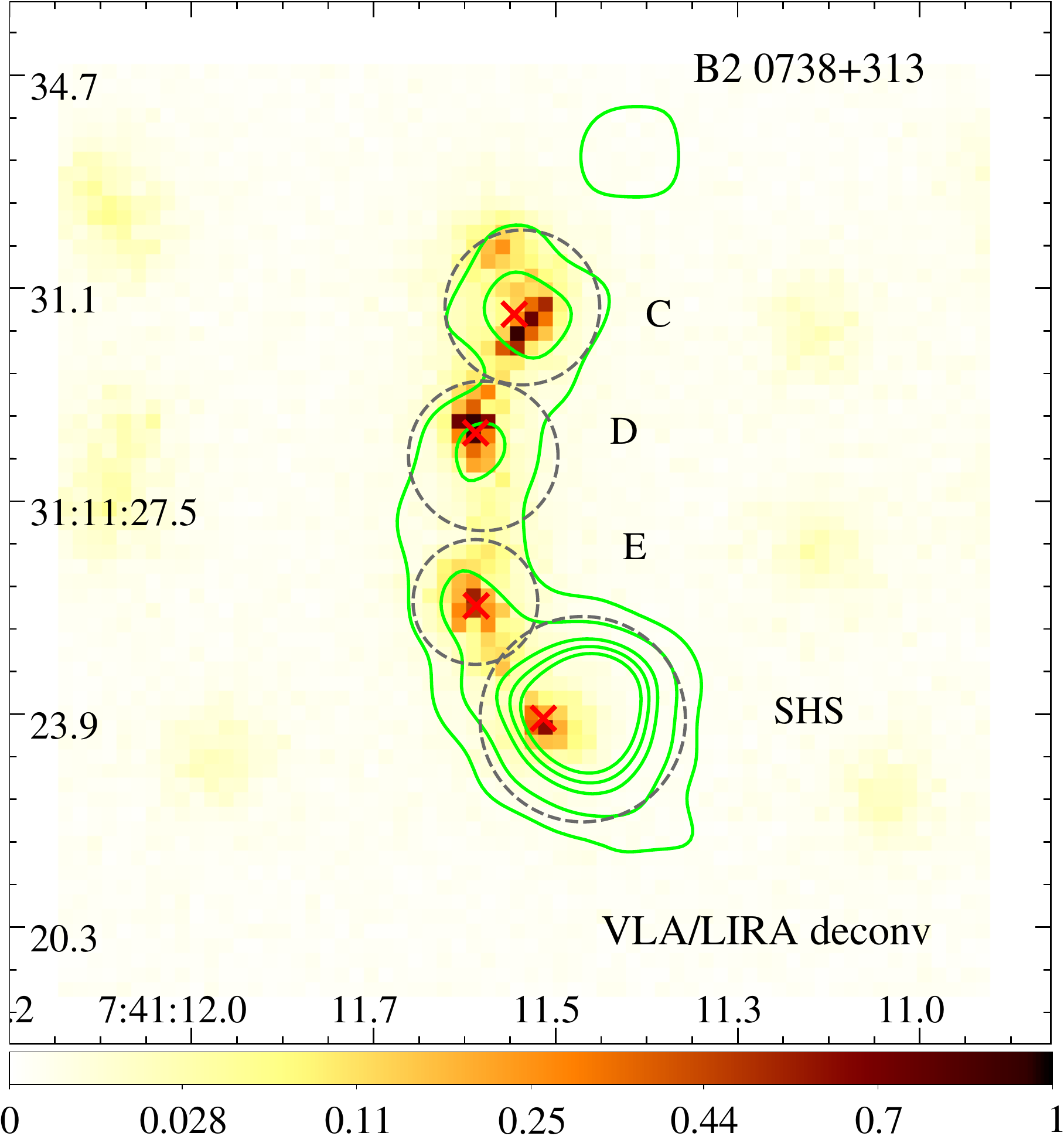}{0.5\textwidth}{(e)}
    }}
    \caption{Same as in Fig. \ref{fig:results-3C9} but for B2 0738+313. The radio contours are given by 0.1, 0.2, 0.4, 0.5, 0.6, 2.0, 100.0, 1000.0, 2000.0 mJy beam$^{-1}$.\label{fig:results-B20738+313}}
\end{figure*}

\begin{figure*}[ht]
    \gridline{
        \fig{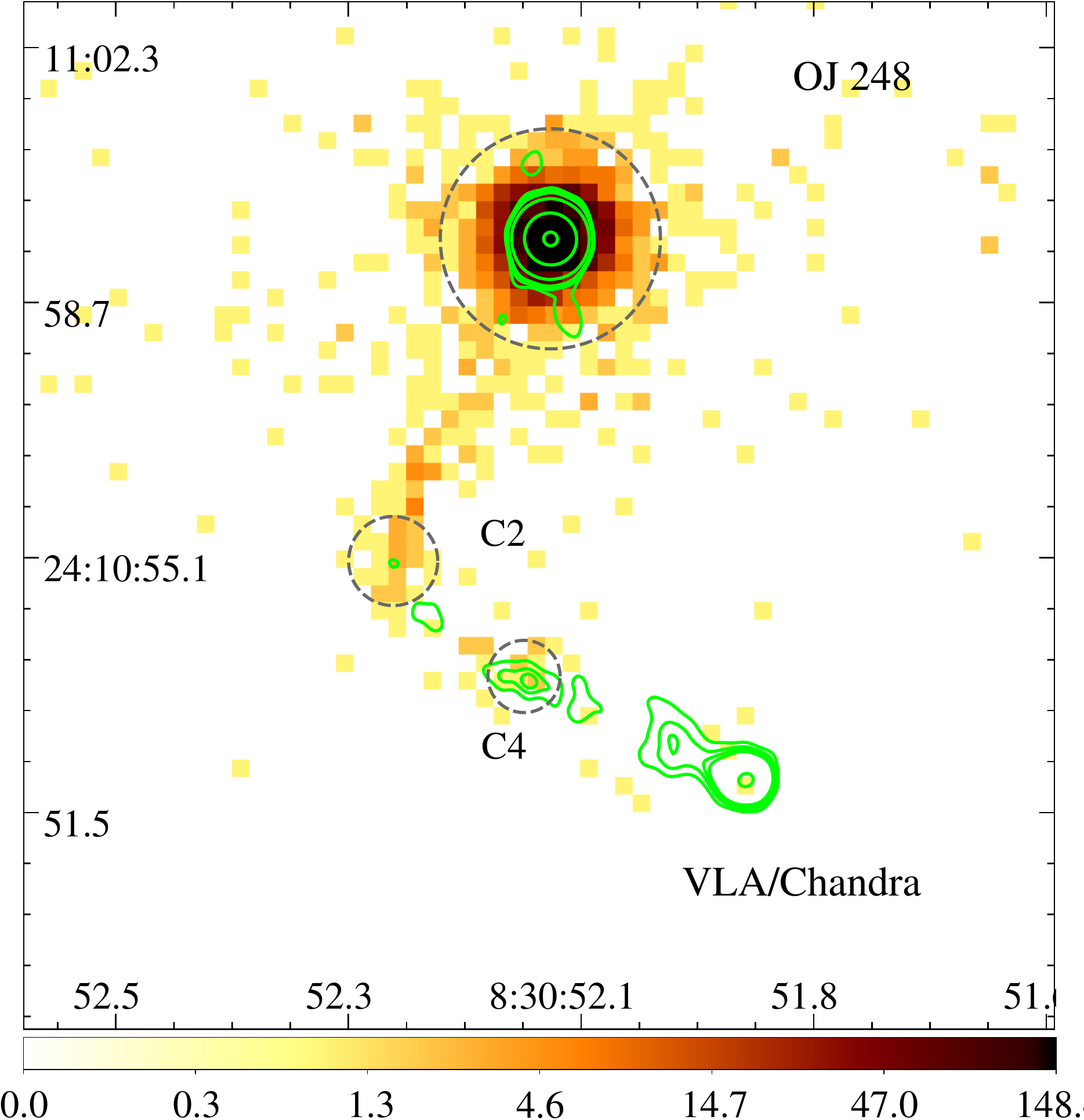}{0.5\textwidth}{(a)}
        \fig{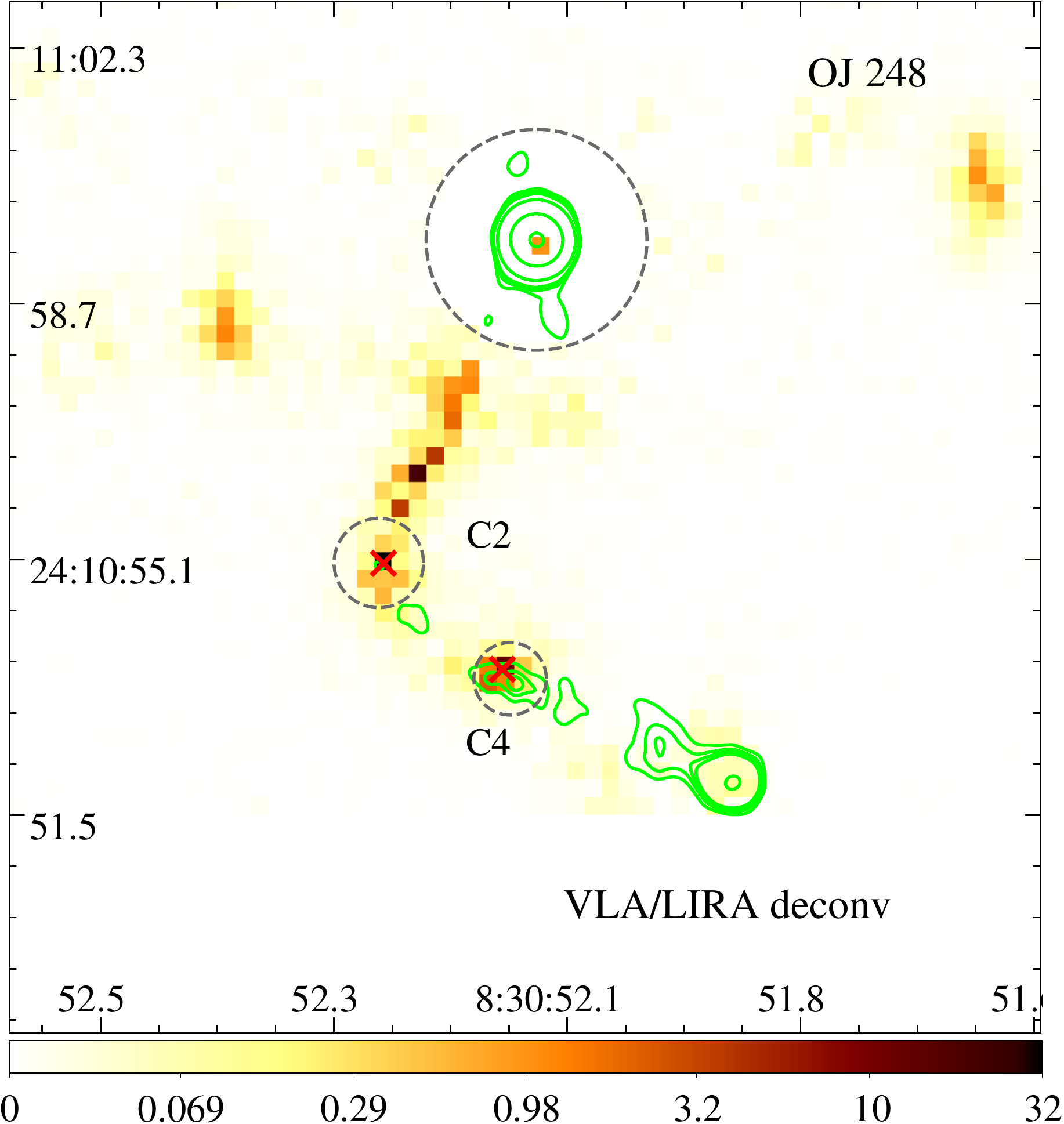}{0.5\textwidth}{(b)}
    }
    \caption{Same as in Fig. \ref{fig:results-3C9} but for OJ 248. The radio contours are given by 0.03, 0.06, 0.08, 0.1, 0.4, 10.0, 100.0 mJy beam$^{-1}$.\label{fig:results-OJ248}}
\end{figure*}

\begin{figure*}[ht]
    \gridline{
        \fig{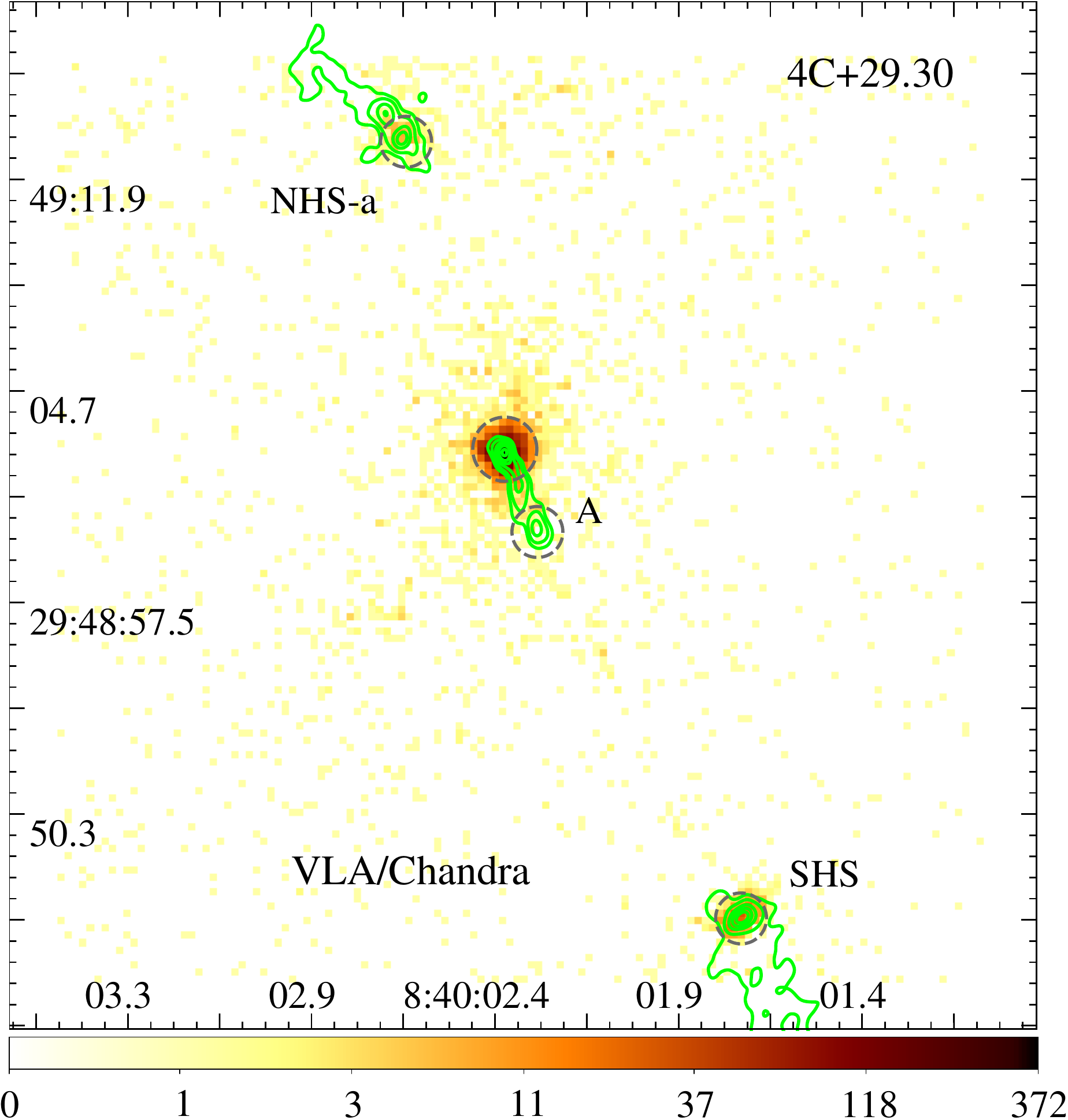}{0.5\textwidth}{(a)}
        \fig{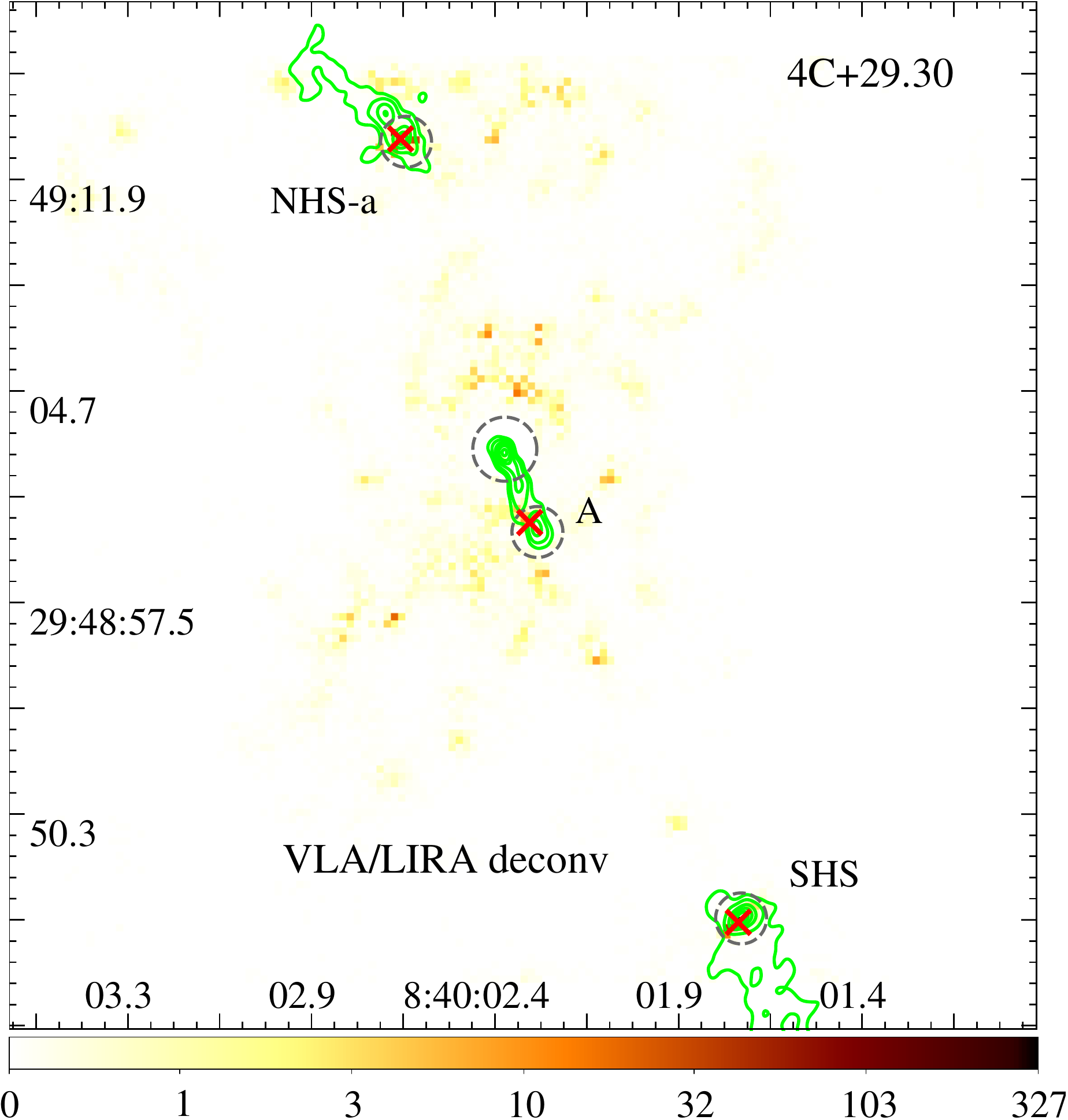}{0.5\textwidth}{(b)}
    }
    \caption{Same as in Fig. \ref{fig:results-3C9} but for 4C +29.30. The radio contours are given by 0.5, 1.0, 2.0, 3.0, 4.0, 6.0 mJy beam$^{-1}$.\label{fig:results-4C+29.30}}
\end{figure*}

\begin{figure*}[ht]
    \gridline{
        \fig{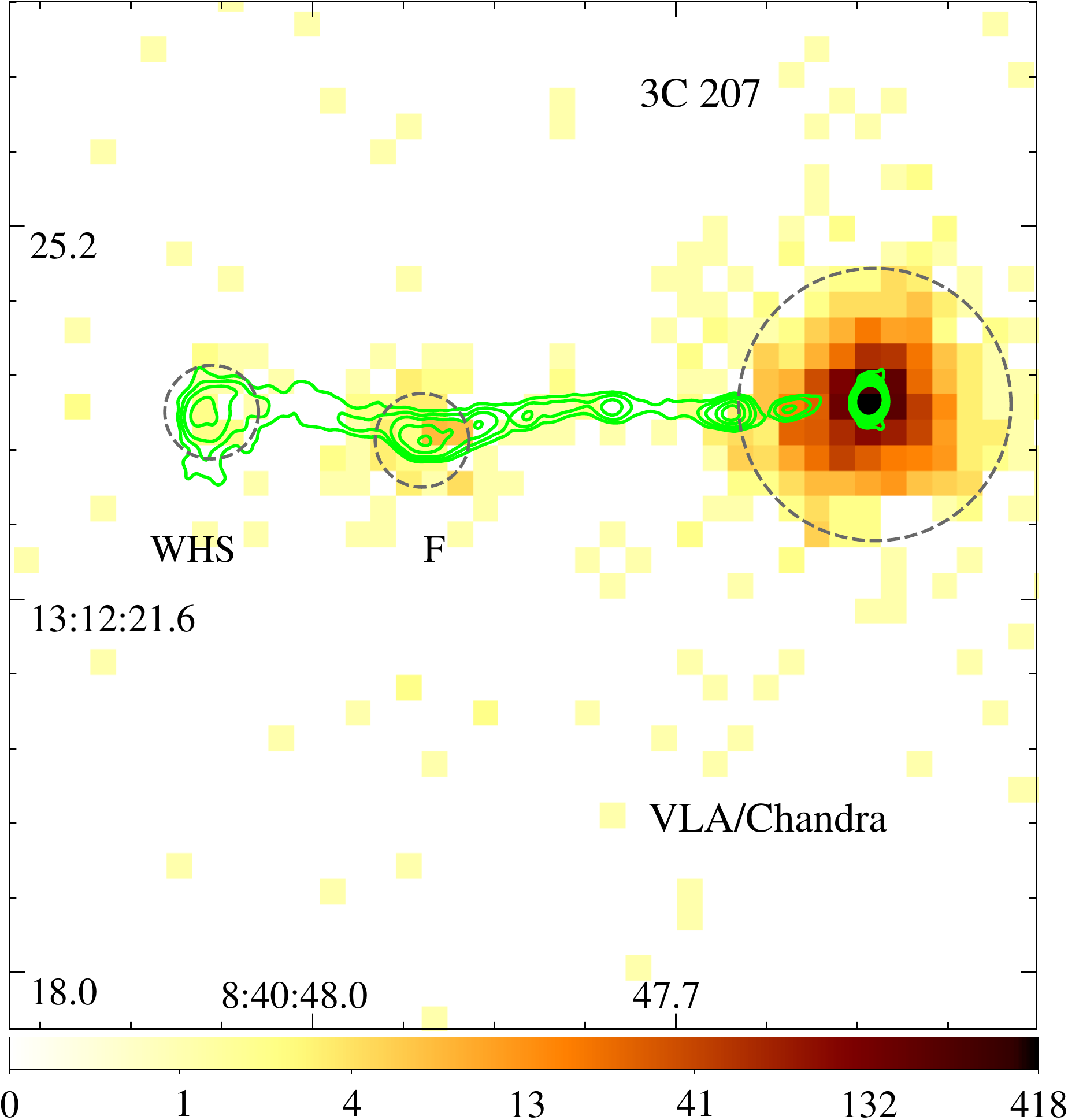}{0.5\textwidth}{(a)}
        \fig{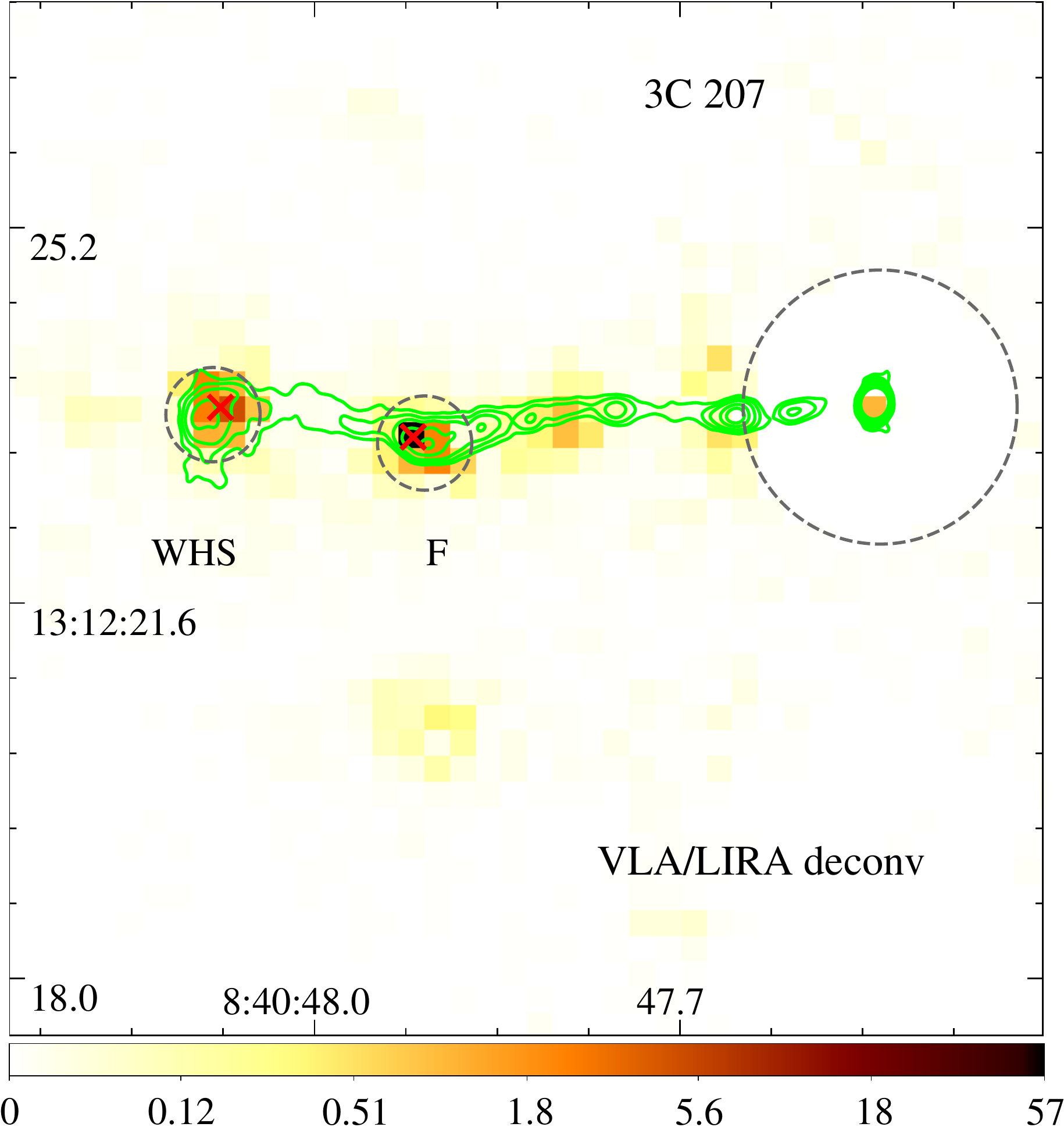}{0.5\textwidth}{(b)}
    }
    \caption{Same as in Fig. \ref{fig:results-3C9} but for 3C 207. The radio contours are given by 0.2, 0.4, 0.8, 2.0, 4.0, 8.0, 20.0 mJy beam$^{-1}$.\label{fig:results-3C207}}
\end{figure*}

\begin{figure*}[ht]
    \gridline{
        \fig{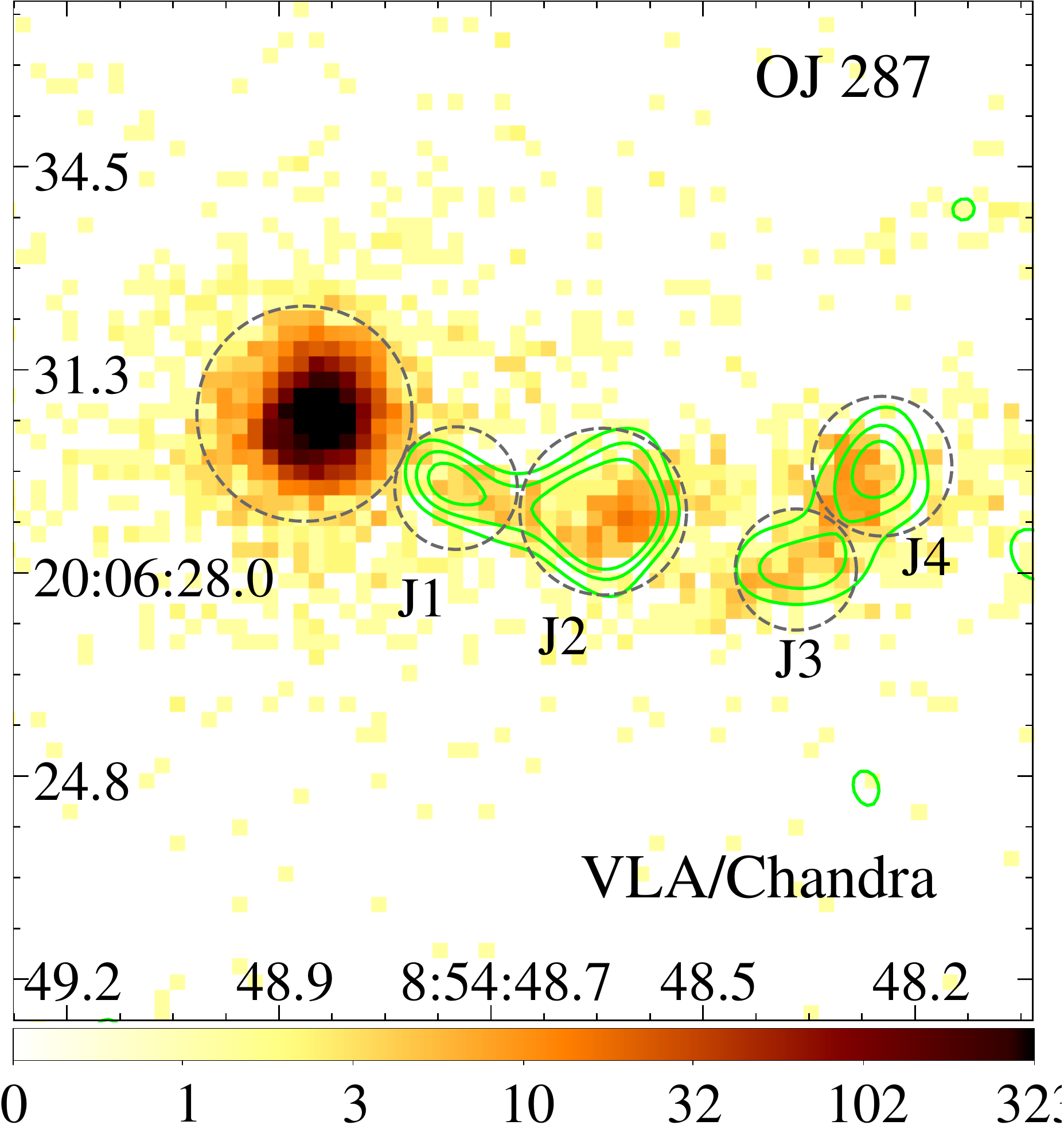}{0.5\textwidth}{(a)}
        \fig{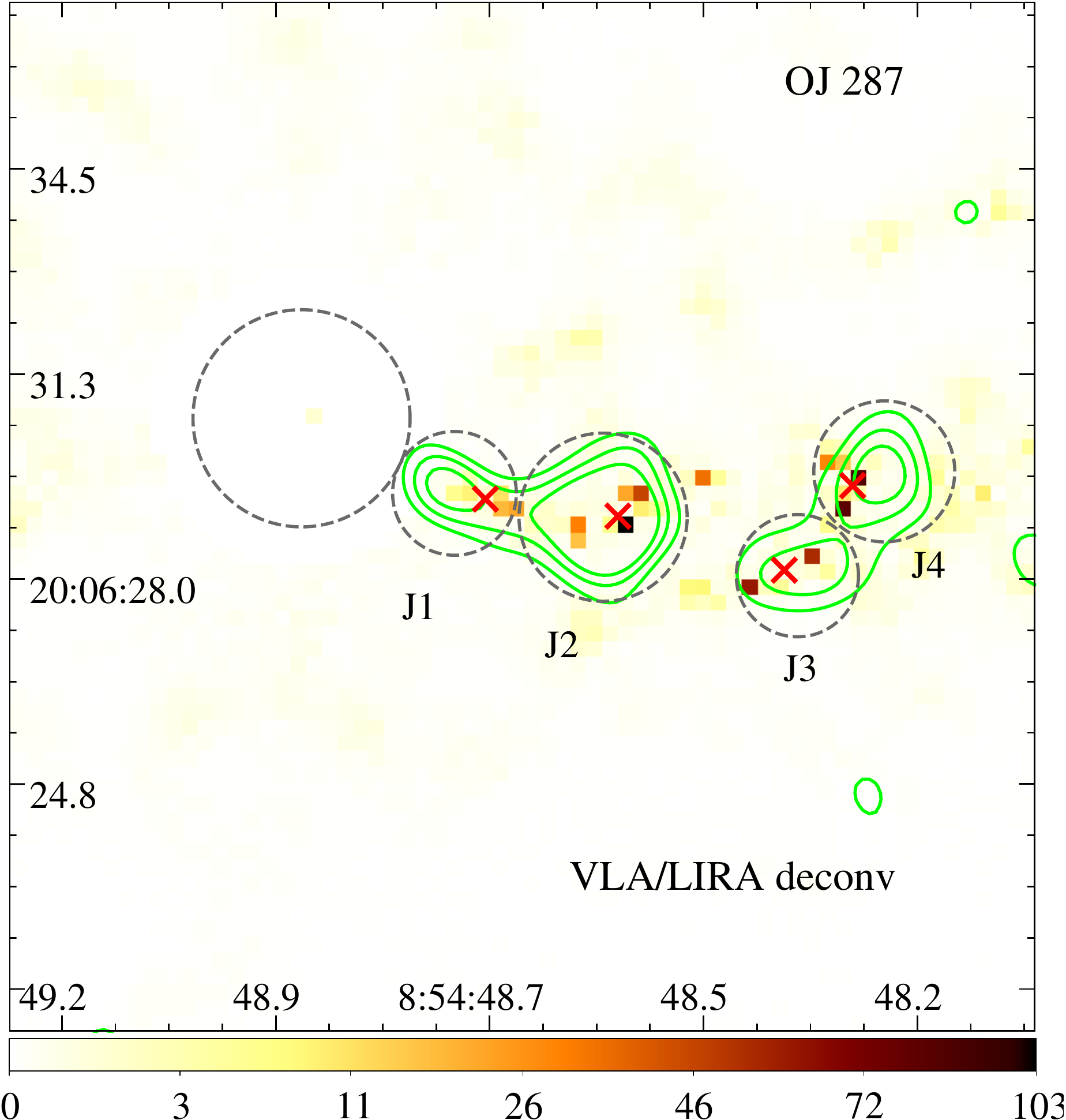}{0.5\textwidth}{(b)}
    }
    \caption{Same as in Fig. \ref{fig:results-3C9} but for OJ 287. The radio contours are given by 0.4, 0.6, 0.8 mJy beam$^{-1}$.\label{fig:results-OJ287}}
\end{figure*}

\begin{figure*}[ht]
    \gridline{
        \fig{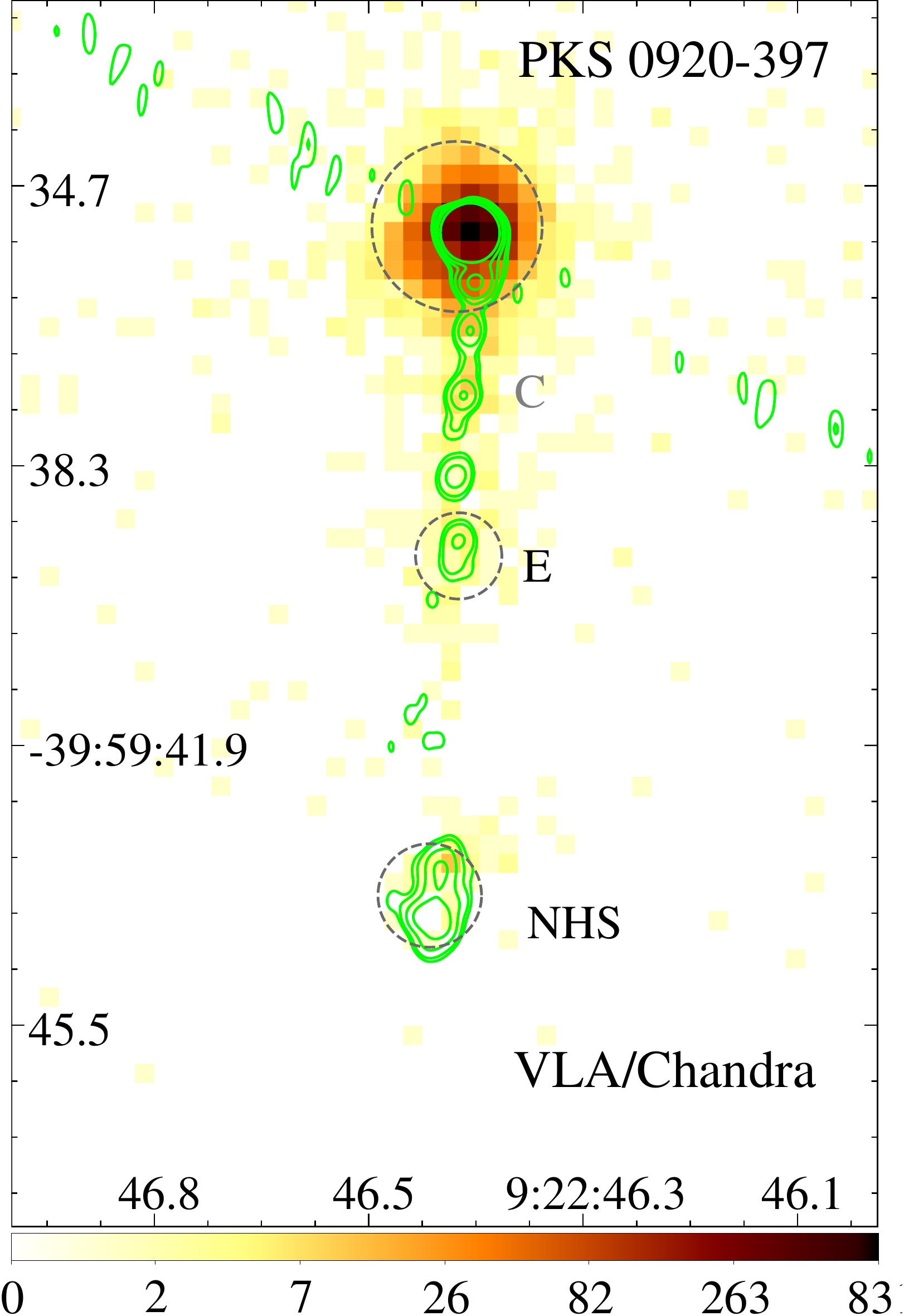}{0.5\textwidth}{(a)}
        \fig{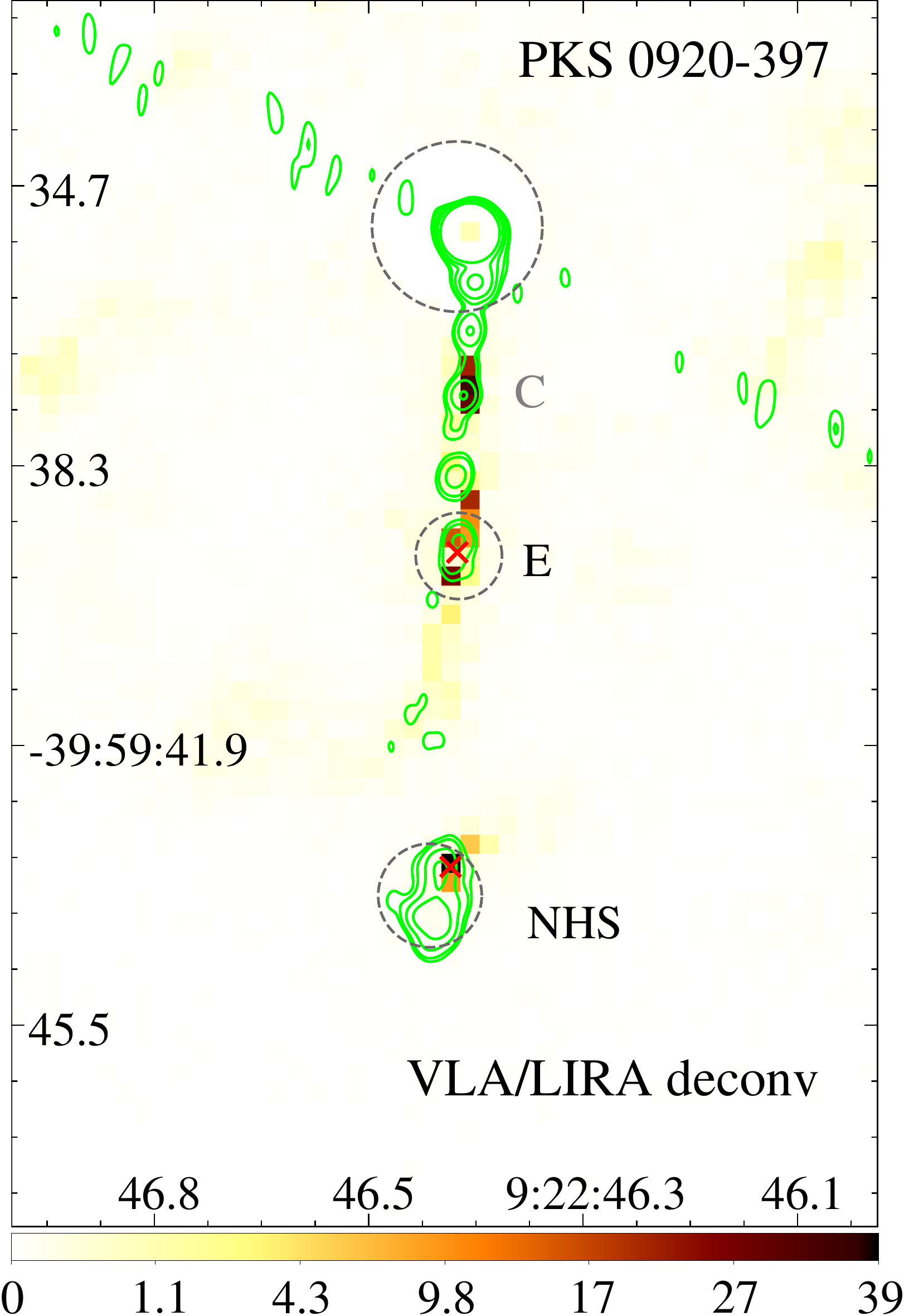}{0.5\textwidth}{(b)}
    }
    \caption{Same as in Fig. \ref{fig:results-3C9} but for PKS 0920-397. The radio contours are given by 0.2, 0.4, 0.8, 2.0, 4.0, 8.0, 20.0 mJy beam$^{-1}$.\label{fig:results-PKS0920-397}}
\end{figure*}

\begin{figure*}[ht]
    \gridline{
        \fig{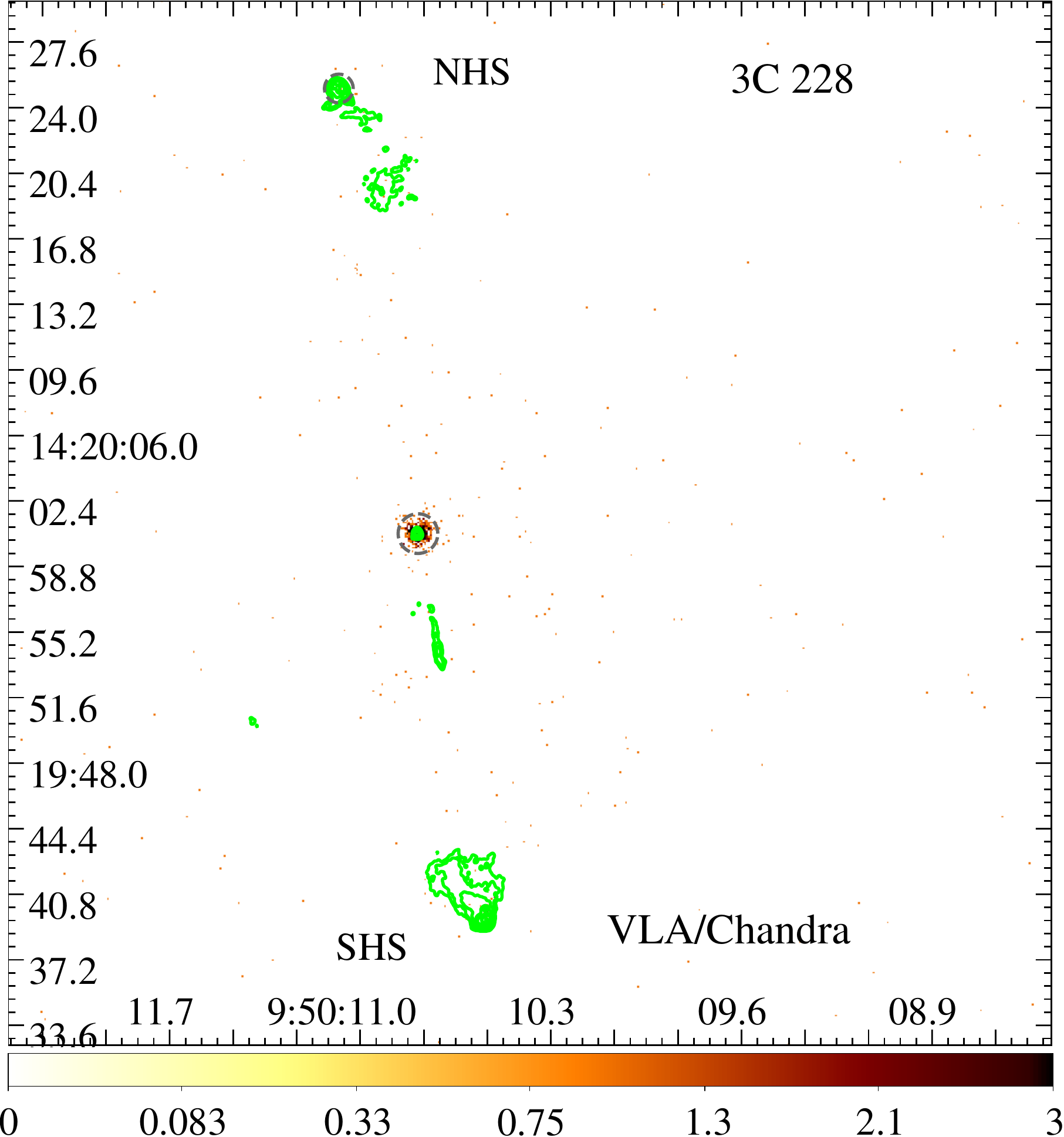}{0.5\textwidth}{(a)}
    }
    \comment{
    \gridline{
        \fig{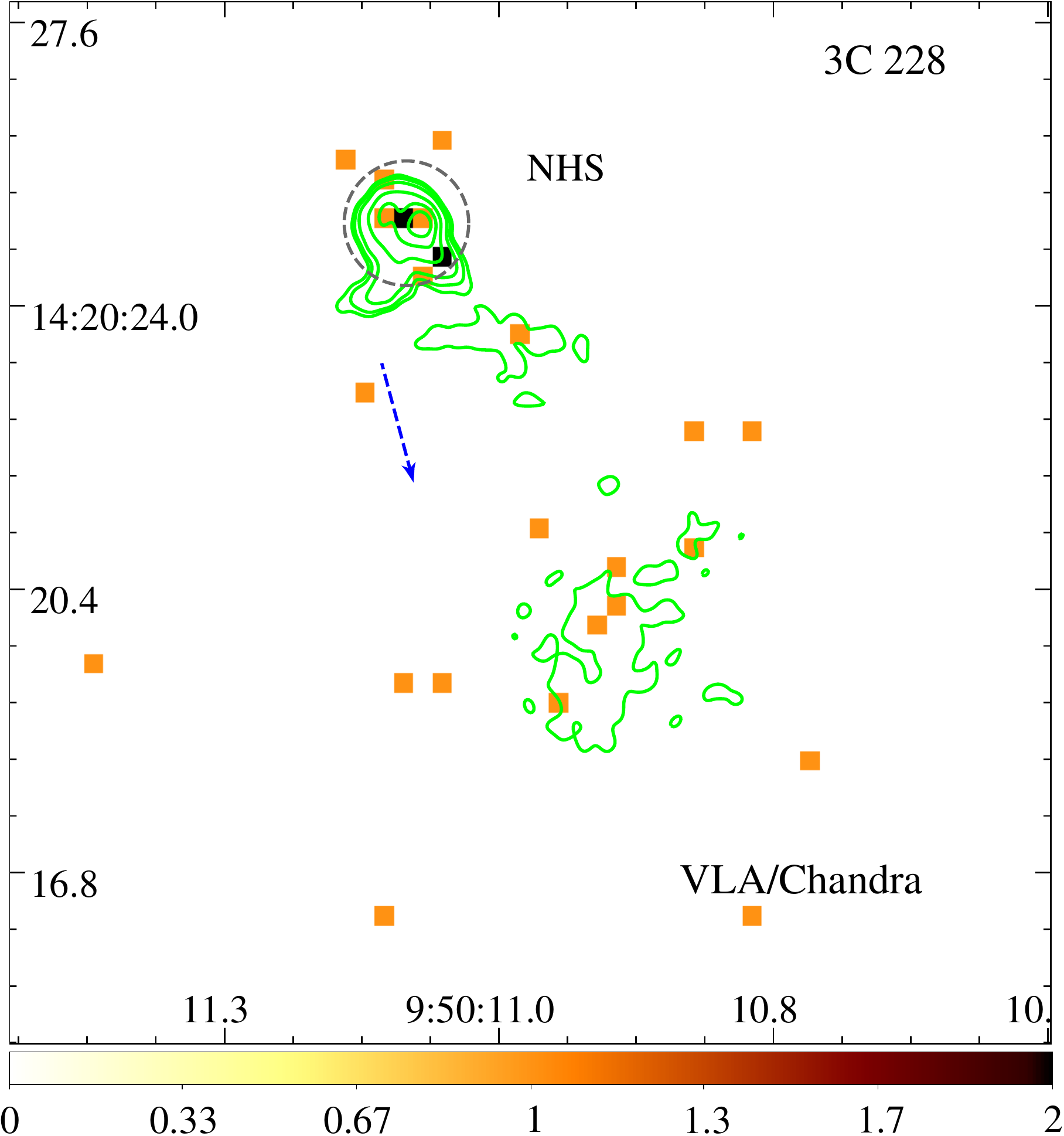}{0.5\textwidth}{(b)}
        \fig{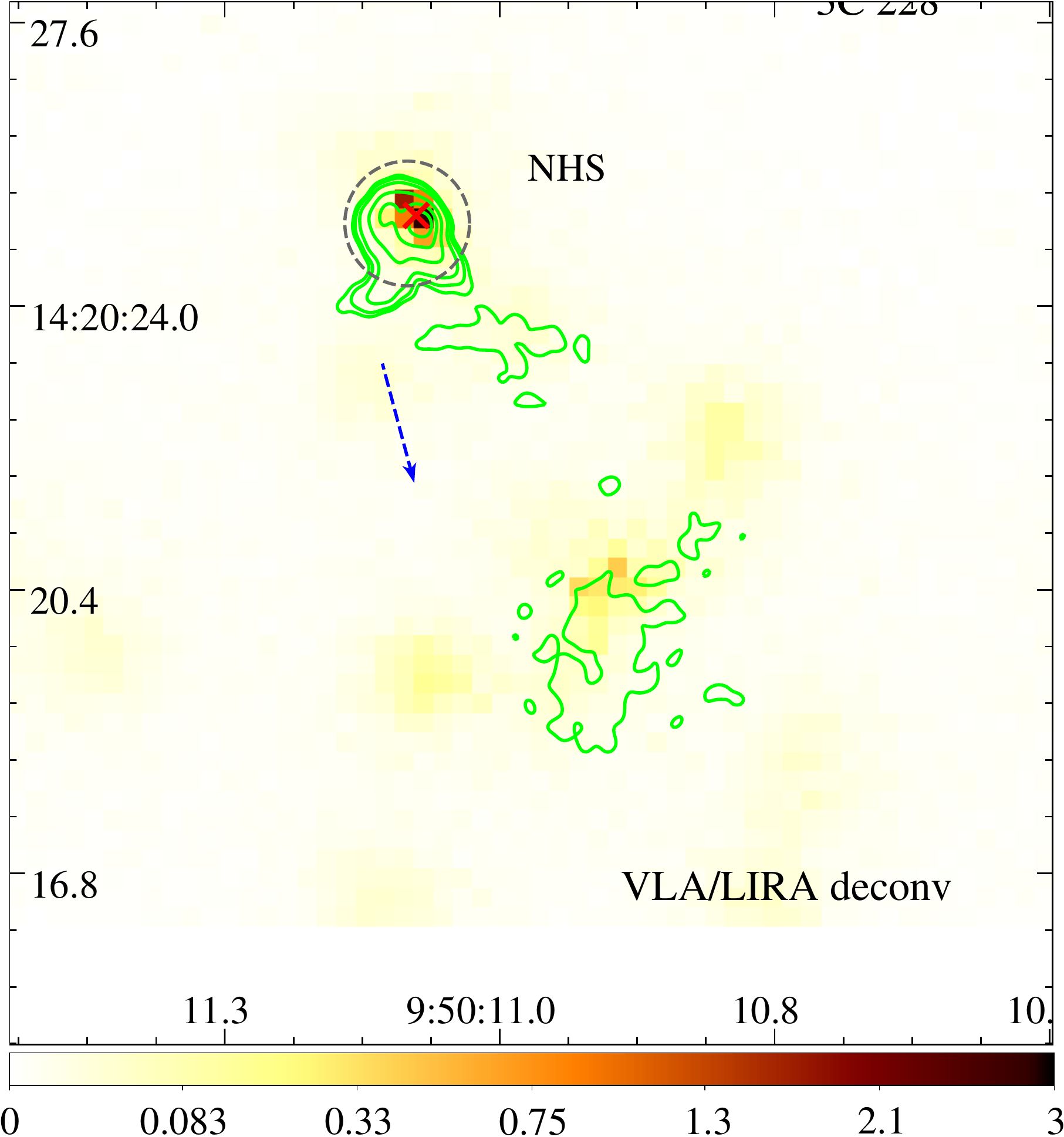}{0.5\textwidth}{(c)}
    }
    }
    \gridline{
        \fig{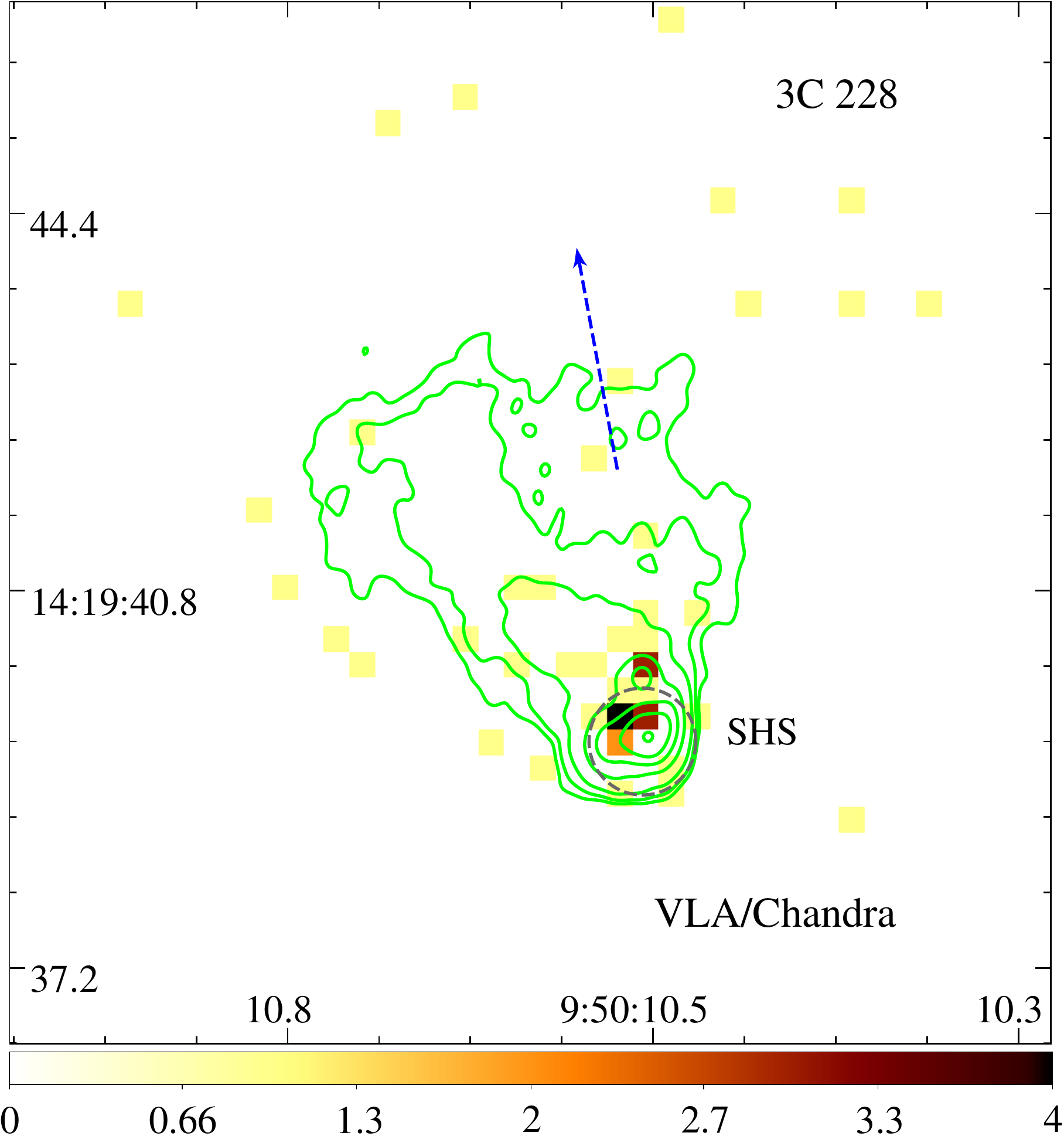}{0.5\textwidth}{(d)}
        \fig{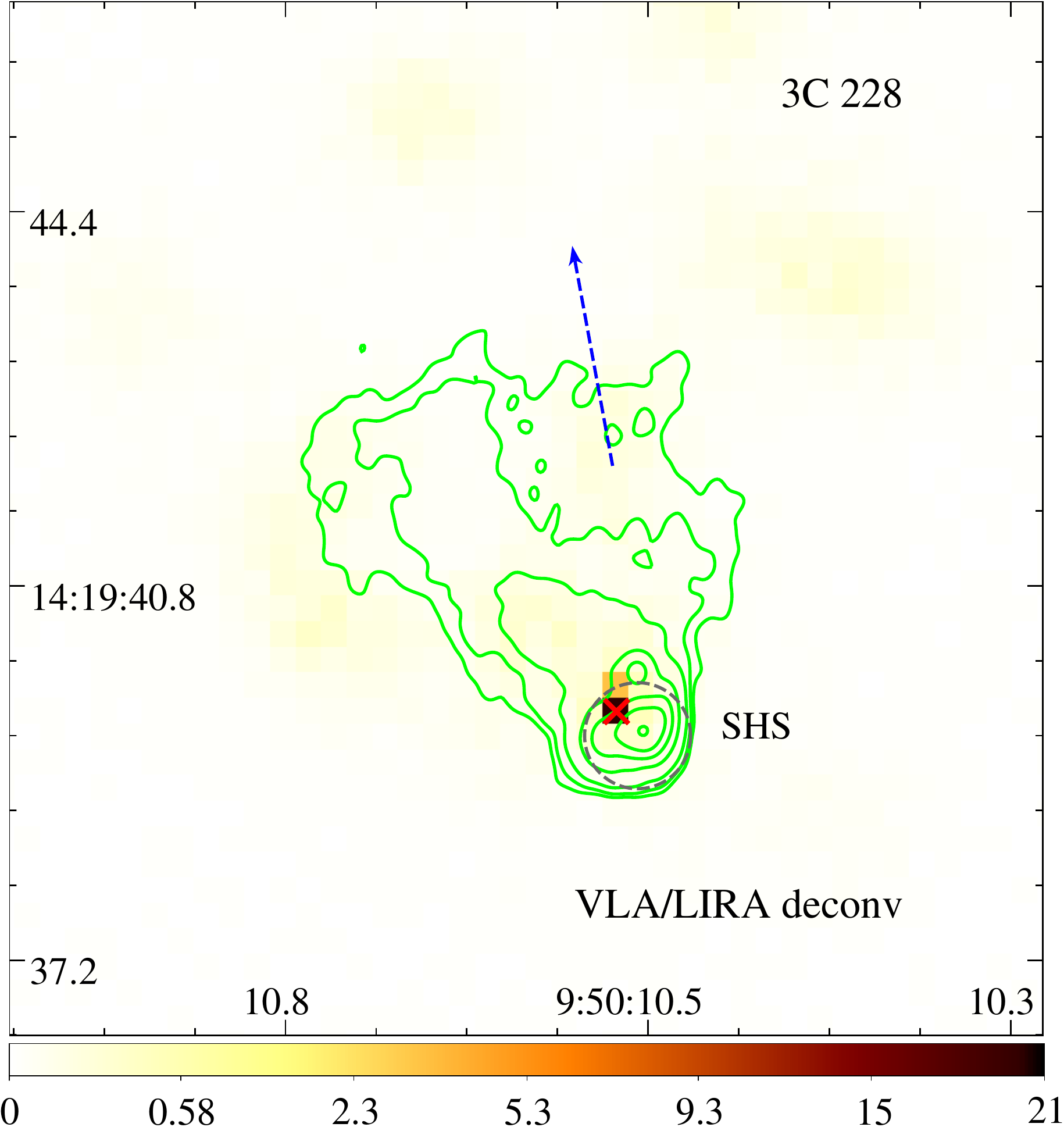}{0.5\textwidth}{(e)}
    }
    \caption{Same as in Fig. \ref{fig:results-3C9} but for 3C 228. (a) shows the full image while (b) and (c) zoom into the southern hotspot (SHS). The radio contours are given by 0.2, 0.4, 1.0, 4.0, 10.0, 20.0, 40.0 mJy beam$^{-1}$.\label{fig:results-3C228}}
\end{figure*}

\begin{figure*}[ht]
    \gridline{
        \fig{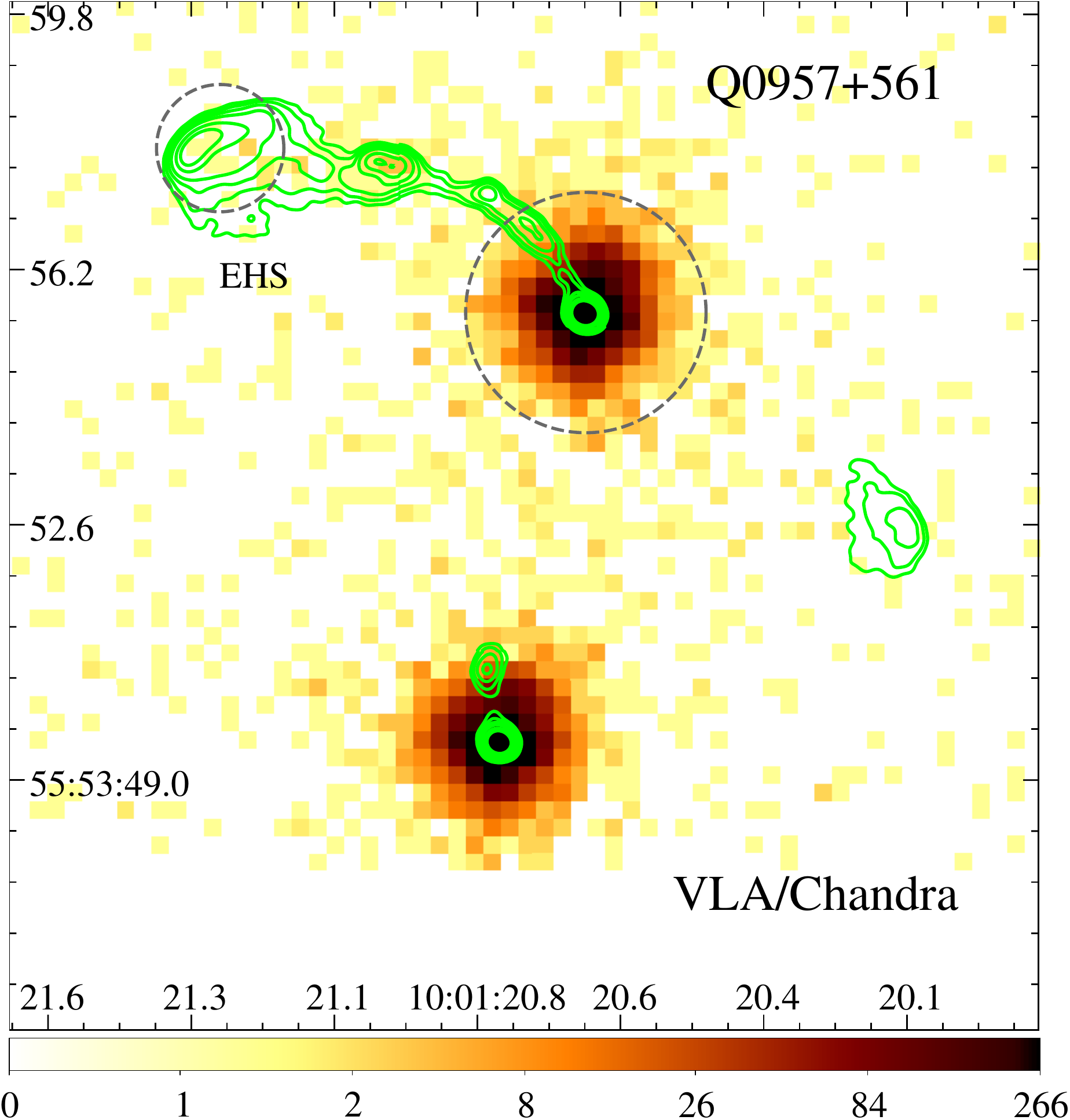}{0.5\textwidth}{(a)}
        \fig{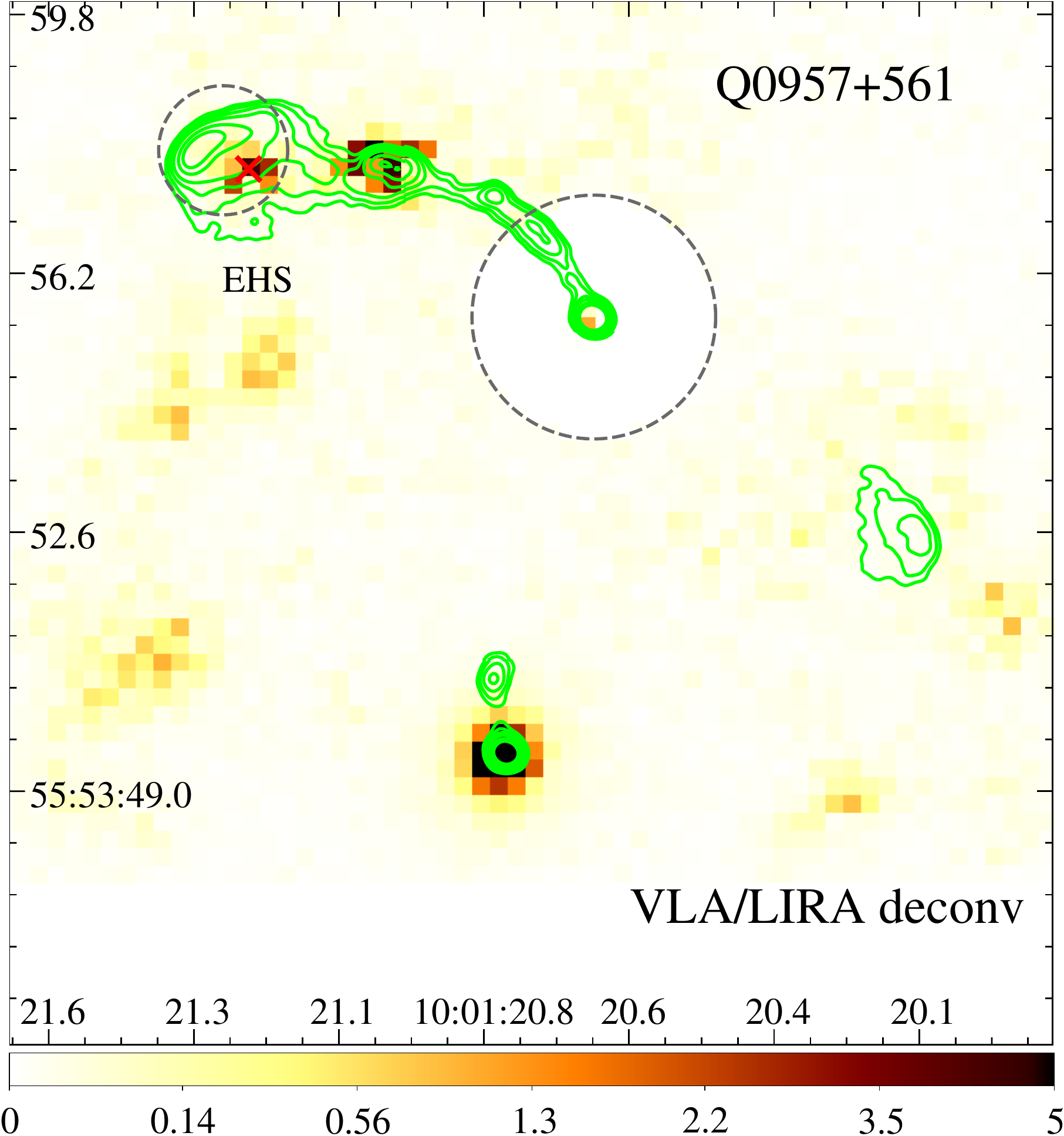}{0.5\textwidth}{(b)}
    }
    \caption{Same as in Fig. \ref{fig:results-3C9} but for Q 0957+561. The radio contours are given by 0.1, 0.2, 0.4, 0.8, 2.0, 3.8 mJy beam$^{-1}$.\label{fig:results-QSO0957+561}}
\end{figure*}

\begin{figure*}[ht]
    \gridline{
        \fig{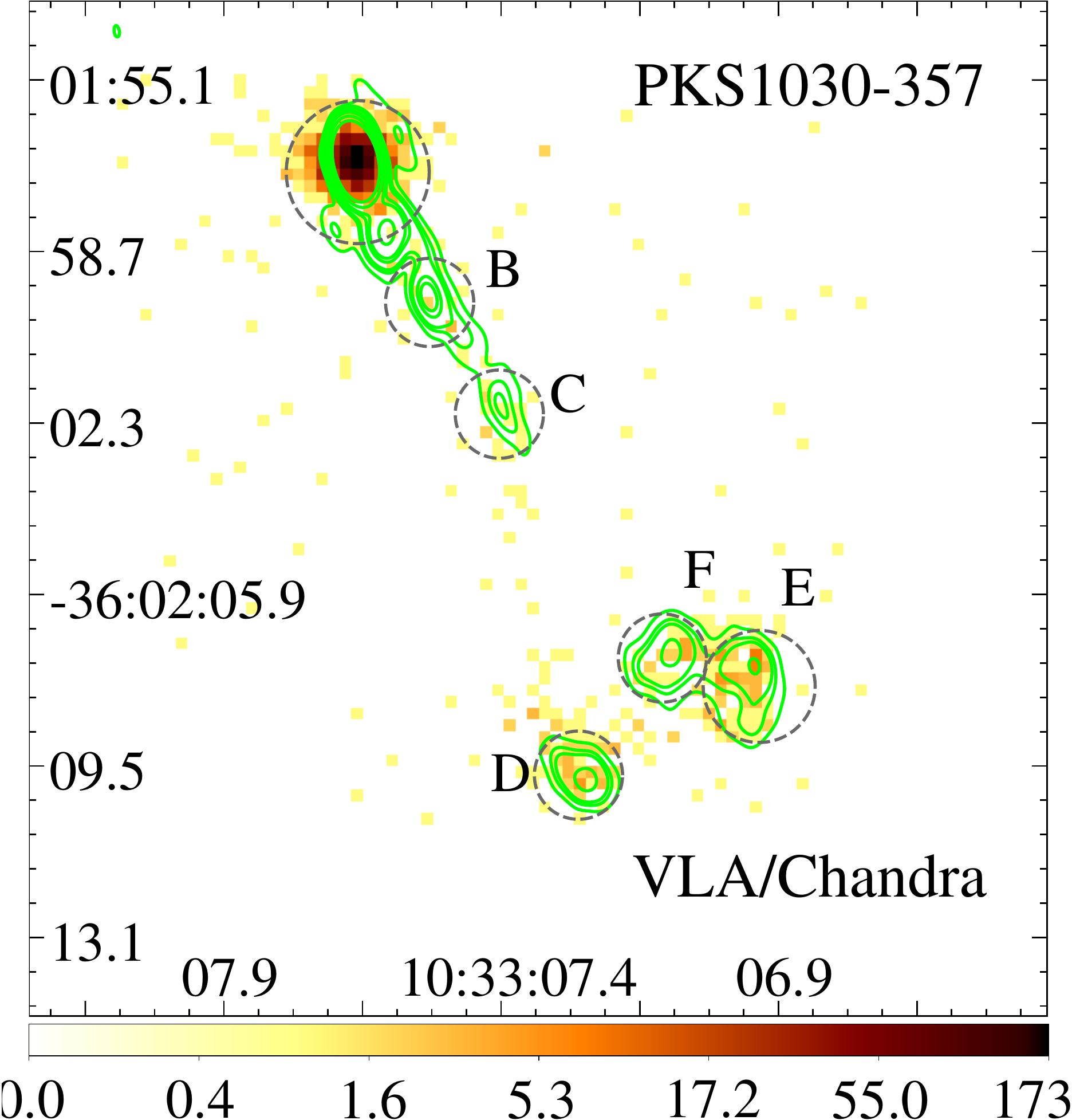}{0.5\textwidth}{(a)}
        \fig{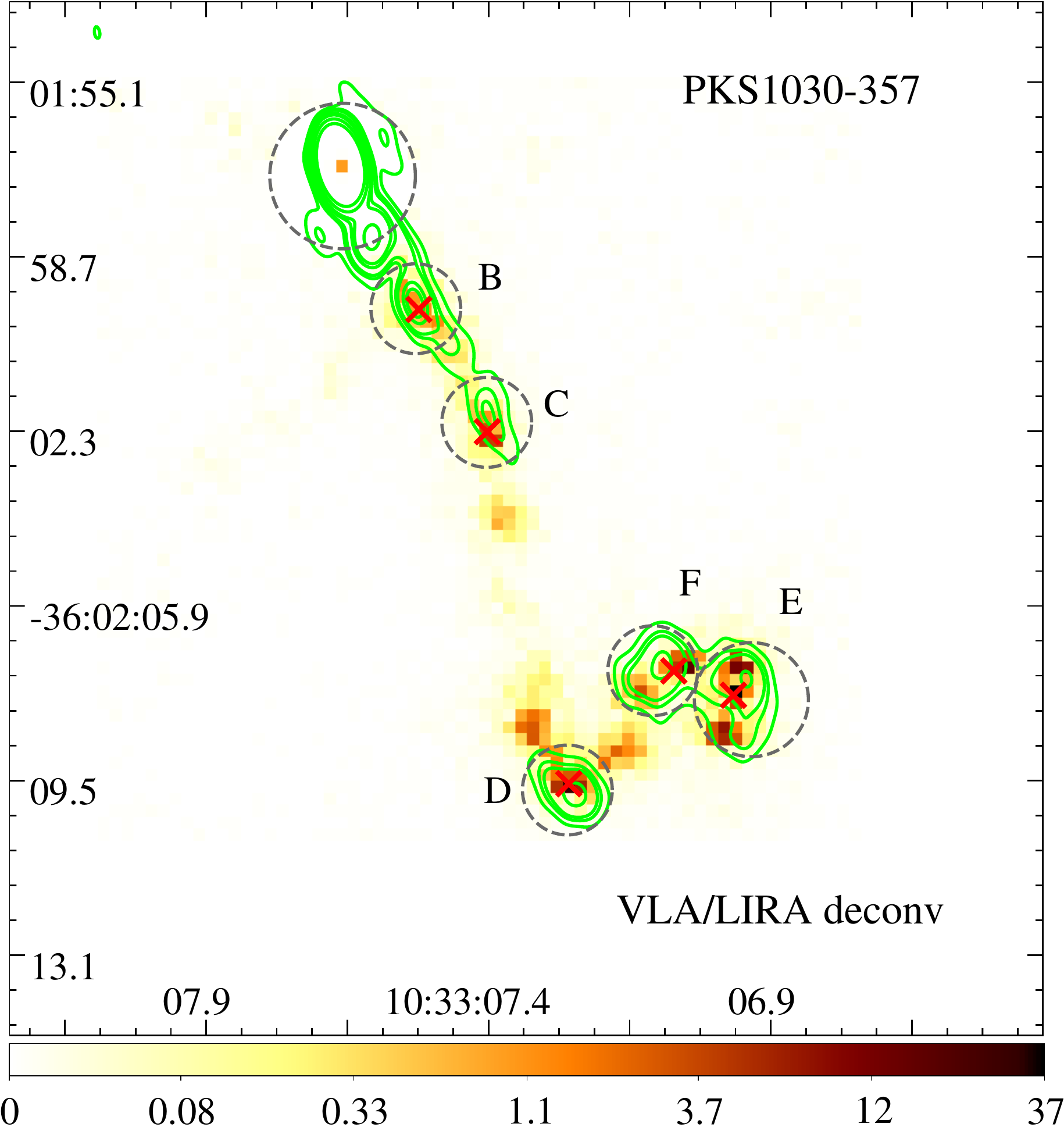}{0.5\textwidth}{(b)}
    }
    \caption{Same as in Fig. \ref{fig:results-3C9} but for PKS 1030-357. The radio contours are given by 0.3, 0.6, 0.8, 1.5, 2.0, 4.0, 8.0 mJy beam$^{-1}$.\label{fig:results-PKS1030-357}}
\end{figure*}

\begin{figure*}[ht]
    \gridline{
        \fig{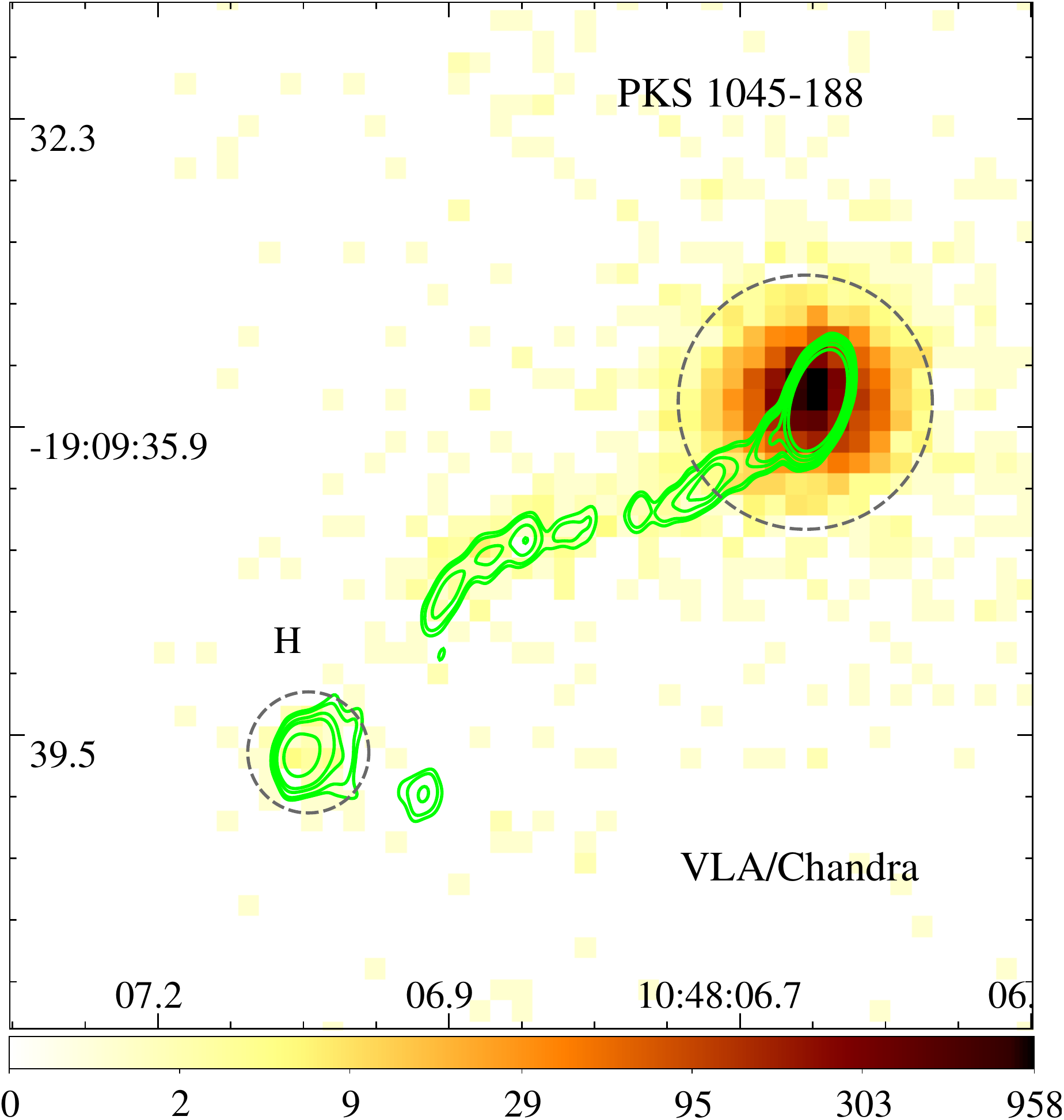}{0.5\textwidth}{(a)}
        \fig{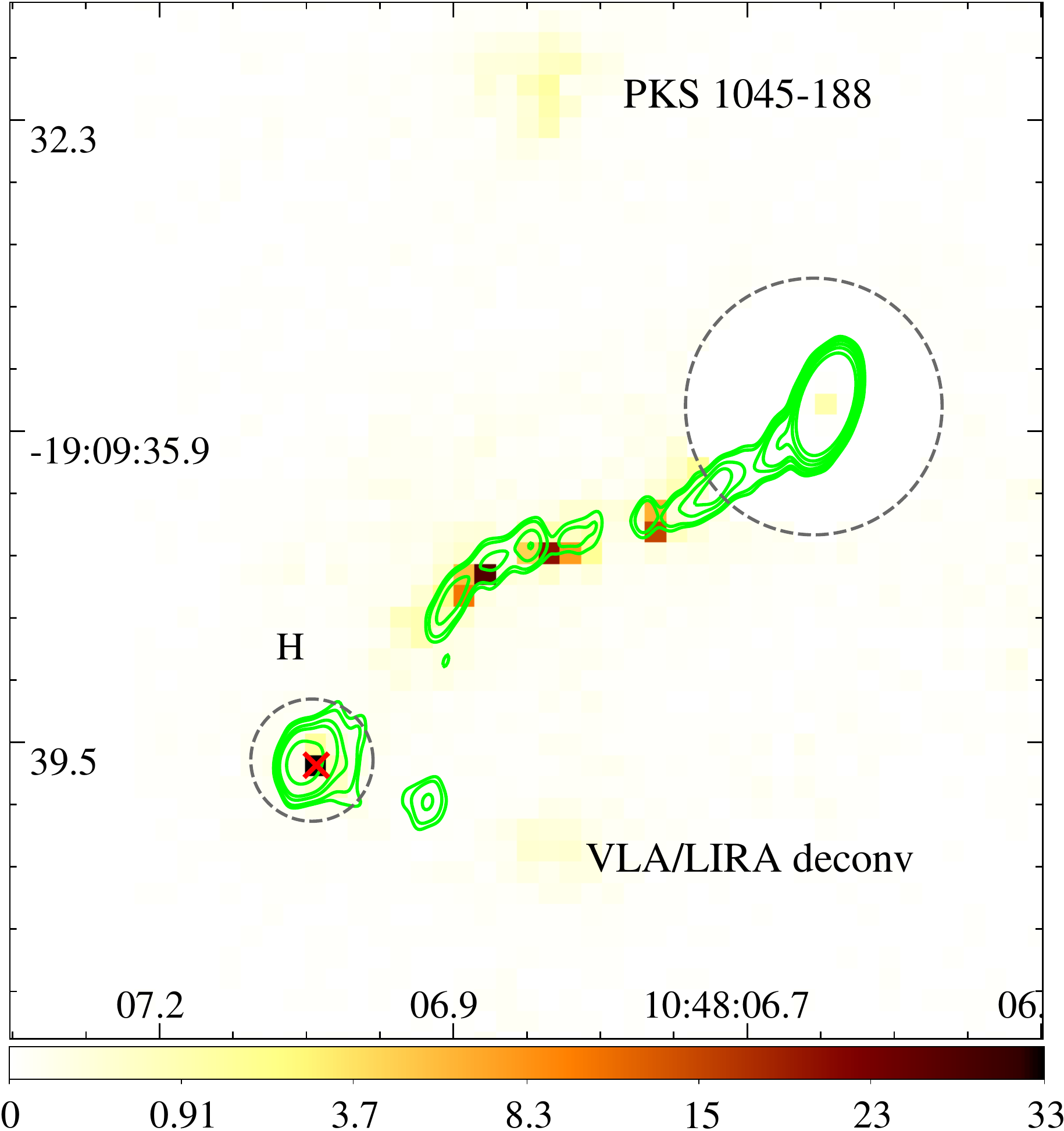}{0.5\textwidth}{(b)}
    }
    \caption{Same as in Fig. \ref{fig:results-3C9} but for PKS 1045-188. The radio contours are given by 1.0, 1.5, 0.8, 2.0, 4.0, 8.0, 20.0 mJy beam$^{-1}$.\label{fig:results-PKS1045-188}}
\end{figure*}

\begin{figure*}[ht]
    \gridline{
        \fig{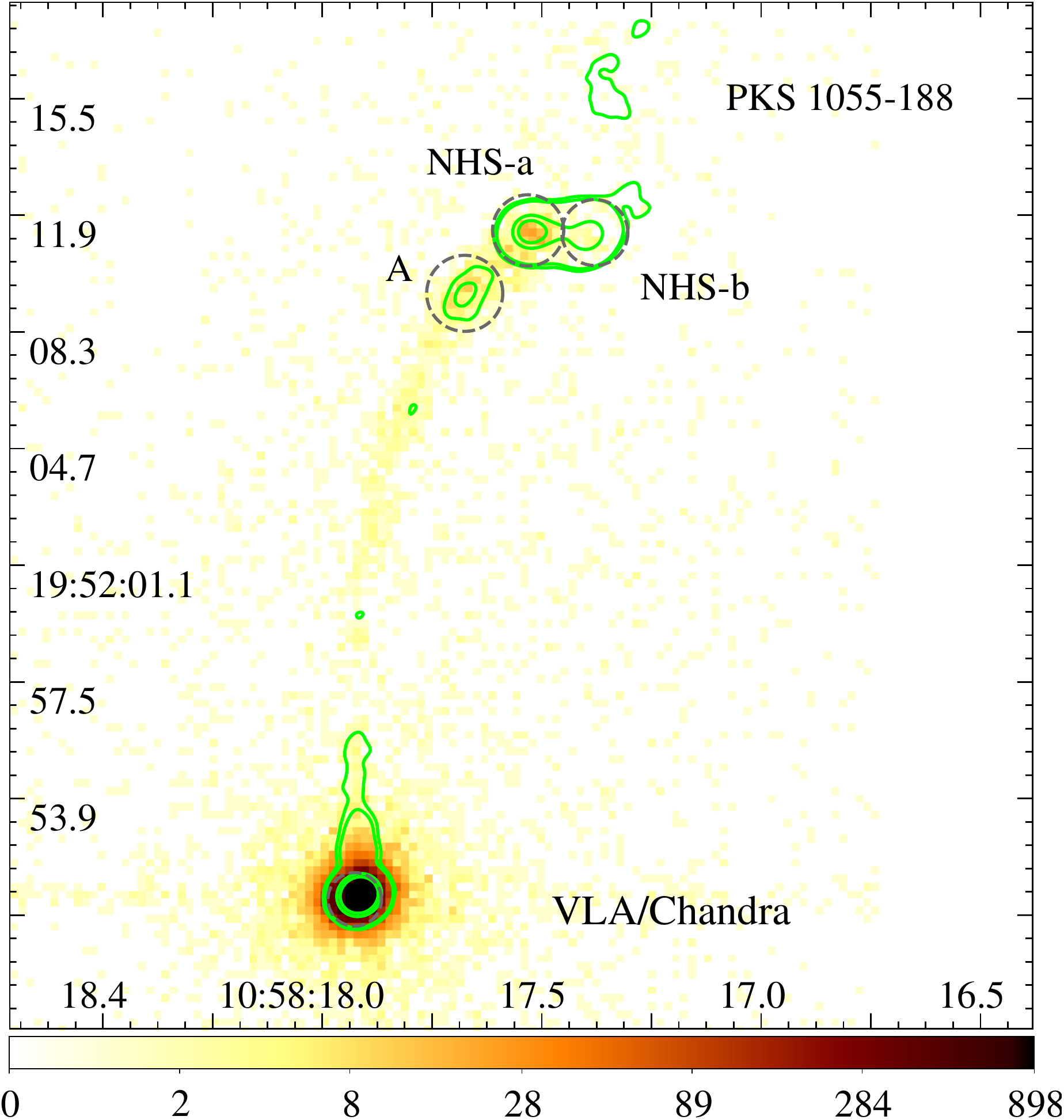}{0.5\textwidth}{(a)}
        \fig{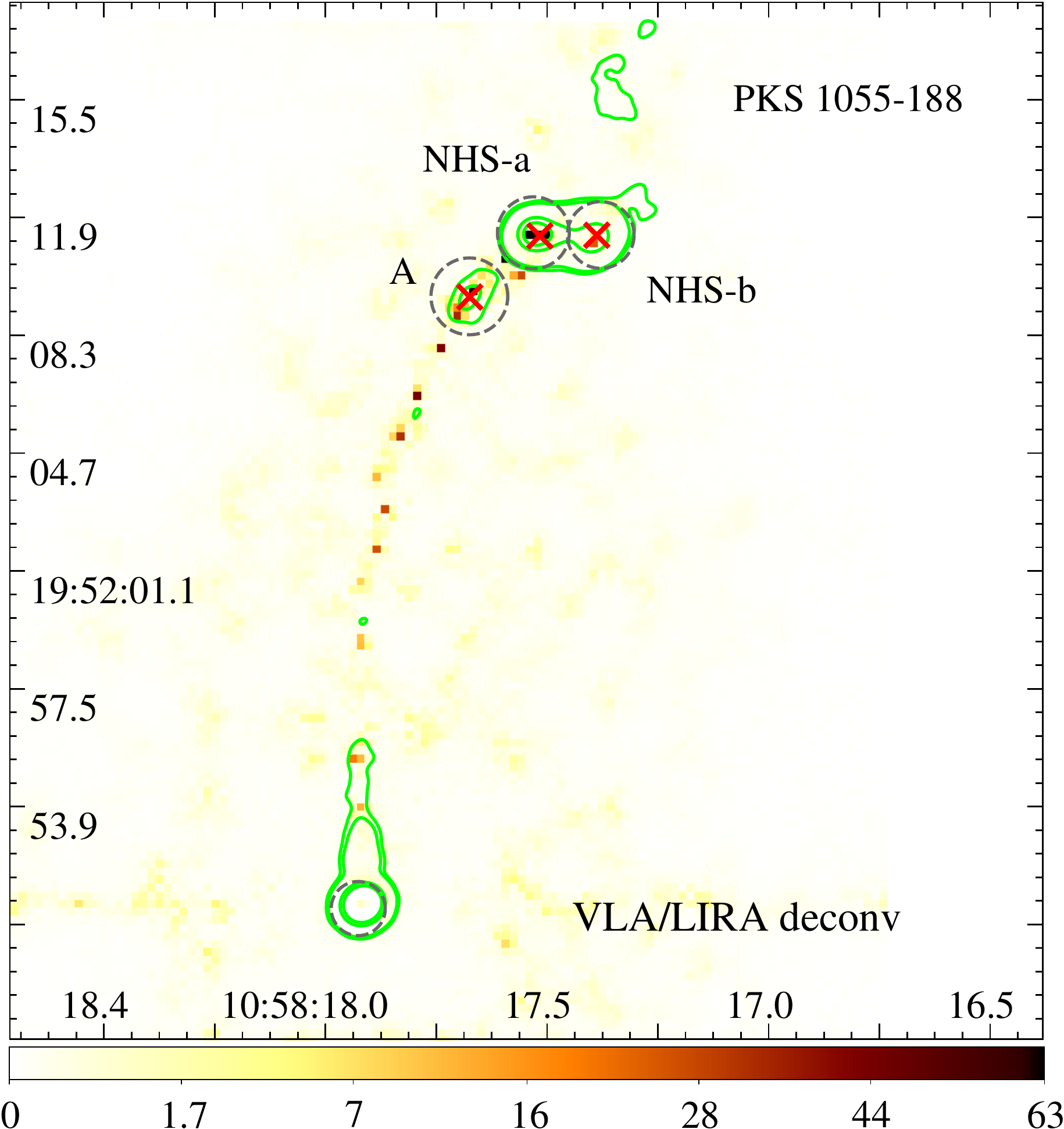}{0.5\textwidth}{(b)}
    }
    \caption{Same as in Fig. \ref{fig:results-3C9} but for PKS 1055+201. The radio contours are given by 2.0, 4.0, 100.0, 200.0 mJy beam$^{-1}$.\label{fig:results-PKS1055+201}}
\end{figure*}

\begin{figure*}[ht]
    \gridline{
        \fig{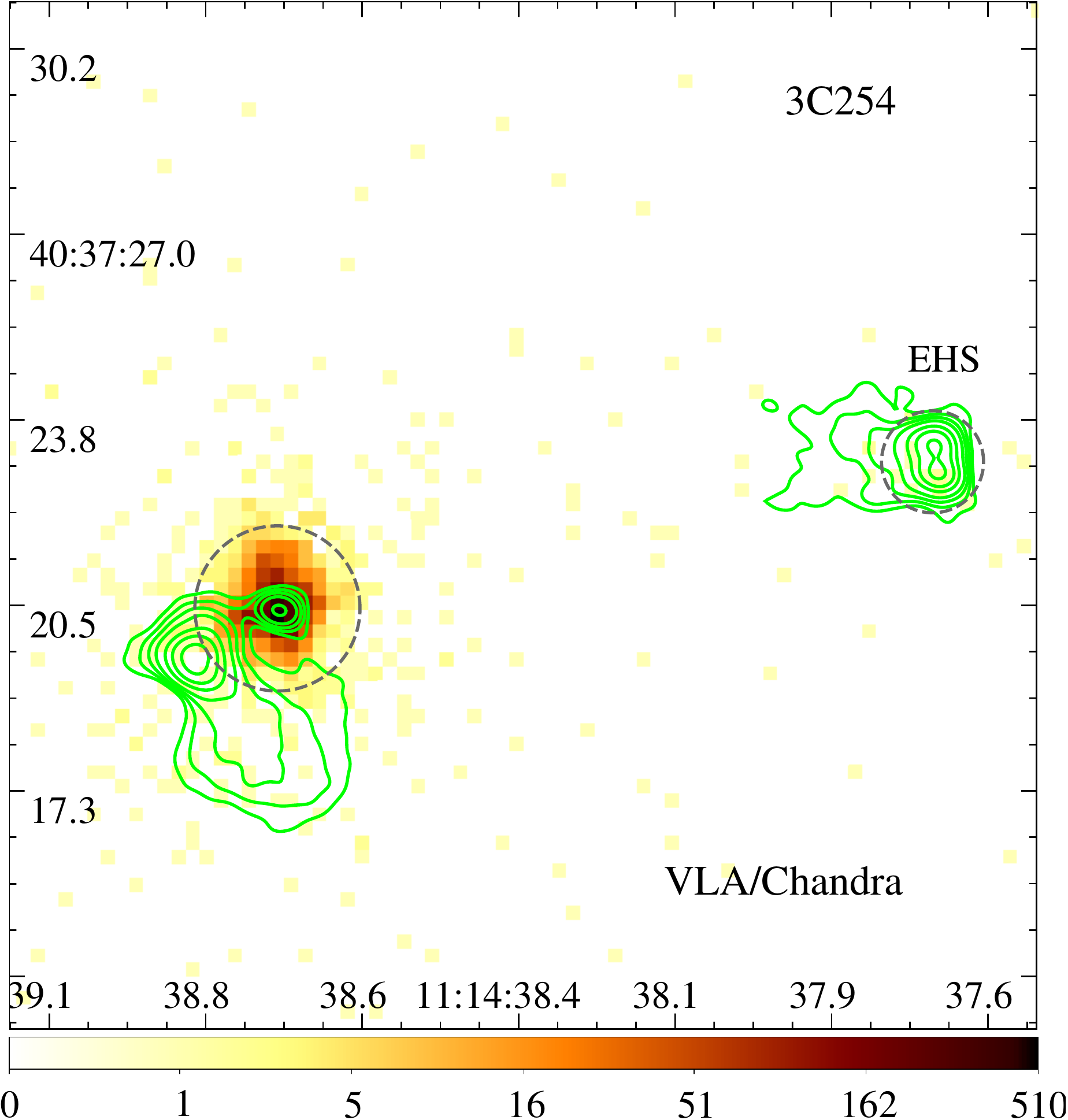}{0.5\textwidth}{(a)}
        \fig{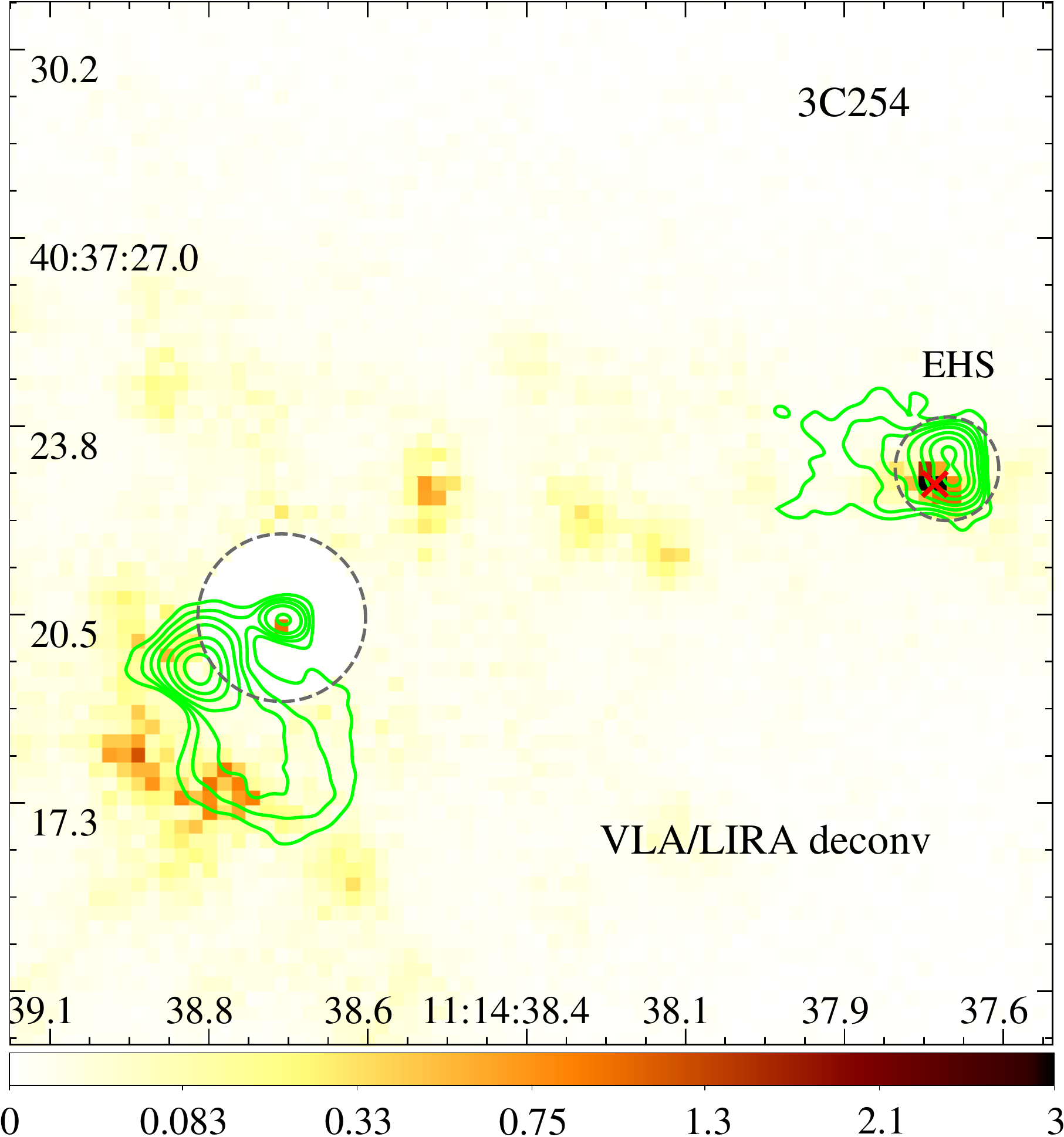}{0.5\textwidth}{(b)}
    }
    \caption{Same as in Fig. \ref{fig:results-3C9} but for 3C 254. The radio contours are given by 0.8, 2.0, 4.0, 8.0, 20.0, 40.0, 80.0 mJy beam$^{-1}$.\label{fig:results-3C254}}
\end{figure*}

\begin{figure*}[ht]
    \gridline{
        \fig{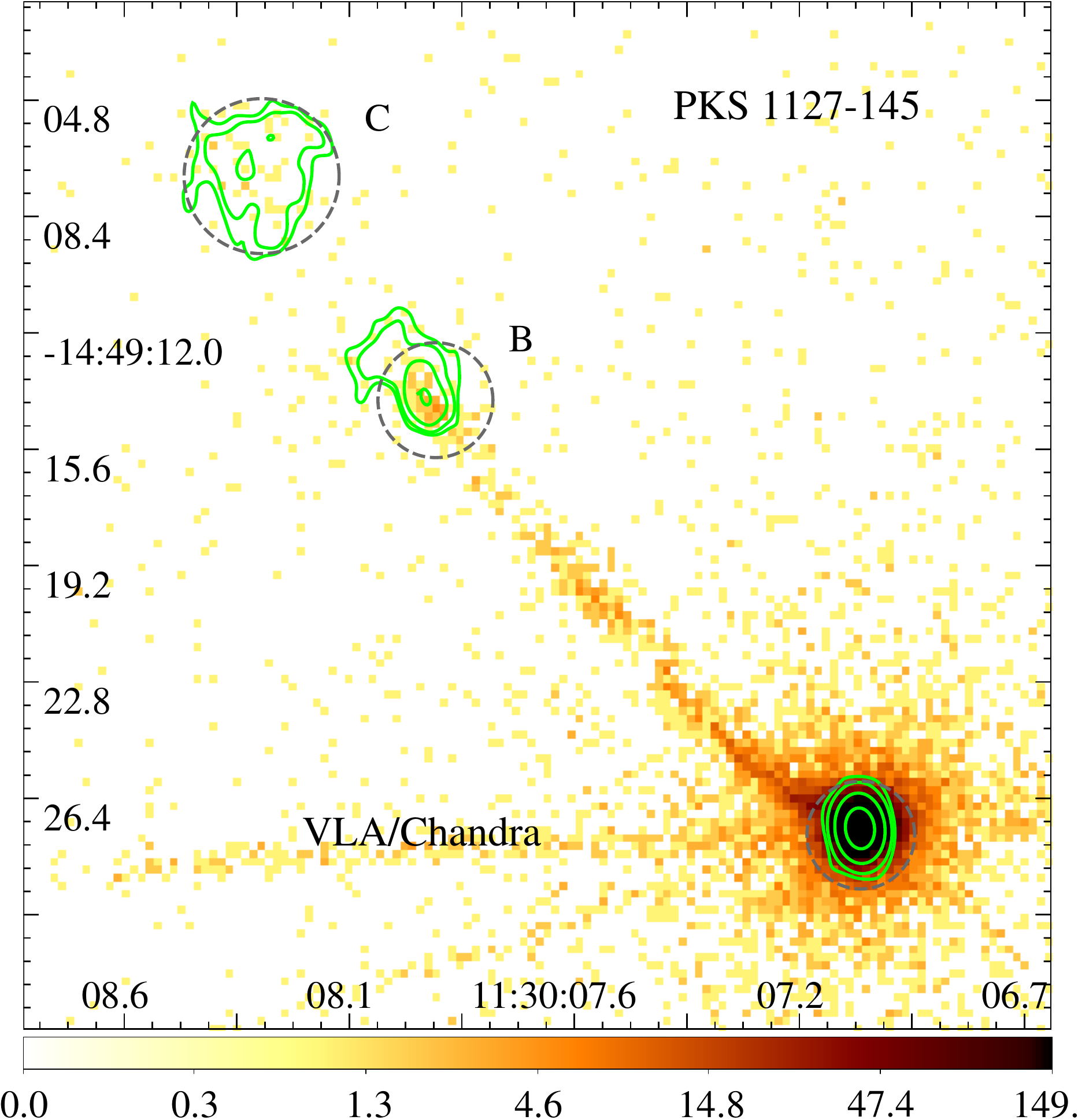}{0.5\textwidth}{(a)}
        \fig{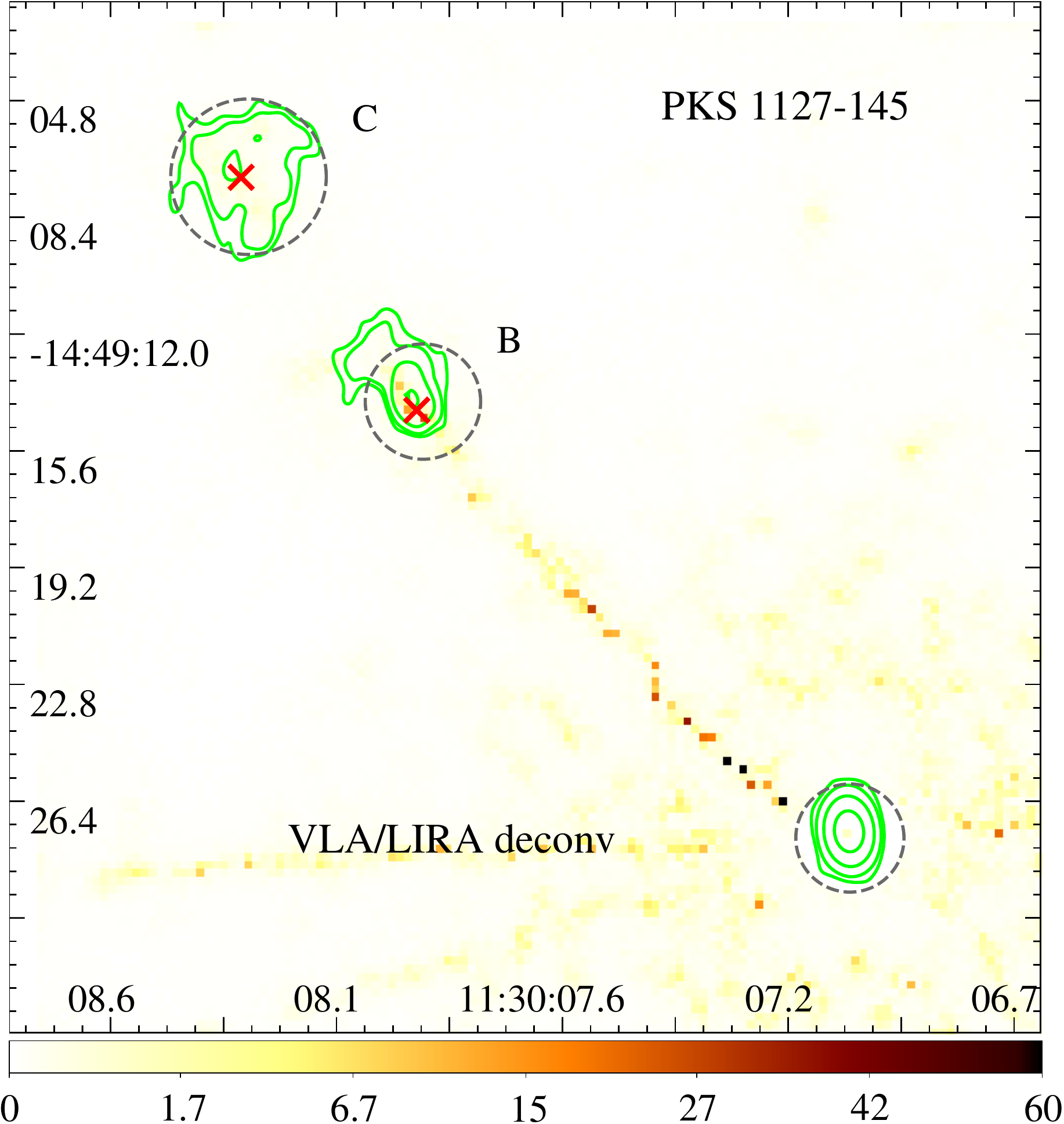}{0.5\textwidth}{(b)}
    }
    \caption{Same as in Fig. \ref{fig:results-3C9} but for PKS 1127-145. The radio contours are given by 0.25, 0.4, 0.85, 2.3, 10.0, 100.0, 1000.0 mJy beam$^{-1}$.\label{fig:results-PKS1127-145}}
\end{figure*}

\begin{figure*}[ht]
    \gridline{
        \fig{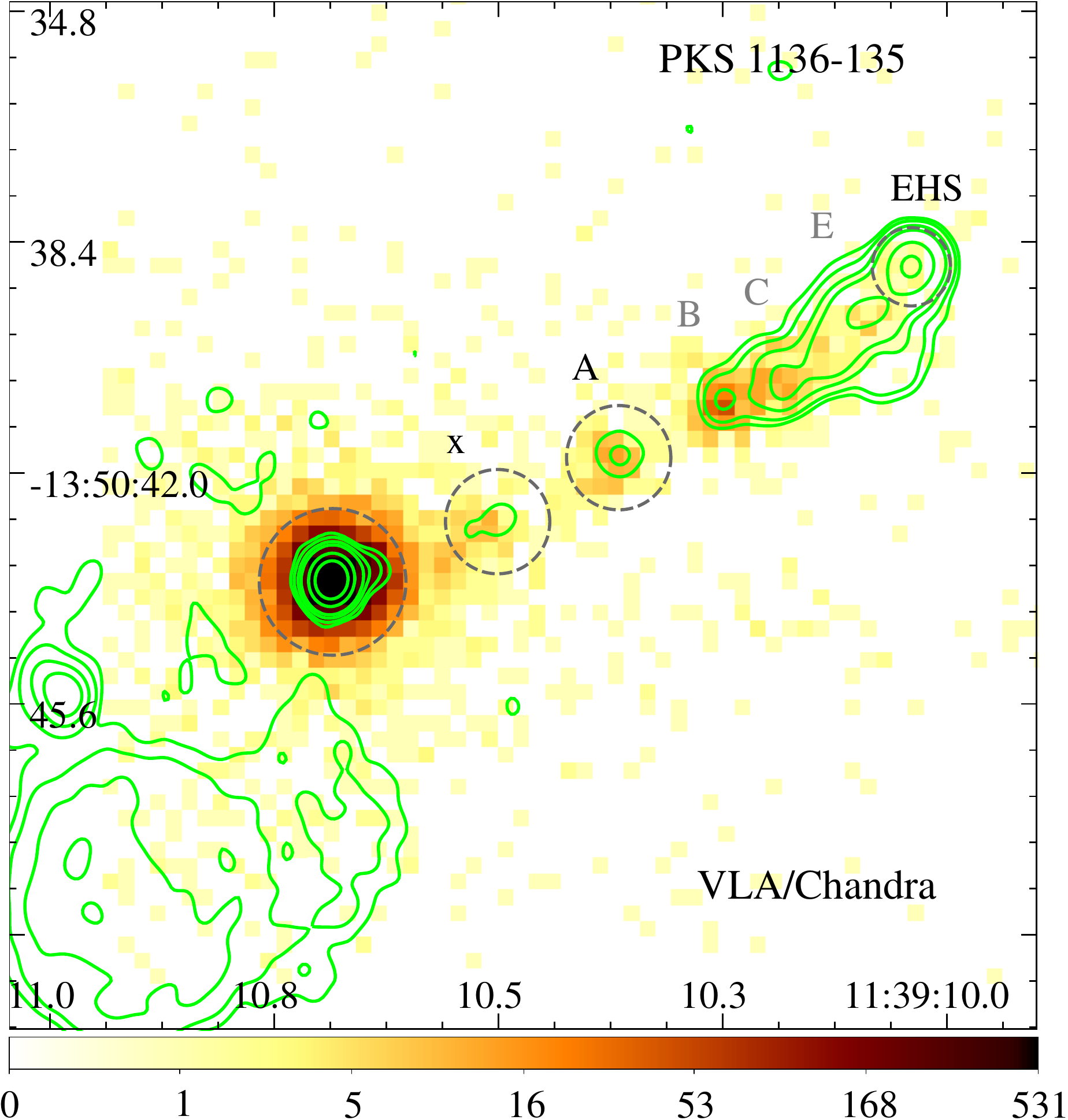}{0.5\textwidth}{(a)}
        \fig{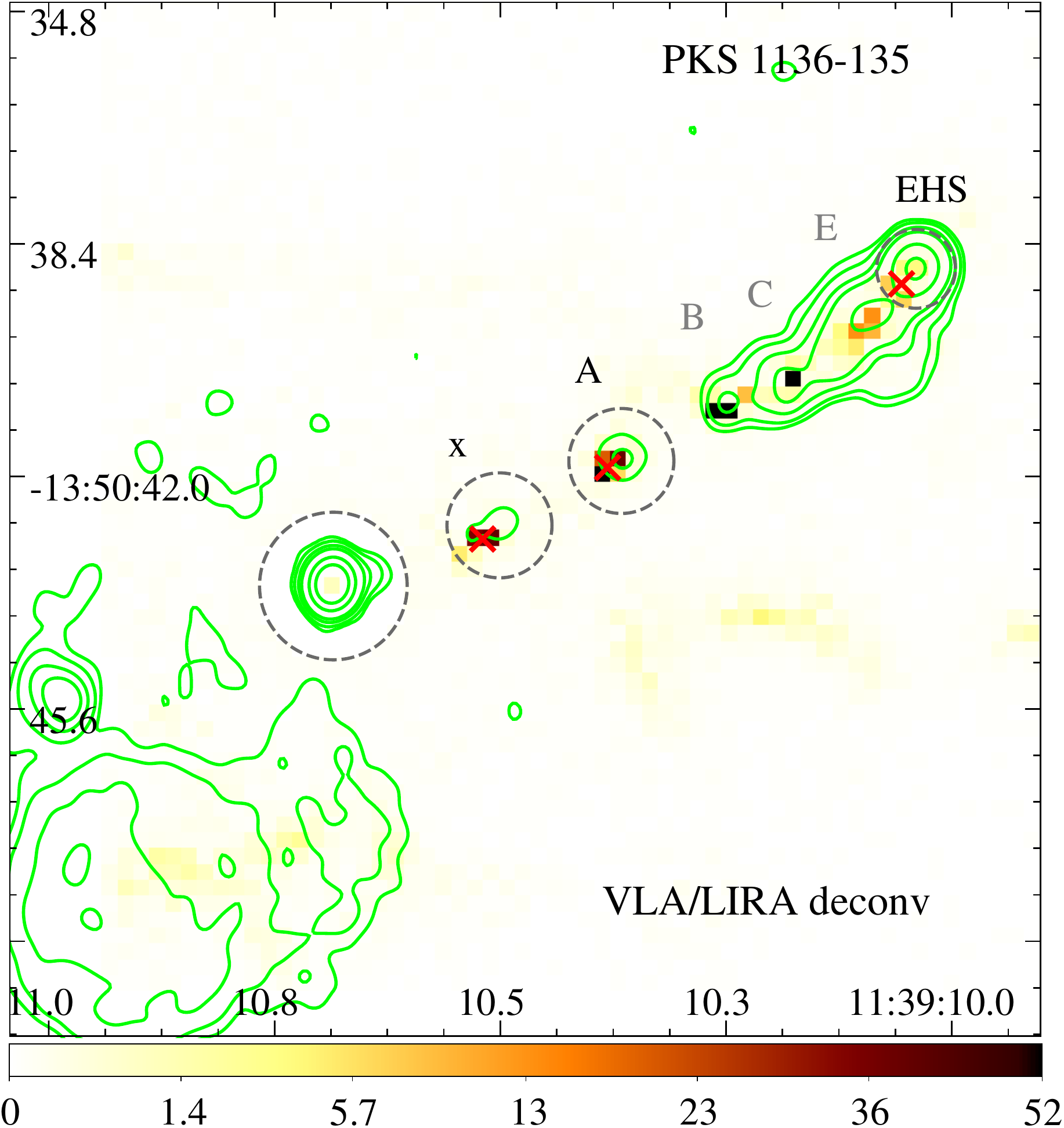}{0.5\textwidth}{(b)}
    }
    \caption{Same as in Fig. \ref{fig:results-3C9} but for PKS 1136-135. The radio contours are given by 0.9, 2.0, 5.0, 10.0, 40.0, 100.0 mJy beam$^{-1}$.\label{fig:results-PKS1136-135}}
\end{figure*}

\begin{figure*}[ht]
    \gridline{
        \fig{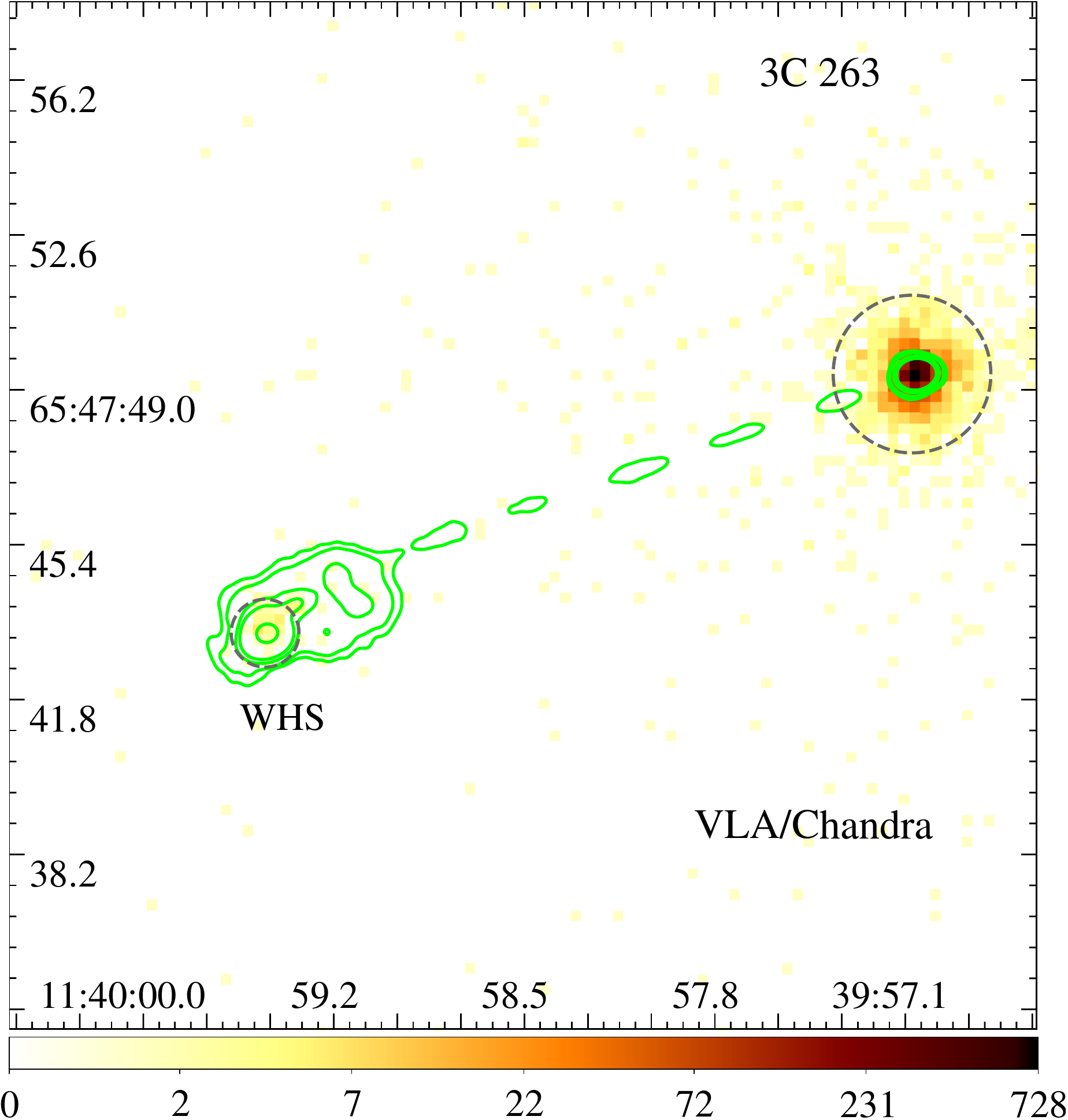}{0.5\textwidth}{(a)}
        \fig{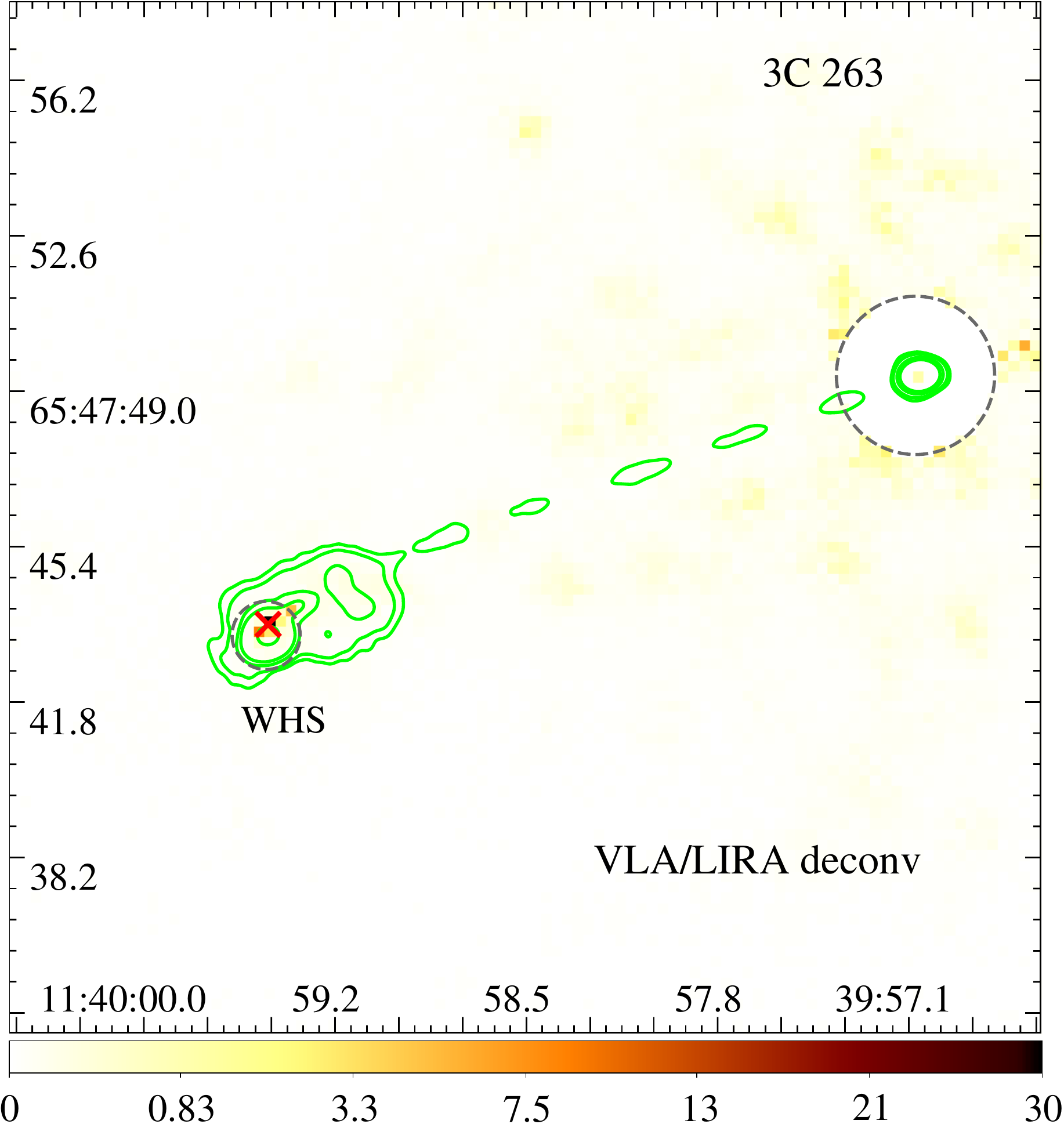}{0.5\textwidth}{(b)}
    }
    \caption{Same as in Fig. \ref{fig:results-3C9} but for 3C 263. The radio contours are given by 0.4, 1.0, 4.0, 10.0, 200.0 mJy beam$^{-1}$.\label{fig:results-3C263}}
\end{figure*}

\begin{figure*}[ht]
    \gridline{
        \fig{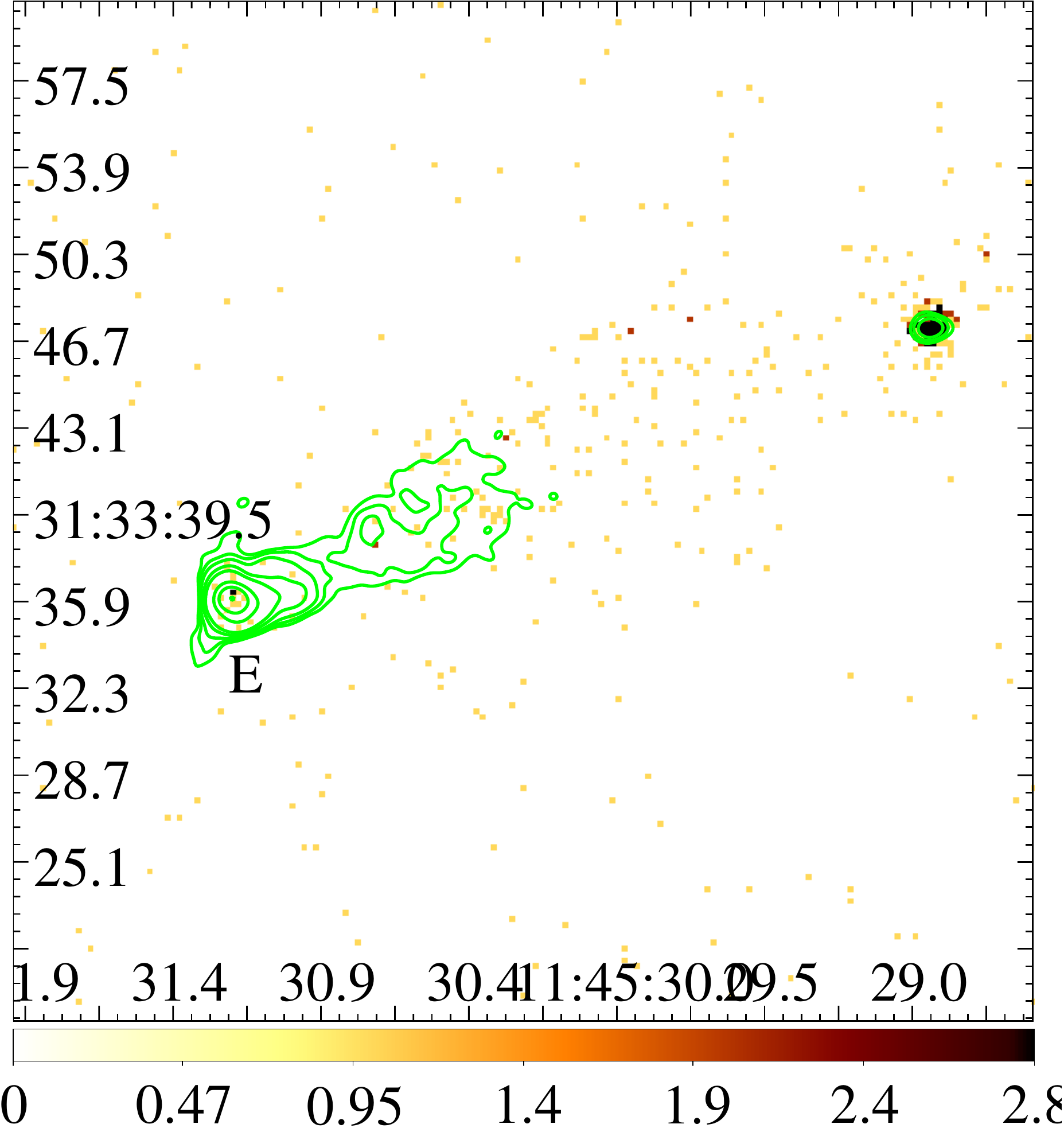}{0.5\textwidth}{(a)}
    }
    \gridline{
        \fig{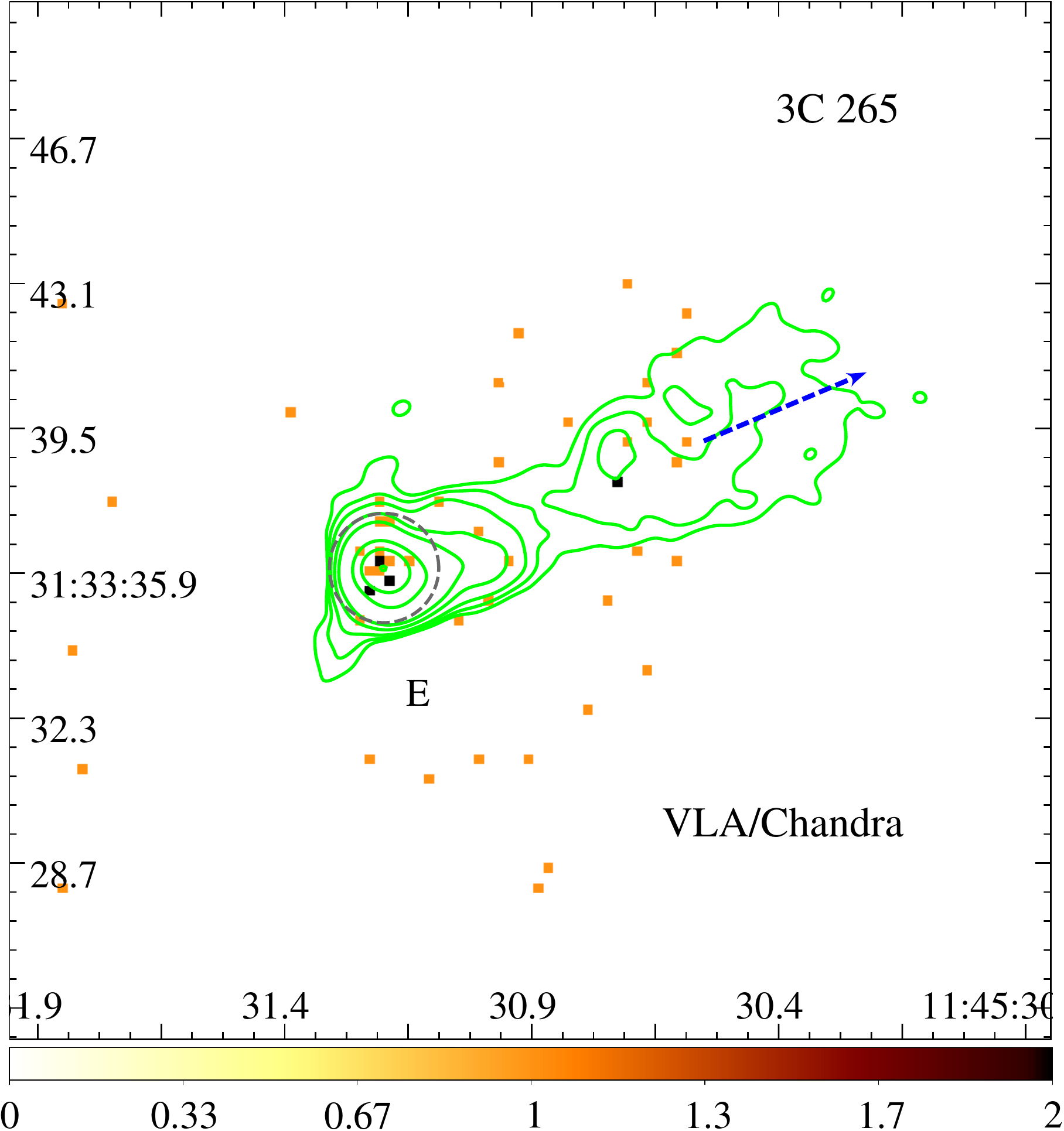}{0.5\textwidth}{(b)}
        \fig{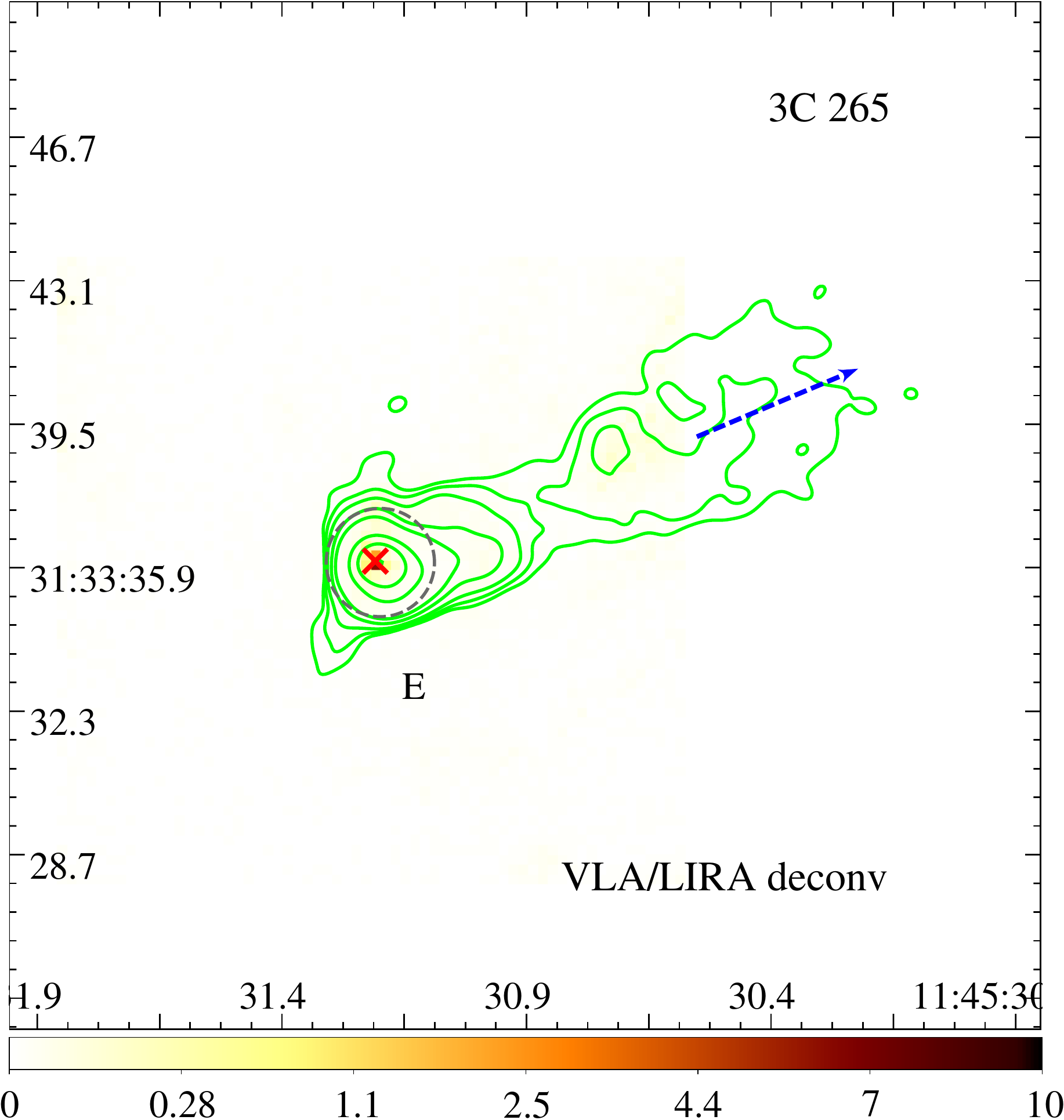}{0.5\textwidth}{(c)}
    }
    \comment{
    \gridline{
        \fig{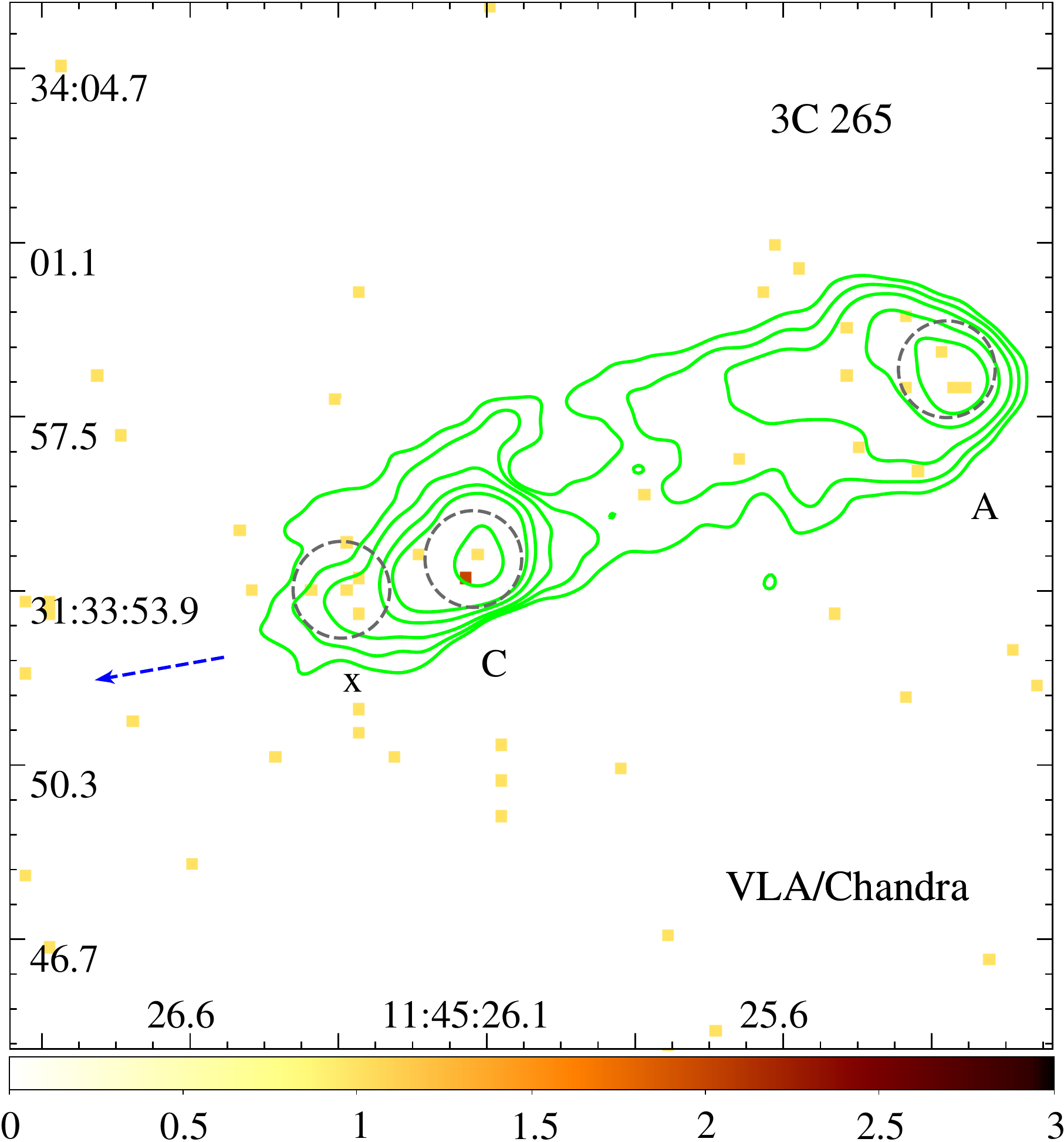}{0.5\textwidth}{(b)}
        \fig{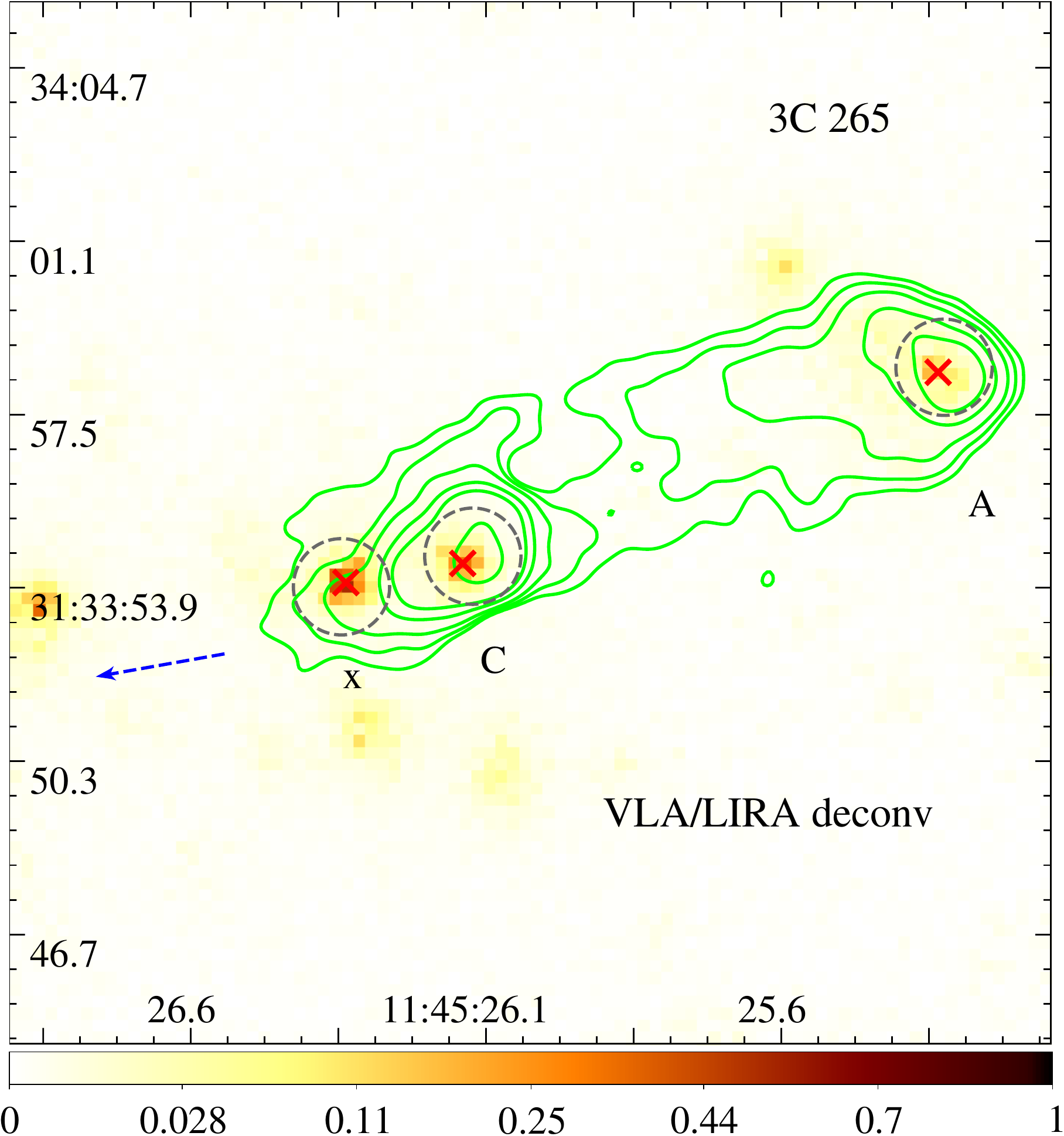}{0.5\textwidth}{(c)}
    }
    }
    \caption{Same as in Fig. \ref{fig:results-3C9} but for 3C 265. (a) shows the full image of eastern side jet. (b) and (c) show the zoomed-in region of the eastern hotspot. The radio contours are given by 0.2, 0.4, 0.8, 2.0, 4.0, 20.0, 40.0, 80.0 mJy beam$^{-1}$.\label{fig:results-3C265}}
\end{figure*}

\begin{figure*}[ht]
    \gridline{
        \fig{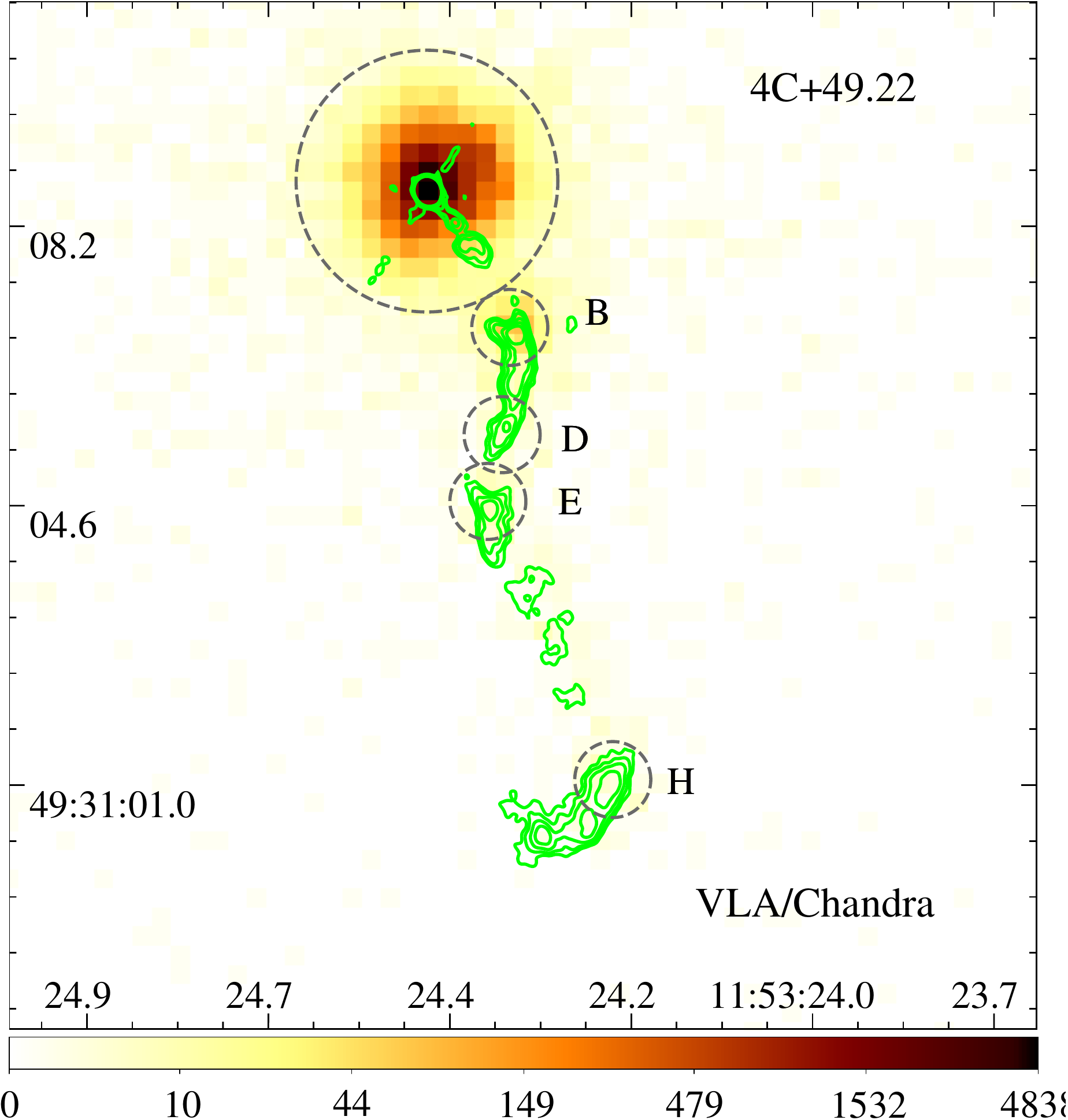}{0.5\textwidth}{(a)}
        \fig{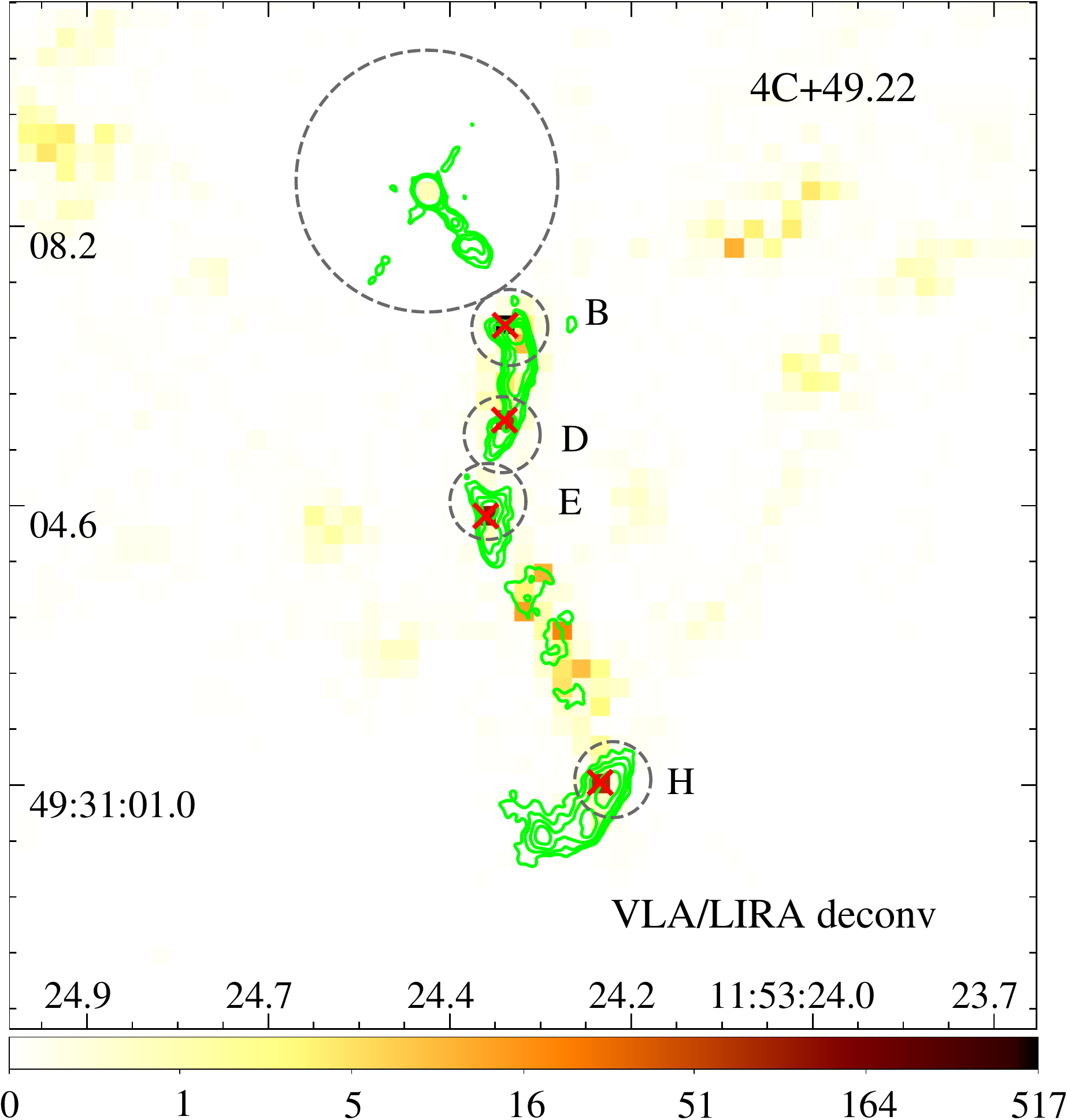}{0.5\textwidth}{(b)}
    }
    \caption{Same as in Fig. \ref{fig:results-3C9} but for 4C +49.22. The radio contours are given by 0.05, 0.1, 0.2, 0.4, 0.8 mJy beam$^{-1}$.\label{fig:results-4C+49.22}}
\end{figure*}

\begin{figure*}[ht]
    \gridline{
        \fig{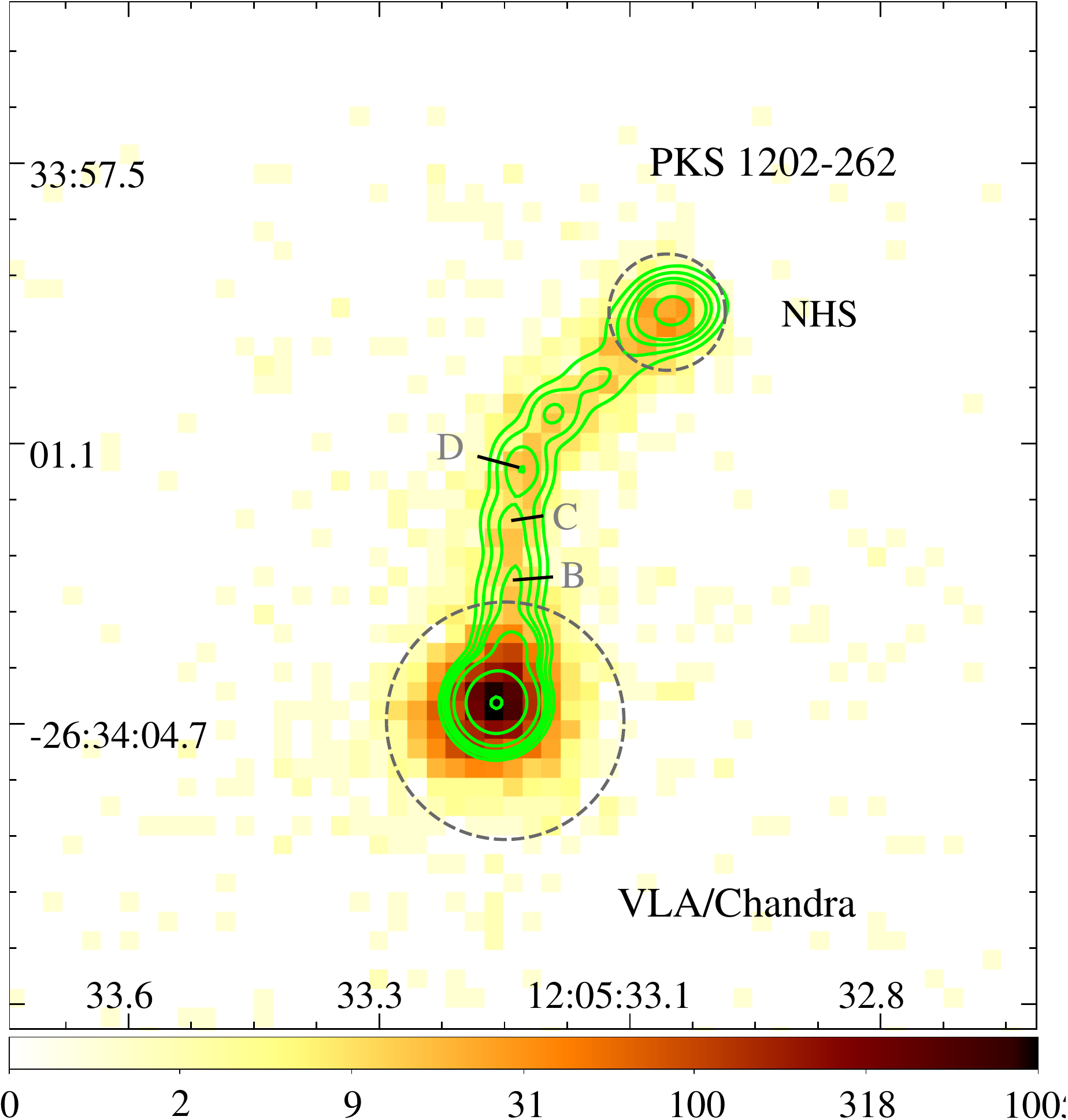}{0.5\textwidth}{(a)}
        \fig{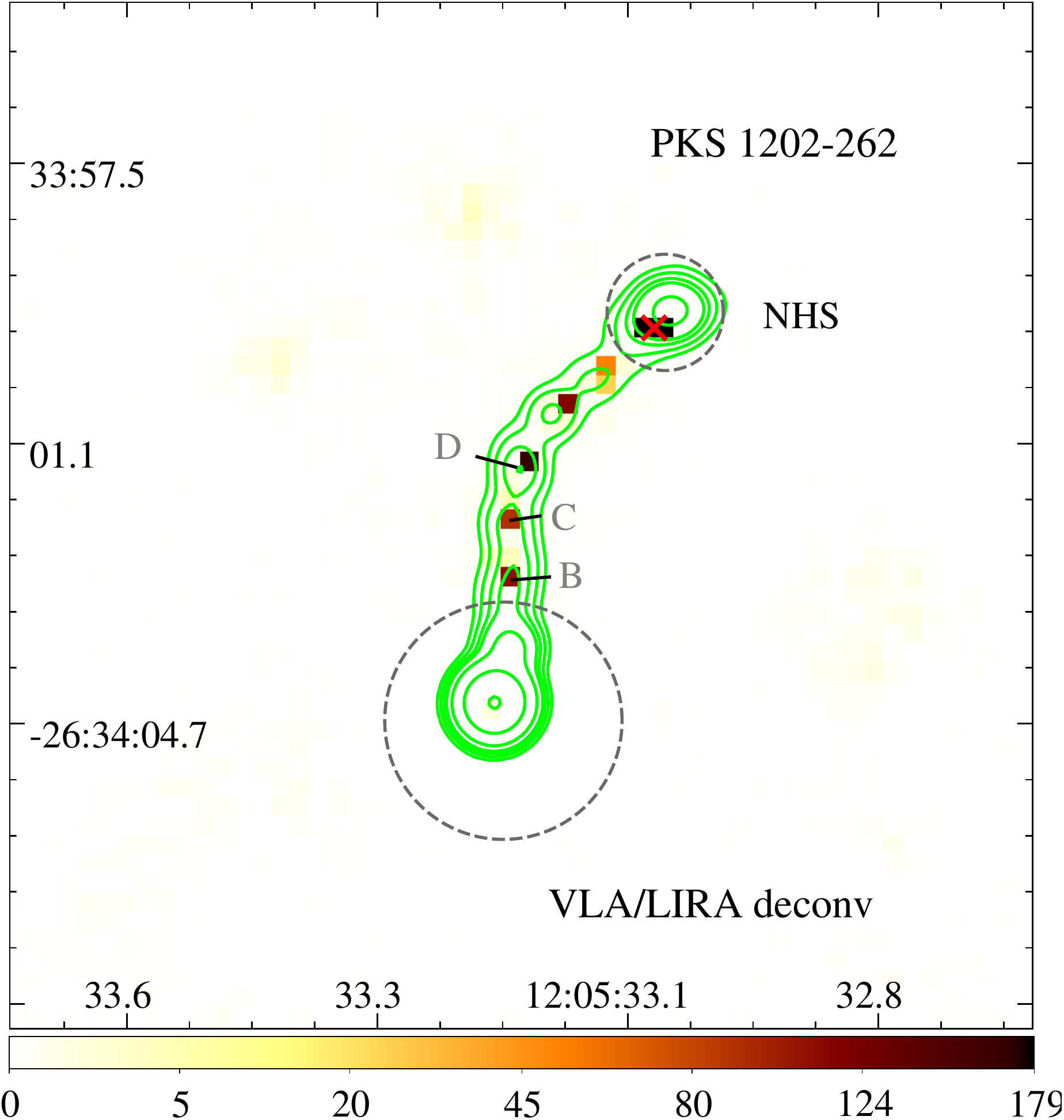}{0.5\textwidth}{(b)}
    }
    \caption{Same as in Fig. \ref{fig:results-3C9} but for PKS 1202-262. The radio contours are given by 1.5, 3.0, 5.4, 8.0, 20.0, 100.0, 500.0 mJy beam$^{-1}$.\label{fig:results-PKS1202-262}}
\end{figure*}

\begin{figure*}[ht]
    \gridline{
        \fig{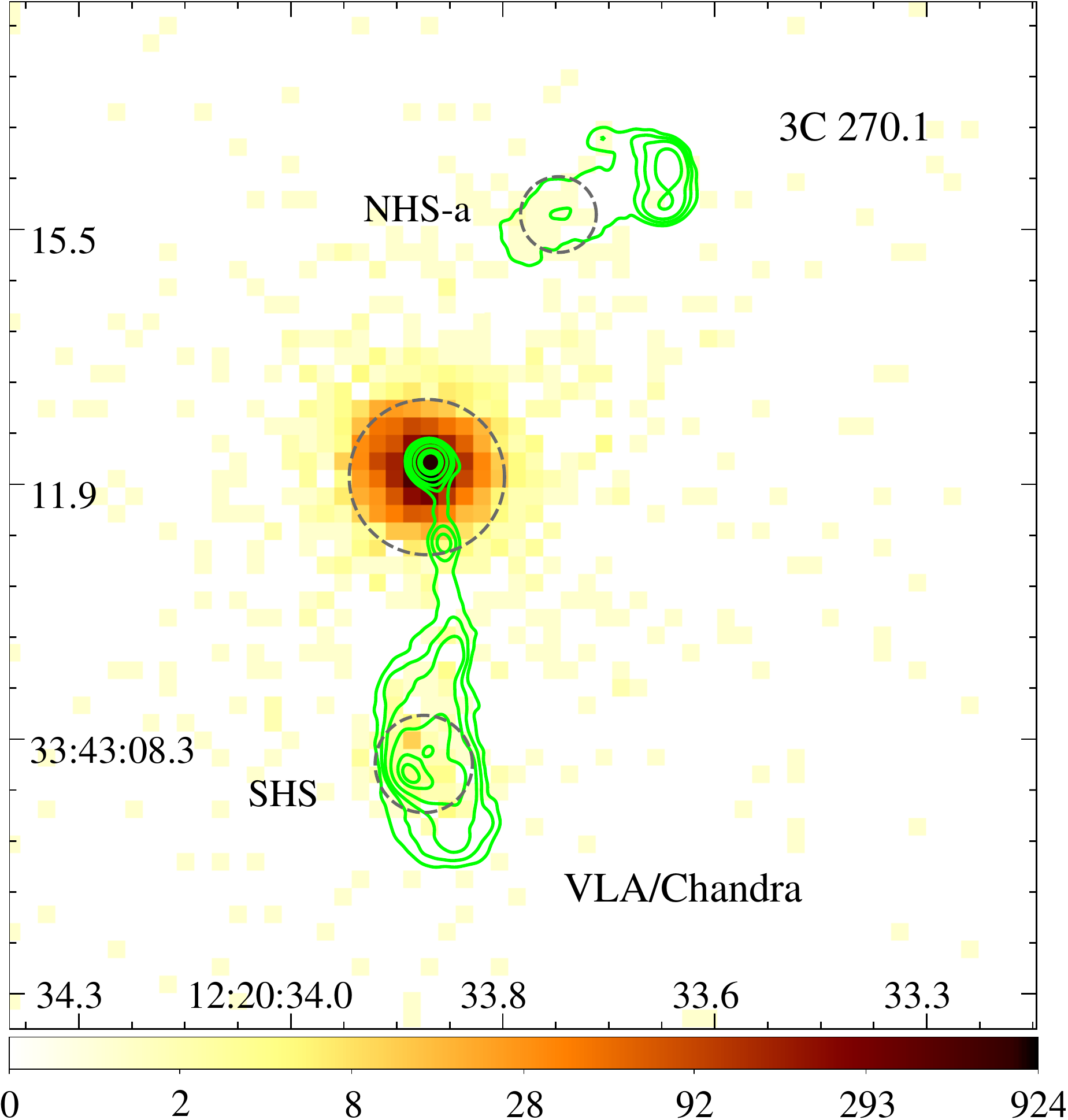}{0.5\textwidth}{(a)}
        \fig{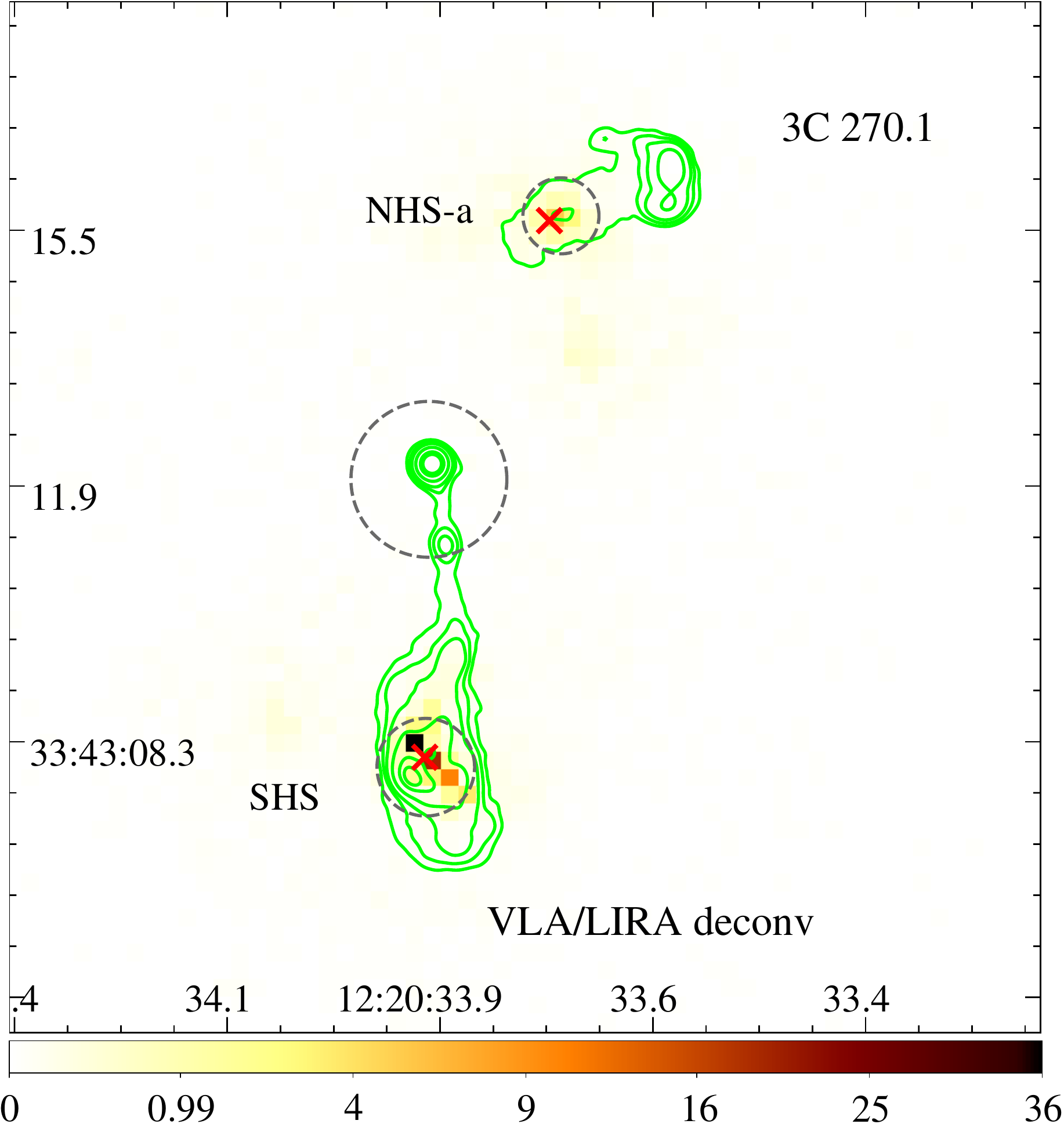}{0.5\textwidth}{(b)}
    }
    \caption{Same as in Fig. \ref{fig:results-3C9} but for 3C 270.1. The radio contours are given by 0.1, 0.36, 1.0, 5.0, 22.0, 40.0 mJy beam$^{-1}$.\label{fig:results-3C270.1}}
\end{figure*}

\begin{figure*}[ht]
    \gridline{
        \fig{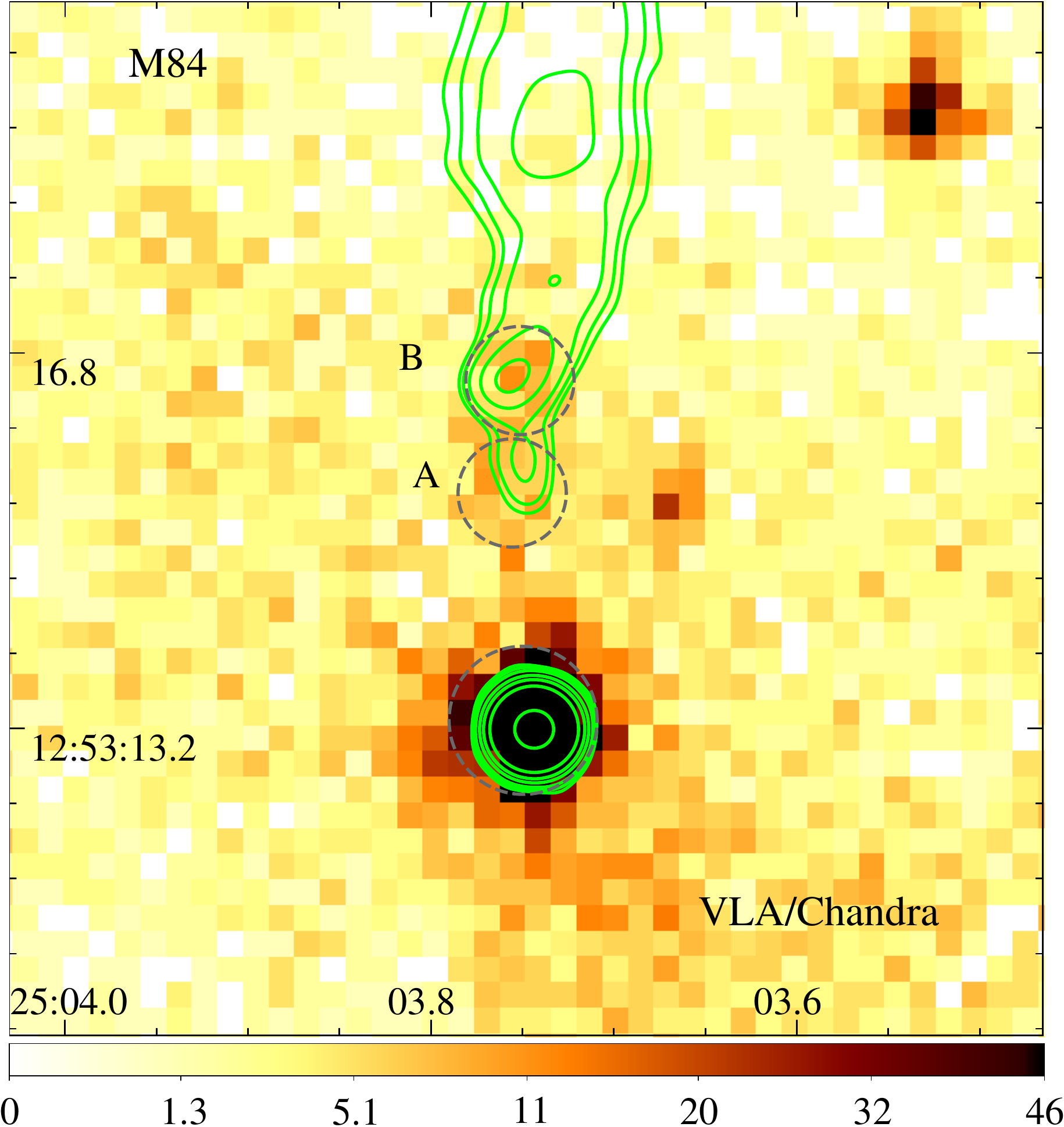}{0.5\textwidth}{(a)}
        \fig{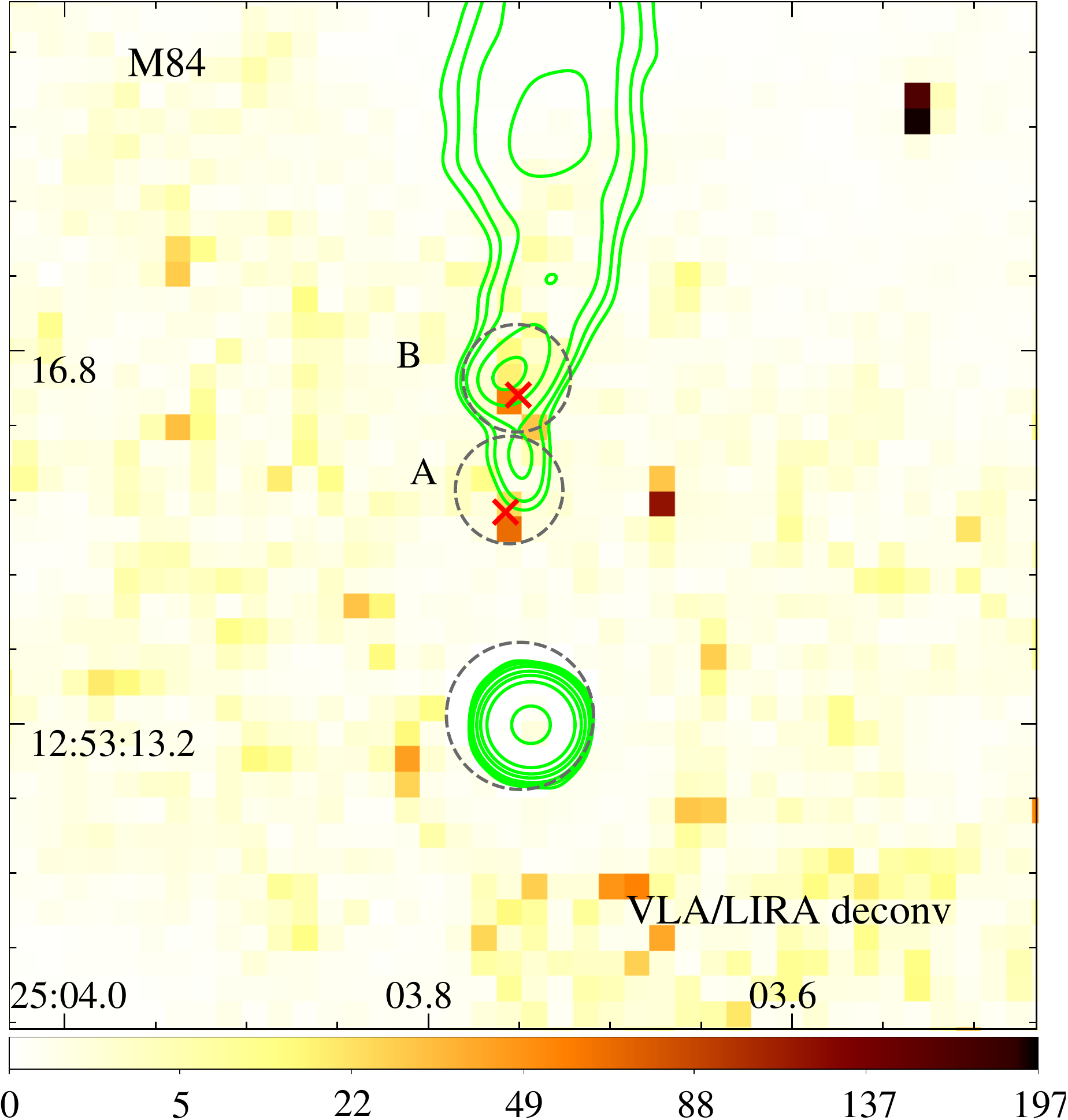}{0.5\textwidth}{(b)}
    }
    \caption{Same as in Fig. \ref{fig:results-3C9} but for M84. The radio contours are given by 0.2, 0.4, 0.8, 2.0, 4.0, 10.0, 100.0 mJy beam$^{-1}$.\label{fig:results-M84}}
\end{figure*}

\begin{figure*}[ht]
    \gridline{
        \fig{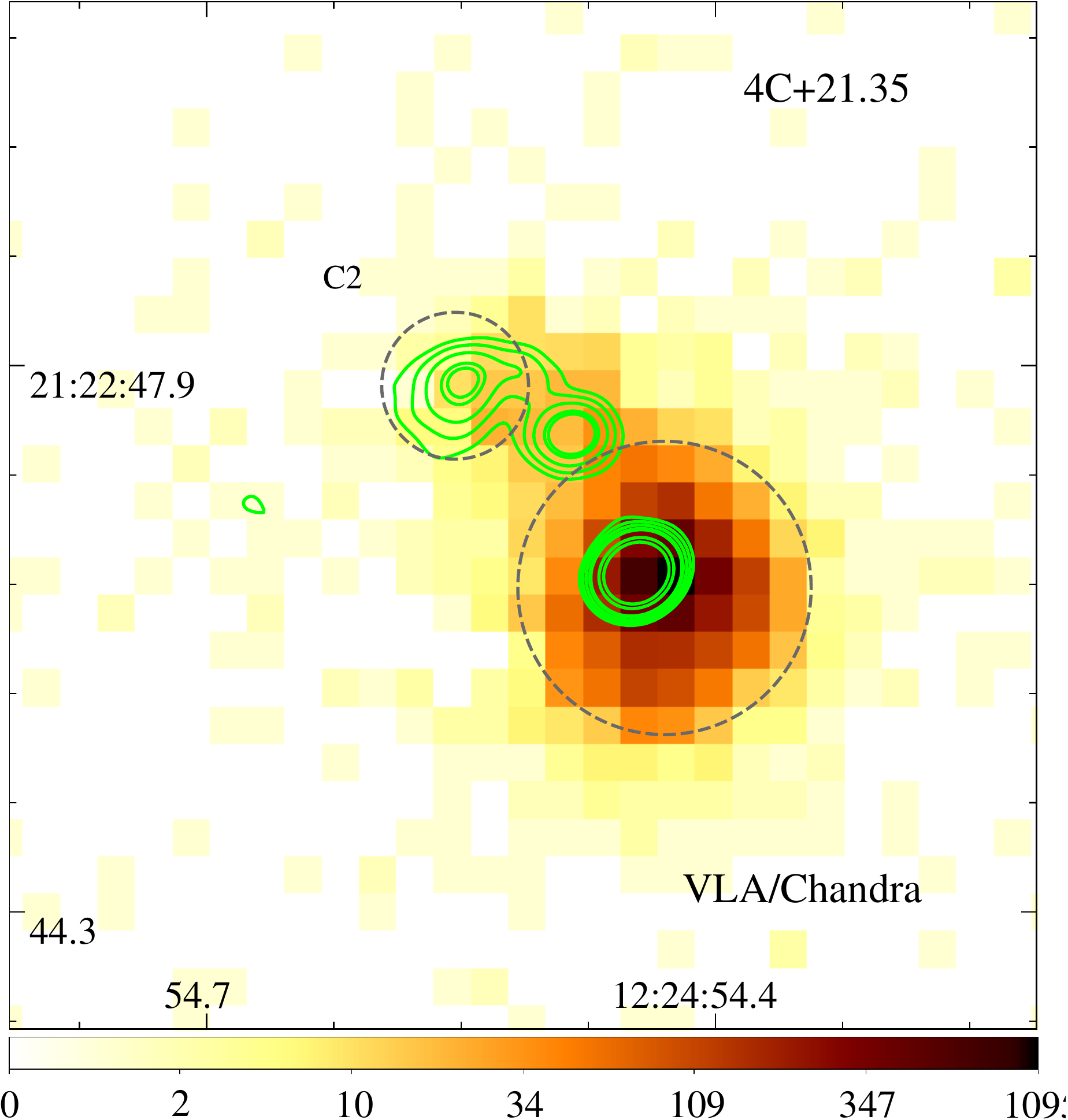}{0.5\textwidth}{(a)}
        \fig{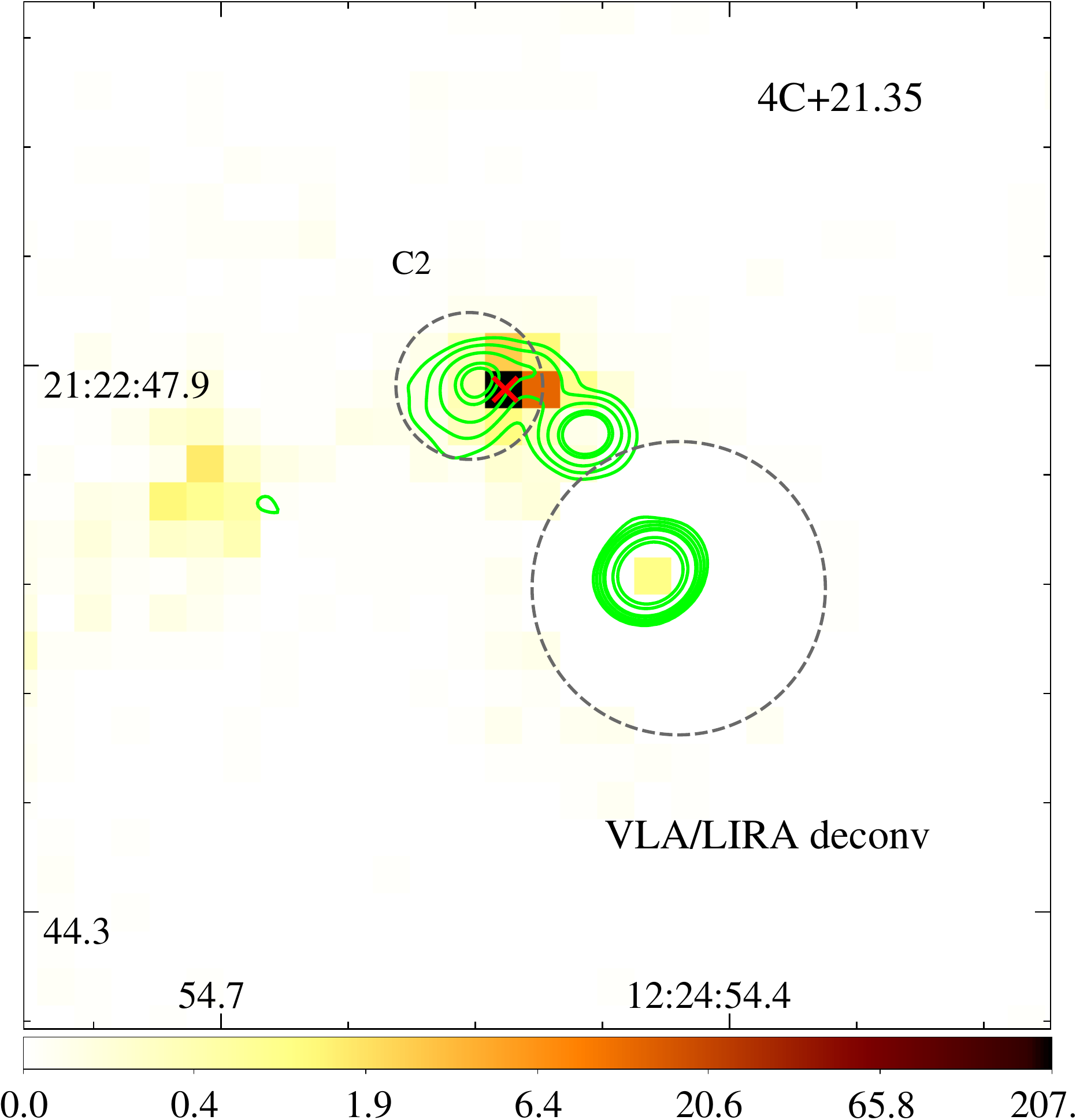}{0.5\textwidth}{(b)}
    }
    \caption{Same as in Fig. \ref{fig:results-3C9} but for 4C +21.35. The radio contours are given by 1.0, 2.0, 4.0, 8.0, 10.0, 40.0, 80.0 mJy beam$^{-1}$.\label{fig:results-4C+21.35}}
\end{figure*}

\begin{figure*}[ht]
    \gridline{
        \fig{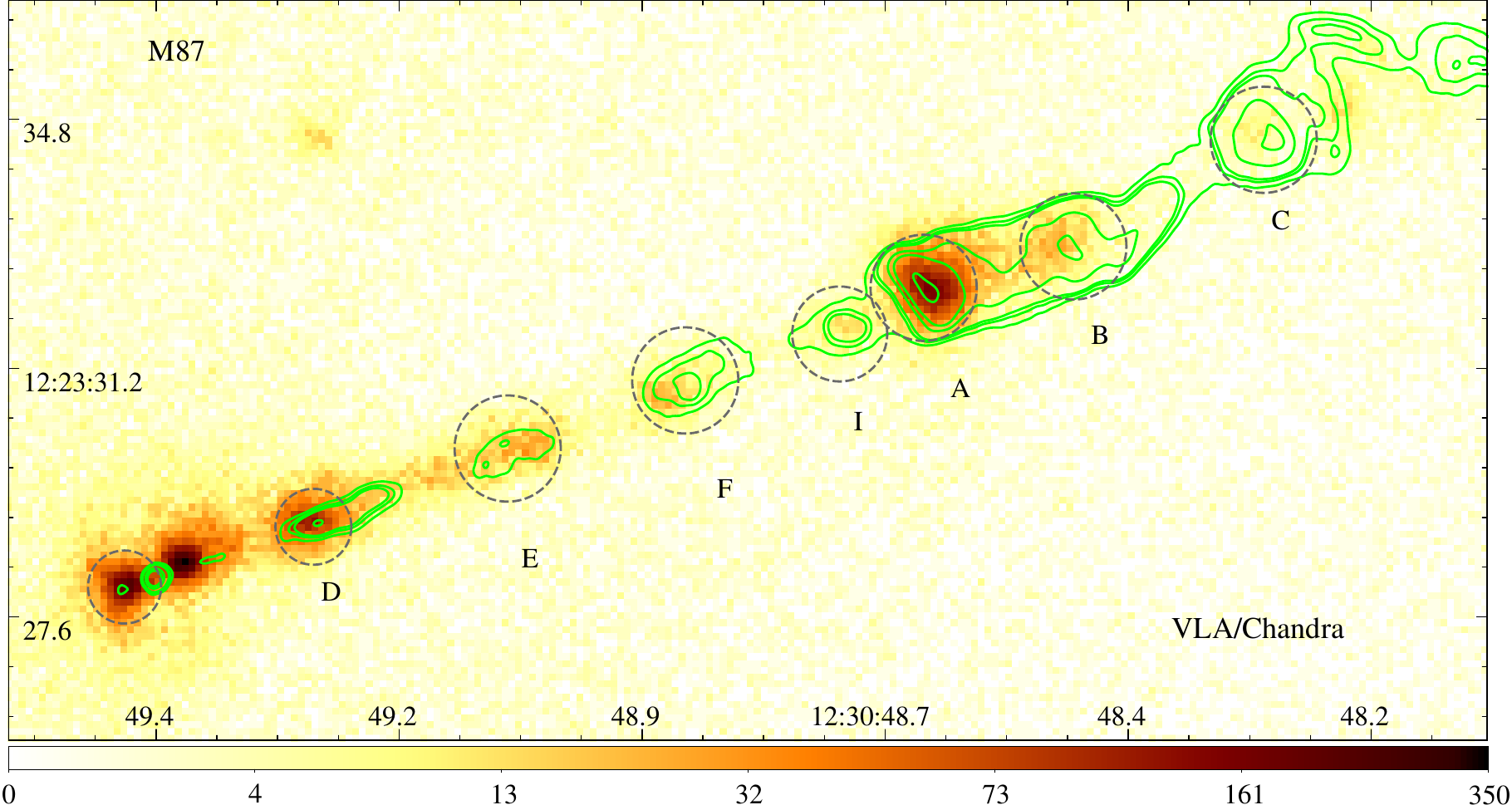}{\textwidth}{(a)}
    }
    \gridline{
        \fig{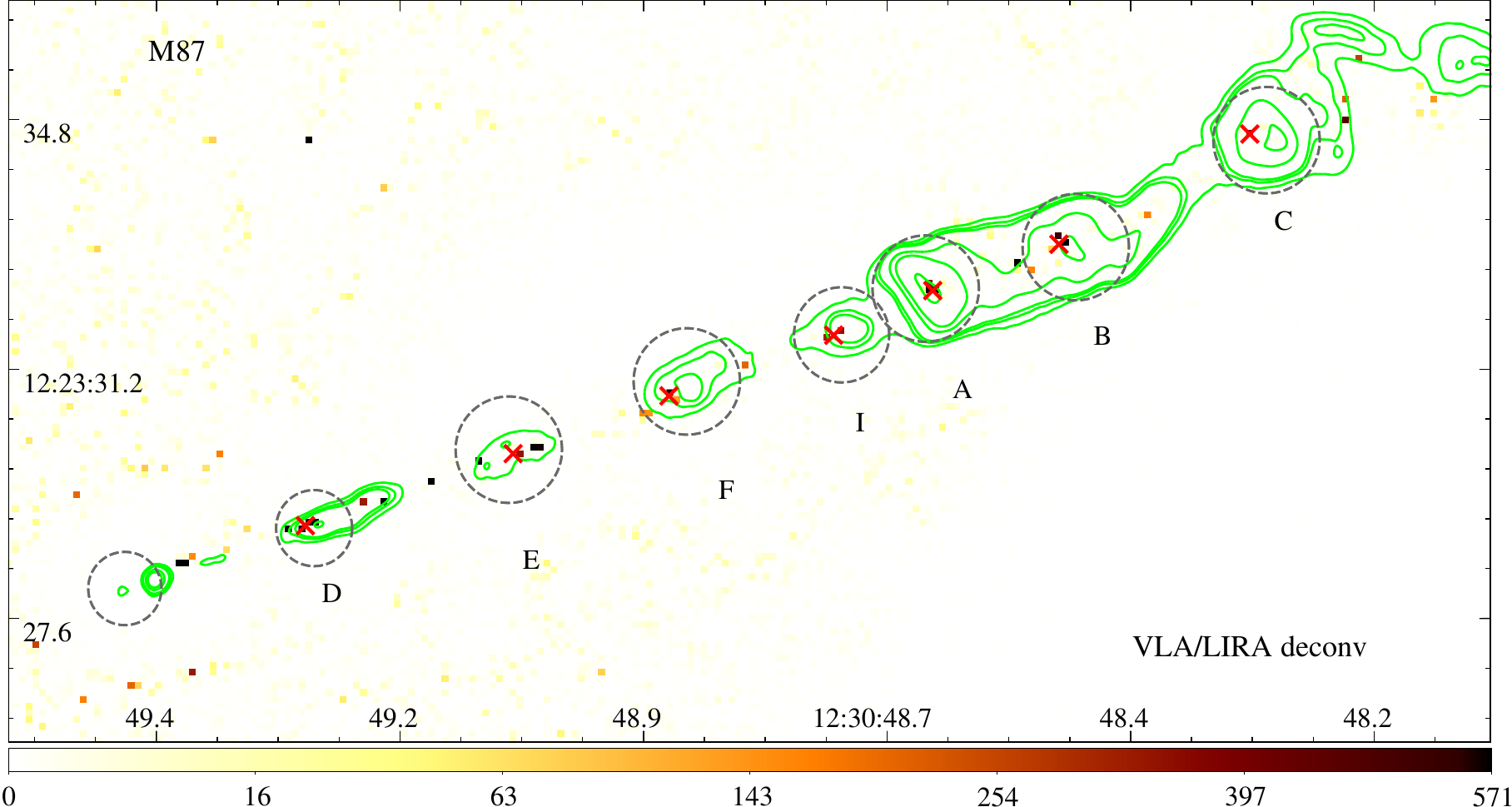}{\textwidth}{(b)}
    }
    \caption{Same as in Fig. \ref{fig:results-3C9} but for M87. The radio contours are given by 5.0, 8.0, 23.0, 35.0, 70.0 mJy beam$^{-1}$.\label{fig:results-M87}}
\end{figure*}

\begin{figure*}[ht]
    \gridline{
        \fig{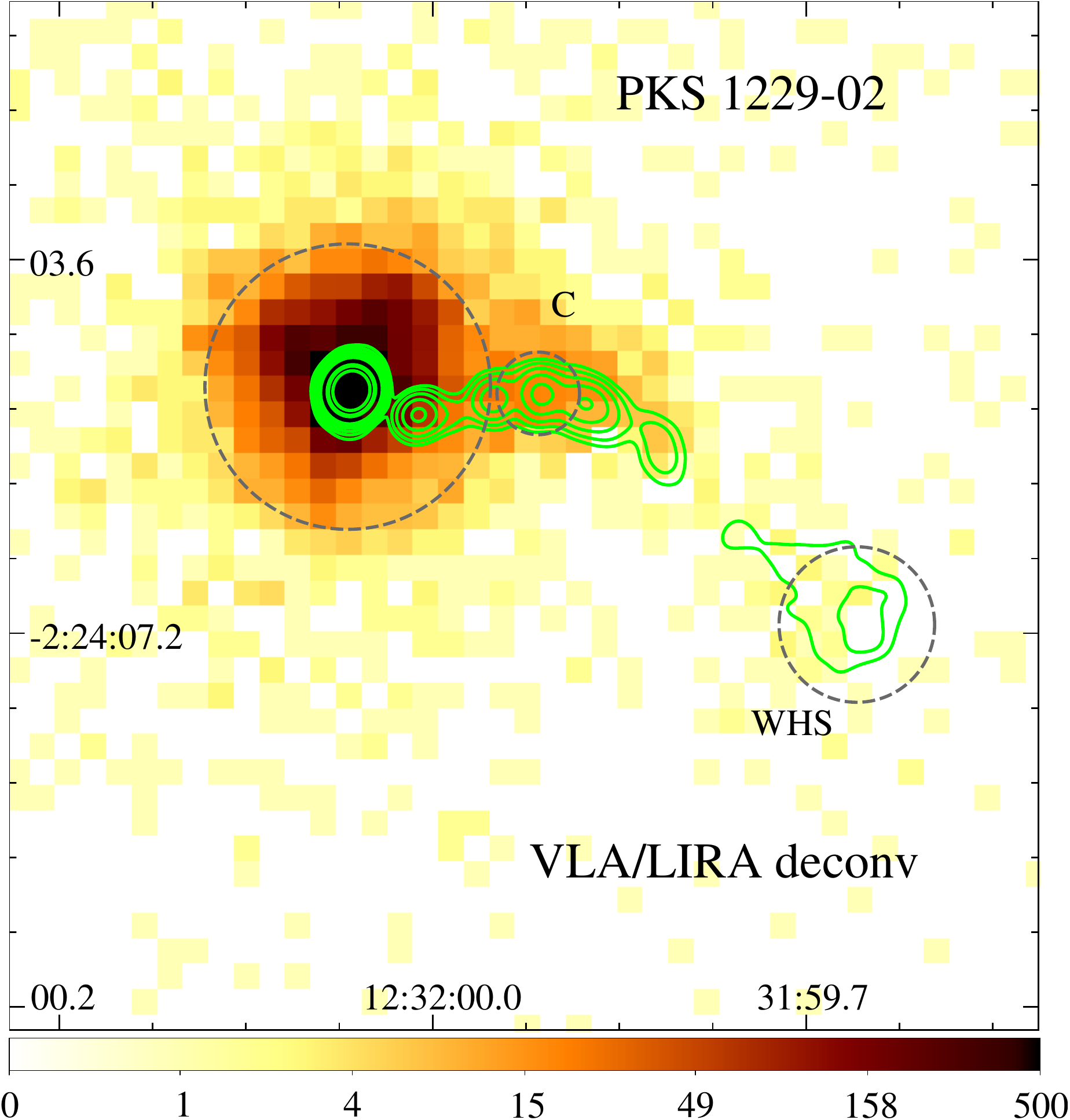}{0.5\textwidth}{(a)}
        \fig{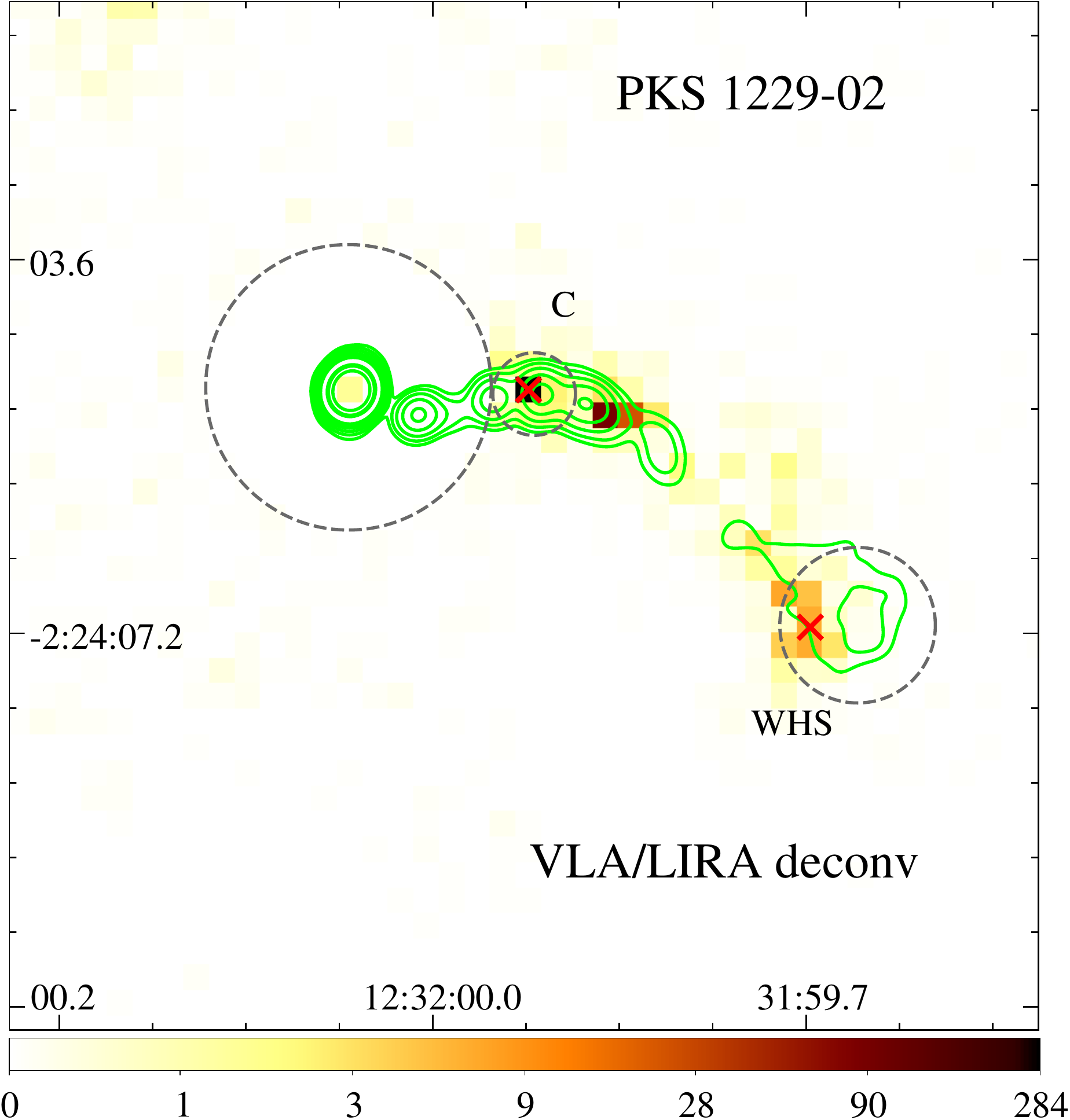}{0.5\textwidth}{(b)}
    }
    \caption{Same as in Fig. \ref{fig:results-3C9} but for PKS 1229-02. The radio contours are given by 1.0, 2.0, 4.0, 8.0, 15.0, 20.0, 80.0, 100.0, 200.0 mJy beam$^{-1}$.\label{fig:results-PKS1229-02}}
\end{figure*}

\begin{figure*}[ht]
    \gridline{
        \fig{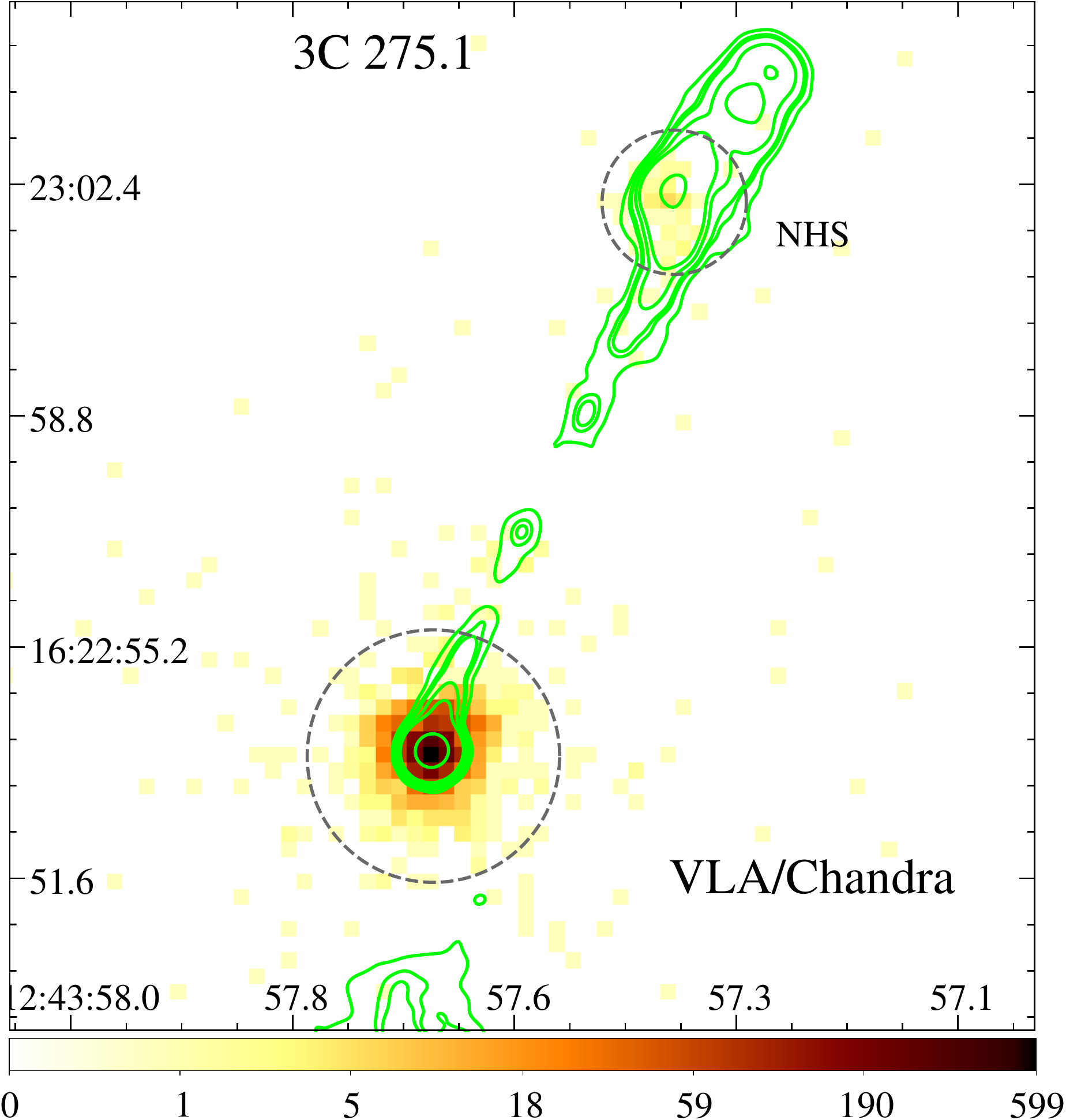}{0.5\textwidth}{(a)}
        \fig{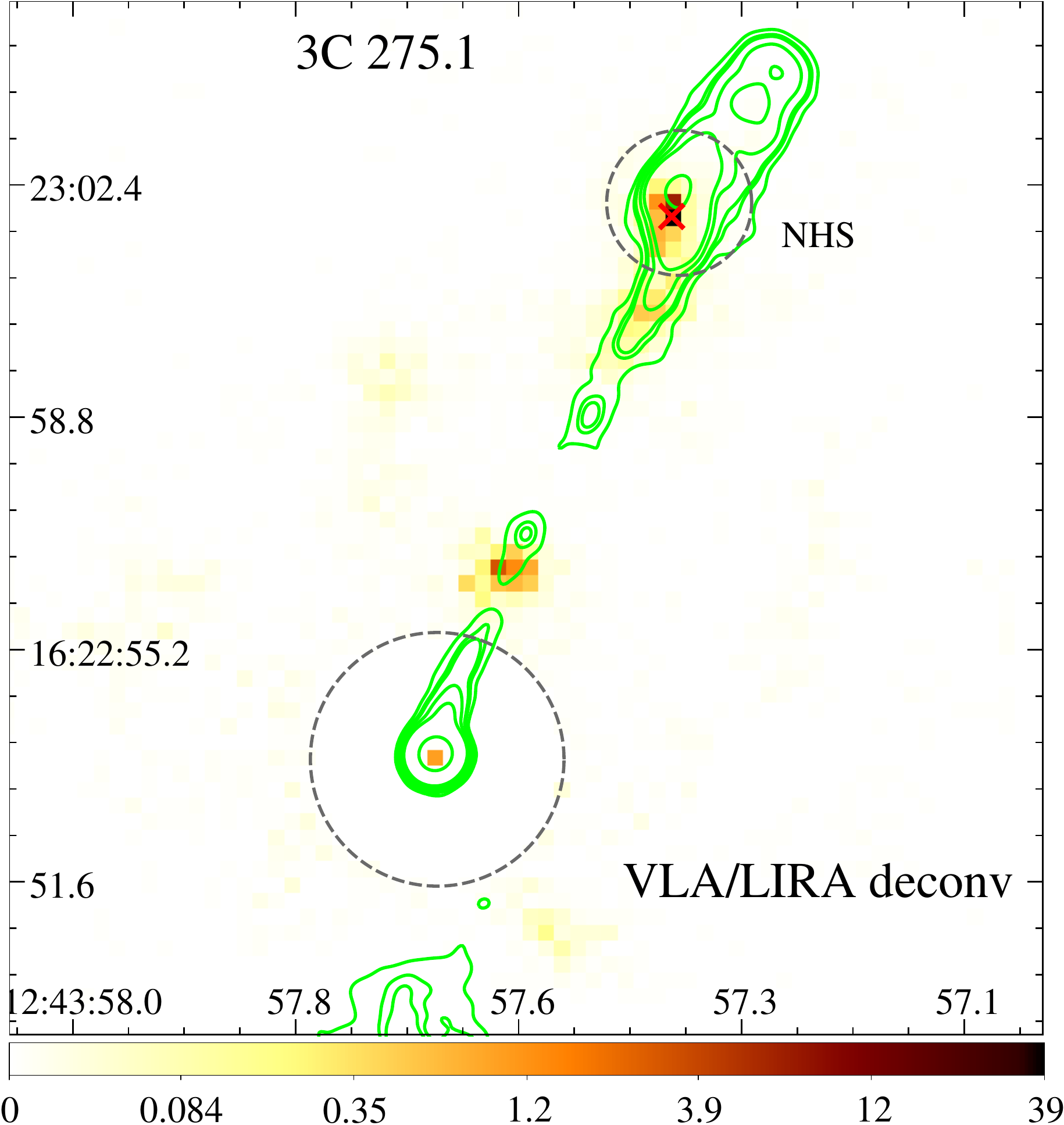}{0.5\textwidth}{(b)}
    }
    \caption{Same as in Fig. \ref{fig:results-3C9} but for 3C 275.1. The radio contours are given by 0.4, 0.8, 1.0, 2.0, 4.0, 70.0 mJy beam$^{-1}$.\label{fig:results-3C275.1}}
\end{figure*}

\begin{figure*}[ht]
    \gridline{
        \fig{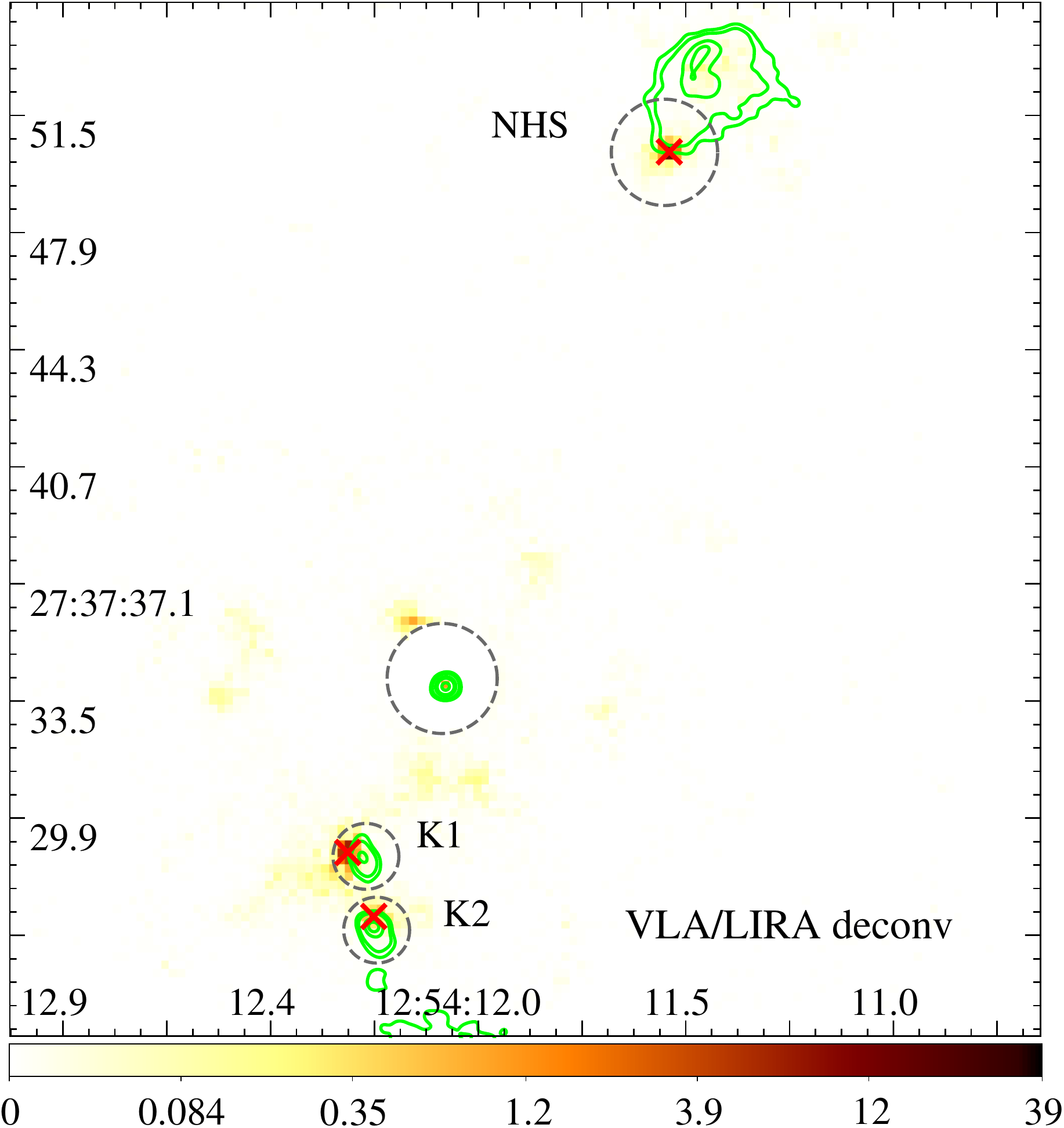}{0.5\textwidth}{(a)}
        \fig{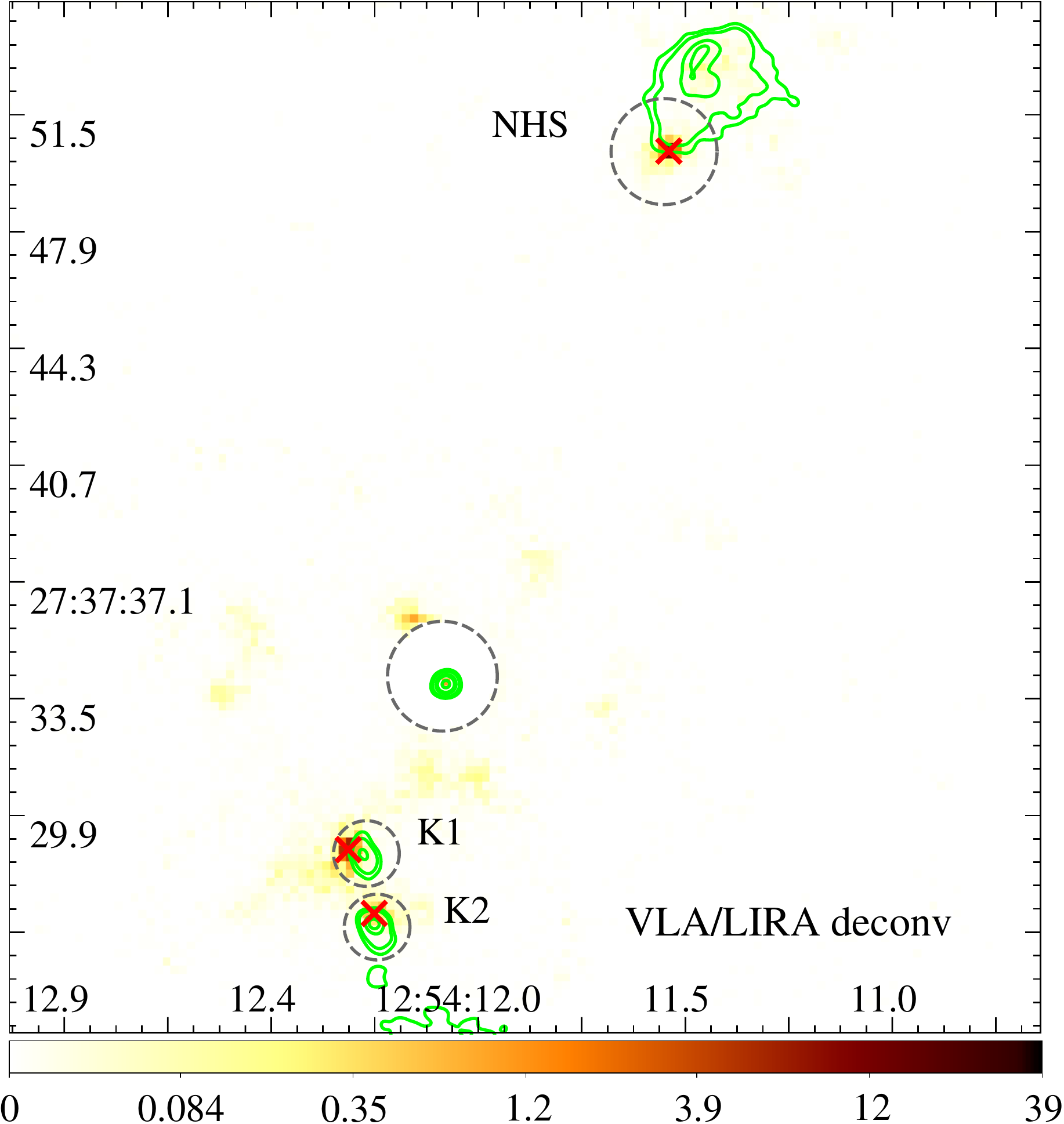}{0.5\textwidth}{(b)}
    }
    \caption{Same as in Fig. \ref{fig:results-3C9} but for 3C 277.3. The radio contours are given by 0.5, 1.0, 3.0, 4.5, 10.0 mJy beam$^{-1}$.\label{fig:results-3C277.3}}
\end{figure*}

\begin{figure*}[ht]
    \gridline{
        \fig{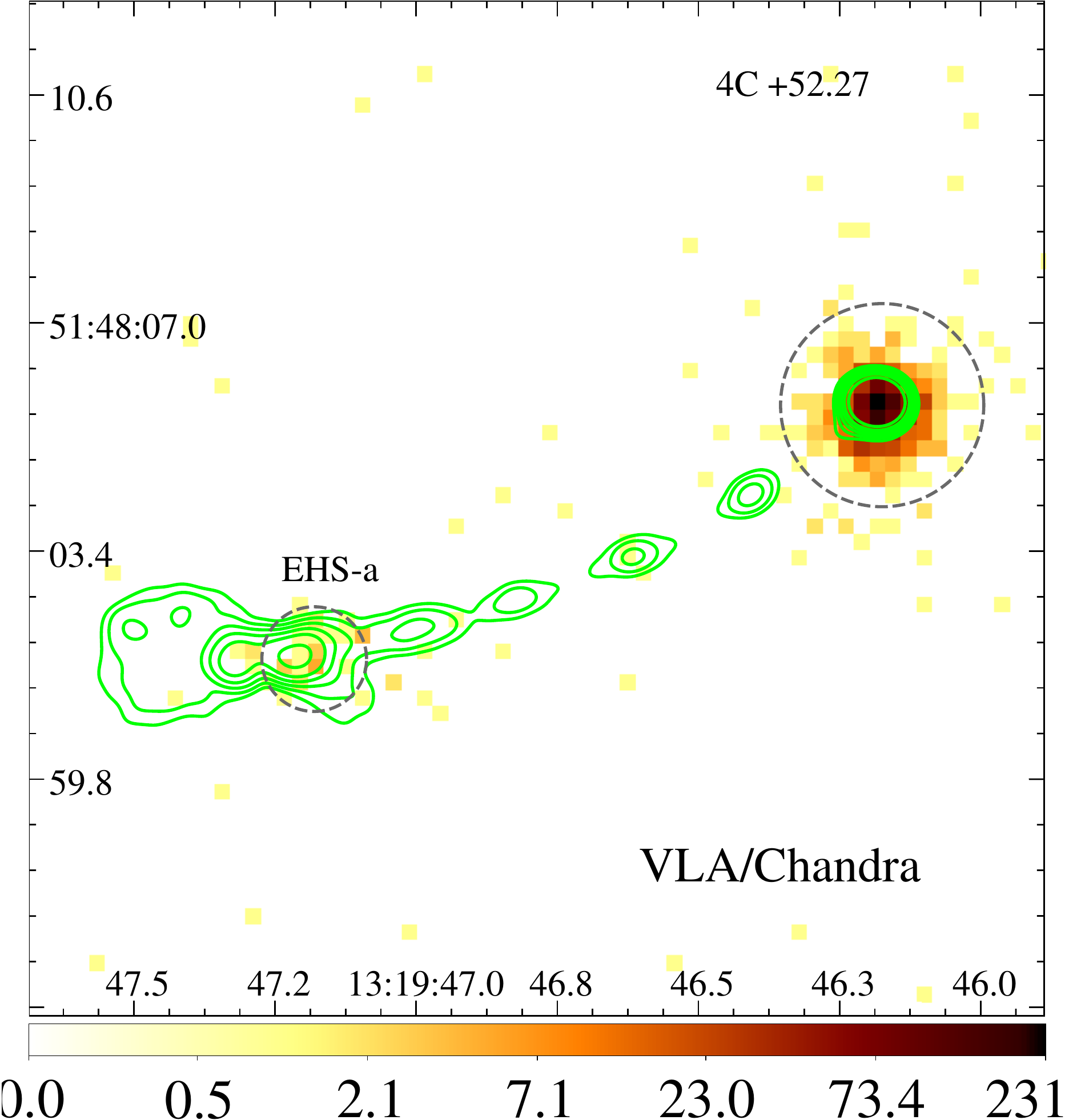}{0.5\textwidth}{(a)}
        \fig{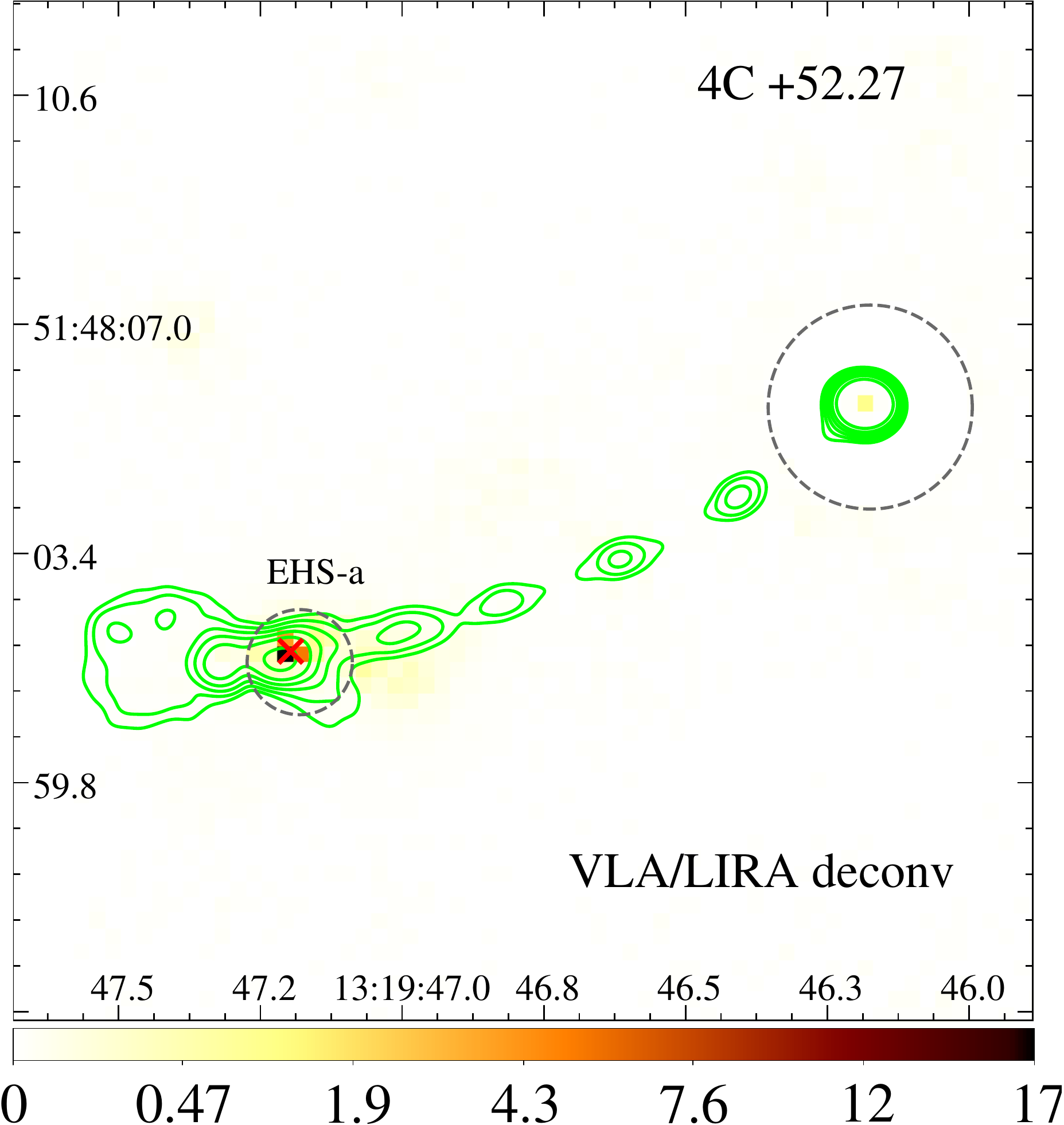}{0.5\textwidth}{(b)}
    }
    \caption{Same as in Fig. \ref{fig:results-3C9} but for 4C +52.27. The radio contours are given by 0.5, 1.0, 2.0, 4.0, 8.0, 20.0 mJy beam$^{-1}$.\label{fig:results-4C+52.27}}
\end{figure*}

\begin{figure*}[ht]
    \gridline{
        \fig{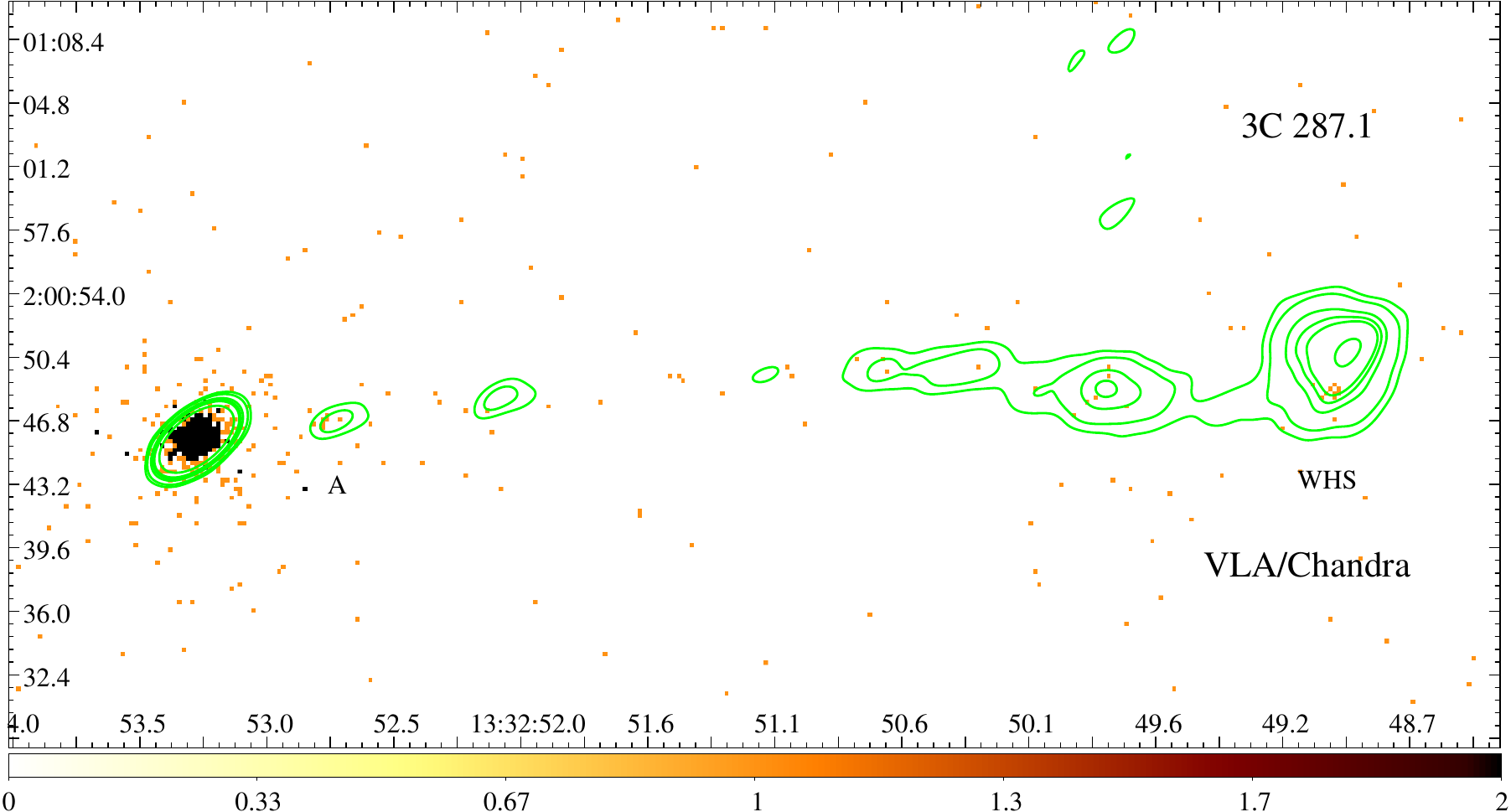}{\textwidth}{(a)}
    }
    \gridline{
        \fig{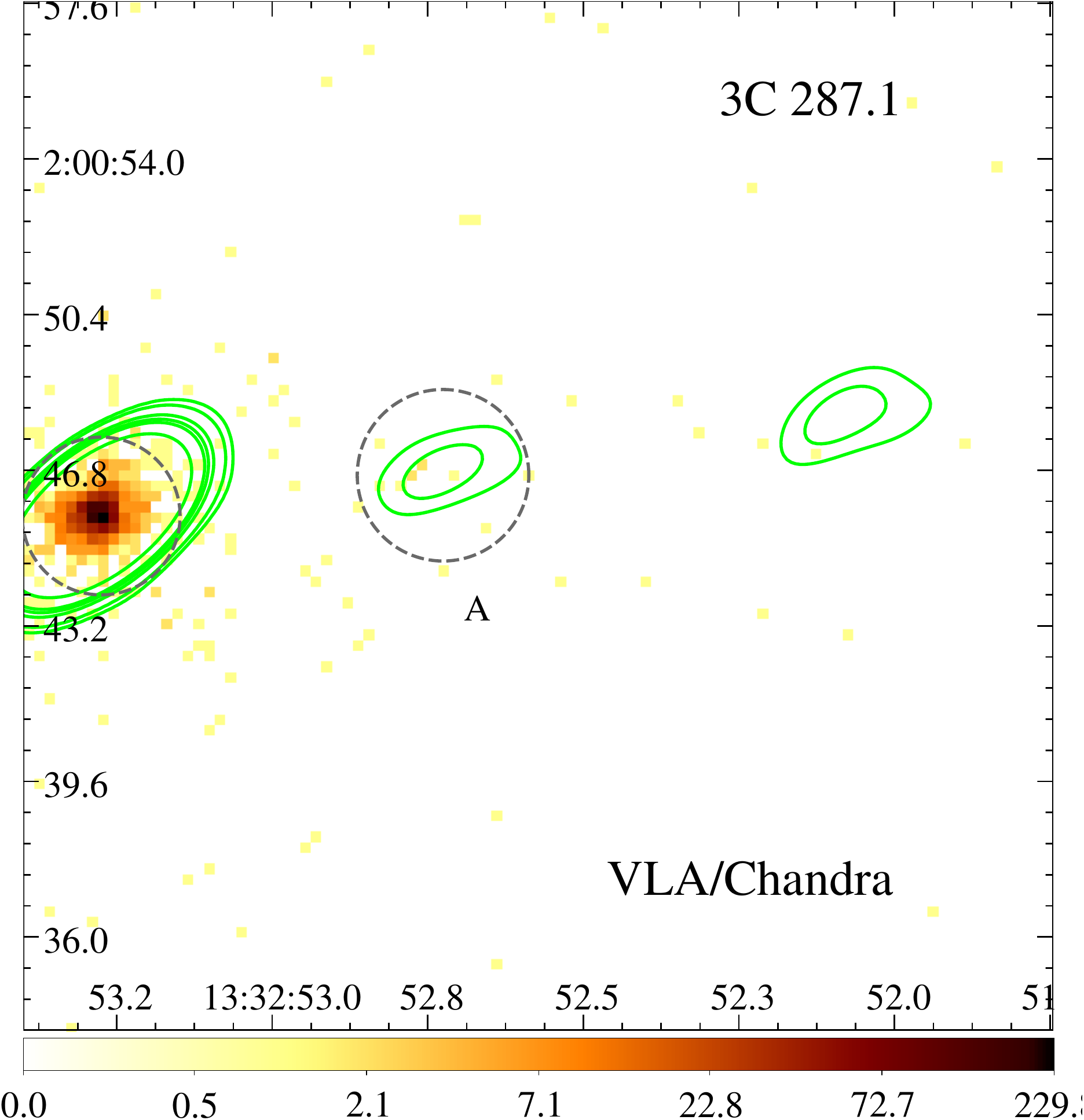}{0.3\textwidth}{(b)}
        \fig{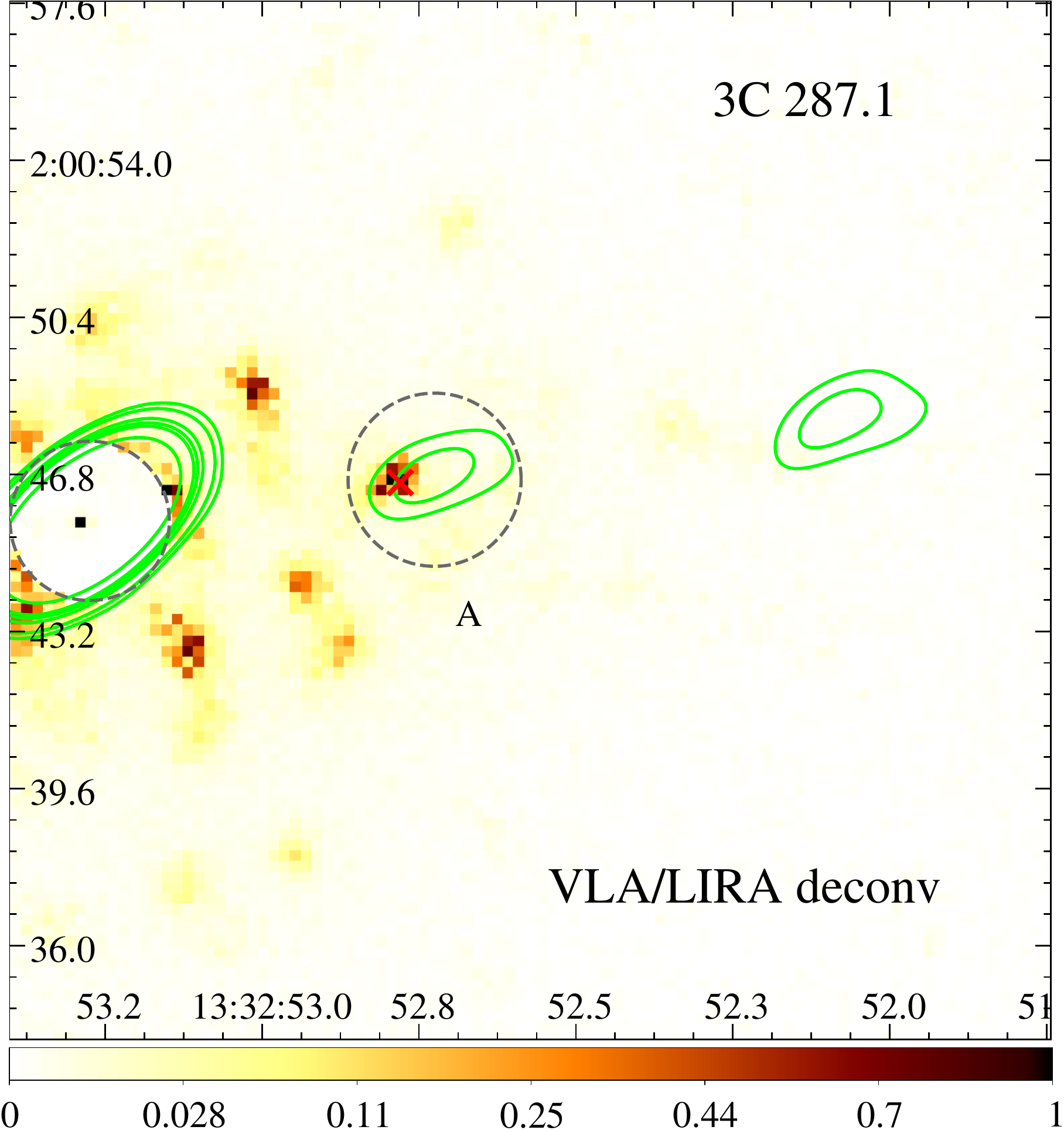}{0.3\textwidth}{(c)}
    }
    \gridline{
        \fig{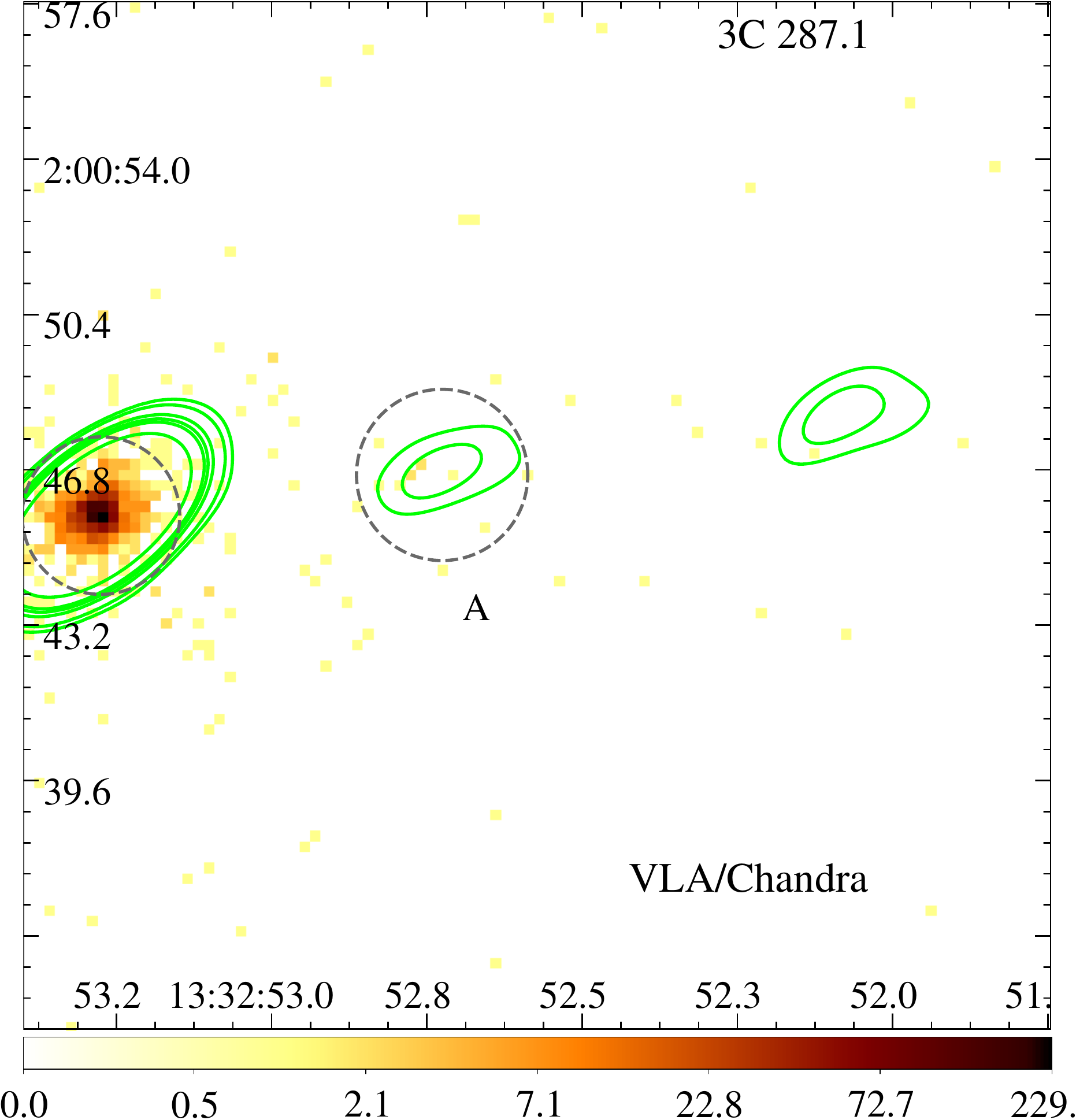}{0.3\textwidth}{(d)}
        \fig{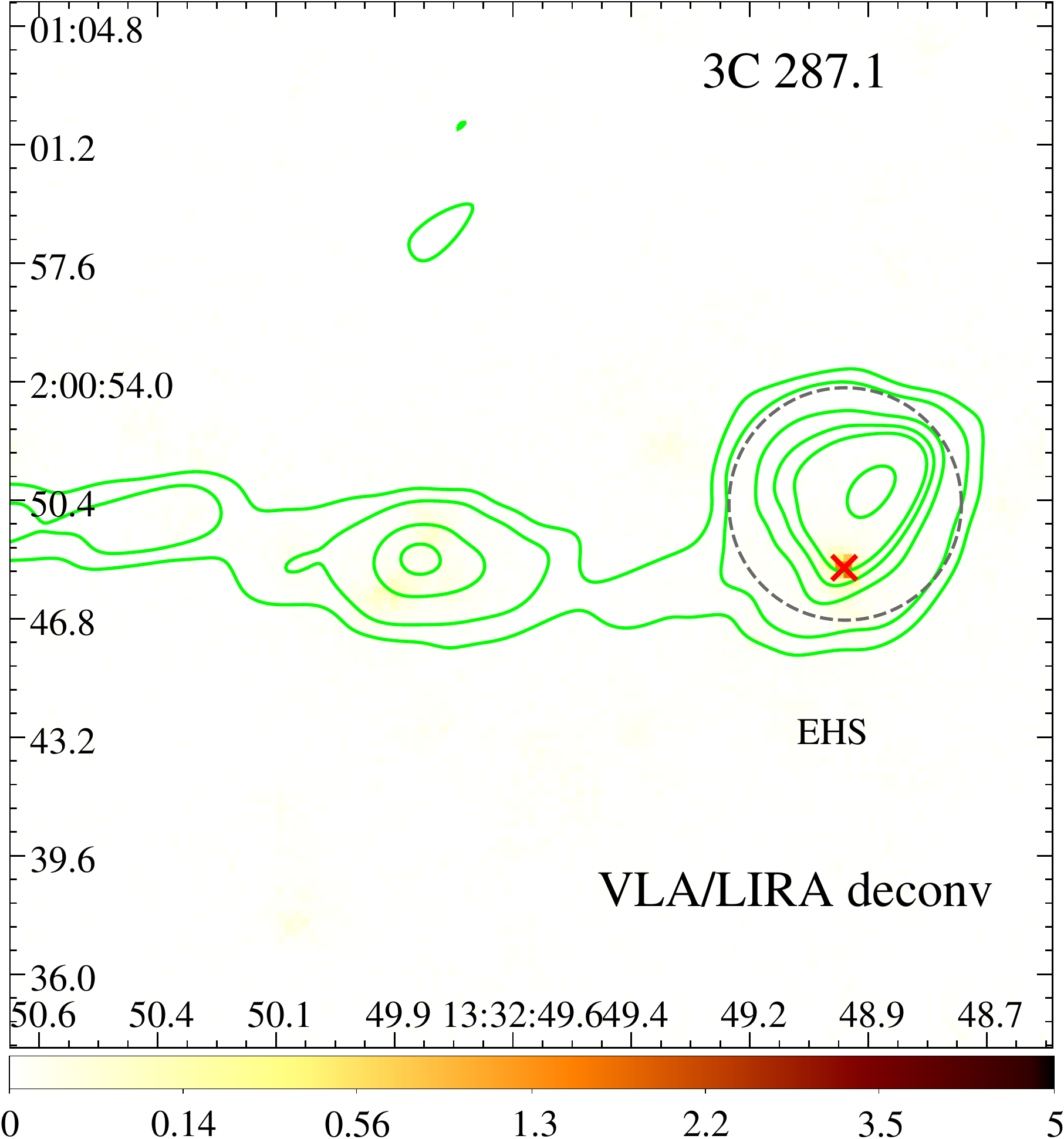}{0.3\textwidth}{(e)}
    }
    \caption{Same as in Fig. \ref{fig:results-3C9} but for 3C 287.1. (a) shows the full image while (b), (c) and (d),(e) show the zoomed-in regions around knot A and eastern hotspot (EHS), respectively. The radio contours are given by 0.8, 1.5, 4.0, 6.0, 8.0, 20.0 mJy beam$^{-1}$.\label{fig:results-3C287.1}}
\end{figure*}

\begin{figure*}[ht]
    \gridline{
        \fig{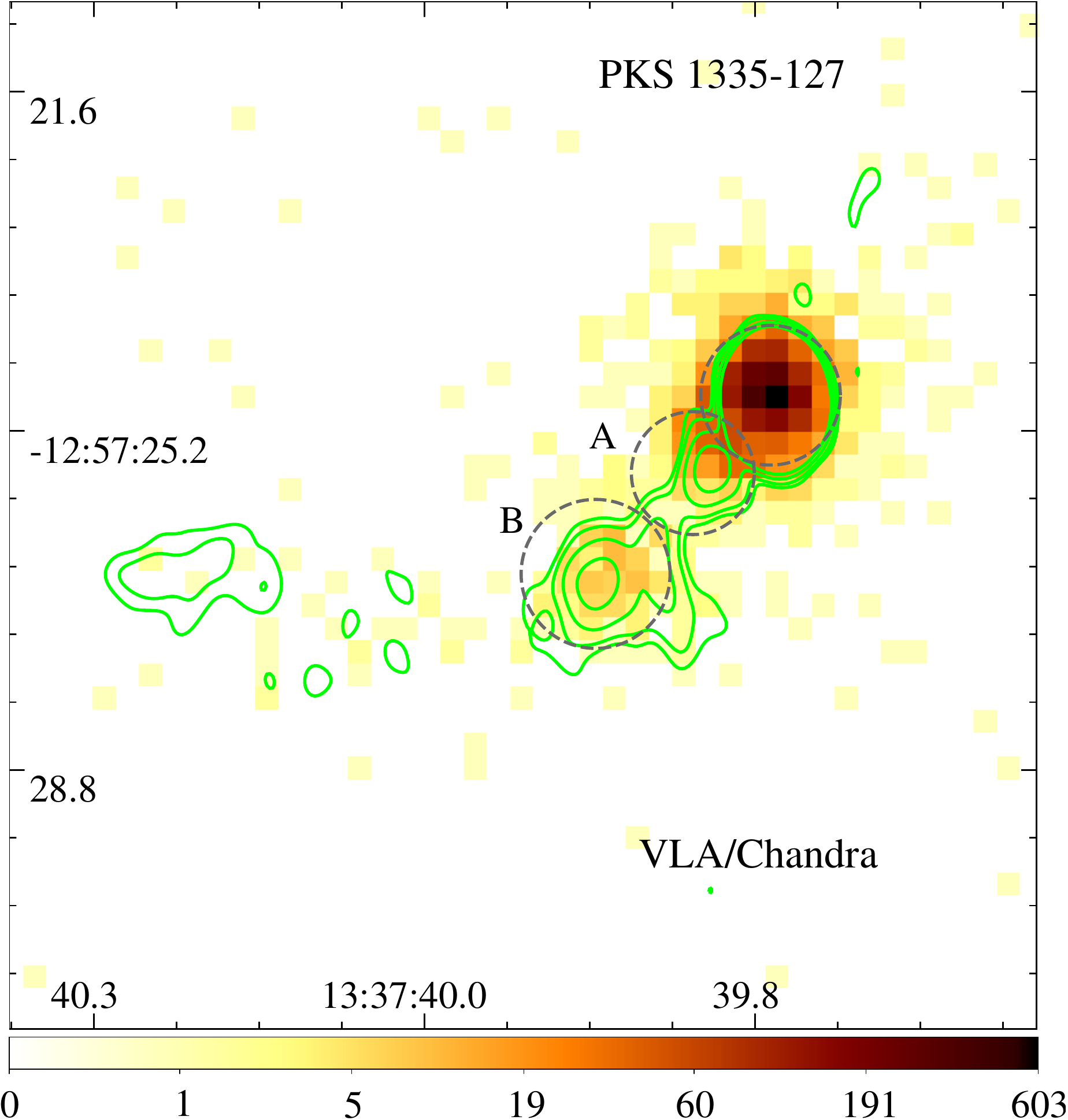}{0.5\textwidth}{(a)}
        \fig{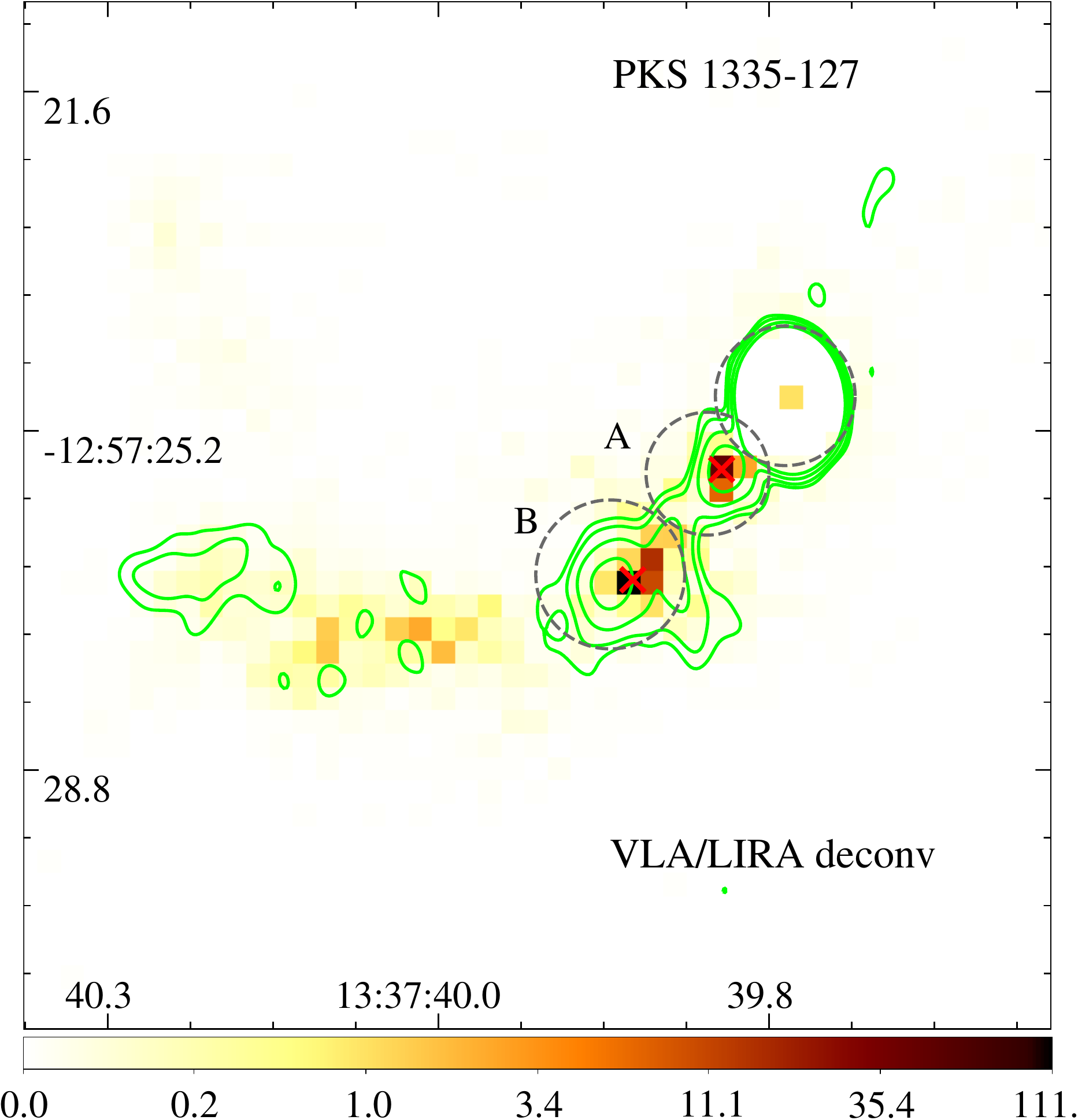}{0.5\textwidth}{(b)}
    }
    \caption{Same as in Fig. \ref{fig:results-3C9} but for PKS 1335-127. The radio contours are given by 1.2, 2.0, 4.0, 8.0 mJy beam$^{-1}$.\label{fig:results-PKS1335-127}}
\end{figure*}

\begin{figure*}[ht]
    \gridline{
        \fig{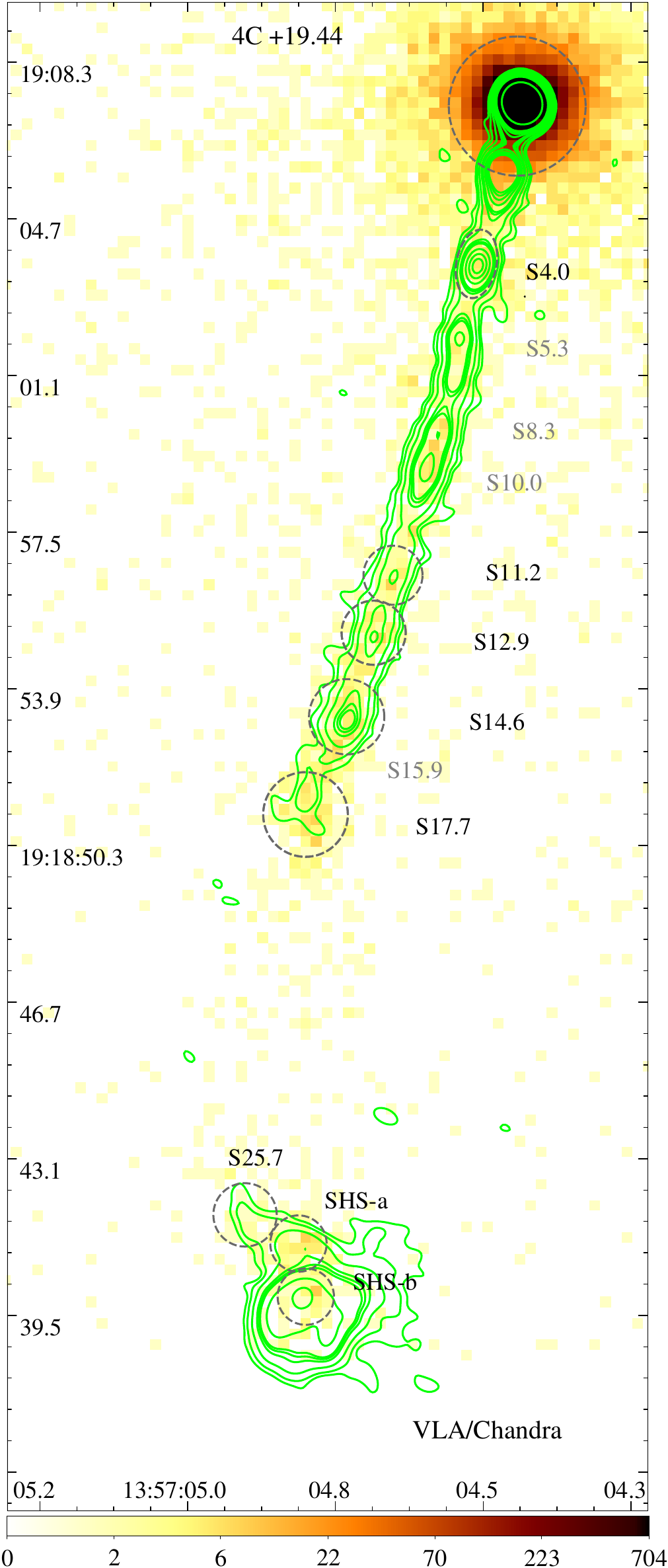}{0.5\textwidth}{(a)}
        \fig{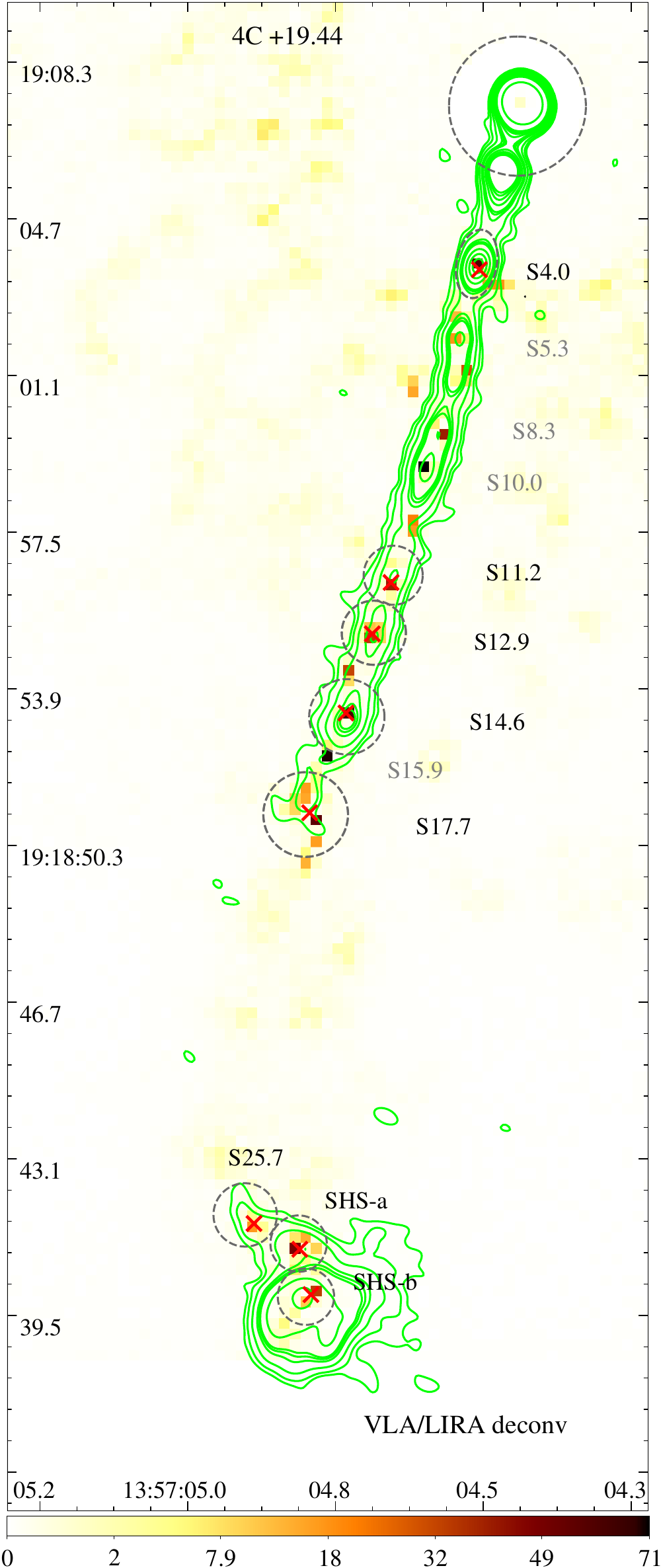}{0.5\textwidth}{(b)}
    }
    \caption{Same as in Fig. \ref{fig:results-3C9} but for 4C+19.44. The radio contours are given by 0.25, 0.4, 0.8, 1.6, 2.3, 2, 4, 6, 8, 10, 60 mJy beam$^{-1}$.\label{fig:results-4C+19.44}}
\end{figure*}

\begin{figure*}[ht]
    \gridline{
        \fig{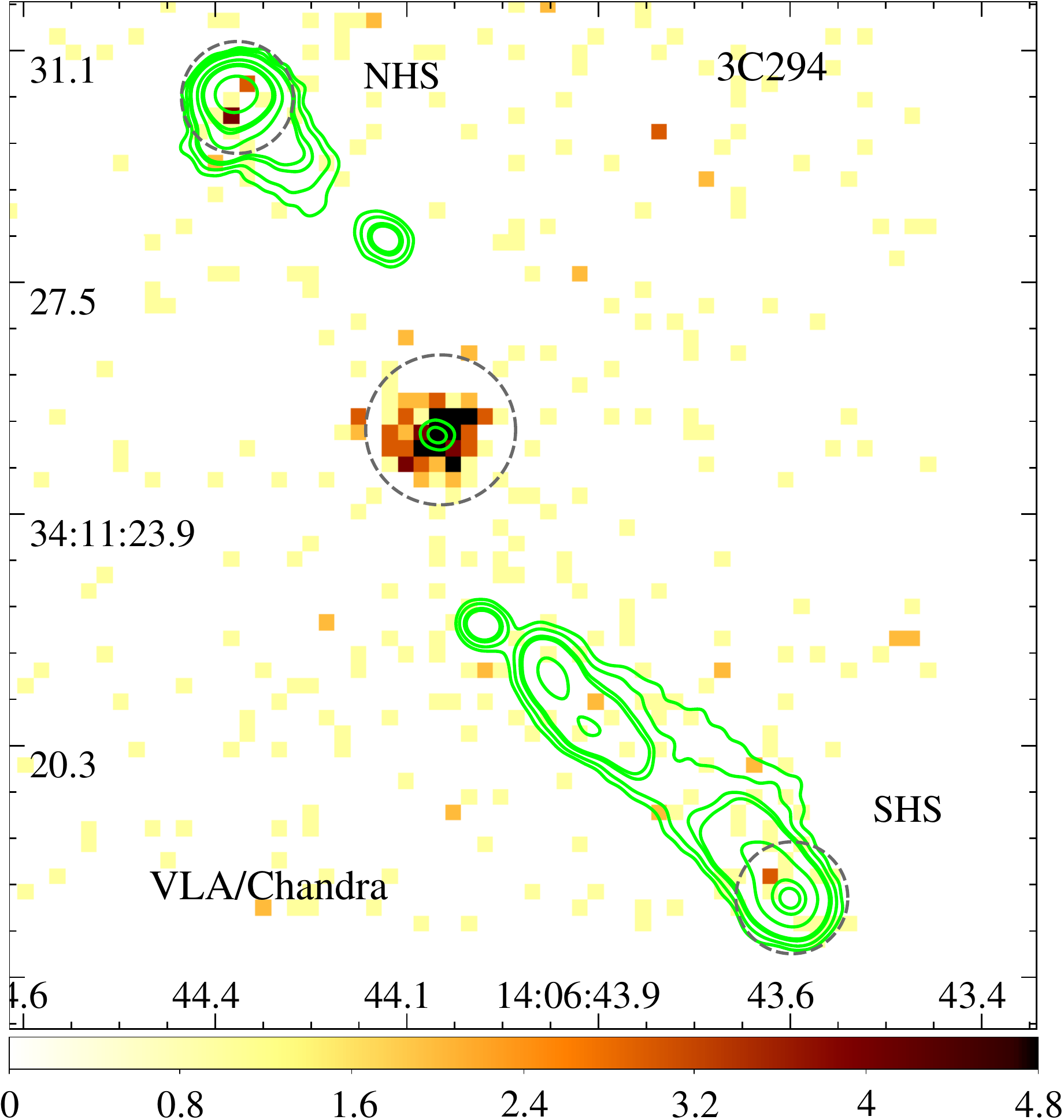}{0.5\textwidth}{(a)}
        \fig{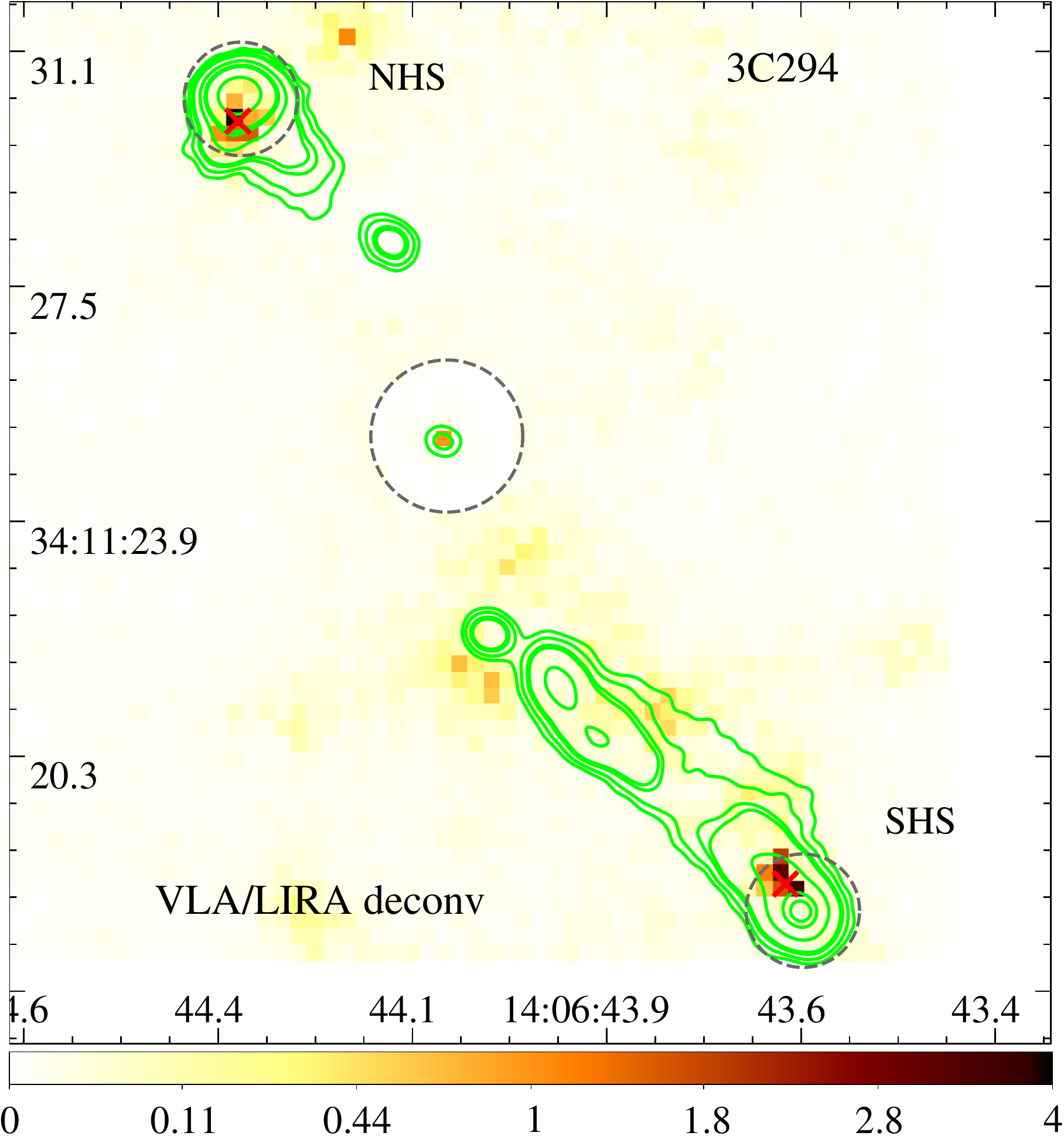}{0.5\textwidth}{(b)}
    }
    \caption{Same as in Fig. \ref{fig:results-3C9} but for 3C 294. The radio contours are given by 0.2, 0.4, 0.8, 1.0, 3.5, 8.0, 10.0, 40.0 mJy beam$^{-1}$.\label{fig:results-3C294}}
\end{figure*}

\begin{figure*}[ht]
    \gridline{
        \fig{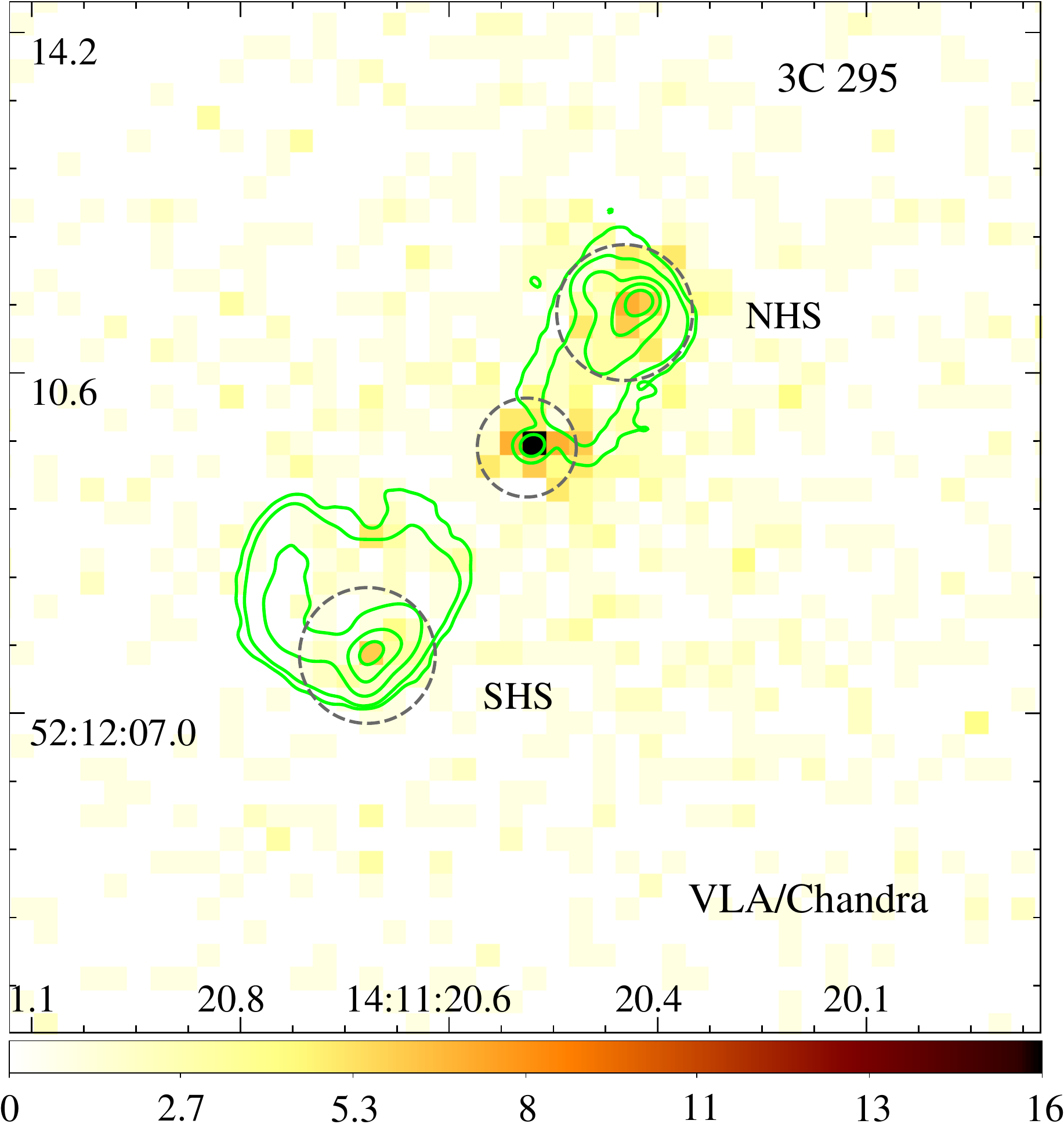}{0.5\textwidth}{(a)}
        \fig{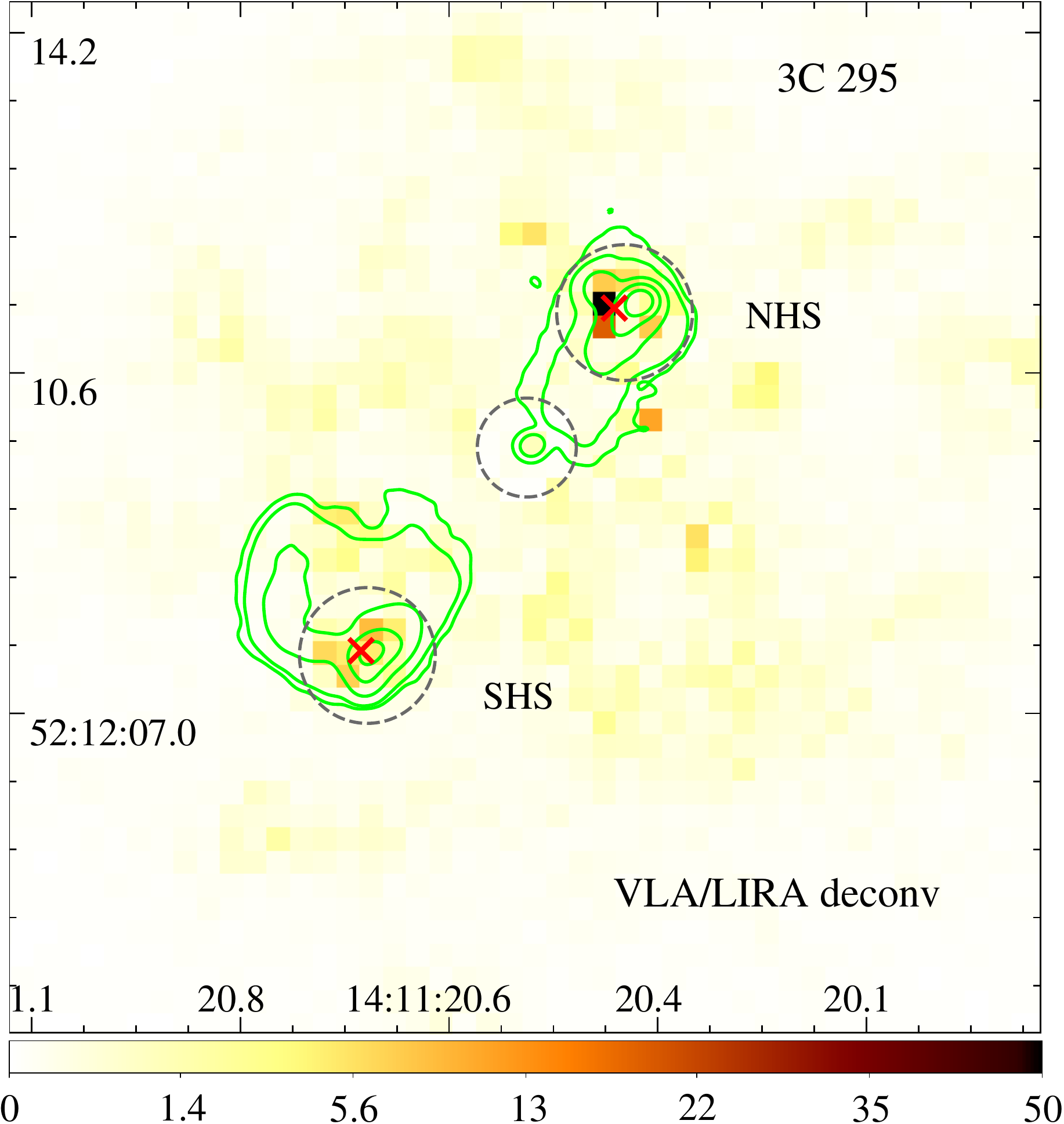}{0.5\textwidth}{(b)}
    }
    \caption{Same as in Fig. \ref{fig:results-3C9} but for 3C 295. The radio contours are given by  0.4, 2.0, 20.0, 100.0, 300.0 mJy beam$^{-1}$.\label{fig:results-3C295}}
\end{figure*}

\begin{figure*}[ht]
    \gridline{
        \fig{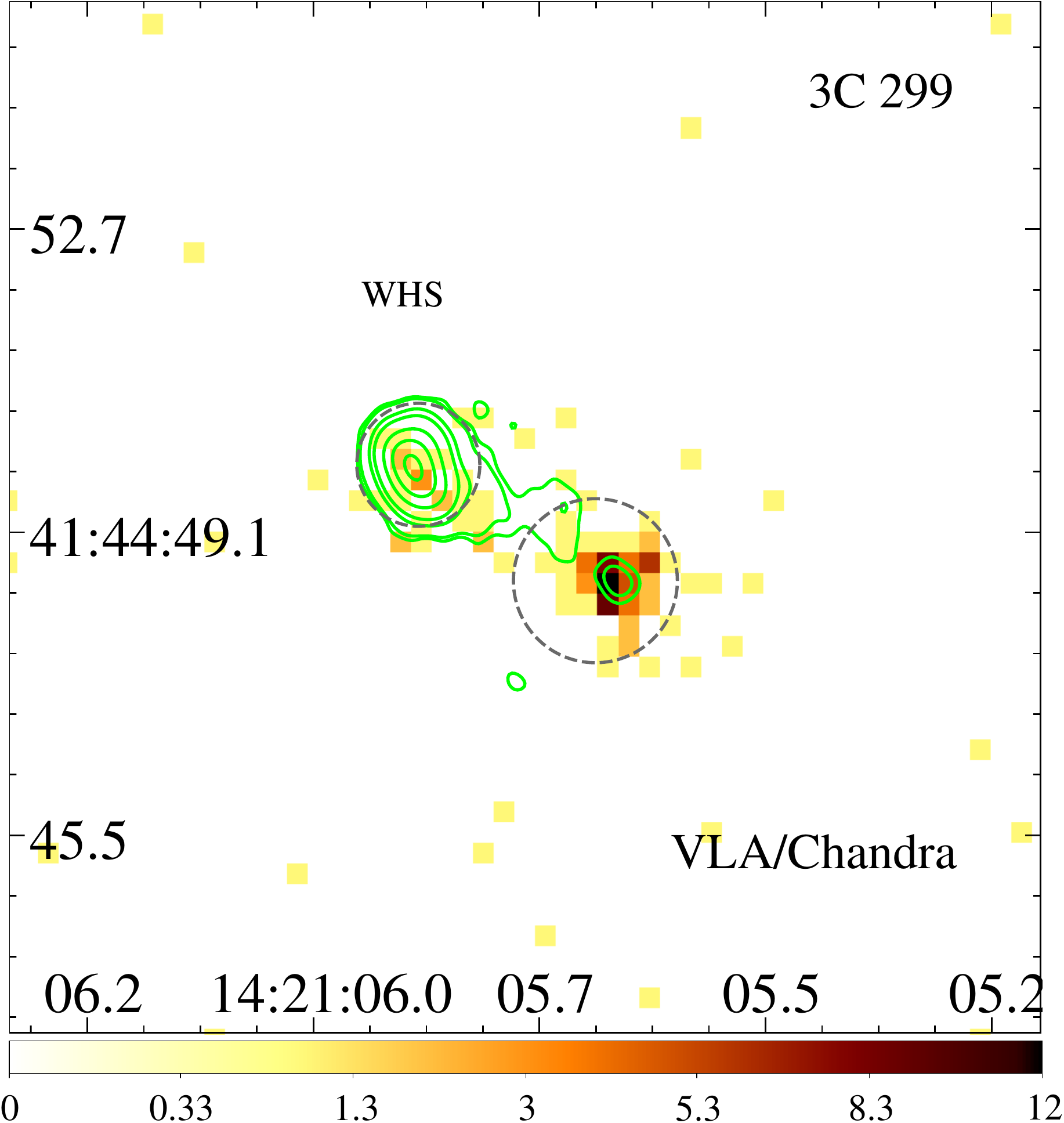}{0.5\textwidth}{(a)}
        \fig{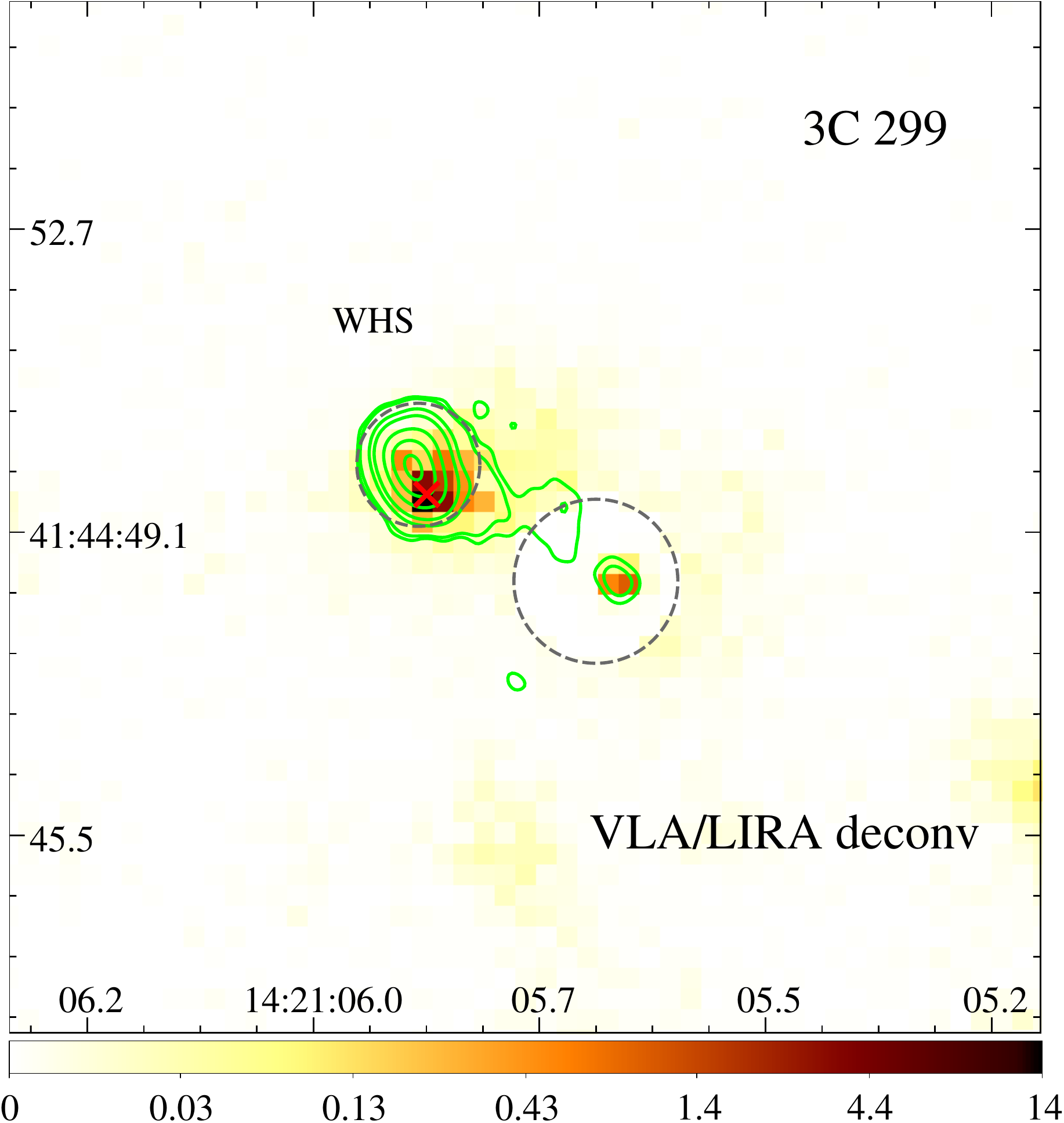}{0.5\textwidth}{(b)}
    }
    \caption{Same as in Fig. \ref{fig:results-3C9} but for 3C 299. The radio contours are given by 0.5, 1.0, 8.0, 20.0, 80.0, 200.0, 400.0 mJy beam$^{-1}$.\label{fig:results-3C299}}
\end{figure*}

\begin{figure*}[ht]
    \gridline{
        \fig{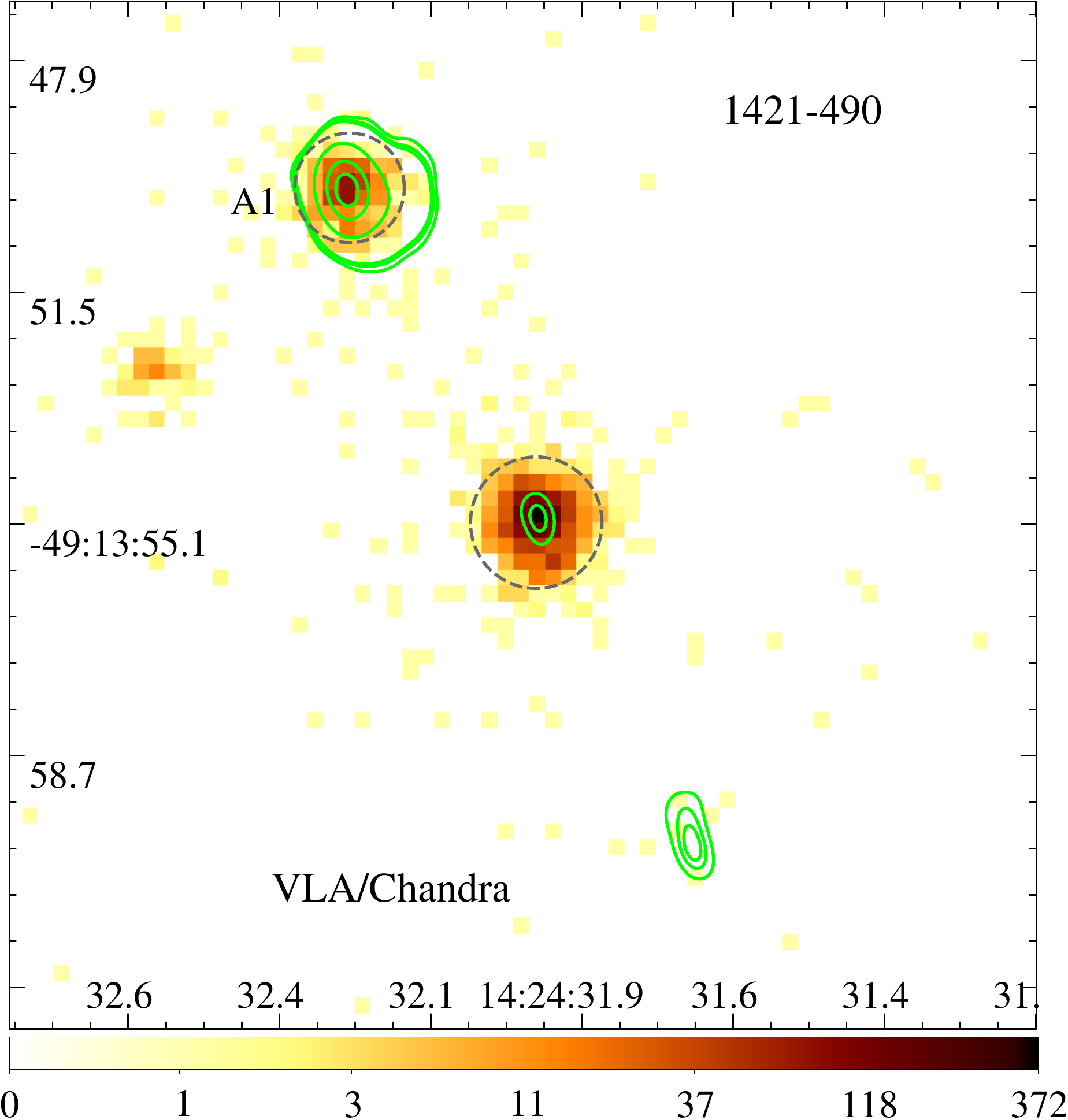}{0.5\textwidth}{(a)}
        \fig{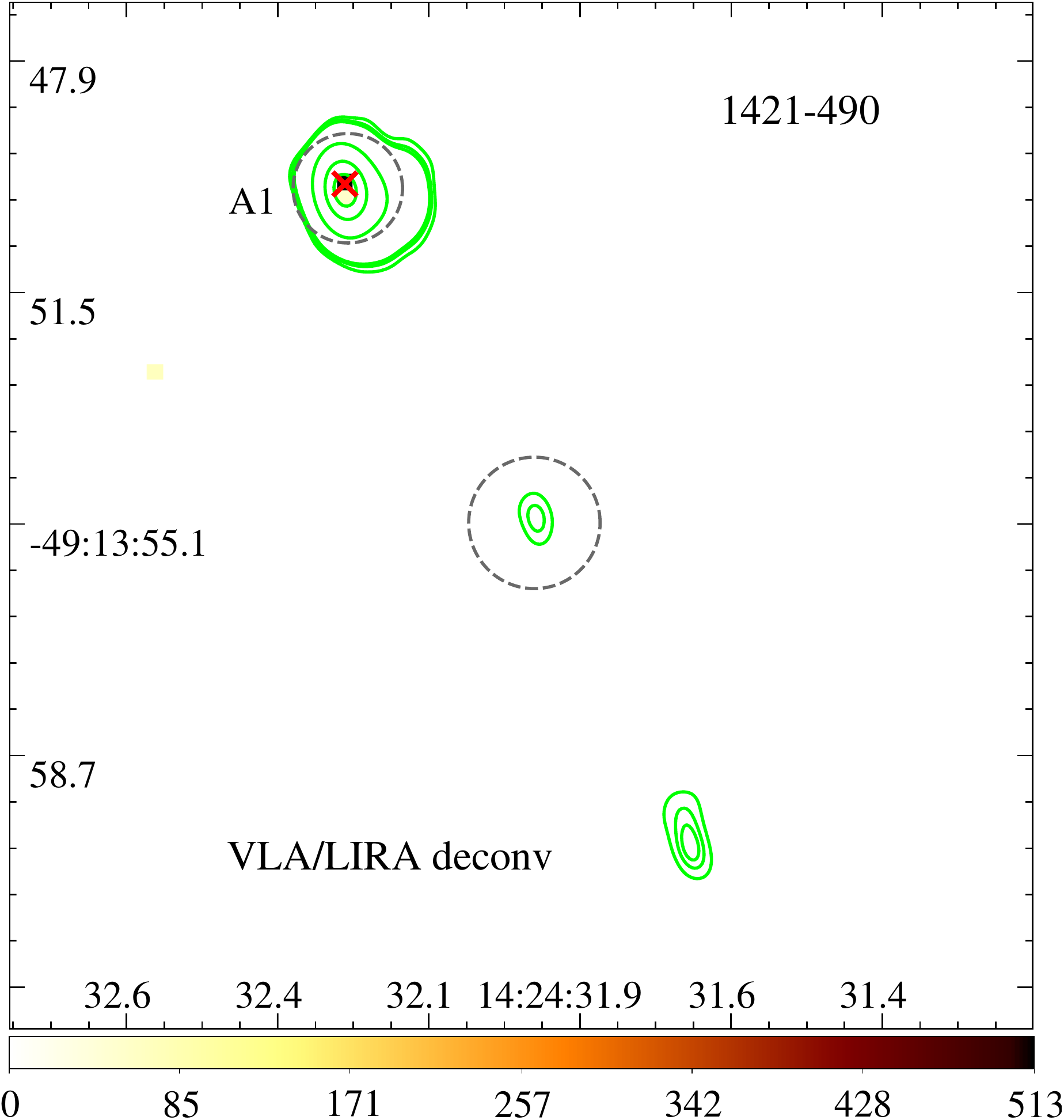}{0.5\textwidth}{(b)}
    }
    \caption{Same as in Fig. \ref{fig:results-3C9} but for PKS 1421-490. The radio contours are given by 5, 8,10, 200, 1000, 2000 mJy beam$^{-1}$.\label{fig:results-1421-490}}
\end{figure*}

\begin{figure*}[ht]
    \gridline{
        \fig{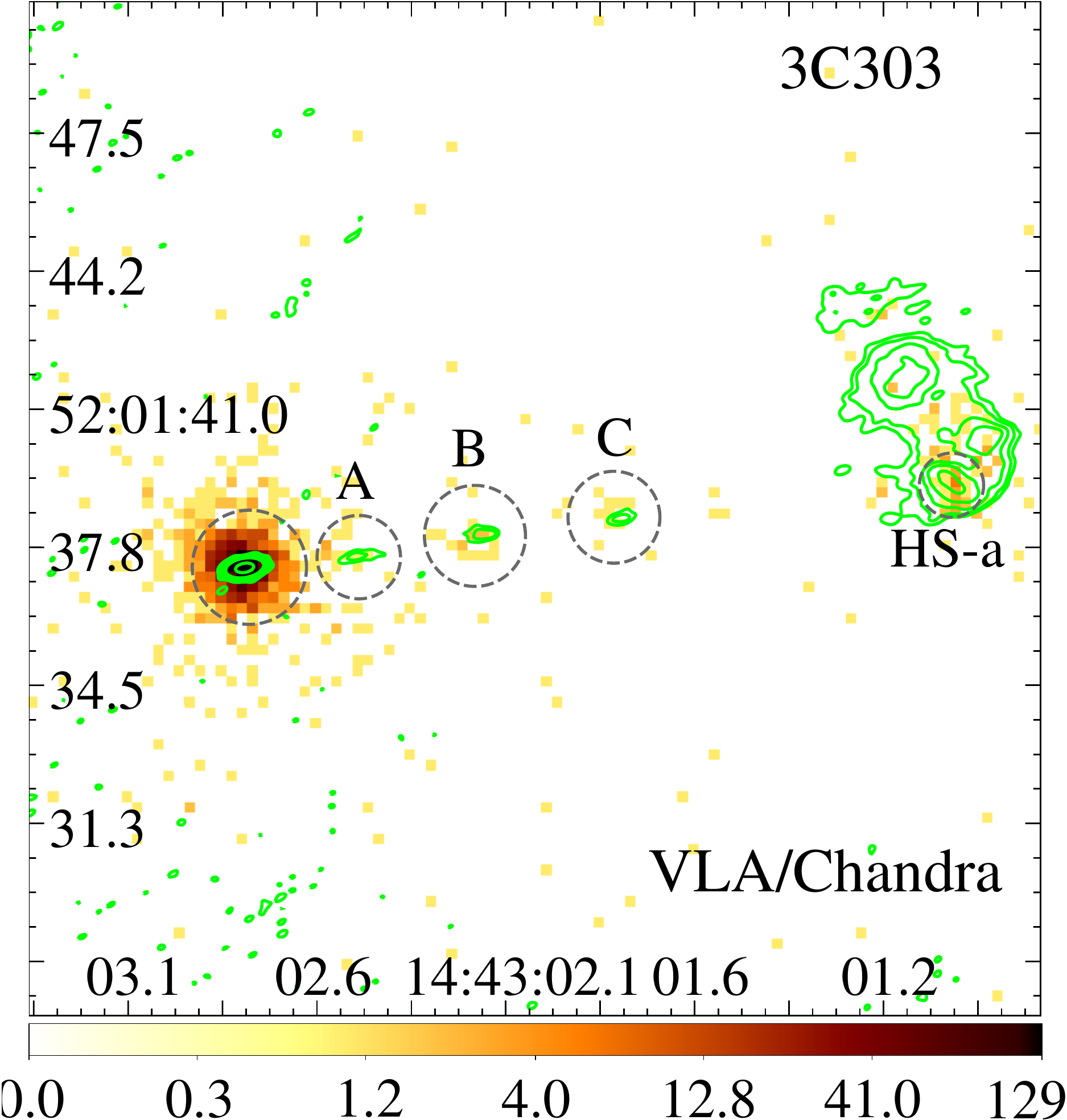}{0.5\textwidth}{(a)}
        \fig{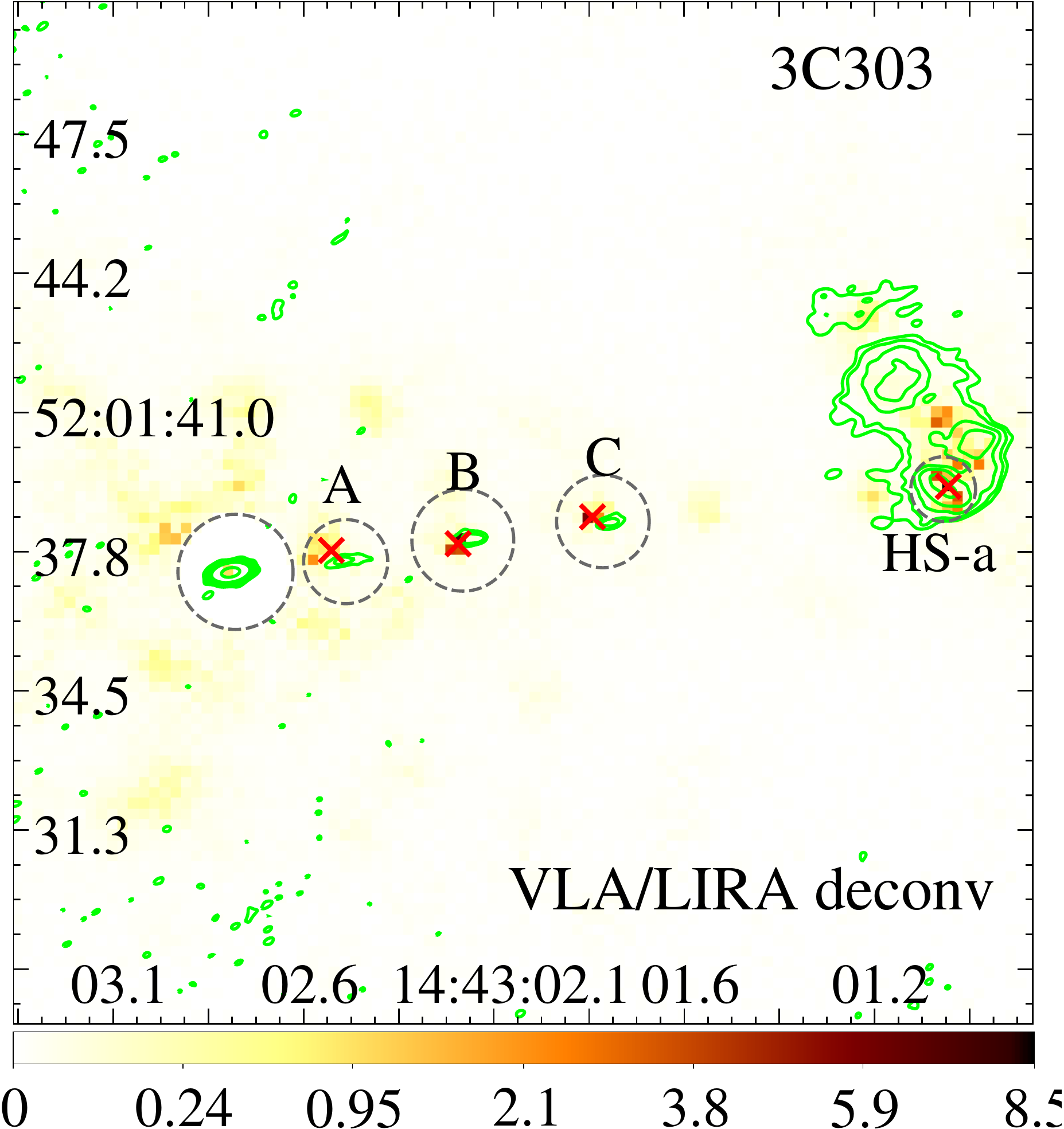}{0.5\textwidth}{(b)}
    }
    \caption{Same as in Fig. \ref{fig:results-3C9} but for 3C 303. The radio contours are given by 0.4, 0.8, 2.0, 4.0, 10.0, 100.0 mJy beam$^{-1}$.\label{fig:results-3C303}}
\end{figure*}

\begin{figure*}[ht]
    \gridline{
        \fig{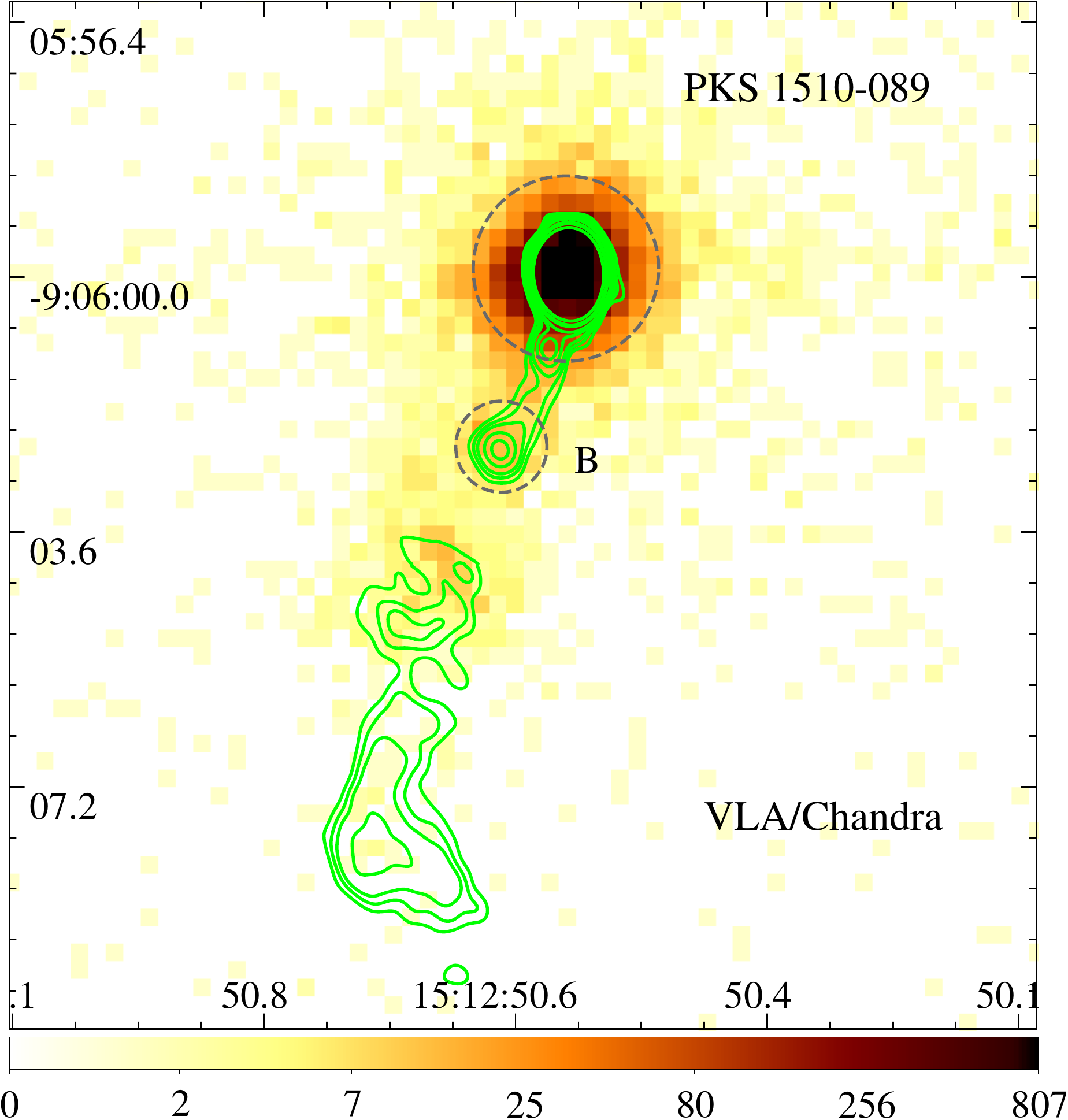}{0.5\textwidth}{(a)}
        \fig{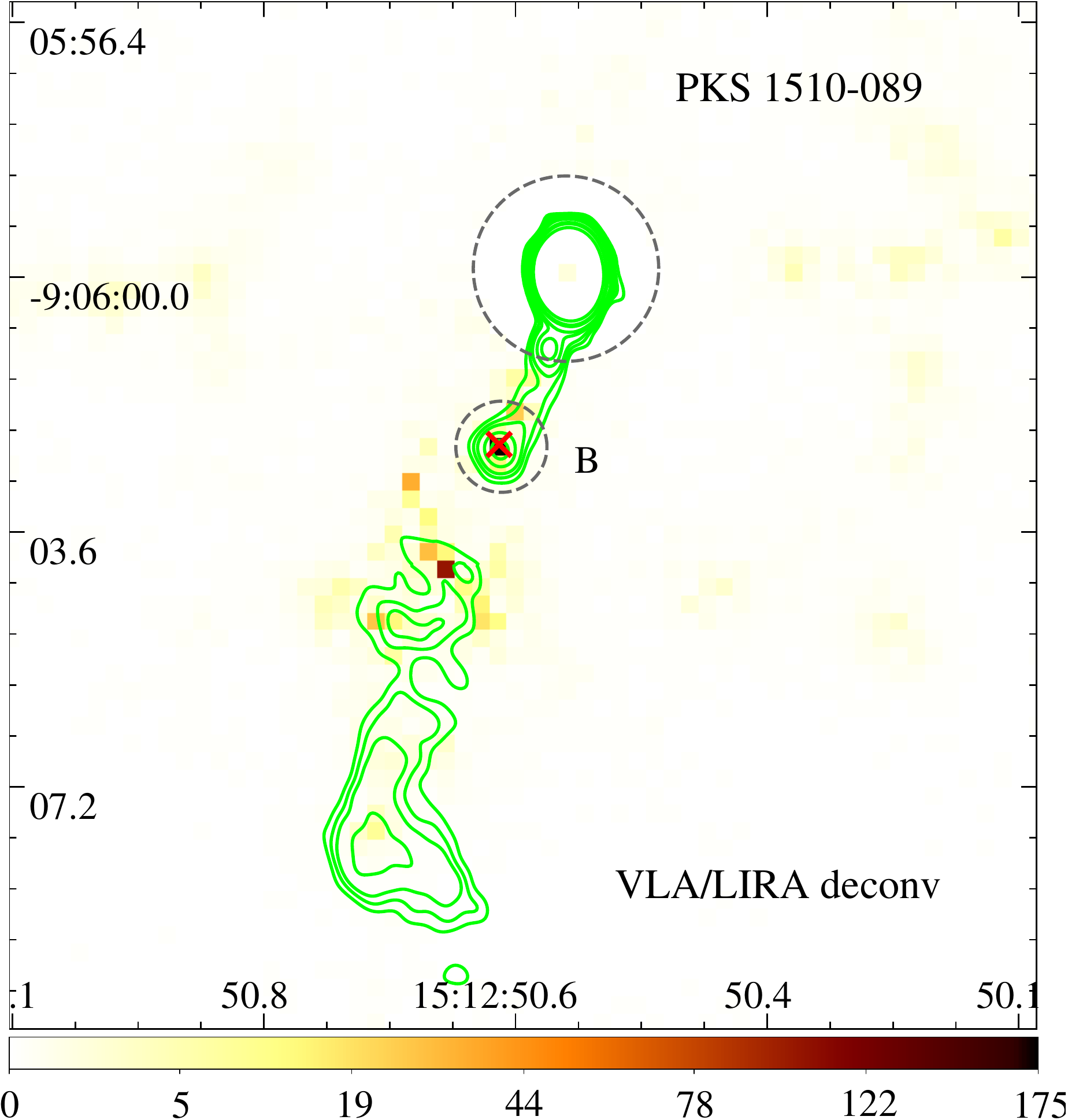}{0.5\textwidth}{(b)}
    }
    \caption{Same as in Fig. \ref{fig:results-3C9} but for PKS 1510-08. The radio contours are given by 1.0, 1.5, 2.0, 3.0, 4.0, 8.0, 10.0, 20.0 mJy beam$^{-1}$.\label{fig:results-PKS1510-089}}
\end{figure*}

\begin{figure*}[ht]
    \gridline{
        \fig{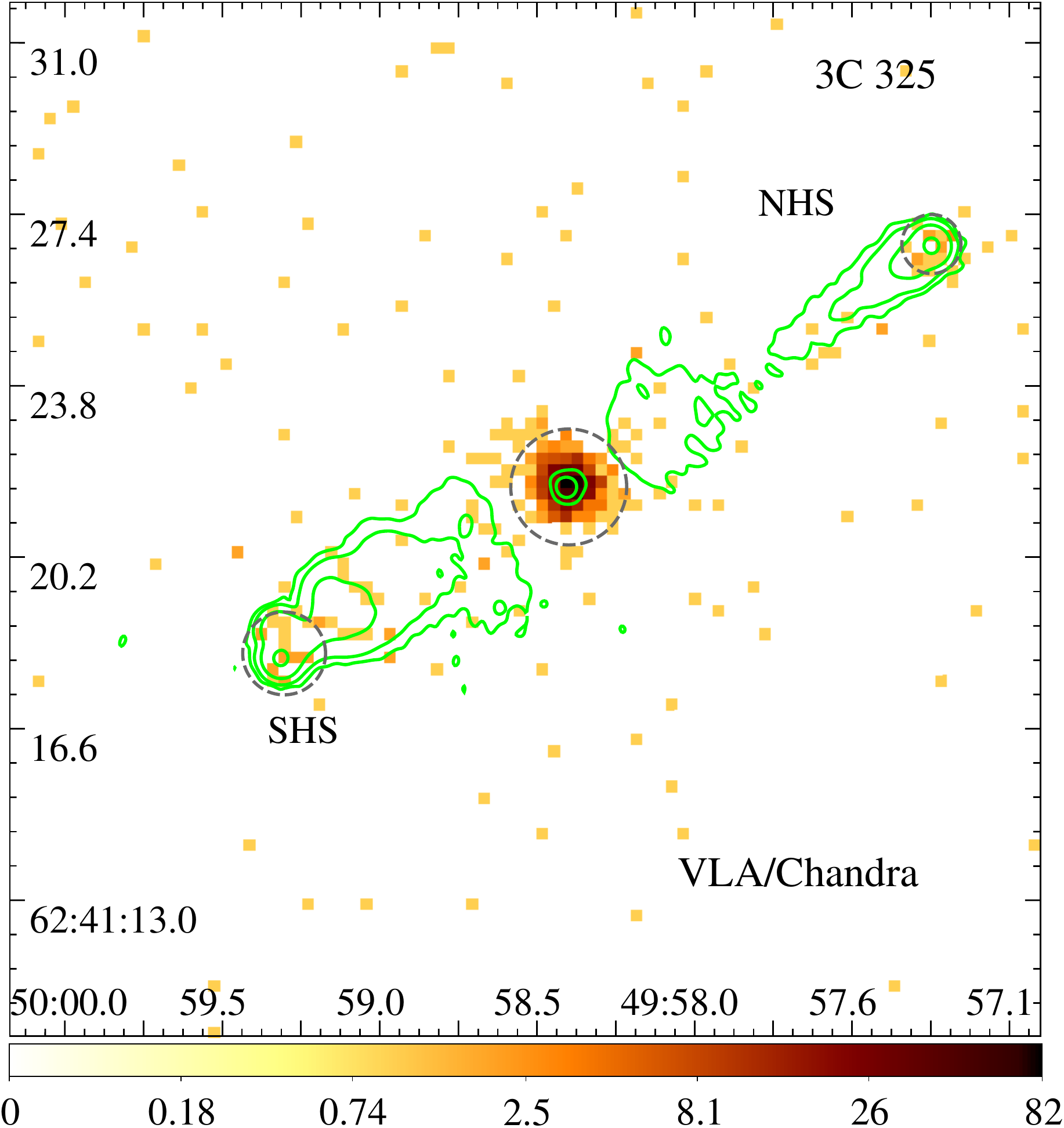}{0.5\textwidth}{(a)}
        \fig{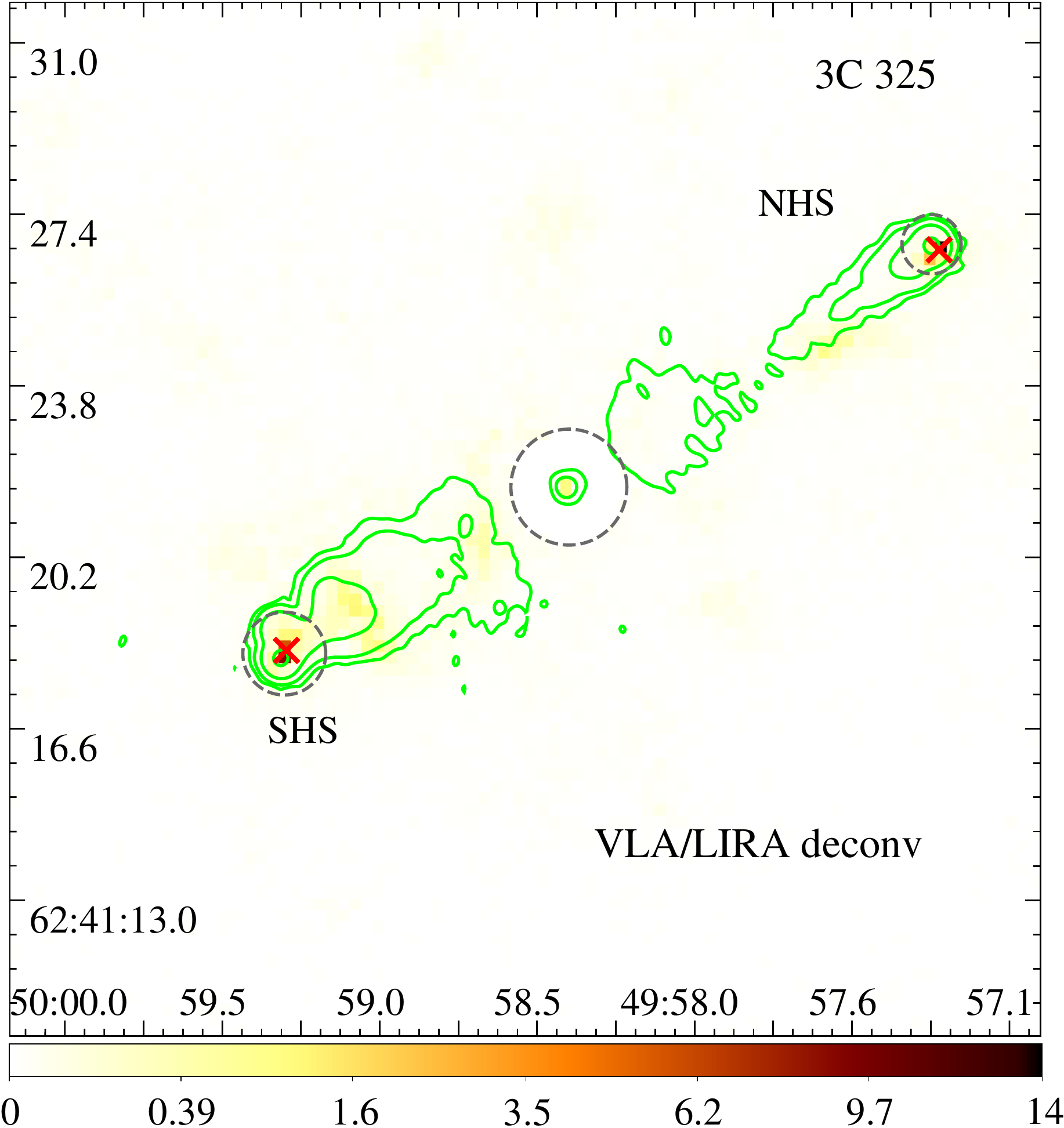}{0.5\textwidth}{(b)}
    }
    \caption{Same as in Fig. \ref{fig:results-3C9} but for 3C 325. The radio contours are given by 0.15, 1.0, 10.0, 130.0 mJy beam$^{-1}$.\label{fig:results-3C325}}
\end{figure*}

\begin{figure*}[ht]
    \gridline{
        \fig{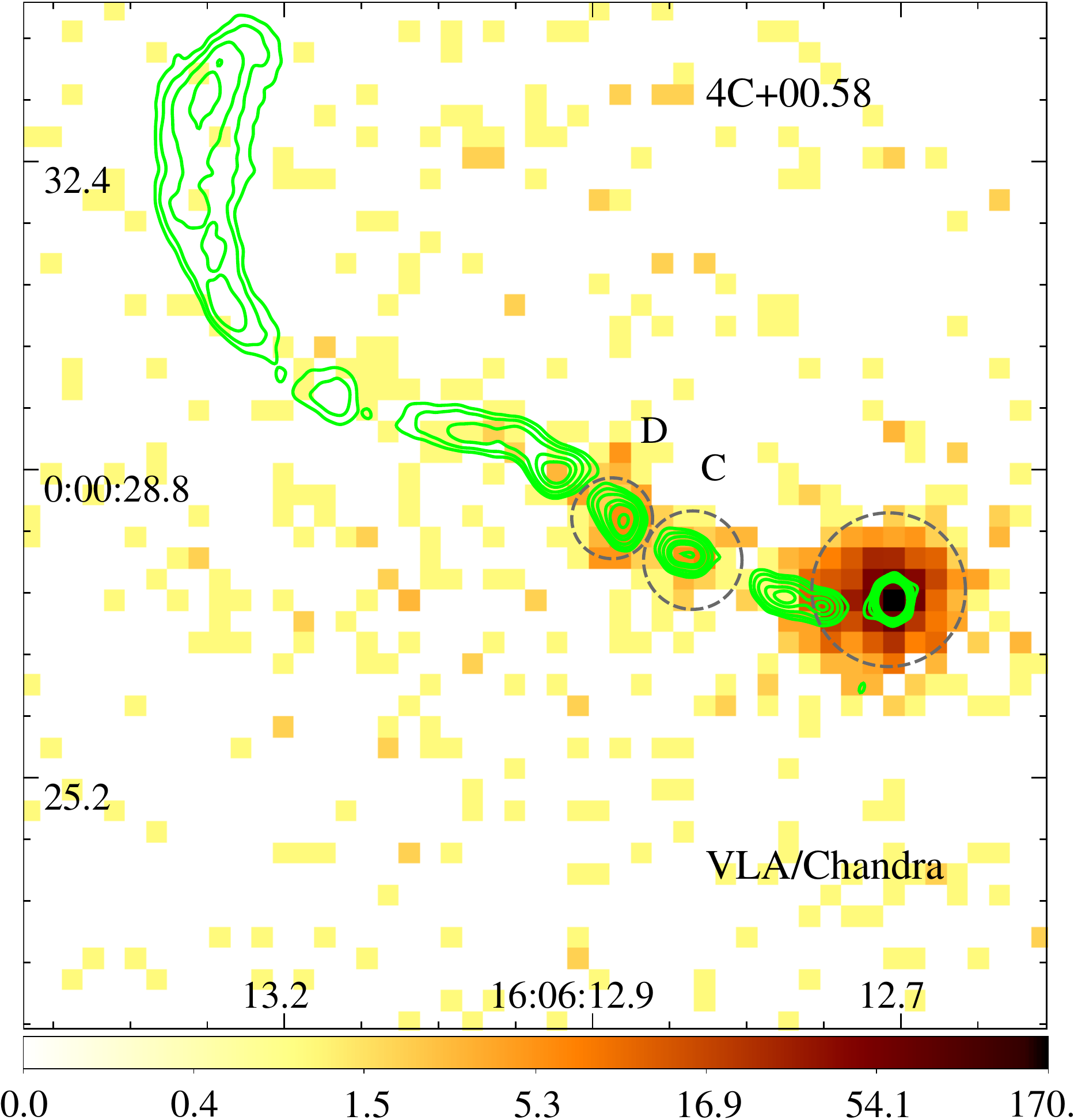}{0.5\textwidth}{(a)}
        \fig{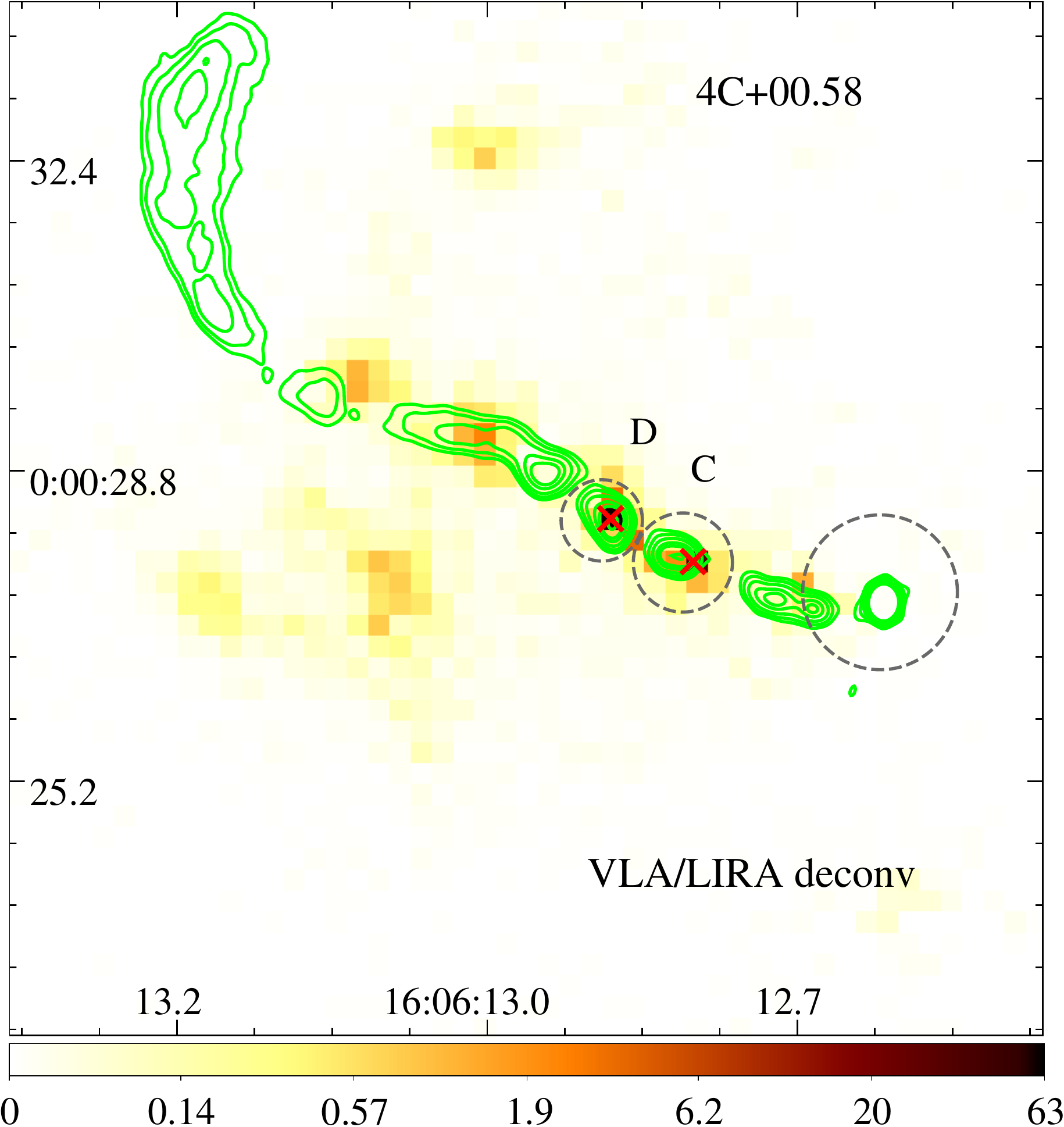}{0.5\textwidth}{(b)}
    }
    \caption{Same as in Fig. \ref{fig:results-3C9} but for 4C +00.58. The radio contours are given by 0.1, 0.2, 0.4, 0.8, 1.2, 2.2 mJy beam$^{-1}$.\label{fig:results-4C+00.58}}
\end{figure*}

\comment{
\begin{figure*}[ht]
    \gridline{
        \fig{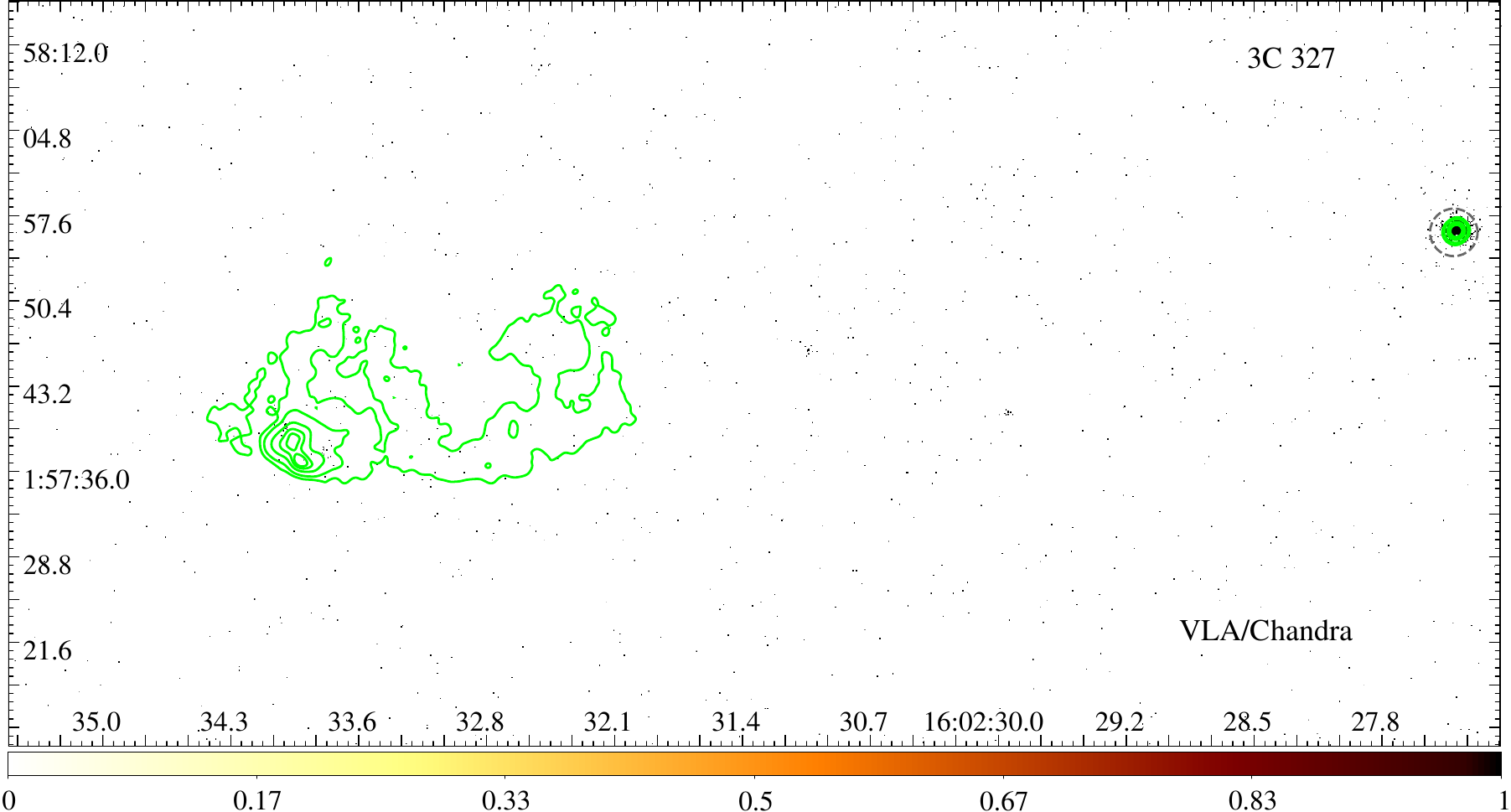}{\textwidth}{(a)}
    }
    \gridline{
        \fig{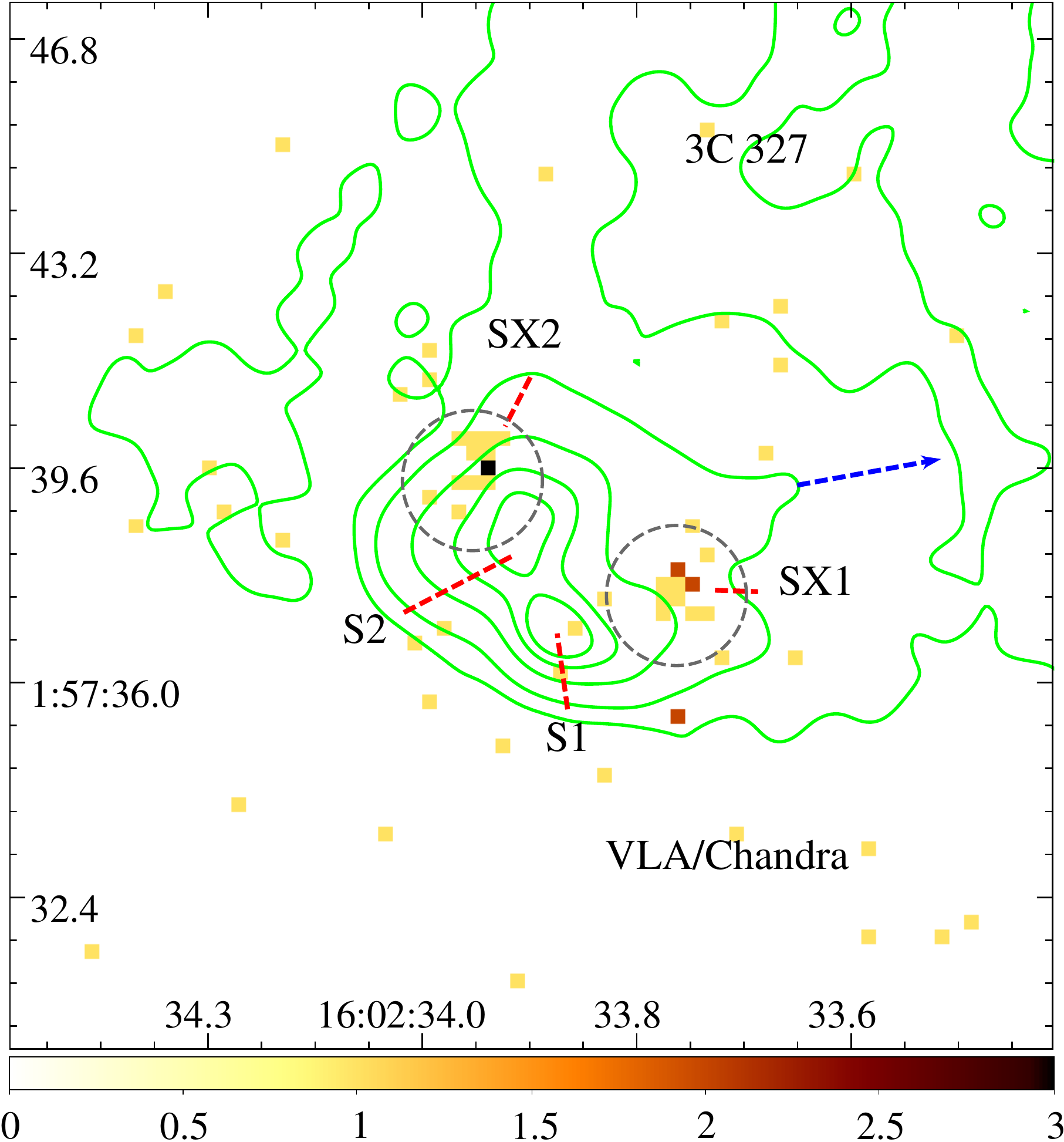}{0.5\textwidth}{(b)}
        \fig{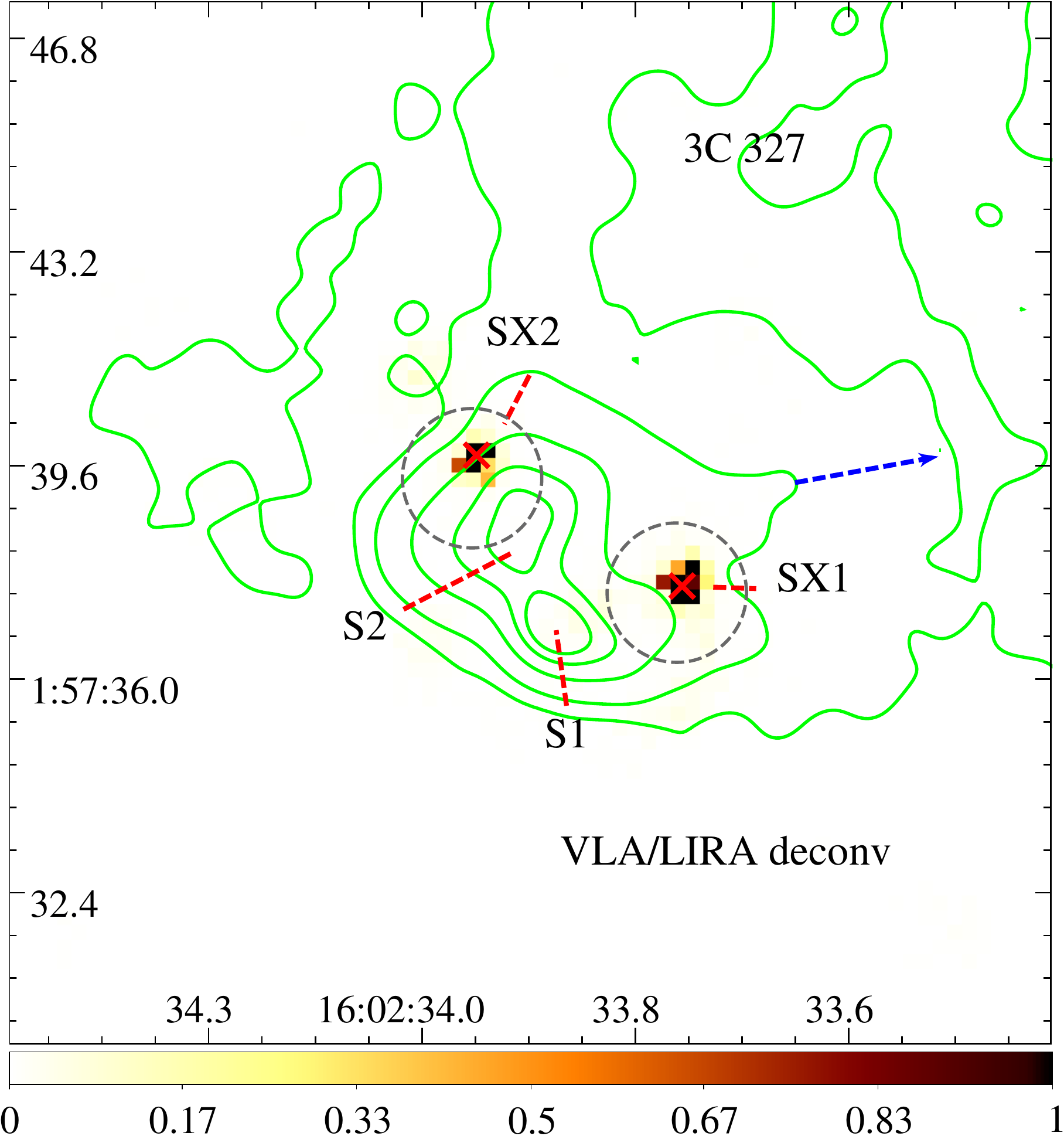}{0.5\textwidth}{(c)}
    }
    \caption{Same as in Fig. \ref{fig:results-3C9} but for 3C 327. The radio contours are given by 0.3, 1.0, 4.0, 8.0, 11.0 mJy beam$^{-1}$.\label{fig:results-3C327}}
\end{figure*}
}
\begin{figure*}[ht]
    \gridline{
        \fig{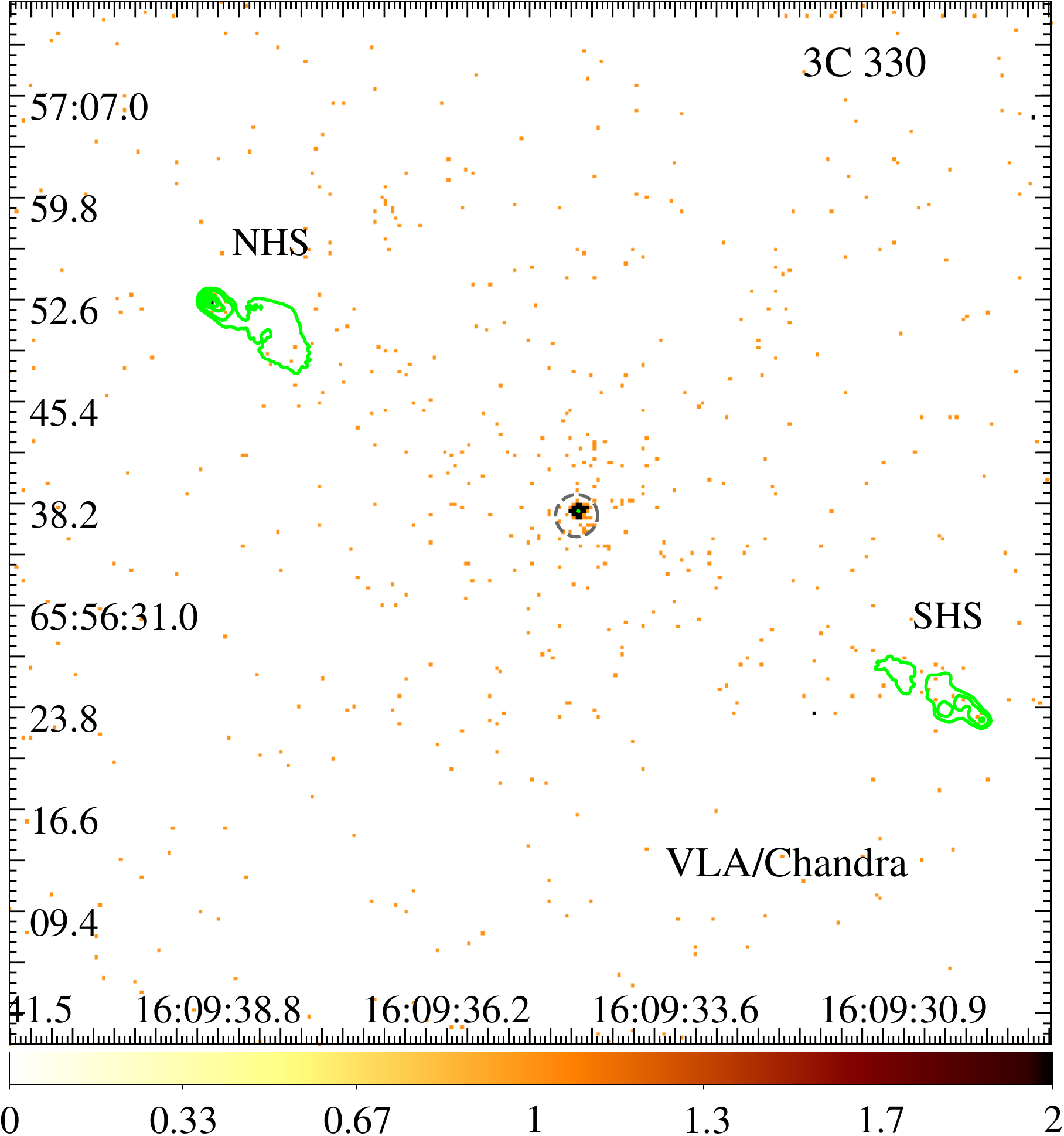}{0.5\textwidth}{(a)}
    }
    \gridline{
        \fig{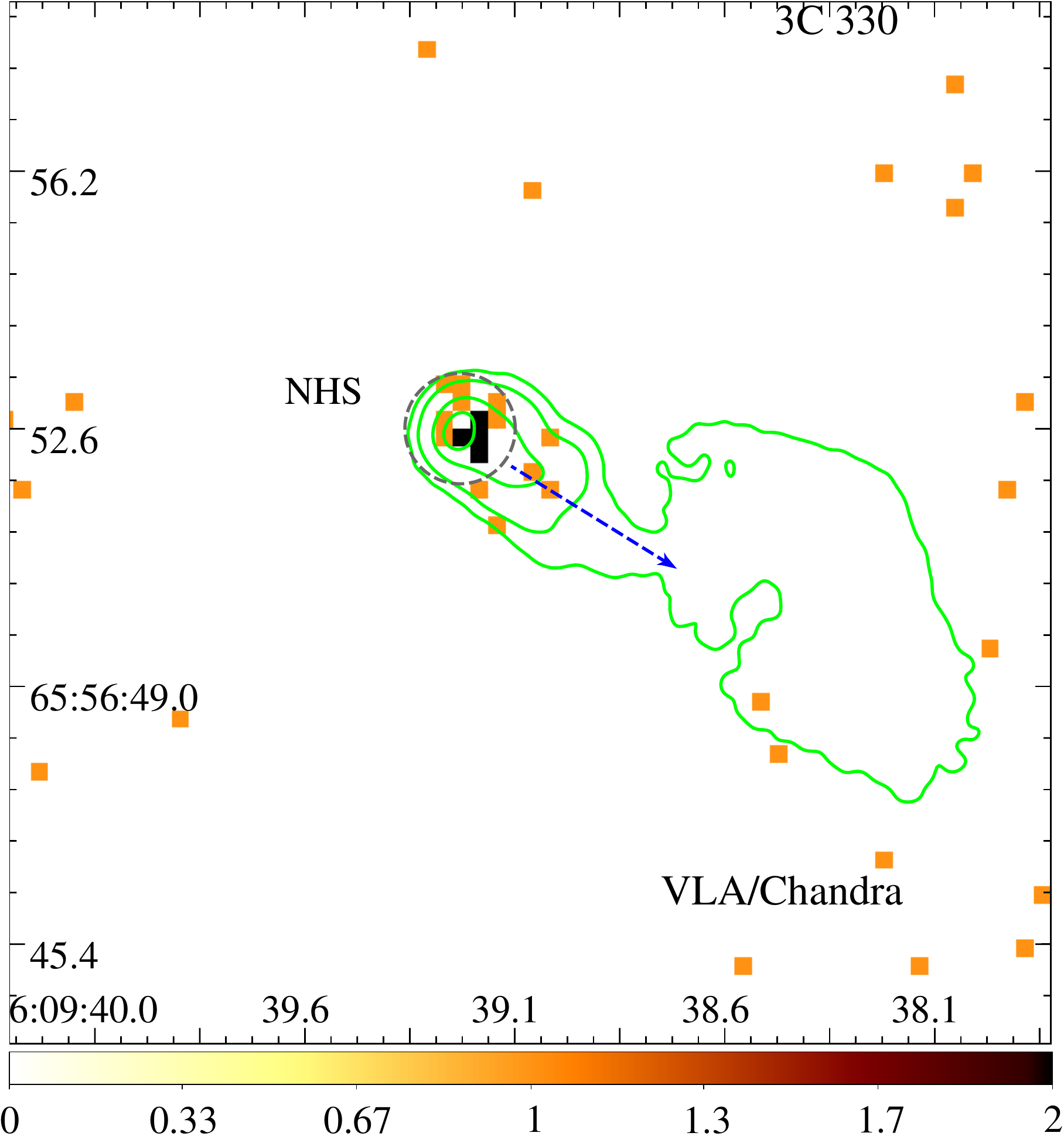}{0.5\textwidth}{(b)}
        \fig{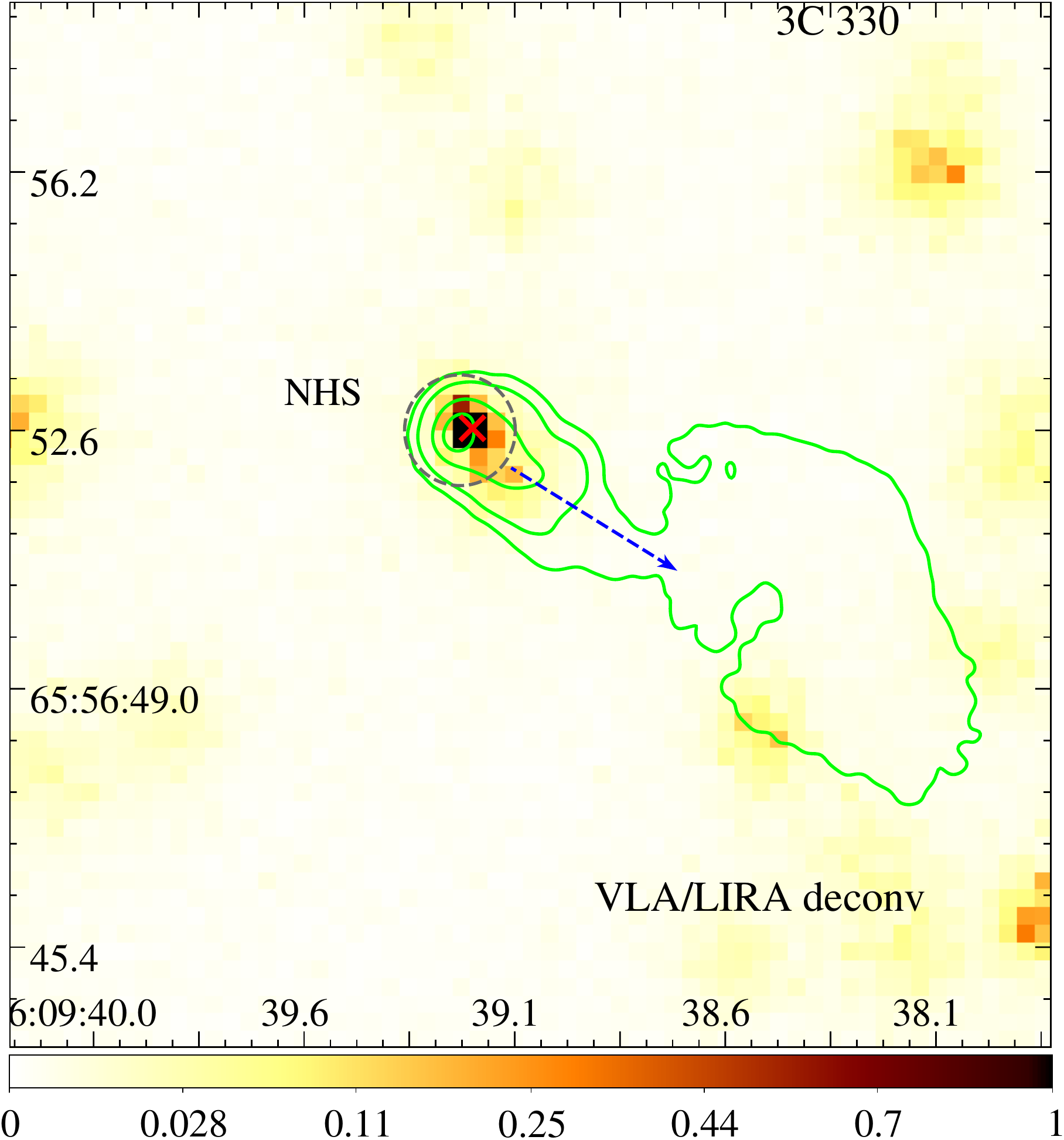}{0.5\textwidth}{(c)}
    }
    \comment{
    \gridline{
        \fig{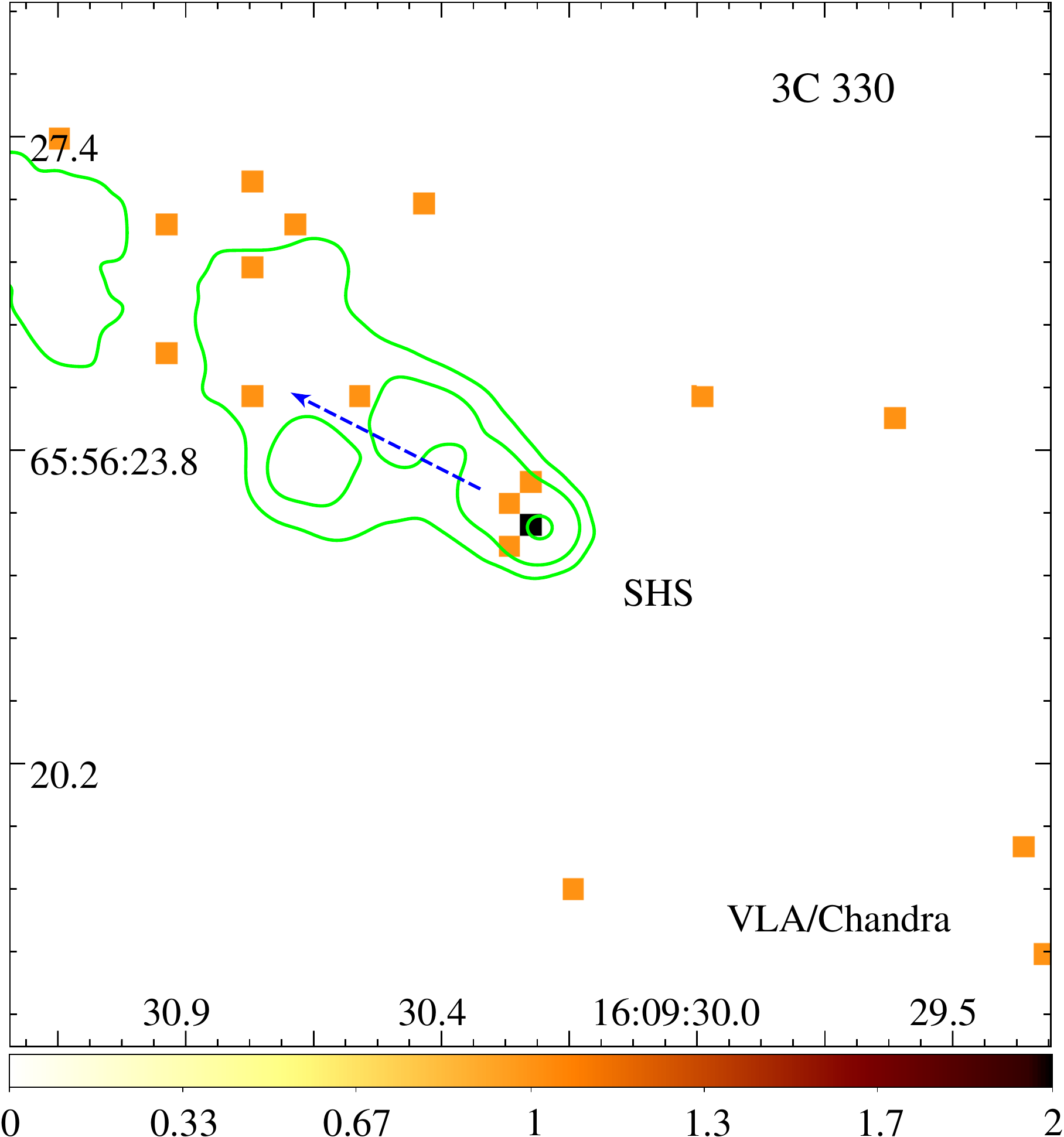}{0.5\textwidth}{(d)}
        \fig{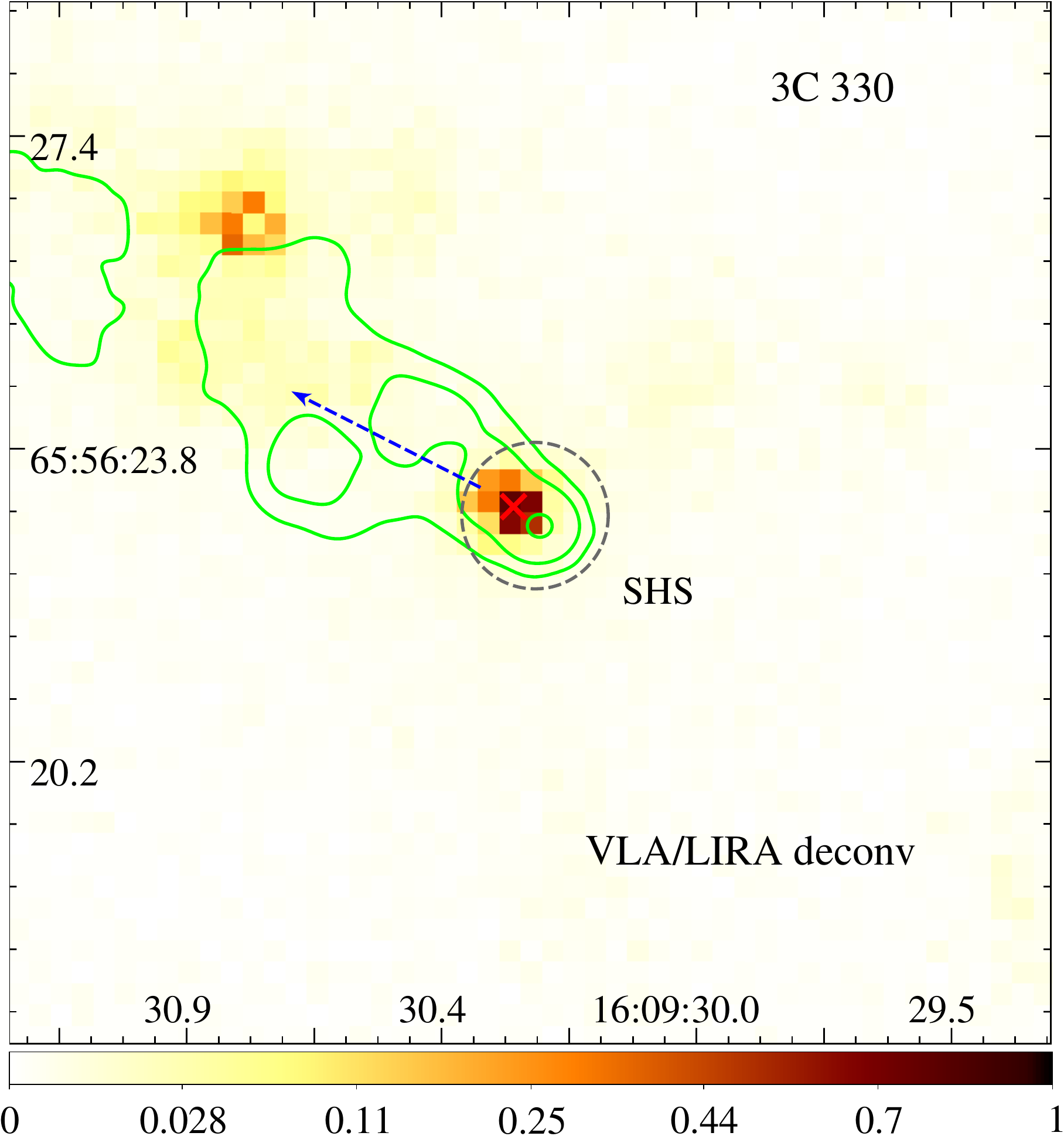}{0.5\textwidth}{(e)}
    }
    }
    \caption{Same as in Fig. \ref{fig:results-3C9} but for 3C 330. (a) shows the full image while (b) and (c) zoomed-in regions around the northern hotspot (NHS).
        The radio contours are given by 0.8, 8.0, 80.0, 200.0 mJy beam$^{-1}$.\label{fig:results-3C330}}
\end{figure*}

\begin{figure*}[ht]
    \gridline{
        \fig{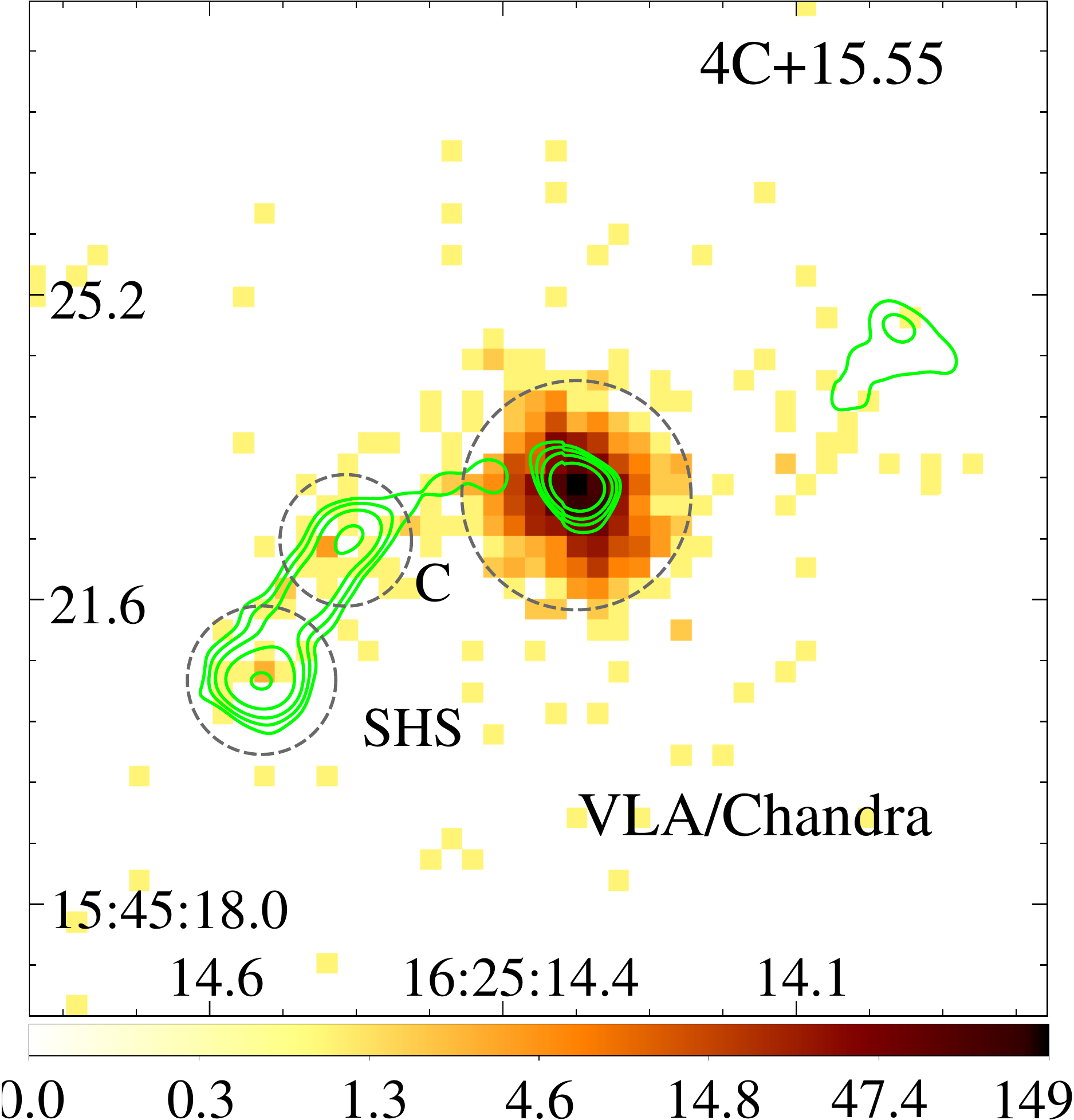}{0.5\textwidth}{(a)}
        \fig{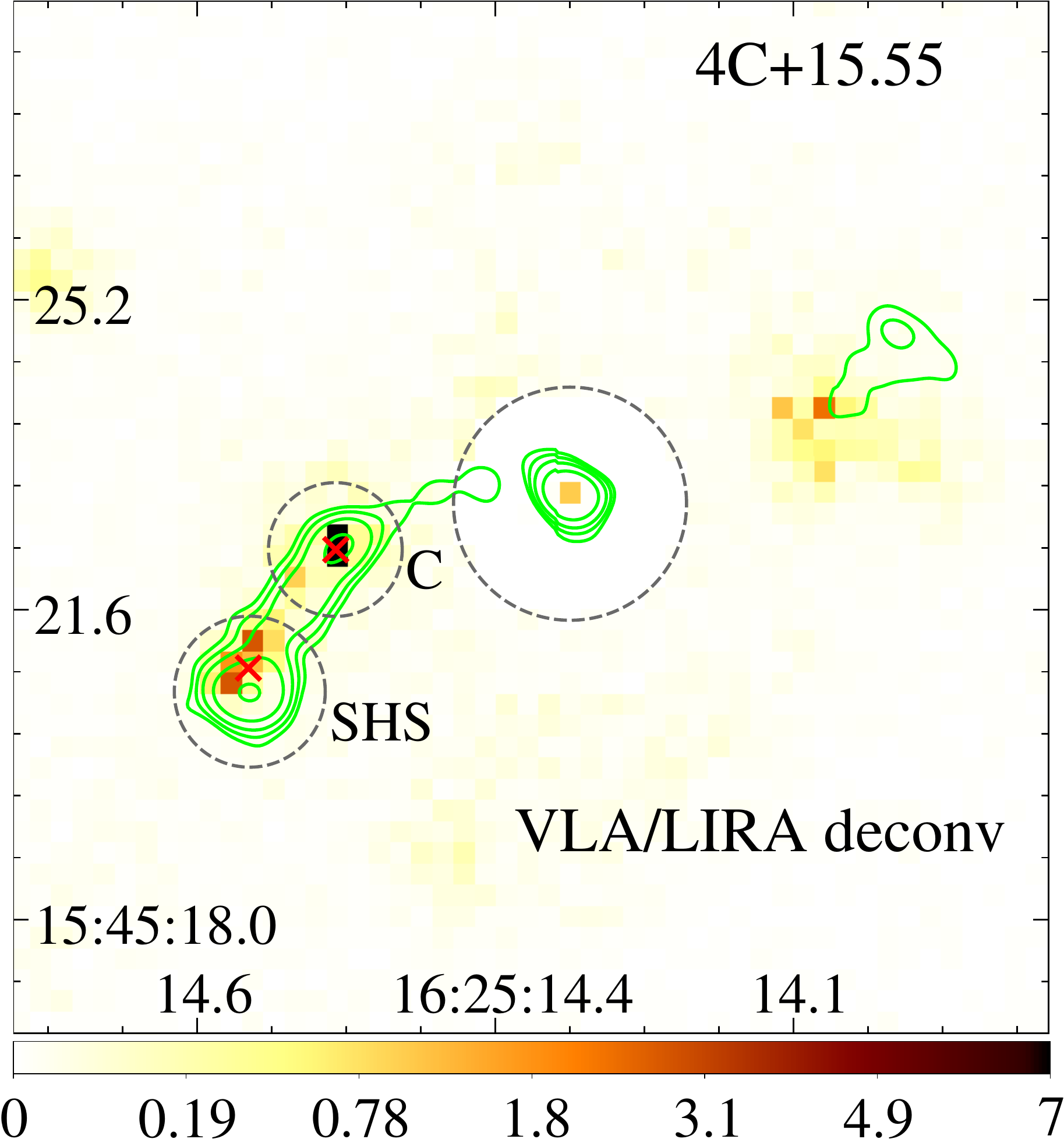}{0.5\textwidth}{(b)}
    }
    \caption{Same as in Fig. \ref{fig:results-3C9} but for 4C +15.55. The radio contours are given by 2.0, 4.0, 8.0, 20.0, 80.0 mJy beam$^{-1}$.\label{fig:results-4C+15.55}}
\end{figure*}

\begin{figure*}[ht]
    \gridline{
        \fig{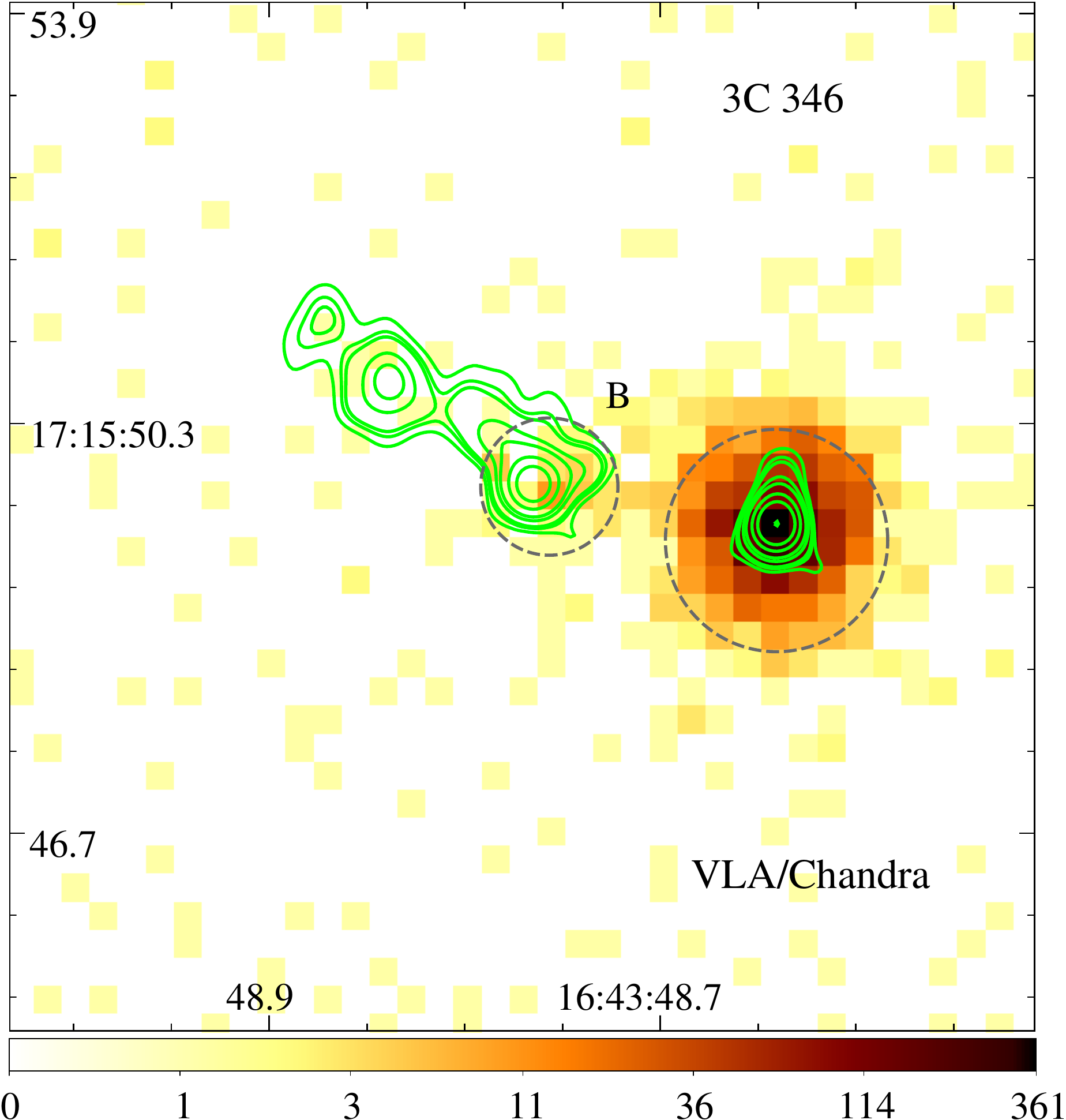}{0.5\textwidth}{(a)}
        \fig{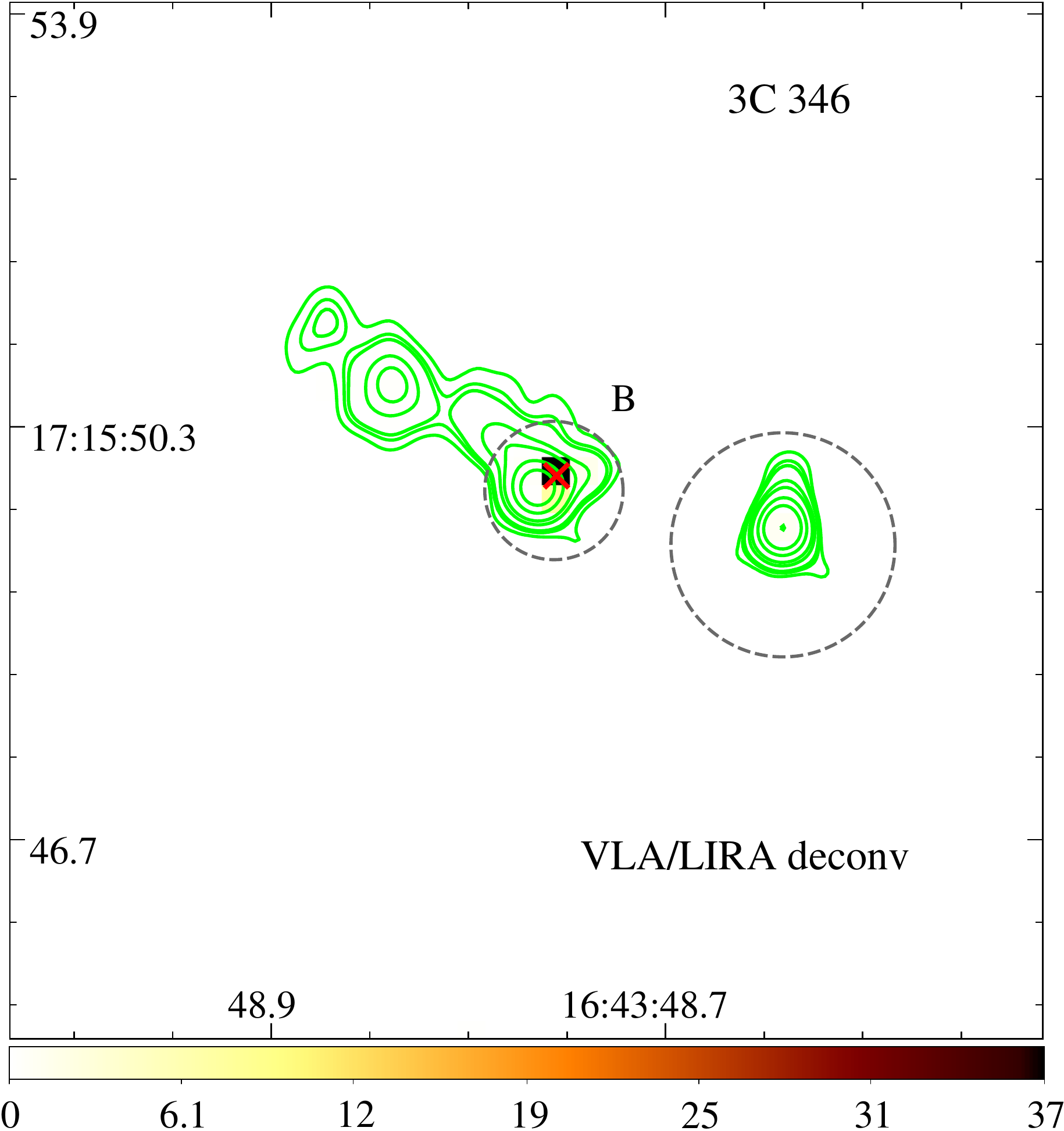}{0.5\textwidth}{(b)}
    }
    \caption{Same as in Fig. \ref{fig:results-3C9} but for 3C 346. The radio contours are given by 5.0, 8.0, 10.0, 20.0, 60.0, 10.0, 30.0, 100.0, 200.0 mJy beam$^{-1}$.\label{fig:results-3C346}}
\end{figure*}

\begin{figure*}[ht]
    \gridline{
        \fig{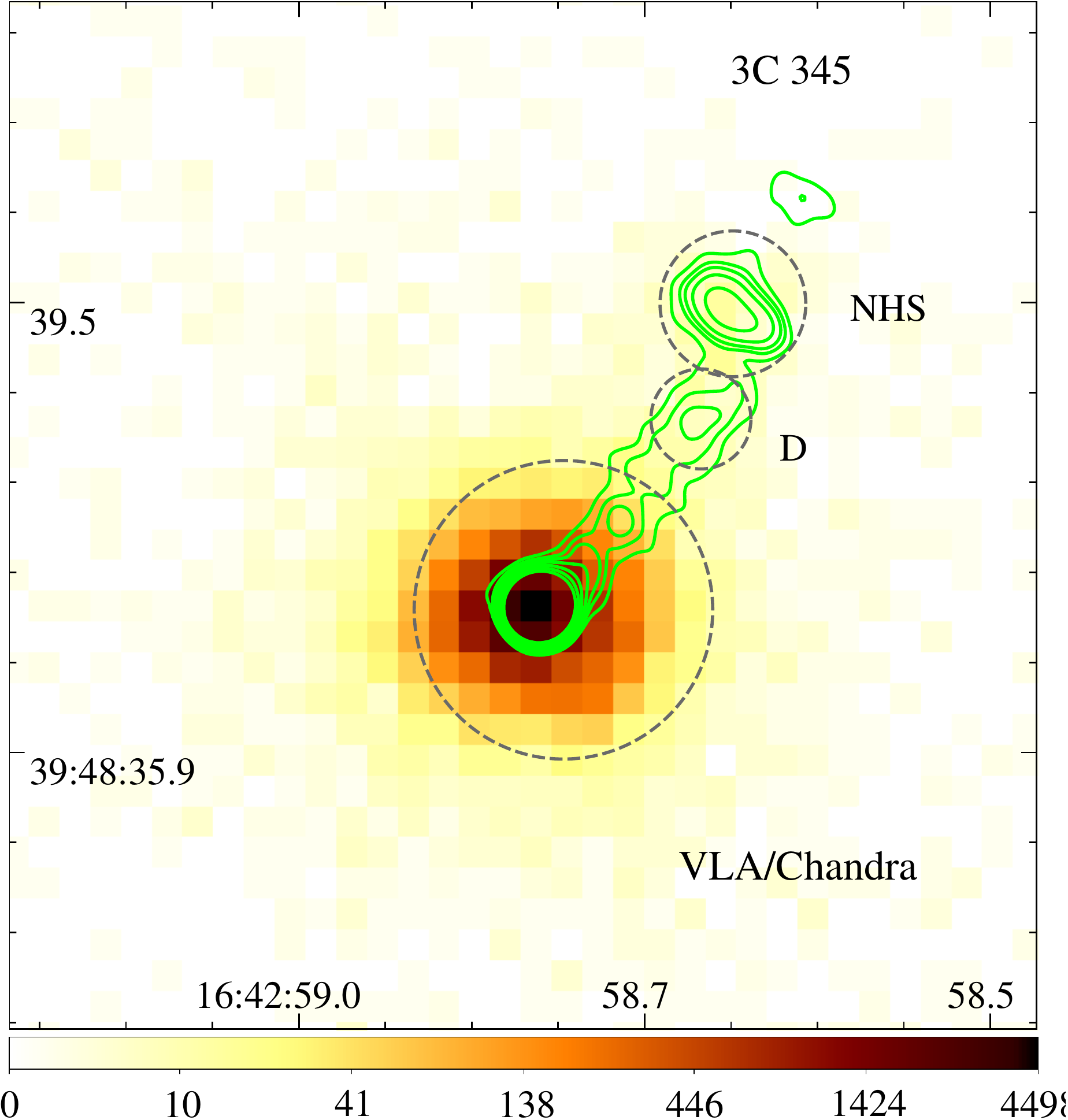}{0.5\textwidth}{(a)}
        \fig{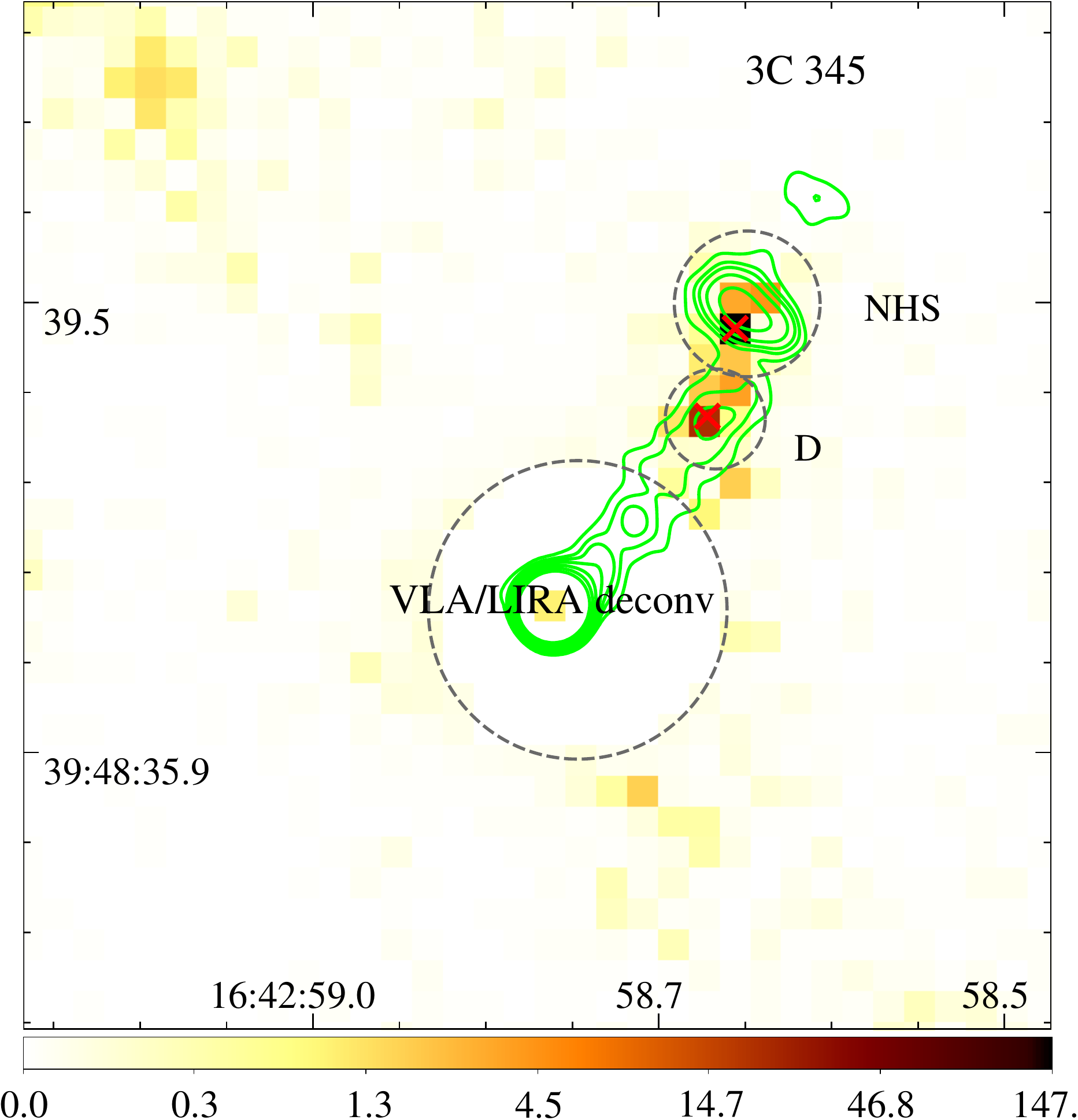}{0.5\textwidth}{(b)}
    }
    \caption{Same as in Fig. \ref{fig:results-3C9} but for 3C 345. The radio contours are given by 2.0, 4.0, 6.0, 10.0, 20.0, 40.0, 80.0 mJy beam$^{-1}$.\label{fig:results-3C345}}
\end{figure*}

\begin{figure*}[ht]
    \gridline{
        \fig{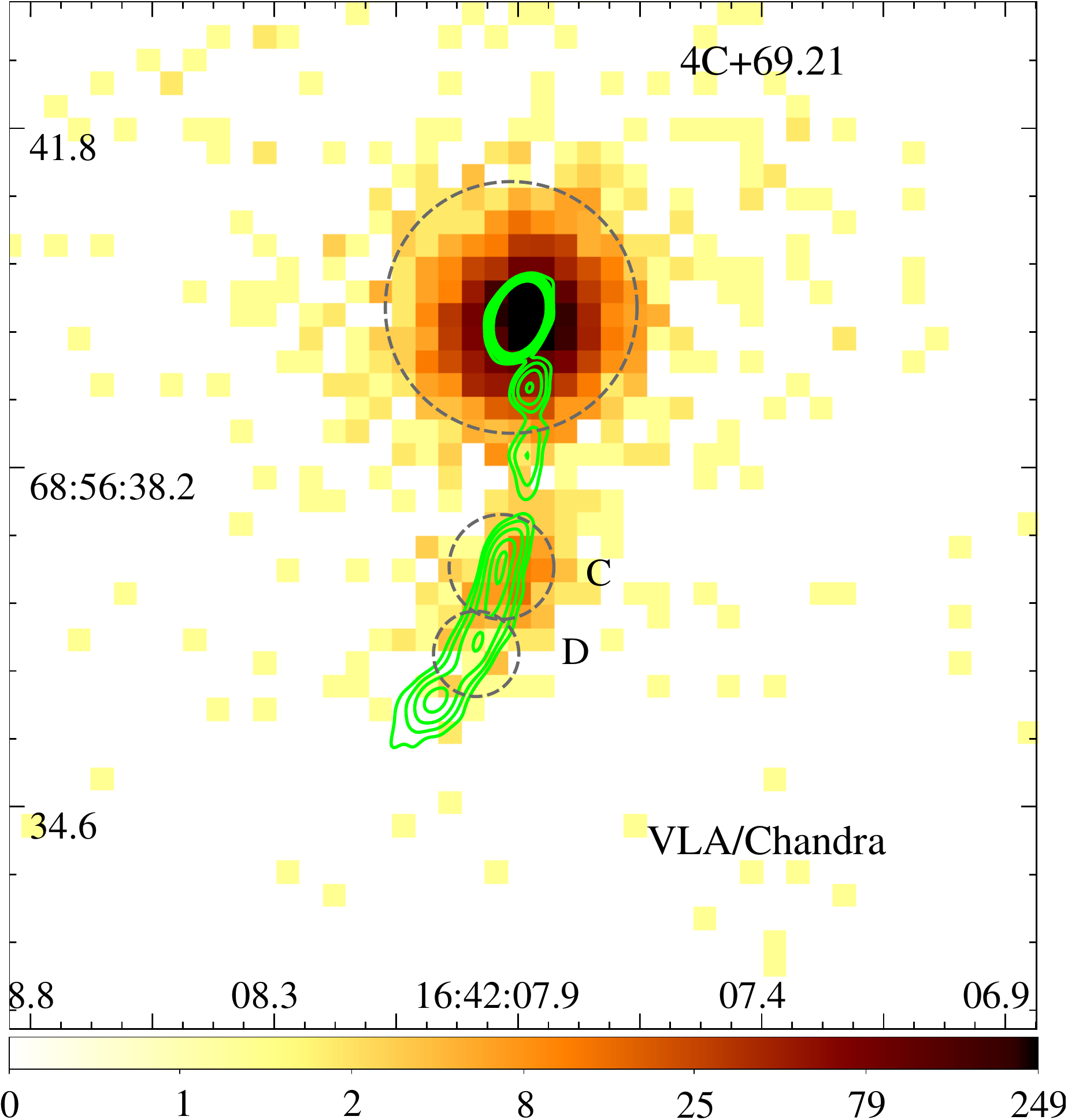}{0.5\textwidth}{(a)}
        \fig{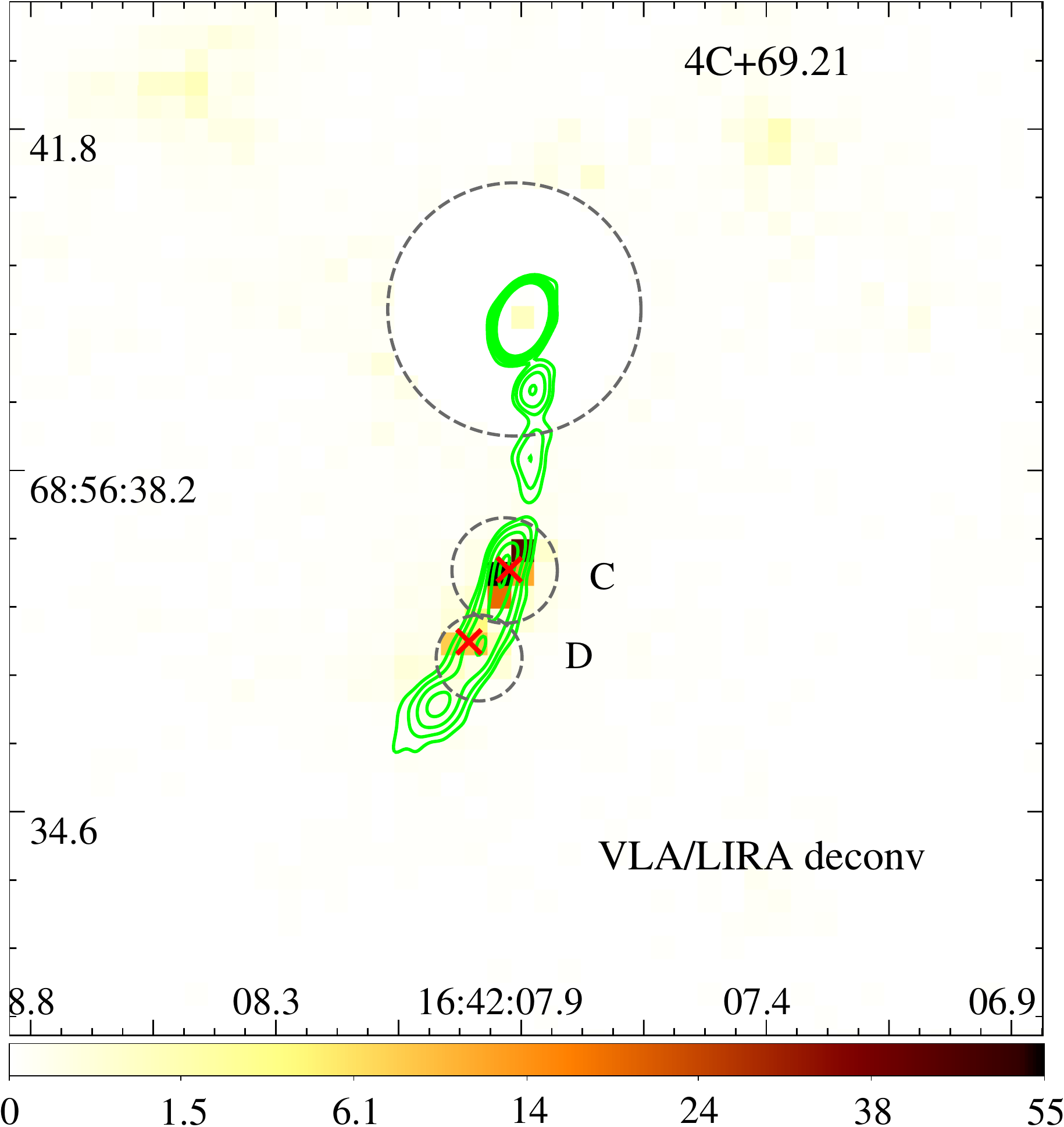}{0.5\textwidth}{(b)}
    }
    \caption{Same as in Fig. \ref{fig:results-3C9} but for 4C +69.21. The radio contours are given by 0.5, 1.0, 2.0, 4.0, 6.0, 8.0 mJy beam$^{-1}$.\label{fig:results-4C+69.21}}
\end{figure*}

\begin{figure*}[ht]
    \gridline{
        \fig{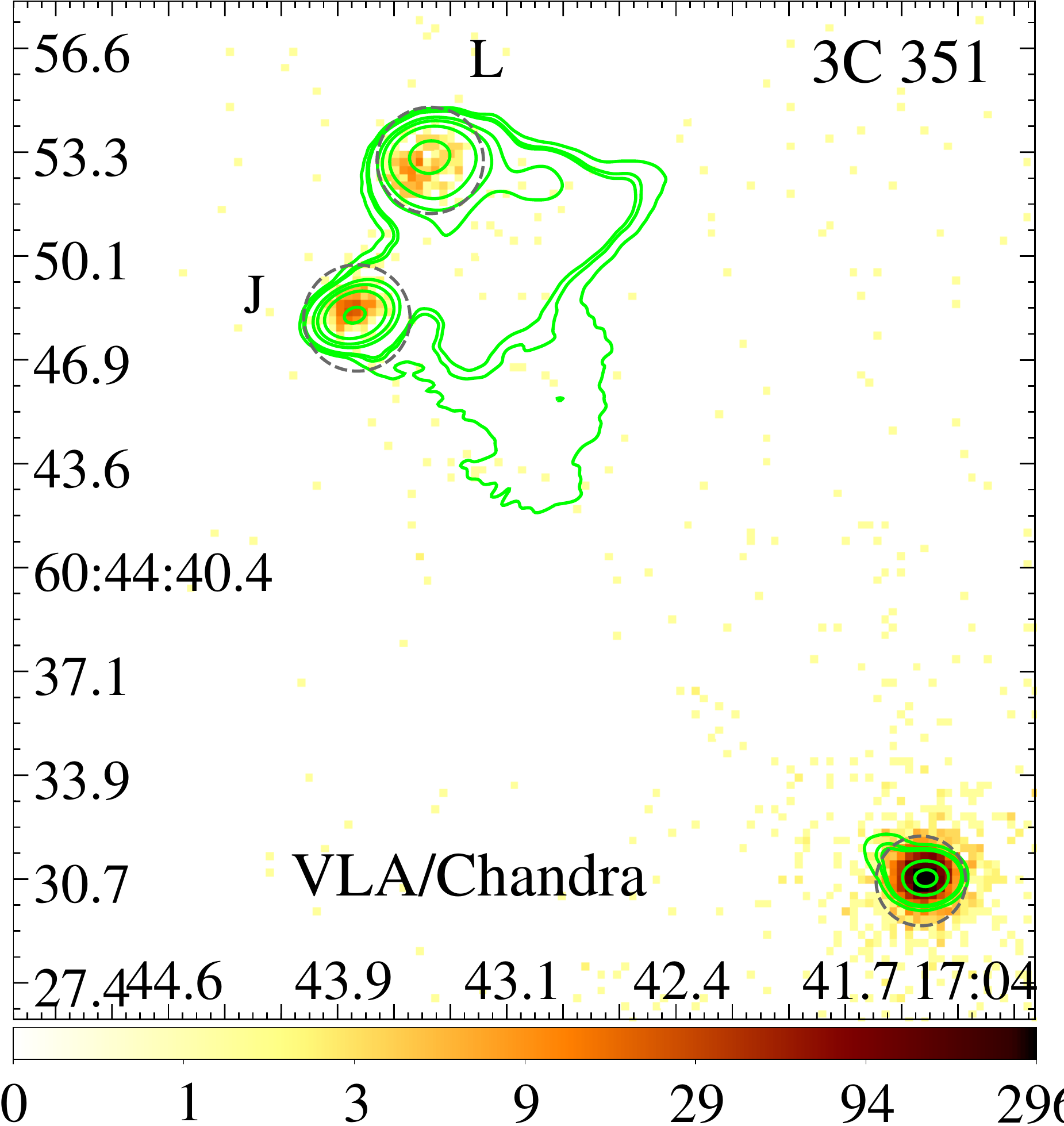}{0.5\textwidth}{(a)}
        \fig{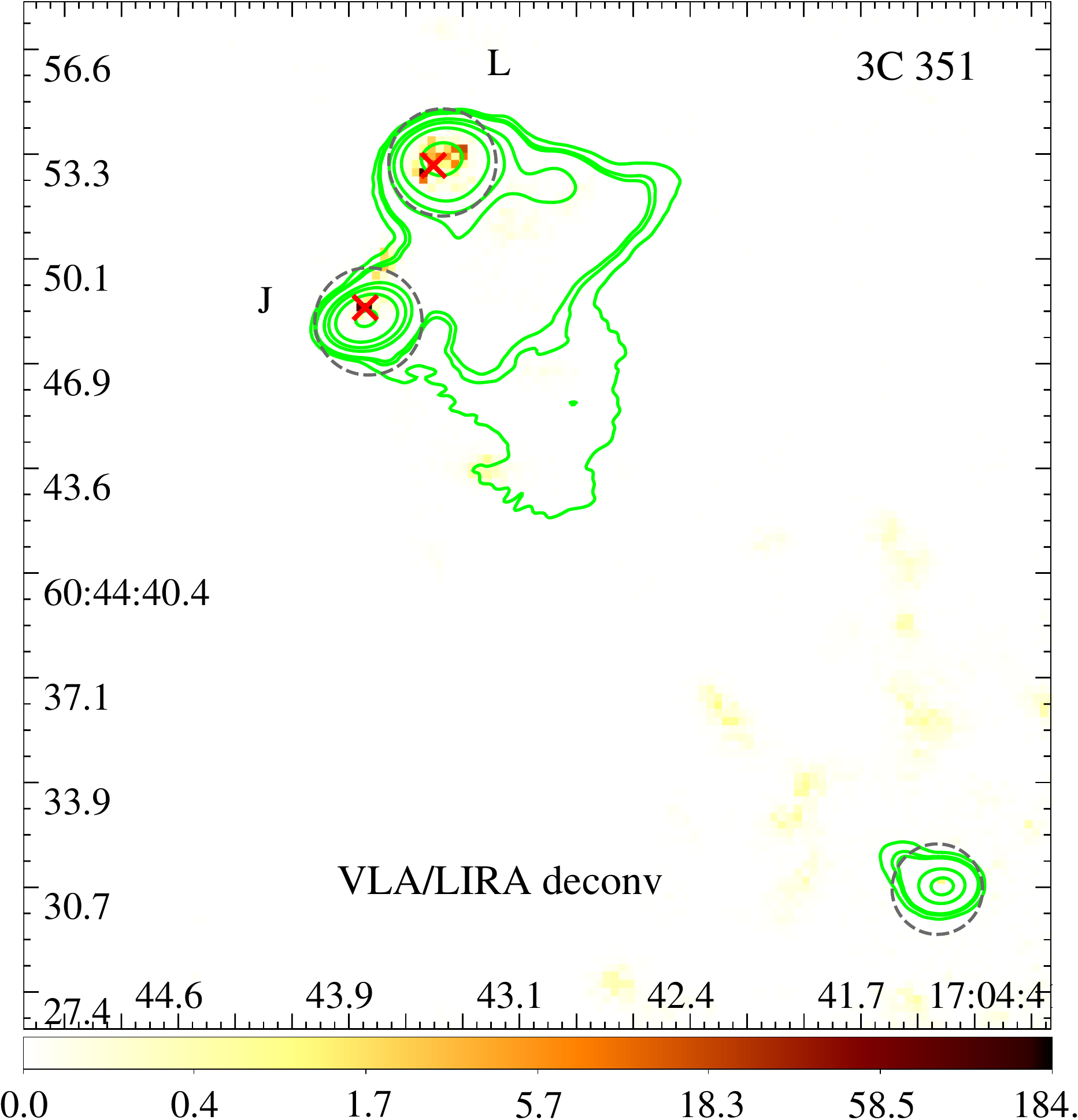}{0.5\textwidth}{(b)}
    }
    \caption{Same as in Fig. \ref{fig:results-3C9} but for 3C 351 . The radio contours are given by 0.4, 0.8, 1.0, 4.0, 8.0, 20.0, 100.0 mJy beam$^{-1}$.\label{fig:results-3C351}}
\end{figure*}

\begin{figure*}[ht]
    \gridline{
        \fig{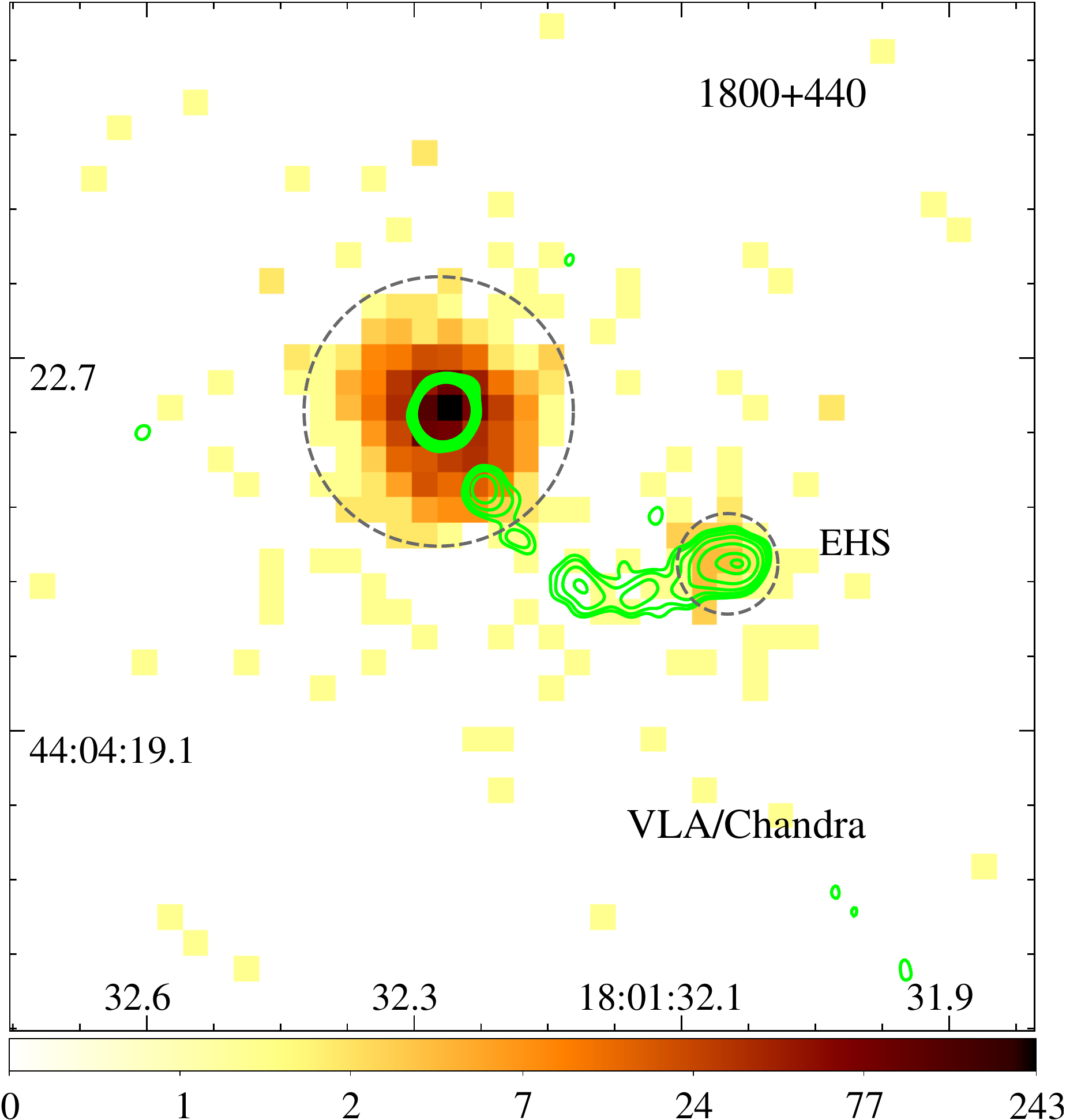}{0.5\textwidth}{(a)}
        \fig{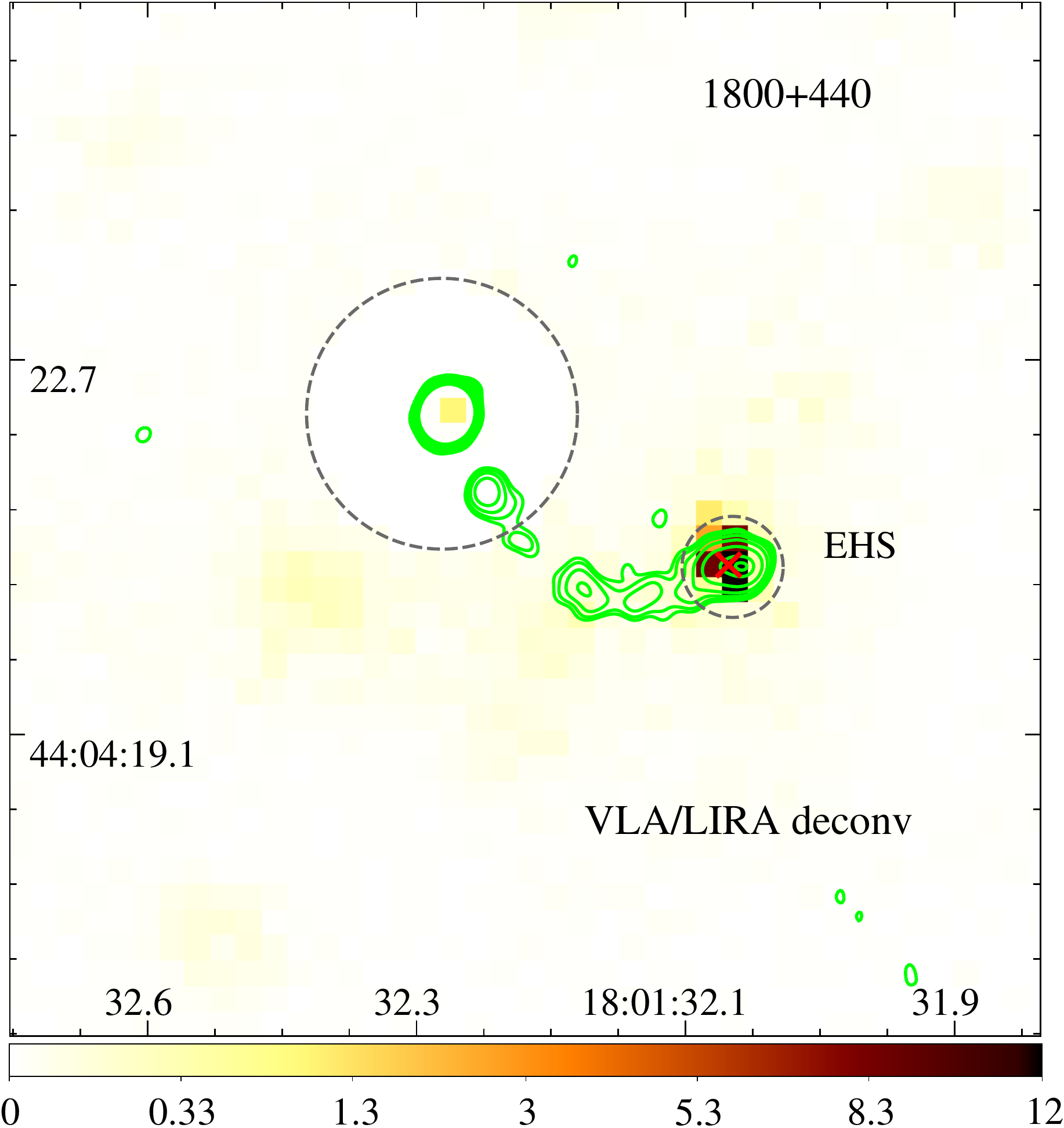}{0.5\textwidth}{(b)}
    }
    \caption{Same as in Fig. \ref{fig:results-3C9} but for 1800+440. The radio contours are given by 0.6, 0.8, 1.2, 2.0, 4.0, 8.0, 10.0 mJy beam$^{-1}$.\label{fig:results-1800+440}}
\end{figure*}

\begin{figure*}[ht]
    \gridline{
        \fig{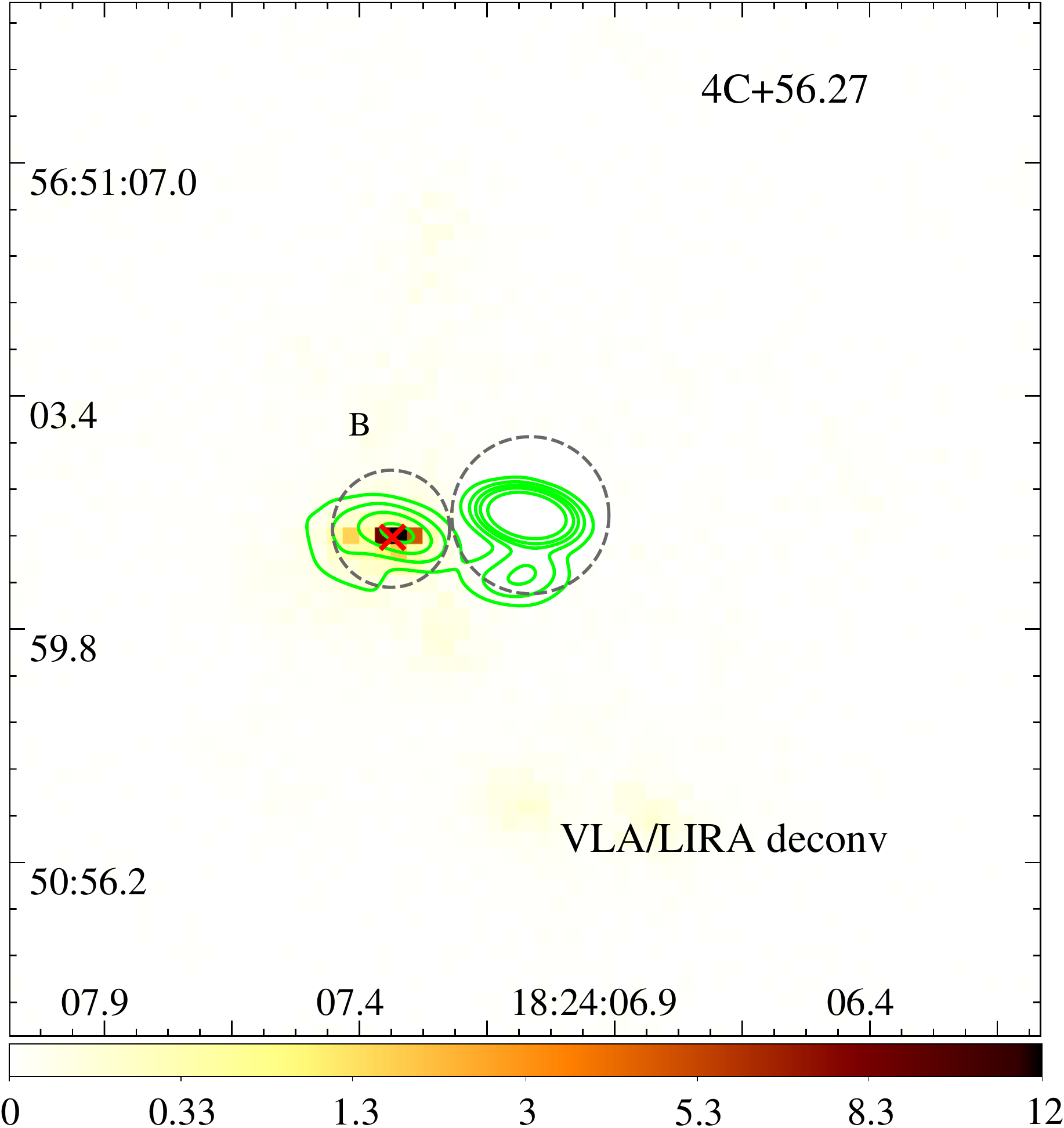}{0.5\textwidth}{(a)}
        \fig{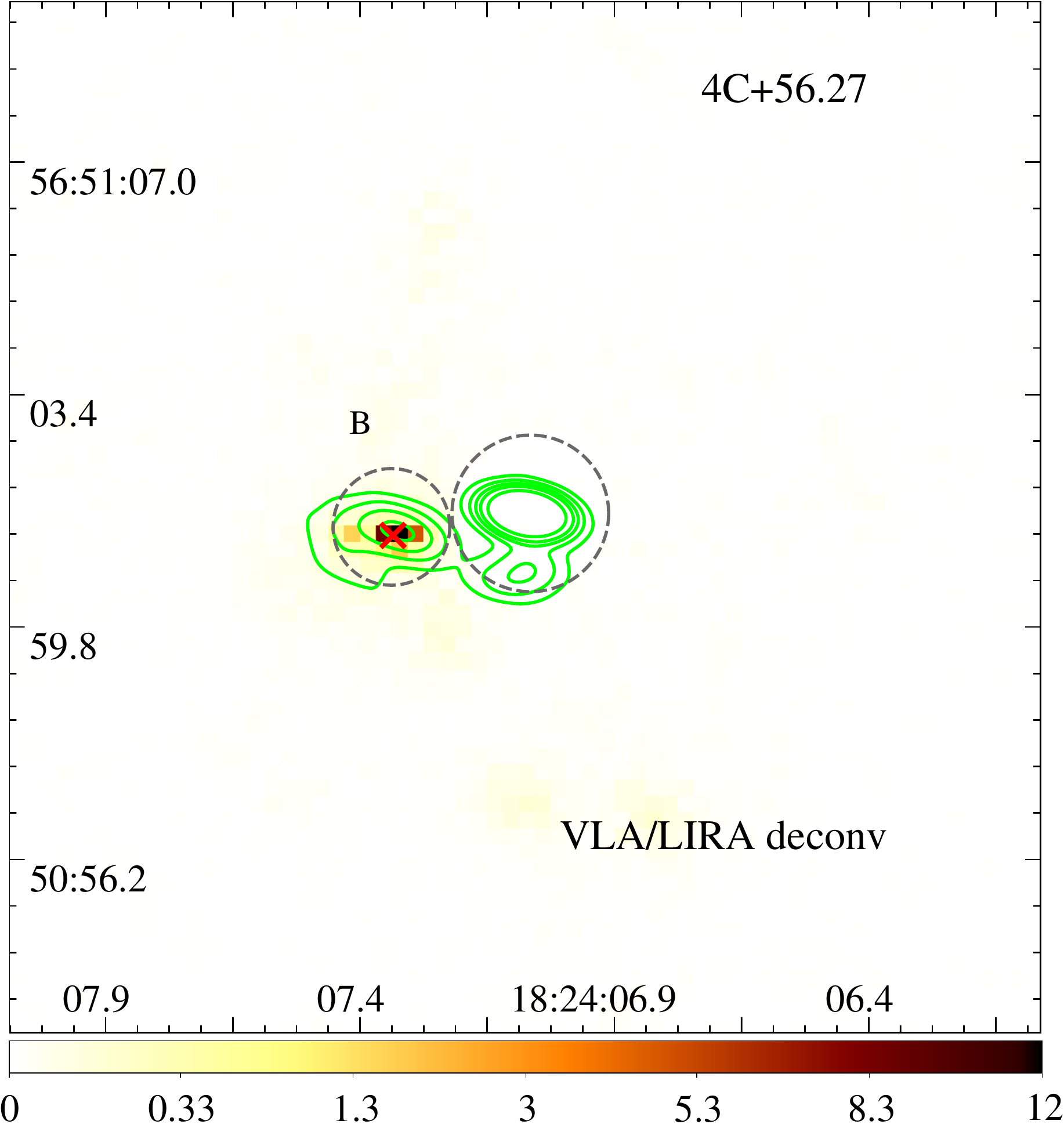}{0.5\textwidth}{(b)}
    }
    \caption{Same as in Fig. \ref{fig:results-3C9} but for 4C +56.27. The radio contours are given by 2.0, 8.0, 20.0, 40.0, 1000.0 mJy beam$^{-1}$.\label{fig:results-4C+56.27}}
\end{figure*}

\begin{figure*}[ht]
    \gridline{
        \fig{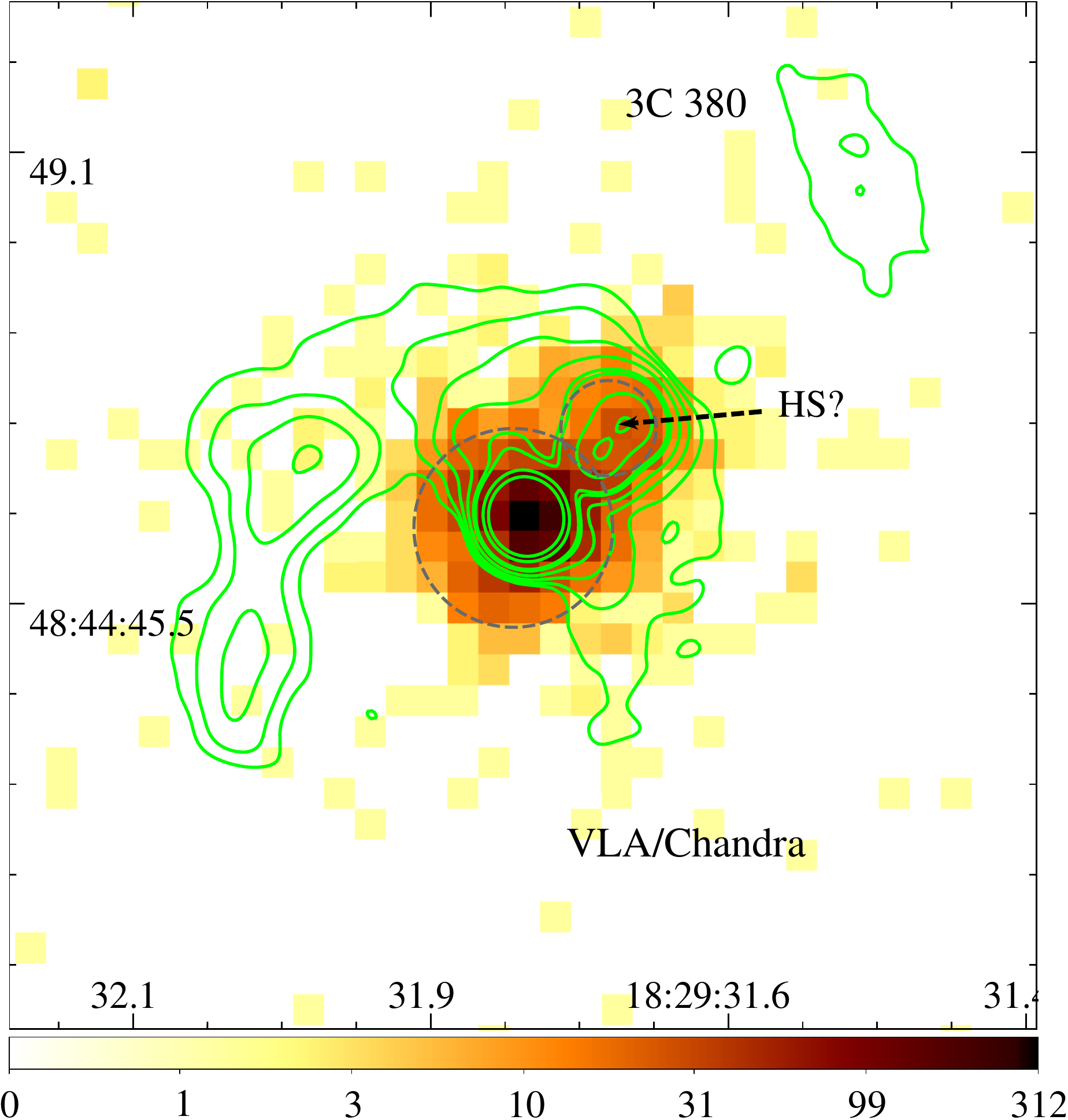}{0.5\textwidth}{(a)}
        \fig{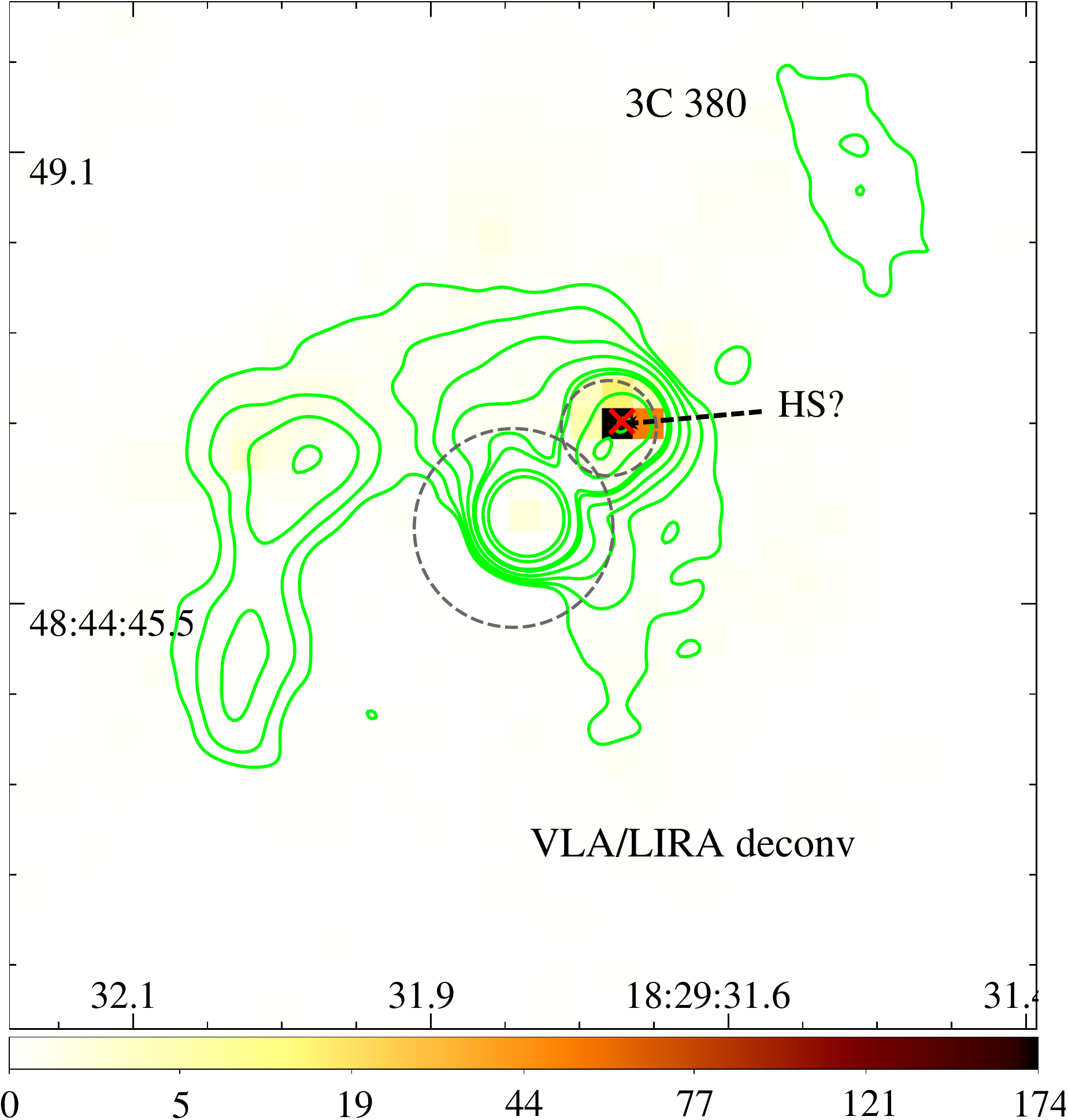}{0.5\textwidth}{(b)}
    }
    \caption{Same as in Fig. \ref{fig:results-3C9} but for 3C 380. The radio contours are given by 5.0, 10.0, 20.0, 40.0, 80.0, 100.0, 250.0, 425.0 mJy beam$^{-1}$.\label{fig:results-3C380}}
\end{figure*}

\begin{figure*}[ht]
    \gridline{
        \fig{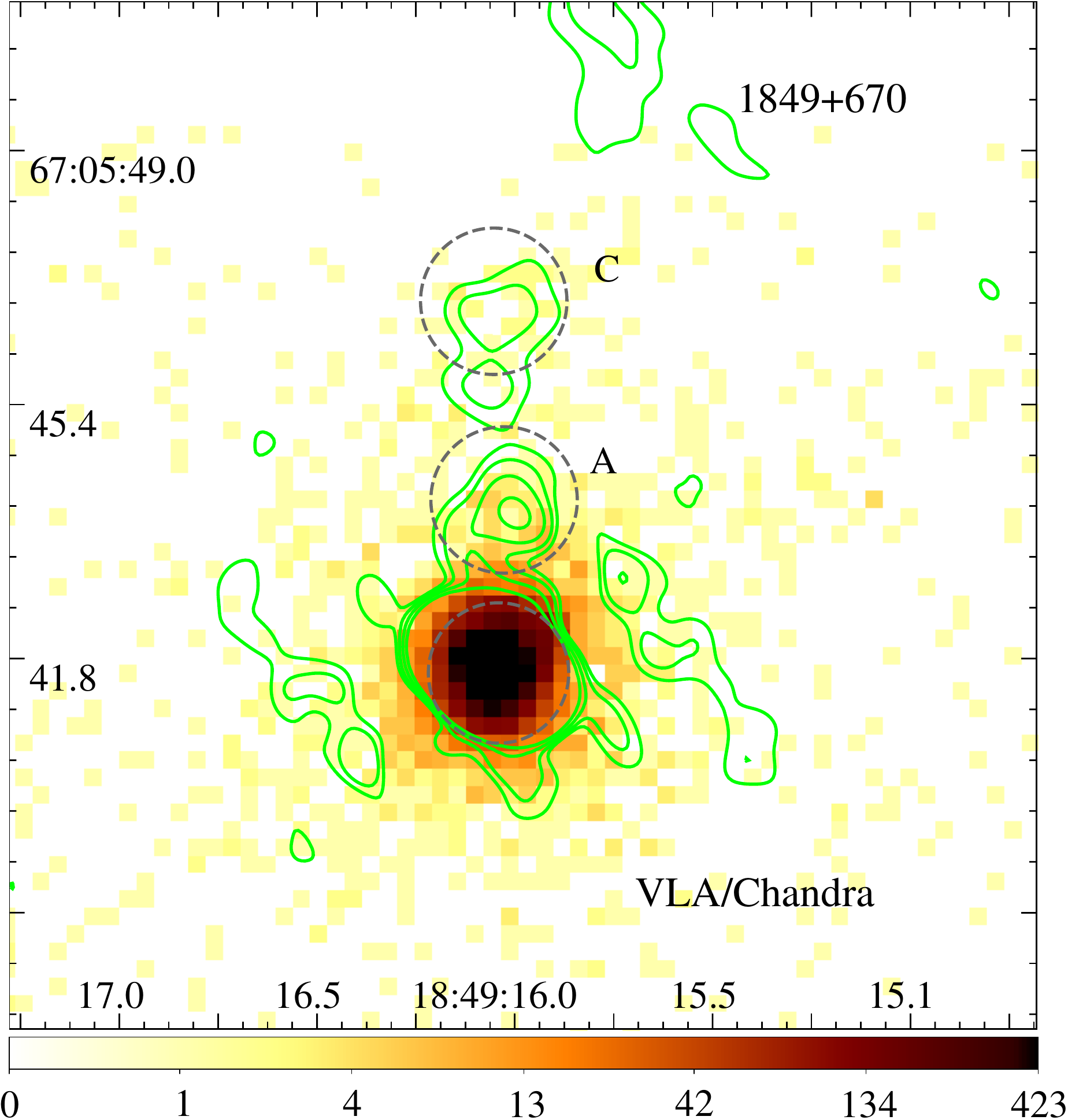}{0.5\textwidth}{(a)}
        \fig{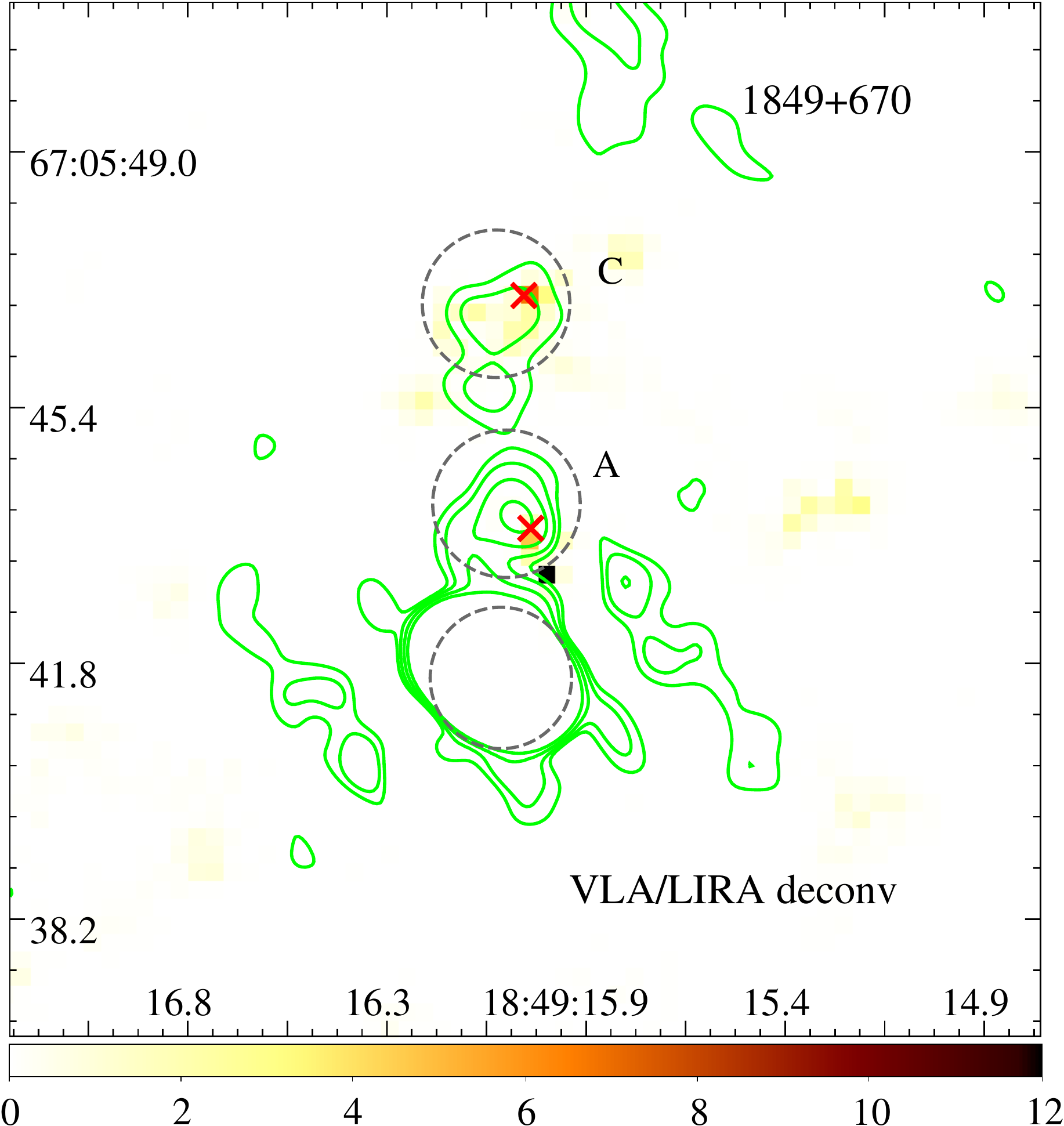}{0.5\textwidth}{(b)}
    }
    \caption{Same as in Fig. \ref{fig:results-3C9} but for 1849+670. The radio contours are given by 0.05, 0.1, 0.2, 0.4 mJy beam$^{-1}$.\label{fig:results-1849+670}}
\end{figure*}

\begin{figure*}[ht]
    \gridline{
        \fig{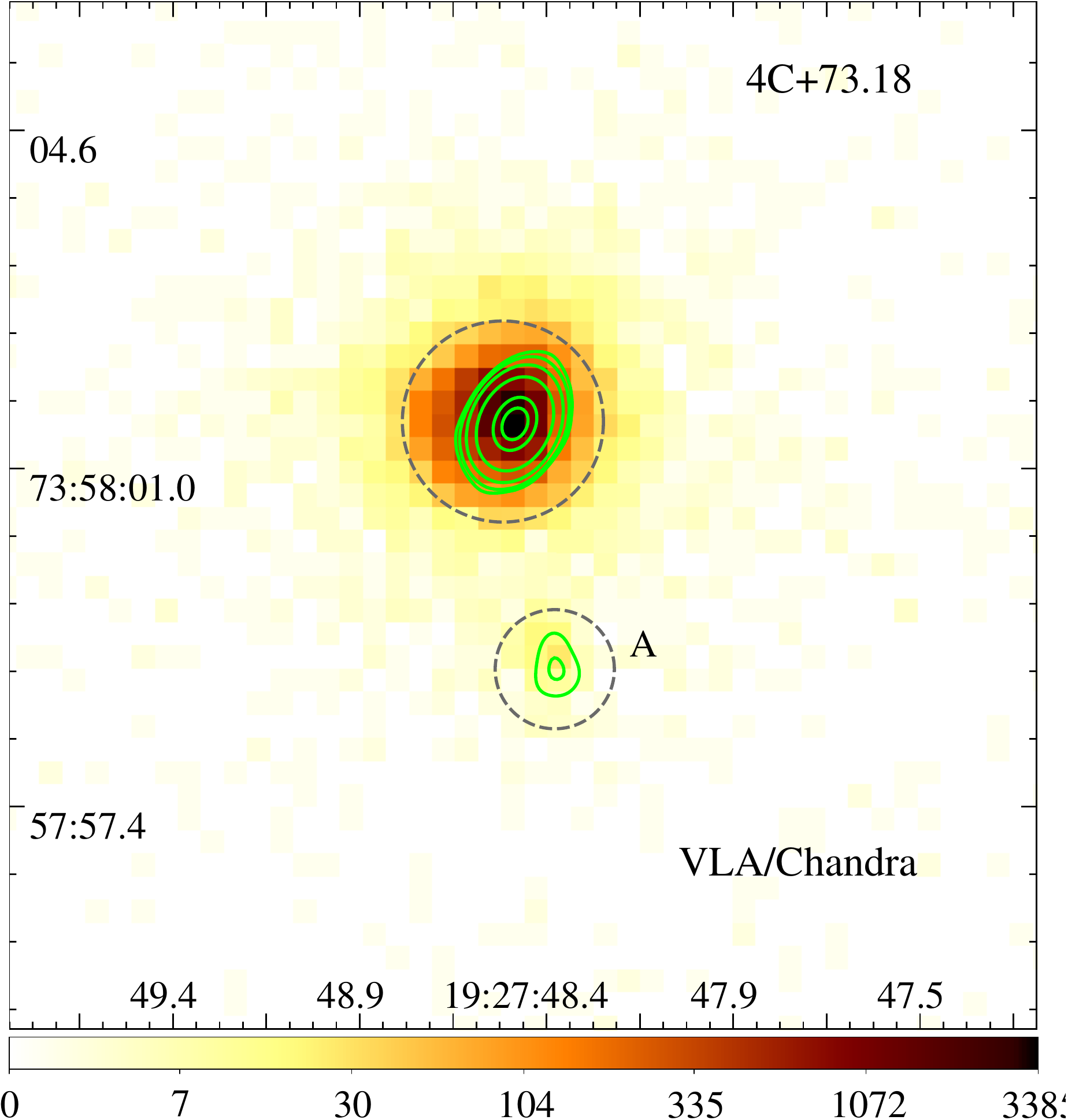}{0.5\textwidth}{(a)}
        \fig{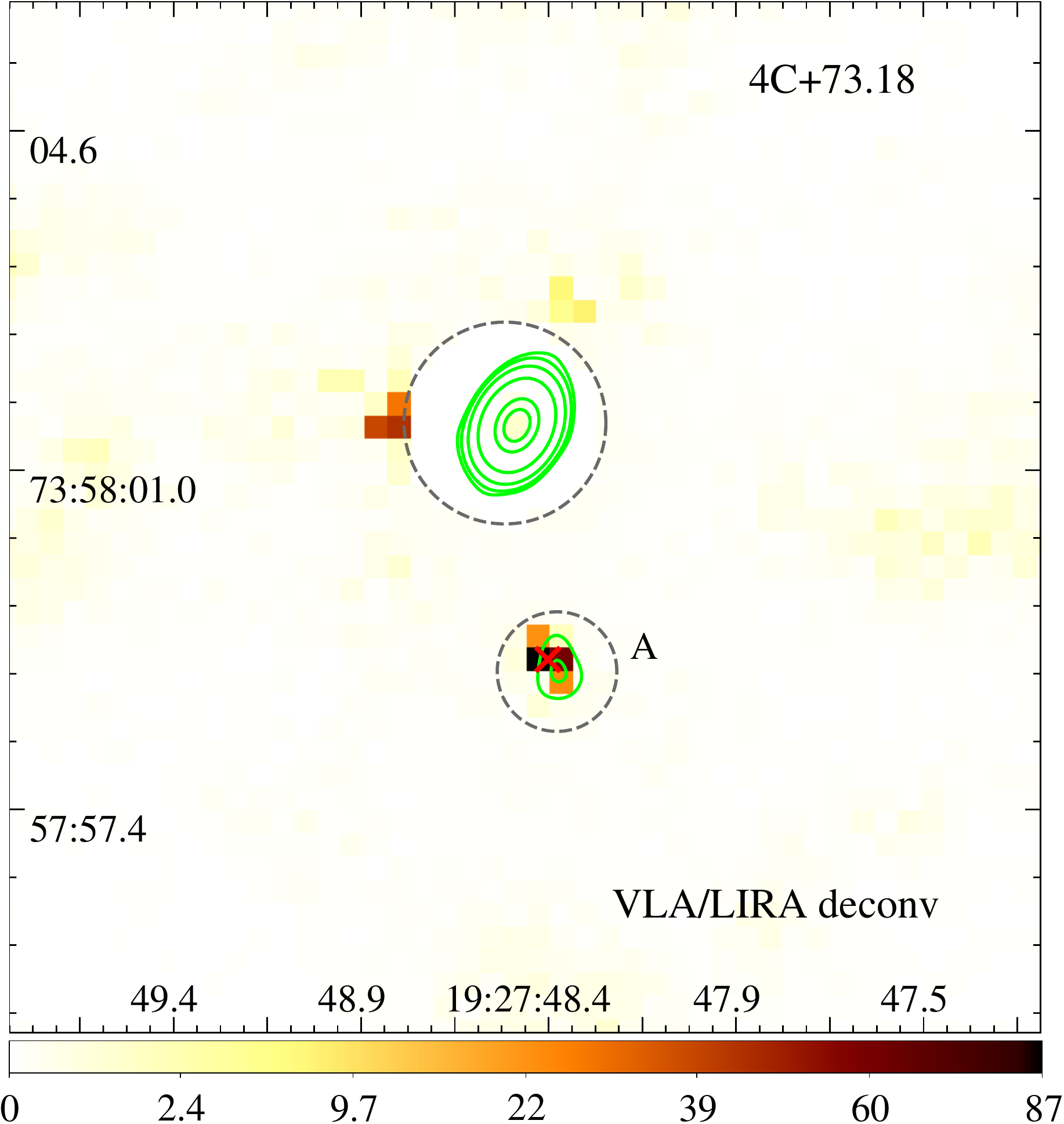}{0.5\textwidth}{(b)}
    }
    \caption{Same as in Fig. \ref{fig:results-3C9} but for 4C +73.18. The radio contours are given by 2.0, 4.0, 15.0, 100.0, 1000.0, 2000.0 mJy beam$^{-1}$.\label{fig:results-4C+73.18}}
\end{figure*}

\begin{figure*}[ht]
    \gridline{
        \fig{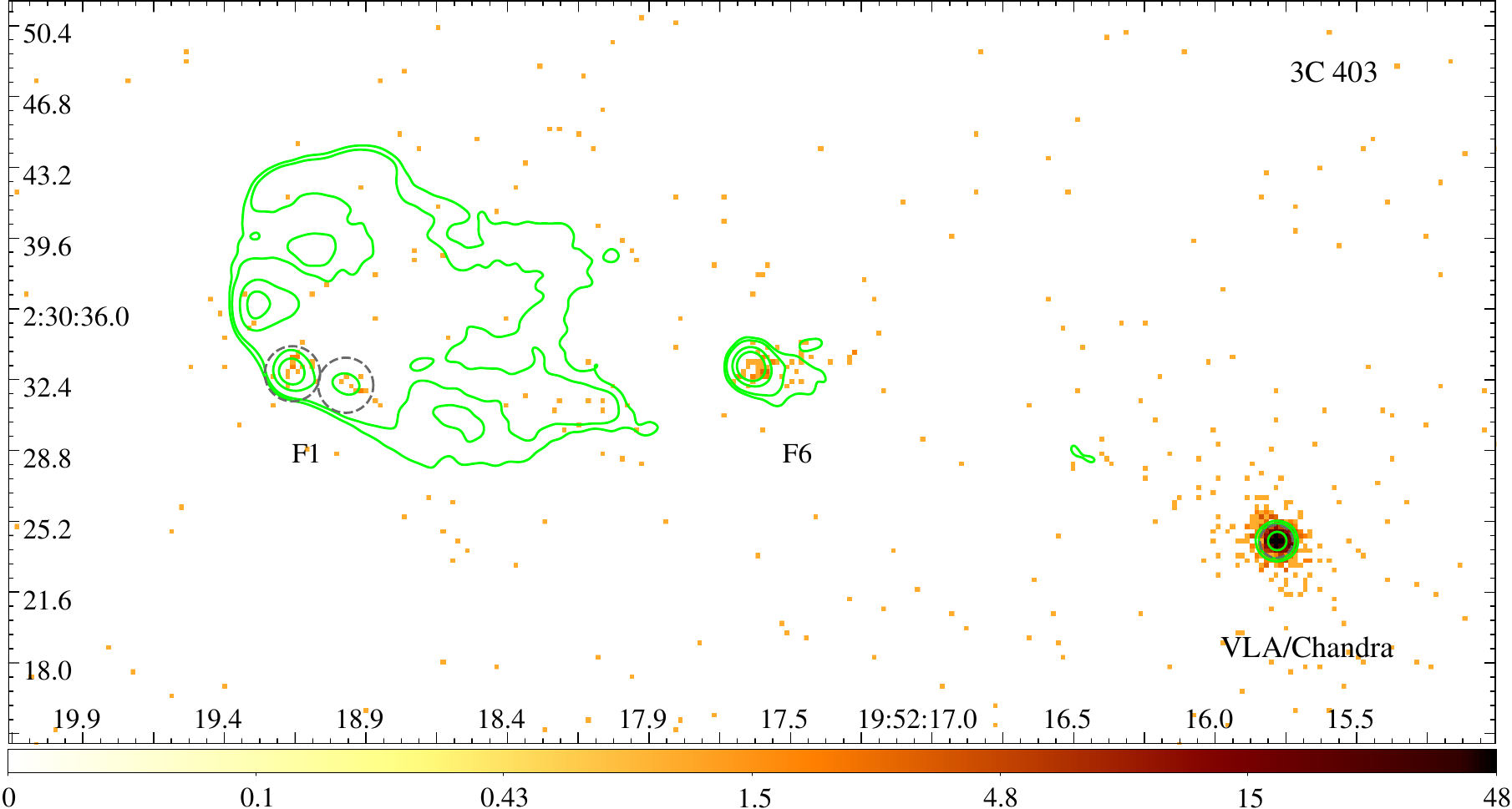}{\textwidth}{(a)}
    }
    \gridline{
        \fig{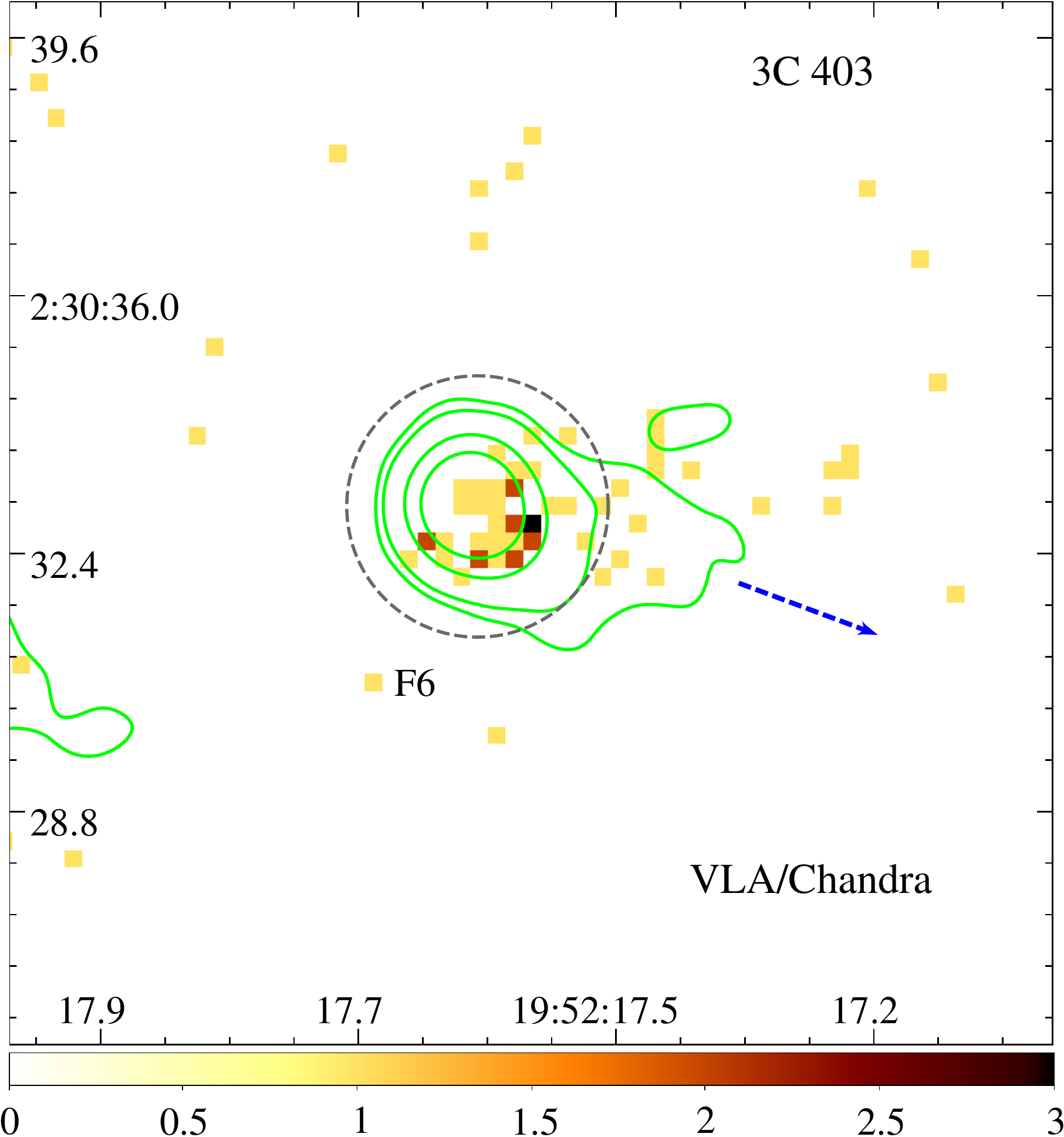}{0.3\textwidth}{(b)}
        \fig{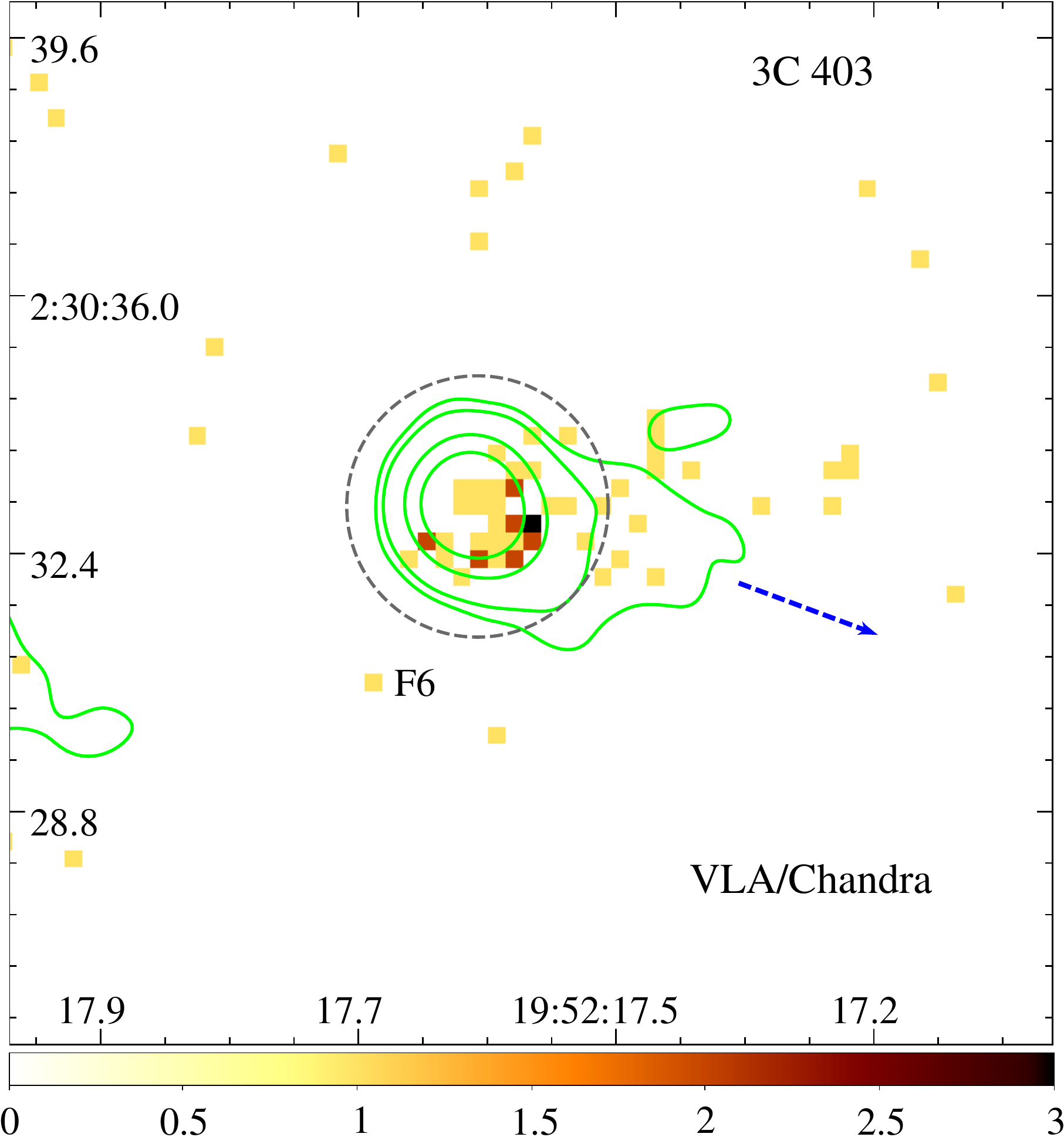}{0.3\textwidth}{(c)}
    }
    \gridline{
        \fig{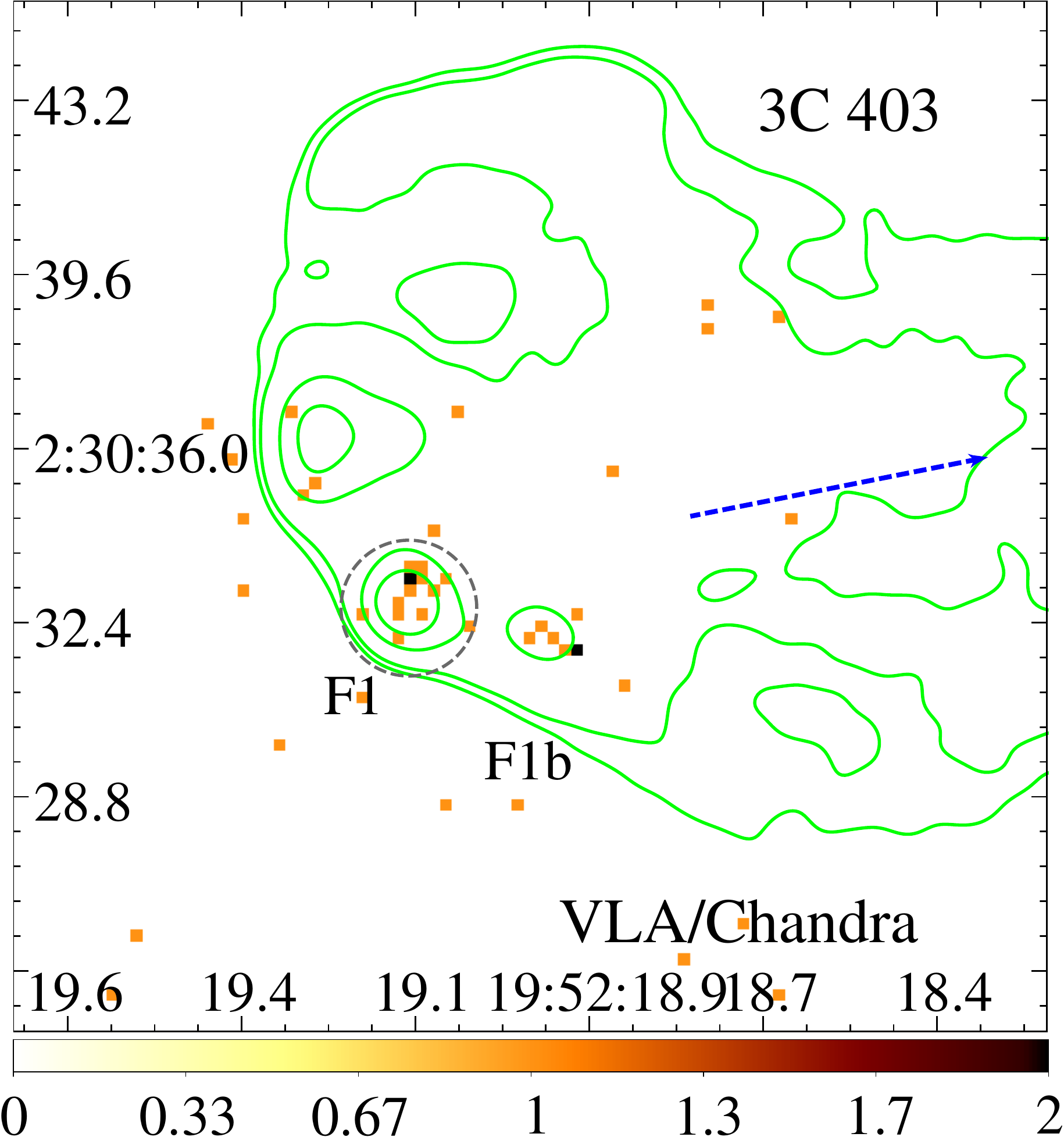}{0.3\textwidth}{(d)}
        \fig{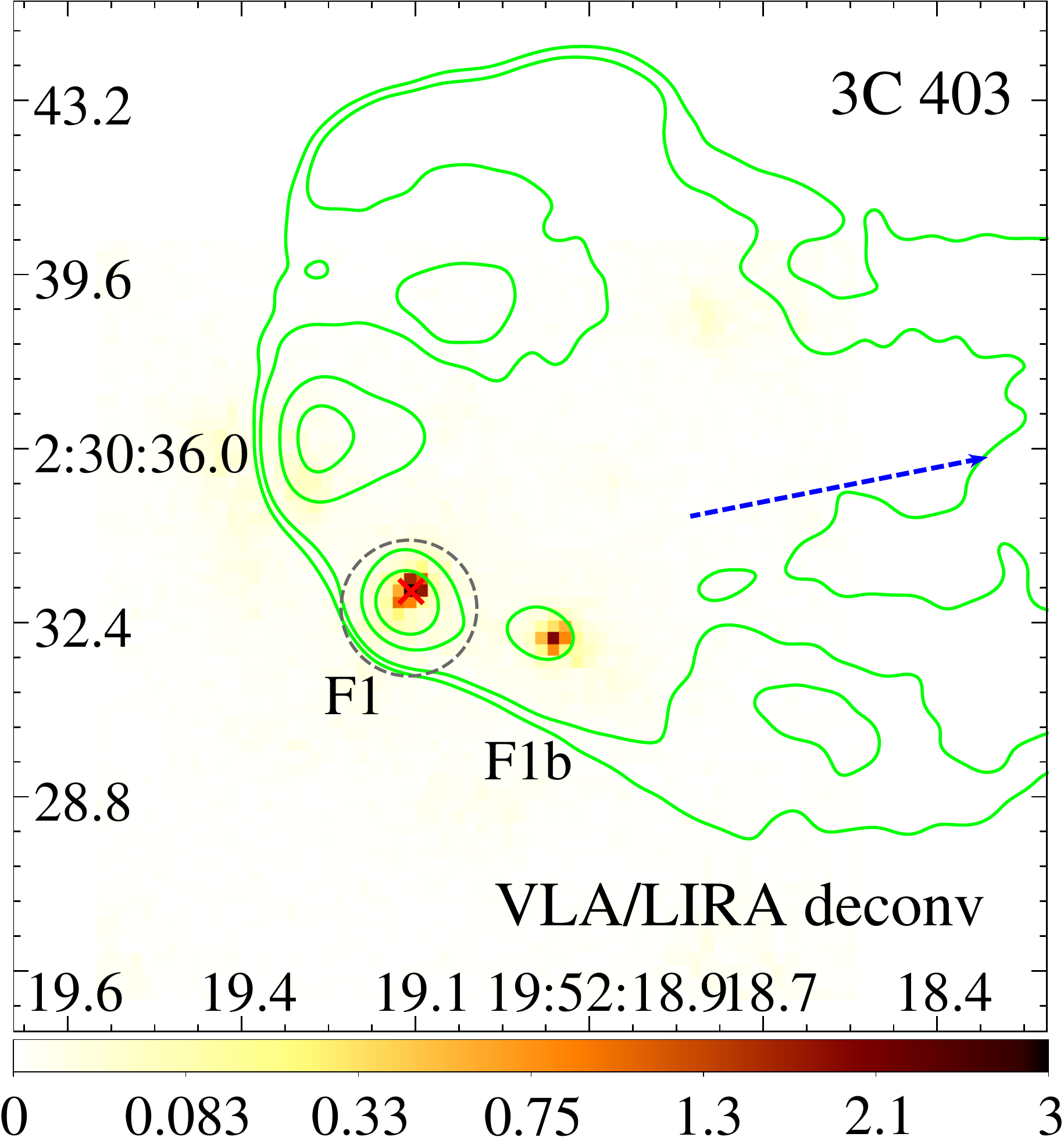}{0.3\textwidth}{(e)}
    }
    \caption{Same as in Fig. \ref{fig:results-3C9} but for 3C 403. (a) shows the full image while (b), (c) and (d),(e) show the zoomed-in regions around knot F6 and the western hotspot, respectively. The dashed-blue arrow in panels (d) and (e) indicate the tentative direction of the jet from knot F6.
        The radio contours are given by 0.4, 0.8, 4.0, 10.0 mJy beam$^{-1}$.\label{fig:results-3C403}}
\end{figure*}

\begin{figure*}[ht]
    \gridline{
        \fig{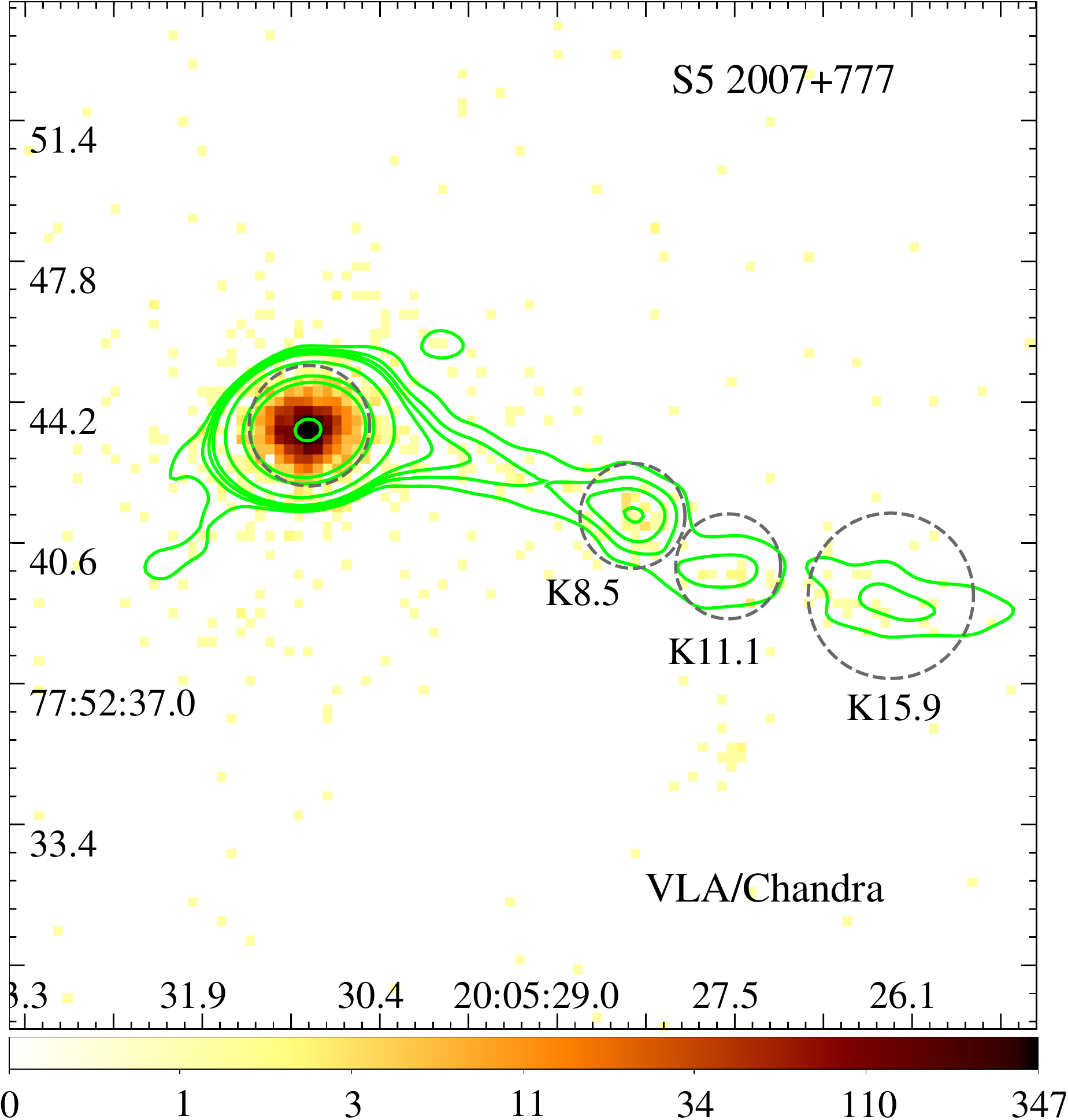}{0.5\textwidth}{(a)}
        \fig{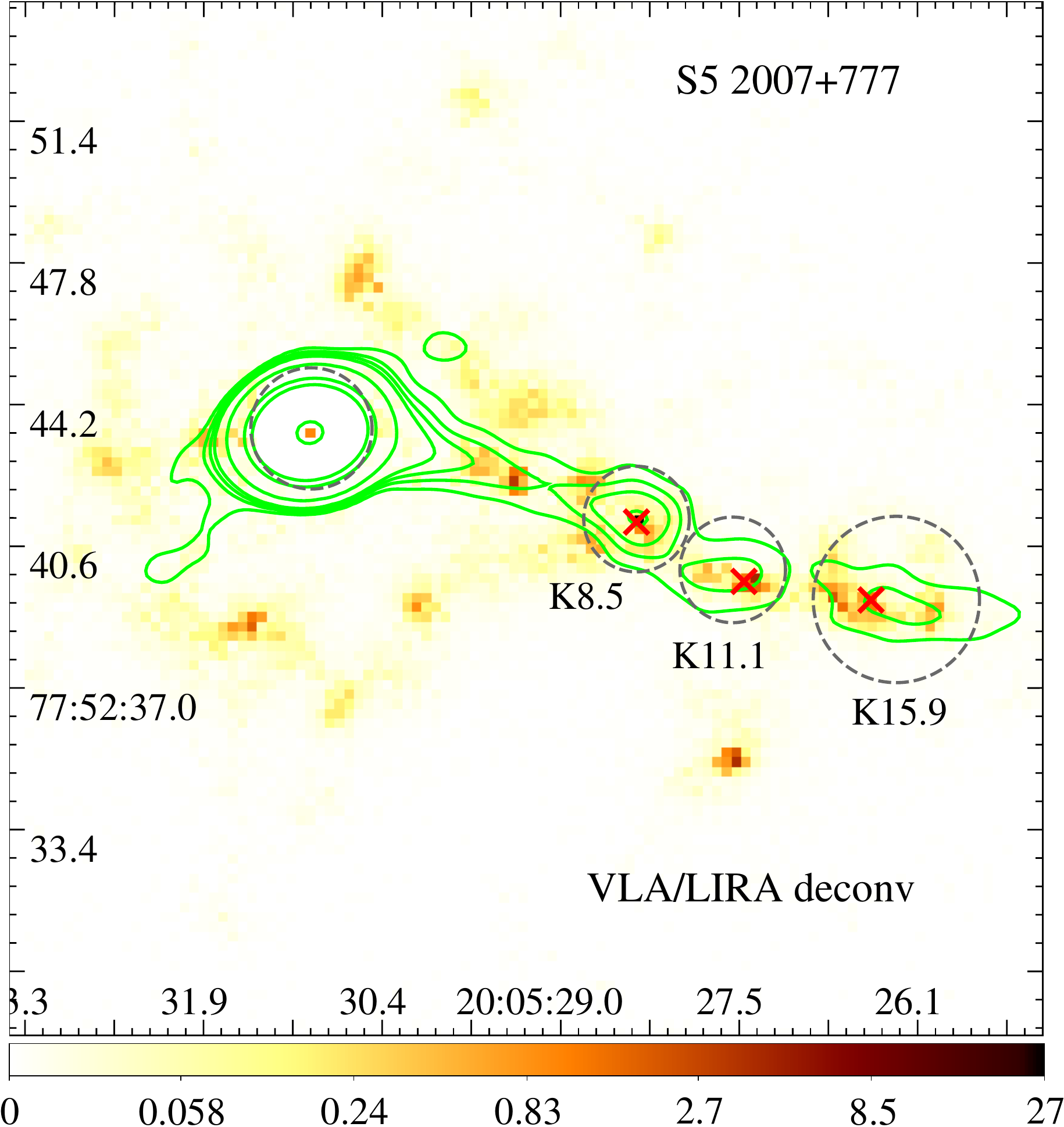}{0.5\textwidth}{(b)}
    }
    \caption{Same as in Fig. \ref{fig:results-3C9} but for S5 2007+777. The radio contours are given by 0.5, 1.0, 1.5, 2.2, 8.0, 50.0, 100.0, 1000.0 mJy beam$^{-1}$.\label{fig:results-S52007+777}}
\end{figure*}

\begin{figure*}[ht]
    \gridline{
        \fig{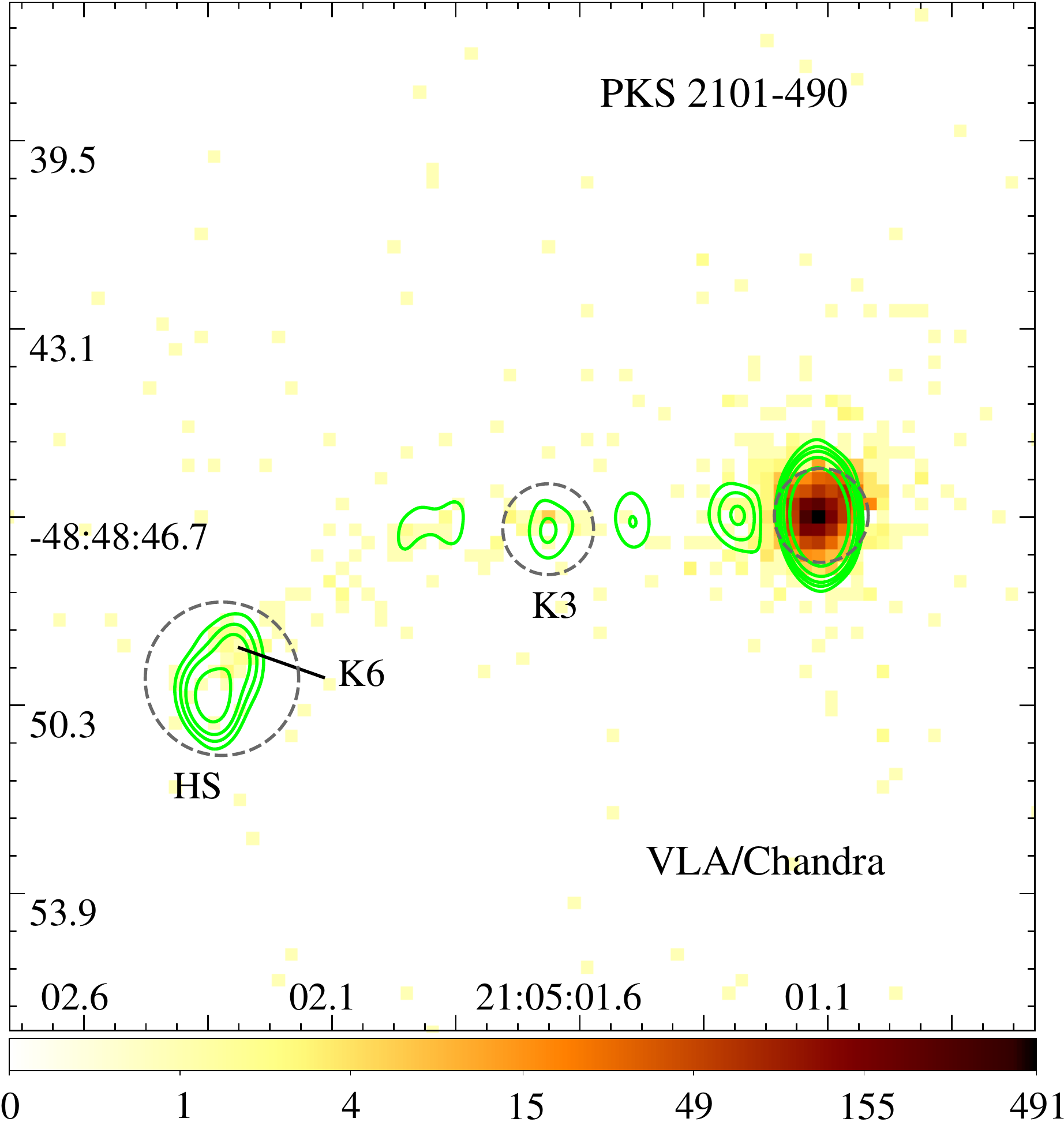}{0.5\textwidth}{(a)}
        \fig{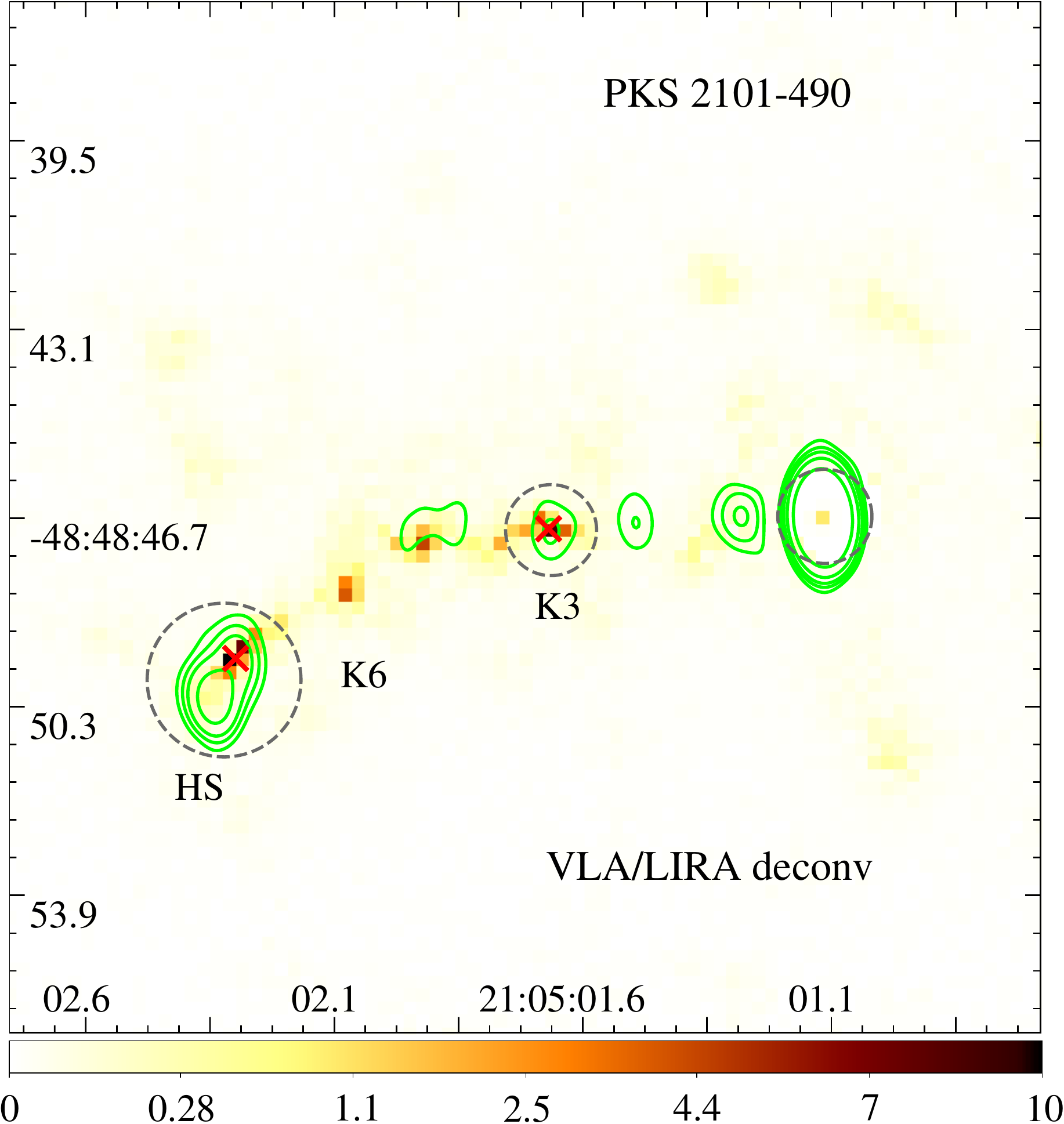}{0.5\textwidth}{(b)}
    }
    \caption{Same as in Fig. \ref{fig:results-3C9} but for PKS 2101-490. The radio contours are given by 1.0, 2.0, 3.5, 8.0, 30.0 mJy beam$^{-1}$.\label{fig:results-PKS2101-490}}
\end{figure*}

\begin{figure*}[ht]
    \gridline{
        \fig{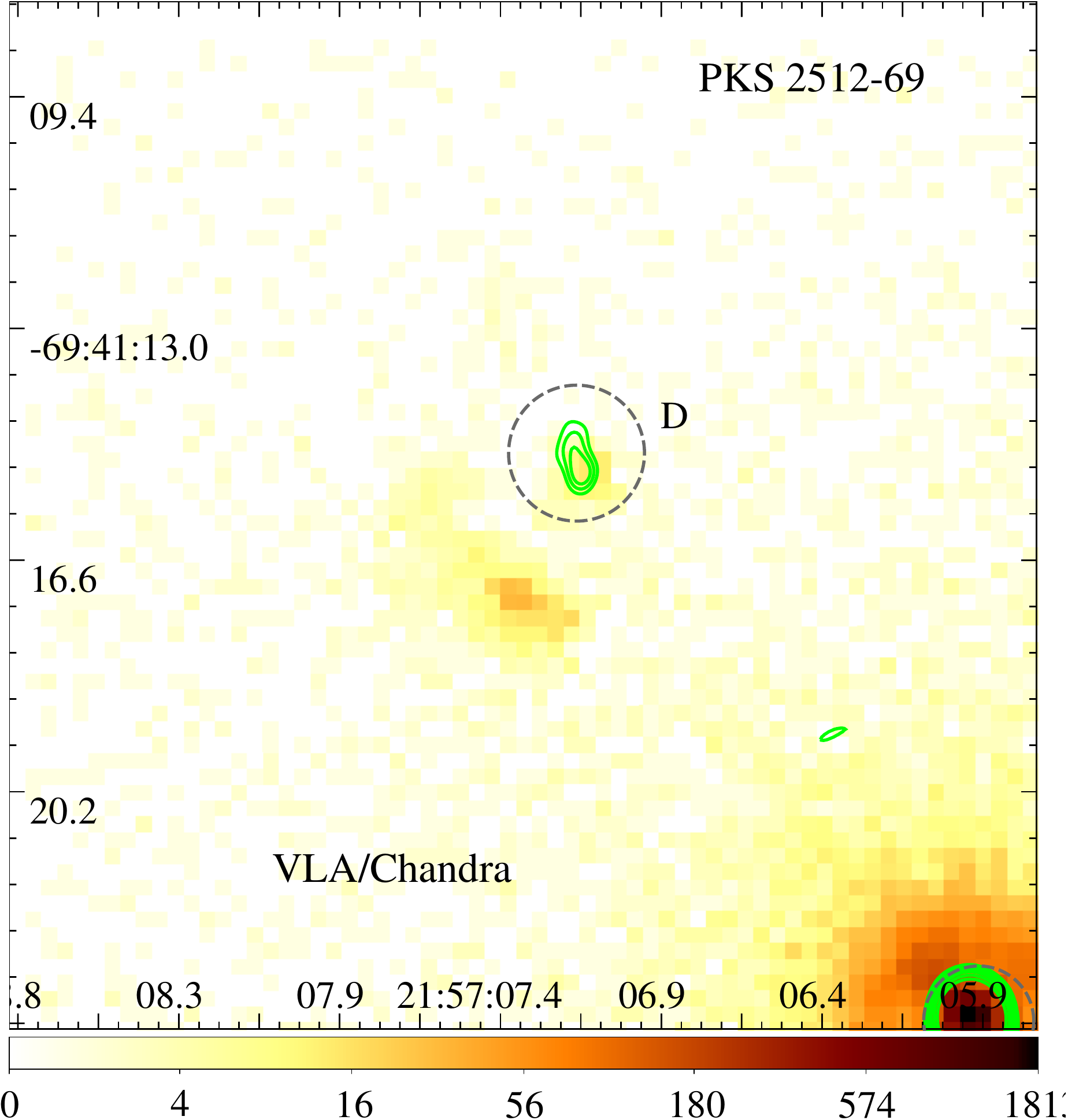}{0.5\textwidth}{(a)}
        \fig{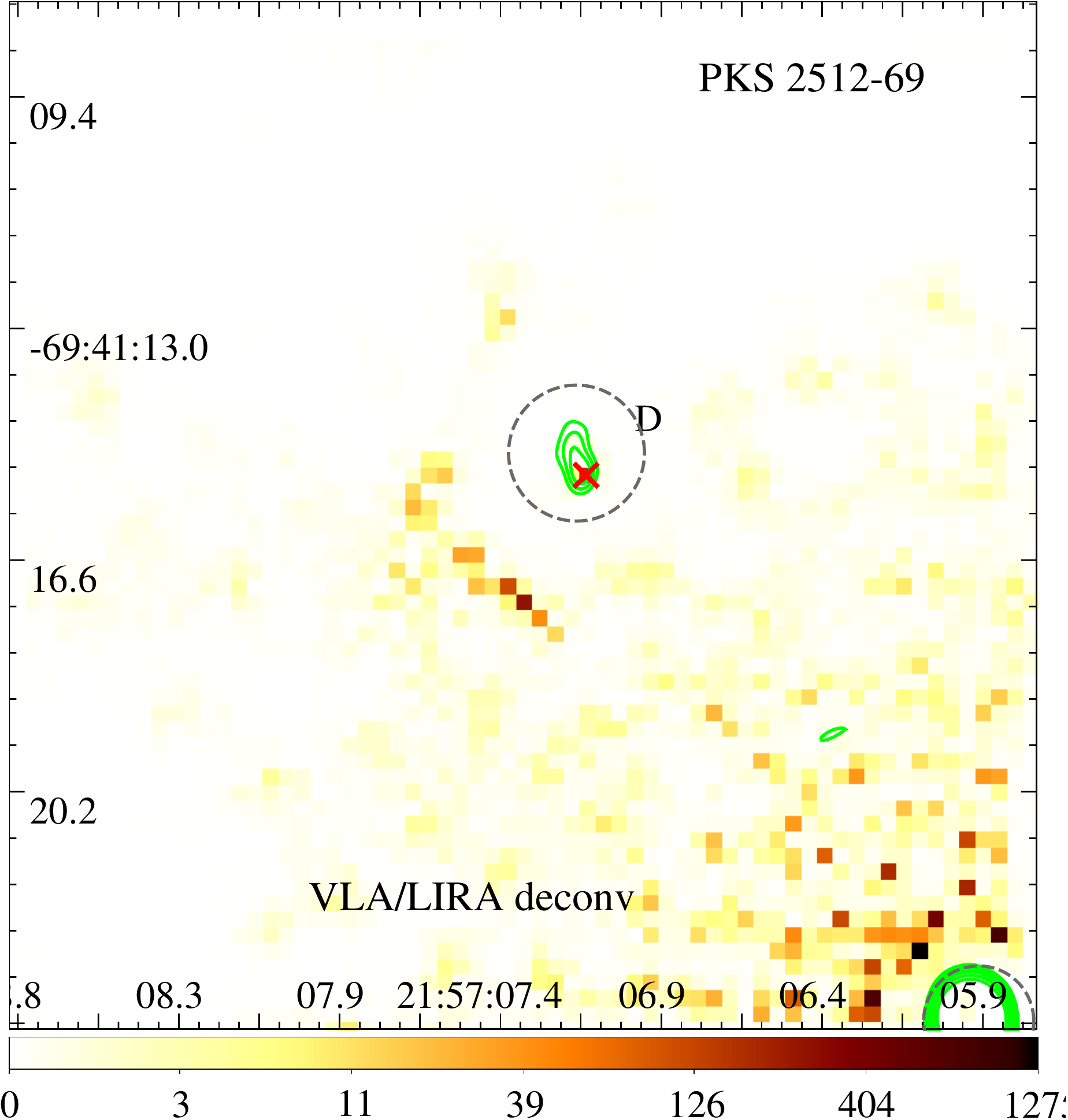}{0.5\textwidth}{(b)}
    }
    \caption{Same as in Fig. \ref{fig:results-3C9} but for PKS 2152-69. The radio contours are given by 2.0, 3.0, 4.0, 8.0, 10.0, 20.0, 40.0 mJy beam$^{-1}$.\label{fig:results-PKS2152-69}}
\end{figure*}

\begin{figure*}[ht]
    \gridline{
        \fig{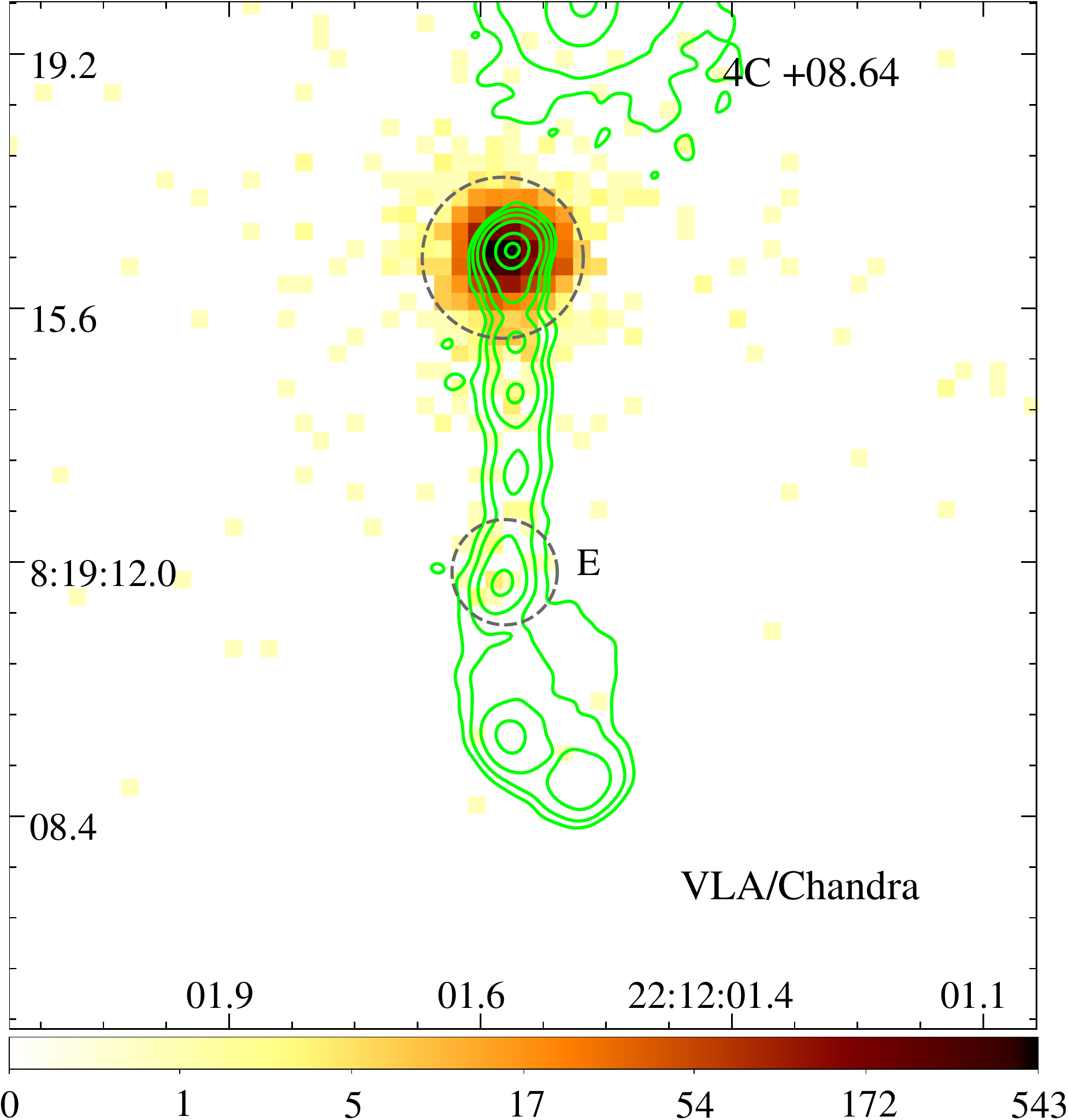}{0.5\textwidth}{(a)}
        \fig{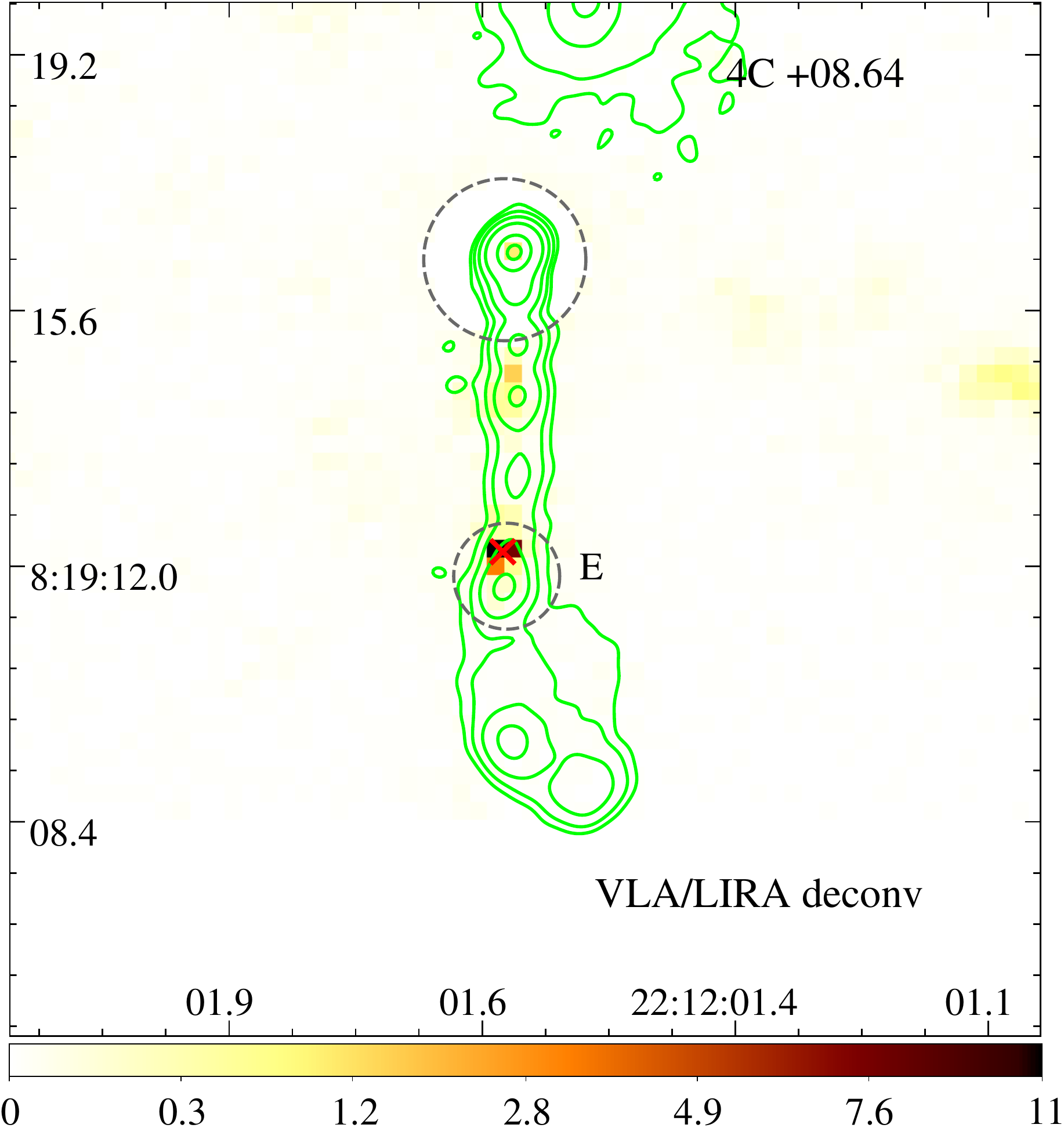}{0.5\textwidth}{(b)}
    }
    \caption{Same as in Fig. \ref{fig:results-3C9} but for 4C +08.64. The radio contours are given by 0.7, 2.0, 6.0, 20.0, 100.0, 200.0 mJy beam$^{-1}$.\label{fig:results-4C+08.64}}
\end{figure*}

\begin{figure*}[ht]
    \gridline{
        \fig{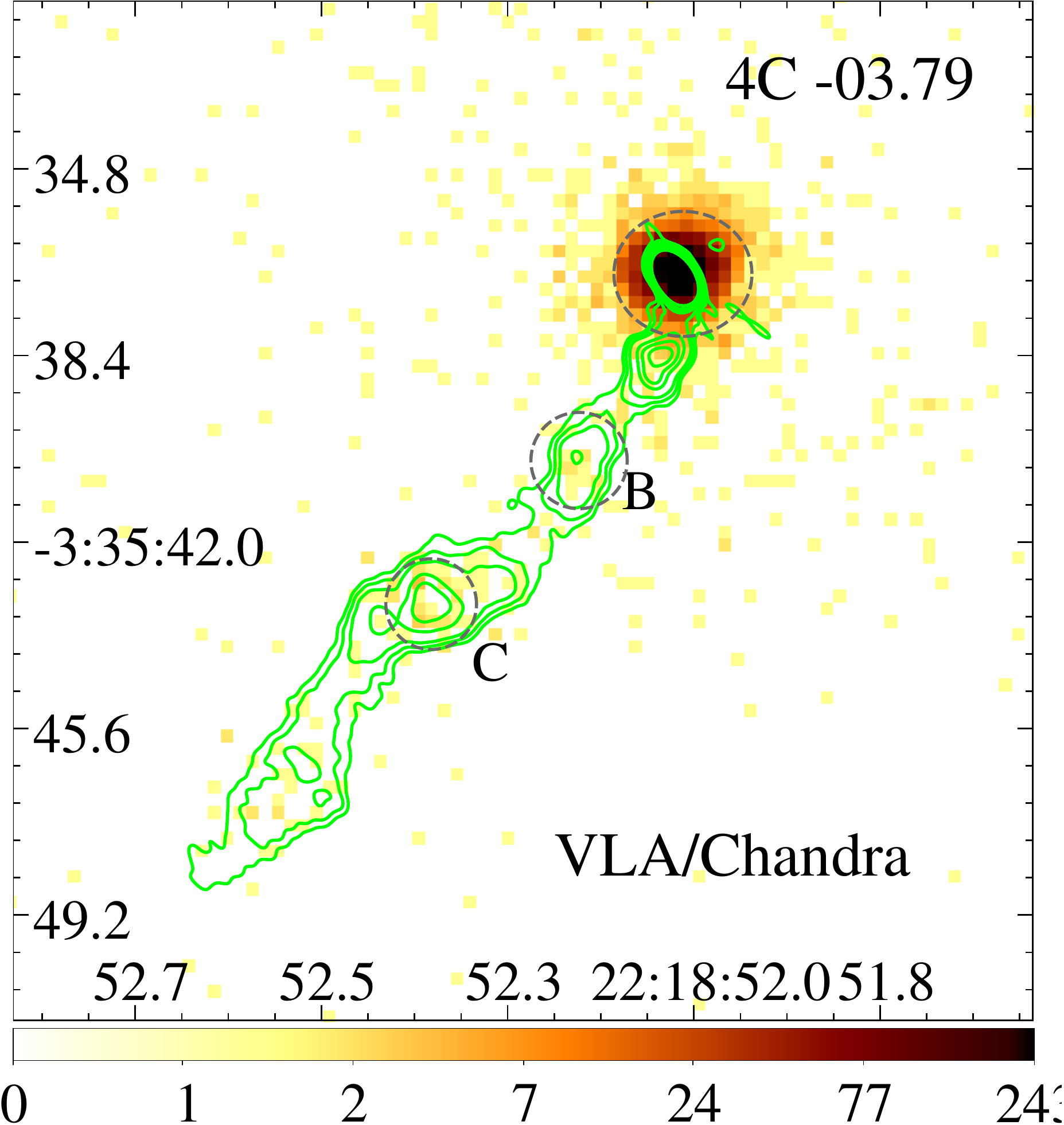}{0.5\textwidth}{(a)}
        \fig{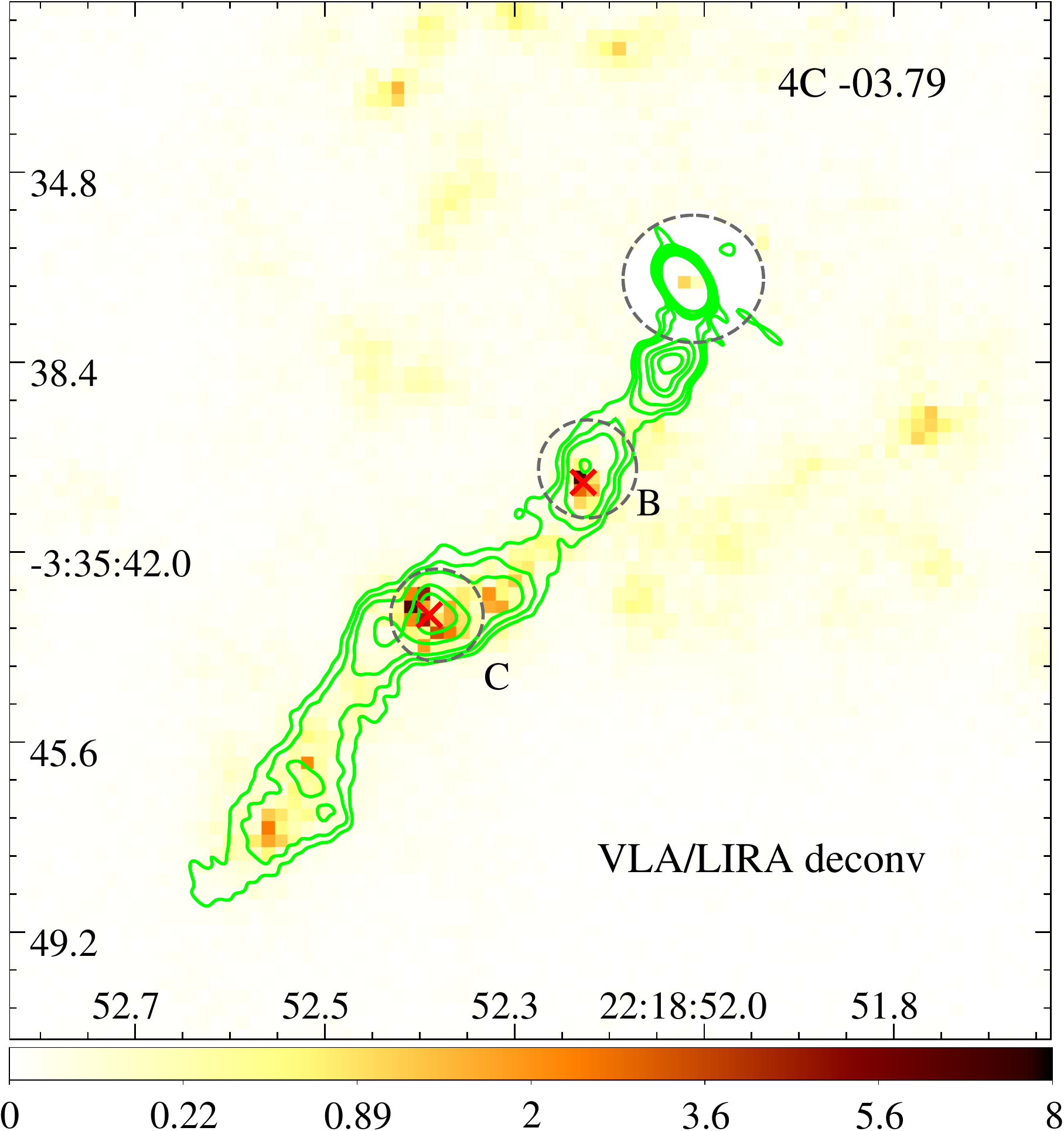}{0.5\textwidth}{(b)}
    }
    \caption{Same as in Fig. \ref{fig:results-3C9} but for 4C -03.79. The radio contours are given by 0.1, 0.2, 0.4, 8.0, 1.2, 2.0, 3.0 mJy beam$^{-1}$.\label{fig:results-4C-03.79}}
\end{figure*}
\comment{

\begin{figure*}[ht]
    \gridline{
        \fig{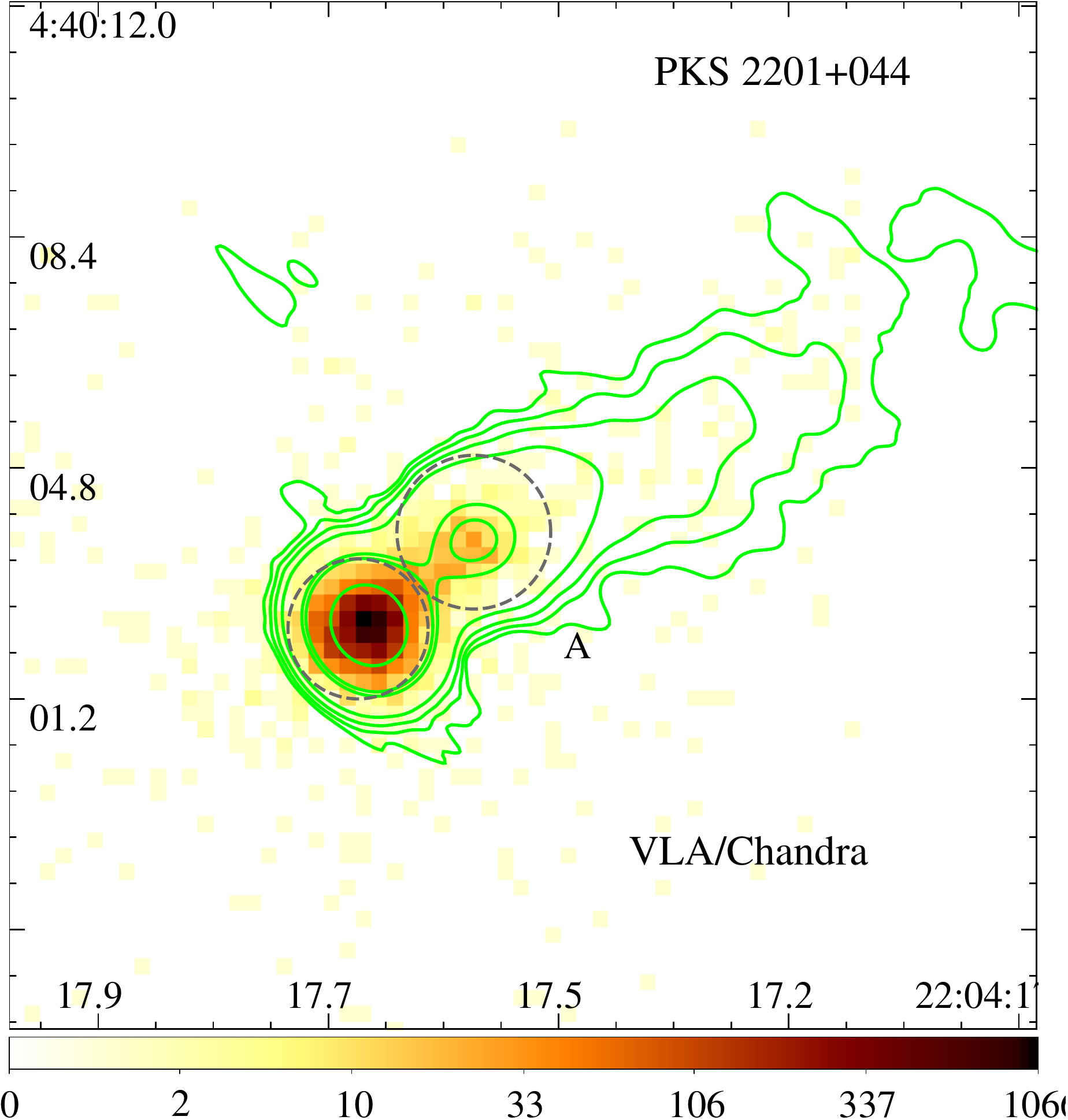}{0.5\textwidth}{(a)}
        \fig{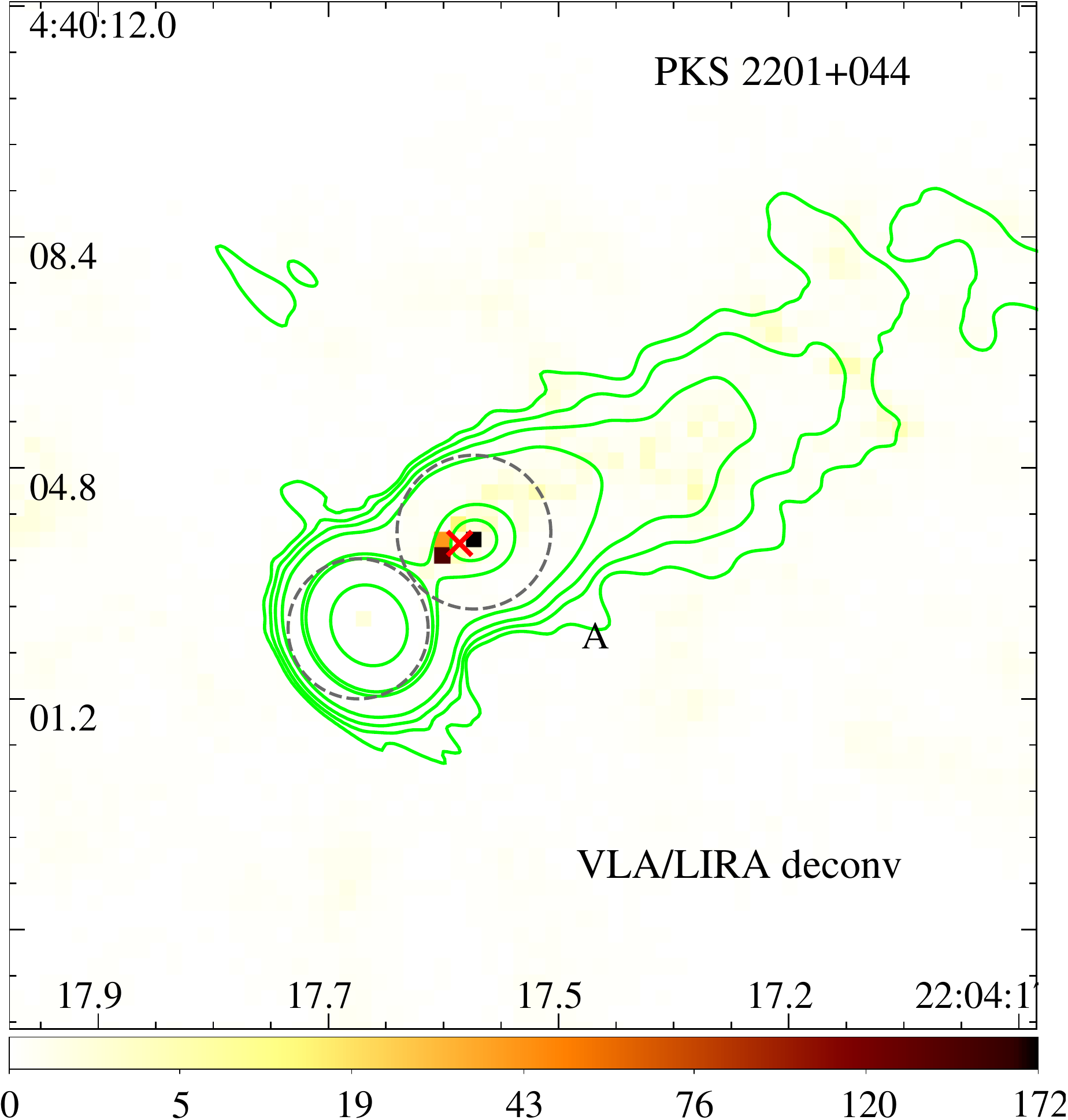}{0.5\textwidth}{(b)}
    }
    \caption{Same as in Fig. \ref{fig:results-3C9} but for PKS 2201+044. The radio contours are given by 0.1, 0.2, 0.4, 1.0, 15.0, 10.0, 100.0 mJy beam$^{-1}$.\label{fig:results-PKS2201+044}}
\end{figure*}
}

\begin{figure*}[ht]
    \gridline{
        \fig{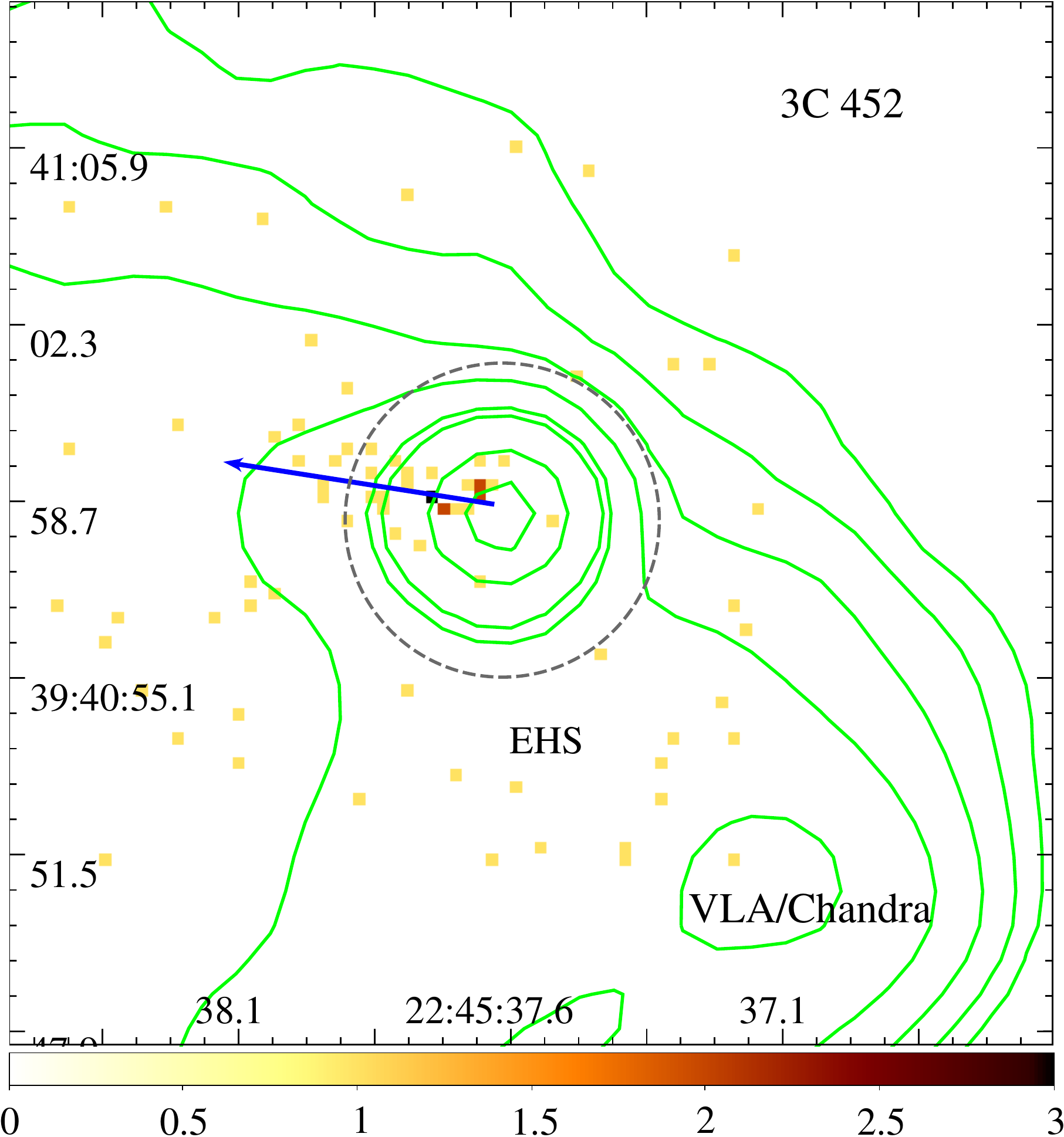}{0.5\textwidth}{(a)}
        \fig{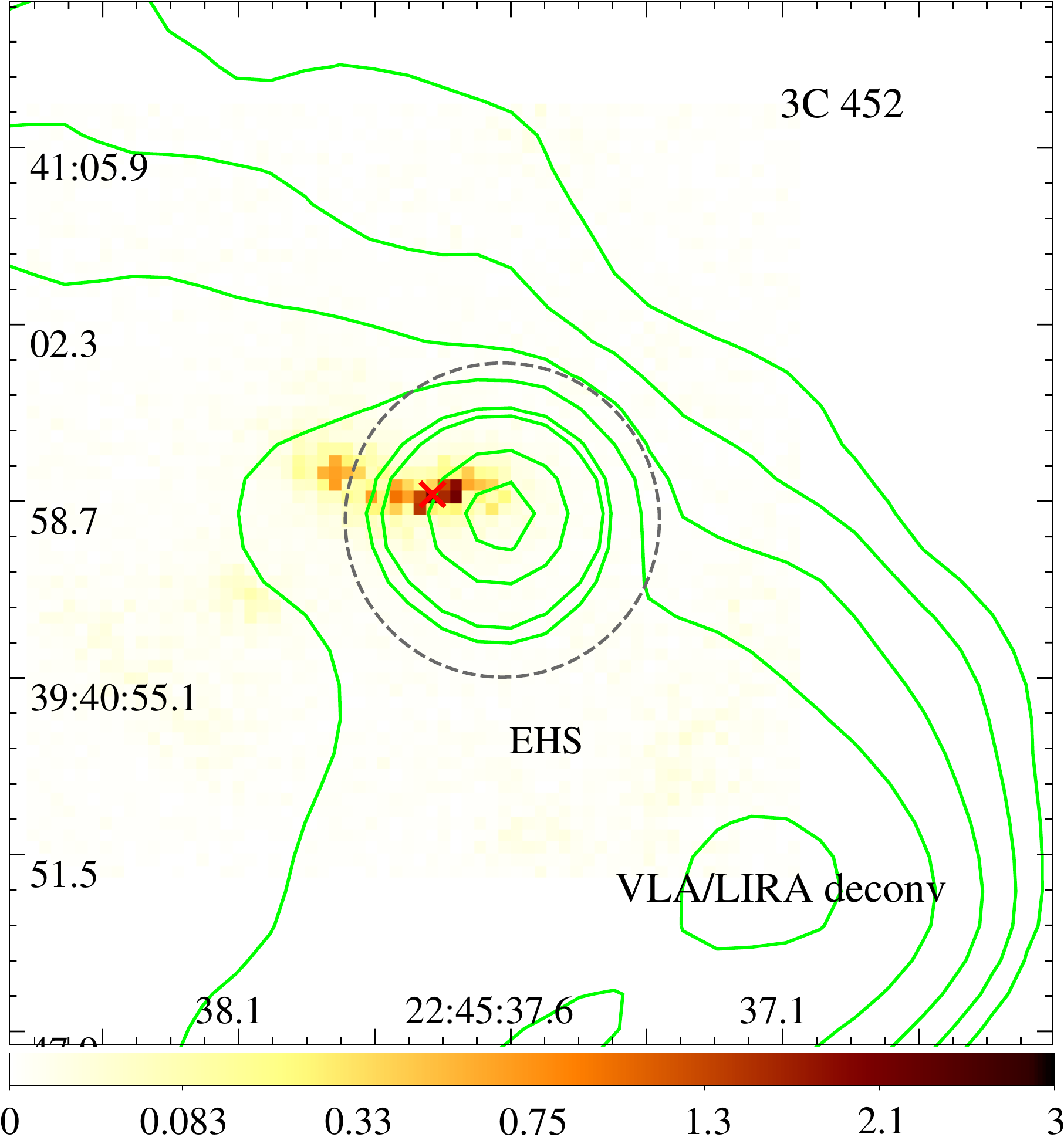}{0.5\textwidth}{(b)}
    }
    \caption{Same as in Fig. \ref{fig:results-3C9} but for 3C 452. The radio contours are given by 0.5, 1.0, 2.0, 4.0, 8.0, 10.0, 20.0, 30.0, 40.0 mJy beam$^{-1}$.\label{fig:results-3C452}}
\end{figure*}

\begin{figure*}[ht]
    \gridline{
        \fig{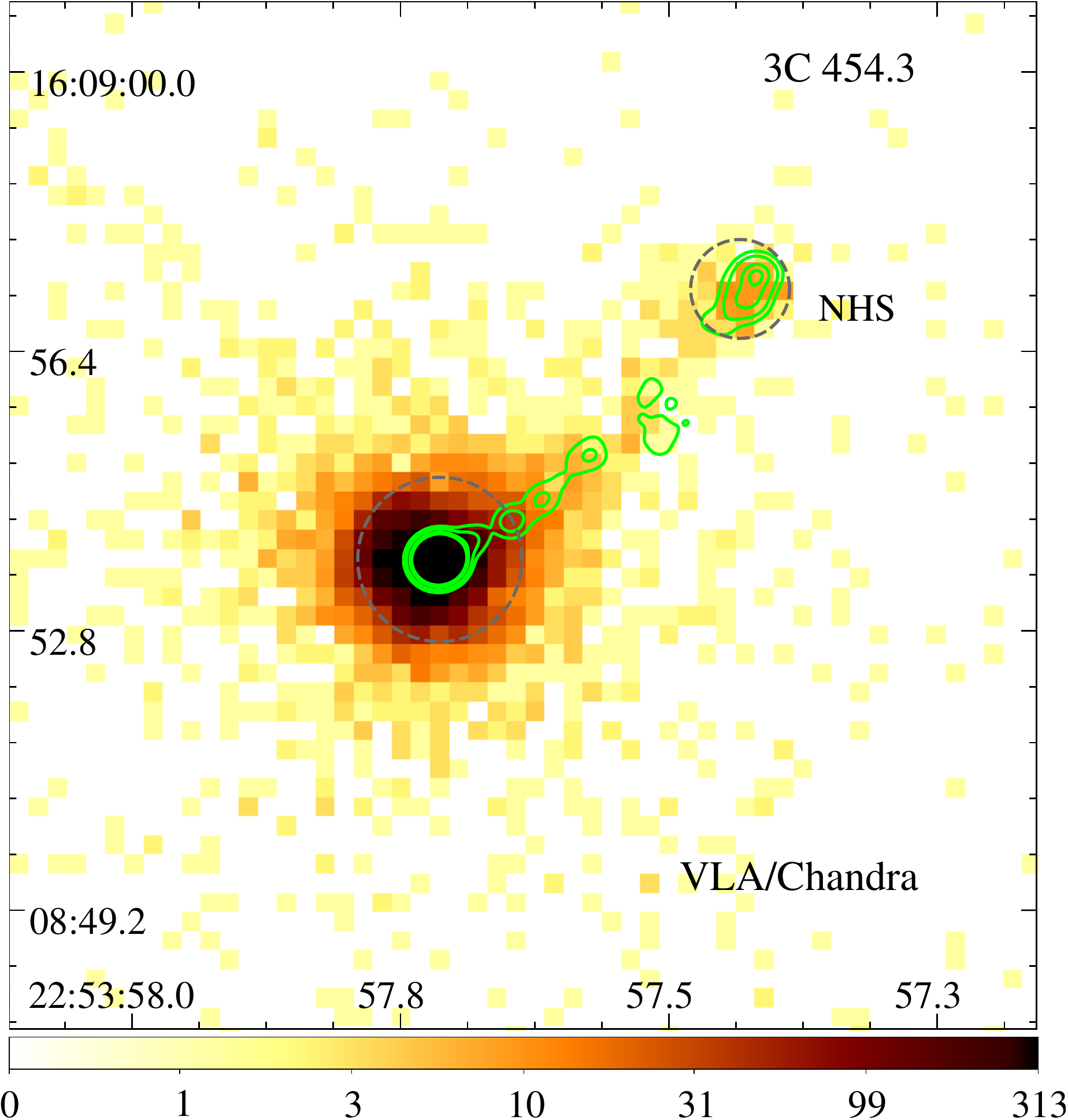}{0.5\textwidth}{(a)}
        \fig{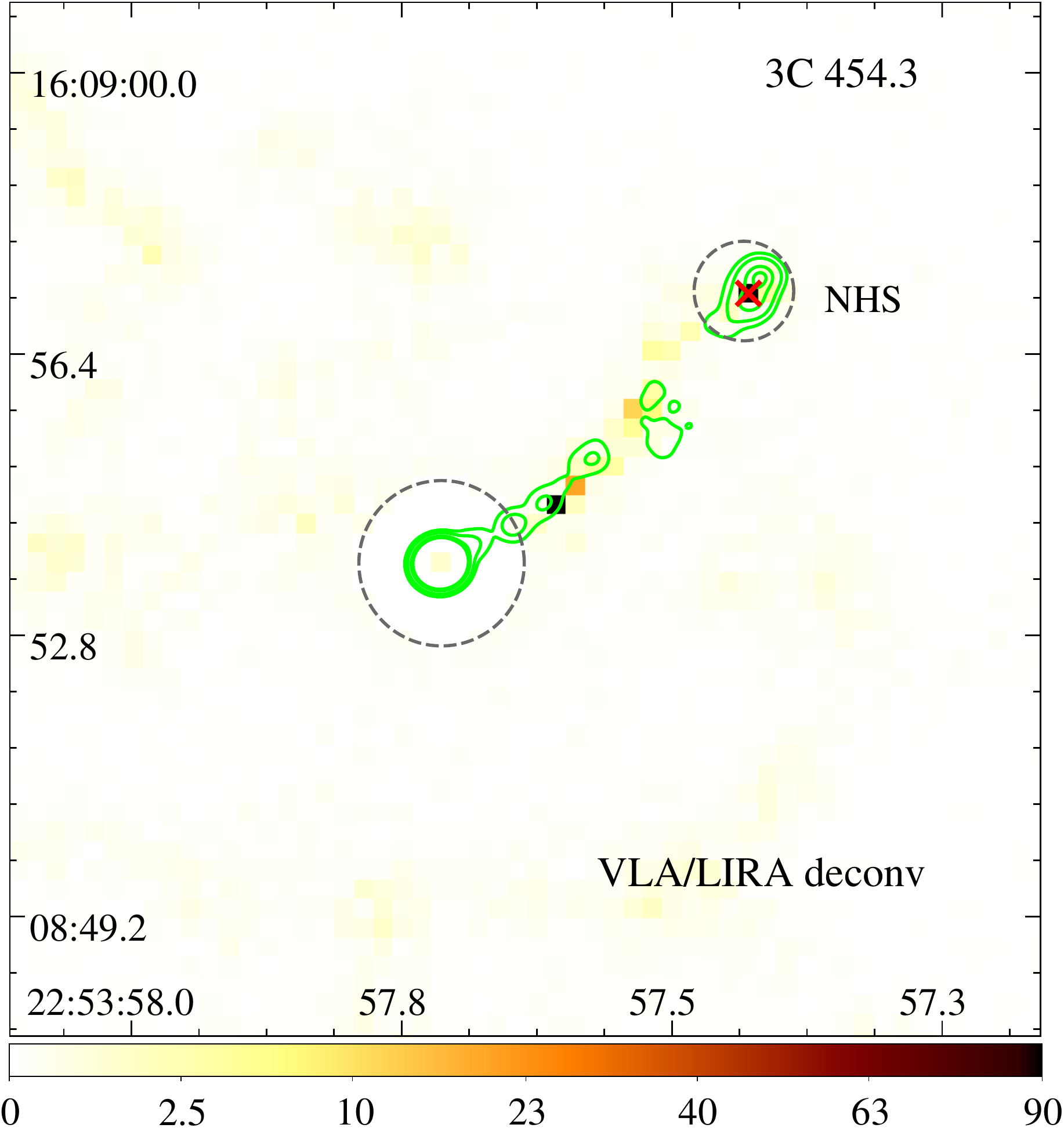}{0.5\textwidth}{(b)}
    }
    \caption{Same as in Fig. \ref{fig:results-3C9} but for 3C 454.3. The radio contours are given by 1.0, 4.0, 20.0, 40.0 mJy beam$^{-1}$.\label{fig:results-3C454.3}}
\end{figure*}


\section{Notes on Individual Sources\label{sec:src-notes}}
\ntitl{3C 9}This is an intermediate redshift LDQ \z{2.01} with a knotty jet.  The jet bends towards west at knot H, which produces the brightest radio feature in the jet, presumably interacting with the large-scale gas clouds \citep{2003MNRAS.338L...7F} and producing an internal shock. This knot shows an Xf-type offset of \ascale{0.29}{8.575}.
\\\\
\ntitl{3C 15}This is an FR I source. The X-ray emission brightens at the so-called \textit{flaring} region \citep[e.g.,][]{hardcastle2001chandra} while the radio peaks \ascale{0.657}{1.4} further downstream at knot B, showing an Xf-type offset. On the other hand, the X-rays coincide with the radio in knot C.
\\\\
\ntitl{3C 31}This is an FR I source. The jet most likely flares and decelerates near the radio knot B \citep{hardcastle2002chandra}. The X-ray cetroid lies \ascale{0.77}{0.34} upstream of the radio peak in knot B. Despite the presence of significant X-ray emission further downstream of knot B, we exclude it from offset analysis due to lack of clear association with the radio morphology.
\\\\
\ntitl{4C+01.02}This is an intermediate redshift \z{2.01}~CDQ. The jet makes an apparent bend to the southwest at knot B where the X-ray centroid and the radio peak coincide. It is possible that the jet presumably aligns closer to our line of sight \citep[][]{2012ApJ...748...81K}. On the other hand, the southern hotspot (SHS) shows a \ascale{0.40}{8.43} Xf-type offset. 
\\\\
\ntitl{3C 47}This is an LDQ. The jet makes a sharp apparent bend to the west at SHS where the X-ray centroid coincides with the radio peak. Although the radio jet travels further down terminating in a fainter peak, it lacks any significant X-ray emission.
\\\\
\ntitl{PKS 0144-522}This is a nearby quasar detected in a recent survey \citep{2018ApJ...856...66M}. The radio map shows a faint jet an inner knot an extension to the southeast of the core. The inner knot, A, shows no offset, while only little X-ray emission is seen beyond this knot.
\\\\
\ntitl{4C+35.03}This is an FR I source. The knot A, unlike the typical knots in FR I jets, shows a \ascale{0.37}{0.72} Rf-type offset. We exclude the remaining  X-ray emission from our analysis due to absence of any radio components.
\\\\
\ntitl{PKS 0208-512}This is a CDQ. The radio and X-ray images show a compact extended structure to the southeast of the core. \citep{2011ApJ...739...65P} use high resolution ATCA images to find no clear X-ray/radio spatial correlation in this structure. In K3, which lies further downstream, the X-ray centroid coincides with the radio peak.
\\\\
\ntitl{3C 66B}This is an FR I source. The jet flares at knot B \citep{2001MNRAS.326.1499H}. We find two distinct peaks in this region, one, upstream of knot B and also downstream of knot x, and, two, \ascale{0.422}{0.43} to its southeast, which is the presumed obstacle (see \ref{subsec:new_detections}).  We exclude the upstream X-ray knot from our analysis due to its unclear association with the two radio peaks x and B. Further downstream, knot C shows a \ascale{0.29}{0.43} Xf-type offset while the X-rays coincide with the radio in knot E.
\\\\
\ntitl{4C+28.07}This is a CDQ. We exclude the inner jet from offset analysis due to the small separation \lsep between the radio knots. Nevertheless, the rough coincidence between the X-ray peaks in averaged LIRA images and their radio counterparts in  knots C and D suggest a Co-s-type offset. We identify an X-ray feature \ascale{0.366}{8.36}~upstream of NHS-a where the jet presumably enters the turbulent hotspot region.
\\\\
\ntitl{3C 78}This a nearby FR I radio galaxy. We present core-subtracted radio map for this source, showing a knotty structure. Due to the proximity of the inner knot to the core, we are unable to detect its X-ray emission in the deconvolved images. However, we find a bright X-ray knot coinciding with its radio counterpart.
\\\\
\ntitl{3C 83.1}This FR I source with a narrow-angle tail morphology. We detect Xf-type offsets in knots E1 and W1 with magnitudes of \ascale{0.725}{0.5} and \ascale{0.65}{0.5}, respectively. 
\\\\
\ntitl{3C 88}This is an FR I source. The jet makes a sharp turn from the northwest to the southwest at knot C. We find a \ascale{1.14}{0.614} Xf-type offset in this knot.
\\\\
\ntitl{PKS 0405-12}This is a CDQ. The X-ray centroid  in the northern hotspot (NHS) coincides with the radio peak.
\\\\
\comment{
\ntitl{3C 109}This is an FR II source. We find a \ascale{0.31}{4.5} Xf-type offset in the southern hotspot (SHS).
\\\\}
\ntitl{PKS 0413-21}This is a CDQ. The radio image shows a bright extended feature, denoted as A, to the southwest of the core. The X-ray centroid of this feature lies \ascale{0.24}{7.45} upstream of its radio peak indicating an Xf-type offset.
\\\\
\textbf{3C 111} (Figures \ref{fig:results-3C111-A}-\ref{fig:results-3C111-D}): This is a nearby FR-II jet with about a 120\as~long knotty jet. To limit the usage of computational resources, we divide this jet into four parts. We find Xf-type offsets in K9, K14, K30, K61 and NHS, while we find no significant evidence for an offset in K45. Although we also find a offset in K97, we indicate it as Amb-type due to uncertainty in the direction of the jet at this location.
\\\\
\ntitl{3C 120}: This is classified as a Seyfert I galaxy, although with an FR I-type jet. We find an Xf-type offset in K4 while the X-ray and radio peaks are within the error limits of each other.
\\\\
\ntitl{3C 123}This is an FR II source. The southern radio hotspot is detected in the X-rays whose centroid lies \ascale{0.37}{4.5} upstream of the radio peak. Although the radio map indicates a precursor-hotspot in the SHS region, we find not X-ray emission.
\\\\
\ntitl{PKS 0454-463}This is a CDQ. The southern jet makes a sharp turn from the southwest to the south at knot A. The X-ray centroid of this knot lies within 0.15\as~of the radio peak.
\\\\
\ntitl{PKS 0605-089}This is a CDQ. The radio image shows a faint inner jet followed by a bend to the southwest before terminating in the western hotspot (WHS). We exclude the inner jet from the offset analysis due to small separation \lsep between the radio knots. WHS shows a \ascale{1.02}{6.9} Xf-type offset.
\\\\
\ntitl{PKS 0637-752}This is a moderately high-redshift CDQ, and is the first ever jet to be detected by \chdr \citep{chartas2000chandra,schwartz2000chandra}. The inner knots are excluded from our analysis due to their close spacing. We find no offsets in knots further down the jet.\\\\
\ntitl{B2 0738+313}This is a CDQ. Knot A, which lies roughly midway between the core and the southern hotspot (SHS) shows a \ascale{0.54}{6.9} Xf-type offset.
\\\\
\ntitl{OJ 248}This is a CDQ. The X-ray image shows a one-sided jet to the south without any radio counterpart. It makes a sharp bend to the east where the radio jet becomes visible. Knot C2, which lies at this bend, shows a Co-s-type offset. We find a \ascale{0.292}{7.9} Xf-type offset in knot C4. 
\\\\
\ntitl{4C +29.30}This is an FR II source. Knot A  shows a \ascale{0.319}{1.2} Xf-type offset, while the two hotspots, NHS-a and SHS, both show no offset.
\\\\
\ntitl{3C 207}This is an LDQ. The jet makes a small projected bend to the southwest at knot F and shows a \ascale{0.35}{7.1} Rf-type offset. This is contrary to many knots at apparent jet-bends in our sample that show Xf-type offsets. The western hotspot shows a \ascale{0.33}{7.1} Xf-type offset.
\\\\
\ntitl{OJ 287}This is a BL Lac type object and is one of the few sources in our sample with Rf and Xf-type offsets in the jet. The two knots in the inner jet, A and B, show \ascale{0.51}{4.5} and \ascale{0.23}{4.5} Rf-type offsets, respectively. On the other hand, two knots in the outer jet, which are detected only in the VLA 1.4 GHz band, show \ascale{0.364}{4.5} and \ascale{0.471}{4.5} Xf-type offsets, respectively.
\\\\
\ntitl{PKS 0920-397}This is a CDQ with a highly aligned jet (\bappb{30.8}) with near periodic knot spacing similar to PKS 0637-752 \citep{godfrey2012periodic}. While our LIRA deconvolution is unable to detect emission from the inner knots, we find no offset in knot C and E. Although we find X-ray emission between knots C and E, its radio association is unclear with the current data. The X-ray and radio jet fades away at knot E before re-emerging at the southern hotspot. The X-ray emission peaks \ascale{0.81}{8.5} of the radio hotspot. 
\\\\
\ntitl{3C 228}This is an FR II source. The radio map shows a precursor hotspot followed by the southern hotspot (SHS). The X-ray centroid lies \ascale{0.31}{6.68} Xf-type offset    upstream of SHS.
\\\\
\textbf{QSO 0957+516} (Figure \ref{fig:results-QSO0957+561}): This is a gravitationally-lensed LDQ. We exclude the knot complex upstream of the western hotspot (EHS) from offset analysis due to the close separation \lsep between the radio knots. We find a \ascale{0.74}{8.5} Xf-type offset in WHS.
\\\\
\ntitl{PKS 1030-357}This is a CDQ. The radio image shows a knotty jet to the southwest where knot B shows a Co-s-type offset, while knot C shows a \ascale{0.41}{8.5} Rf-type offset. The radio jet disappears at knot C before re-appearing at knot D. The situation in knot C is similar to 4C+19.44 where the radio jet disappears while the X-ray emission persists further downstream\citep{Harris-2017}. Knot D shows a \ascale{0.27}{8.5} Xf-type offset where the jet makes a 90\degr~projected-bend towards knot E. Knot E shows a \ascale{0.40}{8.5} Rf-type offset where the jet makes another 90\degr~bend to knot F that shows a \ascale{0.31}{8.5} Xf-type offset. 
\\\\
\ntitl{PKS 1045-188}This is a CDQ. We exclude the inner jet from offset analysis due to the small separation \lsep between the radio knots. Further downstream of the inner jet, the jet makes a 90\degr~projected bend to the east at knot H that shows a \ascale{0.21}{6.64} Xf-type offset.
\\\\
\ntitl{PKS 1055+201}This is an LDQ. The deconvolved X-ray image shows a long knotty jet to the north, although with only extremely faint radio emission,  terminating in a double-hotspot structure perpendicular to the jet's original direction. Knot A and both the hotspots show no offset.
\\\\
\ntitl{3C 254}This is an LDQ. The eastern hotspot shows a two-peak structure oriented in the north-south direction. The X-ray centroid in this hotspot lies \ascale{0.34}{7.3} upstream of the southern radio peak.
\\\\
\ntitl{PKS 1127-145}This is a CDQ and has one of the longest X-ray jets known to date. The X-ray image shows a bright inner jet with two knots but without any significant radio emission, and hence are excluded from are analysis. The two outer knots B and C show \ascale{0.47}{8.3} and \ascale{0.50}{8.3} Xf-type offsets respectively.
\\\\
\ntitl{PKS 1136-135}This is an LDQ. The two knots in the inner jet, x and A, show \ascale{0.37}{6.4} and \ascale{0.281}{6.4} Xf-type offsets, respectively. We exclude the knots in the outer jet from offset analysis due to their close separation \lsep, which, although show roughly coincident X-ray and radio peaks. Hence, we tentatively include them under the Co-s-type category. The eastern hotspot (EHS) shows a \ascale{0.35}{6.4} Xf-type offset.
\\\\
\ntitl{3C 263}This is an LDQ. The radio image shows a knotty radio jet to the southeast although without any X-ray counterpart. The jet terminates in the western hotspot with a \ascale{0.28}{7} Xf-type offset.
\\\\
\ntitl{3C 265}This is an FR II source. Its eastern hotspot (E) shows a Co-s-type offset.
\\\\
\ntitl{4C +49.22}This is a CDQ. The jet bends from the southwest to the southeast at knot B and shows a \ascale{0.197}{4.8} Xf-type offset. Knot D, which lies further down, shows a Co-s-type offset. The jet then turns to the southwest at knot E and shows a Co-s-type offset. Knot H, which lies at another apparent bend in the jet, also shows a Co-s-type offset.
\\\\
\ntitl{PKS 1202-262}This is a CDQ. We exclude knots in the inner jet from offset analysis due to their small separation \lsep, however, the deconvolved X-ray image shows a knotty structure coinciding with radio knots B, C, D. Hence we tentatively include them under the Co-s-type. Although we find two radio peaks further down the jet, they lie in between the radio peaks, making their radio association ambiguous. The eastern hotspot shows a \ascale{0.53}{8.1} Xf-type offset.
\\\\
\ntitl{PKS 1202-262}This is a CDQ. The radio image shows a one-sided jet to the northeast. We exclude the knots from offset analysis due to their close separation \lsep. The northern hotspot (NHS) shows a \ascale{0.28}{7.48} Xf-type offset.
\\\\
\ntitl{3C 270.1}This is a quasar. As mentioned in section \ref{sec:results}, we find a new bar-like structure in the X-rays in the southern hotspot (SHS) region \ascale{0.375}{8.69} upstream of the radio peak. This bar is presumably the location where the jet loses a major portion of its kinetic energy producing  X-rays, similar to 4C +74.26 \citep{erlund2007luminous,erlund2010two}. A faint knot upstream of the northern hotspot, NHS-a, where the jet presumably enters the turbulent hotspot region, shows a Co-s-type offset.
\\\\
\ntitl{M84}This is an FR I source. The radio image shows no radio emission prior to knot A, where the jet presumably flares, and produces a bright knotty jet further down. We find 0.598\as (0.081 kpc) and \ascale{1.366}{0.082} Xf-type offsets in knot A and B, respectively.
\\\\
\ntitl{4C+21.35}This is a CDQ. The radio image shows a knotty jet to the northeast of the core turning to the southeast at knot C2. We find a \ascale{0.19}{5.6} Xf-type offset in knot C2. 
\\\\
\ntitl{M87}This is a well-studied nearby FR-I jet. Because of its close-proximity and brightness, majority of its \chdr observations have many counts. Hence, we considered only one of the observations for our analysis. We find Xf-type offsets in knots B, C, D, F, I while none in knots A and E.
\\\\
\ntitl{PKS 1229-02}This is a CDQ. We exclude knots in the inner jet from offset analysis due to their small separation \lsep. The eastern hotspot shows a \ascale{0.53}{8.1} Xf-type offset.
\\\\
\ntitl{3C 275.1}This is an LDQ. The northern hotspot (NHS) shows a \ascale{0.41}{6.4} Xf-type offset. 
\\\\
\ntitl{3C 277.3}This is an FR II source. A previously cold gas cloud deflects the jet from the southwest to the southeast at knot K1. The X-rays in this knot region are produced by the shock heated cloud \citep[][]{worrall2016x} and hence we exclude K1 from the offset analysis. The jet bends again at knot K2 where it shows a \ascale{0.27}{1.668} Xf-type offset. Similarly, the northern hotspot (NHS) also shows \ascale{3.1}{1.668} Xf-type offset.
\\\\
\ntitl{4C +52.27}This is a CDQ. The radio image shows a knotty jet to the southeast without any X-ray counterpart. The jet terminates in the eastern hotspot with a double-peaked structure oriented along the east-west direction. The western peak, EHS-a, shows a \ascale{0.22}{8.15} Xf-type offset.
\\\\
\ntitl{3C 287.1}This is a quasar. The radio image shows a one-sized knotty jet to the west although with significant emission only from knot A. We measure a \ascale{0.92}{3.619} Xf-type in knot A. The jet shows a sharp bend to the north at the eastern hotspot (WHS). We find a \ascale{2.39}{3.619} Xf-type offset in WHS.
\\\\
\ntitl{PKS 1335-127}This is a CDQ. The inner knot, A, shows a Co-s-type offset. Further downstream of knot A, the jet makes a sharp turn to the east at knot B showing a \ascale{0.22}{6.33} Xf-type offset.
\\\\
\ntitl{4C+19.44}This is CDQ with a 18\as~long knotty jet to the southeast. We find no offsets in the inner jet except at knot S17.7, where the X-rays extend past the radio. As discussed in section \ref{subsec:fr2-offsets}, it is possible that a stationary obstacle interacting presumably produces these X-rays beyond the jet. The jet vanishes at S17.1 and re-emerges at S2.5.7 where X-rays coincide with the radio. Further downstream of this knot, the jet terminates in a double hotspot-like structure (SHS), where X-rays coincide with the radio in SHS-a, while they peak \ascale{0.275}{7.2} upstream of the radio peak in SHS-b.
\\\\
\ntitl{3C 294}This is an FR II source. The radio image shows two hotspots to the north and south of the core.  The  southern hotspot (SHS) shows a \ascale{0.50}{8.5} Xf-type offset. The northern jet produces a radio knot before terminating in the northern hotspot. Although the NHS shows a \ascale{0.40}{8.5} offset, the X-ray centroid  does not lie along the line joining this knot and NHS. However, it lies to the southwest of NHS and aligns with elongated radio structure to the southwest of NHS where the jet presumably enters the jet hotspot region. Hence we indicate it as an Xf-type offset.
\\\\
\ntitl{3C 295}This is an FR II jet, embedded in a bright cluster gas. While the northern hotspot (NHS) shows a \ascale{0.267}{5.7} Xf-type offset, the X-rays coincide with the radio in the southern hotspot (SHS).
\\\\
\ntitl{3C 299}This is an FR II source. We detect Xf-type offsets in both the western (WHS) and the eastern (EHS) hotspot with magnitudes of \ascale{0.35}{5.252} and \ascale{0.33}{5.252}, respectively.
\\\\
\ntitl{1421-490}The northern hotspot (A1) in this source is one of the brightest knots hotspots and is possibly highly-beamed \citep{2009ApJ...695..707G}. The X-ray centroid lies with 0.1\as~of the radio peak.
\\\\
\ntitl{3C 303}This is an FR II source. The radio image shows a knotty jet to the west. Knots A, B and C show \ascale{0.46}{2.5}, \ascale{0.33}{2.5}, \ascale{0.46}{2.5} Xf-type offsets respectively. The jet makes a sharp bend to the north at HS-a showing a Co-s-type offset.
\\\\
\ntitl{PKS 1510-089}This is a CDQ. Knot B shows a Co-s-type offset. We exclude the rest of the jet from offset analysis due to unclear spatial correlation between the X-ray and the radio structures.
\\\\
\ntitl{3C 325}This is a double-lobed FR II source oriented in the northeast-southwest direction. We detect a Co-s-type offset in the northern hotspot (NHS). The southern hotspot shows a double-peaked structure aligned roughly in the north-south direction  and the southern peak shows a Co-s-type offset.
\\\\
\ntitl{4C +00.58} This is an LDQ. The radio image shows a knotty jet to the northwest. The jet bends twice to the northwest at knots C and D, possibly produced by jet-precession \citep{2010ApJ...717L..37H}. Knot C shows a \ascale{0.21}{1.1} Xf-type offset while knot D shows a Co-s-type offset.
\\\\
\ntitl{3C 330}This is an FR II source. In the NHS region, the jet makes a sharp turn to the southwest before terminating in the hotspot showing a \ascale{0.27}{6.4} Xf-type offset.
\\\\
\ntitl{4C +15.55}This is a quasar.  The radio image shows a faint jet to the southwest making a southerly bend at knot C. This knot shows a Co-s-type offset. The jet terminates in the southern hotspot where we detect a \ascale{0.291}{8.659} Xf-type offset.
\\\\
\ntitl{3C346}This is an FR I source. The jet turns from the southwest to the northwest at knot B and shows a \ascale{0.22}{2.74} Xf-type offset.
\\\\
\ntitl{3C 345}This is a CDQ. The radio image shows a knotty jet to the northeast. The inner knots are undetected in the deconvolved images due to their proximity to the core. Knot D shows a Co-s-type offset while the northern hotspot NHS shows a \ascale{0.17}{6.6} Xf-type offset.
\\\\
\ntitl{4C +69.21}This is a CDQ. The radio image shows a knotty jet curving to the southeast. Knots C and D show Co-s-type offsets.
\\\\
\ntitl{3C 351}This is an FR II source. The northern lobe shows a double-peaked hotspot structure. Both the hotspots, J and L, show offsets of \ascale{0.36}{5.1} and \ascale{0.33}{5.1}, respectively. However, due to the uncertain direction of the jet, the nature of these offsets remain ambiguous.
\\\\
\ntitl{1800+440}This is a CDQ. The radio image shows a knotty jet curving from the southeast to the east but without an X-ray counterpart. The jet terminates in the eastern hotspot (EHS) with a Co-s-type offset.
\\\\
\ntitl{4C +56.27}This is a newly identifed X-ray jet \citep{2018ApJ...856...66M}. The radio map shows an inner jet to the south of core that bends to the west at knot B. We find no offset in this knot.
\\\\
\ntitl{3C 380}The radio map shows a bright extended structure to the northwest of the core, while the deconvoled X-ray image shows a bright X-ray peak at the terminus of the jet, coincident with a radio peak. While is unclear whether this structure represents a knot or a hotspot, we tentatively include it in our sample as a hotspot due to its location at the tip of the jet.
\\\\
\ntitl{1849+670}This is a hybrid blazar with an FR I jet to the north. The jet sharply turns to the east at knot A. Although we find a \ascale{0.39}{6.95} offset in knot A, the position angle of the offset lies perpendicular to the direction of the jet and the X-ray centroid closer lies to the edge of the radio knot. Hence, we label this offset as ambiguous. The jet returns to the northward direction after knot A, where we detect another X-ray knot, C, with a Co-s-type offset.
\\\\
\ntitl{4C +73.18}This is a CDQ. The radio image shows a knot, A, with a bright X-ray counterpart. Lower-frequency images presented in \citet{Sambruna_2004} reveal a bend at this knot. Knot A shows a \ascale{0.24}{4.4} Xf-type offset.
\\\\
\ntitl{3C 403}The is an FR II source with a one-sided jet to the northwest. The inner knot, F6, shows a \ascale{0.48}{1.13} Xf-type offset. The western radio lobe shows a north-south oriented double-peaked hotspot structure with the southern peak, F1, showing significant X-ray emission. F1, along with a fainter knot F1b, located downstream of F1 along the tentative direction of the jet, show \ascale{0.30}{1.13} and \ascale{0.36}{1.13} Xf-type offsets, respectively.
\\\\
\ntitl{S5 2007+777}This is a BL Lac object. Knot K8.5 shows a Co-s-type offset. The jet slightly bends to the southeast at knot K11.1 that shows a \ascale{0.529}{4.8} Rf-type offset while knot 15.9, which lies further down the jet, shows a \ascale{0.77}{4.8} Xf-type offset.
\\\\
\ntitl{PKS 2101-490}We exclude all knots except K3 due to lack of clear X-ray/Radio correlation. K3 shows no offset while the X-rays peak \ascale{0.83}{8.1} upstream of the radio peak, while coincident with a faint radio peak K6.
\\\\
\ntitl{PKS 2152-69}This is an FR II source with a one-sided jet to the northwest. The jet interacts with a high-ionization cloud at knot D \citep[][]{worrall2012jet}. This knot shows a \ascale{0.16}{0.56} Xf-type offset.
\\\\
\ntitl{4C +08.64}This is  a CDQ with a one-sided jet to the south. The jet bends to the southwest at knot E where the X-rays centroid lies \ascale{0.48}{5.94} upstream of its radio peak.
\\\\
\comment{
\ntitl{PKS 2201+044}This is a BL Lac object with a jet to the northeast. The inner knot A, shows a \ascale{0.18}{0.535} Xf-type offset.
\\\\}
\ntitl{4C -03.79}This is a hybrid source with a FR I morphology to the south and an FR II to the north.  The radio jet is detected on the southern side. Knot B shows a \ascale{0.37}{7.81} Rf-type offset. The jet bends to the east at knot C and shows a Co-s-type offset.
\\\\
\ntitl{3C 452}This is an FR II source with the X-rays detected in the western hotspot (WHS). The X-ray image shows an extended structure contrary to a point-like appearance in the radio. We find a \ascale{1.37}{1.5} Xf-type offset in WHS.
\\\\
\ntitl{3C 454.3}This is a well-studied blazar. All the knots in the jet are excluded from offset analysis due to their close separation \lsep. We find a \ascale{0.26}{7.7} Xf-type offset in the northern hotspot (NHS).
\\\\\comment{
\ntitl{3C 171}This is an FR II source. Knot B is the location where the jet enters the eastern lobe and shows a Co-s-type offset. The X-ray feature further down knot B, despite lying upstream of the eastern hotspot (EHS), lies on the edge of the radio lobe. Hence, it is possible that a large-scale hot gas cloud may be producing those X-rays rather than EHS.
We detect a \ascale{0.18}{3.894} Rf-type offset in WHS unlike the typical Xf-type offset found in most of the X-ray jets in our sample.
\\\\
\ntitl{3C 327}This is a FR II source. For the eastern hotspot, we adopt the nomenclature for the hotspots from \citet{2007ApJ...669..893H}, where they associate radio hotspots S1 and S2 with X-ray peaks SX1 and SX2 respectively. SX1 lies \ascale{2.40}{1.89} S1 along the tentative direction of the jet. Although SX2-S2 hotspot shows a \ascale{1.46}{1.89} offset, due to the lack of information on the direction of the jet, the type of offset is ambiguous.}
\comment{
\subsection{Ignored sources}
3C171 and 3C305: X-rays are produced by large-scale hot gases.\\
3C371:knots too close\\
3C290.3:pileup and core alignment issue
3C445:large hotspot
3C452:needs full image
}




\acknowledgements
We acknowledge financial support from the National Science Foundation under Grant No. 1714380 and NASA Astrophysics Data Analysis Program Grant No. 80NSSC21K0639.

The scientific results reported in this article are based in part on observations made by the Chandra X-ray Observatory and data obtained from the Chandra Data Archive. This research has made use of software provided by the Chandra X-ray Center (CXC) in the application packages CIAO, ChIPS, and Sherpa. The National Radio Astronomy Observatory is a facility of the National Science Foundation operated under cooperative agreement by Associated Universities, Inc. The Australia Telescope Compact Array is part of the Australia Telescope National Facility which is funded by the Australian Government for operation as a National Facility managed by CSIRO. We acknowledge the Gomeroi people as the traditional owners of the Observatory site. 

\facilities{VLA, EVLA, CXO, ATCA}
\software{CIAO (Fruscione et al. 2006), Sherpa (Freeman et al. 2001, Doe et al. 2007, Burke et al. 2020), ChiPS (Germain et al. 2006), MIRIAD (Sault et al. 1995), CASA (McMullin et al. 2007), ds9 (Joye et al. 2003), LIRA (Esch et al. 2004, Connors et al. 2007, Connors et al. 2011, Stein et al. 2015)}

\bibliography{references}

\end{document}